\documentclass[10pt,letterpaper]{article}
\usepackage{amssymb}
\usepackage[authoryear,round]{natbib}
\usepackage{tikz}
\usetikzlibrary{arrows, calc, decorations.pathmorphing, backgrounds, positioning, fit, petri, automata}
\definecolor{grey1}{rgb}{0.5,0.5,0.5}
\usepackage[top=2.54cm, bottom=2.54cm, left=2.2cm, right=2.2cm]{geometry}
\usepackage{algorithmicx}
\usepackage{algcompatible}
\usepackage{algorithm}

\usepackage{algpseudocode}


\usepackage{changebar}

\IfFileExists{xurl.sty}{\usepackage{xurl}}{
  \usepackage{url}%
  \expandafter\def\expandafter\UrlBreaks\expandafter{\UrlBreaks
    \do\a\do\b\do\c\do\d\do\e\do\f\do\g\do\h\do\i\do\j\do\k\do\l\do\m
    \do\n\do\o\do\p\do\q\do\r\do\s\do\t\do\u\do\v\do\w\do\x\do\y\do\z
    \do\A\do\B\do\C\do\D\do\E\do\F\do\G\do\H\do\I\do\J\do\K\do\L\do\M
    \do\N\do\O\do\P\do\Q\do\R\do\S\do\T\do\U\do\V\do\W\do\X\do\Y\do\Z
    \do0\do1\do2\do3\do4\do5\do6\do7\do8\do9}}
\usepackage{hyperref}
\hypersetup{ 
  colorlinks,
  citecolor=blue,
  linkcolor=red,
  urlcolor=magenta}

\usepackage{mathtools}
\usepackage{graphicx,booktabs}
\usepackage{subfigure}
\usepackage{amsmath,amsthm,bm}
\newcommand{\mathbbm}[1]{\mathbf{#1}}
\usepackage{rotating}
\usepackage{mathrsfs}
\usepackage{chngcntr}
\usepackage{apptools}
\usepackage[titletoc,title]{appendix}
\usepackage{comment}

\usepackage{xcolor}         
\definecolor{grau}{rgb}{0.8,0.8,0.8}

\usepackage{setspace,lscape,longtable}
\usepackage{amsmath,amsthm,bm}
\usepackage{fullpage}  
\usepackage[shortlabels]{enumitem}
\usepackage{titlesec}
\usepackage[english]{babel}
\numberwithin{equation}{section}
\numberwithin{equation}{section}

\newtheorem{theorem}{Theorem}[section]
\newtheorem{corollary}{Corollary}[section]
\newtheorem{lemma}[theorem]{Lemma}
\newtheorem{proposition}{Proposition}[section]
\theoremstyle{remark}
\newtheorem{remark}{Remark}

\newtheorem{assumption}{Assumption}
\newtheorem{example}{Example} 

\DeclareMathOperator*{\argmin}{arg\,min}
\DeclareMathOperator*{\argmax}{arg\,max}

\newcommand{\prob}{{\mathbb{P}}}
\newcommand{\var}{{\mathrm{var}}}
\newcommand{\expect}{\mathbb{E}}
\newcommand{\vect}{\mathrm{vec}}
\newcommand{\vech}{\mathrm{vech}}
\newcommand{\Optilde}{{\widetilde{O}_{p}}}

\newcommand{\transpose}{^{\mathrm{T}}}

\newcommand{\bLambda}{{\bm{\Lambda}}}

\newcommand{\calE}{{\mathcal{E}}}
\newcommand{\calF}{{\mathcal{F}}}

\newcommand{\calH}{{\mathcal{H}}}
\newcommand{\calI}{{\mathcal{I}}}
\newcommand{\calJ}{{\mathcal{J}}}

\newcommand{\calL}{{\mathcal{L}}}

\newcommand{\calN}{{\mathcal{N}}}

\newcommand{\calS}{{\mathcal{S}}}
\newcommand{\calT}{{\mathcal{T}}}
\newcommand{\calU}{{\mathcal{U}}}
\newcommand{\calV}{{\mathcal{V}}}

\newcommand{\calX}{{\mathcal{X}}}

\newcommand{\scrA}{{\mathscr{A}}}

\newcommand{\scrH}{{\mathscr{H}}}
\newcommand{\scrI}{{\mathscr{I}}}
\newcommand{\scrJ}{{\mathscr{J}}}

\newcommand{\scrL}{{\mathscr{L}}}

\newcommand{\scrP}{{\mathscr{P}}}

\newcommand{\balpha}{{\boldsymbol{\alpha}}}
\newcommand{\bT}{{\mathbf{T}}}
\newcommand{\bU}{{\mathbf{U}}}
\newcommand{\bP}{{\mathbf{P}}}

\newcommand{\bY}{{\mathbf{Y}}}
\newcommand{\bv}{{\mathbf{v}}}
\newcommand{\bx}{{\mathbf{x}}}
\newcommand{\bX}{{\mathbf{X}}}
\newcommand{\bA}{{\mathbf{A}}}
\newcommand{\bB}{{\mathbf{B}}}
\newcommand{\bC}{{\mathbf{C}}}
\newcommand{\bD}{{\mathbf{D}}}
\newcommand{\bE}{{\mathbf{E}}}
\newcommand{\bF}{{\mathbf{F}}}
\newcommand{\bG}{{\mathbf{G}}}
\newcommand{\bH}{{\mathbf{H}}}
\newcommand{\bM}{{\mathbf{M}}}
\newcommand{\bN}{{\mathbf{N}}}
\newcommand{\bK}{{\mathbf{K}}}
\newcommand{\bL}{{\mathbf{L}}}
\newcommand{\bJ}{{\mathbf{J}}}
\newcommand{\bR}{{\mathbf{R}}}
\newcommand{\bQ}{{\mathbf{Q}}}
\newcommand{\bS}{{\mathbf{S}}}
\newcommand{\bW}{{\mathbf{W}}}
\newcommand{\bZ}{{\mathbf{Z}}}

\newcommand{\bw}{{\mathbf{w}}}

\newcommand{\bff}{{\mathbf{f}}}
\newcommand{\bg}{{\mathbf{g}}}
\newcommand{\bh}{{\mathbf{h}}}

\newcommand{\bq}{{\mathbf{q}}}
\newcommand{\br}{{\mathbf{r}}}

\newcommand{\bu}{{\mathbf{u}}}
\newcommand{\bs}{{\mathbf{s}}}
\newcommand{\by}{{\mathbf{y}}}
\newcommand{\bz}{{\mathbf{z}}}

\newcommand{\be}{{\mathbf{e}}}
\newcommand{\bb}{{\mathbf{b}}}

\newcommand{\bV}{{\mathbf{V}}}
\newcommand{\bkappa}{{\bm{\kappa}}}
\newcommand{\bgamma}{{\bm{\gamma}}}
\newcommand{\bGamma}{{\bm{\Gamma}}}
\newcommand{\bbeta}{{\bm{\beta}}}
\newcommand{\bDelta}{{\bm{\Delta}}}
\newcommand{\bOmega}{{\bm{\Omega}}}

\newcommand{\bSigma}{{\bm{\Sigma}}}
\newcommand{\bXi}{{\bm{\Xi}}}

\newcommand{\eye}{{\mathbf{I}}}
\newcommand{\one}{{\mathbf{1}}}
\newcommand{\bTheta}{{\bm{\Theta}}}
\newcommand{\btheta}{{\bm{\theta}}}

\newcommand{\bzeta}{{\bm{\zeta}}}

\newcommand{\bPi}{{\bm{\Pi}}}

\newcommand{\bPsi}{{\bm{\Psi}}}
\newcommand{\bphi}{{\bm{\phi}}}
\newcommand{\bpsi}{{\bm{\psi}}}

\newcommand{\zero}{{\bm{0}}}
\newcommand{\eps}{\epsilon}
\newif\ifblinded\blindedfalse 
\date{}

\providecommand{\bc}{{\mathbf{c}}}

\newcommand{\McalRho}{0.967}
\newcommand{\McalRhoLayer}{0.947}

\newcommand{\MouseAlphaInteractPct}{7.7}
\newcommand{\MouseAlphaInteractRankThree}{61}
\newcommand{\MouseAlphaQ}{172}
\newcommand{\MouseAlphaRankMax}{31}
\newcommand{\MouseAlphaRankNeeded}{23}
\newcommand{\MouseAlphaSignalNoise}{22}
\newcommand{\MouseBiasSE}{0.02}
\newcommand{\MouseCCLt}{-14.6}
\newcommand{\MouseCCRt}{-14.3}
\newcommand{\MouseContrast}{0.366}
\newcommand{\MouseContrastSE}{0.0026}

\newcommand{\MouseCvFrac}{10}
\newcommand{\MouseHemiCorr}{0.84}

\newcommand{\MouseMeasVarPct}{10}
\newcommand{\MouseRhoBTBR}{0.74}
\newcommand{\MouseRhoBetweenMax}{0.06}
\newcommand{\MouseRhoBetweenMin}{0.03}
\newcommand{\MouseRhoOtherMax}{0.39}
\newcommand{\MouseRhoOtherMin}{0.36}
\newcommand{\MouseRhoSE}{0.0064}
\newcommand{\MouseSandMed}{1.01}
\newcommand{\MouseSandRange}{0.88--1.26}
\newcommand{\MouseSigma}{2.23}
\newcommand{\MouseVarY}{8.41}
\newcommand{\MouseWaldBetween}{1012}
\newcommand{\MouseWaldWithin}{260}
\newcommand{\MouseWelchT}{16.1}
\newcommand{\MouseZeroPct}{35}

\begin{document}
\addtocontents{toc}{\protect\setcounter{tocdepth}{-1}} 

\ifblinded
\title{\bf Subspace Inference for Multilayer Latent Space Networks: Intrinsic Geometry and One-Step Estimator}
\author{}
\date{}
\maketitle
\else
\title{\bf Subspace Inference for Multilayer Latent Space Networks: Intrinsic Geometry and One-Step Estimator}
\author{Fangzheng Xie\\Department of Statistics, Indiana University
\and 
Hsin-Hsiung Huang\\Dr. Yvette Kanouff School of\\Data, Mathematical, and Statistical Sciences,\\University of Central Florida}
\date{}
\maketitle
\fi

\begin{abstract}
Multilayer networks on a common node set, such as brain connectomes of several animals with one layer per animal, often share a low-dimensional geometry but differ in how strongly each layer expresses it and in node activity. We study a multilayer latent space network model that combines an invariant subspace common to all layers with layer-specific interaction matrices and node-by-layer degree effects. By introducing the generalized Cayley transform of the invariant subspace, we characterize the intrinsic geometry of the model and obtain sharp convergence rates for the maximum likelihood estimator, asymptotic normality and Neyman orthogonality for vertex-level parameters, and asymptotic normality for the interaction matrices with a non-negligible bias. Plug-in estimates of the bias and covariance give debiased confidence regions and tests of equality between layers.
We also propose an easy-to-implement one-step estimator that attains the same vertex-level limits. In a real-world mouse connectome dataset, the inbred strain that lacks a corpus callosum shows about twice the hemispheric segregation of the other three strains.
\end{abstract}

\noindent%

\section{Introduction}
\label{sec:introduction}

Collections of networks observed on a common set of nodes arise whenever the same system is measured more than once. Diffusion-MRI connectomes of several animals registered to one atlas, interactions among the same actors in different contexts, and one network per period on a fixed set of units are typical examples. Two features recur. First, the layers share a low-dimensional geometry. In the mouse connectomes of Section~\ref{sec:mouse}, regions in the same hemisphere and functional system connect preferentially in every brain. Second, the layers differ, both in how strongly they express the shared geometry and in how active each node is. A brain lacking the corpus callosum, the main commissure joining the two hemispheres, should express the hemispheric axis more strongly, and the regions adjacent to the missing tract should lose connectivity. A statistical model for such data should pool the layers to estimate the common geometry, leave the node- and layer-specific heterogeneity unrestricted, and provide calibrated uncertainty for both.

Consider $m$ layers of networks over a fixed set of $n$ vertices labeled as $[n] := \{1,\ldots,n\}$, and suppose $A_{ij}^{(t)}$ denotes the (possibly weighted) edge between vertices $i,j\in[n]$ at the $t$th layer. Such a collection of vertex-aligned networks is referred to as a multilayer network.
We study a multilayer latent space model in which the edges $A_{ij}^{(t)}$, $1\leq i<j\leq n$, $t\in[m]:=\{1,\ldots,m\}$, are independent random variables from a density function $p_A(\cdot;\theta_{ij}^{(t)})$ with respect to some underlying $\sigma$-finite measure, where
\begin{align}
\label{eqn:theta_intro}
\theta_{ij}^{(t)} = \alpha_i^{(t)} + \alpha_j^{(t)} + \bx_i\transpose\bR^{(t)}\bx_j.
\end{align}
The latent positions $\bx_1,\ldots,\bx_n\in\mathbb{R}^d$ are shared by all layers and span an invariant subspace; the symmetric, possibly indefinite, matrices $\bR^{(t)}$ describe how the shared latent positions interact in the $t$th layer; and the degree effects $\alpha_i^{(t)}$ describe the activity of node $i$ in layer $t$, as in the $p_1$ and $\beta$-models \citep{holland1981exponential,yan2016asymptotics,graham2017econometric}. With $m=1$, \eqref{eqn:theta_intro} is the latent space model of \citet{hoff2002latent} with the degree heterogeneity of \citet{ma2020universal}, which has also been studied in \cite{Tian_Sun_He_2026_Bridging}. With an identity link and no degree effects, model \eqref{eqn:theta_intro} collapses to the common subspace independent edge (COSIE) model of \citet{arroyo2021inference} and has been further studied by \cite{Zheng_Tang_2026_COSIELimit} and \cite{xie2024biascorrected}, which is closely related to the multilayer random dot product graph \citep{jones2020multilayer}. \citet{zhang2020flexible} and \citet{macdonald2022latent} studied estimation in related multilayer latent space models. When the latent positions also imply a community structure over the vertex set, the resulting model is referred to as the multilayer stochastic block model, and community detection has been extensively studied \citep{agterberg2025joint,lei2020consistent,lei2023bias,lei_zhang_zhu_2024_AoS_computational,paul2020spectral,xie2024biascorrected}.

Statistical inference for this model is challenging for four reasons. The latent positions are identified only up to an orthogonal transformation and constrained to a scaled Stiefel manifold, so the parameter space is a quotient manifold.  The number of parameters, $n(m+d)+md(d+1)/2$, grows with both the number of vertices $n$ and number of layers $m$. Different components of parameters are informed by very different amounts of data, and this will be made precise using the intrinsic geometry and the Fisher information metric. From the practical side, the likelihood is not concave, which calls for an easy-to-implement computational algorithm. Spectral estimators, which underlie a broad scope of existing network inference literature (see \citealp{athreya2018statistical,chen_chi_fan_ma_2021spectral} for reviews), fail to handle the case when the latent low-rank structure is masked by a nonlinear link function because the mean matrix $\expect\bA^{(t)}$ is then not of low rank.

The interaction matrices raise a further problem. Each $\bR^{(t)}$ is informed by all $n(n-1)/2$ edges of its layer and is estimated at rate $n^{-1}$. Its estimator also depends on the $mn$ degree effects, each estimated at rate $n^{-1/2}$, and on the $nd$ latent coordinates, each estimated at rate $(mn)^{-1/2}$. Unlike the vertex parameters, $\bR^{(t)}$ is in general not asymptotically Neyman orthogonal to these nuisance parameters, so its covariance must account for them. Summed over all edges, the second-order effects of the errors propagated from these nuisance parameters also produce a bias of order $n^{-1}$, the same order as the standard error. This is a network form of the incidental-parameter problem \citep{neyman1948consistent,hahn2004jackknife,fernandezval2016individual}. Without debiased inference, it leads to invalid uncertainty quantification.

\label{sub:contributions}
This article develops likelihood-based inference for model~\eqref{eqn:theta_intro} that addresses these difficulties. Building on the Euclidean representation of low-rank matrices of \citet{xie2023euclidean}, we parameterize the invariant subspace through a generalized Cayley transform and show that the Fisher information under this parameterization is uniformly equivalent to $\mathrm{diag}(n\eye,n^2\eye,mn^2\eye)$ on the degree, interaction, and subspace coordinates (Proposition~\ref{prop:expected_hessian_bound}), which makes precise the different amounts of information noted above. The maximum likelihood estimator (MLE) of the latent positions is consistent and achieves the rate $\{(\log n)/(mn)\}^{1/2}$ under the two-to-infinity norm metric (Theorem~\ref{thm:two_to_infinity_norm_bound_MLE}). Because the vertex-specific parameters are asymptotically Neyman orthogonal to the remaining parameters, the MLE of any bounded linear combination of the parameters of finitely many vertices is asymptotically normal with the oracle variance that would apply if all other parameters were known (Theorem~\ref{thm:asymptotic_normality_MLE}).

For the interaction matrices, we establish the limiting distribution of fixed linear combinations of $n\{\vech(\widehat\bR^{(t)})-\vech(\bR^{(t)}_0)\}$ over a finite set of layers (Theorem~\ref{thm:asymptotic_normality_R}). The limit is normal with the nuisance-adjusted covariance and a non-negligible noncentrality $\bb_n^{(t)}$, which arises from the curvature of the link, of $\theta_{ij}^{(t)}$, and of the constraint manifold. Plug-in estimates of the bias and covariance give debiased confidence regions in a fixed orientation and rotation-invariant Wald tests of the equality of two layers (Corollary~\ref{corr:plug_in_R}), and plug-in inference for the vertex-specific parameters is justified in the same way (Corollary~\ref{corr:plug_in_vertex}). On the computational side, we propose an easy-to-implement one-step estimator that aggregates layer-wise MLEs and takes one vertex-wise Newton step. It attains the same vertex-level limits under the weaker condition $m^2(\log n)^2/n\to0$ (Theorem~\ref{thm:asymptotic_normality_OSE}).

We apply the methodology to 32 mouse connectomes from four strains \citep{wang2020variability}, with a Gaussian working likelihood for log streamline counts. The fit recovers a hemispheric latent axis without using hemisphere labels. The debiased hemispheric interaction coefficient of the BTBR strain, which lacks a corpus callosum \citep{wahlsten2003survey}, is about twice those of the other strains, and the corpus callosum is among the regions with the largest strain differences in degree effects.

\label{sub:related_work}
Our work relates to several lines of research. One-step estimators have a long history in classical asymptotics \citep{lecam1956asymptotic,bickel1975one,vandervaart1998asymptotic}. \citet{xie2023efficient} proposed a one-step procedure for random dot product graphs, and \citet{xie2024entrywise} established entrywise limit theorems under weak signals. For multiple networks with a common subspace, \citet{arroyo2021inference} and \citet{Zheng_Tang_2026_COSIELimit} developed limit theorems, including noncentered limits for score matrices, for spectral estimators under identity links; \citet{levin2017central} studied the omnibus embedding, which \citet{wang2020variability} used for the mouse connectomes. Our analysis combines entrywise perturbation of low-rank structures \citep{abbe2020entrywise,cape2019two} under a nonlinear link function with likelihood theory with a growing number of parameters \citep{portnoy1988asymptotic,he2000parameters,yan2016asymptotics}.
The generalized Cayley transform that reparameterizes the Grassmannian is related to optimization on matrix manifolds 
\citep{edelman1998geometry,absil2008optimization,wen2013feasible}. Note that this transform is introduced purely as a tool for the theoretical analysis, and practitioners would not have access to it since the transform depends on the unknown value of the true parameter.

Estimation and inference for latent space networks and multilayer networks have been recently explored. \cite{He_Sun_Tian_Ying_Feng_2025_AoS,tian2026efficient} propose a one-step estimator and a likelihood-based refinement for efficient analysis of multilayer latent space networks with invariant and partially invariant latent positions, respectively. \cite{Tian_Sun_He_2026_Bridging} focus on the single-layer network model but establish both the asymptotic distributional theory and a computationally tractable warm-started projected gradient descent algorithm with convergence guarantee. 
The closest work is \citet{liu2026unfolding}, who study the estimation and inference of a closely related multilayer latent space network model. Similar to our theory, they establish limiting distributions for the latent positions, correct a second-order bias of their connection-matrix estimator, and test whether two layers share structure. The key difference in the model assumptions is that they assume a low-rank structure over the degree effect parameters $(\alpha_i^{(t)}:1\leq i\leq n,1\leq t\leq m)$. Consequently, the dimension of the parameter space there is linear in $m + n$ when the rank is assumed to be fixed, while our model handles a parameter space with $O(mn)$ free parameters because it requires no structure on the degree effect parameters. 
The unstructured modeling choice of the degree effects is also driven by the needs of our real-world application. Indeed,
in the mouse connectomics data, a low-rank degree effect structure would need rank \MouseAlphaRankNeeded\ of a possible \MouseAlphaRankMax\ to capture the node-by-layer variation that exceeds the estimation noise (Section~\ref{sec:mouse}). 

\label{sub:organization}
The rest of the article is organized as follows. Section~\ref{sec:multilayer_latent_space_network_with_an_invariant_subspace} introduces the model, Section~\ref{sec:euclidean_representation_and_intrinsic_geometry} its intrinsic geometry, Section~\ref{sec:maximum_likelihood_and_one_step_estimator} the MLE, its limit theorems, and
the one-step estimator. 
Section~\ref{sec:mouse} presents the mouse connectome analysis, and Section~\ref{sec:discussion} concludes with a discussion.
Proofs, computational details, simulation studies, and further applications are in the Supplementary Material.

We use the notation $:=$ to assign mathematical definitions. Given an $n$-dimensional vector $\btheta = [\theta_1,\ldots,\theta_n]\transpose$, we let $\mathrm{diag}(\btheta)$ be the diagonal matrix whose $i$th diagonal element is $\theta_i$. Given a positive integer $d\leq n$, let $\mathbb{O}(n, d) := \{\bU\in\mathbb{R}^{n\times d}:\bU\transpose\bU = \eye_d\}$ denote the set of all $n\times d$ Stiefel matrices, where $\eye_d$ is the $d\times d$ identity matrix. When $n = d$, we simply write $\mathbb{O}(d) = \mathbb{O}(d, d)$. Given a positive integer $n$, let $\one_n:=[1,\ldots,1]\transpose\in\mathbb{R}^n$. For an $n\times m$ matrix $\bX = [X_{ij}]_{i\in[n],j\in[m]}$, we denote by $\|\bX\|_{\mathrm{F}} = (\sum_{i = 1}^n\sum_{j = 1}^mX_{ij}^2)^{1/2}$ the Frobenius norm of $\bX$, $\|\bX\|_2$ the spectral norm (i.e., the largest singular value) of $\bX$, $\|\bX\|_{2\to\infty} = \max_{i\in[n]}(\sum_{j = 1}^mX_{ij}^2)^{1/2}$ the two-to-infinity norm of $\bX$, $\|\bX\|_\infty = \max_{i\in[n]}\sum_{j = 1}^m|X_{ij}|$ the matrix infinity norm of $\bX$, and $\|\bX\|_1 := \|\bX\transpose\|_\infty$. For any two matrices $\bA,\bB$ of the same size, let $\langle\bA,\bB\rangle_{\mathrm{F}} = \mathrm{tr}(\bA\transpose\bB)$ denote the Frobenius inner product between $\bA$ and $\bB$.
For any two matrices $\bA$ and $\bB$, $\bA\otimes \bB$ denotes their Kronecker product. 
For a square matrix $\bA$ with all real eigenvalues, we denote by $\lambda_k(\bA)$ the $k$th largest eigenvalue of $\bA$. Given an $n\times m$ matrix $\bA = [A_{ij}]_{n\times m}$, let $\vect(\bA)\in\mathbb{R}^{nm}$ denote the vectorization of $\bA$ that stacks its columns. When $m = n$ and $\bA$ is symmetric, we denote $\vech(\bA) := [A_{11},A_{12},\ldots,A_{1n},A_{22},\ldots,A_{2n},\ldots,A_{nn}]\transpose\in\mathbb{R}^{n(n + 1)/2}$ the half vectorization of $\bA$. 
Let $\be_i(n)$ denote the $i$th standard basis vector in $\mathbb{R}^n$. When the dimension $n$ is clear from the context, we typically write $\be_i = \be_i(n)$. For a given positive integer $d$, we let $\mathrm{Sym}(d)$ denote the set of all $d\times d$ symmetric matrices. 
For two real numbers $a,b\in\mathbb{R}$, let $a\vee b := \max(a,b)$. Given a sequence of random matrices $(\bX_n)_{n = 1}^\infty$ and a deterministic nonnegative sequence $(\eps_n)_{n = 1}^\infty$, we denote by $\bX_n = \Optilde(\eps_n)$ if for any $c > 0$, there exist constants $N_c,K_c,M_c > 0$, such that $\prob(\|\bX_n\|_2 > M_c\eps_n)\leq K_c n^{-c}$ for any $n\geq N_c$. For two symmetric matrices $\bA$ and $\bB$ of the same dimension, we denote by $\bA\preceq \bB$ or $\bB\succeq \bA$ if $\bB - \bA$ is positive semidefinite. For a random variable $X$, its $\psi_1$-Orlicz norm is defined as $\|X\|_{\psi_1} = \inf\{K:\expect e^{|X|/K}\leq 2\}$. For nonnegative deterministic sequences $(a_n)_{n = 1}^\infty, (b_n)_{n = 1}^\infty$, we write $a_n = O(b_n)$ if there exists a constant $C > 0$ not depending on $n$, such that $a_n\leq Cb_n$ for all $n\geq 1$.

\section{Multilayer latent space network with an invariant subspace}
\label{sec:multilayer_latent_space_network_with_an_invariant_subspace}

Let $p_A(\cdot; \theta)$ be a density function with respect to some $\sigma$-finite measure over $\mathbb{R}$ parameterized by $\theta\in\Theta\subset\mathbb{R}$. The multilayer network adjacency matrices $(\bA^{(t)})_{t = 1}^m = ([A_{ij}^{(t)}]_{n\times n})_{t = 1}^m$ have distribution described by
\begin{align}
\label{eqn:model}
A_{ij}^{(t)}\sim p_A(\cdot; \theta_{ij}^{(t)})\text{ independently for all }i<j,t\in[m],\; \theta_{ij}^{(t)} = \alpha_{i}^{(t)} + \alpha_{j}^{(t)} + \bx_i\transpose\bR^{(t)}\bx_j,
\end{align}
and $A_{ij}^{(t)} = A_{ji}^{(t)}$ for $i > j$, $A_{ii}^{(t)} = 0$ for all $i\in[n],t\in[m]$, where $\bX = [\bx_1,\ldots,\bx_n]\transpose$ satisfies $\bX\transpose\one_n = \zero_d$ and $\bX\transpose\bX = n\eye_d$. Denote by $\balpha^{(t)} = [\alpha_1^{(t)},\ldots,\alpha_n^{(t)}]\transpose$. Let $\theta_{0ij}^{(t)}$, $\alpha_{0i}^{(t)}$, $\bx_{0i}$, $\bR_0^{(t)}$, $\balpha_0^{(t)}$ and $\bX_0$ denote the true values of $\theta_{ij}^{(t)}$, $\alpha_i^{(t)}$, $\bx_i$, $\bR^{(t)}$, $\balpha^{(t)}$, and $\bX$, respectively, that correspond to the data generating distribution.

If the centered latent-position matrix has rank $d$, centering and whitening give the stated normalization without changing the edge parameters. For any $\bX$ of full column rank, any $\bc\in\mathbb{R}^d$, and any invertible $\bG\in\mathbb{R}^{d\times d}$, the reparameterization $\bx_i\mapsto\bG\transpose(\bx_i - \bc)$, $\bR^{(t)}\mapsto\bG^{-1}\bR^{(t)}\bG^{-\mathrm{T}}$, $\alpha_i^{(t)}\mapsto\alpha_i^{(t)} + \bc\transpose\bR^{(t)}(\bx_i - \bc) + \bc\transpose\bR^{(t)}\bc/2$ leaves every $\theta_{ij}^{(t)}$ unchanged. The centering condition $\bX\transpose\one_n = \zero_d$ separates the degree heterogeneity from the latent geometry: writing $\bTheta^{(t)} = [\theta_{ij}^{(t)}]_{n\times n}$ (with the diagonal defined by the same formula), we have $(\eye_n - \one_n\one_n\transpose/n)\bTheta^{(t)}(\eye_n - \one_n\one_n\transpose/n) = \bX\bR^{(t)}\bX\transpose$, so the degree parameters are annihilated by double centering and the column space of $(\eye_n - \one_n\one_n\transpose/n)\bTheta^{(t)}(\eye_n - \one_n\one_n\transpose/n)$ is the invariant subspace $\mathrm{Span}(\bX)$ whenever $\bR^{(t)}$ is nonsingular. The condition $\bX\transpose\bX = n\eye_d$ fixes the scale. After this normalization, the latent positions are identified only up to an orthogonal transformation, $(\bX,\bR^{(t)})\mapsto(\bX\bW,\bW\transpose\bR^{(t)}\bW)$ with $\bW\in\mathbb{O}(d)$, while $\balpha^{(t)}$, the projection matrix $\bX\bX\transpose/n$, the inner products $\bx_i\transpose\bx_j$, and the matrices $\bX\bR^{(t)}\bX\transpose$ are invariant under this transformation.

The interaction matrices $(\bR^{(t)})_{t = 1}^m$ are allowed to be indefinite. For two groups located near opposite values on a coordinate, a negative diagonal interaction encourages their cross-group connection relative to their within-group connection. A positive diagonal interaction has the opposite effect. 
Thus, allowing for indefinite interaction matrices can distinguish overall activity from assortative or disassortative structure. 
Degree effects shift all incident parameters and cannot, by themselves, express this contrast. 

There are two distinct inferential targets. An element of $\bR^{(t)}$ is meaningful after a representative of the latent subspace has been fixed. 
In applications, one may choose a reproducible reference orientation, but its estimation and possible eigenvalue multiplicities must be accounted for before applying an entrywise inference procedure that is not rotation invariant. Meanwhile, inference for quantities such as the fitted
parameters and the eigenvalues of each $\bR^{(t)}$ that are invariant to joint orthogonal rotation can proceed without fixing a reference orientation. 

For a fixed layer, each interaction parameter is informed by $O(n^2)$ edges, whereas each degree parameter sees $O(n)$ edges. A shared latent position receives information across $m$ layers. These counts explain why interaction uncertainty is examined at scale $n$ and vertex uncertainty at scale $\sqrt{mn}$, but do not establish a distributional claim: the estimation error in the growing nuisance vector can contribute at the interaction scale. One of the main contributions of this work is the introduction of the generalized Cayley transform of subspaces below in Section \ref{sec:euclidean_representation_and_intrinsic_geometry}, which permits those nuisance contributions to be tracked without counting equivalent rotations as distinct parameters.

Now we introduce two collections of assumptions that describe the behavior of the edge likelihood and the true value of the parameter. A similar set of assumptions has been adopted in \cite{Tian_Sun_He_2026_Bridging} in the context of a single-layer latent space network model.
\begin{assumption}[Edge distributions]
\label{assumption:edge_distribution}
Denote $\ell(\theta; x):=\log p_A(x;\theta)$ the log-density function (viewed as a function of $\theta$).
\begin{enumerate}[(i)]
  \item The support $\mathcal{X} = \{x : p_A(x;\,\theta) > 0\}$ does not depend on $\theta$.

  \item For any $x\in\calX$, $\ell(\cdot; x)$ is three times continuously differentiable. Also, for any constant $b > 0$, there exist constants $\kappa_1(b),\kappa_2(b), \kappa_3(b) > 0$ depending on $b$, such that
  \[
  \kappa_1(b)\leq -\ell''(\theta; x)\leq \kappa_2(b),\;
  |\ell'''(\theta; x)|\leq \kappa_3(b)\;\text{for all }(\theta, x)\in[-b, b]\times\calX.
  \]

  \item For any $\theta\in\Theta$, $\expect_{A\sim p_A(\cdot; \theta)} \{\ell'(\theta; A)\}^2 = -\expect_{A\sim p_A(\cdot; \theta)} \ell''(\theta; A)$.
\end{enumerate}
\end{assumption}

\begin{assumption}[True values of parameters]
\label{assumption:true_parameter}
The true values of the parameters satisfy the following conditions:
\begin{enumerate}[(i)]
  \item There exists a constant $C_1 > 0$ such that $\|\bX_0\|_{2\to\infty}\leq C_1$ and $\max_{t\in[m]}\|\balpha_0^{(t)}\|_\infty\leq C_1$.

  \item There exist constants $c_2,C_2 > 0$, such that $c_2 < |\lambda_k(\bR_0^{(t)})| < C_2$ for all $k\in[d]$, $t\in[m]$.

  \item $\bX_0\transpose\one_n = \zero_d$ and $\bX_0\transpose\bX_0 = n\eye_d$.

  \item $\expect_0\ell'(\theta_{0ij}^{(t)}; A_{ij}^{(t)}) = 0$, and there exists a constant $\sigma > 0$, such that $\|\ell'(\theta_{0ij}^{(t)}; A_{ij}^{(t)})\|_{\psi_1}\leq \sigma$ for all $i,j\in[n]$ and $t\in[m]$.
\end{enumerate}
\end{assumption}

\begin{remark}
\label{remark:assumptions}
Assumption~\ref{assumption:edge_distribution} is standard and can be satisfied by a broad collection of distributions such as canonical exponential family distributions (see Example \ref{example:canonical_families} below). 
Assumption~\ref{assumption:true_parameter} (i)--(ii) imply that $|\theta_{0ij}^{(t)}|\leq 2C_1 + C_1^2C_2$ for all $i,j,t$, so the networks are dense in the sense that all edge parameters are bounded away from the boundary of the parameter space. Assumption~\ref{assumption:true_parameter} (ii) allows $\bR_0^{(t)}$ to be indefinite, which accommodates disassortative structures \citep{rubindelanchy2022statistical}, and allows the layers to express the shared geometry with different strengths and signs; it does, however, require every layer to express every direction of the invariant subspace. The number of layers $m$ may grow slowly with $n$, as seen in Section \ref{sec:maximum_likelihood_and_one_step_estimator}. 
\end{remark}

\begin{example}[Canonical exponential families]
\label{example:canonical_families}
If $p_A(x;\theta) = h(x)\exp\{x\theta - \psi(\theta)\}$ is a natural exponential family with a smooth cumulant function $\psi$, then $\ell''(\theta; x) = -\psi''(\theta)$ does not depend on $x$, and the Fisher weight is $w(\theta) := -\expect_{A\sim p_A(\cdot;\theta)}\ell''(\theta; A) = \psi''(\theta)$. This covers the Bernoulli distribution with the logistic link, $\psi(\theta) = \log(1 + e^\theta)$, for binary networks; the Poisson distribution with the log link, $\psi(\theta) = e^\theta$, for count-valued networks; and the Gaussian distribution $\mathrm{N}(\theta,\sigma^2)$ with known variance and the identity link, for weighted networks. 
Assumption~\ref{assumption:edge_distribution} holds in all three cases, and $\ell'(\theta; A) = A - \psi'(\theta)$ (divided by $\sigma^2$ in the Gaussian case) is sub-exponential with a bounded $\psi_1$-norm when $\theta$ is bounded, so that Assumption~\ref{assumption:true_parameter} (iv) holds under (i)--(ii).
\end{example}

\section{Euclidean representation and intrinsic geometry}
\label{sec:euclidean_representation_and_intrinsic_geometry}

The parameter $\bX$ lives on the set $\{\bX\in\mathbb{R}^{n\times d}:\bX\transpose\one_n = \zero_d,\bX\transpose\bX = n\eye_d\}$, a scaled Stiefel manifold, and only its orbit under the right action of $\mathbb{O}(d)$ is identifiable. To obtain a Euclidean parameterization suited for asymptotic analysis, we use a generalized Cayley transform centered at a reference point. Let $\bU_{0\perp}\in\mathbb{O}(n, n - d - 1)$ span the orthogonal complement of $\mathrm{Span}([\one_n, \bX_0])$ and denote $\bX_{0\perp} = \sqrt{n}\bU_{0\perp}$. Consider the generalized Cayley transform \citep{Bendokat_Zimmermann_Absil_2024_ACM,Jauch_Hoff_Dunson_2020_Bernoulli,Shepard_Brozell_Gidofalvi_2015_JPCA,xie2023euclidean} with $\bX_0$ being the reference:
\begin{align}\label{eqn:Cayley_transform}
\bX:\bB\in\mathbb{R}^{(n - 1 - d)\times d}\mapsto \bX(\bB):=(\bX_0 + \bX_{0\perp}\bB)(\eye_d + \bB\transpose\bB)^{-1/2}.
\end{align}
For every $\bB$, $\bX(\bB)$ satisfies the normalization in \eqref{eqn:model}, and $\bX_0\transpose\bX(\bB) = n(\eye_d + \bB\transpose\bB)^{-1/2}$ is symmetric positive definite; conversely, any normalized $\bX$ with $\bX_0\transpose\bX$ invertible can be written uniquely as $\bX = \bX(\bB)\bW$ with $\bW\in\mathbb{O}(d)$. Hence, \eqref{eqn:Cayley_transform} is a chart of the quotient manifold that covers the open set where $\bX_0\transpose\bX$ is invertible, and the orthogonal ambiguity is resolved by a canonical alignment. This is natural because one expects that a consistent estimator $\widehat{\bX}$ for $\bX_0$ should concentrate around it so that $\bX_0\transpose\widehat{\bX}$ is invertible with high probability. The transform is closely related to the Euclidean representation of low-rank matrices in \cite{xie2023euclidean}, where the reference matrix is taken to be $[\eye_d, \mathbf{O}]\transpose$ without assuming the orthogonality to $\one_n$.
Let
$\bphi = [\balpha^{(1)\mathrm{T}},\ldots,\balpha^{(m)\mathrm{T}},\vech(\bR^{(1)})\transpose,\ldots,\vech(\bR^{(m)})\transpose,\vect(\bB)\transpose]\transpose\in\mathbb{R}^{d_\bphi}$
denote the compact parameter of interest under reparameterization \eqref{eqn:Cayley_transform} of $\bX$ above, where we denote $d_\bphi = m\{n + d(d + 1)/2\} + d_\bB$ and $d_\bB = (n - 1 - d)d$. Likewise, we denote $d_\balpha = mn$ and $d_\br = md(d + 1)/2$. Clearly, $\bX(\mathbf{O}_{(n - 1 - d)\times d}) = \bX_0$, and we denote
$\bphi_0 = [\balpha^{(1)\mathrm{T}}_0,\ldots,\balpha^{(m)\mathrm{T}}_0,\vech(\bR_0^{(1)})\transpose,\ldots,\vech(\bR_0^{(m)})\transpose,\zero_{d_\bB}\transpose]\transpose$
the true value of $\bphi$. Let $\calL(\bphi)$ denote the log-likelihood function
\begin{align*}
\calL(\bphi) = \sum_{t = 1}^m\sum_{i < j}\ell(\alpha_i^{(t)} + \alpha_j^{(t)} + \bx_i(\bB)\transpose\bR^{(t)}\bx_j(\bB); A_{ij}^{(t)})
\end{align*}
under the compact parameterization, where
$\bx_i(\bB):=\bX(\bB)\transpose\be_i$
is the $i$th row of $\bX(\bB)$ (viewed as a column vector). Let $\bG(\bphi)$ denote the Fisher information matrix with respect to $\bphi$ defined by
$\bG(\bphi) = \expect_{\bphi}
\{\partial\calL(\bphi)/\partial\bphi\}
\{\partial\calL(\bphi)/\partial\bphi\}\transpose$.

\begin{proposition}\label{prop:expected_hessian_bound}
Under Assumptions \ref{assumption:edge_distribution}--\ref{assumption:true_parameter}, for any $b > 0$ such that
$\max_{t\in[m],i,j\in[n]}|\theta_{0ij}^{(t)}|\leq b$
and sufficiently large $n$, we have
\begin{align*}
\frac{\kappa_1(b)}{2}
\mathrm{diag}\bigg(n\eye_{d_\balpha}, \frac{n^2\eye_{d_\br}}{2}, c_2^2mn^2\eye_{d_\bB}\bigg)
\preceq \bG(\bphi_0)
\preceq \frac{\kappa_2(b)}{2}
\mathrm{diag}\left(4n\eye_{d_\balpha}, 2n^2\eye_{d_\br}, 2C_2^2mn^2\eye_{d_\bB}\right)
.
\end{align*}
\end{proposition}

Proposition~\ref{prop:expected_hessian_bound} identifies the variance scales of different parameter components via the intrinsic geometry of the model: under the generalized Cayley transform of the invariant subspace,
the magnitude of the Fisher information is equivalent to the block-diagonal matrix $\mathrm{diag}(n\eye_{d_\balpha}, n^2\eye_{d_\br}, mn^2\eye_{d_\bB})$. Each degree parameter $\alpha_i^{(t)}$ is informed by the $n - 1$ edges incident to vertex $i$ in layer $t$, each $\bR^{(t)}$ by all $\binom{n}{2}$ edges of layer $t$, and each coordinate of $\bB$, which moves the latent positions of all vertices simultaneously through $\bX_{0\perp} = \sqrt{n}\bU_{0\perp}$, by the edges of all $m$ layers. The standard errors of the MLE for the three components are therefore $O(n^{-1/2})$ for $\alpha_i^{(t)}$, $O(n^{-1})$ for $\bR^{(t)}$, and $O((mn)^{-1/2})$ for each of the latent positions. The proof exploits the mutual orthogonality, under the Frobenius inner product, of the three components $\delta\balpha^{(t)}\one_n\transpose + \one_n\delta\balpha^{(t)\mathrm{T}}$, $\bX_0\delta\bR^{(t)}\bX_0\transpose$, and $\bX_{0\perp}\delta\bB\bR_0^{(t)}\bX_0\transpose + \bX_0\bR_0^{(t)}\delta\bB\transpose\bX_{0\perp}\transpose$ of the tangent vectors, which is a direct consequence of the normalization in \eqref{eqn:model} and of the construction of $\bX_{0\perp}$.

Denote
$\bpsi = [\balpha^{(1)\mathrm{T}},\ldots,\balpha^{(m)\mathrm{T}},\vech(\bR^{(1)})\transpose,\ldots,\vech(\bR^{(m)})\transpose,\vect(\bX)\transpose]\transpose\in\mathbb{R}^{d_\bpsi}$ the ambient parameter,
where $d_\bpsi = n(m + d) + md(d + 1)/2$ is the dimension of the ambient space. We
refer to the function $\bphi = [\balpha^{(1)\mathrm{T}},\ldots,\balpha^{(m)\mathrm{T}},\vech(\bR^{(1)})\transpose,\ldots,\vech(\bR^{(m)})\transpose,\vect(\bB)\transpose]\transpose\mapsto \bpsi(\bphi)$, where $\bpsi(\bphi) := [\balpha^{(1)\mathrm{T}},\ldots,\balpha^{(m)\mathrm{T}},\vech(\bR^{(1)})\transpose,\ldots,\vech(\bR^{(m)})\transpose,\vect\{\bX(\bB)\}\transpose]\transpose$,
as the natural embedding of the compact parameter $\bphi$ into the ambient parameter space, and denote its generic inverse function by $\bpsi\mapsto \bphi(\bpsi)$, where
\begin{align*}
\bphi(\bpsi)& := [\balpha^{(1)\mathrm{T}},\ldots,\balpha^{(m)\mathrm{T}},\vech(\bW_\bX\transpose\bR^{(1)}\bW_\bX)\transpose,\ldots,\vech(\bW_\bX\transpose\bR^{(m)}\bW_\bX)\transpose,
\\
&\qquad
\vect\{\bX_{0\perp}\transpose\bX(\bX_0\transpose\bX)^{-1}\}\transpose]\transpose,
\quad \bW_\bX = (\bX_0\transpose\bX)^{-1}\{(\bX\transpose\bX_0)^{-1}(\bX_0\transpose\bX)^{-1}\}^{-1/2}
\end{align*}
whenever $\bX_0\transpose\bX$ is invertible.
Note that $\bphi(\bpsi(\bphi)) = \bphi$ always holds for every $\bphi$, whereas
$\bpsi(\bphi(\bpsi)) = \bpsi$ if and only if $\bX_0\transpose\bX$ is symmetric positive definite.


\section{Maximum likelihood estimation and inference}
\label{sec:maximum_likelihood_and_one_step_estimator}

We consider the bounded ambient parameter space of interest
\begin{equation}
\label{eqn:psi_parameter_space}
\begin{aligned}
\Psi &= \bigg\{\bpsi:
\max_{t\in[m]}\|\balpha^{(t)}\|_\infty\vee \max_{t\in[m]}\|\bR^{(t)}\|_{\mathrm{F}}\vee 4\|\bX\|_{2\to\infty}\leq M,
\bX\transpose\bX = n\eye_d, \bX\transpose\one_n = \zero_d
\bigg\}
\end{aligned}
\end{equation}
and define the MLE of $\bpsi$ by
\begin{align*}
\widehat{\bpsi} &= \argmax_{\bpsi\in\Psi}\calL_\bpsi(\bpsi),\quad 
\calL_\bpsi(\bpsi) = \sum_{t = 1}^m\sum_{i < j}\ell(\alpha_i^{(t)} + \alpha_j^{(t)} + \bx_i\transpose\bR^{(t)}\bx_j; A_{ij}^{(t)}).
\end{align*}
Write
$\widehat{\bpsi} = [\widehat{\balpha}^{(1)\mathrm{T}},\ldots,\widehat{\balpha}^{(m)\mathrm{T}},\vech(\widehat{\bR}^{(1)})\transpose,\ldots,\vech(\widehat{\bR}^{(m)})\transpose,\vect(\widehat{\bX})\transpose]\transpose$, $\widehat{\bX} = [\widehat{\bx}_1,\ldots,\widehat{\bx}_n]\transpose$, $\widehat{\balpha}^{(t)} = [\widehat{\alpha}_1^{(t)},\ldots,\widehat{\alpha}_n^{(t)}]\transpose$,
and denote $\widehat{\balpha}_i = [\widehat{\alpha}_i^{(1)},\ldots,\widehat{\alpha}_i^{(m)}]\transpose$. Also, let $\balpha_{0i} = [\alpha_{0i}^{(1)},\ldots,\alpha_{0i}^{(m)}]\transpose$. 
The constant $M$ here is conceptual in the sense that it facilitates the theoretical analysis. In practice, one typically sets it to be a large constant. 

\subsection{Large-sample properties of the vertex-level estimators}
\label{sub:maximum_likelihood_estimator}

In Theorem~\ref{thm:two_to_infinity_norm_bound_MLE}, the factor $m^{-1/2}$ in the two-to-infinity norm bound for $\widehat{\bX}$ quantifies the benefit of pooling the $m$ layers, even though each layer carries its own degree heterogeneity and interaction matrix.

\begin{theorem}
\label{thm:two_to_infinity_norm_bound_MLE}
Suppose Assumptions \ref{assumption:edge_distribution}--\ref{assumption:true_parameter} hold and $m^{12}(\log n)^4/n \to 0$ as $n\to\infty$. Then, there exists a sufficiently large constant $M > 0$ for $\Psi$ in \eqref{eqn:psi_parameter_space}, such that
\begin{align*}
\|\widehat{\bX}\bW_{\widehat{\bX}} - \bX_0\|_{\mathrm{F}} = \Optilde(1),\quad
\|\widehat{\bX}\bW_{\widehat{\bX}} - \bX_0\|_{2\to\infty} = \Optilde\bigg(\sqrt{\frac{\log n}{mn}}\bigg).
\end{align*}
\end{theorem}

To describe the limiting distributions, we define the following matrices:
\begin{align*}
\bSigma_{\bz_i}^\sharp
 & = -\frac{1}{n}\sum_{t = 1}^m\sum_{j\in[n]\backslash\{i\}} \expect_0\ell''(\theta_{0ij}^{(t)}; A_{ij}^{(t)})
\begin{bmatrix}
\be_t(m)\be_t(m)\transpose & \frac{1}{\sqrt{m}}\be_t(m)\bx_{0j}\transpose\bR_0^{(t)}\\
\frac{1}{\sqrt{m}}\bR_0^{(t)}\bx_{0j}\be_t(m)\transpose & \frac{1}{m}\bR_0^{(t)}\bx_{0j}\bx_{0j}\transpose\bR_0^{(t)}
\end{bmatrix},\\
\bSigma_{\bx_i}^\sharp& = \begin{bmatrix}
\mathbf{O}_{d\times m} & \eye_d
\end{bmatrix} (\bSigma_{\bz_i}^\sharp)^{-1}\begin{bmatrix}
  \mathbf{O}_{m\times d} \\ \eye_d
\end{bmatrix}.
\end{align*}
In the above formulas, $\bSigma_{\bz_i}^\sharp$ is the $(\bz_i, \bz_i)$-block of the Fisher information matrix scaled by the standard error rates of $\balpha_i$ and $\bx_i$ informed by Proposition \ref{prop:expected_hessian_bound}, and $\bSigma_{\bx_i}^\sharp$ is the asymptotic covariance matrix of $\sqrt{mn}(\bW_{\widehat{\bX}}\transpose\widehat{\bx}_i - \bx_{0i})$.
In terms of the (unscaled) vertex-$i$ Fisher information matrix
\begin{align*}
\bG_i(\bpsi) := \sum_{t = 1}^m\sum_{j\in[n]\backslash\{i\}}{\expect_\bpsi\{\ell'(\theta_{ij}^{(t)}; A_{ij}^{(t)})\}^2}\begin{bmatrix}\be_t(m)\\ \bR^{(t)}\bx_j\end{bmatrix}\begin{bmatrix}\be_t(m)\\ \bR^{(t)}\bx_j\end{bmatrix}\transpose\in\mathbb{R}^{(m + d)\times(m + d)},
\end{align*}
we have $\bSigma_{\bz_i}^\sharp = \bLambda_n\bG_i(\bpsi_0)\bLambda_n$ with $\bLambda_n = \mathrm{diag}\{n^{-1/2}\eye_m, (mn)^{-1/2}\eye_d\}$. The matrix $\bG_i(\bpsi_0)$ is the Fisher information about $\bz_i = [\balpha_i\transpose,\bx_i\transpose]\transpose$ in the submodel in which all other parameters are known.

\begin{theorem}
\label{thm:asymptotic_normality_MLE}
Suppose Assumptions \ref{assumption:edge_distribution}--\ref{assumption:true_parameter} hold and $m^{12}(\log n)^4/n \to 0$ as $n\to\infty$. Then, there exists a sufficiently large constant $M > 0$ for $\Psi$ in \eqref{eqn:psi_parameter_space}, such that the following hold:
\begin{enumerate}[(i)]
  \item Let $\calI = \{i_1,\ldots,i_q\}\subset[n]$ be a set of fixed vertices, where $q$ is fixed. For each $i\in\calI$, let $\bg_i = [\bg_{\balpha_i}\transpose,\bg_{\bx_i}\transpose]\transpose\in\mathbb{R}^{m + d}$ be a deterministic unit vector with $\bg_{\balpha_i}\in\mathbb{R}^m$ and $\bg_{\bx_i}\in\mathbb{R}^d$. Suppose $\|\bg_i\|_1 = O(1)$ for all $i\in\calI$. Then,
  \begin{align*}
  \frac{\sum_{i\in\calI}\{\sqrt{n}\bg_{\balpha_i}\transpose(\widehat{\balpha}_i - \balpha_{0i})
  + \sqrt{mn}\bg_{\bx_i}\transpose(\bW_{\widehat{\bX}}\transpose\widehat{\bx}_i - \bx_{0i})\} }{\sqrt{\sum_{i\in\calI}\bg_i\transpose (\bSigma_{\bz_i}^\sharp)^{-1}\bg_i}}\overset{\calL}{\to}\mathrm{N}(0, 1)\quad\text{as }n\to\infty.
  \end{align*}

  \item Let $\bF\in\mathrm{Sym}(n)$ be deterministic with $\|\bF\bX_0\|_{\mathrm{F}} = 1$, $\|\bX_0\transpose\bF\bX_0\|_2 = O(1)$, and $\|\vect(\bF)\|_1 = O(1)$. Then
  \begin{align*}
  \frac{\sqrt{mn}\langle\bF, \widehat{\bX}\widehat{\bX}\transpose - \bX_0\bX_0\transpose\rangle_{\mathrm{F}}}{\sqrt{4\sum_{i = 1}^n\be_i\transpose\bF\bX_0
  \bSigma_{\bx_i}^\sharp\bX_0\transpose\bF\be_i}}
  \overset{\calL}{\to}\mathrm{N}(0, 1)\quad\text{as }n\to\infty.
  \end{align*}
\end{enumerate}
\end{theorem}

Part (i) is a central limit theorem for linear combinations of the parameters of a fixed number of vertices with deterministic directions of bounded $\ell_1$-norm. It covers a single degree parameter $\alpha_i^{(t)}$, a contrast $\alpha_i^{(t)} - \alpha_i^{(s)}$ between two layers, and any linear functional of $\bx_i$; since $d$ is fixed, the Cram\'er--Wold device yields joint asymptotic normality of $\sqrt{mn}(\bW_{\widehat{\bX}}\transpose\widehat{\bx}_i - \bx_{0i})$ with covariance $\bSigma_{\bx_i}^\sharp$. The asymptotic variance $\bg_i\transpose(\bSigma_{\bz_i}^\sharp)^{-1}\bg_i$ is the one computed from $\bG_i(\bpsi_0)^{-1}$, so the MLE is oracle efficient for these linear combinations: there is no first-order price for estimating the $n(m + d) + md(d + 1)/2$ nuisance parameters, and in particular no price for the unknown orientation of the subspace. The underlying reason is that the cross blocks between $\bz_i$ and other parameters are asymptotically negligible, so that $\bz_i$ is asymptotically Neyman orthogonal to these high-dimensional nuisance parameters \citep{Cox_Reid_1987_Orthogonality_JRSSB,Jeffreys_1961_Theory,Neyman_1959_Optimal,Chernozhukov_Chetverikov_Demirer_Duflo_Hansen_Newey_Robins_2018_DML_EconometricsJ}. 
Part (ii) concerns the projection matrix onto the invariant subspace, which is rotation invariant; because of the requirement $\|\vect(\bF)\|_1 = O(1)$, it covers
inference for entries of the Gram matrix $\bX_0\bX_0\transpose$ and for linear combinations of a bounded number of such entries.

\begin{remark}[On the growth condition of $m$]
Theorem \ref{thm:asymptotic_normality_MLE} requires that the number of layers $m$ cannot grow too fast with the number of vertices $n$ in the sense that $m^{12}(\log n)^4/n\to 0$ as $n\to\infty$. Although such a growth condition of $m$ is stronger than that in \cite{liu2026unfolding}, we note that the difference is largely structural rather than technical: \cite{liu2026unfolding} assumes that the $n\times m$ degree effect matrix $[\alpha_i^{(t)}]_{n\times m}$ admits a fixed-rank factorization, so that the intrinsic model dimension there is $O(m + n)$; in contrast, our model allows the degree effect matrix to be fully unstructured and possibly of full rank, so that the model dimension is $O(mn)$. We also remark that the condition $m^{12}(\log n)^4/n \to 0$ might not be sharp, but we conjecture that a slowly growing $m$ is necessary for valid inferential theory.
\end{remark}

\subsection{Plug-in vertex-level inference}
\label{sub:plugin_inference}

The covariances in Theorem~\ref{thm:asymptotic_normality_MLE} depend on the unknown parameters. Let $\widehat{\theta}_{ij}^{(t)} = \widehat{\alpha}_i^{(t)} + \widehat{\alpha}_j^{(t)} + \widehat{\bx}_i\transpose\widehat{\bR}^{(t)}\widehat{\bx}_j
$ and define the plug-in estimators of $\bSigma_{\bz_i}^\sharp$ and $\bSigma_{\bx_i}^\sharp$ by
\begin{align*}
\widehat{\bSigma}_{\bz_i}^\sharp
 & = -\frac{1}{n}\sum_{t = 1}^m\sum_{j\in[n]\backslash\{i\}} \expect_{\widehat{\bpsi}}\ell''(\widehat{\theta}_{ij}^{(t)}; A_{ij}^{(t)}) 
\begin{bmatrix}
\be_t(m)\be_t(m)\transpose & \frac{1}{\sqrt{m}}\be_t(m)\widehat{\bx}_{j}\transpose\widehat{\bR}^{(t)}\\ 
\frac{1}{\sqrt{m}}\widehat{\bR}^{(t)}\widehat{\bx}_{j}\be_t(m)\transpose & \frac{1}{m}\widehat{\bR}^{(t)}\widehat{\bx}_{j}\widehat{\bx}_{j}\transpose\widehat{\bR}^{(t)}
\end{bmatrix},\\
\widehat{\bSigma}_{\bx_i}^\sharp& = \begin{bmatrix}
\mathbf{O}_{d\times m} & \eye_d
\end{bmatrix} (\widehat{\bSigma}_{\bz_i}^\sharp)^{-1}\begin{bmatrix}
  \mathbf{O}_{m\times d} \\ \eye_d
\end{bmatrix}.
\end{align*}

Then, Corollary \ref{corr:plug_in_vertex} below shows that the asymptotic normality still holds when the unknown covariance is estimated using the plug-in principle. 
\begin{corollary}
\label{corr:plug_in_vertex}
Suppose Assumptions \ref{assumption:edge_distribution}--\ref{assumption:true_parameter} hold and $m^{12}(\log n)^4/n\to 0$ as $n\to\infty$. Then, with $M$ as in Theorem \ref{thm:asymptotic_normality_MLE}:
\begin{enumerate}[(i)]
\item For $\calI$ and $(\bg_i)_{i\in\calI}$ as in Theorem \ref{thm:asymptotic_normality_MLE}(i),
\begin{align*}
\frac{\sum_{i\in\calI}\{\sqrt{n}\bg_{\balpha_i}\transpose(\widehat{\balpha}_i - \balpha_{0i}) + \sqrt{mn}\bg_{\bx_i}\transpose(\bW_{\widehat{\bX}}\transpose\widehat{\bx}_i - \bx_{0i})\}}{\sqrt{\sum_{i\in\calI}\bg_i\transpose\{\mathrm{diag}(\eye_m, \bW_{\widehat{\bX}}\transpose)\widehat{\bSigma}_{\bz_i}^\sharp\mathrm{diag}(\eye_m, \bW_{\widehat{\bX}})\}^{-1}\bg_i}}\overset{\calL}{\to}\mathrm{N}(0, 1)\quad\text{as }n\to\infty.
\end{align*}
\item For $\bF$ as in Theorem \ref{thm:asymptotic_normality_MLE}(ii),
\begin{align*}
\frac{\sqrt{mn}\langle\bF, \widehat{\bX}\widehat{\bX}\transpose - \bX_0\bX_0\transpose\rangle_{\mathrm{F}}}{\sqrt{4\sum_{i = 1}^n\be_i\transpose\bF\widehat\bX\widehat{\bSigma}_{\bx_i}^\sharp\widehat\bX\transpose\bF\be_i}}\overset{\calL}{\to}\mathrm{N}(0, 1)\quad\text{as }n\to\infty.
\end{align*}
\end{enumerate}
\end{corollary}

Corollary~\ref{corr:plug_in_vertex} turns Theorem~\ref{thm:asymptotic_normality_MLE} into practical procedures. For an estimator $\bar{\bpsi} = (\bar{\balpha},\bar{\bR},\bar{\bX})$, we use the plug-in vertex information $\widehat{\bG}_i = \bG_i(\bar{\bpsi})$ and estimate the covariance of $\bar{\bz}_i = [\bar{\balpha}_i\transpose,\bar{\bx}_i\transpose]\transpose$ by $\widehat{\bG}_i^{-1}$. In particular, $\widehat{\var}(\bar{\alpha}_i^{(t)})$ can be taken as the $t$th diagonal element of $\widehat{\bG}_i^{-1}$, and the covariance of $\bar{\bx}_i$ is estimated by $\widehat{\bC}_i := [\mathbf{O}_{d\times m},\eye_d]\widehat{\bG}_i^{-1}[\mathbf{O}_{d\times m}, \eye_d]\transpose
$. For the MLE, $\bLambda_n\widehat{\bG}_i\bLambda_n=\widehat\bSigma_{\bz_i}^\sharp$, and Corollary~\ref{corr:plug_in_vertex} together with the delta method yields the following procedures for fixed vertices, all of which are invariant or equivariant under the orthogonal transformations that leave the model unchanged:
\begin{enumerate}[(a),leftmargin=*]
\item a $(1 - a)$ confidence interval $\bar{\alpha}_i^{(t)}\pm z_{1 - a/2}\{\widehat{\var}(\bar{\alpha}_i^{(t)})\}^{1/2}$ for $\alpha_{0i}^{(t)}$, and, more generally, $\bg\transpose\bar{\bz}_i\pm z_{1 - a/2}\{\bg\transpose\widehat{\bG}_i^{-1}\bg\}^{1/2}$ for a vector $\bg$ of bounded $\ell_1$-norm, such as $\alpha_{0i}^{(t)} - \alpha_{0i}^{(s)}$;
\item a Wald confidence ellipsoid $\{\bx:(\bar{\bx}_i - \bx)\transpose\widehat{\bC}_i^{-1}(\bar{\bx}_i - \bx)\leq\chi^2_{d,1 - a}\}$ for the latent position $\bW_{\bar{\bX}}\bx_{0i}$ expressed in the coordinate system aligned with $\bar{\bX}$;
\item a confidence interval $\bar{\bx}_i\transpose\bar{\bx}_j\pm z_{1 - a/2}\widehat{\sigma}_{ij}$ for the inner product $\bx_{0i}\transpose\bx_{0j}$, where $\widehat{\sigma}_{ij}^2 = \bar{\bx}_j\transpose\widehat{\bC}_i\bar{\bx}_j + \bar{\bx}_i\transpose\widehat{\bC}_j\bar{\bx}_i$ for $i\neq j$ and $\widehat{\sigma}_{ii}^2 = 4\bar{\bx}_i\transpose\widehat{\bC}_i\bar{\bx}_i$ for the \emph{hub score} $\|\bx_{0i}\|_2^2$ (provided $\bx_{0i}$, respectively $(\bx_{0i},\bx_{0j})$, is bounded away from zero);
\item the test of \emph{structural equivalence} $H_0:\bx_{0i} = \bx_{0j}$, which rejects when $T_{ij} = (\bar{\bx}_i - \bar{\bx}_j)\transpose(\widehat{\bC}_i + \widehat{\bC}_j)^{-1}(\bar{\bx}_i - \bar{\bx}_j) > \chi^2_{d,1 - a}$. Under $H_0$, $T_{ij}\overset{\calL}{\to}\chi^2_d$ as $n\to\infty$ because the estimators of distinct fixed vertices are asymptotically independent by Theorem~\ref{thm:asymptotic_normality_MLE} (i).
\end{enumerate}
For Gaussian edges with unknown variance, $\sigma^2$ is estimated by the residual mean square $\widehat{\sigma}^2$, which does not affect the MLE of $(\balpha,\bR,\bX)$. Since $\widehat\sigma^2$ is consistent and the covariances and bias scale with $\sigma^2$, Slutsky's theorem ensures the limiting distributions still hold. 

\subsection{Inference for the interaction matrices}
\label{sub:R_inference}


In order to establish the limiting distribution of $\widehat{\bR}^{(t)}$, we need an additional assumption that the fourth-order derivative of the log-likelihood is bounded, so that its third-order derivative is Lipschitz.

\begin{assumption}[Third-order Lipschitz property of $\ell$]
\label{assumption:third_order_lipschitz_ell}
For any $x\in\calX$, $\ell(\cdot; x)$ is four times continuously differentiable. Also, for any constant $b > 0$, there exists a constant $\kappa_4(b) > 0$ depending on $b$, such that
$|\ell''''(\theta; x)|\leq \kappa_4(b)$ for all $(\theta, x)\in[-b, b]\times \calX$.
\end{assumption}

Because $\bR^{(t)}$ is identifiable only up to a simultaneous rotation of rows and columns, the limiting distribution is described with regard to $\bW_{\widehat{\bX}}\transpose\widehat{\bR}^{(t)}\bW_{\widehat{\bX}}$.
Let $\widehat\bpsi^{\rm a}$ consist of $\widehat\balpha^{(s)}$, $\widehat\bX\bW_{\widehat\bX}$, and $\widehat\bR_{\rm a}^{(s)}=\bW_{\widehat\bX}\transpose\widehat\bR^{(s)}\bW_{\widehat\bX}$ for every $s$.
Suppose $\bU_\perp\in\mathbb{O}(n, n - 1 - d)$ spans $\mathrm{Span}(\one_n, \bX)^\perp$ and define the following matrices:
\begin{align*}
\scrJ(\bpsi) &= \mathrm{diag}(\eye_{d_\balpha + d_\br}, \eye_d\otimes\bU_\perp),\; \widetilde{\bLambda}_\sharp = \mathrm{diag}\bigg(\frac{\eye_{d_\balpha}}{\sqrt{n}}, \frac{\eye_{d_\br}}{n}, \frac{\eye_{d_\bX}}{\sqrt{mn}}\bigg),\\
\bOmega(\bpsi) &= \scrJ(\bpsi)\bigg\{\scrJ(\bpsi)\transpose\widetilde{\bLambda}_\sharp\expect_\bpsi\frac{\partial\calL_\bpsi}{\partial\bpsi}(\bpsi)\frac{\partial\calL_\bpsi}{\partial\bpsi\transpose}(\bpsi)\widetilde{\bLambda}_\sharp\scrJ(\bpsi)\bigg\}^{-1}\scrJ(\bpsi)\transpose,
\\
\scrP_{\br^{(t)}} &= 
\begin{bmatrix}
\mathbf{O}_{d_r\times d_\balpha} & \be_t(m)\transpose\otimes\eye_{d_r} & \mathbf{O}_{d_r\times d_\bX}  
\end{bmatrix},\;
\scrP_{\bx_i} = 
\begin{bmatrix}
\mathbf{O}_{d\times (d_\balpha + d_\br)} & \eye_d\otimes \be_i(n)\transpose 
\end{bmatrix},
\end{align*}
where $d_\bX=nd$ and we denote $d_r = d(d + 1)/2$ (which is different from $d_{\br} = md(d + 1)/2$). For fixed $(\bX,\bR)$, replacing $\bU_\perp$ by another orthonormal basis leaves $\bOmega$ unchanged. 

\begin{theorem}
\label{thm:asymptotic_normality_R}
Suppose Assumptions \ref{assumption:edge_distribution}--\ref{assumption:third_order_lipschitz_ell} hold and $m^{13}(\log n)^{9}/n\to 0$ as $n\to\infty$. Let $\calT\subset[m]$ be a fixed finite set of layer indices, $d'\in\{1,\ldots,|\calT|d_r\}$ be a fixed integer, $\bC:=(\bC_t)_{t\in\calT}$ be a fixed set of $d'\times d(d + 1)/2$ deterministic matrices not depending on $n$ such that $\sum_{t\in\calT}\bC_t\bC_t\transpose$ is nondegenerate, and write $\scrP_\calT(\bC) = \sum_{t\in\calT}\bC_t\scrP_{\br^{(t)}}$. Then, with $M$ as in Theorem \ref{thm:asymptotic_normality_MLE},
\begin{align*}
\big\{\scrP_{\calT}(\bC)\bOmega(\bpsi_0)\scrP_{\calT}(\bC)\transpose\big\}^{-1/2}\sum_{t\in\calT}\bC_t \Big[n\{\vech(\widehat\bR_{\rm a}^{(t)}) - \vech(\bR_0^{(t)})\} - \bb_n^{(t)}(\bpsi_0)\Big]\overset{\calL}{\to}\mathrm{N}(\zero_{d'}, \eye_{d'})
\end{align*}
as $n\to\infty$, where $\bb_n^{(t)}(\bpsi_0)$ is defined via the bias formula
\begin{equation}
\nonumber
\begin{aligned}
\mathbf{b}_n^{(t)}(\bpsi)
& = \sum_{s = 1}^m\sum_{1\leq i < j\leq n}\bigg\{
\expect_\bpsi \ell''(\theta_{ij}^{(s)}; A_{ij}^{(s)})\ell'(\theta_{ij}^{(s)}; A_{ij}^{(s)}) + \frac{1}{2}\expect_\bpsi \ell'''(\theta_{ij}^{(s)}; A_{ij}^{(s)})
\bigg\}\\
&\quad\times \bigg\{\frac{\partial\theta_{ij}^{(s)}}{\partial\bpsi\transpose}(\bpsi)\widetilde{\bLambda}_\sharp\bOmega(\bpsi)\widetilde{\bLambda}_\sharp\frac{\partial\theta_{ij}^{(s)}}{\partial\bpsi}(\bpsi)\bigg\}\scrP_{\br^{(t)}}\bOmega(\bpsi)\widetilde{\bLambda}_\sharp\frac{\partial\theta_{ij}^{(s)}}{\partial\bpsi}(\bpsi)\\
&\quad + 
\frac{1}{2}\sum_{s = 1}^m\sum_{1\leq i < j\leq n}\expect_\bpsi\ell''(\theta_{ij}^{(s)}; A_{ij}^{(s)})\mathrm{tr}\bigg\{\widetilde{\bLambda}_\sharp\frac{\partial^2\theta_{ij}^{(s)}}{\partial\bpsi\partial\bpsi\transpose}(\bpsi)\widetilde{\bLambda}_\sharp\bOmega(\bpsi)\bigg\}\\
&\quad\times  \scrP_{\br^{(t)}}\bOmega(\bpsi)\widetilde{\bLambda}_\sharp\frac{\partial\theta_{ij}^{(s)}}{\partial\bpsi}(\bpsi)
+ 
\frac{1}{2mn}\vech\left\{\bT(\bpsi)\bR^{(t)} + \bR^{(t)}\bT(\bpsi)\right\},
\end{aligned}
\end{equation}
$\bT(\bpsi) := \sum_{i = 1}^n \scrP_{\bx_i}\bOmega(\bpsi)\scrP_{\bx_i}\transpose$, and
$\scrP_{\calT}(\bC)\bOmega(\bpsi_0)\scrP_{\calT}(\bC)\transpose$ has eigenvalues bounded away from $0$ and $\infty$ uniformly in $n$.
\end{theorem}

The three terms of $\bb_n^{(t)}$ are contributed by different sources. The first term combines the curvature of the link, through $\expect\ell'''$, with the mixed moment $\expect(\ell''\ell')$; the latter vanishes for canonical exponential families but not, for example, for the probit link.
The second term depends on the curvature of $\theta_{ij}^{(s)}$ under the ambient $\bpsi$-parameterization.
The third term is rooted in the curvature of the constraint manifold where $\bX$ lives.
Two features distinguish $\widehat\bR^{(t)}$ from the vertex-level estimators. First, $\bR^{(t)}$ is in general not asymptotically Neyman orthogonal to the nuisance parameters: the rescaled cross-blocks of the Fisher information between $\bR^{(t)}$ and the vertex parameters are sparse but not negligible (see the Supplementary Material). The covariance in Theorem~\ref{thm:asymptotic_normality_R} is therefore the $\bR^{(t)}$ block of the full inverse information $\bOmega(\bpsi_0)$, which by the block-inverse formula dominates, in the Loewner order, the oracle covariance obtained by inverting the $\bR^{(t)}$ block alone. Second, the first two terms of $\bb_n^{(t)}$ sum contributions from all $mn(n - 1)/2$ edges, and the third collects the variances of all $n$ latent positions, so the second-order effects of the nuisance estimation errors accumulate to a shift of the same order as the standard deviation, and $\bb_n^{(t)}(\bpsi_0)$ does not vanish as $n\to\infty$. This is the incidental parameter problem of \cite{neyman1948consistent} in network form. As there, the shift is a second-order effect that orthogonality alone would not remove \citep{Cox_Reid_1987_Orthogonality_JRSSB}: for Gaussian edges, $\expect\ell''' = 0$ and the first term vanishes, yet the second and third terms remain. Valid inference for $\bR^{(t)}$ therefore requires both the nuisance-adjusted covariance and the bias correction.

\begin{corollary}
\label{corr:plug_in_R}
Under the conditions of Theorem~\ref{thm:asymptotic_normality_R},
\[
\big\{\scrP_{\calT}(\bC)\bOmega(\widehat\bpsi^{\rm a})\scrP_{\calT}(\bC)\transpose\big\}^{-1/2}\sum_{t\in\calT}\bC_t\Big[n\{\vech(\widehat\bR_{\rm a}^{(t)}) - \vech(\bR_0^{(t)})\} - \bb_n^{(t)}(\widehat\bpsi^{\rm a})\Big]\overset{\calL}{\to}\mathrm{N}(\zero_{d'}, \eye_{d'})
\]
as $n\to\infty$. 
\end{corollary}

With $\calT=\{t\}$ and $\bC_t=\eye_{d_r}$, Theorem~\ref{thm:asymptotic_normality_R} gives the marginal limiting distribution for a single $\widehat{\bR}^{(t)}$. With $\calT=\{s,t\}$ and $\bC = (\eye_{d_r},-\eye_{d_r})$, it gives the limiting distribution of the contrast $\widehat\bR^{(s)}_{\rm a}-\widehat\bR^{(t)}_{\rm a}$, whose covariance $(\scrP_{\br^{(s)}}-\scrP_{\br^{(t)}})\bOmega(\bpsi_0)(\scrP_{\br^{(s)}}-\scrP_{\br^{(t)}})\transpose$ includes the cross-layer blocks of $\bOmega(\bpsi_0)$. This observation directly yields Wald tests of $H_0:\bR_0^{(s)}=\bR_0^{(t)}$. More generally, stacking successive differences over $\calT$ gives a joint Wald test that all layers in $\calT$ share the same interaction matrix. Because $\bb_n^{(t)}$ and $\bOmega$ transform equivariantly under a simultaneous rotation of the latent coordinates (see the Supplementary Material),
the Wald statistic computed from the unaligned MLE $\widehat\bpsi$ equals the one computed from $\widehat\bpsi^{\rm a}$, so the test does not require the true orientation. 



Likewise, principled plug-in inference procedures can be guided by Corollary \ref{corr:plug_in_R}. For any confidence level $a\in(0, 1)$, a $(1-a)$ confidence region for $\vech(\bR_0^{(t)})$ in the aligned coordinates is centered at $\vech(\widehat\bR^{(t)}_{\rm a})-\bb_n^{(t)}(\widehat\bpsi^{\rm a})/n$, has shape $\scrP_{\br^{(t)}}\bOmega(\widehat\bpsi^{\rm a})\scrP_{\br^{(t)}}\transpose/n^2$, and consists of the points at which the corresponding quadratic form is at most $\chi^2_{d_r,1-a}$.

\subsection{One-step estimator}
\label{sub:one_step_estimator}

Computing the MLE requires iterating over all $n(m + d) + md(d + 1)/2$ parameters until convergence, and any nonlinear optimization algorithm may stop at a local maximum and be sensitive to the initialization because the log-likelihood function is nonconcave. We now describe a one-step estimator that only requires the $m$ layer-wise MLEs, each of which is a single-layer problem with $n + nd + d(d + 1)/2$ parameters, followed by one vertex-wise Newton step. For each $t\in[m]$, let
\begin{align}
\label{eqn:layer_wise_MLE_main}
(\widetilde{\balpha}^{(t)}, \widetilde{\bX}^{(t)}, \overline{\bR}^{(t)}) = \argmax\sum_{i < j}\ell(\alpha_i^{(t)} + \alpha_j^{(t)} + \bx_i\transpose\bR^{(t)}\bx_j; A_{ij}^{(t)})
\end{align}
denote the MLE based on the $t$th layer alone, where the maximum is taken over the parameter space $\Psi$ in \eqref{eqn:psi_parameter_space} with $m = 1$. For the single-layer model, the MLE can be computed by a projected gradient descent algorithm warm-started at a singular value thresholding estimator, with a convergence guarantee \citep{Tian_Sun_He_2026_Bridging}. The estimated subspaces $(\widetilde{\bX}^{(t)})_{t = 1}^m$ are aggregated by the leading eigenvectors of the average of the projection matrices: let $\widetilde{\bU}\in\mathbb{O}(n, d)$ be the eigenvector matrix of $(1/m)\sum_{t = 1}^m\widetilde{\bX}^{(t)}\widetilde{\bX}^{(t)\mathrm{T}}$ corresponding to its $d$ largest eigenvalues, and let $\widetilde{\bX} = [\widetilde{\bx}_1,\ldots,\widetilde{\bx}_n]\transpose = \sqrt{n}\widetilde{\bU}$. The layer-wise interaction matrices are then rotated onto the aggregated subspace, $\widetilde{\bR}^{(t)} = (1/n^2)\widetilde{\bX}\transpose\widetilde{\bX}^{(t)}\overline{\bR}^{(t)}\widetilde{\bX}^{(t)\mathrm{T}}\widetilde{\bX}$. The resulting preliminary estimator is further refined by a single Newton step for each vertex block $[\alpha_i^{(1)},\ldots,\alpha_i^{(m)},\bx_i\transpose]\transpose$ that holds the 
remaining parameters
fixed. 
Formally, for any $t\in[m]$, $i,j\in[n]$, let $\widetilde{\balpha}^{(t)} = [\widetilde{\alpha}_1^{(t)},\ldots,\widetilde{\alpha}_n^{(t)}]\transpose$, $\widetilde{\theta}_{ij}^{(t)} = \widetilde{\alpha}_i^{(t)} + \widetilde{\alpha}_j^{(t)} + \widetilde{\bx}_i\transpose\widetilde{\bR}^{(t)}\widetilde{\bx}_j$, $\widetilde{\balpha}_i = [\widetilde{\alpha}_i^{(1)},\ldots,\widetilde{\alpha}_i^{(m)}]\transpose$, and define the one-step Newton update
    \begin{align*}
    \begin{bmatrix}
      \widehat{\balpha}_i^{(\textsf{OS})}\\
      \widetilde{\bx}_i^{(\textsf{OS})}
    \end{bmatrix}
    & =
     \begin{bmatrix}
      \widetilde{\balpha}_i\\
      \widetilde{\bx}_i
    \end{bmatrix}
     -
     \left\{
     \frac{1}{n}\sum_{t = 1}^m\sum_{j\in[n]\backslash\{i\}}\ell''(\widetilde{\theta}_{ij}^{(t)}; A_{ij}^{(t)})\begin{bmatrix}
       \be_t\be_t\transpose & \be_t\widetilde{\bx}_j\transpose\widetilde{\bR}^{(t)}\\
       \widetilde{\bR}^{(t)}\widetilde{\bx}_j\be_t\transpose & \widetilde{\bR}^{(t)}\widetilde{\bx}_j\widetilde{\bx}_j\transpose\widetilde{\bR}^{(t)}
     \end{bmatrix}
     \right\}^{-1}\\
     &\quad\times \frac{1}{n}\sum_{t = 1}^m\sum_{j\in[n]\backslash\{i\}}\ell'(\widetilde{\theta}_{ij}^{(t)}; A_{ij}^{(t)})\begin{bmatrix}
      \be_t\\
      \widetilde{\bR}^{(t)}\widetilde{\bx}_j
     \end{bmatrix}.
\end{align*}
Then, denote $\widetilde{\bX}^{(\textsf{OS})} = [\widetilde{\bx}_1^{(\textsf{OS})},\ldots,\widetilde{\bx}_n^{(\textsf{OS})}]\transpose$ and define the one-step estimator $\widehat{\bX}^{(\textsf{OS})} = (\eye_n - \one_n\one_n\transpose/n)\widetilde{\bX}^{(\textsf{OS})}$. Denote its rows by $\widehat{\bx}_1^{(\textsf{OS})},\ldots,\widehat{\bx}_n^{(\textsf{OS})}$.

\begin{theorem}\label{thm:asymptotic_normality_OSE}
Suppose Assumptions \ref{assumption:edge_distribution}--\ref{assumption:true_parameter} hold and $m^2(\log n)^2/n \to 0$ as $n\to\infty$.  Then, with the constant $M$ of the parameter space $\Psi$ of the layer-wise MLEs in \eqref{eqn:layer_wise_MLE_main} chosen as in Theorem~\ref{thm:asymptotic_normality_MLE}, the following hold:

\begin{enumerate}[(i)]
  \item Let $\calI = \{i_1,\ldots,i_q\}\subset[n]$ be a set of fixed vertices, where $q$ is fixed. For each $i\in\calI$, let $\bg_i = [\bg_{\balpha_i}\transpose,\bg_{\bx_i}\transpose]\transpose\in\mathbb{R}^{m + d}$ be a deterministic unit vector with $\bg_{\balpha_i}\in\mathbb{R}^m$ and $\bg_{\bx_i}\in\mathbb{R}^d$. Suppose $\|\bg_i\|_1 = O(1)$ for all $i\in\calI$. Then, there exists a random orthogonal matrix $\bW\in\mathbb{O}(d)$, such that
  \begin{align*}
  \frac{\sum_{i\in\calI}\{\sqrt{n}\bg_{\balpha_i}\transpose(\widehat{\balpha}_i^{(\textsf{OS})} - \balpha_{0i})
  + \sqrt{mn}\bg_{\bx_i}\transpose(\bW\transpose\widehat{\bx}_i^{(\textsf{OS})} - \bx_{0i})\} }{\sqrt{\sum_{i\in\calI}\bg_i\transpose (\bSigma_{\bz_i}^\sharp)^{-1}\bg_i}}\overset{\calL}{\to}\mathrm{N}(0, 1)\quad\text{as }n\to\infty.
  \end{align*}

  \item Let $\bF\in\mathrm{Sym}(n)$ be deterministic with $\|\bF\bX_0\|_{\mathrm{F}} = 1$, $\|\bX_0\transpose\bF\bX_0\|_2 = O(1)$, and $\|\vect(\bF)\|_1 = O(1)$. Then
  \begin{align*}
  \frac{\sqrt{mn}\langle\bF, \widehat{\bX}^{(\textsf{OS})}\widehat{\bX}^{(\textsf{OS})\mathrm{T}} - \bX_0\bX_0\transpose\rangle_{\mathrm{F}}}{\sqrt{4\sum_{i = 1}^n\be_i\transpose\bF\bX_0
  \bSigma_{\bx_i}^\sharp\bX_0\transpose\bF\be_i}}
  \overset{\calL}{\to}\mathrm{N}(0, 1)\quad\text{as }n\to\infty.
  \end{align*}
\end{enumerate}
\end{theorem}

Plug-in inference based on the one-step estimator proceeds similarly as in Section~\ref{sub:plugin_inference}. In practice, the one-step estimator and the associated plug-in inference rely on all $m$ layer-wise MLEs being computed successfully, so that the aggregated subspace estimator $\widetilde{\bX}$ is consistent; its finite-sample performance is reported in the Supplementary Material.

\section{Mouse connectomes}
\label{sec:mouse}

\subsection{Data and model}
\citet{wang2020variability} imaged the brains of 32 mice at microscopic resolution with diffusion MRI: eight animals, four of each sex, from each of four inbred strains, C57BL/6J (B6), DBA/2J, CAST/EiJ, and BTBR T$^+$ \emph{Itpr3}$^{tf}$/J (BTBR). Each brain was registered to a common atlas of 166 regions per hemisphere, and tractography produced, for every pair of the $n=332$ regions, a count of reconstructed streamlines. We use the connectomes as distributed with the \texttt{graspologic} package \citep{chung2019graspy}. Each mouse is a layer, so independence across layers is a consequence of the design.
The BTBR strain is a natural test case because it lacks the corpus callosum and has a severely reduced hippocampal commissure \citep{wahlsten2003survey}, so its two hemispheres should be less connected to each other than in the other strains.

The counts are heavy-tailed: \MouseZeroPct\% of region pairs have no streamline, and the largest counts exceed $10^5$. We therefore model $A_{ij}^{(t)}=\log(1+N_{ij}^{(t)})$ with the Gaussian working likelihood of Example~\ref{example:canonical_families}, with $\sigma^2$ estimated by the residual mean square. A binary alternative thresholded by $\mathbf{1}\{N_{ij}^{(t)}>100\}$ is not suitable for likelihood inference: some regions have no such edge in some mice, so they become isolated vertices and their degree effects diverge.
The degree effects $\alpha_i^{(t)}$ absorb differences across regions and mice in overall tract density, including differences in region volume and in tractography sensitivity, while the interaction matrix describes which regions connect more than their degree effects can account for.

\subsection{Dimension and fit}
Table~\ref{tab:mouse_cv} reports held-out error when \MouseCvFrac\% of region pairs are removed from all 32 layers at once, so that a held-out pair is never observed in any mouse. Degree effects alone leave most of the structure unexplained. A shared subspace of dimension $d=3$ halves the error, and the error continues to decrease up to $d=6$. Separate three-dimensional subspaces for each mouse, with 32 times as many latent coordinates, predict about as well as a shared subspace of dimension five. The one-step estimate is slightly less accurate than the MLE at $d=3$, and in one of the three random holdouts, one of its 32 single-mouse fits did not achieve convergence. Prediction alone does not select $d$, and we report the analysis for $d=3$ and check that the conclusions hold for $d=4,5,6$ (see Table~\ref{tab:mouse_d}). The fitted degree effects imply substantial node-by-layer interaction. We fit the two-way decomposition $\widehat\alpha_i^{(t)} = \mu + a_i + b_t + e_{it}$, with region effects $a_i$ and mouse effects $b_t$, by least squares; because the table is complete, this double-centers the $332\times32$ matrix $[\widehat\alpha_i^{(t)}]$ by its row and column means, and the residual matrix $\widehat\bE = [\widehat e_{it}]$ has rank at most \MouseAlphaRankMax. Its squared Frobenius norm is \MouseAlphaInteractPct\% of the total sum of squares $\sum_{i,t}(\widehat\alpha_i^{(t)} - \bar\alpha)^2$, where $\bar\alpha$ is the grand mean. This is small in proportion but about \MouseAlphaSignalNoise\ times the estimation noise $\widehat N = \sum_{i,t}\widehat\var(\widehat\alpha_i^{(t)})$, the sum of the plug-in variances of Section~\ref{sub:plugin_inference}, and $\widehat\bE$ is not of low rank: with singular values $s_1\geq s_2\geq\cdots$, its three leading components account for \MouseAlphaInteractRankThree\% of $\|\widehat\bE\|_{\mathrm{F}}^2$, and the residual $\sum_{l > k}s_l^2$ of its best rank-$k$ approximation falls below $\widehat N$ only for $k\geq\MouseAlphaRankNeeded$.
This is the structure that the unrestricted degree effects of model~\eqref{eqn:model} accommodate.

\begin{table}[t]
\caption{Held-out mean squared error of $\log(1+N_{ij}^{(t)})$ for the mouse connectomes, averaged over three random holdouts of \MouseCvFrac\% of region pairs (removed from all mice). ``Shared'' is model~\eqref{eqn:model}; ``separate'' fits each mouse alone with $d=3$; ``one-step'' is the one-step estimate defined in Section~\ref{sub:one_step_estimator}. The total variance of the response is \MouseVarY.}
\label{tab:mouse_cv}
\centering\small
\begin{tabular}{lcccc}
\toprule
Model & $d$ & Held-out MSE & Splits & Median seconds \\
\midrule
Degree effects only & 0 & 4.787 & 3 & 3 \\
Shared & 2 & 2.522 & 3 & 18 \\
Shared & 3 & 2.287 & 3 & 66 \\
Shared & 4 & 2.099 & 3 & 113 \\
Shared & 5 & 1.947 & 3 & 98 \\
Shared & 6 & 1.846 & 3 & 208 \\
Separate per mouse & 3 & 1.927 & 3 & 85 \\
One-step & 3 & 2.332 & 3 & 178 \\
\bottomrule
\end{tabular}

\end{table}

For the case of $d=3$, the fit converges to an interior stationary point. Three runs of the block Fisher scoring algorithm, one started from the link-transformed spectral estimate and two from perturbations of the best fit (both defined in Section~I.3 of the Supplementary Material), reached the same log-likelihood value. 
The residual variance is $\widehat\sigma^2=\MouseSigma$, which is used as the estimated variance in the Gaussian working model, against a total variance of \MouseVarY. The direction in the fitted latent subspace $\mathrm{Span}(\widehat{\bX})$ that is most correlated with the hemisphere indicator $\bh\in\{\pm1\}^{n}$ has correlation \MouseHemiCorr\ with $\bh$, where the $i$th element of $\bh$ takes value $+1$ if the $i$th region of interest is in the left hemisphere and $-1$ otherwise. Note that hemisphere labels are not used in the fit. To check the Gaussian working variance, we compared the model-based standard errors of the interaction entries with heteroskedasticity-robust standard errors derived from the sandwich covariance matrix $\bOmega(\widehat{\bpsi})\widetilde{\bLambda}_\sharp\bar{\bS}(\widehat{\bpsi})\widetilde{\bLambda}_\sharp\bOmega(\widehat{\bpsi})$, where $\bar{\bS}(\bpsi) = \sum_{t = 1}^m\sum_{i < j}\{\ell'(\theta_{ij}^{(t)}; A_{ij}^{(t)})\}^2(\partial\theta_{ij}^{(t)}/\partial\bpsi)(\partial\theta_{ij}^{(t)}/\partial\bpsi\transpose)$ replaces the expected squared scores in the Fisher information by their empirical versions, and
their ratio has median \MouseSandMed\ and range \MouseSandRange. We report model-based standard errors below since sandwich versions lead to the same conclusions.

\subsection{Hemispheric segregation}
\label{sub:mouse_hemi}
Let $\bh\in\{-1,1\}^n$ indicate the hemisphere of each region. The rotation-invariant coefficient
$\rho_t={\bh\transpose\bX\bR^{(t)}\bX\transpose\bh}/{\|\bX\transpose\bh\|_2^2}$
is the interaction of mouse $t$ along the direction of the latent space most aligned with the hemispheric contrast. A larger $\rho_t$ means that, after adjustment for degree effects, same-hemisphere pairs are more strongly connected relative to cross-hemisphere pairs. Because $\rho_t$ is invariant under $(\bX,\bR^{(s)})\mapsto(\bX\bW,\bW\transpose\bR^{(s)}\bW)$, no reference orientation is needed. We evaluate $\rho_t$ at $\widehat\bX$ and the debiased interaction matrix, whose half-vectorization is $\vech(\widehat\bR^{(t)}) - \bb_n^{(t)}(\widehat\bpsi)/n$, and estimate its variance by the delta method, $\widehat{\var}(\widehat\rho_t) = (\partial\rho_t/\partial\bpsi\transpose)\widetilde{\bLambda}_\sharp\bOmega(\widehat\bpsi)\widetilde{\bLambda}_\sharp(\partial\rho_t/\partial\bpsi)$ with the derivatives evaluated at $\widehat\bpsi$; contrasts of $\rho_t$ across mice are handled in the same way. With $\bX$ held fixed, $\rho_t$ is a fixed linear combination of the entries of $\bR^{(t)}$, so this part of the error, of order $n^{-1}$, is covered by Corollary~\ref{corr:plug_in_R}. The variance also includes the contribution of the estimated subspace through the subspace blocks of $\bOmega(\widehat\bpsi)$, which is heuristically of order $n^{-1}m^{-1/2}$. This contribution rests on a formal first-order expansion rather than on our theorems, so we checked it by a parametric bootstrap at the dimensions of the data (Supplementary Material): the 95\% intervals for $\rho_t$ and for the strain contrast below had empirical coverage \McalRhoLayer\ and \McalRho, and the latter is unchanged when the subspace contribution is omitted.

\begin{figure}[t]
\centering\includegraphics[width=\linewidth]{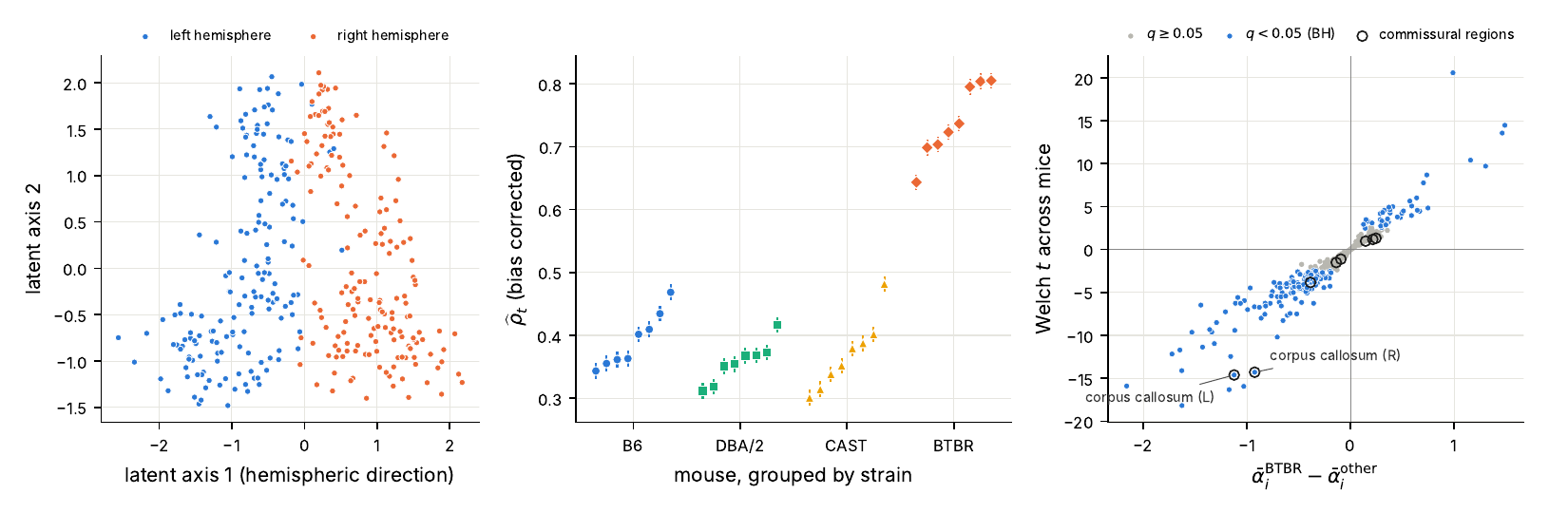}
\caption{Mouse connectomes, shared model with $d=3$. Left: fitted latent positions of the 332 regions on the first two axes, colored by hemisphere; the first axis is the direction most aligned with the hemispheric contrast. Center: debiased hemispheric coefficient $\widehat\rho_t$ for each mouse with 95\% confidence intervals (most intervals are shorter than the symbols). Right: strain contrast of degree effects, BTBR minus the mean of the other strains, for each region against its Welch $t$ statistic across mice; commissural regions are circled and the corpus callosum is labeled.}
\label{fig:mouse}
\end{figure}

Figure~\ref{fig:mouse} (center) shows the estimates. The BTBR mice have $\widehat\rho_t$ near \MouseRhoBTBR, about twice the means of the other three strains, which range from \MouseRhoOtherMin\ to \MouseRhoOtherMax. The within-mouse standard error, about \MouseRhoSE, is an order of magnitude smaller than the spread between mice of the same strain, \MouseRhoBetweenMin--\MouseRhoBetweenMax. For the 32 mice in hand, the BTBR-minus-other contrast, defined as the average of $\widehat\rho_t$ over the eight BTBR mice minus the average over the 24 mice of the other three strains, is \MouseContrast\ with standard error \MouseContrastSE. Treating mice as replicates of their strain, Welch's test of the same contrast gives a $t$-statistic $\MouseWelchT$ ($p<10^{-6}$).
The conclusion does not depend on $d$ (Table~\ref{tab:mouse_d}). It is the expected consequence of a missing corpus callosum: with fewer cross-hemisphere tracts, the hemispheric axis carries more of the connectivity.

\begin{table}[t]
\caption{Sensitivity of the hemispheric coefficient to the latent dimension. Strain-specific means of $\widehat\rho_t$, the BTBR-minus-other contrast with its standard error, Welch's $t$ statistic across mice, and the median absolute estimated bias relative to the standard error.}
\label{tab:mouse_d}
\centering\small
\begin{tabular}{ccccccccc}
\toprule
& \multicolumn{4}{c}{Mean of $\widehat\rho_t$} & \multicolumn{2}{c}{BTBR $-$ other} & Welch & Bias/SE \\
\cmidrule(lr){2-5}\cmidrule(lr){6-7}
$d$ & B6 & DBA/2 & CAST & BTBR & Estimate & SE & $t$ & median \\
\midrule
3 & 0.392 & 0.358 & 0.369 & 0.739 & 0.366 & 0.0026 & 16.1 & 0.018 \\
4 & 0.357 & 0.313 & 0.348 & 0.774 & 0.435 & 0.0025 & 18.1 & 0.009 \\
5 & 0.356 & 0.316 & 0.349 & 0.768 & 0.427 & 0.0024 & 18.6 & 0.007 \\
6 & 0.356 & 0.317 & 0.350 & 0.769 & 0.428 & 0.0023 & 18.6 & 0.007 \\
\bottomrule
\end{tabular}

\end{table}

The estimated bias is small in this application, with a median of \MouseBiasSE\ standard errors for the interaction entries. 
Within-mouse uncertainty is far smaller than variation between mice: Wald statistics for $\bR^{(s)}=\bR^{(t)}$ based on Corollary~\ref{corr:plug_in_R} exceed the Bonferroni-corrected critical value for every pair of mice, with medians \MouseWaldWithin\ for pairs of the same strain and \MouseWaldBetween\ for pairs of different strains, on six degrees of freedom. Under the independent-edge working model, individual connectomes thus differ even within an inbred strain, and differ more between strains.

\subsection{Regional degree effects}
Theorem~\ref{thm:asymptotic_normality_MLE} gives intervals for contrasts of degree effects of a single region across mice. For each region $i\in[332]$, we estimate the BTBR-minus-other degree effect contrast $\bar\alpha_i^{\rm BTBR}-\bar\alpha_i^{\rm other}$ of strain averages. Its sampling variance based on the 32 mice is small: for each region $i$, we compute the ratio of the model-based variance of a single $\widehat\alpha_i^{(t)}$ to the variance of $\widehat\alpha_i^{(t)}$ between mice of the same strain, and the median of this ratio over the 332 regions is \MouseMeasVarPct\%. Strain-level conclusions are therefore driven by biological variation, and we test them with Welch's test across mice and the Benjamini--Hochberg procedure. \MouseAlphaQ\ of the 332 regions have $q<0.05$.
The left and right corpus callosum are among the ten regions with the largest absolute values of the $t$-statistic due to lower degree effect values in BTBR ($t=\MouseCCLt$ and $\MouseCCRt$). Other regions with large absolute values of the $t$-statistic include the lateral ventricles and primary visual cortex (lower in BTBR) and cingulate cortex (higher). The degree effect of a region summarizes the strength of all of its tracts, so a lower value for the corpus callosum, the commissural white matter itself, directly reflects the missing commissure.

\subsection{Limitations}
The model treats region pairs as conditionally independent given the parameters, but tractography counts for pairs sharing a region are computed from overlapping fiber reconstructions. The sandwich check addresses heteroskedasticity but may be insufficient to address this dependence. The common-subspace assumption is an approximation in terms of the held-out prediction performance, as the comparison with separate subspaces shows in Table \ref{tab:mouse_cv} with $d = 3$, so $\bR^{(t)}$ should be read as the interaction of mouse $t$ projected on the shared geometry. Finally, with only eight mice per strain, strain-level inference rests on few replicates: it supports comparisons of a few prespecified functionals such as $\rho_t$, but the analysis of the 332 regional degree effects is exploratory.

\section{Discussion}
\label{sec:discussion}

We have developed likelihood inference for multilayer latent space networks in which a shared invariant subspace coexists with layer-specific interaction matrices and unrestricted node-by-layer degree effects. By introducing the generalized Cayley transform that reparameterizes the latent subspace manifold, we show that the degree effects, the interaction matrices, and the shared latent subspace are informed by different amounts of information under the intrinsic Fisher information geometry.
Leveraging this tool, we further show that the MLE of
the vertex parameters is asymptotically normal and Neyman orthogonal to the remaining parameters, whereas the MLE of an interaction matrix is not Neyman orthogonal to the nuisance parameters and is asymptotically normal with a non-negligible second-order bias, of the order of its standard error, that arises from the estimation of the nuisance parameters and calls for plug-in debiased inference with the nuisance-adjusted covariance.
Application of our proposed methodology to a real-world mouse connectome analysis illustrates what the model offers: a hemispheric axis recovered without using labels in the fit, a rotation-invariant interaction coefficient that separates the strain lacking a corpus callosum with uncertainty quantification, and regional degree effects whose strain differences are among the largest in the corpus callosum.

Several extensions are open. Our growth conditions on $m$ are sufficient and may not be sharp, and simulation studies in the Supplementary Material suggest that, at least for the Gaussian working model, calibration holds well beyond them. The condition of \citet{liu2026unfolding} for their unfolding estimator is weaker, and sharpening ours for the joint MLE is an open nontrivial problem. Inference for a growing number of layers simultaneously, such as a test of homogeneity across all $m$ layers, is not covered by our results. The theory assumes dense layers with bounded
parameters. In sparse binary layers, isolated or nearly isolated vertices make the unrestricted MLE fail to exist, 
and a penalty on the degree effects, with a corresponding sparse-regime theory, is needed. The one-step estimator inherits the numerical instability of the single-layer fits when some of the layer-wise maximum likelihood optimizations do not converge, and a damped or pooled initialization may make it usable for smaller $n$ and comparatively sparse layers. Finally, the model treats the layers as independent, which holds for different animals but not for networks observed repeatedly over time. Dynamic extensions would need their own limit theory.

\section*{Supplementary Material}
The Supplementary Material contains the proofs of all theorems and corollaries, the implementation of the bias formula, computational details, the simulation studies, analyses of a three-layer \emph{C.~elegans} connectome, CDC syphilis surveillance profile networks, and a synthetic partnership network, and a reproducibility package with Python code, the public data with checksums, replicate-level results, and scripts that regenerate every table, figure, and number in the paper.

\section*{Data Availability Statement}
The mouse connectome data of \citet{wang2020variability} are publicly available in the \texttt{graspologic} package \citep{chung2019graspy}. The data and code that reproduce all results are included in the Supplementary Material.

\ifblinded
\else
\section*{Acknowledgments}
H.-H.~Huang acknowledges support from the National Science Foundation under grants DMS-1924792 and DMS-2318925. Both authors acknowledge support from the National Institutes of Health (NIH), National Institute of Allergy and Infectious Diseases (NIAID) R01 award R01AI206603. The authors acknowledge the interactive use of generative AI (Anthropic models
Claude Opus 5.5 and Fable 5.2) for checking the mathematical correctness, including the calculations and proofs of the propositions, theorems, corollaries, and auxiliary lemmas. The authors have verified the mathematical contents and assume full responsibility for the contents of the work. 

\section*{Disclosure statement}
The authors report that there are no competing interests to declare.
\fi

\clearpage
\begin{center}
{\LARGE Supplementary Material}
\end{center}
\bigskip

\noindent 
The Supplementary Material contains two parts: the proofs of the theoretical results in the main text (Sections \ref{sec:intrinsic_geometry_of_multilayer_latent_space_network_model}--\ref{sec:technical_lemmas}), and additional implementation details, including the relevant computation formulas, algorithms, additional numerical examples, and reproducibility (Sections \ref{sec:implementation_audit}--\ref{sec:reproducibility}). 
\begin{itemize}
    \item \textbf{Part I:} Section~A develops the intrinsic geometry and proves Proposition~\ref{prop:expected_hessian_bound}. Sections~B and~C establish the local spectral bounds and the consistency of the MLE with respect to the Frobenius norm metric and the two-to-infinity norm metric, and Section~D proves Theorems~\ref{thm:two_to_infinity_norm_bound_MLE}--\ref{thm:asymptotic_normality_MLE}. Section~E proves Theorem~\ref{thm:asymptotic_normality_R}, Section~F proves the consistency of the plug-in estimators and Corollaries~\ref{corr:plug_in_vertex}--\ref{corr:plug_in_R}, and Section~G proves Theorem~\ref{thm:asymptotic_normality_OSE}. Section~H collects technical lemmas that are used throughout the proof. 

    \item \textbf{Part II:} Sections~I--M contain computational details, additional simulations, details of the mouse analysis, two further applications and a synthetic example, and a description of the reproducibility package.
\end{itemize}

\appendix
\setcounter{table}{0}
\setcounter{figure}{0}
\renewcommand{\thetable}{S\arabic{table}}
\renewcommand{\thefigure}{S\arabic{figure}}
\renewcommand{\theHtable}{S\arabic{table}}   
\renewcommand{\theHfigure}{S\arabic{figure}}

\addtocontents{toc}{\protect\setcounter{tocdepth}{3}} 
\tableofcontents

\allowdisplaybreaks

\paragraph*{Additional notation for the Supplementary Material.} In addition to the notation introduced in Section~\ref{sub:organization} of the main text, the following notation is used throughout the Supplementary Material. For any two matrices $\bA,\bB$ of the same size, let $\bA\circ\bB$ denote the elementwise product between $\bA$ and $\bB$. For indices $i,j\in[n]$, let $\delta_{ij} = \mathbf{1}(i = j)$ denote the Kronecker delta. Let $\bD_d$ denote the duplication matrix such that $\bD_d\vech(\bA) = \vect(\bA)$ for any $\bA\in\mathrm{Sym}(d)$, and $\bK_{d_1d_2}$ denote the commutation matrix such that $\vect(\bM\transpose) = \bK_{d_1d_2}\vect(\bM)$ for any $\bM\in\mathbb{R}^{d_1\times d_2}$. Throughout, we use $C$ to denote constants that may change from line to line that may depend on the model constants but do not change with the varying indices $m,n$. For nonnegative deterministic sequences $(a_n)_{n = 1}^\infty, (b_n)_{n = 1}^\infty$, we write $a_n\lesssim b_n$ or $b_n\gtrsim a_n$ if $a_n = O(b_n)$, and $a_n = o(b_n)$ if $a_n/b_n\to 0$ as $n\to\infty$. Given a sequence of events $(\calE_n)_{n = 1}^\infty$, we say that $\calE_n$ occurs with high probability (w.h.p.), if for any $c > 0$, there exist constants $N_c,K_c > 0$, possibly depending on $c$, such that $\prob(\calE_n) \geq 1 - K_cn^{-c}$ for all $n\geq N_c$. More generally than the $\Optilde(\cdot)$ notation in the main text, given $N$ sequences of random matrices $\{(\bX_{\alpha n})_{n = 1}^\infty:\alpha\in[N]\}$ (where $N$ can possibly grow with $n$) and a deterministic nonnegative sequence $(\eps_n)_{n = 1}^\infty$, we say that $\{(\bX_{\alpha n})_{n = 1}^\infty:\alpha\in[N]\}$ are uniformly $\Optilde(\eps_n)$, if for any $c > 0$, there exist constants $N_c,K_c,M_c > 0$ that only depend on $c$ but do not depend on $\alpha\in[N]$, such that $\prob(\|\bX_{\alpha n}\|_2 > M_c\eps_n)\leq K_c n^{-c}$ for any $n\geq N_c$ and any $\alpha\in[N]$.

\section{Intrinsic geometry of multilayer latent space network model}
\label{sec:intrinsic_geometry_of_multilayer_latent_space_network_model}

This section presents results that characterize the intrinsic geometry of the multilayer latent space network model \eqref{eqn:model} and proves Proposition \ref{prop:expected_hessian_bound} in the main text. We first present the matrix calculus of the generalized Cayley transform \eqref{eqn:Cayley_transform} as well as its Lipschitz properties. 

\subsection{Matrix calculus of the generalized Cayley transform}
\label{sub:jacobian_of_cayley_transform}

We begin with the Jacobian of the generalized Cayley transform. Recall that $\bS(\bB) = \eye_d + \bB\transpose\bB$. We first define the Lyapunov operator $\mathscr{L}$ over $\mathrm{Sym}(d)$ associated with $\bS(\bB)^{1/2}$ as $\mathscr{L}(\bM):=\bS(\bB)^{1/2}\bM + \bM\bS(\bB)^{1/2}$. Theorem VII.2.1 in \cite{bhatia2013matrix} guarantees that $\scrL$ is invertible and its inverse can be obtained via the integral operator described in Lemma \ref{lemma:inverse_Lyapunov_operator_integral}. Denote $B_{pq}$ the $(p, q)$th element of $\bB$. 
\begin{lemma}\label{Lemma:dsqrtB_dB}
For any $(p, q)\in[n - 1 - d]\times [d]$,
\begin{align*}
\frac{\partial}{\partial B_{pq}}\bS(\bB)^{-1/2} = -\bS(\bB)^{-1/2}\scrL^{-1}(\be_q\be_p\transpose\bB + \bB\transpose\be_p\be_q\transpose)\bS(\bB)^{-1/2}.
\end{align*}
\end{lemma}

\begin{proof}[\bf Proof]
By definition, $\bS(\bB)^{-1} = \bS(\bB)^{-1/2}\bS(\bB)^{-1/2}$, so that
\begin{align*}
\frac{\partial}{\partial B_{pq}}\bS(\bB)^{-1}
& = -\bS(\bB)^{-1}\frac{\partial}{\partial B_{pq}}\bS(\bB) \bS(\bB)^{-1}
  = -\bS(\bB)^{-1}(\be_q\be_p\transpose\bB + \bB\transpose\be_p\be_q\transpose) \bS(\bB)^{-1}
\\
& = \frac{\partial}{\partial B_{pq}}\bS(\bB)^{-1/2}\bS(\bB)^{-1/2} + \bS(\bB)^{-1/2}\frac{\partial}{\partial B_{pq}}\bS(\bB)^{-1/2}.
\end{align*}
This implies that
\begin{align*}
	\bigg\{\frac{\partial}{\partial B_{pq}}\bS(\bB)^{-1/2}\bigg\}\bS(\bB)^{1/2} + \bS(\bB)^{1/2}\bigg\{\frac{\partial}{\partial B_{pq}}\bS(\bB)^{-1/2}\bigg\}
	= -\bS(\bB)^{-1/2}(\be_q\be_p\transpose\bB + \bB\transpose\be_p\be_q\transpose)\bS(\bB)^{-1/2}.
\end{align*}
This implies that 
\begin{align*}
	\frac{\partial}{\partial B_{pq}}\bS(\bB)^{-1/2}
	& = \scrL^{-1}\left(-\bS(\bB)^{-1/2}(\be_q\be_p\transpose\bB + \bB\transpose\be_p\be_q\transpose)\bS(\bB)^{-1/2}\right)\\
	& = -\bS(\bB)^{-1/2}\scrL^{-1}(\be_q\be_p\transpose\bB + \bB\transpose\be_p\be_q\transpose)\bS(\bB)^{-1/2},
\end{align*}
where the second equality is due to Lemma \ref{lemma:inverse_Lyapunov_operator_integral}. This completes the proof. 
\end{proof}

\begin{proposition}\label{prop:Cayley_transform_jacobian}
The Jacobian of the generalized Cayley transform \eqref{eqn:Cayley_transform} is defined by the following derivatives:
\begin{align*}
\frac{\partial\bX}{\partial B_{pq}}(\bB)
& = \bX_{0\perp}\be_p\be_q\transpose\bS(\bB)^{-1/2}
 - \bX(\bB)\mathscr{L}^{-1}(\bB\transpose\be_p\be_q\transpose + \be_q\be_p\transpose\bB)\bS(\bB)^{-1/2},
\end{align*}
where $(p, q)\in[n - 1 - d]\times [d]$. In particular, at $\bB = \mathbf{O}$, we have
\begin{align}
\label{eqn:dX_dB_zero}
\frac{\partial\bX}{\partial B_{pq}}(\mathbf{O}) = \bX_{0\perp}\be_p\be_q\transpose. 
\end{align}
\end{proposition}

\begin{proof}[\bf Proof]
By definition and Lemma \ref{Lemma:dsqrtB_dB}, 
\begin{align*}
\frac{\partial\bX}{\partial B_{pq}}(\bB)
& = \frac{\partial}{\partial B_{pq}}(\bX_0 + \bX_{0\perp}\bB)\bS(\bB)^{-1/2} + (\bX_0 + \bX_{0\perp}\bB)\frac{\partial}{\partial B_{pq}}\bS(\bB)^{-1/2}\\
& = \bX_{0\perp}\be_p\be_q\transpose\bS(\bB)^{-1/2} -
(\bX_0 + \bX_{0\perp}\bB)\bS(\bB)^{-1/2}\scrL^{-1}(\be_q\be_p\transpose\bB + \bB\transpose\be_p\be_q\transpose)\bS(\bB)^{-1/2}\\
& = \bX_{0\perp}\be_p\be_q\transpose\bS(\bB)^{-1/2} -
\bX(\bB)\scrL^{-1}(\be_q\be_p\transpose\bB + \bB\transpose\be_p\be_q\transpose)\bS(\bB)^{-1/2},
\end{align*}
which is a direct calculation. This completes the proof. 
\end{proof}

\begin{proposition}\label{prop:Cayley_transform_hessian}
The second-order derivatives of the Cayley transform \eqref{eqn:Cayley_transform} are given by
\begin{align*}
\frac{\partial^2\bX}{\partial B_{pq} \partial B_{rs}}(\bB)
& = -\bX_{0\perp}\be_p\be_q\transpose\bS(\bB)^{-1/2}\mathscr{L}^{-1}(\dot{\bS}_{rs})\bS(\bB)^{-1/2}\\
&\quad -\bX_{0\perp}\be_r\be_s\transpose\bS(\bB)^{-1/2}\mathscr{L}^{-1}(\dot{\bS}_{pq})\bS(\bB)^{-1/2}\\
&\quad + \bX(\bB)
\mathscr{L}^{-1}(\dot{\bS}_{pq})\bS(\bB)^{-1/2}\mathscr{L}^{-1}(\dot{\bS}_{rs})\bS(\bB)^{-1/2}\\
&\quad + \bX(\bB)
\mathscr{L}^{-1}(\dot{\bS}_{rs})\bS(\bB)^{-1/2}\mathscr{L}^{-1}(\dot{\bS}_{pq})\bS(\bB)^{-1/2}\\
&\quad - \bX(\bB)\mathscr{L}^{-1}\left(\ddot{\bS}_{pq,rs}
  - \mathscr{L}^{-1}(\dot{\bS}_{pq})\mathscr{L}^{-1}(\dot{\bS}_{rs})
   - \mathscr{L}^{-1}(\dot{\bS}_{rs})\mathscr{L}^{-1}(\dot{\bS}_{pq})
\right)\bS(\bB)^{-1/2},
\end{align*}
where $(p, q), (r, s)\in[n - 1 - d]\times [d]$, $\dot{\bS}_{pq} = \be_q\be_p\transpose\bB + \bB\transpose\be_p\be_q\transpose$, $\ddot{\bS}_{pq,rs} = \delta_{pr}(\be_q\be_s\transpose + \be_s\be_q\transpose)$. In particular, at $\bB = \mathbf{O}$, we have
\begin{align}
\label{eqn:d2X_dB2_zero}
\frac{\partial^2\bX}{\partial B_{pq} \partial B_{rs}}(\mathbf{O}) = -\frac{1}{2}\delta_{pr}\bX_0(\be_q\be_s\transpose + \be_s\be_q\transpose). 
\end{align}
\end{proposition}

\begin{proof}[\bf Proof]
A direct computation and Lemma \ref{Lemma:dsqrtB_dB} yield
\begin{align*}
\frac{\partial^2\bS(\bB)}{\partial B_{pq}\partial B_{rs}}
& = \delta_{pr}(\be_q\be_s\transpose + \be_s\be_q\transpose),\\
\frac{\partial\bS(\bB)^{-1/2}}{\partial B_{pq}}
& = -\bS(\bB)^{-1/2}\mathscr{L}^{-1}(\dot{\bS}_{pq})\bS(\bB)^{-1/2}
\end{align*}
and
\begin{align*}
\frac{\partial^2\bS(\bB)^{-1/2}}{\partial B_{pq}\partial B_{rs}}
& = \bS(\bB)^{-1/2}
\mathscr{L}^{-1}(\dot{\bS}_{pq})\bS(\bB)^{-1/2}\mathscr{L}^{-1}(\dot{\bS}_{rs})\bS(\bB)^{-1/2}\\
&\quad + \bS(\bB)^{-1/2}\mathscr{L}^{-1}(\dot{\bS}_{rs})\bS(\bB)^{-1/2}\mathscr{L}^{-1}(\dot{\bS}_{pq})\bS(\bB)^{-1/2}\\
&\quad - \bS(\bB)^{-1/2}\mathscr{L}^{-1}\left(\ddot{\bS}_{pq,rs}
  - \mathscr{L}^{-1}(\dot{\bS}_{pq})\mathscr{L}^{-1}(\dot{\bS}_{rs})
   - \mathscr{L}^{-1}(\dot{\bS}_{rs})\mathscr{L}^{-1}(\dot{\bS}_{pq})
\right)\bS(\bB)^{-1/2}
\end{align*}
where
\begin{align*}
\dot{\bS}_{pq} = \be_q\be_p\transpose\bB + \bB\transpose\be_p\be_q\transpose\quad \text{and}\quad
\ddot{\bS}_{pq,rs} = \delta_{pr}(\be_q\be_s\transpose + \be_s\be_q\transpose).
\end{align*}
Consequently, we obtain
\begin{align*}
\frac{\partial^2\bX}{\partial B_{pq}\partial B_{rs}}(\bB)
& = -\bX_{0\perp}\be_p\be_q\transpose\bS(\bB)^{-1/2}\mathscr{L}^{-1}(\dot{\bS}_{rs})\bS(\bB)^{-1/2} -\bX_{0\perp}\be_r\be_s\transpose\bS(\bB)^{-1/2}\mathscr{L}^{-1}(\dot{\bS}_{pq})\bS(\bB)^{-1/2}\\
&\quad + (\bX_0 + \bX_{0\perp}\bB)\frac{\partial^2\bS(\bB)^{-1/2}}{\partial B_{pq}\partial B_{rs}}.
\end{align*}
The proof is completed by invoking the second-order derivative formula of $\bS(\bB)^{-1/2}$ into the last term above. 
\end{proof}

\subsection{Lipschitz properties of the generalized Cayley transform}
\label{sub:lipschitz_properties_of_cayley_transform}

We next investigate the Lipschitz properties of the generalized Cayley transform \eqref{eqn:Cayley_transform}, its Jacobian, and Hessian. The following shorthand notations turn out to be convenient for us to leverage. 
For any $\bB\in\mathbb{R}^{(n - 1 - d)\times d}$ and any matrix-valued function $\bY:\bB\in\mathbb{R}^{(n - 1 - d)\times d}\mapsto \bY(\bB)\in \mathbb{R}^{n_1\times n_2}$ with $n_1,n_2\in\mathbb{N}_+$, define the following $\bB$-dependent multilinear maps
\begin{equation}
\label{eqn:dot_notations}
\begin{aligned}
  &\dot{\bY}(\bB)[\cdot]:\delta\bB\in\mathbb{R}^{(n - 1 - d)\times d}\mapsto \sum_{p = 1}^{n - 1 - d}\sum_{q = 1}^d\frac{\partial\bY}{\partial B_{pq}}(\bB)\be_p\transpose\delta\bB\be_q\in\mathbb{R}^{n_1\times n_2},\\
  &\ddot{\bY}(\bB)[\cdot, \cdot]:(\delta\bB_1,\delta\bB_2)\in\mathbb{R}^{(n - 1 - d)\times d}\times\mathbb{R}^{(n - 1 - d)\times d}\\
  &\qquad\qquad\quad\mapsto \sum_{p_1,p_2 = 1}^{n - 1 - d}\sum_{q_1,q_2 = 1}^d\frac{\partial^2\bY}{\partial B_{p_1q_1}\partial B_{p_2q_2}}(\bB)\be_{p_1}\transpose\delta\bB_1\be_{q_1}\be_{p_2}\transpose\delta\bB_2\be_{q_2}\in\mathbb{R}^{n_1\times n_2},\\
  &\dddot{\bY}(\bB)[\cdot, \cdot, \cdot]:(\delta\bB_1,\delta\bB_2, \delta\bB_3)\in\mathbb{R}^{(n - 1 - d)\times d}\times\mathbb{R}^{(n - 1 - d)\times d}\times\mathbb{R}^{(n - 1 - d)\times d}\\
  &\qquad\qquad\quad\mapsto \sum_{p_1,p_2,p_3 = 1}^{n - 1 - d}\sum_{q_1,q_2,q_3 = 1}^d\frac{\partial^3\bY}{\partial B_{p_1q_1}\partial B_{p_2q_2}\partial B_{p_3q_3}}(\bB)\be_{p_1}\transpose\delta\bB_1\be_{q_1}\be_{p_2}\transpose\delta\bB_2\be_{q_2}\be_{p_3}\transpose\delta\bB_3\be_{q_3}\in\mathbb{R}^{n_1\times n_2}.
\end{aligned}
\end{equation}
With $\bS(\bB) = \eye_d + \bB\transpose\bB$, we immediately obtain
\[
	\dot{\bS}(\bB)[\delta\bB] = \delta\bB\transpose\bB + \bB\transpose\delta\bB,\quad \ddot{\bS}(\bB)[\delta\bB_1,\delta\bB_2] = \delta\bB_1\transpose\delta\bB_2 + \delta\bB_2\transpose\delta\bB_1,\quad \dddot{\bS}(\bB)\equiv\mathbf{O}.
\]

\begin{lemma}\label{lemma:integral_formula_matrix_squarerootinverse}
For any $\bB,\delta\bB_1,\delta\bB_2,\delta\bB_3\in\mathbb{R}^{(n - 1 - d)\times d}$, we have the following integral formula for the derivatives of $\bN(\bB):=\bS(\bB)^{-1/2}$:
\begin{align*}
  \dot{\bN}(\bB)[\delta\bB_1]
  & = -\frac{2}{\pi}\int_0^\infty\bR_\tau\dot{\bS}[\delta\bB_1]\bR_\tau\mathrm{d}\tau,\\
  \ddot{\bN}(\bB)[\delta\bB_1, \delta\bB_2]
  & = \frac{2}{\pi}\int_0^\infty\left(\bR_\tau\dot{\bS}[\delta\bB_1]\bR_\tau\dot{\bS}[\delta\bB_2]\bR_\tau + \bR_\tau\dot{\bS}[\delta\bB_2]\bR_\tau\dot{\bS}[\delta\bB_1]\bR_\tau - \bR_\tau\ddot{\bS}[\delta\bB_1,\delta\bB_2]\bR_\tau\right)\mathrm{d}\tau,\\
  \dddot{\bN}(\bB)[\delta\bB_1, \delta\bB_2,\delta\bB_3]
  & = \frac{2}{\pi}\int_0^\infty\bigg(-\sum_{\iota\in\calS_3}\bR_\tau\dot{\bS}[\delta\bB_{\iota(1)}]\bR_\tau\dot{\bS}[\delta\bB_{\iota(2)}]\bR_\tau\dot{\bS}[\delta\bB_{\iota(3)}]\bR_\tau\bigg)\mathrm{d}\tau\\
  &\quad + \frac{2}{\pi}\int_0^\infty\bigg(\sum_{(j,k,l)\in\{(1,2,3),(1,3,2),(2,3,1)\}}\bR_\tau\ddot{\bS}[\delta\bB_j,\delta\bB_k]\bR_\tau\dot{\bS}[\delta\bB_l]\bR_\tau\bigg)\mathrm{d}\tau\\
  &\quad + \frac{2}{\pi}\int_0^\infty\bigg(\sum_{(j,k,l)\in\{(1,2,3),(1,3,2),(2,3,1)\}}\bR_\tau\dot{\bS}[\delta\bB_l]\bR_\tau\ddot{\bS}[\delta\bB_j,\delta\bB_k]\bR_\tau\bigg)\mathrm{d}\tau,
\end{align*}
where $\bR_\tau:=\{\tau^2\eye_d + \bS(\bB)\}^{-1}$, $\calS_3$ denotes the set of all permutations over $\{1,2,3\}$, and the integrals apply elementwise. Note that the dependency of $\bR_\tau$ and $\dot{\bS}$ on $\bB$ is hidden on the right-hand sides of the equations. Furthermore, we have
\begin{align*}
  \|\dot\bN(\bB)[\delta\bB_1]\|_2&\leq \|\bB\|_2\|\delta\bB_1\|_2,\\
  \|\ddot\bN(\bB)[\delta\bB_1,\delta\bB_2]\|_2&\leq (3\|\bB\|_2^2 + 1)\|\delta\bB_1\|_2\|\delta\bB_2\|_2,\\
  \|\dddot\bN(\bB)[\delta\bB_1,\delta\bB_2,\delta\bB_3]\|_2
  &\leq (15\|\bB\|_2^3 + 9\|\bB\|_2)\|\delta\bB_1\|_2\|\delta\bB_2\|_2\|\delta\bB_3\|_2,
\end{align*}
and when $\|\bB\|_2\leq 1$, there exists some absolute constant $C > 0$ such that
\begin{align*}
  \|\ddddot\bN(\bB)[\delta\bB_1,\delta\bB_2,\delta\bB_3,\delta\bB_4]\|_2
  &\leq C\|\delta\bB_1\|_2\|\delta\bB_2\|_2\|\delta\bB_3\|_2\|\delta\bB_4\|_2,
\end{align*}
\end{lemma}

\begin{proof}[\bf Proof]
The key is the integral representation $\bN = (2/\pi)\int_0^\infty(\tau^2\eye_d + \bS)^{-1}\mathrm{d}\tau$ (see, e.g., \cite{higham2008functions}), which follows from $\int_0^\infty(\tau^2 + \lambda)^{-1}\mathrm{d}\tau = \pi/(2\sqrt{\lambda})$ applied to the eigenvalues of $\bS(\bB)\succeq\eye_d$. By dominated convergence theorem, 
\begin{align*}
  \dot{\bN}[\delta\bB_1]
  & = \frac{\mathrm{d}}{\mathrm{d}s}\bN(\bB + s\delta\bB_1)\mathrel{\bigg|_{s = 0}} = \frac{2}{\pi}\int_0^\infty\frac{\mathrm{d}}{\mathrm{d}s}(\tau^2\eye_d + \bS(\bB + s\delta\bB_1))^{-1}\mathrel{\bigg|_{s = 0}}\mathrm{d}\tau = -\frac{2}{\pi}\int_0^\infty\bR_\tau\dot{\bS}[\delta\bB_1]\bR_\tau\mathrm{d}\tau.
\end{align*}
For $\ddot{\bN}$, we differentiate $\dot{\bN}(\bB + s\delta\bB_2)[\delta\bB_1]$ at $s = 0$ to obtain
\begin{align*}
  \ddot{\bN}[\delta\bB_1, \delta\bB_2]
  & = \frac{\mathrm{d}}{\mathrm{d}s}\dot{\bN}(\bB + s\delta\bB_2)[\delta\bB_1]\mathrel{\bigg|_{s = 0}}\\
  & = -\frac{2}{\pi}\int_0^\infty\frac{\mathrm{d}}{\mathrm{d}s}\bR_\tau(\bB + s\delta\bB_2)\dot{\bS}(\bB + s\delta\bB_2)[\delta\bB_1]\bR_\tau(\bB + s\delta\bB_2)\mathrel{\bigg|_{s = 0}}\mathrm{d}\tau\\
  & = \frac{2}{\pi}\int_0^\infty\left(\bR_\tau\dot{\bS}[\delta\bB_1]\bR_\tau\dot{\bS}[\delta\bB_2]\bR_\tau + \bR_\tau\dot{\bS}[\delta\bB_2]\bR_\tau\dot{\bS}[\delta\bB_1]\bR_\tau - \bR_\tau\ddot{\bS}[\delta\bB_1,\delta\bB_2]\bR_\tau\right)\mathrm{d}\tau,
\end{align*}
where we have applied $(\mathrm{d}/\mathrm{d}s)\bR_\tau(\bB + s\delta\bB_j)|_{s = 0} = -\bR_\tau\dot{\bS}[\delta\bB_j]\bR_\tau$ repeatedly for $j\in\{1,2,3\}$, $(\mathrm{d}/\mathrm{d}s)\dot{\bS}(\bB + s\delta\bB_2)[\delta\bB_1] = \ddot{\bS}(\bB)[\delta\bB_1, \delta\bB_2]$, and $(\mathrm{d}/\mathrm{d}s)\ddot{\bS} = \mathbf{O}$. 
For $\dddot{\bN}$, we again differentiate $\ddot{\bN}$. The sum over $\iota\in\calS_3$ is obtained by replacing exactly one of the $\bR_\tau$ factors in $\bR_\tau\dot{\bS}[\delta\bB_j]\bR_\tau\dot{\bS}[\delta\bB_k]\bR_\tau$ ($(j, k)\in\{(1,2),(2,1)\}$) by $-\bR_\tau\dot{\bS}[\delta\bB_3]\bR_\tau$. For the remaining two sums in the second and third lines, the cases with $k = 3$ are contributed by replacing exactly one of the $\dot{\bS}[\delta\bB_j]$ ($j\in\{1,2\}$) factors in $\bR_\tau\dot{\bS}[\delta\bB_1]\bR_\tau\dot{\bS}[\delta\bB_2]\bR_\tau$ and $\bR_\tau\dot{\bS}[\delta\bB_2]\bR_\tau\dot{\bS}[\delta\bB_1]\bR_\tau$ by $\ddot{\bS}[\delta\bB_j, \delta\bB_3]$, and the cases with $l = 3$ are contributed by replacing exactly one of the $\bR_\tau$ factors in $-\bR_\tau\ddot{\bS}[\delta\bB_1, \delta\bB_2]\bR_\tau$ by $-\bR_\tau\dot{\bS}[\delta\bB_3]\bR_\tau$. The same bookkeeping applies to $\ddddot\bN = (\mathrm d/\mathrm ds)\dddot\bN(\bB + s\delta\bB_4)[\delta\bB_1,\delta\bB_2,\delta\bB_3]|_{s = 0}$, which is an integral of a sum of finitely many terms, each of which is a product that can be categorized to one of the following cases:
\begin{itemize}
   \item four $\dot\bS$ factors;

   \item one $\ddot\bS$ and two $\dot\bS$ factors;

   \item two $\ddot\bS$ factors.
\end{itemize} 
Each term is also associated with a coefficient that is bounded by an absolute constant when $\|\bB\|_2\leq 1$, which is sufficient to obtain the desired bound on $\|\ddddot{\bN}\|_2$. 
For the (unconditional) spectral norms of $\dot{\bN},\ddot{\bN},\dddot{\bN}$, we apply the integral identities
\[
  \int_0^\infty\frac{\mathrm{d}\tau}{(\tau^2 + 1)^k} = \frac{(2k - 3)!!\pi}{2(2k - 2)!!}
\]
and the facts that $\|\bR_\tau\|_2\leq (\tau^2 + 1)^{-1}$, $\|\dot{\bS}[\delta\bB_1]\|_2\leq 2\|\bB\|_2\|\delta\bB_1\|_2$, and $\|\ddot{\bS}[\delta\bB_1, \delta\bB_2]\|_2\leq 2\|\delta\bB_1\|_2\|\delta\bB_2\|_2$. 
The proof is thus completed. 
\end{proof}

\begin{lemma}
\label{lemma:Lipschitz_Cayley_transform}
For any $\delta\bB_1,\delta\bB_2,\delta\bB_3\in\mathbb{R}^{(n - 1 - d)\times d}$ and $\bB\in\mathbb{R}^{(n - 1 - d)\times d}$ with $\|\bB\|_2 = \beta$, we have
\begin{align*}
\|\dot{\bX}(\bB)[\delta\bB_1]\|_2&\leq \sqrt{n}(1 + \beta\sqrt{1 + \beta^2})\|\delta\bB_1\|_2,\\
\|\ddot{\bX}(\bB)[\delta\bB_1,\delta\bB_2]\|_2&\leq \sqrt{n}\{2\beta + (3\beta^2 + 1)\sqrt{1 + \beta^2}\}\|\delta\bB_1\|_2\|\delta\bB_2\|_2,\\
\|\dddot{\bX}(\bB)[\delta\bB_1,\delta\bB_2,\delta\bB_3]\|_2&\leq \sqrt{n}\{(9\beta^2 + 3) + (15\beta^3 + 9\beta)\sqrt{1 + \beta^2}\}\|\delta\bB_1\|_2\|\delta\bB_2\|_2\|\delta\bB_3\|_2
\end{align*}
and
\begin{align*}
\|\dot{\bX}(\bB)[\delta\bB_1]\|_{2\to\infty}&\leq \beta(\|\bX_0\|_{2\to\infty} + \|\bX_{0\perp}\bB\|_{2\to\infty})\|\delta\bB_1\|_2 + \|\bX_{0\perp}\delta\bB_1\|_{2\to\infty},\\
\|\ddot{\bX}(\bB)[\delta\bB_1,\delta\bB_2]\|_{2\to\infty}&\leq \beta\|\bX_{0\perp}\delta\bB_1\|_{2\to\infty}\|\delta\bB_2\|_2 + \beta\|\bX_{0\perp}\delta\bB_2\|_{2\to\infty}\|\delta\bB_1\|_2\\
&\quad + (3\beta^2 + 1)(\|\bX_0\|_{2\to\infty} + \|\bX_{0\perp}\bB\|_{2\to\infty})\|\delta\bB_1\|_2\|\delta\bB_2\|_2,\\
\|\dddot{\bX}(\bB)[\delta\bB_1,\delta\bB_2,\delta\bB_3]\|_{2\to\infty}&\leq 
(3\beta^2 + 1)\left(\|\bX_{0\perp}\delta\bB_1\|_{2\to\infty}\|\delta\bB_2\|_2\|\delta\bB_3\|_2 + \|\bX_{0\perp}\delta\bB_2\|_{2\to\infty}\|\delta\bB_1\|_2\|\delta\bB_3\|_2\right)\\
&\quad + (3\beta^2 + 1)\|\bX_{0\perp}\delta\bB_3\|_{2\to\infty}\|\delta\bB_1\|_2\|\delta\bB_2\|_2\\
&\quad + (15\beta^3 + 9\beta)(\|\bX_0\|_{2\to\infty} + \|\bX_{0\perp}\bB\|_{2\to\infty})\|\delta\bB_1\|_2\|\delta\bB_2\|_2\|\delta\bB_3\|_2.
\end{align*}
Also, when $\beta\leq 1$, there exists an absolute constant $C > 0$, such that
\begin{align*}
\|\ddddot{\bX}(\bB)[\delta\bB_1,\delta\bB_2,\delta\bB_3,\delta\bB_4]\|_2
&\leq C\sqrt{n}\prod_{j = 1}^4\|\delta\bB_j\|_2,\\
\|\ddddot{\bX}(\bB)[\delta\bB_1,\delta\bB_2,\delta\bB_3,\delta\bB_4]\|_{2\to\infty}
&\leq C\beta\sum_{j = 1}^4\|\bX_{0\perp}\delta\bB_j\|_{2\to\infty}
  \prod_{k\in[4]\backslash\{j\}}\|\delta\bB_k\|_2\\
&\quad + C(\|\bX_0\|_{2\to\infty} + \|\bX_{0\perp}\bB\|_{2\to\infty})
  \prod_{j = 1}^4\|\delta\bB_j\|_2.
\end{align*}
\end{lemma}

\begin{proof}[\bf Proof]
By definition, $\bX(\bB) = (\bX_0 + \bX_{0\perp}\bB)\bN(\bB)$, so
\begin{align*}
\dot{\bX}(\bB)[\delta\bB_1]
& = \frac{\mathrm{d}\bX}{\mathrm{d}s}(\bB + s\delta\bB_1)\mathrel{\bigg|_{s = 0}}
  = \bX_{0\perp}\delta\bB_1 \bN(\bB) + (\bX_0 + \bX_{0\perp}\bB)\dot{\bN}(\bB)[\delta\bB_1].
\end{align*}
By Lemma \ref{lemma:integral_formula_matrix_squarerootinverse}, 
\[
\|\dot{\bX}(\bB)[\delta\bB_1]\|_2 \leq \|\bX_{0\perp}\|_2\|\delta\bB_1\|_2\|\bN(\bB)\|_2 + \sqrt{n}\|\eye_d + \bB\transpose\bB\|_2^{1/2}\|\dot{\bN}(\bB)[\delta\bB_1]\|_2\leq \sqrt{n}(1 + \beta\sqrt{1 + \beta^2})\|\delta\bB_1\|_2
\]
and
\begin{align*}
\|\dot{\bX}(\bB)[\delta\bB_1]\|_{2\to\infty}
& \leq \|\bX_{0\perp}\delta\bB_1\|_{2\to\infty}\|\bN(\bB)\|_2 + (\|\bX_0\|_{2\to\infty} + \|\bX_{0\perp}\bB\|_{2\to\infty})\|\dot{\bN}(\bB)[\delta\bB_1]\|_2\\
& \leq (\|\bX_0\|_{2\to\infty} + \|\bX_{0\perp}\bB\|_{2\to\infty})\|\bB\|_2\|\delta\bB_1\|_2 + \|\bX_{0\perp}\delta\bB_1\|_{2\to\infty}.
\end{align*}
For $\ddot{\bX}$, we have
\begin{align*}
  \ddot{\bX}(\bB)[\delta\bB_1, \delta\bB_2]
  & = \frac{\mathrm{d}\dot\bX}{\mathrm{d}s}(\bB + s\delta\bB_2)[\delta\bB_1]\mathrel{\bigg|_{s = 0}}\\
  & = \bX_{0\perp}\delta\bB_1\dot{\bN}(\bB)[\delta\bB_2] + \bX_{0\perp}\delta\bB_2\dot{\bN}(\bB)[\delta\bB_1] + (\bX_0 + \bX_{0\perp}\bB)\ddot{\bN}(\bB)[\delta\bB_1, \delta\bB_2].
\end{align*}
Invoking Lemma \ref{lemma:integral_formula_matrix_squarerootinverse}, we obtain
\begin{align*}
\|  \ddot{\bX}(\bB)[\delta\bB_1, \delta\bB_2]
\|_2
&\leq \sqrt{n}\{2\beta + (3\beta^2 + 1)\sqrt{1 + \beta^2}\}\|\delta\bB_1\|_2\|\delta\bB_2\|_2
\end{align*}
and
\begin{align*}
\|  \ddot{\bX}(\bB)[\delta\bB_1, \delta\bB_2]\|_{2\to\infty}
&\leq \beta\left(\|\bX_{0\perp}\delta\bB_1\|_{2\to\infty} \|\delta\bB_2\|_2 + \|\bX_{0\perp}\delta\bB_2\|_{2\to\infty} \|\delta\bB_1\|_2\right)\\
&\quad + (3\beta^2 + 1)(\|\bX_0\|_{2\to\infty} + \|\bX_{0\perp}\bB\|_{2\to\infty})\|\delta\bB_1\|_2\|\delta\bB_2\|_2.
\end{align*}
The same reasoning, together with
\[
\dddot{\bX}(\bB)[\delta\bB_1,\delta\bB_2,\delta\bB_3] = \sum_{(j,k,l)}\bX_{0\perp}\delta\bB_j\ddot{\bN}(\bB)[\delta\bB_k,\delta\bB_l] + (\bX_0 + \bX_{0\perp}\bB)\dddot{\bN}(\bB)[\delta\bB_1,\delta\bB_2,\delta\bB_3],
\]
where the sum is over $(j,k,l)\in\{(1,2,3),(2,1,3),(3,1,2)\}$, leads to the spectral norm and the two-to-infinity norm bounds for $\dddot{\bX}$.
Also, we have 
\begin{align*}
\ddddot\bX(\bB)[\delta\bB_1,\delta\bB_2,\delta\bB_3,\delta\bB_4]
& = \sum_{j = 1}^4\bX_{0\perp}\delta\bB_j\dddot\bN(\bB)[\delta\bB_{a(j)}, \delta\bB_{b(j)}, \delta\bB_{c(j)}]\\
&\quad + (\bX_0 + \bX_{0\perp}\bB)\ddddot\bN(\bB)[\delta\bB_1, \delta\bB_2, \delta\bB_3, \delta\bB_4],
\end{align*}
where $(a(1), b(1), c(1)) = (2, 3, 4)$, $(a(2), b(2), c(2)) = (1, 3, 4)$, $(a(3), b(3), c(3)) = (1, 2, 4)$, $(a(4), b(4), c(4)) = (1, 2, 3)$. 
The proof is therefore completed. 
\end{proof}

\subsection{Gradient and Hessian calculations}
\label{sub:gradient_hessian}

This subsection calculates the gradients and Hessians of $\theta_{ij}^{(t)} = \alpha_i^{(t)} + \alpha_j^{(t)} + \bx_i(\bB)\transpose\bR^{(t)}\bx_j(\bB)$ and the log-likelihood function $\calL(\bphi)$ with respect to the compact parameter $\bphi$. Let $\bTheta^{(t)} = [\theta_{ij}^{(t)}]_{n\times n}$. Recall that $\theta_{0ij}^{(t)}$ denote the true value of $\theta_{ij}^{(t)}$ and we denote $\bTheta^{(t)}_0 = [\theta_{0ij}^{(t)}]_{n\times n}$. 
Denote
\begin{align}
\ell'(\bTheta^{(t)}) = 
\begin{bmatrix}
\label{eqn:ell_prime_Theta_t}
0 & \ell'(\theta_{12}^{(t)}; A_{12}^{(t)}) & \ell'(\theta_{13}^{(t)}; A_{13}^{(t)}) & \ldots & \ell'(\theta_{1n}^{(t)}; A_{1n}^{(t)})\\
\ell'(\theta_{21}^{(t)}; A_{21}^{(t)}) & 0 & \ell'(\theta_{23}^{(t)}; A_{23}^{(t)})
& \ldots & \ell'(\theta_{2n}^{(t)}; A_{2n}^{(t)})\\
\ell'(\theta_{31}^{(t)}; A_{31}^{(t)}) & \ell'(\theta_{32}^{(t)}; A_{32}^{(t)}) & 0 & \ldots &\ell'(\theta_{3n}^{(t)}; A_{3n}^{(t)})\\
\vdots & \vdots & \vdots && \vdots\\
\ell'(\theta_{n1}^{(t)}; A_{n1}^{(t)}) & \ell'(\theta_{n2}^{(t)}; A_{n2}^{(t)}) & \ell'(\theta_{n3}^{(t)}; A_{n3}^{(t)}) & \ldots & 0
\end{bmatrix}	
\end{align}
 the $n\times n$ matrix whose $(i, j)$th element is $\ell'(\theta_{ij}^{(t)}; A_{ij}^{(t)})$ if $i \neq j$ and $0$ otherwise. $\ell''(\bTheta^{(t)})$ is defined analogously. The derivative can be specialized to the following scenarios (up to terms free of $\theta$):
 \begin{itemize}
     \item The canonical exponential family with binary edges, giving log-likelihood $\ell(\theta; A) = A\theta-\log(1+e^\theta)$, with $\ell'(\theta; A) = A - e^\theta/(1 + e^\theta)$. 

     \item The canonical exponential family with Poisson edges, giving likelihood
     $\ell(\theta; A) = A\theta-e^\theta$ with $\ell'(\theta; A) = A - e^\theta$. 
     
     \item The canonical exponential family with Gaussian edges, giving likelihood $\ell(\theta; A) = -(A-\theta)^2/(2\sigma^2)$ and $\ell'(\theta; A) = (A - \theta)/\sigma^2$. 
     
     \item The probit model leads to likelihood $\ell(\theta; A) = A\log\Phi(\theta)+(1-A)\log\Phi(-\theta)$, where $\Phi(\cdot)$ is the cumulative distribution function (CDF) of $\mathrm{N}(0, 1)$. The derivative is $\ell'(\theta; A) = A\phi(\theta)/\{\Phi(\theta)\Phi(-\theta)\} - \phi(\theta)/\Phi(-\theta)$, where $\phi(\cdot)$ is the density of $\mathrm{N}(0, 1)$. 
 \end{itemize}

\noindent $\blacksquare$ \textbf{Gradient of $\theta_{ij}^{(t)}$. }We have
\begin{equation}
\label{eqn:dtheta}
\begin{aligned}
\frac{\partial\theta_{ij}^{(t)}}{\partial\alpha_{k}^{(s)}} &= \delta_{st}(\delta_{ik} + \delta_{jk}),\\
\frac{\partial\theta_{ij}^{(t)}}{\partial\vech(\bR^{(s)})} &= \delta_{st}\bD_d\transpose\{\bx_j(\bB)\otimes\bx_i(\bB)\},\\
\frac{\partial\theta_{ij}^{(t)}}{\partial B_{kl}} &= \be_i\transpose\frac{\partial\bX}{\partial B_{kl}}(\bB)\bR^{(t)}\bx_j(\bB) + \bx_i(\bB)\transpose\bR^{(t)}\bigg\{\frac{\partial\bX}{\partial B_{kl}}(\bB)\bigg\}\transpose\be_j.
\end{aligned}
\end{equation}
In particular, at $\bB = \mathbf{O}$, by \eqref{eqn:dX_dB_zero} we have
\begin{align}
\label{eqn:dtheta_dB_zero}
\frac{\partial\theta_{ij}^{(t)}}{\partial B_{kl}}\mathrel{\Bigg|_{\bB = \mathbf{O}}} = \be_i\transpose(\bX_{0\perp}\be_k\be_l\transpose\bR^{(t)}\bX_0\transpose + \bX_0\bR^{(t)}\be_l\be_k\transpose\bX_{0\perp}\transpose)\be_j.
\end{align}

\noindent $\blacksquare$ \textbf{Gradient of $\calL(\bphi)$. }The above gradient of $\theta_{ij}^{(t)}$ enables us to compute the gradient of the log-likelihood function as follows:
\begin{align}\label{eqn:grad_alpha}
\nabla_{\balpha^{(t)}}\calL(\bphi) = \frac{1}{2}\sum_{i \neq j}\ell'(\theta_{ij}^{(t)};A_{ij}^{(t)})\frac{\partial\theta_{ij}^{(t)}}{\partial\balpha^{(t)}}
= \frac{1}{2}\sum_{i \neq j}\ell'(\theta_{ij}^{(t)};A_{ij}^{(t)})(\be_i + \be_j) = \ell'(\bTheta^{(t)})\one_n.
\end{align}
A direct calculation also yields
\begin{align}\label{eqn:grad_R}
\nabla_{\vech(\bR^{(t)})}\calL(\bphi)
= \frac{1}{2}\sum_{i\neq j}\ell'(\theta_{ij}^{(t)}; A_{ij}^{(t)})\frac{\partial\theta_{ij}^{(t)}}{\partial\vech(\bR^{(t)})}
= \frac{1}{2}\bD_d\transpose\vect\{\bX(\bB)\transpose\ell'(\bTheta^{(t)})\bX(\bB)\}
\end{align}
and
\begin{align*}
\nabla_{B_{pq}}\calL(\bphi)
& = \frac{1}{2}\sum_{t = 1}^m\sum_{i\neq j}\ell'(\theta_{ij}^{(t)}; A_{ij}^{(t)})\frac{\partial\theta_{ij}^{(t)}}{\partial B_{pq}}
 = \frac{1}{2}\sum_{t = 1}^m\sum_{i\neq j}\ell'(\theta_{ij}^{(t)}; A_{ij}^{(t)})\bigg\{\frac{\partial\bx_i(\bB)\transpose}{\partial B_{pq}}\bR^{(t)}\bx_j(\bB) + \bx_i(\bB)\transpose\bR^{(t)}\frac{\partial\bx_j(\bB)}{\partial B_{pq}}\bigg\}\\
& = \sum_{t = 1}^m\mathrm{tr}\bigg\{\bR^{(t)}\sum_{i\neq j}\bx_j(\bB)\ell'(\theta_{ij}^{(t)}; A_{ij}^{(t)})\frac{\partial\bx_i(\bB)\transpose}{\partial B_{pq}}\bigg\}
  = \sum_{t = 1}^m\mathrm{tr}\bigg\{\bR^{(t)}\bX(\bB)\transpose\ell'(\bTheta^{(t)})\frac{\partial \bX(\bB)}{\partial B_{pq}}\bigg\}.
\end{align*}
By Proposition \ref{prop:Cayley_transform_jacobian}, we have
\begin{equation}
\label{eqn:grad_B}
\begin{aligned}
\nabla_{B_{pq}}\calL(\bphi)& = \sum_{t = 1}^m\be_p\transpose\bX_{0\perp}\transpose\ell'(\bTheta^{(t)})\bX(\bB)\bR^{(t)}\bS(\bB)^{-1/2}\be_q\\
&\quad - \sum_{t = 1}^m\mathrm{tr}\bigg\{\bR^{(t)}\bX(\bB)\transpose\ell'(\bTheta^{(t)})\bX(\bB)\mathscr{L}^{-1}(\bB\transpose\be_p\be_q\transpose + \be_q\be_p\transpose\bB)\bS(\bB)^{-1/2}\bigg\}.
\end{aligned}
\end{equation}
In particular,
\begin{align}
\label{eqn:grad_B_zero}
\nabla_{B_{pq}}\calL(\bphi)\mathrel{\bigg|_{\bB = \mathbf{O}}} = \sum_{t = 1}^m\be_p\transpose\bX_{0\perp}\transpose\ell'(\bTheta^{(t)})\bX_0\bR^{(t)}\be_q.
\end{align}

\noindent $\blacksquare$ \textbf{Hessian of $\theta_{ij}^{(t)}$. }We next proceed to the calculations of the Hessians of $\theta_{ij}^{(t)}$. The following second-order derivatives of $\theta_{ij}^{(t)}$ are zero for any $s,t\in[m]$:
\[
\frac{\partial^2\theta_{ij}^{(t)}}{\partial\balpha^{(s)}\partial\balpha^{(t)\mathrm{T}}},\quad 
\frac{\partial^2\theta_{ij}^{(t)}}{\partial\vech(\bR^{(s)})\partial\vech(\bR^{(t)})\transpose},\quad 
\frac{\partial^2\theta_{ij}^{(t)}}{\partial\balpha^{(s)}\partial\vech(\bR^{(t)})\transpose},\quad 
\frac{\partial^2\theta_{ij}^{(t)}}{\partial\balpha^{(s)}\partial\vect(\bB)\transpose}.
\]
The two nonzero blocks of the second-order derivatives of $\theta_{ij}^{(t)}$ are:
\begin{equation}
\label{eqn:ddtheta}
\begin{aligned}
\frac{\partial^2\theta_{ij}^{(t)}}{\partial\vech(\bR^{(s)})\partial B_{kl}} &= \delta_{st}\bD_d\transpose\bigg[\bigg\{\frac{\partial\bX(\bB)}{\partial B_{kl}}\bigg\}\transpose\be_j\otimes\bx_i(\bB) + \bx_j(\bB) \otimes \bigg\{\frac{\partial\bX(\bB)}{\partial B_{kl}}\bigg\}\transpose\be_i\bigg],\\
\frac{\partial^2\theta_{ij}^{(t)}}{\partial B_{kl}\partial B_{pq}} &= \be_i\transpose\frac{\partial^2\bX}{\partial B_{kl}\partial B_{pq}}(\bB)\bR^{(t)}\bx_j(\bB) + \be_i\transpose\frac{\partial\bX}{\partial B_{pq}}(\bB)\bR^{(t)}\bigg\{\frac{\partial\bX(\bB)}{\partial B_{kl}}\bigg\}\transpose\be_j\\
&\quad + \be_i\transpose\frac{\partial\bX}{\partial B_{kl}}(\bB)\bR^{(t)}\bigg\{\frac{\partial\bX(\bB)}{\partial B_{pq}}\bigg\}\transpose\be_j + \bx_i(\bB)\transpose\bR^{(t)}\bigg\{\frac{\partial^2\bX}{\partial B_{kl}\partial B_{pq}}(\bB)\bigg\}\transpose\be_j.
\end{aligned}
\end{equation}
In particular, at $\bphi = \bphi_0$, we have
\begin{equation}
\label{eqn:ddtheta_dRdB_zero}
\frac{\partial^2\theta_{ij}^{(t)}}{\partial\vech(\bR^{(t)})\partial\vect(\bB)\transpose}(\bphi_0)
 = \bD_d\transpose\left\{\eye_d\otimes (\bx_{0i}\be_j\transpose + \bx_{0j}\be_i\transpose)\bX_{0\perp}\right\}
\end{equation}
and
\begin{equation}
\label{eqn:ddtheta_dBdB_zero}
\begin{aligned}
\frac{\partial^2\theta_{ij}^{(t)}}{\partial\vect(\bB)\partial\vect(\bB)\transpose}(\bphi_0)
& = \bR_0^{(t)}\otimes\{\bX_{0\perp}\transpose(\be_i\be_j\transpose + \be_j\be_i\transpose)\bX_{0\perp}\}\\
&\quad - \frac{1}{2}\left\{\bR_0^{(t)}(\bx_{0i}\bx_{0j}\transpose + \bx_{0j}\bx_{0i}\transpose) + (\bx_{0i}\bx_{0j}\transpose + \bx_{0j}\bx_{0i}\transpose)\bR_0^{(t)}\right\}\otimes\eye_{n - 1 - d}.
\end{aligned}
\end{equation}

\noindent $\blacksquare$ \textbf{Hessian of $\calL(\bphi)$. }We finally work on the calculation of the Hessian of the log-likelihood $\calL$. 

\noindent \underline{\it Hessian of $\calL$: $\balpha^{(s)}$-$\balpha^{(t)}$ block:} This block vanishes when $s\neq t$. For the case when $s = t$, we obtain:
\begin{align}
\label{eqn:hessian_alpha_alpha}
\frac{\partial^2\calL}{\partial\balpha^{(t)}\partial\balpha^{(t)\mathrm{T}}}(\bphi)
& = \frac{1}{2}\sum_{i\neq j}\ell''(\theta_{ij}^{(t)}; A_{ij}^{(t)})\bigg(\frac{\partial\theta_{ij}^{(t)}}{\partial\balpha^{(t)}}\bigg)\bigg(\frac{\partial\theta_{ij}^{(t)}}{\partial\balpha^{(t)}}\bigg)\transpose = 
\ell''(\bTheta^{(t)}) + \mathrm{diag}(\ell''(\bTheta^{(t)})\one_n).
\end{align}

\noindent\underline{\it Hessian of $\calL$: $\vech(\bR^{(s)})$-$\vech(\bR^{(t)})$ block:}
The block vanishes when $s\neq t$. When $s = t$, we have
\begin{align*}
\frac{\partial^2\calL}{\partial\vech(\bR^{(t)})\partial\vech(\bR^{(t)})\transpose}(\bphi)
& = \frac{1}{2}\sum_{i\neq j}\ell''(\theta_{ij}^{(t)}; A_{ij}^{(t)})\bigg\{\frac{\partial\theta_{ij}^{(t)}}{\partial\vech(\bR^{(t)})}\bigg\}\bigg\{\frac{\partial\theta_{ij}^{(t)}}{\partial\vech(\bR^{(t)})}\bigg\}\transpose\\
& = \frac{1}{2}\sum_{i\neq j}\ell''(\theta_{ij}^{(t)}; A_{ij}^{(t)})\bD_d\transpose\{\bx_j(\bB)\otimes\bx_i(\bB)\}\{\bx_j(\bB)\otimes\bx_i(\bB)\}\transpose\bD_d\\
& = \frac{1}{2}\bD_d\transpose(\bX(\bB)\otimes\bX(\bB))\transpose\mathrm{diag}(\vect(\ell''(\bTheta^{(t)})))(\bX(\bB)\otimes\bX(\bB))\bD_d.
\end{align*}
Namely,
\begin{align}\label{eqn:hessian_R_R}
\frac{\partial^2\calL}{\partial\vech(\bR^{(t)})\partial\vech(\bR^{(t)})\transpose}(\bphi)
& = \frac{1}{2}\bD_d\transpose(\bX(\bB)\otimes\bX(\bB))\transpose\mathrm{diag}(\vect(\ell''(\bTheta^{(t)})))(\bX(\bB)\otimes\bX(\bB))\bD_d.
\end{align}

\noindent\underline{\it Hessian of $\calL$: $\balpha^{(s)}$-$\vech(\bR^{(t)})$ block:}
The block vanishes when $s\neq t$. For the case where $s = t$, we have
\begin{align*}
\frac{\partial^2\calL}{\partial\balpha^{(t)}\partial\vech(\bR^{(t)})\transpose}(\bphi)
& = \frac{1}{2}\sum_{i\neq j}\ell''(\theta_{ij}^{(t)}; A_{ij}^{(t)})\bigg(\frac{\partial\theta_{ij}^{(t)}}{\partial\balpha^{(t)}}\bigg)
\bigg\{\frac{\partial\theta_{ij}^{(t)}}{\partial\vech(\bR^{(t)})}\bigg\}\transpose\\
& = \frac{1}{2}\sum_{i\neq j}\ell''(\theta_{ij}^{(t)}; A_{ij}^{(t)})(\be_i + \be_j)\{\bx_j(\bB)\otimes \bx_i(\bB)\}\transpose\bD_d\\
& = \frac{1}{2}(\one_n\transpose\otimes\eye_n + \eye_n\otimes \one_n\transpose)\mathrm{diag}(\vect(\ell''(\bTheta^{(t)})))\{\bX(\bB)\otimes\bX(\bB)\}\bD_d.
\end{align*}
Namely,
\begin{align}
\label{eqn:hessian_alpha_R}
\frac{\partial^2\calL}{\partial\balpha^{(t)}\partial\vech(\bR^{(t)})\transpose}(\bphi)
 = \frac{1}{2}(\one_n\transpose\otimes\eye_n + \eye_n\otimes \one_n\transpose)\mathrm{diag}(\vect(\ell''(\bTheta^{(t)})))\{\bX(\bB)\otimes\bX(\bB)\}\bD_d.
\end{align}

\noindent\underline{\it Hessian of $\calL$: $\balpha^{(t)}$-$\vect(\bB)$ block:}
Let $\partial\bTheta^{(t)}/\partial B_{pq}$ be the $n\times n$ matrix whose $(i, j)$th element is $\partial\theta_{ij}^{(t)}/\partial B_{pq}$. Then,
\begin{align*}
\frac{\partial^2\calL}{\partial\balpha^{(t)}\partial B_{pq}}(\bphi)
& = \frac{1}{2}\sum_{i\neq j}\ell''(\theta_{ij}^{(t)}; A_{ij}^{(t)})\frac{\partial\theta_{ij}^{(t)}}{\partial\balpha^{(t)}}\frac{\partial\theta_{ij}^{(t)}}{\partial B_{pq}}\\
& = \frac{1}{2}\sum_{i\neq j}\ell''(\theta_{ij}^{(t)}; A_{ij}^{(t)})(\be_i + \be_j)\bigg\{\frac{\partial\bx_i(\bB)\transpose}{\partial B_{pq}}\bR^{(t)}\bx_j(\bB) + \bx_i(\bB)\transpose\bR^{(t)}\frac{\partial\bx_j(\bB)}{\partial B_{pq}}\bigg\}\\
& = \bigg\{\ell''(\bTheta^{(t)})\circ \frac{\partial\bTheta^{(t)}}{\partial B_{pq}}\bigg\}\one_n. 
\end{align*}
Namely,
\begin{align}\label{eqn:hessian_alpha_B}
\frac{\partial^2\calL}{\partial\balpha^{(t)}\partial B_{pq}}(\bphi)
 = \bigg\{\ell''(\bTheta^{(t)})\circ \frac{\partial\bTheta^{(t)}}{\partial B_{pq}}\bigg\}\one_n. 
\end{align}
In particular, at $\bB = \mathbf{O}$, we have
\begin{align}
\label{eqn:hessian_alpha_B_zero}
\frac{\partial^2\calL}{\partial\balpha^{(t)}\partial B_{pq}}(\bphi){\mathrel{\Bigg|_{\bB = \mathbf{O}}}}
 = \left\{\ell''(\bTheta^{(t)})\circ (\bX_{0\perp}\be_p\be_q\transpose\bR^{(t)}\bX_0\transpose + \bX_0\bR^{(t)}\be_q\be_p\transpose\bX_{0\perp}\transpose)\right\}\one_n. 
\end{align}

\noindent\underline{\it Hessian of $\calL$: $\vech(\bR^{(t)})$-$\vect(\bB)$ block:}
By definition, we have
\begin{align*}
\frac{\partial^2\calL}{\partial\vech(\bR^{(t)})\partial B_{pq}}(\bphi)
& = \frac{1}{2}\sum_{i\neq j}\ell''(\theta_{ij}^{(t)}; A_{ij}^{(t)})\frac{\partial\theta_{ij}^{(t)}}{\partial\vech(\bR^{(t)})}\frac{\partial\theta_{ij}^{(t)}}{\partial B_{pq}} + \frac{1}{2}\sum_{i\neq j}\ell'(\theta_{ij}^{(t)}; A_{ij}^{(t)})\frac{\partial^2\theta_{ij}^{(t)}}{\partial\vech(\bR^{(t)})\partial B_{pq}}\\
& = \frac{1}{2}\sum_{i\neq j}\ell''(\theta_{ij}^{(t)}; A_{ij}^{(t)})\bD_d\transpose\{\bx_j(\bB)\otimes\bx_i(\bB)\}\frac{\partial\theta_{ij}^{(t)}}{\partial B_{pq}}\\
&\quad + \frac{1}{2}\sum_{i\neq j}\ell'(\theta_{ij}^{(t)}; A_{ij}^{(t)})\bD_d\transpose\bigg[\bigg\{\frac{\partial\bX(\bB)}{\partial B_{pq}}\bigg\}\transpose\be_j\otimes\bx_i(\bB) + \bx_j(\bB) \otimes \bigg\{\frac{\partial\bX(\bB)}{\partial B_{pq}}\bigg\}\transpose\be_i\bigg]\\
& = \frac{1}{2}\bD_d\transpose\vect\bigg[\bX(\bB)\transpose\bigg\{\ell''(\bTheta^{(t)})\circ \frac{\partial\bTheta^{(t)}}{\partial B_{pq}}\bigg\}\bX(\bB)\bigg]\\
&\quad + \frac{1}{2}\bD_d\transpose\vect\bigg[\bigg\{\frac{\partial\bX}{\partial B_{pq}}(\bB)\bigg\}\transpose\ell'(\bTheta^{(t)})\bX(\bB) + \bX(\bB)\transpose\ell'(\bTheta^{(t)})\bigg\{\frac{\partial\bX}{\partial B_{pq}}(\bB)\bigg\}\bigg].
\end{align*}
Namely, 
\begin{equation}
\begin{aligned}
\label{eqn:hessian_R_B}
\frac{\partial^2\calL}{\partial\vech(\bR^{(t)})\partial B_{pq}}(\bphi)
& = 
\frac{1}{2}\bD_d\transpose\vect\bigg[\bX(\bB)\transpose\bigg\{\ell''(\bTheta^{(t)})\circ \frac{\partial\bTheta^{(t)}}{\partial B_{pq}}\bigg\}\bX(\bB)\bigg]\\
&\quad + \frac{1}{2}\bD_d\transpose\vect\bigg[\bigg\{\frac{\partial\bX}{\partial B_{pq}}(\bB)\bigg\}\transpose\ell'(\bTheta^{(t)})\bX(\bB) + \bX(\bB)\transpose\ell'(\bTheta^{(t)})\bigg\{\frac{\partial\bX}{\partial B_{pq}}(\bB)\bigg\}\bigg].
\end{aligned}
\end{equation}
In particular, when $\bB = \mathbf{O}$, we obtain
\begin{equation}
\begin{aligned}
\label{eqn:hessian_R_B_zero}
\frac{\partial^2\calL}{\partial\vech(\bR^{(t)})\partial B_{pq}}(\bphi)\mathrel{\Bigg|_{\bB = \mathbf{O}}}
& = 
\frac{1}{2}\bD_d\transpose\vect\bigg[\bX_0\transpose\bigg\{\ell''(\bTheta^{(t)})\circ (\bX_{0\perp}\be_p\be_q\transpose\bR^{(t)}\bX_0\transpose + \bX_0\bR^{(t)}\be_q\be_p\transpose\bX_{0\perp}\transpose)\bigg\}\bX_0\bigg]\\
&\quad + \frac{1}{2}\bD_d\transpose\vect\bigg\{\be_q\be_p\transpose\bX_{0\perp}\transpose\ell'(\bTheta^{(t)})\bX_0 + \bX_0\transpose\ell'(\bTheta^{(t)})\bX_{0\perp}\be_p\be_q\transpose\bigg\}.
\end{aligned}
\end{equation}

\noindent\underline{\it Hessian of $\calL$: $\vect(\bB)$-$\vect(\bB)$ block:}
For any $(p, q),(r, s)\in[n - 1 - d]\times [d]$, we first compute
\begin{equation}
\label{eqn:d2Theta_dB2}
\begin{aligned}
\frac{\partial^2\bTheta^{(t)}}{\partial B_{pq}\partial B_{rs}}
& = \frac{\partial^2\bX}{\partial B_{pq}\partial B_{rs}}(\bB)\bR^{(t)}\bX(\bB)\transpose + \frac{\partial\bX}{\partial B_{rs}}(\bB)\bR^{(t)}\frac{\partial\bX\transpose}{\partial B_{pq}}(\bB)\\
&\quad + \frac{\partial\bX}{\partial B_{pq}}(\bB)\bR^{(t)}\frac{\partial\bX\transpose}{\partial B_{rs}}(\bB) + \bX(\bB)\bR^{(t)}\frac{\partial^2\bX\transpose}{\partial B_{pq}\partial B_{rs}}(\bB)
\end{aligned}
\end{equation}
and
\begin{equation}
\label{eqn:d2Theta_dB2_zero}
\begin{aligned}
\frac{\partial^2\bTheta^{(t)}}{\partial B_{pq}\partial B_{rs}}\mathrel{\Bigg|_{\bB = \mathbf{O}}}
& = -\frac{1}{2}\delta_{pr}\bX_0(\be_q\be_s\transpose + \be_s\be_q\transpose)\bR^{(t)}\bX_0\transpose + \bX_{0\perp}\be_r\be_s\transpose\bR^{(t)}\be_q\be_p\transpose\bX_{0\perp}\transpose\\
&\quad + \bX_{0\perp}\be_p\be_q\transpose\bR^{(t)}\be_s\be_r\transpose\bX_{0\perp}\transpose - \frac{1}{2}\delta_{pr}\bX_0\bR^{(t)}(\be_q\be_s\transpose + \be_s\be_q\transpose)\bX_0\transpose.
\end{aligned}
\end{equation}
Then, it follows that
\begin{equation}
\label{eqn:hessian_B_B}
\begin{aligned}
\frac{\partial^2\calL}{\partial B_{pq} \partial B_{rs}}(\bphi)
& = \frac{1}{2}\sum_{t = 1}^m\sum_{i\neq j}\ell''(\theta_{ij}^{(t)}; A_{ij}^{(t)})\frac{\partial\theta_{ij}^{(t)}}{\partial B_{pq}}\frac{\partial\theta_{ij}^{(t)}}{\partial B_{rs}}
 + \frac{1}{2}\sum_{t = 1}^m\sum_{i\neq j}\ell'(\theta_{ij}^{(t)}; A_{ij}^{(t)})\frac{\partial^2\theta_{ij}^{(t)}}{\partial B_{pq}\partial B_{rs}}\\
& = \frac{1}{2}\sum_{t = 1}^m\one_n\transpose\bigg\{\ell''(\bTheta^{(t)})\circ \frac{\partial\bTheta^{(t)}}{\partial B_{pq}}\circ\frac{\partial\bTheta^{(t)}}{\partial B_{rs}} + \ell'(\bTheta^{(t)})\circ\frac{\partial^2\bTheta^{(t)}}{\partial B_{pq}\partial B_{rs}}\bigg\}\one_n.
\end{aligned}
\end{equation}

\subsection{Proof of Proposition \ref{prop:expected_hessian_bound}}
\label{sub:proof_of_expected_hessian_bound}

Denote by $\bH(\bphi) = \expect_0\nabla_\bphi^2\calL(\bphi)$ the expected Hessian of the log-likelihood, and the three diagonal blocks by
\begin{align*}
\bH_{\balpha\balpha}(\bphi) &= \expect_0\bigg\{\mathrm{diag}\bigg(\frac{\partial^2\calL}{\partial\balpha^{(t)}\partial\balpha^{(t)\mathrm{T}}}(\bphi):t\in[m]\bigg)\bigg\},\\
\bH_{\bR\bR}(\bphi) &= \expect_0\bigg\{\mathrm{diag}\bigg(\frac{\partial^2\calL}{\partial\vech(\bR^{(t)})\partial\vech(\bR^{(t)})\transpose}(\bphi):t\in[m]\bigg)\bigg\},\\
\bH_{\bB\bB}(\bphi) &= \expect_0\bigg\{\frac{\partial^2\calL}{\partial\vect(\bB)\partial\vect(\bB)\transpose}(\bphi)\bigg\}
\end{align*}
Likewise, denote the off-diagonal blocks of the expected Hessian by
\begin{align*}
\bH_{\balpha\bR}(\bphi) &= \expect_0\bigg\{\mathrm{diag}\bigg(\frac{\partial^2\calL}{\partial\balpha^{(t)}\partial\vech(\bR^{(t)})\transpose}(\bphi):t\in[m]\bigg)\bigg\},\\
\bH_{\balpha\bB}(\bphi) &= \expect_0\bigg\{\bigg[\frac{\partial^2\calL}{\partial\vect(\bB)\partial\balpha^{(1)\mathrm{T}}}(\bphi),\ldots,\frac{\partial^2\calL}{\partial\vect(\bB)\partial\balpha^{(m)\mathrm{T}}}(\bphi)\bigg]\transpose\bigg\},\\
\bH_{\bR\bB}(\bphi) &= \expect_0\bigg\{\bigg[\frac{\partial^2\calL}{\partial\vect(\bB)\partial\vech(\bR^{(1)})\transpose}(\bphi),\ldots,\frac{\partial^2\calL}{\partial\vect(\bB)\partial\vech(\bR^{(m)})\transpose}(\bphi)\bigg]\transpose\bigg\}.
\end{align*}
It is clear that the Fisher information matrix is the negative expected Hessian since Assumption \ref{assumption:true_parameter} requires $\expect_0\ell'(\theta_{0ij}^{(t)}; A_{ij}^{(t)}) = 0$. The proof below works on $\bH$ directly. 
\begin{proof}[\bf Proof of Proposition \ref{prop:expected_hessian_bound}]
Let $\delta\balpha^{(t)} = [\delta\alpha_1^{(t)},\ldots,\delta\alpha_n^{(t)}]\transpose\in\mathbb{R}^{n}$, $\delta\br^{(t)}\in\mathbb{R}^{d(d + 1)/2}$, and $\delta\bB\in\mathbb{R}^{(n - 1 - d)\times d}$ be arbitrary, where $t$ ranges over $[m]$. Denote by
\[
\delta\bphi = \begin{bmatrix}
  \delta\balpha\\\delta\br\\\vect(\delta\bB)
\end{bmatrix},\text{ where } 
\delta\balpha = \begin{bmatrix}
  \delta \balpha^{(1)} \\ \vdots \\ \delta \balpha^{(m)}
\end{bmatrix} \text{ and } \delta\br = \begin{bmatrix}
  \delta \br^{(1)} \\ \vdots \\ \delta \br^{(m)}
\end{bmatrix}.
\] 
For $\bH_{\balpha\balpha}$, let $\bH_{\balpha^{(t)}\balpha^{(t)}}(\bphi)$ denote the $t$th diagonal block of $\bH_{\balpha\balpha}(\bphi)$. For $\bH_{\bR\bR}$, let $\bH_{\bR^{(t)}\bR^{(t)}}(\bphi)$ denote the $t$th diagonal block of $\bH_{\bR\bR}(\bphi)$. 
Also, denote by
\begin{align*}
\bH_{\balpha^{(t)}\bR^{(t)}}
& = \expect_0\bigg\{\frac{\partial^2\calL}{\partial\balpha^{(t)}\partial\vech(\bR^{(t)})\transpose}(\bphi)\bigg\},\;
\bH_{\balpha^{(t)}\bB}
 = \expect_0\bigg\{\frac{\partial^2\calL}{\partial\balpha^{(t)}\partial\vect(\bB)\transpose}(\bphi)\bigg\},\\
\bH_{\bR^{(t)}\bB}
& = \expect_0\bigg\{\frac{\partial^2\calL}{\partial\vech(\bR^{(t)})\partial\vect(\bB)\transpose}(\bphi)\bigg\}.
\end{align*}
By \eqref{eqn:hessian_alpha_alpha}, 
\begin{align*}
&-\delta\balpha^{(t)\mathrm{T}}\bH_{\balpha^{(t)}\balpha^{(t)}}(\bphi_0)\delta\balpha^{(t)}\\
&\quad = -\sum_{i\neq j}\expect_0\ell''(\theta_{0ij}^{(t)}; A_{ij}^{(t)})\delta\alpha^{(t)}_i\delta\alpha^{(t)}_j - \sum_{i = 1}^n\sum_{j\in[n]\backslash\{i\}}\expect_0\ell''(\theta_{0ij}^{(t)}; A_{ij}^{(t)})(\delta\alpha^{(t)}_i)^2\\
&\quad = -\sum_{i\neq j}\expect_0\ell''(\theta_{0ij}^{(t)}; A_{ij}^{(t)})\delta\alpha^{(t)}_i\delta\alpha^{(t)}_j - \frac{1}{2}\sum_{i = 1}^n\sum_{j\in[n]\backslash\{i\}}\expect_0\ell''(\theta_{0ij}^{(t)}; A_{ij}^{(t)})\{(\delta\alpha^{(t)}_i)^2 + (\delta\alpha^{(t)}_j)^2\}\\
&\quad = -\frac{1}{2}\sum_{j\neq i}\expect_0\ell''(\theta_{0ij}^{(t)}; A_{ij}^{(t)})(\delta\alpha^{(t)}_i + \delta\alpha^{(t)}_j)^2. 
\end{align*}
Let $\delta\bR^{(t)}\in\mathrm{Sym}(d)$ satisfy $\vect(\delta\bR^{(t)}) = \bD_d\delta\br^{(t)}$. 
By \eqref{eqn:hessian_R_R}, we have
\begin{align*}
-\delta\br^{(t)\mathrm{T}}\bH_{\bR^{(t)}\bR^{(t)}}(\bphi_0)\delta\br^{(t)}
& = -\frac{1}{2}\delta\br^{(t)\mathrm{T}}\bD_d\transpose(\bX_0\otimes\bX_0)\transpose\mathrm{diag}(\vect(\expect_0\ell''(\bTheta_0^{(t)})))(\bX_0\otimes\bX_0)\bD_d\delta\br^{(t)}\\
& = -\frac{1}{2}\vect(\delta\bR^{(t)})\transpose\sum_{i\neq j}\expect_0\ell''(\theta_{0ij}^{(t)}; A_{ij}^{(t)})(\bx_{0j}\otimes\bx_{0i})(\bx_{0j}\otimes\bx_{0i})\transpose\vect(\delta\bR^{(t)})\\
& = -\frac{1}{2} \sum_{i\neq j}\expect_0\ell''(\theta_{0ij}^{(t)}; A_{ij}^{(t)})(\bx_{0i}\transpose\delta\bR^{(t)}\bx_{0j})^2.
\end{align*}
For $\bH_{\bB\bB}$, by \eqref{eqn:hessian_B_B}, for any $\delta\bB = [\delta B_{pq}]_{(n - 1 - d)\times d}\in\mathbb{R}^{(n - 1 - d)\times d}$, we have
\begin{align*}
-\vect(\delta\bB)\transpose\bH_{\bB\bB}(\bphi_0)\vect(\delta\bB)
& = -\frac{1}{2}\sum_{p, r = 1}^{n - 1 - d}\sum_{q, s = 1}^d\sum_{t = 1}^m\one_n\transpose\bigg\{\expect_0\ell''(\bTheta^{(t)}_0)\circ \frac{\partial\bTheta^{(t)}}{\partial B_{pq}}(\bphi_0)\circ\frac{\partial\bTheta^{(t)}}{\partial B_{rs}}(\bphi_0)\bigg\}\one_n\delta B_{pq}\delta B_{rs}\\
& = -\frac{1}{2}\sum_{p, r = 1}^{n - 1 - d}\sum_{q, s = 1}^d\sum_{t = 1}^m\sum_{i\neq j}\expect_0\ell''(\theta_{0ij}^{(t)}; A_{ij}^{(t)})\frac{\partial\theta_{ij}^{(t)}}{\partial B_{pq}}(\bphi_0)\frac{\partial\theta_{ij}^{(t)}}{\partial B_{rs}}(\bphi_0)\delta B_{pq}\delta B_{rs}\\
& = -\frac{1}{2}\sum_{t = 1}^m\sum_{i\neq j}\expect_0\ell''(\theta_{0ij}^{(t)}; A_{ij}^{(t)})\bigg\{\sum_{p = 1}^{n - 1 - d}\sum_{q = 1}^d\frac{\partial\theta_{ij}^{(t)}}{\partial B_{pq}}(\bphi_0)\delta B_{pq}\bigg\}^2\\
& = -\frac{1}{2}\sum_{t = 1}^m\sum_{i\neq j}\expect_0\ell''(\theta_{0ij}^{(t)}; A_{ij}^{(t)})\left\{
\be_i\transpose(\bX_{0\perp}\delta\bB\bR_0^{(t)}\bX_0\transpose + \bX_0\bR_0^{(t)}\delta\bB\transpose\bX_{0\perp}\transpose)\be_j
\right\}^2.
\end{align*}
Then, by \eqref{eqn:hessian_alpha_R}, \eqref{eqn:hessian_alpha_B}, and \eqref{eqn:hessian_R_B}, we have
\begin{align*}
\delta\balpha^{(t)\mathrm{T}}\bH_{\balpha^{(t)}\bR^{(t)}}(\bphi_0)\delta\br^{(t)}
& = \frac{1}{2}\sum_{i\neq j}\expect_0\{\ell''(\theta_{0ij}^{(t)}; A_{ij}^{(t)})\}(\delta\alpha_i^{(t)} + \delta\alpha_j^{(t)})\{\bx_{0j}\otimes\bx_{0i}\}\transpose\bD_d\delta\br^{(t)}\\
& = \frac{1}{2}\sum_{i\neq j}\expect_0\{\ell''(\theta_{0ij}^{(t)}; A_{ij}^{(t)})\}(\delta\alpha_i^{(t)} + \delta\alpha_j^{(t)})(\bx_{0i}\transpose\delta\bR^{(t)}\bx_{0j}),\\
\delta\balpha^{(t)\mathrm{T}}\bH_{\balpha^{(t)}\bB}(\bphi_0)\vect(\delta\bB)
& = \frac{1}{2}\sum_{i\neq j}\expect_0\{\ell''(\theta_{0ij}^{(t)}; A_{ij}^{(t)})\}(\delta\alpha_i^{(t)} + \delta\alpha_j^{(t)})(\be_i\transpose\bX_{0\perp}\delta\bB\bR^{(t)}_0\bx_{0j} + \bx_{0i}\transpose\bR_0^{(t)}\delta\bB\transpose\bX_{0\perp}\transpose\be_j),\\
\delta\br^{(t)\mathrm{T}}\bH_{\bR^{(t)}\bB}(\bphi_0)\vect(\delta\bB)
& = \frac{1}{2}\sum_{i\neq j}\expect_0\{\ell''(\theta_{0ij}^{(t)}; A_{ij}^{(t)})\}(\bx_{0i}\transpose\delta\bR^{(t)}\bx_{0j})\be_i\transpose(\bX_{0\perp}\delta\bB\bR_0^{(t)}\bX_0\transpose + \bX_0\bR^{(t)}_0\delta\bB\transpose\bX_{0\perp}\transpose)\be_j.
\end{align*}
Combined with Assumption \ref{assumption:edge_distribution} (recall that $|\theta_{0ij}^{(t)}|\leq b$), we obtain
\begin{align*}
-\delta\bphi\transpose\bH(\bphi_0)\delta\bphi
& \geq \frac{\kappa_1(b)}{2}\sum_{t = 1}^m\sum_{i \neq j}\left\{\delta\alpha_i^{(t)} + \delta\alpha_j^{(t)} + \bx_{0i}\transpose\delta\bR^{(t)}\bx_{0j} + \be_i\transpose(\bX_{0\perp}\delta\bB\bR_0^{(t)}\bX_0\transpose + \bX_0\bR^{(t)}_0\delta\bB\transpose\bX_{0\perp}\transpose)\be_j\right\}^2\\
& = \frac{\kappa_1(b)}{2}\sum_{t = 1}^m\left\|
  (\delta\balpha^{(t)}\one_n\transpose + \one_n\delta\balpha^{(t)\mathrm{T}}) + \bX_0\delta\bR^{(t)}\bX_0\transpose + (\bX_{0\perp}\delta\bB\bR_0^{(t)}\bX_0\transpose + \bX_0\bR_0^{(t)}\delta\bB\transpose\bX_{0\perp}\transpose)
\right\|_{\mathrm{F}}^2\\
&\quad - \frac{\kappa_1(b)}{2}\sum_{t = 1}^m\sum_{i = 1}^n(2\delta\alpha_i^{(t)} + \bx_{0i}\transpose\delta\bR^{(t)}\bx_{0i} + 2\be_i\transpose\bX_{0\perp}\delta\bB\bR_0^{(t)}\bX_0\transpose\be_i)^2
\end{align*}
and
\begin{align*}
-\delta\bphi\transpose\bH(\bphi_0)\delta\bphi
& \leq \frac{\kappa_2(b)}{2}\sum_{t = 1}^m\sum_{i \neq j}\left\{\delta\alpha_i^{(t)} + \delta\alpha_j^{(t)} + \bx_{0i}\transpose\delta\bR^{(t)}\bx_{0j} + \be_i\transpose(\bX_{0\perp}\delta\bB\bR_0^{(t)}\bX_0\transpose + \bX_0\bR^{(t)}_0\delta\bB\transpose\bX_{0\perp}\transpose)\be_j\right\}^2\\
& \leq \frac{\kappa_2(b)}{2}\sum_{t = 1}^m\left\|
  (\delta\balpha^{(t)}\one_n\transpose + \one_n\delta\balpha^{(t)\mathrm{T}}) + \bX_0\delta\bR^{(t)}\bX_0\transpose + (\bX_{0\perp}\delta\bB\bR_0^{(t)}\bX_0\transpose + \bX_0\bR_0^{(t)}\delta\bB\transpose\bX_{0\perp}\transpose)
\right\|_{\mathrm{F}}^2.
\end{align*}
For the first terms in the above two inequalities, observe the mutual orthogonality among $\delta\balpha^{(t)}\one_n\transpose + \one_n\delta\balpha^{(t)\mathrm{T}}$, $\bX_0\delta\bR^{(t)}\bX_0\transpose$, and 
$\bX_{0\perp}\delta\bB\bR_0^{(t)}\bX_0\transpose + \bX_0\bR_0^{(t)}\delta\bB\transpose\bX_{0\perp}\transpose$ under the Frobenius inner product since $\bX_0\transpose\one_n$, $\bX_0\transpose\bX_{0\perp}$, and $\bX_{0\perp}\transpose\one_n$ are zero, implying that
\begin{align*}
&\left\|
  (\delta\balpha^{(t)}\one_n\transpose + \one_n\delta\balpha^{(t)\mathrm{T}}) + \bX_0\delta\bR^{(t)}\bX_0\transpose + (\bX_{0\perp}\delta\bB\bR_0^{(t)}\bX_0\transpose + \bX_0\bR_0^{(t)}\delta\bB\transpose\bX_{0\perp}\transpose)
\right\|_{\mathrm{F}}^2\\
&\quad = 
\|\delta\balpha^{(t)}\one_n\transpose + \one_n\delta\balpha^{(t)\mathrm{T}}\|_{\mathrm{F}}^2 + \|\bX_0\delta\bR^{(t)}\bX_0\transpose\|_{\mathrm{F}}^2 + \|\bX_{0\perp}\delta\bB\bR_0^{(t)}\bX_0\transpose + \bX_0\bR_0^{(t)}\delta\bB\transpose\bX_{0\perp}\transpose\|_{\mathrm{F}}^2\\
&\quad = 2(\one_n\transpose\delta\balpha^{(t)})^2 + 2n\|\delta\balpha^{(t)}\|_2^2 + n^2\|\delta\bR^{(t)}\|_{\mathrm{F}}^2 + 2n^2\|\delta\bB\bR_0^{(t)}\|_{\mathrm{F}}^2\\
&\quad \geq 2n\|\delta\balpha^{(t)}\|_2^2 + n^2\|\delta\br^{(t)}\|_2^2 + 2c_2^2n^2\|\delta\bB\|_{\mathrm{F}}^2
\end{align*}
and 
\begin{align*}
&\left\|
  (\delta\balpha^{(t)}\one_n\transpose + \one_n\delta\balpha^{(t)\mathrm{T}}) + \bX_0\delta\bR^{(t)}\bX_0\transpose + (\bX_{0\perp}\delta\bB\bR_0^{(t)}\bX_0\transpose + \bX_0\bR_0^{(t)}\delta\bB\transpose\bX_{0\perp}\transpose)
\right\|_{\mathrm{F}}^2\\
&\quad \leq 4n\|\delta\balpha^{(t)}\|_2^2 + 2n^2\|\delta\br^{(t)}\|_2^2 + 2C_2^2n^2\|\delta\bB\|_{\mathrm{F}}^2.
\end{align*}
For the second term in the first inequality, by Cauchy-Schwarz inequality, we have 
\begin{align*}
&\sum_{i = 1}^n(2\delta\alpha_i^{(t)} + \bx_{0i}\transpose\delta\bR^{(t)}\bx_{0i} + 2\be_i\transpose\bX_{0\perp}\delta\bB\bR_0^{(t)}\bX_0\transpose\be_i)^2\\
&\quad \leq 8\|\delta\balpha^{(t)}\|_2^2 + 4\|\bX_0\|_{2\to\infty}^2\sum_{i = 1}^n\|\bx_{0i}\|_2^2\|\delta\bR^{(t)}\|_{\mathrm{F}}^2 + 16\sum_{i = 1}^n\|\be_i\transpose\bX_{0\perp}\delta\bB\|_2^2\|\bR_0^{(t)}\|_2^2\|\bX_0\|_{2\to\infty}^2\\
&\quad \leq O(1)\|\delta\balpha^{(t)}\|_2^2 + O(nd)\|\delta\br^{(t)}\|_2^2 + O(n)\|\delta\bB\|_{\mathrm{F}}^2. 
\end{align*}
Hence, we conclude that for sufficiently large $n$,
\begin{align*}
- \bH(\bphi_0)
&\succeq \frac{\kappa_1(b)}{2}
\begin{bmatrix}
  n\eye_{mn} & & \\ 
& n^2/2\eye_{md(d + 1)/2} & \\
& & c_2^2mn^2\eye_{d_\bB}
\end{bmatrix}
\end{align*}
and
\begin{align*}
-\bH(\bphi_0)
&\preceq \frac{\kappa_2(b)}{2}
\begin{bmatrix}
  4n\eye_{mn} & & \\ 
& 2n^2\eye_{md(d + 1)/2} & \\
& & 2C_2^2mn^2\eye_{d_\bB}
\end{bmatrix}.
\end{align*}
The proof is thus completed.
\end{proof}

\subsection{Spectral properties of the sample Hessian at truth}
\label{sub:spectral_properties_of_the_sample_hessian_at_truth}
We slightly extend the spectral bound of the expected Hessian in Proposition \ref{prop:expected_hessian_bound} to the sample Hessian evaluated at the truth $\bphi = \bphi_0$ below, the proof of which is a simple application of the matrix Bernstein's inequality (Lemma \ref{lemma:matrix_Bernstein}; also see Theorem 6.2 in \cite{tropp2012user}). 
\begin{proposition}\label{prop:hessian_bound}
Under Assumptions \ref{assumption:edge_distribution}--\ref{assumption:true_parameter}, for any $b > 0$ such that 
\[
\max_{t\in[m],i,j\in[n]}|\theta_{0ij}^{(t)}|\leq b
\]
and sufficiently large $n$, we have
\begin{align*}
\frac{\kappa_1(b)}{4}
\begin{bmatrix}
  n\eye_{mn} & & \\ 
& n^2/2\eye_{md(d + 1)/2} & \\
& & c_2^2mn^2\eye_{d_\bB}
\end{bmatrix}
\preceq - \nabla^2\calL(\bphi_0)
\preceq \kappa_2(b)
\begin{bmatrix}
  4n\eye_{mn} & & \\ 
& 2n^2\eye_{md(d + 1)/2} & \\
& & 2C_2^2mn^2\eye_{d_\bB}
\end{bmatrix}
\end{align*}
occurs w.h.p.
\end{proposition}

\begin{proof}[\bf Proof]
By the calculation in Section \ref{sub:gradient_hessian}, we decompose $\nabla^2\calL(\bphi_0) = \bGamma(\bphi_0) + \bDelta$, where
\begin{align}
\label{eqn:Gamma_matrix}
\bGamma(\bphi_0) & := \frac{1}{2}\sum_{t = 1}^m\sum_{i\neq j}\ell''(\theta_{0ij}^{(t)}; A_{ij}^{(t)})\bigg\{\frac{\partial\theta_{ij}^{(t)}}{\partial\bphi}(\bphi_0)\bigg\}\bigg\{\frac{\partial\theta_{ij}^{(t)}}{\partial\bphi}(\bphi_0)\bigg\}\transpose,\\
\bDelta  &:= \frac{1}{2}\sum_{t = 1}^m\sum_{i\neq j}\ell'(\theta_{0ij}^{(t)}; A_{ij}^{(t)})\frac{\partial^2\theta_{ij}^{(t)}}{\partial\bphi\partial\bphi\transpose}(\bphi_0).\nonumber
\end{align}
Since Assumption \ref{assumption:edge_distribution} gives $\kappa_1(b)\leq -\ell''(\theta_{0ij}^{(t)}; A_{ij}^{(t)})\leq \kappa_2(b)$ for every realization of $A_{ij}^{(t)}$, and the proof of Proposition \ref{prop:expected_hessian_bound} uses $\expect_0\ell''(\theta_{0ij}^{(t)}; A_{ij}^{(t)})$ only through these bounds (both $-\bH(\bphi_0)$ and $-\bGamma(\bphi_0)$ have the quadratic form $\frac{1}{2}\sum_{t = 1}^m\sum_{i\neq j}w_{ij}^{(t)}\{\delta\bphi\transpose\partial\theta_{ij}^{(t)}(\bphi_0)/\partial\bphi\}^2$ with $w_{ij}^{(t)} = -\expect_0\ell''(\theta_{0ij}^{(t)}; A_{ij}^{(t)})$ and $w_{ij}^{(t)} = -\ell''(\theta_{0ij}^{(t)}; A_{ij}^{(t)})$, respectively), the two-sided bound in Proposition \ref{prop:expected_hessian_bound} holds with $\bG(\bphi_0)$ replaced by $-\bGamma(\bphi_0)$ for every realization of $(\bA^{(t)})_{t = 1}^m$. Clearly, $\bDelta$ has only three non-zero blocks:
\begin{align*}
\bDelta_{\bB\bR}
& = [\bDelta_{\bB\bR^{(1)}},\ldots,\bDelta_{\bB\bR^{(m)}}] = \bDelta_{\bR\bB}\transpose,\quad
\bDelta_{\bB\bR^{(t)}} = \frac{1}{2}\sum_{i\neq j}\ell'(\theta_{0ij}^{(t)}; A_{ij}^{(t)})\frac{\partial^2\theta_{ij}^{(t)}}{\partial\vect(\bB)\partial\vech(\bR^{(t)})\transpose}(\bphi_0),\\
\bDelta_{\bB\bB}
& = \frac{1}{2}\sum_{t = 1}^m\sum_{i\neq j}\ell'(\theta_{0ij}^{(t)}; A_{ij}^{(t)})\frac{\partial^2\theta_{ij}^{(t)}}{\partial\vect(\bB)\partial\vect(\bB)\transpose}(\bphi_0).
\end{align*}
Next, we will invoke the matrix Bernstein's inequality (Lemma \ref{lemma:matrix_Bernstein}; also see Theorem 6.2 in \cite{tropp2012user}) to control $\|\bDelta_{\bB\bB}\|_2$ and $\|\bDelta_{\bB\bR}\|_2$. First note that each of the summands in $\bDelta_{\bB\bB}$ and $\bDelta_{\bB\bR}$ is a scalar sub-exponential random variable times a deterministic matrix. Then, by the relation between the $\psi_1$-Orlicz norm of a sub-exponential random variable and its $p$th moment, the constant $R$ in Lemma \ref{lemma:matrix_Bernstein} becomes $\sigma$ times the maximum of the spectral norms of these deterministic matrices, and $\bA_k$'s are $2\sigma$ times these matrices. More specifically, leveraging \eqref{eqn:ddtheta_dRdB_zero} and \eqref{eqn:ddtheta_dBdB_zero}, we compute
\begin{align*}
R_{\bB\bB} &= \max_{t\in[m],i,j\in[n]}\bigg\|\frac{\partial^2\theta_{ij}^{(t)}}{\partial\vect(\bB)\partial\vect(\bB)\transpose}(\bphi_0)\bigg\|_2
\leq 2(n + \|\bX_0\|_{2\to\infty}^2)\max_{t\in[m]}\|\bR_0^{(t)}\|_2 = O(n),\\
R_{\bR\bB} &= \sqrt{m}\max_{t\in[m],i,j\in[n]}\bigg\|\frac{\partial^2\theta_{ij}^{(t)}}{\partial\vech(\bR^{(t)})\partial\vect(\bB)\transpose}(\bphi_0)\bigg\|_2
\leq 2\sqrt{2}m^{1/2}\|\bX_0\|_{2\to\infty}\|\bX_{0\perp}\|_2\leq O(\sqrt{mn})
\end{align*}
for the $R$ parameters and
\begin{align*}
\sigma^2_{\bB\bB}
& = \bigg\|\sum_{t = 1}^m\sum_{i < j}\bigg\{\frac{\partial^2\theta_{ij}^{(t)}}{\partial\vect(\bB)\partial\vect(\bB)\transpose}(\bphi_0)\bigg\}^2\bigg\|_2 = O(mn^3),\\
\sigma^2_{\bB\bR}
& \leq \bigg\|\sum_{i < j}\bigg[\frac{\partial^2\theta_{ij}^{(s)}}{\partial\vech(\bR^{(s)})\partial\vect(\bB)\transpose}(\bphi_0)\frac{\partial^2\theta_{ij}^{(t)}}{\partial\vect(\bB)\partial\vech(\bR^{(t)})\transpose}(\bphi_0)\bigg]_{m\times m}\bigg\|_2\\
&\quad + \bigg\|\sum_{t = 1}^m\sum_{i < j}\frac{\partial^2\theta_{ij}^{(t)}}{\partial\vect(\bB)\partial\vech(\bR^{(t)})\transpose}(\bphi_0)\frac{\partial^2\theta_{ij}^{(t)}}{\partial\vech(\bR^{(t)})\partial\vect(\bB)\transpose}(\bphi_0)\bigg\|_2\\
&\leq 2mn^2\max_{i,j\in[n]}\bigg\|\frac{\partial^2\theta_{ij}^{(t)}}{\partial\vech(\bR^{(t)})\partial\vect(\bB)\transpose}(\bphi_0)\bigg\|_2^2 = O(mn^3)
\end{align*}
for the $\sigma^2$ parameters by operator Cauchy-Schwarz inequality, the facts that $\partial^2\theta_{ij}^{(t)}/\partial\vech(\bR^{(t)})\partial\vect(\bB)\transpose$ does not depend on $\bR^{(t)}$, and $\|\one_m\one_m\transpose\|_2 = m$. By Lemma \ref{lemma:matrix_Bernstein}, we obtain
\begin{align*}
\|\bDelta_{\bB\bB}\|_2 = \Optilde(m^{1/2}n^{3/2}\sqrt{\log n}), \quad 
\|\bDelta_{\bR\bB}\|_2 = \Optilde(m^{1/2}n^{3/2}\sqrt{\log n}).
\end{align*}
Namely, 
\begin{align*}
&\left\|\begin{bmatrix}
\frac{1}{\sqrt{n}}\eye_{mn} & &\\
& \frac{1}{n}\eye_{md(d + 1)/2} &\\
& & \frac{1}{m^{1/2}n}\eye_{d_\bB}
\end{bmatrix}\bDelta\begin{bmatrix}
\frac{1}{\sqrt{n}}\eye_{mn} & &\\
& \frac{1}{n}\eye_{md(d + 1)/2} &\\
& & \frac{1}{m^{1/2}n}\eye_{d_\bB}
\end{bmatrix}\right\|_2\\
&\quad = 
\left\|
\begin{bmatrix}
    \mathbf{O} & \mathbf{O} & \mathbf{O}\\
    \mathbf{O} & \mathbf{O} & \frac{1}{\sqrt{m}n^2}\bDelta_{\bR\bB}\\
    \mathbf{O} & \frac{1}{\sqrt{m}n^2}\bDelta_{\bB\bR} & \frac{1}{mn^2}\bDelta_{\bB\bB}
  \end{bmatrix}
\right\|_2\leq \frac{2}{\sqrt{m}n^2}\|\bDelta_{\bR\bB}\|_2 + \frac{1}{mn^2}\|\bDelta_{\bB\bB}\|_2 = \Optilde\bigg(\sqrt{\frac{\log n}{n}}\bigg).
\end{align*}
The proof is then completed by applying the two-sided bound for $-\bGamma(\bphi_0)$ obtained above, since $-\nabla^2\calL(\bphi_0) = -\bGamma(\bphi_0) - \bDelta$. 
\end{proof}

\section{Local spectral bound of the sample Hessian}
\label{sec:local_spectral_properties_of_the_sample_hessian}

Section \ref{sec:intrinsic_geometry_of_multilayer_latent_space_network_model} above characterizes the intrinsic geometry (Fisher information metric) of the multilayer latent space network model. The analyses there are mostly deterministic or static. This section takes a step further and investigates the local spectral bounds of the sample Hessian of the log-likelihood $\calL(\bphi)$ with respect to the compact parameter $\bphi$. 

We first describe the neighborhood of interest for this section. To this end, we introduce the following native-equivalent rescaled norm $\|\cdot\|_\sharp$. For any $\bphi\in\mathbb{R}^{d_\bphi}$, we define
\begin{align}\label{eqn:native_norm}
\|\bphi\|_\sharp = \bigg\{n\sum_{t = 1}^m\|\balpha^{(t)}\|_2^2 + n^2\sum_{t = 1}^m\|\vech(\bR^{(t)})\|_2^2 + mn^2\|\bB\|_{\mathrm{F}}^2\bigg\}^{1/2}.
\end{align}
which, by Proposition~\ref{prop:expected_hessian_bound}, is equivalent up to constants to the Fisher information metric induced by $\bG(\bphi_0)$.
For any $\epsilon, M > 0$, consider the following restricted neighborhood of the truth $\bphi_0$:
\begin{align}
\label{eqn:restricted_neigbhorhood}
\calN(\epsilon, M) := \bigg\{\bphi\in\mathbb{R}^{d_\bphi}:\|\bphi - \bphi_0\|_\sharp \leq \epsilon, \max_{t\in[m]}\|\balpha^{(t)}\|_\infty\leq M, \max_{t\in[m]}\|\br^{(t)}\|_2\leq M, \|\bX_{0\perp}\bB\|_{2\to\infty}\leq M/2\bigg\}
\end{align}
and denote by the native-norm rescaling matrix
\begin{align}
\label{eqn:native_norm_rescaling_matrix}
\bLambda_\sharp: = \mathrm{diag}\bigg(\frac{1}{\sqrt{n}}\eye_{d_\balpha}, \frac{1}{n}\eye_{d_\br}, \frac{1}{\sqrt{m}n}\eye_{d_\bB}\bigg).
\end{align}
The goal of this section is to prove Proposition \ref{prop:local_hessian_bound}, a spectral bound of the sample Hessian locally around $\bphi$ in a restricted neighborhood described in \eqref{eqn:restricted_neigbhorhood} above. Note that the power of Proposition \ref{prop:local_hessian_bound} lies in the fact that it does not require any specific rate of the shrinking radius $\eps_n$ as long as $\eps_n \to 0$ as $n\to\infty$. Therefore, even inside a neighborhood $\calN(\sqrt{m}n\eps_n, M)$ of $\bphi_0$ with a sub-optimal convergence rate $\eps_n$, the log-likelihood function is still strongly concave. This feature will be useful for establishing the uniform norm consistency of the MLE in Section \ref{sec:consistency_MLE}. 
\begin{proposition}\label{prop:local_hessian_bound}
Under Assumptions \ref{assumption:edge_distribution}--\ref{assumption:true_parameter} and the additional condition that $m = O(n^2)$, for any $M > 0$ and any $(\epsilon_n)_{n = 1}^\infty\subset(0, +\infty)$ with $\lim_{n\to\infty}\eps_n = 0$, there exist constants $K_1(M), K_2(M) > 0$ depending on $M$ such that
\begin{align*}
K_1(M)\leq \inf_{\bphi\in\calN(m^{1/2}n\epsilon_n, M)}\lambda_{\min}\left\{-\bLambda_\sharp\nabla^2\calL(\bphi)\bLambda_\sharp\right\}\leq \sup_{\bphi\in\calN(m^{1/2}n\epsilon_n, M)}\lambda_{\max}\left\{-\bLambda_\sharp\nabla^2\calL(\bphi)\bLambda_\sharp\right\}\leq K_2(M)
\end{align*}
occurs w.h.p. 
\end{proposition}

\begin{proof}[\bf Proof]
Observe that for any $\delta\bphi\in\mathbb{R}^{d_\bphi}$, we have
\begin{align*}
  \delta\bphi\transpose\nabla^2\calL(\bphi)\delta\bphi
  & = \frac{1}{2}\sum_{t = 1}^m\sum_{i\neq j}\ell''(\theta_{ij}^{(t)}; A_{ij}^{(t)})\delta\bphi\transpose\bigg\{\frac{\partial\theta_{ij}^{(t)}}{\partial\bphi}(\bphi)\bigg\}\bigg\{\frac{\partial\theta_{ij}^{(t)}}{\partial\bphi}(\bphi)\bigg\}\transpose\delta\bphi\\
  &\quad + \frac{1}{2}\sum_{t = 1}^m\sum_{i\neq j}\ell'(\theta_{ij}^{(t)}; A_{ij}^{(t)})\delta\bphi\transpose\frac{\partial^2\theta_{ij}^{(t)}}{\partial\bphi\partial\bphi\transpose}(\bphi)\delta\bphi
\end{align*}
and that the first term satisfies
\begin{align*}
\kappa_1(2M + 2M^3)\delta\bphi\transpose\bigg\{\frac{\partial\theta_{ij}^{(t)}}{\partial\bphi}(\bphi)\bigg\}\bigg\{\frac{\partial\theta_{ij}^{(t)}}{\partial\bphi}(\bphi)\bigg\}\transpose\delta\bphi
&\leq 
-\ell''(\theta_{ij}^{(t)}; A_{ij}^{(t)})\delta\bphi\transpose\bigg\{\frac{\partial\theta_{ij}^{(t)}}{\partial\bphi}(\bphi)\bigg\}\bigg\{\frac{\partial\theta_{ij}^{(t)}}{\partial\bphi}(\bphi)\bigg\}\transpose\delta\bphi\\
&\leq \kappa_2(2M + 2M^3)\delta\bphi\transpose\bigg\{\frac{\partial\theta_{ij}^{(t)}}{\partial\bphi}(\bphi)\bigg\}\bigg\{\frac{\partial\theta_{ij}^{(t)}}{\partial\bphi}(\bphi)\bigg\}\transpose\delta\bphi
\end{align*}
by Assumption \ref{assumption:edge_distribution}. Therefore, it is sufficient to provide sharp bounds for the following three quantities uniformly over $\bphi\in\calN(m^{1/2}n\epsilon_n, M)$:
\begin{align}
\label{eqn:gram_term_local_spectral_hessian}
&\text{Gram term: }G(\bphi, \delta\bphi) = \sum_{t = 1}^m\sum_{i\neq j}\delta\bphi\transpose\bigg\{\frac{\partial\theta_{ij}^{(t)}}{\partial\bphi}(\bphi)\bigg\}\bigg\{\frac{\partial\theta_{ij}^{(t)}}{\partial\bphi}(\bphi)\bigg\}\transpose\delta\bphi,\\
\label{eqn:score_term_local_spectral_hessian}
&\text{Score term: }J(\bphi, \delta\bphi) = \sum_{t = 1}^m\sum_{i\neq j}\ell'(\theta_{0ij}^{(t)}; A_{ij}^{(t)})\delta\bphi\transpose\frac{\partial^2\theta_{ij}^{(t)}}{\partial\bphi\partial\bphi\transpose}(\bphi)\delta\bphi,\\
\label{eqn:drift_term_local_spectral_hessian}
&\text{Drift term: }D(\bphi, \delta\bphi) = \sum_{t = 1}^m\sum_{i\neq j}\{\ell'(\theta_{ij}^{(t)}; A_{ij}^{(t)}) - \ell'(\theta_{0ij}^{(t)}; A_{ij}^{(t)})\}\delta\bphi\transpose\frac{\partial^2\theta_{ij}^{(t)}}{\partial\bphi\partial\bphi\transpose}(\bphi)\delta\bphi.
\end{align}
Then, the proof of Proposition \ref{prop:local_hessian_bound} is completed by Lemmas \ref{lemma:Gram_perturbation}--\ref{lemma:drift_perturbation} below. 
\end{proof}


\subsection{Local perturbation of the Gram term}
\label{sub:local_perturbation_of_the_gram_term}

\begin{lemma}\label{lemma:Gram_perturbation}
Under the conditions of Proposition \ref{prop:local_hessian_bound}, there exists a constant $C_G$ depending on $M$, $C_1,c_2,C_2$ in Assumption \ref{assumption:true_parameter}, such that 
$|G(\bphi, \delta\bphi) - G(\bphi_0, \delta\bphi)|\leq C_G\epsilon_n G(\bphi_0, \delta\bphi)$ for any $\bphi\in\calN(m^{1/2}n\epsilon_n, M)$ and $\delta\bphi\in\mathbb{R}^{d_\bphi}$, where $G(\bphi, \delta\bphi)$ is defined in \eqref{eqn:gram_term_local_spectral_hessian} and satisfies $\|\delta\bphi\|_\sharp^2\lesssim G(\bphi_0, \delta\bphi)\lesssim \|\delta\bphi\|_\sharp^2$.
\end{lemma}

\begin{proof}[\bf Proof]
It is sufficient to show that there exists some constant $C_\Delta > 0$ depending on $M,C_1,c_2,C_2$, such that
\begin{align}
\label{eqn:Gram_perturbation_inequality}
  \sum_{t = 1}^m\sum_{i\neq j}\bigg[\bigg\{\frac{\partial\theta_{ij}^{(t)}}{\partial\bphi}(\bphi)\bigg\}\transpose\delta\bphi - \bigg\{\frac{\partial\theta_{ij}^{(t)}}{\partial\bphi}(\bphi_0)\bigg\}\transpose\delta\bphi\bigg]^2\leq C_\Delta\epsilon_n^2 G(\bphi_0, \delta\bphi). 
\end{align}
To see this, by Cauchy-Schwarz inequality, we first have
\begin{align*}
  |G(\bphi, \delta\bphi) - G(\bphi_0, \delta\bphi)|
  &\leq \sqrt{C_\Delta}\epsilon_n\sqrt{G(\bphi_0, \delta\bphi)}\sqrt{2G(\bphi, \delta\bphi) + 2G(\bphi_0, \delta\bphi)}.
\end{align*}
With $y = G(\bphi, \delta\bphi)/G(\bphi_0, \delta\bphi)$, we have $(y - 1)^2\leq C_\Delta\epsilon_n^2(2y + 2)$, so that 
\[
y^2 - (2 + 2C_\Delta\epsilon_n^2)y + (1 - 2C_\Delta\epsilon_n^2)\leq 0.
\]
This implies that
\[
|y - (1 + C_\Delta\epsilon_n^2)|\leq \sqrt{4C_\Delta\epsilon_n^2 + C_\Delta^2\epsilon_n^4},
\]
and therefore,
\begin{align*}
|y - 1|\leq C_\Delta\epsilon_n^2 + \sqrt{4C_\Delta\epsilon_n^2 + C_\Delta^2\epsilon_n^4}\leq C_G\epsilon_n
\end{align*}
for some constant $C_G$ depending on $C_\Delta$ for sufficiently large $n$. This implies the assertion of the lemma. Hence, it is now sufficient to focus on proving \eqref{eqn:Gram_perturbation_inequality}. By Cauchy-Schwarz inequality and the fact that $\partial\theta_{ij}^{(t)}/\partial\balpha^{(t)}$ is constant, we obtain
\[
  \sum_{t = 1}^m\sum_{i\neq j}\bigg[\bigg\{\frac{\partial\theta_{ij}^{(t)}}{\partial\bphi}(\bphi)\bigg\}\transpose\delta\bphi - \bigg\{\frac{\partial\theta_{ij}^{(t)}}{\partial\bphi}(\bphi_0)\bigg\}\transpose\delta\bphi\bigg]^2
  \leq 2\Sigma_\bR + 8\Sigma_\bB,
\]
where
\begin{align*}
  \Sigma_\bR &= \sum_{t = 1}^m\sum_{i\neq j}\{\be_i\transpose\bX(\bB)\delta\bR^{(t)}\bX(\bB)\transpose\be_j - \bx_{0i}\transpose\delta\bR^{(t)}\bx_{0j}\}^2,\\
  \Sigma_\bB &= \sum_{t = 1}^m\sum_{i\neq j}\{\be_i\transpose\dot{\bX}(\bB)[\delta\bB]\bR^{(t)}\bX(\bB)\transpose\be_j - \be_i\transpose\dot{\bX}(\mathbf{O})[\delta\bB]\bR_0^{(t)}\bx_{0j}\}^2.
\end{align*}
For $\Sigma_\bR$, we have
\begin{align*}
\Sigma_\bR
&\leq 3\sum_{t = 1}^m\sum_{i\neq j}[\{\be_i\transpose\bX(\bB) - \bx_{0i}\transpose\}\delta\bR^{(t)}\{\bX(\bB)\transpose\be_j - \bx_{0j}\}]^2 + 3\sum_{t = 1}^m\sum_{i\neq j}[\bx_{0i}\transpose\delta\bR^{(t)}\{\bX(\bB)\transpose\be_j - \bx_{0j}\}]^2\\
&\quad + 3\sum_{t = 1}^m\sum_{i\neq j}[\{\be_i\transpose\bX(\bB) - \bx_{0i}\transpose\}\delta\bR^{(t)}\bx_{0j}]^2\\
&\leq 6\{\|\bX(\bB) - \bX_0\|_2^4 + 2n\|\bX(\bB) - \bX_0\|_2^2\}\|\delta\br\|_2^2.
\end{align*}
Since $\|\bphi - \bphi_0\|_\sharp < m^{1/2}n\epsilon_n$, we obtain $\|\bB\|_2 < \eps_n$ and $\|\br - \br_0\|_2 < m^{1/2}\eps_n$ immediately, where $\br = [\vech(\bR^{(1)})\transpose,\ldots,\vech(\bR^{(m)})\transpose]\transpose$ and $\br_0 = [\vech(\bR_0^{(1)})\transpose,\ldots,\vech(\bR_0^{(m)})\transpose]\transpose$. 
By mean-value theorem and Lemma \ref{lemma:Lipschitz_Cayley_transform}, we have 
\[
\|\bX(\bB) - \bX_0\|_2\leq \int_0^1\|\dot{\bX}(s\bB)[\bB]\|_2\mathrm{d}s\leq 2\sqrt{n}\|\bB\|_2
\]
for sufficiently large $n$, so that $\Sigma_\bR\leq 144n^2\eps_n^2\|\delta\br\|_2^2$, where the integral applies elementwise. For convenience, we denote by $\dot{\bx}_i = \dot{\bX}(\bB)[\delta\bB]\transpose\be_i$ and $\bx_j = \bX(\bB)\transpose\be_j$, hiding the dependency on $\bB$ and $\delta\bB$, and write $\dot{\bx}_{0i} := \dot{\bX}(\mathbf{O})[\delta\bB]\transpose\be_i$. Then, 
\begin{align*}
\dot{\bx}_i\transpose\bR^{(t)}\bx_j - \dot{\bx}_{0i}\transpose\bR_0^{(t)}\bx_{0j}
& = (\dot{\bx}_i - \dot{\bx}_{0i})\transpose\bR_0^{(t)}\bx_{0j} + \dot{\bx}_{0i}\transpose(\bR^{(t)} - \bR_0^{(t)})\bx_{0j} + \dot{\bx}_{0i}\transpose\bR^{(t)}_0(\bx_j - \bx_{0j})\\
&\quad + (\dot{\bx}_i - \dot{\bx}_{0i})\transpose(\bR^{(t)} - \bR_0^{(t)})\bx_{0j} + (\dot{\bx}_i - \dot{\bx}_{0i})\transpose\bR_0^{(t)}(\bx_{j} - \bx_{0j})\\
&\quad + \dot{\bx}_{0i}\transpose(\bR^{(t)} - \bR_0^{(t)})(\bx_j - \bx_{0j}) + (\dot{\bx}_i - \dot{\bx}_{0i})\transpose(\bR^{(t)} - \bR_0^{(t)})(\bx_j - \bx_{0j}). 
\end{align*}
By mean-value theorem and Lemma \ref{lemma:Lipschitz_Cayley_transform} again, 
\[
\|\dot{\bX}(\bB)[\delta\bB] - \dot{\bX}(\mathbf{O})[\delta\bB]\|_2\leq \int_0^1\|\ddot{\bX}(s\bB)[\delta\bB, \bB]\|_2\mathrm{d}s\leq 2\sqrt{n}\eps_n\|\delta\bB\|_2.
\]
It follows immediately that
\begin{align*}
  \Sigma_\bB&\leq 7\sum_{t = 1}^m\|\dot{\bX}(\bB)[\delta\bB] - \dot{\bX}(\mathbf{O})[\delta\bB]\|_2^2\\
  &\quad\quad\times(\|\bR_0^{(t)}\|_{\mathrm{F}}^2\|\bX_0\|_2^2 + \|\bR^{(t)} - \bR_0^{(t)}\|_2^2\|\bX_0\|_2^2 + \|\bR^{(t)} - \bR_0^{(t)}\|_{\mathrm{F}}^2\|\bX(\bB) - \bX_0\|_2^2 + \|\bR_0^{(t)}\|_2^2\|\bX(\bB) - \bX_0\|_2^2)\\
  &\quad + 7\sum_{t = 1}^m\|\bR^{(t)} - \bR_0^{(t)}\|_2^2(\|\dot{\bX}(\mathbf{O})[\delta\bB]\|_2^2\|\bX_0\|_2^2 + \|\dot{\bX}(\mathbf{O})[\delta\bB]\|_2^2\|\bX(\bB) - \bX_0\|_2^2)\\
  &\quad + 7\sum_{t = 1}^m\|\dot{\bX}(\mathbf{O})[\delta\bB]\|_2^2\|\bR_0^{(t)}\|_{\mathrm{F}}^2\|\bX(\bB) - \bX_0\|_2^2 = O(mn^2\eps_n^2\|\delta\bB\|_{\mathrm{F}}^2).
\end{align*}
Hence, we conclude that $2\Sigma_\bR + 8\Sigma_\bB \leq C\eps_n^2\|\delta\bphi\|_\sharp^2$ for some constant $C > 0$. On the other hand, the proof of Proposition \ref{prop:expected_hessian_bound} clearly shows that $\|\delta\bphi\|_\sharp^2\lesssim G(\bphi_0, \delta\bphi) \lesssim \|\delta\bphi\|_\sharp^2$. The proof of \eqref{eqn:Gram_perturbation_inequality} is thus completed. 
\end{proof}

\subsection{Uniform bound of the score term}
\label{sub:uniform_bound_of_the_score_term}

\begin{lemma}\label{lemma:score_uniform_bound}
Under the conditions of Proposition \ref{prop:local_hessian_bound}, let
\[
\bM_J(\bphi) := \sum_{t = 1}^m\sum_{i\neq j}\ell'(\theta_{0ij}^{(t)}; A_{ij}^{(t)})\frac{\partial^2\theta_{ij}^{(t)}}{\partial\bphi\partial\bphi\transpose}(\bphi),
\]
so that $J(\bphi, \delta\bphi) = \delta\bphi\transpose\bM_J(\bphi)\delta\bphi$ for the score term \eqref{eqn:score_term_local_spectral_hessian}. Then
\[
\sup_{\bphi\in\calN(m^{1/2}n\epsilon_n, M)}\|\bLambda_\sharp\{\bM_J(\bphi) - \bM_J(\bphi_0)\}\bLambda_\sharp\|_2 = \Optilde\bigg(\eps_n\sqrt{\frac{\log n}{n}}\bigg)
\quad\text{and}\quad
\|\bLambda_\sharp\bM_J(\bphi_0)\bLambda_\sharp\|_2 = \Optilde\bigg(\sqrt{\frac{\log n}{n}}\bigg).
\]
In particular, 
\[
\sup_{\bphi\in\calN(m^{1/2}n\epsilon_n, M)}\sup_{\delta\bphi\neq\zero}\frac{|J(\bphi, \delta\bphi)|}{\|\delta\bphi\|_\sharp^2} = \Optilde\bigg(\sqrt{\frac{\log n}{n}}\bigg).
\]
\end{lemma}

\begin{proof}[\bf Proof]
By the computation in Section \ref{sub:gradient_hessian}, for any $\delta\bphi\in\mathbb{R}^{d_\bphi}$, we have
\begin{align*}
  \delta\bphi\transpose\frac{\partial^2\theta_{ij}^{(t)}}{\partial\bphi\partial\bphi\transpose}(\bphi)\delta\bphi
  & = 2\dot{\bx}_i\transpose\delta\bR^{(t)}\bx_j + 2\bx_i\transpose\delta\bR^{(t)}\dot{\bx}_j + 2\dot{\bx}_i\transpose\bR^{(t)}\dot{\bx}_j + \ddot{\bx}_i\transpose\bR^{(t)}\bx_j + \bx_i\transpose\bR^{(t)}\ddot{\bx}_j,
\end{align*}
where we take $\delta\bR^{(t)}\in\mathrm{Sym}(d)$ such that $\vect(\delta\bR^{(t)}) = \bD_d\delta\br^{(t)}$ and denote $\bx_i,\dot{\bx}_i,\ddot{\bx}_i$ the $i$th rows of $\bX(\bB),\dot{\bX}(\bB)[\delta\bB],\ddot{\bX}(\bB)[\delta\bB, \delta\bB]$ (defined in \eqref{eqn:dot_notations}), respectively, viewed as column vectors. Note that we suppress the dependency of $\bx_i$ on $\bB$ and $\dot{\bx}_i,\ddot{\bx}_i$ on $\bB$ and $\delta\bB$ for notation convenience here. Also, denote by $\bx_{0\perp i} = \bX_{0\perp}\transpose\be_i$ the $i$th row of $\bX_{0\perp}$ (viewed as a column vector), $\Delta\bR^{(t)} = \bR^{(t)} - \bR_0^{(t)}$, and $\bN := \bN(\bB)$, $\dot{\bN} := \dot{\bN}(\bB)[\delta\bB]$, $\ddot{\bN} := \ddot{\bN}(\bB)[\delta\bB, \delta\bB]$, where the notation \eqref{eqn:dot_notations} is applied to $\bN(\bB) = \bS(\bB)^{-1/2}$.

Since $\bX(\bB) = (\bX_0 + \bX_{0\perp}\bB)\bN$, differentiating as in the proof of Lemma \ref{lemma:Lipschitz_Cayley_transform} gives, for every $i\in[n]$,
\[
\bx_i = \bA_{\bx}\bx_{0i} + \bC_{\bx}\transpose\bx_{0\perp i},\quad
\dot{\bx}_i = \bA_{\dot{\bx}}\bx_{0i} + \bC_{\dot{\bx}}\transpose\bx_{0\perp i},\quad
\ddot{\bx}_i = \bA_{\ddot{\bx}}\bx_{0i} + \bC_{\ddot{\bx}}\transpose\bx_{0\perp i},
\]
where
\[
\bA_{\bx} = \bN,\quad \bC_{\bx} = \bB\bN,\quad \bA_{\dot{\bx}} = \dot{\bN},\quad \bC_{\dot{\bx}} = \delta\bB\bN + \bB\dot{\bN},\quad \bA_{\ddot{\bx}} = \ddot{\bN},\quad \bC_{\ddot{\bx}} = 2\delta\bB\dot{\bN} + \bB\ddot{\bN}
\]
do not depend on $i$ and depend on $(\bB, \delta\bB)$ only. At $\bB = \mathbf{O}$, by \eqref{eqn:dX_dB_zero} and \eqref{eqn:d2X_dB2_zero},
\[
\bA_{\bx}|_{\bB = \mathbf{O}} = \eye_d,\quad
\bC_{\bx}|_{\bB = \mathbf{O}} = \mathbf{O},\quad
\bA_{\dot{\bx}}|_{\bB = \mathbf{O}} = \mathbf{O},\quad
\bC_{\dot{\bx}}|_{\bB = \mathbf{O}} = \delta\bB,\quad
\bA_{\ddot{\bx}}|_{\bB = \mathbf{O}} = -\delta\bB\transpose\delta\bB,\quad
\bC_{\ddot{\bx}}|_{\bB = \mathbf{O}} = \mathbf{O}.
\]
Moreover, for $\|\bB\|_2\leq 1$, by Lemma \ref{lemma:integral_formula_matrix_squarerootinverse}, 
\begin{align*}
&\|\bA_{\bx} - \bA_{\bx}|_{\bB = \mathbf{O}}\|_2\lesssim \|\bB\|_2^2,\quad
\|\bC_{\bx} - \bC_{\bx}|_{\bB = \mathbf{O}}\|_2\lesssim \|\bB\|_2,\\
&\|\bA_{\dot{\bx}} - \bA_{\dot{\bx}}|_{\bB = \mathbf{O}}\|_2\lesssim \|\bB\|_2\|\delta\bB\|_2,\quad
\|\bC_{\dot{\bx}} - \bC_{\dot{\bx}}|_{\bB = \mathbf{O}}\|_2\lesssim \|\bB\|_2^2\|\delta\bB\|_2,\\
&\|\bA_{\ddot{\bx}} - \bA_{\ddot{\bx}}|_{\bB = \mathbf{O}}\|_2\lesssim \|\bB\|_2^2\|\delta\bB\|_2^2,\quad
\|\bC_{\ddot{\bx}} - \bC_{\ddot{\bx}}|_{\bB = \mathbf{O}}\|_2\lesssim \|\bB\|_2\|\delta\bB\|_2^2.
\end{align*}
By the formula for $\delta\bphi\transpose\{\partial^2\theta_{ij}^{(t)}/\partial\bphi\partial\bphi\transpose\}(\bphi)\delta\bphi$ and the representations of $\bx_i,\dot{\bx}_i,\ddot{\bx}_i$ above,
\[
\delta\bphi\transpose\frac{\partial^2\theta_{ij}^{(t)}}{\partial\bphi\partial\bphi\transpose}(\bphi)\delta\bphi = \sum_{(a, \bW^{(t)}, b)}(\bA_a\bx_{0i} + \bC_a\transpose\bx_{0\perp i})\transpose\bW^{(t)}(\bA_b\bx_{0j} + \bC_b\transpose\bx_{0\perp j}),
\]
where the sum runs over the five triples
\[
(a, \bW^{(t)}, b)\in\{(\dot{\bx}, 2\delta\bR^{(t)}, \bx),\ (\bx, 2\delta\bR^{(t)}, \dot{\bx}),\ (\dot{\bx}, 2\bR^{(t)}, \dot{\bx}),\ (\ddot{\bx}, \bR^{(t)}, \bx),\ (\bx, \bR^{(t)}, \ddot{\bx})\}.
\]
Denote by
\[
\bK_{\bx\bx}^{(t)} := \sum_{(a, \bW^{(t)}, b)}\bA_a\bW^{(t)}\bA_b,\quad
\bK_{\bx\bx_\perp}^{(t)} := \sum_{(a, \bW^{(t)}, b)}\bA_a\bW^{(t)}\bC_b\transpose,\quad
\bK_{\bx_\perp\bx_\perp}^{(t)} := \sum_{(a, \bW^{(t)}, b)}\bC_a\bW^{(t)}\bC_b\transpose,
\]
which are deterministic and depend on $(\bphi, \delta\bphi)$. Since
\[
\sum_{i\neq j}\ell'(\theta_{0ij}^{(t)}; A_{ij}^{(t)})\bx_{0i}\transpose\bK\bx_{0j} = \mathrm{tr}\{\bK\bX_0\transpose\ell'(\bTheta_0^{(t)})\bX_0\},\quad
\sum_{i\neq j}\ell'(\theta_{0ij}^{(t)}; A_{ij}^{(t)})\bx_{0i}\transpose\bK\bx_{0\perp j} = \mathrm{tr}\{\bK\bX_{0\perp}\transpose\ell'(\bTheta_0^{(t)})\bX_0\},
\]
and $\sum_{i\neq j}\ell'(\theta_{0ij}^{(t)}; A_{ij}^{(t)})\bx_{0\perp i}\transpose\bK\bx_{0\perp j} = \mathrm{tr}\{\bK\bX_{0\perp}\transpose\ell'(\bTheta_0^{(t)})\bX_{0\perp}\}$ for any conformable matrix $\bK$, where $\ell'(\bTheta_0^{(t)})$ is defined in \eqref{eqn:ell_prime_Theta_t} with $\bTheta^{(t)}$ replaced by $\bTheta_0^{(t)}$, we obtain
\begin{align*}
J(\bphi, \delta\bphi) = \sum_{t = 1}^m\Big[\mathrm{tr}\{\bK_{\bx\bx}^{(t)}\bX_0\transpose\ell'(\bTheta_0^{(t)})\bX_0\} + 2\mathrm{tr}\{\bK_{\bx\bx_\perp}^{(t)}\bX_{0\perp}\transpose\ell'(\bTheta_0^{(t)})\bX_0\} + \mathrm{tr}\{\bK_{\bx_\perp\bx_\perp}^{(t)}\bX_{0\perp}\transpose\ell'(\bTheta_0^{(t)})\bX_{0\perp}\}\Big].
\end{align*}

Each of $\bK_{\bx\bx}^{(t)}, \bK_{\bx\bx_\perp}^{(t)}, \bK_{\bx_\perp\bx_\perp}^{(t)}$ is a sum of five summands, which we bound separately below. For $\bK_{\bx\bx}^{(t)}$:
\begin{itemize}
  \item $(\dot{\bx}, 2\delta\bR^{(t)}, \bx)$: $\bA_{\dot{\bx}}(2\delta\bR^{(t)})\bA_{\bx}$, which vanishes at $\bphi_0$ and has spectral norm $O(\|\bB\|_2\|\delta\bB\|_2\|\delta\bR^{(t)}\|_2)$;
  \item $(\bx, 2\delta\bR^{(t)}, \dot{\bx})$: $\bA_{\bx}(2\delta\bR^{(t)})\bA_{\dot{\bx}}$, which vanishes at $\bphi_0$ and has spectral norm $O(\|\bB\|_2\|\delta\bB\|_2\|\delta\bR^{(t)}\|_2)$;
  \item $(\dot{\bx}, 2\bR^{(t)}, \dot{\bx})$: $\bA_{\dot{\bx}}(2\bR^{(t)})\bA_{\dot{\bx}}$, which vanishes at $\bphi_0$ and has spectral norm $O(\|\bB\|_2^2\|\delta\bB\|_2^2)$;
  \item $(\ddot{\bx}, \bR^{(t)}, \bx)$: the difference is 
  \begin{align*}
  &\bA_{\ddot{\bx}}\bR^{(t)}\bA_{\bx} - (-\delta\bB\transpose\delta\bB)\bR_0^{(t)}\eye_d\\
  &\quad = (\bA_{\ddot{\bx}} - \bA_{\ddot{\bx}}|_{\bB = \mathbf{O}})\bR^{(t)}\bA_{\bx} - \delta\bB\transpose\delta\bB\{\Delta\bR^{(t)}\bA_{\bx} + \bR_0^{(t)}(\bA_{\bx} - \bA_{\bx}|_{\bB = \mathbf{O}})\},
  \end{align*}
  whose spectral norm is $O(\|\bB\|_2^2\|\delta\bB\|_2^2 + \|\delta\bB\|_2^2\|\Delta\bR^{(t)}\|_2)$;
  \item $(\bx, \bR^{(t)}, \ddot{\bx})$: the transpose of the previous one; its spectral norm is $O(\|\bB\|_2^2\|\delta\bB\|_2^2 + \|\delta\bB\|_2^2\|\Delta\bR^{(t)}\|_2)$.
\end{itemize}
For $\bK_{\bx\bx_\perp}^{(t)}$:
\begin{itemize}
  \item $(\dot{\bx}, 2\delta\bR^{(t)}, \bx)$: $\bA_{\dot{\bx}}(2\delta\bR^{(t)})\bC_{\bx}\transpose$, which vanishes at $\bphi_0$ and has spectral norm $O(\|\bB\|_2^2\|\delta\bB\|_2\|\delta\bR^{(t)}\|_2)$;
  \item $(\bx, 2\delta\bR^{(t)}, \dot{\bx})$: the difference is
  \begin{align*}
  \bA_{\bx}(2\delta\bR^{(t)})\bC_{\dot{\bx}}\transpose - \eye_d(2\delta\bR^{(t)})\delta\bB\transpose
  = \bA_{\bx}(2\delta\bR^{(t)})(\bC_{\dot{\bx}} - \bC_{\dot{\bx}}|_{\bB = \mathbf{O}})\transpose + (\bA_{\bx} - \bA_{\bx}|_{\bB = \mathbf{O}})(2\delta\bR^{(t)})\delta\bB\transpose,
  \end{align*}
  whose spectral norm is $O(\|\bB\|_2^2\|\delta\bB\|_2\|\delta\bR^{(t)}\|_2)$;
  \item $(\dot{\bx}, 2\bR^{(t)}, \dot{\bx})$: $\bA_{\dot{\bx}}(2\bR^{(t)})\bC_{\dot{\bx}}\transpose$, which vanishes at $\bphi_0$ and has spectral norm $O(\|\bB\|_2\|\delta\bB\|_2^2)$;
  \item $(\ddot{\bx}, \bR^{(t)}, \bx)$: $\bA_{\ddot{\bx}}\bR^{(t)}\bC_{\bx}\transpose$, which vanishes at $\bphi_0$ and has spectral norm $O(\|\bB\|_2\|\delta\bB\|_2^2)$;
  \item $(\bx, \bR^{(t)}, \ddot{\bx})$: $\bA_{\bx}\bR^{(t)}\bC_{\ddot{\bx}}\transpose$, which vanishes at $\bphi_0$ and has spectral norm $O(\|\bB\|_2\|\delta\bB\|_2^2)$.
\end{itemize}
For $\bK_{\bx_\perp\bx_\perp}^{(t)}$:
\begin{itemize}
  \item $(\dot{\bx}, 2\delta\bR^{(t)}, \bx)$: $\bC_{\dot{\bx}}(2\delta\bR^{(t)})\bC_{\bx}\transpose$, which vanishes at $\bphi_0$ and has spectral norm $O(\|\bB\|_2\|\delta\bB\|_2\|\delta\bR^{(t)}\|_2)$;
  \item $(\bx, 2\delta\bR^{(t)}, \dot{\bx})$: $\bC_{\bx}(2\delta\bR^{(t)})\bC_{\dot{\bx}}\transpose$, which vanishes at $\bphi_0$ and has spectral norm $O(\|\bB\|_2\|\delta\bB\|_2\|\delta\bR^{(t)}\|_2)$;
  \item $(\dot{\bx}, 2\bR^{(t)}, \dot{\bx})$: the difference is
  \begin{align*}
  &\bC_{\dot{\bx}}(2\bR^{(t)})\bC_{\dot{\bx}}\transpose - \delta\bB(2\bR_0^{(t)})\delta\bB\transpose\\
  &\quad = (\bC_{\dot{\bx}} - \bC_{\dot{\bx}}|_{\bB = \mathbf{O}})(2\bR^{(t)})\bC_{\dot{\bx}}\transpose + \delta\bB(2\bR^{(t)})(\bC_{\dot{\bx}} - \bC_{\dot{\bx}}|_{\bB = \mathbf{O}})\transpose + \delta\bB(2\Delta\bR^{(t)})\delta\bB\transpose,
  \end{align*}whose spectral norm is $O(\|\bB\|_2^2\|\delta\bB\|_2^2 + \|\delta\bB\|_2^2\|\Delta\bR^{(t)}\|_2)$;
  \item $(\ddot{\bx}, \bR^{(t)}, \bx)$: $\bC_{\ddot{\bx}}\bR^{(t)}\bC_{\bx}\transpose$, which vanishes at $\bphi_0$ and has spectral norm $O(\|\bB\|_2^2\|\delta\bB\|_2^2)$;
  \item $(\bx, \bR^{(t)}, \ddot{\bx})$: $\bC_{\bx}\bR^{(t)}\bC_{\ddot{\bx}}\transpose$, which vanishes at $\bphi_0$ and has spectral norm $O(\|\bB\|_2^2\|\delta\bB\|_2^2)$.
\end{itemize}
Adding up the five summands, since $\|\bB\|_2\leq 1$, for any $\bK^{(t)}\in\{\bK_{\bx\bx}^{(t)}, \bK_{\bx\bx_\perp}^{(t)}, \bK_{\bx_\perp\bx_\perp}^{(t)}\}$, we have
\[
\|\bK^{(t)}(\bphi) - \bK^{(t)}(\bphi_0)\|_2\lesssim \|\bB\|_2\|\delta\bB\|_2^2 + \|\delta\bB\|_2^2\|\Delta\bR^{(t)}\|_2 + \|\bB\|_2\|\delta\bB\|_2\|\delta\bR^{(t)}\|_2.
\]
For $\bphi\in\calN(m^{1/2}n\eps_n, M)$, we have $\|\bB\|_2 < \eps_n$ and $\sum_{t = 1}^m\|\Delta\bR^{(t)}\|_{\mathrm{F}}^2 < 2m\eps_n^2$, and hence, by Cauchy-Schwarz inequality, $\sum_{t = 1}^m\|\Delta\bR^{(t)}\|_2\leq\sqrt{2}m\eps_n$ and $\sum_{t = 1}^m\|\delta\bR^{(t)}\|_2\leq\sum_{t = 1}^m\|\delta\bR^{(t)}\|_{\mathrm{F}}\leq\sqrt{2m}\|\delta\br\|_2$, so that for any $\bK^{(t)}\in\{\bK_{\bx\bx}^{(t)}, \bK_{\bx\bx_\perp}^{(t)}, \bK_{\bx_\perp\bx_\perp}^{(t)}\}$, 
\[
\sup_{\bphi\in\calN(m^{1/2}n\eps_n, M)}\sum_{t = 1}^m\|\bK^{(t)}(\bphi) - \bK^{(t)}(\bphi_0)\|_2\lesssim \eps_n\Big(m\|\delta\bB\|_{\mathrm{F}}^2 + \sqrt{m}\|\delta\bB\|_{\mathrm{F}}\|\delta\br\|_2\Big)\lesssim \frac{\eps_n}{n^2}\|\delta\bphi\|_\sharp^2.
\]

By Assumption \ref{assumption:true_parameter} and Lemma \ref{lemma:matrix_Bernstein}, $\|\ell'(\bTheta_0^{(t)})\|_2 = \Optilde(\sqrt{n\log n})$. Hence, 
\[  
\|\bX_{0\perp}\transpose\ell'(\bTheta_0^{(t)})\bX_{0\perp}\|_2\leq n\|\ell'(\bTheta_0^{(t)})\|_2 = \Optilde(n^{3/2}\sqrt{\log n}),\quad
\|\bX_0\transpose\ell'(\bTheta_0^{(t)})\bX_{0\perp}\|_2 = \Optilde(n^{3/2}\sqrt{\log n}).
\]
By Lemma \ref{lemma:matrix_Bernstein}, $\bX_0\transpose\ell'(\bTheta_0^{(t)})\bX_0 = \sum_{i < j}\ell'(\theta_{0ij}^{(t)}; A_{ij}^{(t)})(\bx_{0i}\bx_{0j}\transpose + \bx_{0j}\bx_{0i}\transpose) = \Optilde(n\sqrt{\log n})$. These bounds hold uniformly over $t\in[m]$ by Lemma \ref{lemma:uniform_Optilde}, since $m = O(n^2)$.

By von-Neumann trace inequality, $|\mathrm{tr}(\bK\bW)|\leq 5d\|\bK\|_2\|\bW\|_2$ and the bounds obtained above, simultaneously for all 
$\delta\bphi\in\mathbb{R}^{d_\bphi}$,
\begin{align*}
&\sup_{\bphi\in\calN(m^{1/2}n\eps_n, M)}|J(\bphi, \delta\bphi) - J(\bphi_0, \delta\bphi)|\\
&\quad\leq 5d\sup_{\bphi\in\calN(m^{1/2}n\eps_n, M)}\sum_{t = 1}^m\Big\{\|\bK_{\bx\bx}^{(t)}(\bphi) - \bK_{\bx\bx}^{(t)}(\bphi_0)\|_2 + 2\|\bK_{\bx\bx_\perp}^{(t)}(\bphi) - \bK_{\bx\bx_\perp}^{(t)}(\bphi_0)\|_2\\
&\quad\quad + \|\bK_{\bx_\perp\bx_\perp}^{(t)}(\bphi) - \bK_{\bx_\perp\bx_\perp}^{(t)}(\bphi_0)\|_2\Big\}\Optilde(n^{3/2}\sqrt{\log n})\\
&\quad = \Optilde\bigg(\eps_n\sqrt{\frac{\log n}{n}}\bigg)\|\delta\bphi\|_\sharp^2 .
\end{align*}
This implies that
\begin{align*}
  \sup_{\bphi\in\calN(m^{1/2}n\eps_n, M)}\|\bLambda_\sharp\{\bM_J(\bphi) - \bM_J(\bphi_0)\}\bLambda_\sharp\|_2
  & = \sup_{\bphi\in\calN(m^{1/2}n\eps_n, M)}\sup_{\delta\bphi\neq\zero}\frac{|J(\bphi, \delta\bphi) - J(\bphi_0, \delta\bphi)|}{\|\delta\bphi\|_\sharp^2}\\
  & = \Optilde\bigg(\eps_n\sqrt{\frac{\log n}{n}}\bigg).
\end{align*}
Since $\bM_J(\bphi) - \bM_J(\bphi_0)$ is symmetric, this is the first assertion. Finally, $\bM_J(\bphi_0) = 2\bDelta$ with $\bDelta$ defined in the proof of Proposition \ref{prop:hessian_bound}, and the proof there gives $\|\bLambda_\sharp\bDelta\bLambda_\sharp\|_2 = \Optilde(\sqrt{\log n/n})$. The proof is thus completed.
\end{proof}

\subsection{Local perturbation of the drift term}
\label{sub:local_perturbation_of_the_drift_term}

\begin{lemma}\label{lemma:drift_perturbation}
Under the conditions of Proposition \ref{prop:local_hessian_bound}, there exists a constant $C_G$ depending on $M$, $C_1,c_2,C_2$ in Assumption \ref{assumption:true_parameter}, such that 
$|D(\bphi, \delta\bphi)|\lesssim \eps_n\|\delta\bphi\|_\sharp^2$.
\end{lemma}

\begin{proof}[\bf Proof]
By definition, for any $\theta_{ij}^{(t)} = \theta_{ij}^{(t)}(\bphi)$ where $\bphi\in\calN(m^{1/2}n\eps_n, M)$, 
\begin{align*}
  |\theta_{ij}^{(t)} - \theta_{0ij}^{(t)}|
  & \leq |\alpha_i^{(t)} - \alpha_{0i}^{(t)}| + |\alpha_j^{(t)} - \alpha_{0j}^{(t)}|
  + |\bx_i(\bB)\transpose\Delta\bR^{(t)}\bx_j(\bB)| + |\bx_i(\bB)\transpose\bR_0^{(t)}\bx_j(\bB) - \bx_{0i}\transpose\bR_0^{(t)}\bx_{0j}|,
\end{align*}
where we follow the notations in the proof of Lemma \ref{lemma:score_uniform_bound}. It follows from mean-value theorem and Lemma \ref{lemma:Lipschitz_Cayley_transform} that
\begin{align*}
\sum_{t = 1}^m\sum_{i\neq j}(\theta_{ij}^{(t)} - \theta_{0ij}^{(t)})^2
&\lesssim n\|\balpha - \balpha_0\|_2^2 + \sum_{t = 1}^m\|\bX\Delta\bR^{(t)}\bX\transpose\|_{\mathrm{F}}^2 + \sum_{t = 1}^m\|\{\bX(\bB) - \bX_0\}\bR_0^{(t)}\bX_0\transpose\|_{\mathrm{F}}^2\lesssim \|\bphi - \bphi_0\|_\sharp^2. 
\end{align*}
Similarly, by Lemma \ref{lemma:Lipschitz_Cayley_transform},
\begin{align*}
\sum_{t = 1}^m\sum_{i\neq j}\bigg(\delta\bphi\transpose\frac{\partial^2\theta_{ij}^{(t)}}{\partial\bphi\partial\bphi\transpose}\delta\bphi\bigg)^2
&\lesssim \sum_{t = 1}^m\|\dot{\bX}(\bB)[\delta\bB]\delta\bR^{(t)}\bX\transpose\|_{\mathrm{F}}^2 + \sum_{t = 1}^m\|\dot{\bX}(\bB)[\delta\bB]\bR^{(t)}\dot{\bX}(\bB)\transpose\|_{\mathrm{F}}^2\\
&\quad + \sum_{t = 1}^m\|\ddot{\bX}(\bB)[\delta\bB, \delta\bB]\bR^{(t)}\bX\transpose\|_{\mathrm{F}}^2\\
&\lesssim mn^2\|\delta\bB\|_2^4 + n^2\|\delta\bB\|_2^2\|\delta\br\|_2^2 \lesssim \frac{\|\delta\bphi\|_\sharp^4}{mn^2}.
\end{align*}
Therefore, by mean-value theorem, Cauchy-Schwarz inequality, and Assumption \ref{assumption:edge_distribution}, we have
\begin{align*}
  |D(\bphi, \delta\bphi)|
  &\leq \kappa_2(2M + 2M^3)\bigg\{\sum_{t = 1}^m\sum_{i\neq j}(\theta_{ij}^{(t)} - \theta_{0ij}^{(t)})^2\bigg\}^{1/2}\bigg\{\sum_{t = 1}^m\sum_{i\neq j}\bigg(\delta\bphi\transpose\frac{\partial^2\theta_{ij}^{(t)}}{\partial\bphi\partial\bphi\transpose}\delta\bphi\bigg)^2\bigg\}^{1/2} \lesssim \eps_n\|\delta\bphi\|_\sharp^2.
\end{align*}
The proof is thus completed. 
\end{proof}

\section{Consistency of the maximum likelihood estimator}
\label{sec:consistency_MLE}

The main goal of this section is to establish the consistency and convergence rates of the MLE defined in Section \ref{sec:maximum_likelihood_and_one_step_estimator} of the main text under both the native-equivalent rescaled norm and the uniform metric in Propositions \ref{prop:consistency_native_norm} and \ref{prop:stationarity_uniqueness_uniform_consistency_MLE} below. 

\begin{proposition}[Native norm consistency]
\label{prop:consistency_native_norm}
Suppose Assumptions \ref{assumption:edge_distribution}--\ref{assumption:true_parameter} hold and $m = O(n^2)$. Let $\widehat{\bpsi}$ be the MLE defined in Section \ref{sec:maximum_likelihood_and_one_step_estimator} of the main text. Then, with $M > 4C_1\vee\sqrt{d}C_2$ and $M = O(1)$, $\bX_0\transpose\widehat{\bX}$ is invertible w.h.p., and there exists an $n$-dependent random $\bW_{\widehat{\bX}} = (\bX_0\transpose\widehat{\bX})^{-1}\{(\bX_0\transpose\widehat{\bX})^{-\mathrm{T}}(\bX_0\transpose\widehat{\bX})^{-1}\}^{-1/2}\in\mathbb{O}(d)$, such that
\begin{align*}
&\sum_{t = 1}^m\|\widehat{\balpha}^{(t)} - \balpha_0^{(t)}\|_2^2 = \Optilde(m),\; \sum_{t = 1}^m\|\bW\transpose_{\widehat{\bX}}\widehat{\bR}^{(t)}\bW_{\widehat{\bX}} - \bR_0^{(t)}\|_{\mathrm{F}}^2 = \Optilde(m/n),\\
&\|\widehat{\bX}\bW_{\widehat{\bX}} - \bX_0\|_2 = \Optilde(1),\; 
\|\widehat{\bB}\|_2 = \Optilde(n^{-1/2}),\quad\text{where }\widehat{\bB} = \bX_{0\perp}\transpose\widehat{\bX}(\bX_0\transpose\widehat{\bX})^{-1}.
\end{align*}
\end{proposition}

\begin{proposition}[Uniform norm consistency]
\label{prop:stationarity_uniqueness_uniform_consistency_MLE}
Suppose Assumptions \ref{assumption:edge_distribution}--\ref{assumption:true_parameter} hold and $m^{12}(\log n)^4/n \to 0$ as $n\to\infty$. Let $\widehat{\bpsi}$ be the MLE defined in Section \ref{sec:maximum_likelihood_and_one_step_estimator} of the main text and $\widehat{\bphi} = \argmax_{\bphi:\bpsi(\bphi)\in\Psi}\calL(\bphi)$. Then, with $M > 4C_1\vee\sqrt{d}C_2$ and $M = O(1)$, we have,
\begin{align*}
\widehat{\bphi}\text{ is unique, }\widehat{\bphi} = \bphi(\widehat{\bpsi}),\text{ and }\frac{\partial\calL}{\partial\bphi}(\widehat{\bphi}) = \zero_{d_\bphi}\text{ w.h.p.}
\end{align*}
Also, with $\widehat{\balpha}^{(t)}$, $\widehat{\bR}^{(t)}$, $\widehat{\bX} =\bX(\widehat{\bB})$ attached to $\widehat{\bphi}$, we have
\begin{align*}
  \max_{t\in[m]}\left(\|\widehat{\balpha}^{(t)} - \balpha_0^{(t)}\|_\infty + \|\widehat{\bR}^{(t)} - \bR_0^{(t)}\|_{\mathrm{F}}\right)
   + \|\widehat{\bX} - \bX_0\|_{2\to\infty} = \Optilde\bigg(\frac{m^{2}\sqrt{\log n} + m^{3/2}\log n}{\sqrt{n}}\bigg).
\end{align*}
\end{proposition}

\subsection{Consistency under native norm}
\label{sub:consistency_under_native_norm}

\begin{proof}[\bf Proof of Proposition \ref{prop:consistency_native_norm}]
Let $\widehat{\bTheta}^{(t)} = \widehat{\balpha}^{(t)}\one_n\transpose + \one_n\widehat{\balpha}^{(t)\mathrm{T}} + \widehat{\bX}\widehat{\bR}^{(t)}\widehat{\bX}\transpose$ and let $\widehat{\theta}_{ij}^{(t)}$ denote the $(i, j)$th element of $\widehat{\bTheta}^{(t)}$. By Taylor's theorem, 
\begin{align*}
\sum_{t = 1}^m\sum_{i < j}\ell(\widehat{\theta}_{ij}^{(t)}; A_{ij}^{(t)})
& = \sum_{t = 1}^m\sum_{i < j}\ell(\theta_{0ij}^{(t)}; A_{ij}^{(t)}) + \sum_{t = 1}^m\sum_{i < j}\ell'(\theta_{0ij}^{(t)}; A_{ij}^{(t)})(\widehat{\theta}_{ij}^{(t)} - \theta_{0ij}^{(t)})\\ 
&\quad + \frac{1}{2}
\sum_{t = 1}^m\sum_{i < j}\ell''(\widetilde{\theta}_{ij}^{(t)}; A_{ij}^{(t)})(\widehat{\theta}_{ij}^{(t)} - \theta_{0ij}^{(t)})^2,
\end{align*}
where $\widetilde{\theta}_{ij}^{(t)}$ lies between $\theta_{0ij}^{(t)}$ and $\widehat{\theta}_{ij}^{(t)}$. Since $\widehat{\bTheta}^{(t)}$ is the MLE, it follows from Assumption \ref{assumption:edge_distribution} that
\begin{align*}
  2\sum_{t = 1}^m\sum_{i < j}\ell'(\theta_{0ij}^{(t)}; A_{ij}^{(t)})(\widehat{\theta}_{ij}^{(t)} - \theta_{0ij}^{(t)})&\geq \frac{\kappa_1(2M + 2M^3)}{2}\sum_{t = 1}^m\sum_{i \neq j} (\widehat{\theta}_{ij}^{(t)} - \theta_{0ij}^{(t)})^2\\
  & = \frac{\kappa_1(2M + 2M^3)}{2}\sum_{t = 1}^m\|\widehat{\bTheta}^{(t)} - \bTheta_0^{(t)}\|_{\mathrm{F}}^2 - |O(mn)|.
\end{align*}
On the other hand, by von-Neumann trace inequality, Cauchy-Schwarz inequality, and Lemma 4.3 of \cite{10.1214/11-AOP734},
\begin{align*}
  2\sum_{t = 1}^m\sum_{i < j}\ell'(\theta_{0ij}^{(t)}; A_{ij}^{(t)})(\widehat{\theta}_{ij}^{(t)} - \theta_{0ij}^{(t)})
  \lesssim \sum_{t = 1}^m\|\ell'(\bTheta_0^{(t)})\|_2\|\widehat{\bTheta}^{(t)} - \bTheta_0^{(t)}\|_{\mathrm{F}}\leq \Optilde(\sqrt{mn})\bigg(\sum_{t = 1}^m\|\widehat{\bTheta}^{(t)} - \bTheta_0^{(t)}\|_{\mathrm{F}}^2\bigg)^{1/2}.
\end{align*}
Combining the two inequalities above, we further obtain
\begin{align*}
\sum_{t = 1}^m\|\widehat{\bTheta}^{(t)} - \bTheta_0^{(t)}\|_{\mathrm{F}}^2 - B_n\bigg(\sum_{t = 1}^m\|\widehat{\bTheta}^{(t)} - \bTheta_0^{(t)}\|_{\mathrm{F}}^2\bigg)^{1/2} - C_n\leq 0,
\end{align*}
where $B_n \geq 0$ with $B_n = \Optilde(\sqrt{mn})$ and $C_n \geq 0$ with $C_n = O(mn)$. This implies that
\[
\bigg(\sum_{t = 1}^m\|\widehat{\bTheta}^{(t)} - \bTheta_0^{(t)}\|_{\mathrm{F}}^2\bigg)^{1/2}\leq \frac{B_n + \sqrt{B_n^2 + 4C_n}}{2} = \Optilde(\sqrt{mn}). 
\]
Exploiting the structure of $\|\widehat{\bTheta}^{(t)} - \bTheta_0^{(t)}\|_{\mathrm{F}}^2$, we then have
\begin{align*}
\|\widehat{\bTheta}^{(t)} - \bTheta_0^{(t)}\|_{\mathrm{F}}^2
& \geq \|\widehat{\bX}\widehat{\bR}^{(t)}\widehat{\bX}\transpose - \bX_0\bR_0^{(t)}\bX_0\transpose\|_{\mathrm{F}}^2 + 2n\|\widehat{\balpha}^{(t)} - \balpha_0^{(t)}\|_2^2.
\end{align*}
This immediately yields $\sum_{t = 1}^m\|\widehat{\balpha}^{(t)} - \balpha_0^{(t)}\|_2^2 = \Optilde(m)$. 
For the first term, simple algebra shows that
\begin{align*}
  \|\widehat{\bX}(\widehat{\bR}^{(t)})^2\widehat{\bX}\transpose - \bX_0(\bR_0^{(t)})^2\bX_0\transpose\|_{\mathrm{F}}
  \lesssim \|\widehat{\bX}\widehat{\bR}^{(t)}\widehat{\bX}\transpose - \bX_0\bR_0^{(t)}\bX_0\transpose\|_{\mathrm{F}}
\end{align*}
and
\begin{align*}
&\sum_{t = 1}^m\|\widehat{\bX}(\widehat{\bR}^{(t)})^2\widehat{\bX}\transpose - \bX_0(\bR_0^{(t)})^2\bX_0\transpose\|_{\mathrm{F}}^2\\
&\quad = \left\|
(\eye_m\otimes \widehat{\bX})\mathrm{diag}((\widehat{\bR}^{(1)})^2,\ldots,(\widehat{\bR}^{(m)})^2)(\eye_m\otimes \widehat{\bX})\transpose - (\eye_m\otimes \bX_0)\mathrm{diag}(({\bR}_0^{(1)})^2,\ldots,({\bR}_0^{(m)})^2)(\eye_m\otimes{\bX}_0)\transpose
\right\|_{\mathrm{F}}^2.
\end{align*}
Then, by Davis-Kahan theorem, 
\begin{align*}
\bigg\|\sin\Theta\bigg(\eye_m\otimes\frac{\widehat{\bX}}{\sqrt{n}}, \eye_m\otimes\frac{\bX_0}{\sqrt{n}}\bigg)\bigg\|_{\mathrm{F}}^2
&\lesssim \frac{\sum_{t = 1}^m\|\widehat{\bX}(\widehat{\bR}^{(t)})^2\widehat{\bX}\transpose - \bX_0(\bR_0^{(t)})^2\bX_0\transpose\|_{\mathrm{F}}^2}{n^2} = \Optilde\bigg(\frac{m}{n}\bigg).
\end{align*}
This implies that $\|\sin\Theta(n^{-1/2}\widehat{\bX}, n^{-1/2}\bX_0)\|_{\mathrm{F}} = \Optilde(n^{-1/2})$, so that $\widehat{\bX}\transpose\bX_0$ is invertible w.h.p.. 
\begin{align*}
\bW_{\widehat{\bX}}& = (\bX_0\transpose\widehat{\bX})^{-1}\{(\bX_0\transpose\widehat{\bX})^{-\mathrm{T}}(\bX_0\transpose\widehat{\bX})^{-1}\}^{-1/2},\; 
\widehat{\bB} = \bX_{0\perp}\transpose\widehat{\bX}(\bX_0\transpose\widehat{\bX})^{-1}.
\end{align*}
Then, a direct calculation shows that
$\bX(\widehat{\bB}) = (\bX_0 + \bX_{0\perp}\widehat{\bB})(\eye_d + \widehat{\bB}\transpose\widehat{\bB})^{-1/2} = \widehat{\bX}\bW_{\widehat{\bX}}$. 
Because a simultaneous rotation $\bQ$ applied to $\bQ\transpose\bR^{(t)}\bQ$ for all $t$ and $\bX\bQ$ does not change the optimality of the problem, we may assume $\bW_{\widehat{\bX}} = \eye_d$ without loss of generality. Note that $\widehat{\bB}$ is not accessible for the practitioners since it depends on the unknown $\bX_0$, and we only introduce it here for the sake of consistency analysis. This further implies that $\|(\eye_d + \widehat{\bB}\transpose\widehat{\bB})^{-1/2} - \eye_d\|_{\mathrm{F}} = \Optilde(1/n)$, and hence, $\|\widehat{\bB}\|_2 = \Optilde(n^{-1/2})$ by leveraging the singular value decomposition of $\widehat{\bB}$. Hence, by Lemma \ref{lemma:Lipschitz_Cayley_transform}, 
\[
\|\widehat{\bX} - \bX_0\|_2\leq \int_0^1\|\dot{\bX}(s\widehat{\bB})[\widehat{\bB}]\|_2\mathrm{d}s = \Optilde(1).
\]
Also, simple algebra shows that
\begin{align*}
\sum_{t = 1}^m\|\widehat{\bR}^{(t)} - \bR_0^{(t)}\|_{\mathrm{F}}^2
&\leq \frac{3}{n^4}\sum_{t = 1}^m\left\{\|(\widehat{\bX} - \bX_0)\transpose\widehat{\bTheta}^{(t)}\widehat{\bX}\|_{\mathrm{F}}^2 +
\|\bX_0\transpose(\widehat{\bTheta}^{(t)} - \bTheta_0^{(t)})\widehat{\bX}\|_{\mathrm{F}}^2 +
\|\bX_0\transpose\bTheta_0^{(t)}(\widehat{\bX} - \bX_0)\|_{\mathrm{F}}^2
\right\}\\
& = \Optilde(m/n).
\end{align*}
The proof is completed by combining the above results. 
\end{proof}

\subsection{Proof architecture of the uniform norm consistency}
\label{sub:proof_sketch_of_uniform_norm_consistency}

This subsection first outlines the proof architecture of Proposition \ref{prop:stationarity_uniqueness_uniform_consistency_MLE}. Denote by
\begin{align}
\label{eqn:Lambda_sharp_tilde}
\widetilde{\bLambda}_{\sharp}& = \mathrm{diag}\bigg(\frac{1}{\sqrt{n}}\eye_{d_\balpha}, \frac{1}{n}\eye_{d_\br}, \frac{1}{\sqrt{mn}}\eye_{d_\bX}\bigg),\\
\label{eqn:J(B)}
\bJ(\bB)& = \mathrm{diag}\bigg\{\eye_{d_\balpha + d_\br}, \frac{\partial\vect(\bX)}{\partial\vect(\bB)\transpose}(\bB)\bigg\},\quad\bJ_0 = \bJ(\mathbf{O}), \\
\label{eqn:J_n(B)}
\bJ_n(\bB)& = \mathrm{diag}\bigg\{\eye_{d_\balpha + d_\br}, \frac{1}{\sqrt{n}}\frac{\partial\vect(\bX)}{\partial\vect(\bB)\transpose}(\bB)\bigg\},\quad
\bJ_{0n} = \bJ_n(\mathbf{O}),
\end{align}
where we recall that $d_\bX = nd$. Note that $\bJ(\bB) = \partial\bpsi/\partial\bphi\transpose$, where $\bpsi$ and $\bphi$ are the ambient parameter and compact parameter, and $\bJ_0\bLambda_\sharp = \widetilde{\bLambda}_\sharp\bJ_{0n}$ (recall the definition of $\bLambda_\sharp$ in \eqref{eqn:native_norm_rescaling_matrix}). Now we define $\breve{\bphi}$ to be
\[
  \breve{\bphi} = \argmin_{\bphi\in\calN(\sqrt{m}n\eps_n, M)}\left\|\mathrm{diag}\bigg(\frac{1}{\sqrt{n}}\eye_{d_\balpha}, \frac{1}{n}\eye_{d_\br}, \frac{1}{\sqrt{m}n^{3/2}}\eye_{d_\bX}\bigg)\bJ_0
  \frac{\partial\calL}{\partial\bphi}(\bphi)\right\|_\infty,\quad\eps_n = 
  \frac{m^2(\log n)^2}{\sqrt{n}},
\]
where we recall $\bLambda_\sharp$ is defined in \eqref{eqn:native_norm_rescaling_matrix}. We will argue that $\breve{\bphi}$ above is achieved inside the interior of $\calN(\sqrt{m}n\eps_n, M)$ and coincides with the MLE $\widehat{\bphi}:=\bphi(\widehat{\bpsi})$ w.h.p., where $\bpsi\mapsto \bphi$ denotes the inverse of the natural embedding $\bphi\mapsto \bpsi$ of the compact parameter $\bphi$ into the ambient parameter space. 

For any matrix-valued function $\bY:\bphi\in\mathbb{R}^{d_\bphi}\mapsto\bY(\bphi)\in\mathbb{R}^{n_1\times n_2}$, define the integral operator
\[
(\scrI \bY)(\bphi) = \int_0^1\bY(\bphi_0 + \tau({\bphi} - \bphi_0))\mathrm{d}\tau. 
\]
By the fundamental theorem of calculus and the fact that $\calN(\sqrt{m}n\eps_n, M)$ is convex,
\begin{align*}
\frac{\partial\calL}{\partial\bphi}(\breve{\bphi}) - \frac{\partial\calL}{\partial\bphi}(\bphi_0)
& = \int_0^1 \frac{\mathrm{d}}{\mathrm{d}\tau}\frac{\partial\calL}{\partial\bphi}(\bphi_0 + \tau(\breve{\bphi} - \bphi_0))\mathrm{d}\tau = 
\int_0^1 \frac{\partial^2\calL}{\partial\bphi\partial\bphi\transpose}(\bphi_0 + \tau(\breve{\bphi} - \bphi_0))(\breve{\bphi} - \bphi_0)\mathrm{d}\tau\\
& = \bigg(\scrI\frac{\partial^2\calL}{\partial\bphi\partial\bphi\transpose}\bigg)(\breve{\bphi})(\breve{\bphi} - \bphi_0),
\end{align*}
implying that
\begin{align*}
\|\bJ_0(\breve{\bphi} - \bphi_0)\|_\infty
& = \bigg\|\bJ_0\bigg\{\bigg(\scrI\frac{\partial^2\calL}{\partial\bphi\partial\bphi\transpose}\bigg)(\breve{\bphi})\bigg\}^{-1}(\bJ_0\transpose\bJ_0)^{-1}\bJ_0\transpose\bJ_0\bigg\{\frac{\partial\calL}{\partial\bphi}(\breve{\bphi}) - \frac{\partial\calL}{\partial\bphi}(\bphi_0)\bigg\}\bigg\|_\infty\\ 
& \leq \left\|
\bJ_0\bigg\{\bigg(\scrI\frac{\partial^2\calL}{\partial\bphi\partial\bphi\transpose}\bigg)(\breve{\bphi})\bigg\}^{-1}
\bJ_0\transpose\widetilde{\bLambda}_\sharp^{-1}
\right\|_\infty\\
&\quad\times \left\|
\mathrm{diag}\bigg(
\frac{1}{\sqrt{n}}\eye_{d_\balpha}, \frac{1}{n}\eye_{d_\br}, \frac{1}{\sqrt{m}n^{3/2}}\eye_{d_\bX}\bigg)\bJ_0
\bigg\{\frac{\partial\calL}{\partial\bphi}(\breve{\bphi}) - \frac{\partial\calL}{\partial\bphi}(\bphi_0)\bigg\}\right\|_\infty\\
&\leq 
2\left\|
\widetilde{\bLambda}_\sharp\bJ_{0n}\bigg\{\bigg(\bLambda_\sharp\scrI\frac{\partial^2\calL}{\partial\bphi\partial\bphi\transpose}\bLambda_\sharp\bigg)(\breve{\bphi})\bigg\}^{-1}
\bJ_{0n}\transpose
\right\|_\infty
\left\|
\mathrm{diag}\bigg(
\frac{\eye_{d_\balpha}}{\sqrt{n}}, \frac{\eye_{d_\br}}{n}, \frac{\eye_{d_\bX}}{\sqrt{m}n^{3/2}}\bigg)\bJ_0
\frac{\partial\calL}{\partial\bphi}(\bphi_0)\right\|_\infty
\end{align*}
by the minimization property of $\breve{\bphi}$ and triangle inequality. The second factor is $\Optilde(\sqrt{\log n})$ by Lemma \ref{lemma:score_uniform_convergence} below. Note that an error bound on $\|\bJ_0(\breve{\bphi} - \bphi_0)\|_\infty$ automatically implies the uniform error bound on $\balpha^{(t)}$ and $\bR^{(t)}$. For the uniform error bound on $\bX$, note that
\begin{align*}
\|\bX(\bB) - \bX_0\|_{2\to\infty}&\leq \|\bX_0\|_{2\to\infty}\|(\eye_d + \bB\transpose\bB)^{-1/2} - \eye_d\|_2 + \sqrt{d}\|(\eye_d\otimes\bX_{0\perp})\vect(\bB)\|_\infty\\
& \leq O(\|\bB\|_2^2) + \sqrt{d}\|\bJ_0(\bphi - \bphi_0)\|_\infty = O(n^{-1/2}) + \sqrt{d}\|\bJ_0(\bphi - \bphi_0)\|_\infty,
\end{align*}
so that a uniform error bound on $\bX$ can be obtained via $\|\bJ_0(\breve{\bphi} - \bphi_0)\|_\infty$, provided this error bound dominates $n^{-1/2}$. 
The focus is on the first factor.
Let
\begin{align}
\label{eqn:alphai_zi}
\balpha_i = [\alpha_i^{(1)},\ldots,\alpha_i^{(m)}]\transpose\quad\text{and}\quad
\bz_i = [\balpha_i\transpose, \bx_i\transpose]\transpose
\end{align}
be the parameter associated with vertex $i$, including the degree correction parameter $(\alpha_i^{(t)})$ across layers $t\in[m]$ and the latent position $\bx_i$. 
For any generic ambient parameter $\bpsi$, recall the log-likelihood function under the $\bpsi$-parameterization defined Section \ref{sec:maximum_likelihood_and_one_step_estimator} in the main text
and define the following matrix-valued functions 
\begin{align}
\bGamma_{\bZ\bZ}(\bpsi)& = \bigg[
\sum_{t = 1}^m\sum_{a < b}\ell''(\theta_{ab}^{(t)}; A_{ab}^{(t)})
\frac{\partial\theta_{ab}^{(t)}}{\partial\bz_i}
\frac{\partial\theta_{ab}^{(t)}}{\partial\bz_j\transpose}
\bigg]_{n\times n},\nonumber\\
\label{eqn:GammaZZ_sharp}
\bGamma_{\bZ\bZ}^\sharp(\bpsi) &= \bigg\{\eye_n\otimes\mathrm{diag}\bigg(\frac{\eye_m}{\sqrt{n}}, \frac{\eye_d}{\sqrt{mn}}\bigg)\bigg\}\bGamma_{\bZ\bZ}(\bpsi)\bigg\{\eye_n\otimes\mathrm{diag}\bigg(\frac{\eye_m}{\sqrt{n}}, \frac{\eye_d}{\sqrt{mn}}\bigg)\bigg\},\\
\bGamma_{\bZ\bR}(\bpsi)& = \bigg[
\sum_{s = 1}^m\sum_{a < b}\ell''(\theta_{ab}^{(s)}; A_{ab}^{(s)})
\frac{\partial\theta_{ab}^{(s)}}{\partial\bz_i}\frac{\partial\theta_{ab}^{(s)}}{\partial\vech(\bR^{(t)})\transpose}
\bigg]_{n\times m},\nonumber
\\
\label{eqn:GammaZR_sharp}
\bGamma_{\bZ\bR}^\sharp(\bpsi) & = \bGamma_{\bR\bZ}^\sharp(\bpsi)\transpose =\bigg\{\eye_n\otimes\mathrm{diag}\bigg(\frac{\eye_m}{\sqrt{n}}, \frac{\eye_d}{\sqrt{mn}}\bigg)\bigg\}\bGamma_{\bZ\bR}(\bpsi)\frac{\eye_{d_\br}}{n},\\
\label{eqn:GammaRR_sharp}
\bGamma_{\bR\bR}^\sharp(\bpsi) &= \frac{\eye_{d_\br}}{n}\bigg[
\sum_{s = 1}^m\sum_{a < b}\ell''(\theta_{ab}^{(s)}; A_{ab}^{(s)})
\frac{\partial\theta_{ab}^{(s)}}{\partial\vech(\bR^{(t)})}
\frac{\partial\theta_{ab}^{(s)}}{\partial\vech(\bR^{(u)})\transpose}
\bigg]_{m\times m}\frac{\eye_{d_\br}}{n},\\
\label{eqn:Gamma_sharp}
\bGamma_{\bpsi\bpsi}^\sharp(\bpsi) &=
\begin{bmatrix}
\bGamma_{\bZ\bZ}^\sharp(\bpsi) & \bGamma_{\bZ\bR}^\sharp(\bpsi)\\
\bGamma_{\bR\bZ}^\sharp(\bpsi) & \bGamma_{\bR\bR}^\sharp(\bpsi)
\end{bmatrix}.
\end{align}
In other words, $\bGamma_{\bpsi\bpsi}^\sharp$ is the sample Fisher information matrix with respect to $\bpsi$ after being rescaled by $\widetilde{\bLambda}_\sharp$ up to a simultaneous permutation of rows and columns. Clearly, a direct calculation yields
\begin{align*}
\bGamma_{\bZ\bZ}^\sharp(\bpsi)
& = \frac{1}{n}\sum_{t = 1}^m\sum_{i\neq j} \ell''(\theta_{ij}^{(t)}; A_{ij}^{(t)})\be_i\be_i\transpose\otimes 
\begin{bmatrix}
\be_t(m)\be_t(m)\transpose & \frac{1}{\sqrt{m}}\be_t(m)\bx_{j}\transpose\bR^{(t)}\\ 
\frac{1}{\sqrt{m}}\bR^{(t)}\bx_{j}\be_t(m)\transpose & \frac{1}{m}\bR^{(t)}\bx_{j}\bx_{j}\transpose\bR^{(t)}
\end{bmatrix}\\
&\quad + 
\frac{1}{n}\sum_{t = 1}^m\sum_{i\neq j} \ell''(\theta_{ij}^{(t)}; A_{ij}^{(t)})\be_i\be_j\transpose\otimes 
\begin{bmatrix}
\be_t(m)\be_t(m)\transpose & \frac{1}{\sqrt{m}}\be_t(m)\bx_{i}\transpose\bR^{(t)}\\ 
\frac{1}{\sqrt{m}}\bR^{(t)}\bx_{j}\be_t(m)\transpose & \frac{1}{m}\bR^{(t)}\bx_{j}\bx_{i}\transpose\bR^{(t)}
\end{bmatrix}
,\\
\bGamma_{\bZ\bR}^\sharp(\bpsi)
& = \frac{1}{2n^{3/2}}\sum_{t = 1}^m\sum_{i\neq j} \ell''(\theta_{ij}^{(t)}; A_{ij}^{(t)})\left\{\be_i\otimes\begin{bmatrix}
\be_t(m) \\ 
\frac{1}{\sqrt{m}}\bR^{(t)}\bx_{j}
\end{bmatrix}
+
\be_j\otimes\begin{bmatrix}
\be_t(m) \\ 
\frac{1}{\sqrt{m}}\bR^{(t)}\bx_{i}
\end{bmatrix}
\right\}\{\be_t(m)\transpose\otimes(\bx_{j}\otimes\bx_{i})\transpose\bD_d\},\\
\bGamma_{\bR\bR}^\sharp(\bpsi)
& = \sum_{t = 1}^m\be_t(m)\be_t(m)\transpose\otimes\frac{1}{2n^2}\sum_{i\neq j} \ell''(\theta_{ij}^{(t)}; A_{ij}^{(t)})
\bD_d\transpose(\bx_{j}\bx_{j}\transpose\otimes \bx_{i}\bx_{i}\transpose)\bD_d.
\end{align*}
Note that there exists a permutation matrix $\bPi\in\mathbb{R}^{d_\bpsi\times d_\bpsi}$, such that 
\begin{align}
\label{eqn:permutation_matrix_psi}
\bpsi = \begin{bmatrix}
\balpha^{(1)}\\
\vdots\\
\balpha^{(m)}\\
\vech(\bR^{(1)})\\
\vdots\\
\vech(\bR^{(m)})\\
\vect(\bX)
\end{bmatrix}
 = \bPi \begin{bmatrix}
\bz_1\\
\vdots\\
\bz_n\\
\vech(\bR^{(1)})\\
\vdots\\
\vech(\bR^{(m)})
\end{bmatrix}.
\end{align}
Now define the following matrix-valued functions
\begin{align}
\label{eqn:Deltazi_sharp}
\bDelta_{\bz_i}^\sharp(\bpsi)
 & = \frac{1}{n}\sum_{t = 1}^m\sum_{j\in[n]\backslash\{i\}} \ell''(\theta_{ij}^{(t)}; A_{ij}^{(t)}) 
\begin{bmatrix}
\be_t(m)\be_t(m)\transpose & \frac{1}{\sqrt{m}}\be_t(m)\bx_{j}\transpose\bR^{(t)}\\ 
\frac{1}{\sqrt{m}}\bR^{(t)}\bx_{j}\be_t(m)\transpose & \frac{1}{m}\bR^{(t)}\bx_{j}\bx_{j}\transpose\bR^{(t)}
\end{bmatrix},\\
\label{eqn:DeltaZ_sharp}
\bDelta_\bZ^\sharp(\bpsi)& = \sum_{i = 1}^n\be_i\be_i\transpose\otimes\bDelta_{\bz_i}^\sharp(\bpsi),\\
\label{eqn:Deltapsi_sharp}
\bDelta_\bpsi^\sharp(\bpsi)& = \mathrm{diag}\{\bDelta_\bZ^\sharp(\bpsi), \bGamma_{\bR\bR}^\sharp(\bpsi)\},\\
\label{eqn:Deltaphi_sharp}
\bDelta_\bphi^\sharp(\bphi)& = \bJ_n(\bB)\transpose\bPi \bDelta_\bpsi^\sharp(\bpsi(\bphi))\bPi\transpose\bJ_n(\bB),
\end{align}
where $\bpsi(\bphi)$ denotes the natural embedding $\bphi\mapsto \bpsi$ of the compact parameter $\bphi$ into the ambient parameter space. 
We will argue that
\[
\left\|
\widetilde{\bLambda}_\sharp\bJ_{0n}\bigg\{\bLambda_\sharp\bigg(\scrI\frac{\partial^2\calL}{\partial\bphi\partial\bphi\transpose}\bigg)(\breve{\bphi})\bLambda_\sharp\bigg\}^{-1}
\bJ_{0n}\transpose
\right\|_\infty\approx
\left\|
\widetilde{\bLambda}_\sharp\bJ_{0n}\{(\scrI \bDelta_\bphi^\sharp)(\breve{\bphi})\}^{-1}
\bJ_{0n}\transpose
\right\|_\infty\quad\text{w.h.p.}
\]
This will be made precise and rigorous in Section \ref{sub:uniform_perturbation_analysis_of_the_jacobian_matrix}. Then, we write by a simple algebra
\begin{align*}
\bLambda_\sharp\frac{\partial^2\calL}{\partial\bphi\partial\bphi\transpose}(\bphi)\bLambda_\sharp
& = \bJ_n(\bB)\transpose\bPi\bGamma_{\bpsi\bpsi}^\sharp(\bpsi(\bphi))\bPi\transpose\bJ_n(\bB) + \frac{1}{2}\bLambda_\sharp\sum_{t = 1}^m\sum_{i\neq j}\ell'(\theta_{ij}^{(t)}; A_{ij}^{(t)})\frac{\partial^2\theta_{ij}^{(t)}}{\partial\bphi\partial\bphi\transpose}\bLambda_\sharp\\
& = \bJ_n(\bB)\transpose\bPi\bDelta_\bpsi^\sharp(\bpsi(\bphi))\bPi\transpose\bJ_n(\bB) + \bJ_n(\bB)\transpose\bPi\{\bGamma_{\bpsi\bpsi}^\sharp(\bpsi(\bphi)) - \bDelta_\bpsi^\sharp(\bpsi(\bphi))\}\bPi\transpose\bJ_n(\bB)\\
&\quad + \frac{1}{2}\bLambda_\sharp\sum_{t = 1}^m\sum_{i\neq j}\ell'(\theta_{ij}^{(t)}; A_{ij}^{(t)})\frac{\partial^2\theta_{ij}^{(t)}}{\partial\bphi\partial\bphi\transpose}\bLambda_\sharp. 
\end{align*}
Section \ref{sub:uniform_perturbation_analysis_of_the_hessian_off_diagonal_blocks} handles the second term on the right-hand side of the above decomposition, while Section \ref{sub:uniform_perturbation_analysis_of_the_hessian_score_term} deals with the third term. The complete proof of Proposition \ref{prop:stationarity_uniqueness_uniform_consistency_MLE} is detailed in Section \ref{sub:proof_of_uniform_consistency_mle}.

\subsection{Uniform norm error bound of the score function}
\label{sub:uniform_norm_error_bound_of_the_score_function}

\begin{lemma}
\label{lemma:score_uniform_convergence}
Suppose Assumptions \ref{assumption:edge_distribution}--\ref{assumption:true_parameter} hold and $m = O(n^2)$. Then
\begin{align*}
\max_{t\in[m]}\bigg\|\frac{\partial\calL}{\partial\balpha^{(t)}}(\bphi_0)\bigg\|_\infty
& = \Optilde(\sqrt{n\log n}),\\
\bigg\|(\eye_d\otimes\bX_{0\perp})\frac{\partial\calL}{\partial\vect(\bB)}(\bphi_0)\bigg\|_\infty
& = \Optilde(\sqrt{m}n^{3/2}\sqrt{\log n}),\\
\max_{t\in[m]}\bigg\|\frac{\partial\calL}{\partial\vech(\bR^{(t)})}(\bphi_0)\bigg\|_\infty
& = \Optilde(n\sqrt{\log n}).
\end{align*}
\end{lemma}

\begin{proof}[\bf Proof]
By the calculation in Section \ref{sub:gradient_hessian}, 
\begin{align*}
\max_{t\in[m]}\bigg\|\frac{\partial\calL}{\partial\balpha^{(t)}}(\bphi_0)\bigg\|_\infty
& = \max_{t\in[m], i\in[n]}\bigg|\sum_{j\in[n]\backslash\{i\}}\ell'(\theta_{0ij}^{(t)}; A_{ij}^{(t)})\bigg| = \Optilde(\sqrt{n\log n}),\\
\max_{t\in[m]}\bigg\|\frac{\partial\calL}{\partial\vech(\bR^{(t)})}(\bphi_0)\bigg\|_\infty
& \leq \max_{t\in[m]}\frac{1}{2}\|\bD_d\|_2\bigg\|\sum_{i\neq j}\ell'(\theta_{0ij}^{(t)}; A_{ij}^{(t)})\vect(\bx_{0i}\bx_{0j}\transpose)\bigg\|_{\mathrm{F}} = \Optilde(n\sqrt{\log n})
\end{align*}
by Bernstein's inequality and union bounds. For the second assertion, we compute
\begin{align*}
&(\eye_d\otimes\bX_{0\perp})\frac{\partial\calL}{\partial\vect(\bB)}(\bphi_0)\\
&\quad  = \frac{1}{2}\sum_{t = 1}^m\sum_{i\neq j}\ell'(\theta_{0ij}^{(t)}; A_{ij}^{(t)})\{\bR_0^{(t)}\bx_{0j} \otimes (n\eye_n - \one_n\one_n\transpose - \bX_0\bX_0\transpose)\be_i
+ \bR_0^{(t)}\bx_{0i} \otimes (n\eye_n - \one_n\one_n\transpose - \bX_0\bX_0\transpose)\be_j
\}\\
&\quad = n\sum_{t = 1}^m\sum_{j = 1}^n\bR_0^{(t)}\bx_{0j} \otimes \sum_{i\in[n]\backslash\{j\}}\ell'(\theta_{0ij}^{(t)}; A_{ij}^{(t)})\be_i
 - \sum_{t = 1}^m\sum_{i\neq j}\ell'(\theta_{0ij}^{(t)}; A_{ij}^{(t)})\bR_0^{(t)}\bx_{0j}\otimes\bX_0\bx_{0i}\\
 &\qquad
 - \sum_{t = 1}^m\sum_{i\neq j}\ell'(\theta_{0ij}^{(t)}; A_{ij}^{(t)})\bR_0^{(t)}\bx_{0j}\otimes\one_n.
\end{align*}
Therefore, by Bernstein's inequality and a union bound,
\begin{align*}
\bigg\|(\eye_d\otimes\bX_{0\perp})\frac{\partial\calL}{\partial\vect(\bB)}(\bphi_0)\bigg\|_\infty
& \leq \max_{k\in[d],i\in[n]}
\bigg|n\sum_{t = 1}^m\sum_{j\in[n]\backslash\{i\}}\be_k\transpose\bR_0^{(t)}\bx_{0j}\ell'(\theta_{0ij}^{(t)}; A_{ij}^{(t)})\bigg|\\
&\quad + \max_{k\in[d],l\in[n]}\bigg|\sum_{t = 1}^m\sum_{i\neq j}\be_k\transpose\bR_0^{(t)}\bx_{0j}\bx_{0l}\transpose\bx_{0i}\ell'(\theta_{0ij}^{(t)}; A_{ij}^{(t)})\bigg|\\
&\quad + \max_{k\in[d]}\bigg|\sum_{t = 1}^m\sum_{i\neq j}\be_k\transpose\bR_0^{(t)}\bx_{0j}\ell'(\theta_{0ij}^{(t)}; A_{ij}^{(t)})\bigg|\\
& = \Optilde(\sqrt{m}n^{3/2}\sqrt{\log n}).
\end{align*}
The proof is therefore completed.
\end{proof}

\subsection{Uniform perturbation analysis of the Hessian: Off-diagonal blocks}
\label{sub:uniform_perturbation_analysis_of_the_hessian_off_diagonal_blocks}

\begin{lemma}\label{lemma:off_diagonal_term_perturbation}
Suppose Assumptions \ref{assumption:edge_distribution}--\ref{assumption:true_parameter} hold. Then, for $M,K = O(1)$ and any sequences $(a_n)_{n = 1}^\infty$, $(b_n)_{n = 1}^\infty$, $(c_n)_{n = 1}^\infty$, we have
\begin{align*}
&\sup_{\|\bB\|_2\leq K, \bpsi\in\Psi}\bigg\|\mathrm{diag}(a_n\eye_{d_\balpha}, b_n\eye_{d_\br}, c_n\eye_{d_\bX})\bJ_{0n}\bJ_n(\bB)\transpose
\bPi\{\bGamma_{\bpsi\bpsi}^\sharp(\bpsi) - \bDelta_{\bpsi}^\sharp(\bpsi)\}\bPi\transpose
\bJ_n(\bB)\bJ_{0n}\transpose
\bigg\|_{2\to\infty}\\
&\quad = \Optilde\bigg\{\frac{(a_n \vee c_n)}{\sqrt{n}} + b_n\bigg\}.
\end{align*}
\end{lemma}

\begin{proof}[\bf Proof]
By definition, we have
\begin{align*}
\bGamma_{\bpsi\bpsi}^\sharp(\bpsi) - \bDelta_{\bpsi}^\sharp(\bpsi) 
& = \begin{bmatrix}
  \bGamma_{\bZ\bZ}^\sharp(\bpsi) - \bDelta_{\bZ}^\sharp(\bpsi) &   \bGamma_{\bZ\bR}^\sharp(\bpsi)\\
  \bGamma_{\bR\bZ}^\sharp(\bpsi) &   \mathbf{O}
\end{bmatrix},\\
\bGamma_{\bZ\bZ}^\sharp(\bpsi) - \bDelta_{\bZ}^\sharp(\bpsi) & = \frac{1}{n}\sum_{t = 1}^m\sum_{i\neq j} \ell''(\theta_{ij}^{(t)}; A_{ij}^{(t)})\be_i\be_j\transpose\otimes 
\begin{bmatrix}
\be_t(m)\be_t(m)\transpose & \frac{1}{\sqrt{m}}\be_t(m)\bx_{i}\transpose\bR^{(t)}\\ 
\frac{1}{\sqrt{m}}\bR^{(t)}\bx_{j}\be_t(m)\transpose & \frac{1}{m}\bR^{(t)}\bx_{j}\bx_{i}\transpose\bR^{(t)}
\end{bmatrix}.
\end{align*}
Note that $\sup_{\bpsi\in\Psi}\|\bGamma_{\bZ\bR}^\sharp(\bpsi)\|_{2\to\infty} = \Optilde(\sqrt{1/n})$ and $\sup_{\bpsi\in\Psi}\|\bGamma_{\bR\bZ}^\sharp(\bpsi)\|_{2\to\infty} = \Optilde(1)$. Also, 
\begin{align}
\label{eqn:Jn(B)J0n_perturbation}
\sup_{\|\bB\|_2\leq K, \|\bX_{0\perp}\bB\|_{2\to\infty}\leq M}\|\bJ_n(\bB)\bJ_{0n}\transpose\|_2 = \Optilde(1),\;
\sup_{\|\bB\|_2\leq K, \|\bX_{0\perp}\bB\|_{2\to\infty}\leq M}
\|\bJ_{0n}\bJ_n(\bB)\transpose\|_\infty = \Optilde(1).  
\end{align}
Indeed, it is sufficient to consider $\bJ_\bX(\bB) = n^{-1/2}\partial\vect(\bX)/\partial\vect(\bB)\transpose$ and $\bJ_\bX(\mathbf{O})$. Clearly, by the formula of $\bJ_\bX(\bB)$ in Proposition \ref{prop:Cayley_transform_jacobian}, we have $\|\bJ_\bX(\bB)\transpose\bJ_\bX(\bB) - \eye\|_2 = O(\|\bB\|_2)$, which gives the first bound in \eqref{eqn:Jn(B)J0n_perturbation}, and
\begin{align*}
\sup_\bB\|\bJ_{0n}\{\bJ_n(\bB) - \bJ_{0n}\}\transpose\|_\infty 
&\leq \sup_\bB\|\bN - \eye_d\|_\infty\bigg\|\frac{\bX_{0\perp}\bX_{0\perp}\transpose}{n}\bigg\|_\infty + \sup_\bB \frac{O(d)}{n}\|\eye_d\otimes\bX_{0\perp}\bB\|_{\infty}\|\eye_d\otimes\bX(\bB)\transpose\|_\infty\\
& = O(1),
\end{align*}
where the supremum is taken over $\|\bB\|_2\leq K$ and $\|\bX_{0\perp}\bB\|_{2\to\infty}\leq M$, and $\bN = (\eye_d + \bB\transpose\bB)^{-1/2}$. 
It is now sufficient to work on the off-diagonal part of $\bGamma_{\bZ\bZ}^\sharp(\bpsi)$. By construction, we have
\begin{align*}
& \sup_{\bpsi\in\Psi} \left\|\bGamma_{\bZ\bZ}^\sharp(\bpsi) - \bDelta_{\bZ}^\sharp(\bpsi)\right\|_{2\to\infty}\\
&\quad \leq  \sup_{\bpsi\in\Psi}\frac{1}{\sqrt{n}}\max_{i,j}\Bigg\|\sum_{t = 1}^m\ell''(\theta_{ij}^{(t)}; A_{ij}^{(t)})\begin{bmatrix}
\be_t(m)\be_t(m)\transpose & \frac{1}{\sqrt{m}}\be_t(m)\bx_{i}\transpose\bR^{(t)}\\ 
\frac{1}{\sqrt{m}}\bR^{(t)}\bx_{j}\be_t(m)\transpose & \frac{1}{m}\bR^{(t)}\bx_{j}\bx_{i}\transpose\bR^{(t)}
\end{bmatrix}\Bigg\|_{2\to\infty} = O(n^{-1/2}). 
\end{align*}
The proof is completed by combining the above analyses. 
\end{proof}

\subsection{Uniform perturbation analysis of the Hessian: Score term}
\label{sub:uniform_perturbation_analysis_of_the_hessian_score_term}

\begin{lemma}\label{lemma:score_term_perturbation}
Suppose Assumptions \ref{assumption:edge_distribution}--\ref{assumption:true_parameter} hold and $m = O(n^2)$. Then, there exists a $d_\bpsi\times d_\bpsi$ random matrix $\bXi(\bphi)$ depending on $\bphi$ of the form $\bXi(\bphi) = \mathrm{diag}(\mathbf{O}_{(d_\balpha + d_\br)\times (d_\balpha + d_\br)}, \bXi_\bX(\bphi))$, where $\bXi_\bX(\bphi)$ is a $\bphi$-dependent $d_\bX\times d_\bX$ random matrix, such that for $M,\eps_n = O(1)$, and any sequences $(a_n)_{n = 1}^\infty, (b_n)_{n = 1}^\infty, (c_n)_{n = 1}^\infty$, we have
\begin{align*}
&\sup_{\bphi\in\calN(\sqrt{m}n\eps_n, M)}\bigg\|\mathrm{diag}(a_n\eye_{d_\balpha}, b_n\eye_{d_\br}, c_n\eye_{d_\bX})\bigg\{\bJ_{0n}\bLambda_\sharp\sum_{t = 1}^m\sum_{i\neq j}\ell'({\theta}_{ij}^{(t)}; A_{ij}^{(t)})\frac{\partial^2\theta_{ij}^{(t)}}{\partial\bphi\partial\bphi\transpose}\bLambda_\sharp\bJ_{0n}\transpose - \bXi(\bphi)\bigg\}\bigg\|_{2\to\infty}\\
&\quad = \Optilde\bigg\{c_n\bigg(\frac{1}{\sqrt{n}} + \sqrt{\frac{\log n}{mn}}\bigg) + b_n\bigg(\eps_n\log n + \sqrt{\frac{\log n}{mn}}\bigg)\bigg\},\\
&\sup_{\bphi\in\calN(\sqrt{m}n\eps_n, M)}\|\bXi (\bphi)\|_\infty = \sup_{\bphi\in\calN(\sqrt{m}n\eps_n, M)}\|\bXi_\bX(\bphi)\|_\infty = \Optilde(\eps_n\log n). 
\end{align*}
\end{lemma}

\begin{proof}[\bf Proof]
First note that $\bphi\in\calN(\sqrt{m}n\eps_n, M)$ automatically implies the following error bounds:
\begin{align*}
&\|\bB\|_2 = O(\eps_n),\quad \|\bX(\bB) - \bX_0\|_{\mathrm{F}} = O(\sqrt{n}\eps_n),\quad \sum_{t = 1}^m\sum_{i\neq j}(\theta_{ij}^{(t)} - \theta_{0ij}^{(t)})^2 = O(mn^2\eps_n^2),\\
&n^2\sum_{t = 1}^m\|\bR^{(t)} - \bR_0^{(t)}\|_{\mathrm{F}}^2 = O(mn^2\eps_n^2),\quad \max_{t\in[m]}(\|\balpha^{(t)}\|_\infty + \|\bR^{(t)}\|_{\mathrm{F}}) + \|\bX(\bB)\|_{2\to\infty} = O(1).
\end{align*}
By the calculations in Section \ref{sub:gradient_hessian}, 
\begin{align*}
&\mathrm{diag}(a_n\eye_{d_\balpha}, b_n\eye_{d_\br}, c_n\eye_{d_\bX})\bJ_{0n}\bLambda_\sharp\sum_{t = 1}^m\sum_{i\neq j}\bigg\{\ell'({\theta}_{ij}^{(t)}; A_{ij}^{(t)})\frac{\partial^2\theta_{ij}^{(t)}}{\partial\bphi\partial\bphi\transpose}({\bphi})
-
\ell'(\theta_{0ij}^{(t)}; A_{ij}^{(t)})\frac{\partial^2\theta_{ij}^{(t)}}{\partial\bphi\partial\bphi\transpose}(\bphi_0)
\bigg\}\bLambda_\sharp\bJ_{0n}\transpose\\
&\quad = 
\mathrm{diag}(a_n\eye_{d_\balpha}, b_n\eye_{d_\br}, c_n\eye_{d_\bX})\bJ_{0n}\bJ_n(\bB)\transpose\sum_{t = 1}^m\sum_{i\neq j}
\widetilde{\bLambda}_\sharp
\Bigg\{
\ell'({\theta}_{ij}^{(t)}; A_{ij}^{(t)})\\
&\qquad\times \begin{bmatrix}
\mathbf{O} & \mathbf{O} & \mathbf{O}\\
\mathbf{O} & \mathbf{O} & \be_t\otimes\bD_d\transpose\{\eye_d\otimes({\bx}_i\be_j\transpose + {\bx}_j\be_i\transpose)\}\\
\mathbf{O} & \be_t\transpose\otimes\{\eye_d\otimes(\be_j{\bx}_i\transpose + \be_i{\bx}_j\transpose)\}\bD_d& 
\bR^{(t)} \otimes (\be_i\be_j\transpose + \be_j\be_i\transpose)
\end{bmatrix}\\
&\qquad - \ell'(\theta_{0ij}^{(t)}; A_{ij}^{(t)})\\
&\qquad\times \left.\begin{bmatrix}
\mathbf{O} & \mathbf{O} & \mathbf{O}\\
\mathbf{O} & \mathbf{O} & \be_t\otimes\bD_d\transpose\{\eye_d\otimes({\bx}_{0i}\be_j\transpose + {\bx}_{0j}\be_i\transpose)\}\\
\mathbf{O} & \be_t\transpose\otimes\{\eye_d\otimes(\be_j{\bx}_{0i}\transpose + \be_i{\bx}_{0j}\transpose)\}\bD_d& 
\bR_0^{(t)}\otimes(\be_i\be_j\transpose + \be_j\be_i\transpose)
\end{bmatrix}\right\}\\
&\qquad\times \widetilde{\bLambda}_\sharp\bJ_n(\bB)\bJ_{0n}\transpose\\
&\qquad + \left\{\frac{2c_n}{mn^2}\bJ_{0n}\sum_{t = 1}^m\sum_{i\neq j}\ell'({\theta}_{ij}^{(t)}; A_{ij}^{(t)})
\begin{bmatrix}
\mathbf{O} & \mathbf{O} & \mathbf{O}\\
\mathbf{O} & \mathbf{O} & \mathbf{O}\\
\mathbf{O} & \mathbf{O} & \sum_{k = 1}^d\frac{\partial^2x_{ik}(\bB)}{\partial\vect(\bB)\partial\vect(\bB)\transpose}\be_k\transpose\bR^{(t)}\bx_j
\end{bmatrix}
\bJ_{0n}\transpose \right.\\
&\qquad - \left.\frac{2c_n}{mn^2}\bJ_{0n}\sum_{t = 1}^m\sum_{i\neq j}\ell'(\theta_{0ij}^{(t)}; A_{ij}^{(t)})
\begin{bmatrix}
\mathbf{O} & \mathbf{O} & \mathbf{O}\\
\mathbf{O} & \mathbf{O} & \mathbf{O}\\
\mathbf{O} & \mathbf{O} & \sum_{k = 1}^d\frac{\partial^2x_{ik}(\mathbf{O})}{\partial\vect(\bB)\partial\vect(\bB)\transpose}\be_k\transpose\bR_0^{(t)}\bx_{0j}
\end{bmatrix}
\bJ_{0n}\transpose\right\}\\
&\qquad + \mathrm{diag}(a_n\eye_{d_\balpha}, b_n\eye_{d_\br}, c_n\eye_{d_\bX})\bJ_{0n}\{\bJ_n(\bB) - \bJ_{0n}\}\transpose\sum_{t = 1}^m\sum_{i\neq j}
\widetilde{\bLambda}_\sharp
\ell'(\theta_{0ij}^{(t)}; A_{ij}^{(t)})\\
&\qquad\times \begin{bmatrix}
\mathbf{O} & \mathbf{O} & \mathbf{O}\\
\mathbf{O} & \mathbf{O} & \be_t\otimes\bD_d\transpose\{\eye_d\otimes({\bx}_{0i}\be_j\transpose + {\bx}_{0j}\be_i\transpose)\}\\
\mathbf{O} & \be_t\transpose\otimes\{\eye_d\otimes(\be_j{\bx}_{0i}\transpose + \be_i{\bx}_{0j}\transpose)\}\bD_d& 
\bR_0^{(t)}\otimes(\be_i\be_j\transpose + \be_j\be_i\transpose)
\end{bmatrix}\\
&\qquad\times \widetilde{\bLambda}_\sharp{\bJ}_n(\bB)\bJ_{0n}\transpose\\
&\qquad + \mathrm{diag}(a_n\eye_{d_\balpha}, b_n\eye_{d_\br}, c_n\eye_{d_\bX})\bJ_{0n} \bJ_{0n} \transpose\sum_{t = 1}^m\sum_{i\neq j}
\widetilde{\bLambda}_\sharp
\ell'(\theta_{0ij}^{(t)}; A_{ij}^{(t)})\\
&\qquad\times \begin{bmatrix}
\mathbf{O} & \mathbf{O} & \mathbf{O}\\
\mathbf{O} & \mathbf{O} & \be_t\otimes\bD_d\transpose\{\eye_d\otimes({\bx}_{0i}\be_j\transpose + {\bx}_{0j}\be_i\transpose)\}\\
\mathbf{O} & \be_t\transpose\otimes\{\eye_d\otimes(\be_j{\bx}_{0i}\transpose + \be_i{\bx}_{0j}\transpose)\}\bD_d& 
\bR_0^{(t)}\otimes(\be_i\be_j\transpose + \be_j\be_i\transpose)
\end{bmatrix}\\
&\qquad\times \widetilde{\bLambda}_\sharp\{\bJ_n(\bB) - \bJ_{0n}\}\bJ_{0n}\transpose,
\end{align*}
where $x_{ik}(\bB)$ denotes the $(i, k)$th element of $\bX(\bB)$. Below, we bound each of the four terms separately. 

\noindent $\blacksquare$ \textbf{First term.} Note that $\|\bJ_{0n}\bJ_n(\bB)\transpose\|_{\infty} = O(1)$ and $\|\bJ_n(\bB)\bJ_{0n}\transpose\|_{2} = O(1)$ uniformly over $\bphi\in\calN(\sqrt{m}n\eps_n, M)$ by \eqref{eqn:Jn(B)J0n_perturbation}. The three non-zero blocks have the following forms (up to left and right multiplications of $\bJ_{0n}\bJ_n(\bB)\transpose$ and $\bJ_n(\bB)\bJ_{0n}\transpose$):
\begin{align*}
  &\frac{2c_n}{mn}\sum_{t = 1}^m\sum_{i\neq j}\{\ell'(\theta_{ij}^{(t)}; A_{ij}^{(t)}) - \ell'(\theta_{0ij}^{(t)}; A_{ij}^{(t)})\}(\bR^{(t)}\otimes \be_i\be_j\transpose) + \frac{2c_n}{mn}\sum_{t = 1}^m\sum_{i\neq j}\ell'(\theta_{0ij}^{(t)}; A_{ij}^{(t)})\{(\bR^{(t)} - \bR_0^{(t)})\otimes \be_i\be_j\transpose\},\\
    &\frac{c_n}{\sqrt{m}n^{3/2}}\sum_{t = 1}^m\sum_{i\neq j}\{\ell'(\theta_{ij}^{(t)}; A_{ij}^{(t)}) - \ell'(\theta_{0ij}^{(t)}; A_{ij}^{(t)})\}\be_t\transpose\otimes\{\eye_d\otimes(\be_j\bx_i\transpose + \be_i\bx_j\transpose)\}\bD_d\\
  &\quad + \frac{c_n}{\sqrt{m}n^{3/2}}\sum_{t = 1}^m\sum_{i\neq j}  \ell'(\theta_{0ij}^{(t)}; A_{ij}^{(t)}) \be_t\transpose\otimes\{\eye_d\otimes(\be_j\bx_i\transpose + \be_i\bx_j\transpose - \be_j\bx_{0i}\transpose - \be_i\bx_{0j}\transpose)\}\bD_d\\
  &\frac{b_n}{\sqrt{m}n^{3/2}}\sum_{t = 1}^m\sum_{i\neq j}\{\ell'(\theta_{ij}^{(t)}; A_{ij}^{(t)}) - \ell'(\theta_{0ij}^{(t)}; A_{ij}^{(t)})\}\be_t\otimes\bD_d\transpose\{\eye_d\otimes(\be_j\bx_i\transpose + \be_i\bx_j\transpose)\transpose\}\\
  &\quad + \frac{b_n}{\sqrt{m}n^{3/2}}\sum_{t = 1}^m\sum_{i\neq j}  \ell'(\theta_{0ij}^{(t)}; A_{ij}^{(t)}) \be_t\otimes\bD_d\transpose\{\eye_d\otimes(\be_j\bx_i\transpose + \be_i\bx_j\transpose - \be_j\bx_{0i}\transpose - \be_i\bx_{0j}\transpose)\transpose\}. 
\end{align*}
By the Lipschitz continuity of $\ell'$ and Assumption \ref{assumption:edge_distribution}, $|\ell'(\theta_{ij}^{(t)}; A_{ij}^{(t)}) - \ell'(\theta_{0ij}^{(t)}; A_{ij}^{(t)})|\leq \kappa_2(M)|\theta_{ij}^{(t)} - \theta_{0ij}^{(t)}|$, so that the first terms in these three blocks have two-to-infinity norms bounded as
\begin{align*}
  &\sup_{\bphi\in\calN(\sqrt{m}n\eps_n, M)}\bigg\|\frac{2c_n}{mn}\sum_{t = 1}^m\sum_{i\neq j}\{\ell'(\theta_{ij}^{(t)}; A_{ij}^{(t)}) - \ell'(\theta_{0ij}^{(t)}; A_{ij}^{(t)})\}(\bR^{(t)}\otimes\be_i\be_j\transpose)\bigg\|_{2\to\infty}\\
  &\quad = 2\max_{k\in[d],i\in[n]}\sup_{\bphi}\bigg\|\frac{c_n}{mn}\sum_{t = 1}^m\sum_{j\in [n]\backslash\{i\}}\{\ell'(\theta_{ij}^{(t)}; A_{ij}^{(t)}) - \ell'(\theta_{0ij}^{(t)}; A_{ij}^{(t)})\}(\be_k\transpose\bR^{(t)} \otimes\be_j\transpose)\bigg\|_{2\to\infty}\\
  &\quad\lesssim \max_{k\in[d],i\in[n]}\sup_{\bphi}\bigg\|\frac{c_n}{mn}\sum_{t = 1}^m\sum_{j = 1}^n|\ell'(\theta_{ij}^{(t)}; A_{ij}^{(t)}) - \ell'(\theta_{0ij}^{(t)}; A_{ij}^{(t)})|\|\be_k\transpose\bR^{(t)}\|_2\be_j\transpose\bigg\|_{2\to\infty}\\
  &\quad = \max_{k\in[d],i\in[n]}\sup_{\bphi}\frac{c_n}{mn}\bigg[\sum_{j = 1}^n\bigg\{\sum_{t = 1}^m|\ell'(\theta_{ij}^{(t)}; A_{ij}^{(t)}) - \ell'(\theta_{0ij}^{(t)}; A_{ij}^{(t)})|\|\be_k\transpose\bR^{(t)}\|_2\bigg\}^2\bigg]^{1/2}\\
  &\quad\lesssim \max_{k\in[d],i\in[n]}\frac{c_n}{m^{1/2}n}\sup_{\bphi}\bigg\{\sum_{j = 1}^n \sum_{t = 1}^m(\theta_{ij}^{(t)} - \theta_{0ij}^{(t)})^2\bigg\}^{1/2} = O\bigg(\frac{c_n}{\sqrt{n}}\bigg),
\end{align*}
where we have used Cauchy-Schwarz inequality and the fact that $\sup_\bphi\max_{t,i,j}|\theta_{ij}^{(t)}| = O(1)$ in the last inequality, and similarly,
\begin{align*}
  &\sup_{\bphi}\bigg\|\frac{c_n}{\sqrt{m}n^{3/2}}\sum_{t = 1}^m\sum_{i\neq j}\{\ell'(\theta_{ij}^{(t)}; A_{ij}^{(t)}) - \ell'(\theta_{0ij}^{(t)}; A_{ij}^{(t)})\}\be_t\transpose\otimes\{\eye_d\otimes(\be_j\bx_i\transpose + \be_i\bx_j\transpose)\}\bD_d\bigg\|_{2\to\infty}\\
  &\quad \leq\frac{4c_n}{\sqrt{m}n^{3/2}}\max_{k\in[d],j\in[n]}\sup_{\bphi}\bigg\|\sum_{t = 1}^m\sum_{i\in[n]\backslash\{j\}}\{\ell'(\theta_{ij}^{(t)}; A_{ij}^{(t)}) - \ell'(\theta_{0ij}^{(t)}; A_{ij}^{(t)})\}\be_t\transpose\otimes (\eye_d\otimes\be_j\bx_i\transpose)\bigg\|_{2\to\infty} = O\bigg(\frac{c_n}{\sqrt{n}}\bigg),\\
  &\sup_\bphi\bigg\|\frac{b_n}{\sqrt{m}n^{3/2}}\sum_{t = 1}^m\sum_{i\neq j}\{\ell'(\theta_{ij}^{(t)}; A_{ij}^{(t)}) - \ell'(\theta_{0ij}^{(t)}; A_{ij}^{(t)})\}\be_t\otimes\bD_d\transpose\{\eye_d\otimes(\be_j\bx_i\transpose + \be_i\bx_j\transpose)\transpose\}\bigg\|_{2\to\infty}\\
  &\quad\leq \frac{2b_n}{\sqrt{m}n^{3/2}}\max_{k\in[d(d + 1)/2],t\in[m]}\sup_\bphi\bigg\|
  \sum_{i\neq j}\{\ell'(\theta_{ij}^{(t)}; A_{ij}^{(t)}) - \ell'(\theta_{0ij}^{(t)}; A_{ij}^{(t)})\}\be_k\transpose\bD_d\transpose(\eye_d\otimes\bx_i\be_j\transpose)
  \bigg\|_{2\to\infty}\\
  &\quad \lesssim \frac{b_n}{\sqrt{m}n^{3/2}}\sup_\bphi \bigg\{\sum_{j = 1}^n\bigg(\sum_{i = 1}^n|\theta_{ij}^{(t)} - \theta_{0ij}^{(t)}|\bigg)^2\bigg\}^{1/2} = O(b_n\eps_n). 
\end{align*}
For the second terms in these three nonzero blocks, since $\{\ell'(\theta_{0ij}^{(t)}; A_{ij}^{(t)}):1\leq i < j\leq n,t\in[m]\}$ are uniformly sub-exponential random variables, then they satisfy $\max_{t,i,j}|\ell'(\theta_{0ij}^{(t)}; A_{ij}^{(t)})| = \Optilde(\log n)$ by Lemma \ref{lemma:uniform_Optilde}. We then have
\begin{align*}
&\sup_{\bphi\in\calN(\sqrt{m}n\eps_n, M)}\bigg\|\frac{2c_n}{mn}\sum_{t = 1}^m\sum_{i\neq j}\ell'(\theta_{0ij}^{(t)}; A_{ij}^{(t)})\{(\bR^{(t)} - \bR_0^{(t)})\otimes\be_i\be_j\transpose\}\bigg\|_{2\to\infty}\\
&\quad = 2\max_{k\in[d],i\in[n]}\sup_\bphi\frac{c_n}{mn}\bigg\|\sum_{t = 1}^m\sum_{j\in[n]\backslash\{i\}}\ell'(\theta_{0ij}^{(t)}; A_{ij}^{(t)})\{\be_k\transpose(\bR^{(t)} - \bR_0^{(t)})\otimes\be_j\transpose\}\bigg\|_{2}\\
&\quad\lesssim \max_{k\in[d],i\in[n]}\sup_\bphi\frac{c_n}{mn}\bigg[\sum_{j\in[n]\backslash\{i\}}\bigg\{\sum_{t = 1}^m|\ell'(\theta_{0ij}^{(t)}; A_{ij}^{(t)})|\|\be_k\transpose(\bR^{(t)} - \bR_0^{(t)})\|_2\bigg\}^2\bigg]^{1/2}\\
&\quad\leq \max_{t,i,j}|\ell'(\theta_{0ij}^{(t)}; A_{ij}^{(t)})|\max_{k\in[d],i\in[n]}\sup_\bphi\frac{c_n}{mn}\bigg[n\bigg\{\sum_{t = 1}^m\|\be_k\transpose(\bR^{(t)} - \bR_0^{(t)})\|_2\bigg\}^2\bigg]^{1/2}\\
&\quad\lesssim \Optilde(\log n)\max_{i\in[n]}\frac{c_n}{mn}\sup_\bphi\bigg(mn\sum_{t = 1}^m\|\bR^{(t)} - \bR_0^{(t)}\|_{\mathrm{F}}^2\bigg)^{1/2}  = \Optilde\bigg(\frac{c_n\eps_n\log n}{\sqrt{n}}\bigg),
\end{align*}
where we have applied Cauchy-Schwarz inequality in the last inequality above, and similarly, 
\begin{align*}
&\sup_\bphi\bigg\|\frac{c_n}{\sqrt{m}n^{3/2}}\sum_{t = 1}^m\sum_{i\neq j}  \ell'(\theta_{0ij}^{(t)}; A_{ij}^{(t)}) \be_t\transpose\otimes\{\eye_d\otimes(\be_j\bx_i\transpose + \be_i\bx_j\transpose - \be_j\bx_{0i}\transpose - \be_i\bx_{0j}\transpose)\}\bD_d\bigg\|_{2\to\infty}\\
&\quad\leq \frac{4c_n}{\sqrt{m}n^{3/2}}\max_{k\in[d],j\in[n]}\bigg\|
\sum_{t = 1}^m\sum_{i\in[n]\backslash\{j\}}\ell'(\theta_{0ij}^{(t)}; A_{ij}^{(t)}) \be_t\transpose\otimes\{\be_k\transpose\otimes(\bx_i\transpose - \bx_{0i}\transpose)\} 
\bigg\|_2\\
&\quad\leq \max_{t,i,j}|\ell'(\theta_{0ij}^{(t)}; A_{ij}^{(t)})|\frac{c_n}{\sqrt{m}n^{3/2}}\max_{j\in[n]}\sup_\bphi\bigg[
m\bigg\{\sum_{i\in[n]\backslash\{j\}}\|\bx_i\transpose - \bx_{0i}\transpose\|_2
\bigg\}^2\bigg]^{1/2}\\
&\quad \leq\Optilde(\log n)\frac{c_n}{\sqrt{m}n^{3/2}}\bigg(mn\sum_{i = 1}^n\|\bx_i - \bx_{0i}\|_2^2\bigg)^{1/2} = \Optilde\bigg(\frac{c_n\eps_n\log n}{\sqrt{n}}\bigg),\\
&\sup_\bphi\bigg\|\frac{b_n}{\sqrt{m}n^{3/2}}\sum_{t = 1}^m\sum_{i\neq j}  \ell'(\theta_{0ij}^{(t)}; A_{ij}^{(t)}) \be_t\otimes\bD_d\transpose\{\eye_d\otimes(\bx_{0i}\be_j\transpose - \bx_i\be_j\transpose + \bx_{0j}\be_i\transpose - \bx_j\be_i\transpose)\}\bigg\|_{2\to\infty}\\
&\quad\leq \frac{2b_n}{\sqrt{m}n^{3/2}}\max_{t\in[m],k\in[d(d + 1)/2]}\sup_\bphi\bigg\|\sum_{i\neq j}\ell'(\theta_{0ij}^{(t)}; A_{ij}^{(t)})\be_k\transpose\bD_d\transpose\{\eye_d\otimes(\bx_i - \bx_{0i})\be_j\transpose\}\bigg\|_2\\
&\quad\lesssim \max_{t,i,j}|\ell'(\theta_{0ij}^{(t)}; A_{ij}^{(t)})|\frac{b_n}{\sqrt{m}n^{3/2}}\max_{t\in[m]}\bigg[ n \bigg\{\sum_{i = 1}^n \|\bx_i - \bx_{0i}\|\bigg\}^2\bigg]^{1/2}\\
&\quad\lesssim \Optilde(\log n)\frac{b_n}{\sqrt{m}n^{3/2}}\bigg(n^2  \sum_{i = 1}^n \|\bx_i - \bx_{0i}\|^2\bigg)^{1/2} = \Optilde\bigg(\frac{b_n\eps_n\log n}{\sqrt{m}}\bigg).
\end{align*}

\noindent$\blacksquare$ \textbf{Second term. }For the second term, we show below that, for a carefully designed $\bphi$-dependent matrix $\bXi_\bB(\bphi)$ satisfying $\sup_{\bphi\in\calN(\sqrt{m}n\eps_n, M)}\|\bXi_\bB(\bphi)\|_\infty = \Optilde(\eps_n\log n)$, one has
\begin{align*}
&\sup_{\bphi\in\calN(\sqrt{m}n\eps_n, M)}\bigg\|\frac{c_n}{mn^2}\sum_{t = 1}^m\sum_{i\neq j}\bigg(\eye_d\otimes\frac{\bX_{0\perp}}{\sqrt{n}}\bigg)\\
&\quad\times\sum_{k = 1}^d\bigg\{\ell'(\theta_{ij}^{(t)}; A_{ij}^{(t)})\frac{\partial^2x_{ik}(\bB)}{\partial\vect(\bB)\partial\vect(\bB)\transpose}\be_k\transpose\bR^{(t)}\bx_j - \ell'(\theta_{0ij}^{(t)};A_{ij}^{(t)})\frac{\partial^2x_{ik}(\mathbf{O})}{\partial\vect(\bB)\partial\vect(\bB)\transpose}\be_k\transpose\bR_0^{(t)}\bx_{0j}\bigg\}
\\
&\quad\times
\bigg(\eye_d\otimes\frac{\bX_{0\perp}\transpose}{\sqrt{n}}\bigg) - c_n\bXi_\bB(\bphi)\bigg\|_{2\to\infty}
\end{align*}
can be sharply bounded as needed. Then, one can directly take $\bXi_\bX(\bphi) = 2\bXi_\bB(\bphi)$. Recall that the inverse Lyapunov operator $\mathscr{L}^{-1}$ is linear and satisfies $\mathscr{L}^{-1}(\bM) = \int_0^\infty e^{-\tau\bS^{1/2}}\bM e^{-\tau\bS^{1/2}}\mathrm{d}\tau$ for any symmetric $\bM\in\mathbb{R}^{d\times d}$ by Lemma \ref{lemma:inverse_Lyapunov_operator_integral}, where we denote ${\bS} = \eye_d + {\bB}\transpose{\bB}$. Alternatively, one also has $\vect(\mathscr{L}^{-1}(\bM)) = (\bS^{1/2}\otimes \eye_d + \eye_d \otimes {\bS}^{1/2})^{-1}\vect(\bM)$. Therefore, by the calculations in Section \ref{sub:gradient_hessian}, we have
\begin{align*}
&(\eye_d\otimes \bX_{0\perp})\sum_{k = 1}^d\frac{\partial^2x_{ik}({\bB})}{\partial \vect(\bB)\partial \vect(\bB)\transpose}\be_k\transpose{\bR}^{(t)}{\bx}_j(\eye_d\otimes \bX_{0\perp}\transpose)\\
&\quad = -(\eye_d\otimes \bX_{0\perp}\bX_{0\perp}\transpose\be_i)({\bx}_j\transpose{\bR}^{(t)}{\bS}^{-1/2}\otimes {\bS}^{-1/2})({\bS}^{1/2}\otimes\eye_d + \eye_d\otimes {\bS}^{1/2})^{-1}
(\eye_{d^2} + \bK_{dd})(\eye_d\otimes \bB\transpose\bX_{0\perp}\transpose)\\
&\qquad - (\eye_d\otimes\bX_{0\perp}\bB)(\eye_{d^2} + \bK_{dd})({\bS}^{1/2}\otimes\eye_d + \eye_d\otimes {\bS}^{1/2})^{-1}
({\bS}^{-1/2}{\bR}^{(t)}{\bx}_j\otimes {\bS}^{-1/2})(\eye_d\otimes \be_i\transpose\bX_{0\perp}\bX_{0\perp}\transpose)\\
&\qquad + (\eye_d\otimes\bX_{0\perp}\bB)(\eye_{d^2} + \bK_{dd})({\bS}^{1/2}\otimes\eye_d + \eye_d\otimes {\bS}^{1/2})^{-1}
({\bx}_i{\bx}_j\transpose{\bR}^{(t)}{\bS}^{-1/2}\otimes {\bS}^{-1/2})\\
&\qquad\times ({\bS}^{1/2}\otimes\eye_d + \eye_d\otimes {\bS}^{1/2})^{-1}(\eye_{d^2} + \bK_{dd})(\eye_d\otimes \bB\transpose\bX_{0\perp}\transpose)\\
&\qquad + (\eye_d\otimes\bX_{0\perp}\bB)(\eye_{d^2} + \bK_{dd})({\bS}^{1/2}\otimes\eye_d + \eye_d\otimes {\bS}^{1/2})^{-1}({\bS}^{-1/2}{\bR}^{(t)}{\bx}_j{\bx}_i\transpose\otimes {\bS}^{-1/2})\\
&\qquad\times ({\bS}^{1/2}\otimes\eye_d + \eye_d\otimes {\bS}^{1/2})^{-1}(\eye_{d^2} + \bK_{dd})(\eye_d\otimes \bB\transpose\bX_{0\perp}\transpose) \\
&\qquad - \int_0^\infty \bigg\{e^{-\tau{\bS}^{1/2}}({\bx}_i{\bx}_j\transpose{\bR}^{(t)}{\bS}^{-1/2} + {\bS}^{-1/2}{\bR}^{(t)}{\bx}_j{\bx}_i\transpose)e^{-\tau{\bS}^{1/2}}\otimes \bX_{0\perp}\bX_{0\perp}\transpose\bigg\}\mathrm{d}\tau \\
&\qquad + (\eye_d \otimes \bX_{0\perp}\bB)(\eye_{d^2} + \bK_{dd})({\bS}^{1/2}\otimes \eye_d + \eye_d \otimes {\bS}^{1/2})^{-1}\\
&\qquad\times 
\int_0^\infty \bigg\{e^{-\tau{\bS}^{1/2}}({\bx}_i{\bx}_j\transpose{\bR}^{(t)}{\bS}^{-1/2} + {\bS}^{-1/2}{\bR}^{(t)}{\bx}_j{\bx}_i\transpose) e^{-\tau{\bS}^{1/2}}\otimes \eye_d\bigg\}\mathrm{d}\tau\\
&\qquad\times ({\bS}^{1/2}\otimes \eye_d + \eye_d \otimes {\bS}^{1/2})^{-1}
(\eye_{d^2} + \bK_{dd})(\eye_d\otimes \bB\transpose\bX_{0\perp}\transpose).
\end{align*}
Define
\begin{align*}
\bXi_{\bB}^{(1)}(\bphi)
& = -\frac{1}{mn^2}\sum_{t = 1}^m\sum_{i\neq j}\{\ell'(\theta_{ij}^{(t)}; A_{ij}^{(t)}) - \ell'(\theta_{0ij}^{(t)}; A_{ij}^{(t)})\}\\
&\quad \times \bigg[(\eye_d\otimes\bX_{0\perp}\bB)(\eye_{d^2} + \bK_{dd})({\bS}^{1/2}\otimes\eye_d + \eye_d\otimes {\bS}^{1/2})^{-1}
({\bS}^{-1/2}{\bR}^{(t)}{\bx}_j\otimes {\bS}^{-1/2})\bigg(\eye_d\otimes \frac{\be_i\transpose\bX_{0\perp}\bX_{0\perp}\transpose}{n}\bigg)\\
&\quad + \int_0^\infty \bigg\{e^{-\tau{\bS}^{1/2}}({\bx}_i{\bx}_j\transpose{\bR}^{(t)}{\bS}^{-1/2} + {\bS}^{-1/2}{\bR}^{(t)}{\bx}_j{\bx}_i\transpose)e^{-\tau{\bS}^{1/2}}\otimes \frac{\bX_{0\perp}\bX_{0\perp}\transpose}{n}\bigg\}\mathrm{d}\tau\bigg].
\end{align*}
Since $(\be_k\transpose\otimes\be_l\transpose)\bXi_\bB^{(1)}(\bphi)$ is a sum over $(t, i, j)$ of
rows of the form 
\[
\{\ell'(\theta_{ij}^{(t)}; A_{ij}^{(t)}) -
\ell'(\theta_{0ij}^{(t)}; A_{ij}^{(t)})\}\bigg\{\bu_{klj}^{(t)}\bigg(\eye_d\otimes\be_i\transpose
\frac{\bX_{0\perp}\bX_{0\perp}\transpose}{n}\bigg) + \bv_{klij}^{(t)}\otimes\frac{\be_l\transpose\bX_{0\perp}\bX_{0\perp}\transpose}{n}
\bigg\},
\]
where
$\max_{t,k,i,j,l}\|\bu_{klj}^{(t)}\|_\infty\vee\|\bv\|_\infty = O(1)$ and
$\max_{l\in[n]}\|\be_l\transpose\bX_{0\perp}\bX_{0\perp}\transpose\|_\infty/n  = O(1)$, Cauchy-Schwarz inequality implies
\begin{align*}
\sup_\bphi\|\bXi_{\bB}^{(1)}(\bphi)\|_{\infty}
&\lesssim \sup_\bphi\frac{1}{mn^2}\sum_{t = 1}^m\sum_{i\neq j}
  |\ell'(\theta_{ij}^{(t)}; A_{ij}^{(t)}) - \ell'(\theta_{0ij}^{(t)}; A_{ij}^{(t)})|\\
&\lesssim \sup_\bphi\frac{1}{mn^2}\sqrt{mn^2}\bigg(\sum_{t = 1}^m\sum_{i\neq j}
  |\theta_{ij}^{(t)} - \theta_{0ij}^{(t)}|^2\bigg)^{1/2} = O(\eps_n).
\end{align*}
By Cauchy-Schwarz inequality, the formula above also implies that
\begin{align*}
&\sup_{\bphi\in\calN(\sqrt{m}n\eps_n, M)}\bigg\|\frac{c_n}{mn^2}\sum_{t = 1}^m\sum_{i\neq j}\{\ell'(\theta_{ij}^{(t)}; A_{ij}^{(t)}) - \ell'(\theta_{0ij}^{(t)}; A_{ij}^{(t)})\}\\
&\quad\times \bigg(\eye_d\otimes \frac{\bX_{0\perp}}{\sqrt{n}}\bigg)\sum_{k = 1}^d\frac{\partial^2x_{ik}({\bB})}{\partial \vect(\bB)\partial \vect(\bB)\transpose}\be_k\transpose{\bR}^{(t)}{\bx}_j\bigg(\eye_d\otimes \frac{\bX_{0\perp}\transpose}{\sqrt{n}}\bigg) - c_n\bXi_\bB^{(1)}(\bphi)\bigg\|_{2\to\infty}\\
&\quad\lesssim \sup_\bphi \bigg\|\frac{c_n}{mn^2}\sum_{t = 1}^m\sum_{i\neq j}\{\ell'(\theta_{ij}^{(t)}; A_{ij}^{(t)}) - \ell'(\theta_{0ij}^{(t)}; A_{ij}^{(t)})\}\\
&\qquad\times (\eye_d\otimes\be_i)({\bx}_j\transpose{\bR}^{(t)}{\bS}^{-1/2}\otimes {\bS}^{-1/2})({\bS}^{1/2}\otimes\eye_d + \eye_d\otimes {\bS}^{1/2})^{-1}
(\eye_{d^2} + \bK_{dd})(\eye_d\otimes \bB\transpose\bX_{0\perp}\transpose)\bigg\|_{2\to\infty}\\
&\qquad + \sup_\bphi \bigg\|\frac{c_n}{mn^2}\sum_{t = 1}^m\sum_{i\neq j}\{\ell'(\theta_{ij}^{(t)}; A_{ij}^{(t)}) - \ell'(\theta_{0ij}^{(t)}; A_{ij}^{(t)})\}\bigg\{\eye_d\otimes\bigg(\frac{\one_n}{n} + \frac{\bX_0\bx_{0i}}{n}\bigg)\bigg\}\\
&\qquad\times ({\bx}_j\transpose{\bR}^{(t)}{\bS}^{-1/2}\otimes {\bS}^{-1/2})({\bS}^{1/2}\otimes\eye_d + \eye_d\otimes {\bS}^{1/2})^{-1}
(\eye_{d^2} + \bK_{dd})(\eye_d\otimes \bB\transpose\bX_{0\perp}\transpose)\bigg\|_{2\to\infty}\\
&\qquad + \sup_\bphi\frac{c_n}{mn^3}\sum_{t = 1}^m\sum_{i\neq j}|\ell'(\theta_{ij}^{(t)}; A_{ij}^{(t)}) - \ell'(\theta_{0ij}^{(t)}; A_{ij}^{(t)})| \|\bX_{0\perp}\bB\|_{2\to\infty} \|\bB\transpose\bX_{0\perp}\transpose\|_2 \\
&\quad\lesssim \sup_\bphi\max_{i\in[n]}\frac{c_n}{mn^2}\sum_{t = 1}^m\sum_{j = 1}^n|\ell'(\theta_{ij}^{(t)}; A_{ij}^{(t)}) - \ell'(\theta_{0ij}^{(t)}; A_{ij}^{(t)})|\|\bB\transpose\bX_{0\perp}\transpose\|_2\\
&\qquad + \sup_\bphi \frac{c_n}{mn^3}(1 + \|\bX_0\|_{2\to\infty}^2)\sum_{t = 1}^m\sum_{i\neq j}|\ell'(\theta_{ij}^{(t)}; A_{ij}^{(t)}) - \ell'(\theta_{0ij}^{(t)}; A_{ij}^{(t)})|\|\bB\transpose\bX_{0\perp}\transpose\|_2\\
&\qquad + \sup_\bphi\frac{c_n}{m^{1/2}n^{3/2}}\bigg(\sum_{t = 1}^m\sum_{i\neq j}|\theta_{ij}^{(t)} - \theta_{0ij}^{(t)}|^2\bigg)^{1/2}\|\bB\|_2\\
&\quad\lesssim \sup_\bphi\max_{i\in[n]}\frac{c_n}{m^{1/2}n}\bigg(\sum_{t = 1}^m\sum_{j = 1}^n|\theta_{ij}^{(t)} - \theta_{0ij}^{(t)}|^2\bigg)^{1/2}\|\bB\|_2
\\&\qquad
 + \sup_\bphi \frac{c_n}{m^{1/2}n^{3/2}}(1 + \|\bX_0\|_{2\to\infty}^2)\bigg(\sum_{t = 1}^m\sum_{i\neq j}|\theta_{ij}^{(t)} - \theta_{0ij}^{(t)}|^2\bigg)^{1/2}\|\bB\|_2 = O\bigg(\frac{c_n\eps_n}{\sqrt{n}}\bigg). 
\end{align*}
Note that when taking supremum over $\bphi\in\calN(\sqrt{m}n\eps_n, M)$, 
\begin{align*}
&\sup_{\bphi}
\|\bS^{1/2} - \eye_d\|_2 = O(\eps_n^2),\quad
\sup_{\bphi}
\|\bS^{-1/2} - \eye_d\|_2 = O(\eps_n^2),\quad
\sup_{\bphi}
\|e^{-\tau\bS^{1/2}} - e^{-\tau\eye_d}\|_2\leq e^{-\tau}\{1 - e^{-\tau O(\eps_n^2)}\},
\\
&\sup_\bphi\sum_{t,i,j}\int_0^\infty\bigg\|e^{-\tau{\bS}^{1/2}}{\bx}_i{\bx}_j\transpose{\bR}^{(t)}{\bS}^{-1/2}e^{-\tau{\bS}^{1/2}}\ell'(\theta_{0ij}^{(t)}; A_{ij}^{(t)}) - e^{-\tau\eye_d}\bx_{0i}\bx_{0j}\transpose\bR_0^{(t)}e^{-\tau\eye_d}\ell'(\theta_{0ij}^{(t)}; A_{ij}^{(t)})\bigg\|_2\mathrm{d}\tau\\
&\quad\leq \sup_\bphi\sum_{t,i,j}\Optilde(\log n)\int_0^\infty\|e^{-\tau{\bS}^{1/2}} - e^{-\tau\eye_d}\|\|\bx_i\|_2\|\bx_j\|_2\|\bR^{(t)}\|_2\|e^{-\tau{\bS}^{1/2}}\|_2\mathrm{d}\tau\\
&\qquad + \sup_\bphi\sum_{t,i,j}\Optilde(\log n)\int_0^\infty\|e^{-\tau\eye_d}\|(\|\bx_i - \bx_{0i}\|_2\|\bx_j\|_2 + \|\bx_{0i}\|_2\|\bx_j - \bx_{0j}\|_2)\|\bR^{(t)}\|_2\|e^{-\tau{\bS}^{1/2}}\|_2\mathrm{d}\tau\\
&\qquad + \sup_\bphi\sum_{t,i,j}\Optilde(\log n)\int_0^\infty\|e^{-\tau\eye_d}\|\|\bx_{0i}\|_2\|\bx_{0j}\|_2\|\bR^{(t)} - \bR_0^{(t)}\|_{\mathrm{F}}\|e^{-\tau{\bS}^{1/2}}\|_2\mathrm{d}\tau\\
&\qquad + \sup_\bphi\sum_{t,i,j}\Optilde(\log n)\int_0^\infty\|e^{-\tau\eye_d}\|\|\bx_{0i}\|_2\|\bx_{0j}\|_2\|\bR_0^{(t)}\|_2\|e^{-\tau{\bS}^{1/2}} - e^{-\tau\eye_d}\|_2\mathrm{d}\tau\\
&\quad\leq 
 \sup_\bphi \Optilde(mn^2\log n)\int_0^\infty e^{-2\tau}\left\{1 - e^{-\tau O(\eps_n^2)}\right\}\mathrm{d}\tau + \sup_\bphi \Optilde(mn\log n)\sum_{i}\int_0^\infty e^{-2\tau} \|\bx_i - \bx_{0i}\|_2 \mathrm{d}\tau\\
&\qquad
 + \sup_\bphi \Optilde(n^2\log n)\sum_{t}\int_0^\infty e^{-2\tau}\|\bR^{(t)} - \bR_0^{(t)}\|_{\mathrm{F}} \mathrm{d}\tau\\
&\quad = \Optilde(mn^2\eps_n\log n).
\end{align*}
Then, by taking
\begin{align*}
\bXi_\bB^{(2)}(\bphi)
& = -\frac{1}{mn^2}\sum_{t = 1}^m\sum_{i\neq j}\int_0^\infty\bigg\{e^{-\tau\bS^{1/2}}(\bx_i\bx_j\transpose\bR^{(t)}\bS^{-1/2} + \bS^{-1/2}\bR^{(t)}\bx_j\bx_i\transpose)e^{-\tau\bS^{1/2}}\ell'(\theta_{0ij}^{(t)}; A_{ij}^{(t)})\otimes \frac{\bX_{0\perp}\bX_{0\perp}\transpose}{n}\bigg\}\mathrm{d}\tau\\
&\quad + \frac{1}{mn^2}\sum_{t = 1}^m\sum_{i\neq j}\int_0^\infty\bigg\{e^{-\tau\eye_d}(\bx_{0i}\bx_{0j}\transpose\bR_0^{(t)} + \bR_0^{(t)}\bx_{0j}\bx_{0i}\transpose)e^{-\tau\eye_d}\ell'(\theta_{0ij}^{(t)}; A_{ij}^{(t)})\otimes \frac{\bX_{0\perp}\bX_{0\perp}\transpose}{n}\bigg\}\mathrm{d}\tau,
\end{align*}
we have $\sup_\bphi\|\bXi_\bB^{(2)}(\bphi)\|_\infty = \Optilde(\eps_n\log n)$. Also, by the formula of 
\[
  (\eye_d\otimes \bX_{0\perp})\sum_{k = 1}^d\frac{\partial^2x_{ik}({\bB})}{\partial \vect(\bB)\partial \vect(\bB)\transpose}\be_k\transpose{\bR}^{(t)}{\bx}_j(\eye_d\otimes \bX_{0\perp}\transpose),
\]
the fact that
\begin{align*}
&\max_{k\in[d],l\in[n]} 
\sum_{t = 1}^m\sum_{j = 1}^n\bigg\|\sum_{i\in[n]\backslash\{j\}}\ell'(\theta_{0ij}^{(t)}; A_{ij}^{(t)})\bigg(\be_k\transpose\otimes \frac{\be_l\transpose\bX_{0\perp}\bX_{0\perp}\transpose\be_i}{n}\bigg)\bigg\|_2\\
&\quad\leq \max_{k\in[d],l\in[n]} 
\sum_{t = 1}^m\sum_{j = 1}^n\sum_{i\in[n]\backslash\{j\}}\bigg\|\ell'(\theta_{0ij}^{(t)}; A_{ij}^{(t)})\bigg\{\be_k\transpose\otimes \bigg(\be_l\transpose\be_i - \frac{1 + \bx_{0l}\transpose\bx_{0i}}{n}\bigg)\bigg\}\bigg\|_2\\
&\quad\leq O(mn) + \max_{l\in[n]}\sum_{t = 1}^m\sum_{j\in[n]\backslash\{l\}}\{|\ell'(\theta_{0lj}^{(t)}; A_{lj}^{(t)})| - \expect_0|\ell'(\theta_{0lj}^{(t)}; A_{lj}^{(t)})|\}\\
&\qquad + \max_{l\in[n]}\sum_{t = 1}^m\sum_{i\neq j}\{|\ell'(\theta_{0ij}^{(t)}; A_{ij}^{(t)})| - \expect_0|\ell'(\theta_{0ij}^{(t)}; A_{ij}^{(t)})|\}\bigg(\frac{1 + |\bx_{0l}\transpose\bx_{0i}|}{n}\bigg)\\
&\quad = O(mn) + \Optilde(\sqrt{mn\log n}) + \Optilde(\sqrt{m\log n}) = \Optilde(mn)
\end{align*}
by Bernstein's inequality, and the result $\|\ell'(\bTheta_0^{(t)})\|_2 = \Optilde(\sqrt{n\log n})$ by matrix Bernstein's inequality (Lemma \ref{lemma:matrix_Bernstein} and Theorem 6.2 in \cite{tropp2012user}), we further compute the following upper bound:
\begin{align*}
&\sup_{\bphi\in\calN(\sqrt{m}n\eps_n, M)}\bigg\|\frac{c_n}{mn^2}\sum_{t = 1}^m\sum_{i\neq j}\bigg(\eye_d\otimes\frac{\bX_{0\perp}}{\sqrt{n}}\bigg)\\
&\quad\times\sum_{k = 1}^d\ell'(\theta_{0ij}^{(t)}; A_{ij}^{(t)})\bigg\{\frac{\partial^2x_{ik}(\bB)}{\partial\vect(\bB)\partial\vect(\bB)\transpose}\be_k\transpose\bR^{(t)}\bx_j - \frac{\partial^2x_{ik}(\mathbf{O})}{\partial\vect(\bB)\partial\vect(\bB)\transpose}\be_k\transpose\bR_0^{(t)}\bx_{0j}\bigg\}
\\
&\quad\times
\bigg(\eye_d\otimes\frac{\bX_{0\perp}\transpose}{\sqrt{n}}\bigg) - c_n\bXi_\bB^{(2)}(\bphi)\bigg\|_{2\to\infty}\\
&\quad\lesssim \sup_\bphi\frac{c_n}{mn^3}\bigg\|
\sum_{t = 1}^m\sum_{i\neq j}\ell'(\theta_{0ij}^{(t)}; A_{ij}^{(t)})(\eye_d\otimes \bX_{0\perp}\bX_{0\perp}\transpose\be_i)({\bx}_j\transpose{\bR}^{(t)}{\bS}^{-1/2}\otimes {\bS}^{-1/2})\bigg\|_{2\to\infty} \|\bB\transpose\bX_{0\perp}\transpose
\|_2\\
&\qquad + \sup_\bphi\frac{c_n}{mn^3}\bigg\|
\sum_{t = 1}^m\sum_{i\neq j}\ell'(\theta_{0ij}^{(t)}; A_{ij}^{(t)})(\eye_d\otimes\bX_{0\perp}\bB)(\eye_{d^2} + \bK_{dd})({\bS}^{1/2}\otimes\eye_d + \eye_d\otimes {\bS}^{1/2})^{-1}\\
&\qquad\times 
({\bS}^{-1/2}{\bR}^{(t)}{\bx}_j\otimes {\bS}^{-1/2})(\eye_d\otimes \be_i\transpose\bX_{0\perp}\bX_{0\perp}\transpose)\bigg\|_{2\to\infty}\\
&\qquad + \sup_\bphi\frac{c_n}{mn^3}\|\bX_{0\perp}\bB\|_{2\to\infty}
\bigg\|\bigg\{\sum_{t = 1}^m\sum_{i\neq j}\ell'(\theta_{0ij}^{(t)}; A_{ij}^{(t)}){\bx}_i{\bx}_j\transpose{\bR}^{(t)}{\bS}^{-1/2}\bigg\}\otimes {\bS}^{-1/2}\bigg\|_2\|\bB\transpose\bX_{0\perp}\transpose\|_2
\\
&\qquad + \sup_\bphi\frac{c_n}{mn^3}\|\bX_{0\perp}\bB\|_{2\to\infty}\bigg\|\bigg\{
\sum_{t = 1}^m\sum_{i\neq j}\ell'(\theta_{0ij}^{(t)}; A_{ij}^{(t)}){\bS}^{-1/2}{\bR}^{(t)}{\bx}_j{\bx}_i\transpose\bigg\}\otimes {\bS}^{-1/2}\bigg\|_2 \|\bB\transpose\bX_{0\perp}\transpose\|_{2} \\
&\qquad + \sup_\bphi\frac{c_n}{mn^3}\|\bX_{0\perp}\bB\|_{2\to\infty}\\
&\qquad\times 
\bigg\|
\sum_{t = 1}^m\sum_{i\neq j}\ell'(\theta_{0ij}^{(t)}; A_{ij}^{(t)})\int_0^\infty e^{-\tau{\bS}^{1/2}}({\bx}_i{\bx}_j\transpose{\bR}^{(t)}{\bS}^{-1/2} + {\bS}^{-1/2}{\bR}^{(t)}{\bx}_j{\bx}_i\transpose) e^{-\tau{\bS}^{1/2}}\mathrm{d}\tau\bigg\|_2\|\bB\transpose\bX_{0\perp}\transpose\|_{2}\\
&\quad\lesssim \sup_\bphi\frac{c_n}{mn^2}\max_{k\in[d],l\in[n]} 
\sum_{t = 1}^m\sum_{j = 1}^n\bigg\|\sum_{i\in[n]\backslash\{j\}}\ell'(\theta_{0ij}^{(t)}; A_{ij}^{(t)})\bigg(\be_k\transpose\otimes \frac{\be_l\transpose\bX_{0\perp}\bX_{0\perp}\transpose\be_i}{n}\bigg)\bigg\|_2\|{\bx}_j\transpose{\bR}^{(t)}\|_2  \|\bB\transpose\bX_{0\perp}\transpose
\|_2\\
&\qquad + \sup_\bphi\frac{c_n}{mn^3}\|\bX_{0\perp}\bB\|_{2\to\infty}
\sum_{t = 1}^m\sum_{j = 1}^n
\|{\bR}^{(t)}\|_2\|{\bx}_j\|_2\bigg\|\sum_{i\in[n]\backslash\{j\}}\ell'(\theta_{0ij}^{(t)}; A_{ij}^{(t)})\be_i\bigg\|_2\|\bX_{0\perp}\bX_{0\perp}\transpose\|_2\\
&\qquad + \sup_\bphi\frac{c_n}{mn^3}\|\bX_{0\perp}\bB\|_{2\to\infty}
\sum_{t = 1}^m\|\bX(\bB)\|_2\|\ell'(\bTheta_0^{(t)})\|_2\|\bX(\bB)\|_2\|{\bR}^{(t)}\|_2\|\bB\transpose\bX_{0\perp}\transpose\|_2
\\
&\quad = \Optilde\bigg(c_n\sqrt{\frac{\log n}{n}}\bigg).
\end{align*}
Now let $\bXi_\bB(\bphi) = \bXi_\bB^{(1)}(\bphi) + \bXi_\bB^{(2)}(\bphi)$. Clearly, we obtain directly that  $\sup_\bphi\|\bXi_\bB(\bphi)\|_\infty = \Optilde(\eps_n\log n)$ and
\begin{align*}
&\sup_{\bphi\in\calN(\sqrt{m}n\eps_n, M)}\bigg\|\frac{c_n}{mn^2}\sum_{t = 1}^m\sum_{i\neq j}\bigg(\eye_d\otimes\frac{\bX_{0\perp}}{\sqrt{n}}\bigg)\\
&\quad\times\sum_{k = 1}^d\bigg\{\ell'(\theta_{ij}^{(t)}; A_{ij}^{(t)})\frac{\partial^2x_{ik}(\bB)}{\partial\vect(\bB)\partial\vect(\bB)\transpose}\be_k\transpose\bR^{(t)}\bx_j - \ell'(\theta_{0ij}^{(t)};A_{ij}^{(t)})\frac{\partial^2x_{ik}(\mathbf{O})}{\partial\vect(\bB)\partial\vect(\bB)\transpose}\be_k\transpose\bR_0^{(t)}\bx_{0j}\bigg\}
\\
&\quad\times
\bigg(\eye_d\otimes\frac{\bX_{0\perp}\transpose}{\sqrt{n}}\bigg) - c_n\bXi_\bB(\bphi)\bigg\|_{2\to\infty}\\
&\quad\leq \sup_{\bphi\in\calN(\sqrt{m}n\eps_n, M)}\bigg\|\frac{c_n}{mn^2}\sum_{t = 1}^m\sum_{i\neq j}\bigg(\eye_d\otimes\frac{\bX_{0\perp}}{\sqrt{n}}\bigg)\{\ell'(\theta_{ij}^{(t)}; A_{ij}^{(t)}) - \ell'(\theta_{0ij}^{(t)}; A_{ij}^{(t)})\}\\
&\quad\quad\times\sum_{k = 1}^d\bigg\{\frac{\partial^2x_{ik}(\bB)}{\partial\vect(\bB)\partial\vect(\bB)\transpose}\be_k\transpose\bR^{(t)}\bx_j \bigg\}
\bigg(\eye_d\otimes\frac{\bX_{0\perp}\transpose}{\sqrt{n}}\bigg) - c_n\bXi_\bB^{(1)}(\bphi) \bigg\|_{2\to\infty}\\
&\quad\quad + \sup_{\bphi\in\calN(\sqrt{m}n\eps_n, M)}\bigg\|\frac{c_n}{mn^2}\sum_{t = 1}^m\sum_{i\neq j}\bigg(\eye_d\otimes\frac{\bX_{0\perp}}{\sqrt{n}}\bigg)\\
&\qquad\times\sum_{k = 1}^d\ell'(\theta_{0ij}^{(t)}; A_{ij}^{(t)})\bigg\{\frac{\partial^2x_{ik}(\bB)}{\partial\vect(\bB)\partial\vect(\bB)\transpose}\be_k\transpose\bR^{(t)}\bx_j - \frac{\partial^2x_{ik}(\mathbf{O})}{\partial\vect(\bB)\partial\vect(\bB)\transpose}\be_k\transpose\bR_0^{(t)}\bx_{0j}\bigg\}
\\
&\qquad\times
\bigg(\eye_d\otimes\frac{\bX_{0\perp}\transpose}{\sqrt{n}}\bigg) - c_n\bXi_\bB^{(2)}(\bphi)\bigg\|_{2\to\infty} = \Optilde\bigg(c_n\sqrt{\frac{\log n}{n}}\bigg).
\end{align*}

\noindent $\blacksquare$ \textbf{Score term at $\bphi = \bphi_0$ and the remaining terms. }By \eqref{eqn:Jn(B)J0n_perturbation} and the block diagonality of $\bJ_n(\bB),\bJ_{0n}$, we will work on
\begin{align*}
&\bigg\|
\mathrm{diag}(a_n\eye_{d_\balpha}, b_n\eye_{d_\br}, c_n\eye_{d_\bX})
\widetilde{\bLambda}_\sharp
\sum_{t = 1}^m\sum_{i\neq j}\ell'(\theta_{0ij}^{(t)}; A_{ij}^{(t)})\\
&\quad\times \begin{bmatrix}
\mathbf{O} & \mathbf{O} & \mathbf{O}\\
\mathbf{O} & \mathbf{O} & \be_t\otimes\bD_d\transpose\{\eye_d\otimes({\bx}_{0i}\be_j\transpose + {\bx}_{0j}\be_i\transpose)\}\\
\mathbf{O} & \be_t\transpose\otimes\{\eye_d\otimes(\be_j{\bx}_{0i}\transpose + \be_i{\bx}_{0j}\transpose)\}\bD_d& 
\bR_0^{(t)}\otimes(\be_i\be_j\transpose + \be_j\be_i\transpose)
\end{bmatrix}\\
&\quad\times \widetilde{\bLambda}_\sharp
\bigg\|_{2\to\infty} = \Optilde\bigg(\frac{c_n\sqrt{\log n}}{n} + (b_n\vee c_n)\sqrt{\frac{\log n}{mn}}\bigg)
\end{align*}
directly and that implies the necessary error bounds for the remaining terms. Note that this term does not depend on $\bphi$. Therefore, we can apply matrix Bernstein's inequality to obtain
\begin{align*}
&\bigg\|\frac{c_n}{mn}\sum_{t = 1}^m\sum_{i\neq j}\ell'(\theta_{0ij}^{(t)}; A_{ij}^{(t)})\bR_0^{(t)}\otimes (\be_i\be_j\transpose + \be_j\be_i\transpose)\bigg\|_{2\to\infty}\\
&\quad\leq \frac{2c_n}{mn}\max_{k\in[d],i\in[n]}\bigg\|\sum_{t = 1}^m\sum_{j\in[n]\backslash\{i\}}\ell'(\theta_{0ij}^{(t)}; A_{ij}^{(t)})\be_k\transpose\bR_0^{(t)}\be_j\transpose\bigg\|_2\\
&\quad = \frac{c_n}{mn}\Optilde\bigg\{\max_{t,j}\|\be_k\transpose\bR_0^{(t)}\otimes\be_j\transpose\|_2\log n + \bigg(\sum_{t = 1}^m\sum_{j = 1}^n\|\be_k\transpose\bR_0^{(t)}\otimes\be_j\transpose\|_2^2\bigg)^{1/2}\sqrt{\log n}\bigg\} = \Optilde\bigg(c_n\sqrt{\frac{\log n}{mn}}\bigg)\\
&\bigg\|\frac{c_n}{\sqrt{m}n^{3/2}}\sum_{t = 1}^m\sum_{i\neq j}\ell'(\theta_{0ij}^{(t)}; A_{ij}^{(t)})\be_t\transpose\otimes\{\eye_d\otimes(\be_j\bx_{0i}\transpose + \be_i\bx_{0j}\transpose)\}\bD_d\bigg\|_{2\to\infty}\\
&\quad\leq \frac{2c_n}{\sqrt{m}n^{3/2}}\max_{k\in[d],i\in[n]}\bigg\|\sum_{t = 1}^m\sum_{j\in[n]\backslash\{i\}}\ell'(\theta_{0ij}^{(t)}; A_{ij}^{(t)})\be_t\transpose\otimes (\be_k\transpose\otimes\bx_{0j}\transpose)\bD_d\bigg\|_2\\
&\quad = \frac{c_n}{\sqrt{m}n^{3/2}}\Optilde\bigg\{\max_{t,j}\|\be_t\transpose\otimes(\be_k\transpose\otimes\bx_{0j}\transpose)\|_2\log n + \bigg(\sum_{t = 1}^m\sum_{j = 1}^n\|\be_t\transpose\otimes (\be_k\transpose\otimes\bx_{0j}\transpose)\|_2^2\bigg)^{1/2}\sqrt{\log n}\bigg\}\\
&\quad = \Optilde\bigg(\frac{c_n\sqrt{\log n}}{n}\bigg),
\\
&\bigg\|\frac{b_n}{\sqrt{m}n^{3/2}}\sum_{t = 1}^m\sum_{i\neq j}\ell'(\theta_{0ij}^{(t)}; A_{ij}^{(t)})\be_t\otimes\bD_d\transpose\{\eye_d\otimes(\bx_{0i}\be_j\transpose + \bx_{0j}\be_i\transpose)\}\bigg\|_{2\to\infty}\\
&\quad\leq \frac{2b_n}{\sqrt{m}n^{3/2}}\max_{t\in[m],k\in[d(d + 1)/2]}\bigg\|\sum_{i\neq j}\ell'(\theta_{0ij}^{(t)}; A_{ij}^{(t)})\be_k\transpose\bD_d\transpose(\eye_d\otimes\bx_{0i}\be_j\transpose)\bigg\|_2\\
&\quad = \frac{b_n}{\sqrt{m}n^{3/2}}\Optilde\bigg\{\max_{i,j}\|\eye_d\otimes\bx_{0i}\be_j\transpose\|_2\log n + \bigg(\sum_{i\neq j}\|\be_k\transpose\bD_d\transpose(\eye_d\otimes \bx_{0i}\be_j\transpose)\|_2^2\bigg)^{1/2}\sqrt{\log n}\bigg\}\\
&\quad = \Optilde\bigg(\frac{b_n\sqrt{\log n}}{\sqrt{mn}}\bigg),\\
&\bigg\|\frac{c_n}{mn^2}\sum_{t = 1}^m\sum_{i\neq j}\bigg(\eye_d\otimes\frac{\bX_{0\perp}}{\sqrt{n}}\bigg)\sum_{k = 1}^d\ell'(\theta_{0ij}^{(t)}; A_{ij}^{(t)})\frac{\partial^2 x_{ik}(\mathbf{O})}{\partial\vect(\bB)\partial\vect(\bB)\transpose}\be_k\transpose\bR_0^{(t)}\bx_{0j}\bigg(\eye_d\otimes\frac{\bX_{0\perp}\transpose}{\sqrt{n}}\bigg)\bigg\|_{2\to\infty}\\
&\quad = \Optilde\bigg(\frac{c_n\sqrt{\log n}}{\sqrt{m}n}\bigg)\bigg\|\frac{\bX_{0\perp}\bX_{0\perp}\transpose}{n}\bigg\|_{2\to\infty} = \Optilde\bigg(\frac{c_n\sqrt{\log n}}{\sqrt{m}n}\bigg).
\end{align*}
The proof is completed by combining the above analyses.
\end{proof}

\subsection{Uniform perturbation analysis of the Jacobian matrix}
\label{sub:uniform_perturbation_analysis_of_the_jacobian_matrix}

The goal of this subsection is to show that the (rescaled) Jacobian matrix \eqref{eqn:J_n(B)}, which has a small co-dimension in the sense that the orthogonal complement of its column space is low-dimensional and delocalized, is roughly the same as the identity operator. We first establish the following Lemma.  

\begin{lemma}
\label{lemma:jacobian_perturbation}
Suppose Assumption \ref{assumption:true_parameter} holds. Define $\bJ_n^+(\bB) = \{\bJ_n(\bB)\transpose\bJ_n(\bB)\}^{-1}\bJ_n(\bB)\transpose$, where we recall the definition of $\bJ_n(\bB)$ in \eqref{eqn:J_n(B)}. Assume $\|\bB\|_2 = O(n^{-1/4})$, and $\|\bX_{0\perp}\bB\|_{2\to\infty} = O(1)$. 
\begin{enumerate}[(i)]
  \item Then
  \begin{align*}
  &\|\{\bJ_n(\bB) - \bJ_{0n}\}\bJ_{0n}\transpose\|_\infty = O(\|\bB\|_2),\quad
  \|\bJ_{0n}\{\bJ_n(\bB) - \bJ_{0n}\}\transpose\|_\infty = O(1),\\
  &\|\{\bJ_n(\bB) - \bJ_{0n}\}\bJ_n^+(\bB)\|_\infty = O(\|\bB\|_2),\quad 
  \|\bJ^+_n(\bB)\transpose\{\bJ_n(\bB) - \bJ_{0n}\}\transpose\|_\infty = O(1).
  \end{align*}

  \item Let $\bH(\tau):\tau\in[0, 1]\to \mathbb{R}^{d_\bpsi \times d_\bpsi}$ be a positive definite matrix-valued function for all $\tau\in[0, 1]$. Define $\bar{\bH} = \int_{[0, 1]}\bH(\tau)\mathrm{d}\tau$ and $\widehat{\bJ}_n(\bB) = \bar{\bH}^{-1}\int_{[0, 1]}\bH(\tau)\bJ_n(\tau\bB)\mathrm{d}\tau$, where the integral is with regard to any probability measure on $[0, 1]$. 
  Suppose $\|\bar{\bH}^{-1}\|_2 = O(1)$, $\sup_\tau\|\bH(\tau)\|_2 = O(1)$, $n^{-1/2}\sup_\tau\|\bar{\bH}^{-1}\bH(\tau)\|_\infty \to 0$ as $n\to\infty$. 
  Then, for any standard basis vector $\be_i$, $i\in [d_\bpsi]$, 
  \begin{align*}
  \|\be_{i}\transpose\bJ_{0n}\{\widehat{\bJ}_n(\bB)\transpose\bar{\bH}\widehat{\bJ}_n(\bB)\}^{-1}\bJ_{0n}\transpose\|_\infty
  &\lesssim \bigg\{1 + \sup_{\tau\in[0, 1]}\|\bH(\tau)\bar{\bH}^{-1}\|_\infty\bigg\}\bigg(1 + \max_{j\in[d_\bX]}\|\be_{j + d_\balpha + d_\br}\transpose\bar{\bH}^{-1}\|_\infty\bigg)\\
  &\quad\times\bigg\{\|\be_{i}\transpose\bar{\bH}^{-1}\|_\infty + \sup_{\tau\in[0, 1]}\|\bar{\bH}^{-1}\bH(\tau)\|_\infty\|\bB\|_2\|\bar{\bH}^{-1}\|_\infty\bigg\}.
  \end{align*}
\end{enumerate}
\end{lemma}

\begin{proof}[\bf Proof]

Let $\bV_\bB$ denote the matrix with orthonormal columns spanning the orthogonal complement of the column space of $\widehat{\bJ}_n(\bB)$ and $\widehat{\bJ}_n^+(\bB) = \{\widehat{\bJ}_n(\bB)\transpose\widehat{\bJ}_n(\bB)\}^{-1}\widehat{\bJ}_n(\bB)\transpose$. We will directly prove item (ii), and the proof process automatically proves item (i) in Step 3. 

\noindent $\blacksquare$ \textbf{Step 1.} We first show that $\bV_\bB$ is semi-delocalized: 
\[
\max_{i\in [d_\bX]}\|\be_{i + d_\balpha + d_\br}\transpose\bV_\bB\|_2 = O(n^{-1/2}),\quad\|\be_{i}\transpose\bV_\bB\|_2 = 0\text{ if }i\leq d_\balpha + d_\br. 
\]
Denote ${\bJ}_\bX(\bB) = n^{-1/2}\partial\vect\{\bX(\bB)\}/\partial\vect(\bB)\transpose$ and $\tilde{\bV}_\bB$ the orthonormal frame spanning the orthogonal complement of the column space of $\widehat{\bJ}_\bX(\bB) = \int_{[0, 1]}[\bar{\bH}^{-1}\bH(\tau)]_{\bX\bX}\bJ_\bX(\tau\bB)\mathrm{d}\tau$, where $[\bar{\bH}^{-1}\bH(\tau)]_{\bX\bX}\in\mathbb{R}^{d_\bX\times d_\bX}$ denotes the $(\bX,\bX)$-block of $\bar{\bH}^{-1}\bH(\tau)$. Likewise, denote $[\bar{\bH}^{-1}\bH(\tau)]_{\balpha\br\balpha\br}\in\mathbb{R}^{(d_\balpha + d_\br)\times (d_\balpha + d_\br)}$, $[\bar{\bH}^{-1}\bH(\tau)]_{\balpha\br,\bX}\in\mathbb{R}^{(d_\balpha + d_\br)\times d_\bX}$ the remaining blocks of $\bar{\bH}^{-1}\bH(\tau)$. 
For the second assertion, by definition, we have
\begin{align*}
\widehat{\bJ}_n(\bB)& = \bar{\bH}^{-1}\int_{[0, 1]}\bH(\tau)\mathrm{diag}\{\eye_{d_\balpha + d_\br},\bJ_\bX(\tau\bB)\}\mathrm{d}\tau
  = \begin{bmatrix}
  \eye_{d_\balpha + d_\br} & \bar{\bD}_{\balpha\br,\bX}\\
  \mathbf{O} & \widehat{\bJ}_\bX(\bB)
 \end{bmatrix}
\end{align*}
for some matrix $\bar{\bD}_{\balpha\br,\bX} = \int_{[0, 1]}[\bar{\bH}^{-1}\bH(\tau)]_{\balpha\br,\bX}\{\bJ_\bX(\tau\bB) - \bJ_\bX(\mathbf{O})\}\mathrm{d}\tau$. Note that $\int_{[0, 1]}[\bar{\bH}^{-1}\bH(\tau)]_{\bX\bX}\mathrm{d}\tau = \eye_{d_\bX}$. For any $\bu\in\mathbb{R}^{d_\balpha + d_\br}$ and $\bv\in\mathbb{R}^{d_\bX}$, $\widehat{\bJ}_n(\bB)\transpose[\bu\transpose,\bv\transpose]\transpose = \zero$ implies $\bu = 0$. Hence, $\be_i\transpose\bV_\bB = \zero$ for any $i\in [d_\balpha + d_\br]$ since $\widehat{\bJ}_n(\bB)\transpose\bV_\bB = \mathbf{O}$. 

Now it is sufficient to focus on the first assertion, and it is sufficient to work with $\|\tilde{\bV}_\bB\|_{2\to\infty}$. Below, we show that $\vect(\bZ)\perp\mathrm{Span}\{\widehat{\bJ}_\bX(\bB)\}$ if and only if 
\[
\bX_{0\perp}\transpose \int_{[0, 1]}\bZ_\tau\bN_\tau\mathrm{d}\tau = \bB\bPsi(\bZ)
\] 
for a matrix $\bZ = [Z_{pq}]_{n\times d}$, where we define $\bZ_\tau\in\mathbb{R}^{n\times d}$ satisfying $\vect(\bZ_\tau) = [\bar{\bH}^{-1}\bH(\tau)]_{\bX\bX}\transpose\vect(\bZ)$ and 
\begin{align*}
  \bPsi(\bZ)&:=
  \int_{[0, 1]}\tau 
  \mathscr{L}^{-1}(\bN_\tau(\bZ_\tau\transpose\bX_\tau + \bX_\tau\transpose\bZ_\tau)\bN_\tau)\mathrm{d}\tau,\;\bX_\tau = \bX_0 + \tau\bX_{0\perp}\bB,\;
  \bN_\tau = (\eye_d + \tau^2\bB\transpose\bB)^{-1/2},
\end{align*}
and $\bPsi(\cdot)$ satisfies $\|\bPsi(\bZ)\|_2\lesssim \sqrt{n}\sup_{\tau}\|\bar{\bH}^{-1}\bH(\tau)\|_2\|\bZ\|_{\mathrm{F}}$. 
By the properties of the inverse Lyapunov operator, $\|\mathscr{L}^{-1}(\bW)\|_2\leq \|\bW\|_2/2$ and $\mathscr{L}^{-1}$ is self-adjoint, namely, $\langle\mathscr{L}^{-1}(\bW),\bM\rangle_{\mathrm{F}} = \langle\bW,\mathscr{L}^{-1}(\bM)\rangle_{\mathrm{F}}$, by the integral formula $\mathscr{L}^{-1}(\bW) = \int_0^\infty e^{-u\bS_\tau^{1/2}}\bW e^{-u\bS_\tau^{1/2}}\mathrm{d}u$, where $\bS_\tau = \eye_d + \tau^2\bB\transpose\bB$ (see Lemma \ref{lemma:inverse_Lyapunov_operator_integral} or \cite{Bhatia2007,Rosenblum1956}), so the upper bound on $\|\bPsi(\cdot)\|_2$ naturally follows. 
So $\vect(\bZ)\perp\mathrm{Span}\{\widehat{\bJ}_\bX(\bB)\}$ if and only if $\vect(\bZ)\perp \widehat{\bJ}_\bX(\bB)\vect(\bDelta)$ for any test direction $\bDelta\in\mathbb{R}^{(n - 1 - d)\times d}$, or equivalently, $\langle\vect(\bZ),  \int_{[0, 1]}[\bar{\bH}^{-1}\bH(\tau)]_{\bX\bX}\vect\{\dot{\bX}(\tau\bB)[\bDelta]\}\mathrm{d}\tau\rangle = 0$. 
By Proposition \ref{prop:Cayley_transform_jacobian}, this is further equivalent to 
\begin{align*}
0& = \bigg\langle\vect(\bZ), \sum_{p = 1}^{n - 1 - d}\sum_{q = 1}^d \int_{[0, 1]}[\bar{\bH}^{-1}\bH(\tau)]_{\bX\bX}\vect\bigg\{\frac{\partial\bX}{\partial B_{pq}}(\tau\bB)\Delta_{pq}\bigg\}
\mathrm{d}\tau
\bigg\rangle
\\
& = \bigg\langle \vect(\bZ),  \int_{[0, 1]}[\bar{\bH}^{-1}\bH(\tau)]_{\bX\bX}\vect(\bX_{0\perp}\bDelta\bN_\tau)\mathrm{d}\tau\bigg\rangle
\\
&\quad -
 \sum_{p = 1}^{n - 1 - d}\sum_{q = 1}^d\bigg\langle\vect(\bZ),  \int_{[0, 1]}[\bar{\bH}^{-1}\bH(\tau)]_{\bX\bX}\vect\left\{\bX(\tau\bB)\mathscr{L}^{-1}(\tau\bB\transpose\be_p\be_q\transpose + \tau\be_q\be_p\transpose\bB)\bN_\tau\Delta_{pq}\right\}\mathrm{d}\tau\bigg\rangle
 \\
& = \int_{[0, 1]}\langle(\bN_\tau\otimes\bX_{0\perp}\transpose)\vect(\bZ_\tau), \vect(\bDelta)\rangle
 \mathrm{d}\tau -
\bigg\langle\vect(\bZ),  \int_{[0, 1]}[\bar{\bH}^{-1}\bH(\tau)]_{\bX\bX}\vect\left(\bX_\tau\bN_\tau\mathscr{L}^{-1}_\tau\bN_\tau\right)\mathrm{d}\tau\bigg\rangle
\\
& =  \int_{[0, 1]}\langle\bX_{0\perp}\transpose\bZ_\tau\bN_\tau, \bDelta\rangle_{\mathrm{F}}\mathrm{d}\tau - 
 \int_{[0, 1]}\bigg\langle\frac{1}{2}(\bX_\tau\transpose\bZ_\tau + \bZ_\tau\transpose\bX_\tau), \bN_\tau\mathscr{L}^{-1}_\tau\bN_\tau\bigg\rangle_{\mathrm{F}}\mathrm{d}\tau\\
& =  \int_{[0, 1]}\langle\bX_{0\perp}\transpose\bZ_\tau\bN_\tau, \bDelta\rangle_{\mathrm{F}}\mathrm{d}\tau
 -
\int_{[0, 1]}\bigg\langle\frac{1}{2}\bN_\tau(\bX_\tau\transpose\bZ_\tau + \bZ_\tau\transpose\bX_\tau)\bN_\tau, \mathscr{L}^{-1}_\tau\bigg\rangle_{\mathrm{F}}\mathrm{d}\tau
 \\
& = \int_{[0, 1]}\langle\bX_{0\perp}\transpose\bZ_\tau\bN_\tau, \bDelta\rangle_{\mathrm{F}}\mathrm{d}\tau -
\int_{[0, 1]}\bigg\langle\scrL^{-1}\bigg\{\frac{1}{2}\bN_\tau(\bX_\tau\transpose\bZ_\tau + \bZ_\tau\transpose\bX_\tau)\bN_\tau\bigg\}, \tau\bB\transpose\bDelta + \tau\bDelta\transpose\bB\bigg\rangle_{\mathrm{F}}\mathrm{d}\tau\\
& = \bigg\langle\bX_{0\perp}\transpose\int_{[0, 1]}\bZ_\tau\bN_\tau\mathrm{d}\tau, \bDelta\bigg\rangle_{\mathrm{F}} -
\bigg\langle\bB\int_{[0, 1]}\tau\scrL^{-1}\left\{\bN_\tau(\bX_\tau\transpose\bZ_\tau + \bZ_\tau\transpose\bX_\tau)\bN_\tau\right\}\mathrm{d}\tau, \bDelta\bigg\rangle_{\mathrm{F}}
\end{align*}
for any test direction $\bDelta$, where we denote $\scrL_\tau^{-1} = \scrL^{-1}(\tau(\bB\transpose\bDelta + \bDelta\transpose\bB))$ and have used the properties of the trace operator and the fact that $\langle \bA, \bB\rangle_{\mathrm{F}} = \langle \bA\transpose, \bB\rangle_{\mathrm{F}}$ for symmetric $\bB$. This is further equivalent to $\bX_{0\perp}\transpose \int_{[0, 1]}\bZ_\tau\bN_\tau\mathrm{d}\tau = \bB\bPsi(\bZ)$, as required. Now observe that 
\begin{align*}
  \|\widehat{\bJ}_\bX(\bB) - \widehat{\bJ}_\bX(\mathbf{O})\|_2
  & = \bigg\| \int_{[0, 1]}[\bar{\bH}^{-1}\bH(\tau)]_{\bX\bX}\{\bJ_\bX(\tau\bB) - \bJ_\bX(\mathbf{O})\}\mathrm{d}\tau\bigg\|_2\\
  &\leq \sup_\tau\|\bar{\bH}^{-1} \bH(\tau)\|_2\bigg\{\frac{\|(\bN_\tau - \eye_d)\otimes\bX_{0\perp}\|_2}{\sqrt{n}} + \frac{\|\eye_d\otimes(\bX_0 + \tau\bX_{0\perp}\bB)\|_2\|\eye_d\otimes\tau\bB\|_2}{\sqrt{n}}\bigg\}\\
  & = O\bigg(\sup_\tau\|\bar{\bH}^{-1} \bH(\tau)\|_2\|\bB\|_2\bigg) = O(n^{-1/4})
\end{align*}
and $\widehat{\bJ}_\bX(\mathbf{O}) = \bJ_\bX(\mathbf{O})$ has singular values being all $1$, it follows that the singular values of $\widehat{\bJ}_\bX(\bB)$ are all bounded away from $0$, so that $\mathrm{dim}\,\mathrm{Span}\{\widehat{\bJ}_\bX(\bB)\} = (n - 1 - d)d$. Now since $\tilde{\bV}_\bB$ has orthonormal columns spanning the orthogonal complement of $\mathrm{Span}\{\widehat{\bJ}_\bX(\bB)\}$, and it has a small co-dimension $d(d + 1)$, it is only sufficient to focus on $\|\bZ\|_{2\to\infty}$ for a matrix $\bZ$ such that $\vect(\bZ)\in\mathrm{Span}\{\widehat{\bJ}_\bX(\bB)\}^\perp$ where we can treat $\vect(\bZ)$ as a generic column of $\tilde{\bV}_\bB$ having unit length $\|\bZ\|_{\mathrm{F}} = 1$. Note that
\[
\int_{[0, 1]} \vect(\bZ_\tau)\mathrm{d}\tau
 = \int_{[0, 1]} [\bar{\bH}^{-1}\bH(\tau)]_{\bX\bX}\transpose\vect(\bZ)\mathrm{d}\tau =\vect(\bZ)\Longrightarrow \int_{[0, 1]}\bZ_\tau\mathrm{d}\tau = \bZ.
\]
Therefore, by the above derivation, we can always decompose $\bZ$ as
\begin{align*}
\|\bZ\|_{2\to\infty}
&\leq \frac{1}{n}\|[\one_n, \bX_0]\|_{2\to\infty}\|([\one_n, \bX_0])\transpose\bZ\|_2 + \frac{1}{n}\|\bX_{0\perp}(\bX_{0\perp}\transpose\bZ)\|_{2\to\infty}\\
&\leq \frac{1}{n}\|[\one_n, \bX_0]\|_{2\to\infty}\|([\one_n, \bX_0])\transpose\bZ\|_2 + \frac{1}{n}\bigg\|\bX_{0\perp}\bigg(\bX_{0\perp}\transpose \int_{[0, 1]}\bZ_\tau\bN_\tau\mathrm{d}\tau\bigg)\bigg\|_{2\to\infty}\\
&\quad + \frac{1}{n}\bigg\|\bX_{0\perp}\bigg\{\bX_{0\perp}\transpose \int_{[0, 1]}\bZ_\tau(\bN_\tau - \eye_d)\mathrm{d}\tau\bigg\}\bigg\|_{2\to\infty}\\
&\leq O\bigg(\frac{1}{\sqrt{n}}\bigg)\|\bZ\|_{\mathrm{F}} + \frac{1}{n}\|\bX_{0\perp}\bB\bPsi(\bZ)\|_{2\to\infty} + \frac{1}{n}\|\bX_{0\perp}\bX_{0\perp}\transpose\|_{2\to\infty}\sup_\tau\|\bar{\bH}^{-1}\bH(\tau)\|_2\|\bZ\|_{\mathrm{F}}\|\bN_\tau - \eye_d\|_2\\
&\leq O(n^{-1/2}) + \frac{1}{n}\|\bX_{0\perp}\bB\|_{2\to\infty}\|\bPsi(\bZ)\|_2 + O\bigg(\sup_\tau\frac{\|\bar{\bH}^{-1}\bH(\tau)\|_2}{\sqrt{n}}\bigg) = O(n^{-1/2}).
\end{align*}
Hence, $\|\tilde{\bV}_\bB\|_{2\to\infty}$ should satisfy the same $O(n^{-1/2})$ upper bound. 

\vspace*{1ex}
\noindent $\blacksquare$ \textbf{Step 2.} We show the matrix identity
\begin{align*}
  &\bJ_{0n}\{\widehat{\bJ}_n(\bB)\transpose\bar\bH\widehat{\bJ}_n(\bB)\}^{-1}\bJ_{0n}\transpose\\
  &\quad = 
  \left[\eye - \{\widehat{\bJ}_n(\bB) - \bJ_{0n}\}\widehat{\bJ}^+_n(\bB)\right]\left[\widehat{\bJ}_n(\bB)\{\widehat{\bJ}_n(\bB)\transpose\bar\bH\widehat{\bJ}_n(\bB)\}^{-1}\widehat{\bJ}_n(\bB)\transpose\right]\left[\eye - \{\widehat{\bJ}_n(\bB) - \bJ_{0n}\}\widehat{\bJ}_n^+(\bB)\right]\transpose.
\end{align*}
As a result, for any standard basis vector $\be_{i}$ with $i\in[d_\bpsi]$, 
\begin{align*}
  &\|\be_{i}\transpose\bJ_{0n}\{\widehat{\bJ}_n(\bB)\transpose\bar\bH\widehat{\bJ}_n(\bB)\}^{-1}\bJ_{0n}\transpose\|_\infty\\
  &\quad\leq 
  \|\be_{i}\transpose\widehat{\bJ}_n(\bB)\{\widehat{\bJ}_n(\bB)\transpose\bar\bH\widehat{\bJ}_n(\bB)\}^{-1}\widehat{\bJ}_n(\bB)\transpose\|_\infty 
  \left[1 + \|\widehat{\bJ}_n^+(\bB)\transpose\{\widehat{\bJ}_n(\bB) - \bJ_{0n}\}\transpose\|_\infty\right]\\
  &\qquad + 
  \|\{\widehat{\bJ}_n(\bB) - \bJ_{0n}\}\widehat{\bJ}_n^+(\bB)\|_\infty
  \| \widehat{\bJ}_n(\bB)\{\widehat{\bJ}_n(\bB)\transpose\bar\bH\widehat{\bJ}_n(\bB)\}^{-1}\widehat{\bJ}_n(\bB)\transpose\|_\infty \\
  &\qquad\times \left[1 + \|\widehat{\bJ}_n^+(\bB)\transpose\{\widehat{\bJ}_n(\bB) - \bJ_{0n}\}\transpose\|_\infty\right].
\end{align*}
This can be done by simply observing that
\begin{align*}
\bJ_{0n} & = \widehat{\bJ}_n(\bB) - \{\widehat{\bJ}_n(\bB) - \bJ_{0n}\} = \widehat{\bJ}_n(\bB) - \{\widehat{\bJ}_n(\bB) - \bJ_{0n}\}\{\widehat{\bJ}_n(\bB)\transpose\widehat{\bJ}_n(\bB)\}^{-1}\widehat{\bJ}_n(\bB)\transpose\widehat{\bJ}_n(\bB)\\
& = \widehat{\bJ}_n(\bB) - \{\widehat{\bJ}_n(\bB) - \bJ_{0n}\}\widehat{\bJ}_n^+(\bB)\widehat{\bJ}_n(\bB) = \left[\eye - \{\widehat{\bJ}_n(\bB) - \bJ_{0n}\}\widehat{\bJ}_n^+(\bB)\right]\widehat{\bJ}_n(\bB).
\end{align*}

\noindent $\blacksquare$ \textbf{Step 3.} We show the Jacobian perturbation analysis satisfies:
\begin{align*}
&\|\{\widehat{\bJ}_n(\bB) - \bJ_{0n}\}\widehat{\bJ}_n^+(\bB)\|_\infty = O\bigg(\sup_{\tau\in[0, 1]}\|\bar{\bH}^{-1} \bH(\tau)\|_\infty\|\bB\|_2\bigg),\\
&\|\widehat{\bJ}_n^+(\bB)\transpose\{\widehat{\bJ}_n(\bB) - \bJ_{0n}\}\transpose\|_\infty = O\bigg(1 + \sup_{\tau\in[0, 1]}\|\bH(\tau) \bar{\bH}^{-1}\|_\infty\bigg),\\
&\|\{\widehat{\bJ}_n(\bB) - \bJ_{0n}\}\bJ_{0n}\transpose\|_\infty = O\bigg(\sup_{\tau\in[0, 1]}\|\bar{\bH}^{-1} \bH(\tau)\|_\infty\|\bB\|_2\bigg),\\
& 
\|\bJ_{0n}\{\widehat{\bJ}_n(\bB) - \bJ_{0n}\}\transpose\|_\infty = O\bigg(\sup_{\tau\in[0, 1]}\|\bH(\tau)\bar{\bH}^{-1}\|_\infty\bigg).
\end{align*}
Note that when $\mathrm{d}\tau$ corresponds to the point-mass at $\tau = 1$, this recovers item (i) in the statement of the Lemma, since then $\bar{\bH}^{-1}\bH(\tau)\equiv\eye$ and every factor involving $\bar{\bH}^{-1}$ and $\bH(\tau)$ below equals one. A direct calculation shows that
\begin{align*}
&\widehat{\bJ}_n^+(\bB) = \begin{bmatrix}
  \eye_{d_\balpha + d_\br}
  & -\bar{\bD}_{\balpha\br,\bX}\widehat{\bJ}_\bX^+(\bB)\\
  \mathbf{O} & \widehat{\bJ}_\bX^+(\bB)
\end{bmatrix},\quad
\{\widehat{\bJ}_n(\bB) - \bJ_{0n}\}\bJ_{0n}\transpose = \begin{bmatrix}
  \mathbf{O} & \bar{\bD}_{\balpha\br,\bX}\bJ_\bX(\mathbf{O})\transpose\\
  \mathbf{O}
  & \{\widehat{\bJ}_\bX(\bB) - \bJ_\bX(\mathbf{O})\}\bJ_\bX(\mathbf{O})\transpose
\end{bmatrix},\\
&\{\widehat{\bJ}_n(\bB) - \bJ_{0n}\}\widehat{\bJ}_n^+(\bB) = \begin{bmatrix}
    \mathbf{O} & \bar{\bD}_{\balpha\br, \bX}\widehat{\bJ}_\bX^+(\bB)\\
    \mathbf{O} & \{\widehat{\bJ}_\bX(\bB) - \bJ_\bX(\mathbf{O})\}\widehat{\bJ}_\bX^+(\bB)
  \end{bmatrix},
\end{align*}
where $\widehat{\bJ}_\bX^+(\bB)
= \{\widehat{\bJ}_\bX(\bB)\transpose\widehat{\bJ}_\bX(\bB)\}^{-1}
\widehat{\bJ}_\bX(\bB)\transpose$.
The four matrices in question are therefore supported on two blocks each,
and it is sufficient to bound the blocks involving
$\widehat{\bJ}_\bX(\bB)$, $\widehat{\bJ}_\bX(\mathbf{O})$, $\widehat{\bJ}_\bX^+(\bB)$,
and $\bar{\bD}_{\balpha\br,\bX}$.
To see this, we directly compute the Jacobian from Proposition \ref{prop:Cayley_transform_jacobian} and write
\begin{align*}
\frac{\partial\vect(\bX)}{\partial\vect(\bB)\transpose}(\tau\bB)
& = \bN_\tau\otimes\bX_{0\perp}\\
&\quad - \{\eye_d \otimes (\bX_0 + \tau\bX_{0\perp}\bB)\}(\bN_\tau\otimes\bN_\tau)(\eye_d\otimes\bS_\tau^{1/2} + \bS_\tau^{1/2}\otimes\eye_d)^{-1}(\eye_{d^2} + \bK_{dd})(\eye_d \otimes \tau\bB\transpose),
\end{align*}
where we follow the notations in Step 1. By the derivation in Step 1 and a triangle inequality, we also have $\|\widehat{\bJ}_\bX(\bB)\transpose\widehat{\bJ}_\bX(\bB) - \eye\|_2 = O(\sup_\tau\|\bar{\bH}^{-1}\bH(\tau)\|_2\|\bB\|_2)$, so that $\|\widehat{\bJ}_\bX^+(\bB)\|_2\leq 2$ for large $n$. Namely, for any test matrix $\bZ\in\mathbb{R}^{n\times d}$ with $\|\bZ\|_{\max}\leq 1$, we have the following:
\begin{align*}
&\|\{\widehat{\bJ}_\bX(\bB) - \widehat{\bJ}_\bX(\mathbf{O})\}\widehat{\bJ}_\bX^+(\bB)\vect(\bZ)\|_\infty\\
&\quad \leq \int_{[0, 1]}\|[\bar{\bH}^{-1}\bH(\tau)]_{\bX\bX}\{\bJ_\bX(\tau\bB) - \widehat{\bJ}_\bX(\mathbf{O})\}\widehat{\bJ}_\bX^+(\bB)\vect(\bZ)\|_\infty\mathrm{d}\tau\\
&\quad \leq  \sup_{\tau\in[0, 1]}\|\bar{\bH}^{-1} \bH(\tau)\|_\infty\|\{(\bN_\tau - \eye_d)\otimes\eye_n\}\widehat{\bJ}_\bX(\mathbf{O})\widehat{\bJ}_\bX^+(\bB)\vect(\bZ)\|_\infty\\
&\qquad + \sup_{\tau\in[0, 1]}\|\bar{\bH}^{-1} \bH(\tau)\|_\infty\frac{d}{\sqrt{n}}\|\eye_d\otimes(\bX_0 + \tau\bX_{0\perp}\bB)\|_\infty\|(\eye_d\otimes\tau\bB\transpose)\widehat{\bJ}^+_\bX(\bB)\vect(\bZ)\|_\infty\\
&\quad\lesssim \sup_{\tau\in[0, 1]}\|\bar{\bH}^{-1} \bH(\tau)\|_\infty\sup_\tau\|\bN_\tau - \eye_d\|_\infty\|\widehat{\bJ}_\bX(\mathbf{O})\widehat{\bJ}_\bX^+(\bB)\|_\infty\\
&\qquad + \sup_{\tau\in[0, 1]}\|\bar{\bH}^{-1} \bH(\tau)\|_\infty\frac{1}{\sqrt{n}}\sup_\tau\|\eye_d\otimes(\bX_0 + \tau\bX_{0\perp}\bB)\|_\infty\|(\eye_d\otimes\bB\transpose)\widehat{\bJ}^+_\bX(\bB)\vect(\bZ)\|_2\\
&\quad\leq O\bigg(\sup_{\tau\in[0, 1]}\|\bar{\bH}^{-1} \bH(\tau)\|_\infty\|\bB\|_2\bigg)\left\{O(\|\bB\|_2)\|\widehat{\bJ}_\bX(\mathbf{O})\widehat{\bJ}_\bX^+(\bB)\|_\infty + O(1)\right\},\\
&\|\bar{\bD}_{\balpha\br,\bX}\widehat{\bJ}_\bX^+(\bB)\vect(\bZ)\|_\infty\\
&\quad\leq \int_{[0, 1]}\|[\bar{\bH}^{-1}\bH(\tau)]_{\balpha\br,\bX}\{\bJ_\bX(\tau\bB) - \bJ_\bX(\mathbf{O})\}\widehat{\bJ}_\bX^+(\bB)\vect(\bZ)\|_\infty\mathrm{d}\tau\\
&\quad\leq \sup_\tau\|\bar{\bH}^{-1}\bH(\tau)\|_\infty\|\{\bJ_\bX(\tau\bB) - \bJ_\bX(\mathbf{O})\}\widehat{\bJ}_\bX^+(\bB)\vect(\bZ)\|_\infty\\
&\quad\leq O\bigg(\sup_{\tau\in[0, 1]}\|\bar{\bH}^{-1} \bH(\tau)\|_\infty\|\bB\|_2\bigg)\left\{O(\|\bB\|_2)\|\widehat{\bJ}_\bX(\mathbf{O})\widehat{\bJ}_\bX^+(\bB)\|_\infty + O(1)\right\},\\
&\|\{\widehat{\bJ}_\bX(\bB) - \widehat{\bJ}_\bX(\mathbf{O})\}\bJ_\bX(\mathbf{O})\transpose\vect(\bZ)\|_\infty\\
&\quad \leq \int_{[0, 1]}\|[\bar{\bH}^{-1}\bH(\tau)]_{\bX\bX}\{\bJ_\bX(\tau\bB) - \widehat{\bJ}_\bX(\mathbf{O})\}\bJ_\bX(\mathbf{O})\transpose\vect(\bZ)\|_\infty\mathrm{d}\tau\\
&\quad \leq  \sup_{\tau\in[0, 1]}\|\bar{\bH}^{-1} \bH(\tau)\|_\infty\|\{(\bN_\tau - \eye_d)\otimes\eye_n\}\widehat{\bJ}_\bX(\mathbf{O})\bJ_\bX(\mathbf{O})\transpose\vect(\bZ)\|_\infty\\
&\qquad + \sup_{\tau\in[0, 1]}\|\bar{\bH}^{-1} \bH(\tau)\|_\infty\frac{d}{\sqrt{n}}\|\eye_d\otimes(\bX_0 + \tau\bX_{0\perp}\bB)\|_\infty\|(\eye_d\otimes\tau\bB\transpose)\bJ_\bX(\mathbf{O})\transpose\vect(\bZ)\|_\infty\\
&\quad\lesssim \sup_\tau\|\bar{\bH}^{-1} \bH(\tau)\|_\infty\|\bN_\tau - \eye_d\|_\infty\|\widehat{\bJ}_\bX(\mathbf{O})\bJ_\bX(\mathbf{O})\transpose\|_\infty\\
&\qquad + \sup_\tau\|\bar{\bH}^{-1} \bH(\tau)\|_\infty\frac{1}{\sqrt{n}}\sup_\tau\|\eye_d\otimes(\bX_0 + \tau\bX_{0\perp}\bB)\|_\infty\|(\eye_d\otimes\bB\transpose)\bJ_\bX(\mathbf{O})\transpose\vect(\bZ)\|_2\\
&\quad\leq O\bigg(\sup_{\tau\in[0, 1]}\|\bar{\bH}^{-1} \bH(\tau)\|_\infty\|\bB\|_2\bigg)\left\{O(\|\bB\|_2)\|\widehat{\bJ}_\bX(\mathbf{O})\bJ_\bX(\mathbf{O})\transpose\|_\infty + O(1)\right\},
\end{align*}
so that 
\begin{align*}
\|\{\widehat{\bJ}_\bX(\bB) - \widehat{\bJ}_\bX(\mathbf{O})\}\widehat{\bJ}_\bX^+(\bB) \|_\infty&\leq O\bigg(\sup_{\tau\in[0, 1]}\|\bar{\bH}^{-1} \bH(\tau)\|_\infty\|\bB\|_2\bigg)\left\{O(\|\bB\|_2)\|\widehat{\bJ}_\bX(\mathbf{O})\widehat{\bJ}_\bX^+(\bB)\|_\infty + O(1)\right\},\\
\|\bar{\bD}_{\balpha\br,\bX}\widehat{\bJ}_\bX^+(\bB)\|_\infty&\leq O\bigg(\sup_{\tau\in[0, 1]}\|\bar{\bH}^{-1} \bH(\tau)\|_\infty\|\bB\|_2\bigg)\left\{O(\|\bB\|_2)\|\widehat{\bJ}_\bX(\mathbf{O})\widehat{\bJ}_\bX^+(\bB)\|_\infty + O(1)\right\},\\
\|\{\widehat{\bJ}_\bX(\bB) - \widehat{\bJ}_\bX(\mathbf{O})\}\widehat{\bJ}_\bX(\mathbf{O})\transpose \|_\infty&\leq O\bigg(\sup_{\tau\in[0, 1]}\|\bar{\bH}^{-1} \bH(\tau)\|_\infty\|\bB\|_2\bigg)\left\{O(\|\bB\|_2)\|\widehat{\bJ}_\bX(\mathbf{O})\widehat{\bJ}_\bX(\mathbf{O})\transpose\|_\infty + O(1)\right\},
\end{align*}
and
\begin{align*}
\|\{\widehat{\bJ}_n(\bB) - \bJ_{0n}\}\bJ_{0n}\transpose\|_\infty
&\leq \int_{[0, 1]}\|\bar{\bH}^{-1}\bH(\tau)\{\bJ_n(\tau\bB) - \bJ_{0n}\}\bJ_{0n}\transpose\|_\infty\mathrm{d}\tau\\
&\leq \sup_\tau \|\bar{\bH}^{-1} \bH(\tau)\|_\infty \|\{\bJ_\bX(\tau\bB) - \bJ_\bX(\mathbf{O})\}\bJ_\bX(\mathbf{O})\transpose\|_\infty\\
&\leq \sup_\tau \|\bar{\bH}^{-1} \bH(\tau)\|_\infty\bigg\|(\bN_\tau - \eye_d)\otimes\frac{\bX_{0\perp}\bX_{0\perp}\transpose}{n}\bigg\|_\infty\\
&\quad + \sup_\tau \|\bar{\bH}^{-1} \bH(\tau)\|_\infty\frac{d}{n}\|\eye_d\otimes(\bX_0 + \tau\bX_{0\perp}\bB)\|_\infty\|\eye_d\otimes\bB\transpose\bX_{0\perp}\transpose\|_\infty\\
& = O\bigg(\sup_{\tau\in[0, 1]}\|\bar{\bH}^{-1} \bH(\tau)\|_\infty\|\bB\|_2\bigg).
\end{align*}
Also, 
\begin{align*}
\|\widehat{\bJ}_\bX(\mathbf{O})\widehat{\bJ}_\bX^+(\bB)\|_\infty 
&\leq \|\{\widehat{\bJ}_\bX(\bB) - \widehat{\bJ}_\bX(\mathbf{O})\}\widehat{\bJ}_\bX^+(\bB)\|_\infty + \|\widehat{\bJ}_\bX(\bB)\{\widehat{\bJ}_\bX(\bB)\transpose\widehat{\bJ}_\bX(\bB)\}^{-1}\widehat{\bJ}_\bX(\bB)\transpose\|_\infty\\
&\leq O\bigg(\sup_{\tau\in[0, 1]}\|\bar{\bH}^{-1} \bH(\tau)\|_\infty\|\bB\|_2^2\bigg)\|\widehat{\bJ}_\bX(\mathbf{O})\widehat{\bJ}_\bX^+(\bB)\|_\infty + O\bigg(\sup_{\tau\in[0, 1]}\|\bar{\bH}^{-1} \bH(\tau)\|_\infty\|\bB\|_2\bigg)\\
&\quad + \|\widehat{\bJ}_\bX(\bB)\{\widehat{\bJ}_\bX(\bB)\transpose\widehat{\bJ}_\bX(\bB)\}^{-1}\widehat{\bJ}_\bX(\bB)\transpose\|_\infty,\\
\|\widehat{\bJ}_\bX(\mathbf{O})\bJ_\bX(\mathbf{O})\transpose\|_\infty& = \frac{1}{n}\|\eye_d\otimes\bX_{0\perp}\bX_{0\perp}\transpose\|_\infty = O(1),
\end{align*}
so that 
\begin{align*}
\|\widehat{\bJ}_\bX(\mathbf{O})\widehat{\bJ}_\bX^+(\bB)\|_\infty 
&\leq 2\|\widehat{\bJ}_\bX(\bB)\{\widehat{\bJ}_\bX(\bB)\transpose\widehat{\bJ}_\bX(\bB)\}^{-1}\widehat{\bJ}_\bX(\bB)\transpose\|_\infty + O\bigg(\sup_{\tau\in[0, 1]}\|\bar{\bH}^{-1} \bH(\tau)\|_\infty\|\bB\|_2\bigg)\\
&\leq 2 + 2\|\tilde{\bV}_\bB\|_\infty\|\tilde{\bV}_\bB\transpose\|_\infty + O\bigg(\sup_{\tau\in[0, 1]}\|\bar{\bH}^{-1} \bH(\tau)\|_\infty\|\bB\|_2\bigg)\\
& = O\bigg(1 + \sup_{\tau\in[0, 1]}\|\bar{\bH}^{-1} \bH(\tau)\|_\infty\|\bB\|_2\bigg)  
\end{align*} 
for large $n$ by Step 1.
This implies that 
\begin{align*}
  \|\{\widehat{\bJ}_n(\bB) - \bJ_{0n}\}\widehat{\bJ}_n^+(\bB) \|_\infty
  & = O\bigg(\sup_{\tau\in[0, 1]}\|\bar{\bH}^{-1} \bH(\tau)\|_\infty\|\bB\|_2\bigg),\\
  \|\{\widehat{\bJ}_n(\bB) - \bJ_{0n}\}\bJ_{0n}\transpose \|_\infty
  & = O\bigg(\sup_{\tau\in[0, 1]}\|\bar{\bH}^{-1} \bH(\tau)\|_\infty\|\bB\|_2\bigg).
\end{align*}
Similarly, by Lemma \ref{lemma:Lipschitz_Cayley_transform},
\begin{align*}
&\left\|\widehat{\bJ}_\bX^+(\bB)\transpose\{\widehat{\bJ}_\bX(\bB) - \widehat{\bJ}_\bX(\mathbf{O})\}\transpose - \widehat{\bJ}_\bX(\mathbf{O})\{\widehat{\bJ}_\bX(\bB) - \widehat{\bJ}_\bX(\mathbf{O})\}\transpose\right\|_2\\
&\quad \leq \int_{[0, 1]}\|\bar{\bH}^{-1}  \bH(\tau)\|_2^2\|\bJ_\bX(\tau\bB) - \bJ_\bX(\mathbf{O})\|_2^2\mathrm{d}\tau = O(\|\bB\|_2^2),\\
&\sup_{\tau\in[0, 1]}\left\|\widehat{\bJ}_\bX^+(\bB)\transpose\{\bJ_\bX(\tau\bB) - \bJ_\bX(\mathbf{O})\}\transpose - \widehat{\bJ}_\bX(\mathbf{O})\{\bJ_\bX(\tau\bB) - \bJ_\bX(\mathbf{O})\}\transpose\right\|_2\\
&\quad \leq \sup_{\tau\in[0, 1]}\|\bar{\bH}^{-1} \bH(\tau)\|_2\|\bJ_\bX(\tau\bB) - \bJ_\bX(\mathbf{O})\|_2^2 = O(\|\bB\|_2^2),\\
&\left\|\widehat{\bJ}_\bX(\mathbf{O})\{\widehat{\bJ}_\bX(\bB) - \widehat{\bJ}_\bX(\mathbf{O})\}\transpose\right\|_\infty\\
&\quad\lesssim \frac{1}{n}\int_{[0, 1]} \|\eye_d\otimes\tau\bX_{0\perp}\bB\|_\infty\|\eye_d\otimes(\bX_0 + \tau\bX_{0\perp}\bB)\transpose\|_\infty\|\bH(\tau) \bar{\bH}^{-1}\|_\infty\mathrm{d}\tau\\
&\qquad + \frac{1}{n}\int_{[0, 1]}\|(\bN_\tau - \eye_d)\otimes(\bX_{0\perp}\bX_{0\perp}\transpose)\|_\infty
\|\bH(\tau) \bar{\bH}^{-1}\|_\infty\mathrm{d}\tau\\
&\quad\leq \frac{1}{n}\sup_{\tau\in[0, 1]}\|\eye_d\otimes\tau\bX_{0\perp}\bB\|_\infty\|\eye_d\otimes(\bX_0 + \tau\bX_{0\perp}\bB)\transpose\|_\infty\|\bH(\tau) \bar{\bH}^{-1}\|_\infty\\
&\qquad + \frac{1}{n}\sup_{\tau\in[0, 1]}\|(\bN_\tau - \eye_d)\otimes(\bX_{0\perp}\bX_{0\perp}\transpose)\|_\infty\|\bH(\tau) \bar{\bH}^{-1}\|_\infty = O\bigg(\sup_{\tau\in[0, 1]}\|\bH(\tau)\bar{\bH}^{-1}\|_\infty\bigg),\\
&\sup_{\tau\in[0, 1]}\left\|\widehat{\bJ}_\bX(\mathbf{O})\{\bJ_\bX(\tau\bB) - \bJ_\bX(\mathbf{O})\}\transpose\right\|_\infty\\
&\quad\lesssim \frac{1}{n}\sup_{\tau\in[0, 1]}\left\{\|\eye_d\otimes\tau\bX_{0\perp}\bB\|_\infty\|\eye_d\otimes(\bX_0 + \tau\bX_{0\perp}\bB)\transpose\|_\infty +\|(\bN_\tau - \eye_d)\otimes(\bX_{0\perp}\bX_{0\perp}\transpose)\|_\infty\right\} = O(1).
\end{align*}
The fourth error bound above also implies
\begin{align*}
\|\bJ_{0n}\{\widehat{\bJ}_n(\bB) - \bJ_{0n}\}\transpose\|_\infty 
&\leq \int_{[0, 1]}\|\bJ_{0n}[\bar{\bH}^{-1}\bH(\tau)\{\bJ_n(\tau\bB) - \bJ_{0n}\}]\transpose\|_\infty \\
&\leq \sup_\tau\|\bJ_{0n}\{\bJ_n(\tau\bB) - \bJ_{0n}\}\transpose\|_\infty\|\bH(\tau) \bar{\bH}^{-1}\|_\infty \\
&\leq \sup_\tau\|{\bJ}_\bX(\mathbf{O})\{{\bJ}_\bX(\tau\bB) - \bJ_\bX(\mathbf{O})\}\transpose\|_\infty \|\bH(\tau)\bar{\bH}^{-1}\|_\infty\\
& = O\bigg(\sup_{\tau\in[0, 1]}\|\bH(\tau)\bar{\bH}^{-1}\|_\infty\bigg).
\end{align*}
By the relationships among $\|\cdot\|_\infty,\|\cdot\|_{2\to\infty},\|\cdot\|_2$, 
\begin{align*}
  \|\widehat{\bJ}^+_\bX(\bB)\transpose\{\widehat{\bJ}_\bX(\bB) - \widehat{\bJ}_\bX(\mathbf{O})\}\transpose\|_\infty 
  &\leq O\bigg(\sup_{\tau\in[0, 1]}\|\bH(\tau)\bar{\bH}^{-1}\|_\infty + \sqrt{n}\sup_{\tau\in[0, 1]}\|\bar{\bH}^{-1} \bH(\tau)\|_2^2\|\bB\|_2^2\bigg)\\
  & = O\bigg(1 + \sup_{\tau\in[0, 1]}\|\bH(\tau)\bar{\bH}^{-1}\|_\infty\bigg),\\
  \sup_{\tau\in[0, 1]}\|\widehat{\bJ}^+_\bX(\bB)\transpose\{\bJ_\bX(\tau\bB) - \bJ_\bX(\mathbf{O})\}\transpose\|_\infty 
  &\leq O\bigg(1 + \sqrt{n}\sup_{\tau\in[0, 1]}\|\bar{\bH}^{-1}\bH(\tau)\|_2\|\bB\|_2^2\bigg) = O(1).
\end{align*}
The second error bound above further implies that
\begin{align*}
\|\widehat{\bJ}_\bX^+(\bB)\transpose\bar{\bD}_{\balpha\br,\bX}\transpose\|_\infty&
\leq \int_{[0, 1]}\|\widehat{\bJ}_\bX^+(\bB)\transpose\{\bJ_\bX(\tau\bB) - \bJ_\bX(\mathbf{O})\}[\bar{\bH}^{-1}\bH(\tau)]_{\balpha\br,\bX}\transpose\|_\infty\mathrm{d}\tau\\
&\leq \sup_{\tau\in[0, 1]}\|\widehat{\bJ}^+_\bX(\bB)\transpose\{\bJ_\bX(\tau\bB) - \bJ_\bX(\mathbf{O})\}\transpose\|_\infty\|\bH(\tau) \bar{\bH}^{-1}\|_\infty
 = O\bigg(\sup_{\tau\in[0, 1]}\|\bH(\tau) \bar{\bH}^{-1}\|_\infty\bigg).
\end{align*}
Since
\[
\widehat{\bJ}_n^+(\bB)\transpose\{\widehat{\bJ}_n(\bB) - \bJ_{0n}\}\transpose
= \begin{bmatrix}
  \mathbf{O} & \mathbf{O}\\
  \widehat{\bJ}_\bX^+(\bB)\transpose\bar{\bD}_{\balpha\br,\bX}\transpose
  & \widehat{\bJ}_\bX^+(\bB)\transpose
    \{\widehat{\bJ}_\bX(\bB) - \widehat{\bJ}_\bX(\mathbf{O})\}\transpose
\end{bmatrix},
\]
the two bounds above give
\begin{align*}
\|\widehat{\bJ}_n^+(\bB)\transpose\{\widehat{\bJ}_n(\bB) - \bJ_{0n}\}\transpose\|_\infty = O\bigg(1 + \sup_{\tau\in[0, 1]}\|\bH(\tau) \bar{\bH}^{-1}\|_\infty\bigg),
\end{align*}
which completes the proof of Step 3.

\noindent $\blacksquare$ \textbf{Step 4.} We show that the small co-dimension of $\bJ(\bB)$ does not hurt too much:
\[
\widehat{\bJ}_n(\bB)\{\widehat{\bJ}_n(\bB)\transpose\bar{\bH}\widehat{\bJ}_n(\bB)\}^{-1}\widehat{\bJ}_n(\bB)\transpose = \bar{\bH}^{-1} - \bar{\bH}^{-1}\bV_\bB(\bV_\bB\transpose\bar{\bH}^{-1}\bV_\bB)^{-1}\bV_\bB\transpose\bar{\bH}^{-1}
\]
and for any standard basis vector $\be_i$ with $i\in[d_\bpsi]$, 
\[
  \|\be_i\transpose\widehat{\bJ}_n(\bB)\{\widehat{\bJ}_n(\bB)\transpose\bar{\bH}\widehat{\bJ}_n(\bB)\}^{-1}\widehat{\bJ}_n(\bB)\transpose\|_\infty
  \lesssim \bigg(1 + \max_{j\in[d_\bX]}\|\be_{j + d_\balpha + d_\br}\transpose\bar{\bH}^{-1}\|_\infty\bigg)\|\be_i\transpose\bar{\bH}^{-1}\|_\infty.
\] 
The second norm inequality can be obtained by writing 
\begin{align*}
&\|\be_i\transpose\widehat{\bJ}_n(\bB)\{\widehat{\bJ}_n(\bB)\transpose\bar{\bH}\widehat{\bJ}_n(\bB)\}^{-1}\widehat{\bJ}_n(\bB)\transpose\|_\infty\\
&\quad \leq \|\be_i\transpose\bar{\bH}^{-1}\|_\infty + \|\be_i\transpose\bar{\bH}^{-1}\|_\infty\|\tilde{\bV}_\bB\|_\infty\|(\tilde{\bV}_\bB\transpose[\bar{\bH}^{-1}]_{\bX\bX}\tilde{\bV}_\bB)^{-1}\|_\infty\|\tilde{\bV}_\bB\transpose\|_\infty\|\bar{\bH}^{-1}\|_\infty\\
&\quad \leq \|\be_i\transpose\bar{\bH}^{-1}\|_\infty + O({d})\frac{\|\be_i\transpose\bar{\bH}^{-1}\|_{\infty}\|\tilde{\bV}_\bB\|_\infty \|\tilde{\bV}_\bB\transpose\|_\infty \max_{j\in[d_\bX]}\|\be_{j + d_\balpha + d_\br}\transpose\bar{\bH}^{-1}\|_\infty}{\lambda_{\min}({\bV}_\bB\transpose\bar{\bH}^{-1}{\bV}_\bB)}\\
&\quad \lesssim \|\be_i\transpose\bar{\bH}^{-1}\|_\infty + \|\be_i\transpose\bar{\bH}^{-1}\|_{\infty}\|\tilde{\bV}_\bB\|_\infty\|\bar{\bH}\|_2 \|\tilde{\bV}_\bB\transpose\|_\infty \max_{j\in[d_\bX]}\|\be_{j + d_\balpha + d_\br}\transpose\bar{\bH}^{-1}\|_\infty\\
&\quad \lesssim   \bigg(1 +  \max_{j\in[d_\bX]}\|\be_{j + d_\balpha + d_\br}\transpose\bar{\bH}^{-1}\|_\infty\bigg)\|\be_i\transpose\bar{\bH}^{-1}\|_\infty
\end{align*}
since $\bV_\bB = [\mathbf{O},\tilde{\bV}_\bB\transpose]\transpose$, $\lambda_{\min}(\bV_\bB\transpose\bar{\bH}^{-1}\bV_\bB)\geq\lambda_{\min}(\bar{\bH}^{-1}) = 1/\|\bar{\bH}\|_2$, and $\|\tilde{\bV}_\bB\|_\infty\|\tilde{\bV}_\bB\transpose\|_\infty = O(1)$ by Step 1.
So it is sufficient to focus on the first equation. Without loss of generality, by QR factorization, we may replace $\widehat{\bJ}_n(\bB)$ by $\widehat{\bJ}_\bB$ that has the same column space but with orthonormal columns. Observe that the columns spaces of $\bar{\bH}^{1/2}\widehat{\bJ}_\bB$ and $\bar{\bH}^{-1/2}\bV_\bB$ are orthogonal and entire Euclidean space is their direct sum. Therefore, 
\begin{align*}
  \eye = \bar{\bH}^{1/2}\widehat{\bJ}_\bB(\widehat{\bJ}_\bB\transpose\bar{\bH}\widehat{\bJ}_\bB)^{-1}\widehat{\bJ}_\bB\transpose\bar{\bH}^{1/2} + \bar{\bH}^{-1/2}\bV_\bB(\bV_\bB\transpose\bar{\bH}^{-1}\bV_\bB)^{-1}\bV_\bB\transpose\bar{\bH}^{-1/2},
\end{align*}
which implies the target matrix equation. Combining Steps 2--3 and the row bound above, we obtain 
\begin{align*}
\|\be_{i}\transpose\bJ_{0n}\{\widehat{\bJ}_n(\bB)\transpose\bar{\bH}\widehat{\bJ}_n(\bB)\}^{-1}\bJ_{0n}\transpose\|_\infty
&\lesssim \bigg\{1 + \sup_{\tau\in[0, 1]}\|\bH(\tau)\bar{\bH}^{-1}\|_\infty\bigg\}\bigg(1 + \max_{j\in[d_\bX]}\|\be_{j + d_\balpha + d_\br}\transpose\bar{\bH}^{-1}\|_\infty\bigg)\\
&\quad\times\bigg\{\|\be_{i}\transpose\bar{\bH}^{-1}\|_\infty + \sup_{\tau\in[0, 1]}\|\bar{\bH}^{-1}\bH(\tau)\|_\infty\|\bB\|_2\|\bar{\bH}^{-1}\|_\infty\bigg\}
,
\end{align*}
which is item (ii), and the proof is done. 
\end{proof}

Next, we establish an important technical tool that extends the Jacobian perturbation analysis in Lemma \ref{lemma:jacobian_perturbation} to the integral operator case. 
\begin{lemma}
\label{lemma:integrated_jacobian_perturbation}
Suppose Assumption \ref{assumption:true_parameter} holds. 
Let $\bH(\tau):\tau\in[0, 1]\to \mathbb{R}^{d_\bpsi \times d_\bpsi}$ be a positive definite matrix-valued function for all $\tau\in[0, 1]$. Define $\bar{\bH} = \int_{[0, 1]}\bH(\tau)\mathrm{d}\tau$ and $\widehat{\bJ}_n(\bB) = \bar{\bH}^{-1}\int_{[0, 1]}\bH(\tau)\bJ_n(\tau\bB)\mathrm{d}\tau$, where the integral is with regard to any probability measure on $[0, 1]$. 
Suppose $\|\bar{\bH}^{-1}\|_2 = O(1)$, $\sup_\tau\|\bH(\tau)\|_2 = O(1)$, $\|\bB\|_2\sup_\tau\|\bar{\bH}^{-1}\|_\infty^4\|\bH(\tau)\|_\infty^3 \to 0$ as $n\to\infty$. 
Further assume $\|\bB\|_2 = O(n^{-1/4})$, and $\|\bX_{0\perp}\bB\|_{2\to\infty} = O(1)$. Then, for any standard basis vector $\be_i$, $i\in[d_\bpsi]$, 
\begin{align*}
&\bigg\|\be_i\transpose\bJ_{0n}\bigg\{\int_{[0, 1]}\bJ_n(\tau\bB)\transpose\bH(\tau)\bJ_n(\tau\bB)\mathrm{d}\tau\bigg\}^{-1}\bJ_{0n}\transpose\bigg\|_\infty \\
&\quad\lesssim \sup_{\tau\in[0, 1]}\|\bH(\tau)\|_\infty\|\bar{\bH}^{-1}\|_\infty\bigg(1 + \max_{j\in[d_\bX]}\|\be_{j + d_\balpha + d_\br}\transpose\bar{\bH}^{-1}\|_\infty\bigg) \bigg\{\|\be_i\transpose\bar{\bH}^{-1}\|_\infty + \sup_{\tau\in[0, 1]}\|\bH(\tau)\|_\infty\|\bar{\bH}^{-1}\|_\infty^2\|\bB\|_2\bigg\}.
\end{align*}
\end{lemma}

\begin{proof}[\bf Proof]
Denote $\bS_\tau = (\eye_d + \tau^2\bB\transpose\bB)$ and $\bN_\tau = \bS_\tau^{-1/2}$. 

\noindent $\blacksquare$ \textbf{Step 1.} The following the matrix identity holds:
\begin{align*}
    &\int_{[0, 1]} \bJ_n(\tau\bB)\transpose\bH(\tau)\bJ_n(\tau\bB)\mathrm{d}\tau\\
    &\quad = 
  \bigg[\bJ_{0n} - \bar{\bH}^{-1}\int_{[0, 1]}\bH(\tau)\{\bJ_{0n} - \bJ_n(\tau\bB)\}\mathrm{d}\tau\bigg]\transpose\bar{\bH}\bigg[\bJ_{0n} - \bar{\bH}^{-1}\int_{[0, 1]}\bH(\tau)\{\bJ_{0n} - \bJ_n(\tau\bB)\}\mathrm{d}\tau\bigg]\\
  &\qquad + \int_{[0, 1]}\{\bJ_{0n} - \bJ_n(\tau\bB)\}\transpose \bH(\tau)\{\bJ_{0n} - \bJ_n(\tau\bB)\}\mathrm{d}\tau\\
  &\qquad - 
  \bigg[\int_{[0, 1]}\bH(\tau)\{\bJ_{0n} - \bJ_n(\tau\bB)\}\mathrm{d}\tau\bigg]\transpose\bar{\bH}^{-1}\bigg[\int_{[0, 1]}\bH(\tau)\{\bJ_{0n} - \bJ_n(\tau\bB)\}\mathrm{d}\tau\bigg].
\end{align*}
The verification can be done by a direct calculation.

\noindent $\blacksquare$ \textbf{Step 2.} We show that 
\begin{align*}
 &\bigg\|\bJ_{0n}\int_{[0, 1]}\{\bJ_{0n} - \bJ_n(\tau\bB)\}\transpose \bH(\tau)\{\bJ_{0n} - \bJ_n(\tau\bB)\}\mathrm{d}\tau\bJ_{0n}\transpose\\
  &\quad - 
  \bJ_{0n}\bigg[\int_{[0, 1]}\bH(\tau)\{\bJ_{0n} - \bJ_n(\tau\bB)\}\mathrm{d}\tau\bigg]\transpose\bar{\bH}^{-1}\bigg[\int_{[0, 1]}\bH(\tau)\{\bJ_{0n} - \bJ_n(\tau\bB)\}\mathrm{d}\tau\bigg]\bJ_{0n}\transpose\bigg\|_\infty\\
  &\quad = O\bigg(\sup_{\tau\in[0, 1]}\|\bH(\tau)\|_\infty^2\|\bar{\bH}^{-1}\|_\infty\|\bB\|_2\bigg). 
\end{align*}
This is straightforward: by item (i) of Lemma \ref{lemma:jacobian_perturbation},
\begin{align*}
  &\bigg\|\bJ_{0n}\int_{[0, 1]}\{\bJ_{0n} - \bJ_n(\tau\bB)\}\transpose \bH(\tau)\{\bJ_{0n} - \bJ_n(\tau\bB)\}\mathrm{d}\tau\bJ_{0n}\transpose\bigg\|_\infty \\
  &\quad \leq \sup_\tau\|\bJ_{0n}\{\bJ_{0n} - \bJ_n(\tau\bB)\}\transpose\|_\infty\|\bH(\tau)\|_\infty \|\{\bJ_{0n} - \bJ_n(\tau\bB)\}\bJ_{0n}\transpose\|_\infty = O\bigg(\sup_{\tau\in[0, 1]}\|\bH(\tau)\|_\infty\|\bB\|_2\bigg),\\
  &\bigg\|\bJ_{0n}\bigg[\int_{[0, 1]}\bH(\tau)\{\bJ_{0n} - \bJ_n(\tau\bB)\}\mathrm{d}\tau\bigg]\transpose\bar{\bH}^{-1}\bigg[\int_{[0, 1]}\bH(\tau)\{\bJ_{0n} - \bJ_n(\tau\bB)\}\mathrm{d}\tau\bigg]\bJ_{0n}\transpose\bigg\|_\infty \\
  &\quad \leq \sup_\tau\|\bJ_{0n}\{\bJ_{0n} - \bJ_n(\tau\bB)\}\transpose\|_\infty\|\bH(\tau)\|_\infty^2 \|\bar{\bH}^{-1}\|_\infty \|\{\bJ_{0n} - \bJ_n(\tau\bB)\}\bJ_{0n}\transpose\|_\infty\\
  &\quad = O\bigg(\sup_{\tau\in[0, 1]}\|\bH(\tau)\|_\infty^2\|\bar{\bH}^{-1}\|_\infty\|\bB\|_2\bigg). 
\end{align*}


\noindent $\blacksquare$ \textbf{Step 3.} We show that for any standard basis vector $\be_i$ with $i\in[d_\bpsi]$, 
\begin{align*}
&\bigg\|\be_i\transpose\bJ_{0n}\bigg\{\int_{[0, 1]} \bJ_n(\tau\bB)\transpose\bH(\tau)\bJ_{n}(\tau\bB)\mathrm{d}\tau\bigg\}^{-1}\bJ_{0n}\transpose\bigg\|_\infty.
\end{align*}
For any invertible matrices $\bM,\bM^*$ of the same dimension, $\bM^{-1} = (\bM^{*})^{-1} + \bM^{-1}(\bM^* - \bM)(\bM^*)^{-1}$, so that
\[
\|\bJ_{0n}\bM^{-1}\bJ_{0n}\transpose\|_\infty
\leq \|\bJ_{0n}(\bM^{*})^{-1}\bJ_{0n}\transpose\|_\infty + \|\bJ_{0n}\bM^{-1}\bJ_{0n}\transpose\|_\infty\|\bJ_{0n}(\bM^* - \bM)\bJ_{0n}\transpose\|_\infty\|\bJ_{0n}(\bM^*)^{-1}\bJ_{0n}\transpose\|_\infty. 
\]
Denote $\widehat{\bJ}_n(\bB) = \bar{\bH}^{-1}\int_{[0, 1]}\bH(\tau)\bJ_n(\tau\bB)\mathrm{d}\tau$. 
Both $\bM$ and $\bM^*$ below are invertible: $\widehat{\bJ}_n(\bB)$ and every
$\bJ_n(\tau\bB)$ have full column rank by Step 1 of the proof of
Lemma \ref{lemma:jacobian_perturbation}, and $\bH(\tau)$ is positive definite.
By Steps 1--2, by taking
\begin{align*}
\bM &= \int_{[0, 1]}\bJ_n(\tau\bB)\transpose\bH(\tau)\bJ_n(\tau\bB)\mathrm{d}\tau,\\
\bM^* &= \bigg[\bJ_{0n} - \bar{\bH}^{-1}\int_{[0, 1]}\bH(\tau)\{\bJ_{0n} - \bJ_n(\tau\bB)\}\mathrm{d}\tau\bigg]\transpose\bar{\bH}\bigg[\bJ_{0n} - \bar{\bH}^{-1}\int_{[0, 1]}\bH(\tau)\{\bJ_{0n} - \bJ_n(\tau\bB)\}\mathrm{d}\tau\bigg]\\
& = \widehat{\bJ}_n(\bB)\transpose\bar{\bH}\widehat{\bJ}_n(\bB),
\end{align*}
we have 
\begin{align*}
&\bigg\|\be_i\transpose\bJ_{0n}\bigg\{\int_{[0, 1]} \bJ_n(\tau\bB)\transpose\bH(\tau)\bJ_{n}(\tau\bB)\mathrm{d}\tau\bigg\}^{-1}\bJ_{0n}\transpose\bigg\|_\infty\\
&\quad \leq \left\|\be_i\transpose\bJ_{0n}\{
\widehat{\bJ}_n(\bB)\transpose\bar{\bH}\widehat{\bJ}_n(\bB)
\}^{-1}\bJ_{0n}\transpose\right\|_\infty + \bigg\|\be_i\transpose\bJ_{0n}\bigg\{\int \bJ_n(\tau\bB)\transpose\bH(\tau)\bJ_n(\tau\bB)\mathrm{d}\tau\bigg\}^{-1}\bJ_{0n}\transpose\bigg\|_\infty\\
&\qquad\times O\bigg(\sup_{\tau\in[0, 1]}\|\bH(\tau)\|_\infty^2\|\bar{\bH}^{-1}\|_\infty\|\bB\|_2\bigg)
 \left\|\bJ_{0n}\{\widehat{\bJ}_n(\bB)\transpose\bar{\bH}\widehat{\bJ}_n(\bB)
\}^{-1}\bJ_{0n}\transpose\right\|_\infty\\
&\quad \leq \left\|\be_i\transpose\bJ_{0n}\{\widehat{\bJ}_n(\bB)\transpose\bar{\bH}\widehat{\bJ}_n(\bB)\}^{-1}\bJ_{0n}\transpose\right\|_\infty +  \bigg\|\be_{i}\transpose\bJ_{0n}\bigg\{\int_{[0, 1]} \bJ_n(\tau\bB)\transpose\bH(\tau)\bJ_n(\tau\bB)\mathrm{d}\tau\bigg\}^{-1}\bJ_{0n}\transpose\bigg\|_\infty\\
&\qquad\times O\bigg(\|\bB\|_2\sup_{\tau}\|\bH(\tau)\|_\infty^3\|\bar{\bH}^{-1}\|_\infty^4 \bigg),
\end{align*}
where the last step follows from item (ii) of Lemma \ref{lemma:jacobian_perturbation}. Applying item (ii) of Lemma \ref{lemma:jacobian_perturbation} again, we further obtain
\begin{align*}
  &\bigg\|\be_{i}\transpose\bJ_{0n}\bigg\{\int_{[0, 1]} \bJ_n(\tau\bB)\transpose\bH(\tau)\bJ_n(\tau\bB)\mathrm{d}\tau\bigg\}^{-1}\bJ_{0n}\transpose\bigg\|_\infty\\
  &\quad\leq 2\left\|\be_i\transpose\bJ_{0n}\{\widehat{\bJ}_n(\bB)\transpose\bar{\bH}\widehat{\bJ}_n(\bB)\}^{-1}\bJ_{0n}\transpose\right\|_\infty\\
  &\quad \lesssim \sup_{\tau\in[0, 1]}\|\bH(\tau)\|_\infty\|\bar{\bH}^{-1}\|_\infty\bigg(1 + \max_{j\in[d_\bX]}\|\be_{j + d_\balpha + d_\br}\transpose\bar{\bH}^{-1}\|_\infty\bigg) \bigg\{\|\be_i\transpose\bar{\bH}^{-1}\|_\infty + \sup_{\tau\in[0, 1]}\|\bH(\tau)\|_\infty\|\bar{\bH}^{-1}\|_\infty^2\|\bB\|_2\bigg\}.
\end{align*}
This completes the proof. 
\end{proof}

\subsection{Proof of the uniform norm consistency}
\label{sub:proof_of_uniform_consistency_mle}

\begin{proof}[\bf Proof of Proposition \ref{prop:stationarity_uniqueness_uniform_consistency_MLE}]

$\blacksquare$ \textbf{Uniform consistency of $\breve{\bphi}$. }
By Section \ref{sub:proof_sketch_of_uniform_norm_consistency}, we first show that
\begin{align}
\label{eqn:integrated_hessian_inverse_infinity_norm_bound}
 \bigg\|\widetilde{\bLambda}_\sharp\bJ_{0n}\bigg\{\bLambda_\sharp\bigg(\scrI\frac{\partial^2\calL}{\partial\bphi\partial\bphi\transpose}\bigg)(\breve{\bphi})\bLambda_\sharp\bigg\}^{-1}
\bJ_{0n}\transpose\bigg\|_\infty = \Optilde\bigg(\frac{m^{3/2}\sqrt{m \vee \log n}}{\sqrt{n}}\bigg).  
\end{align}
Clearly, $\bphi\in\calN(\sqrt{m}n\eps_n, M)$ implies that
\begin{align*}
\bigg\|\frac{1}{m}\sum_{t = 1}^m\{(\bR^{(t)})^2 - (\bR_0^{(t)})^2\}\bigg\|_2
& = \bigg\|\frac{1}{m}\sum_{t = 1}^m\{
(\bR^{(t)} - \bR_0^{(t)})^2 + 
\bR_0^{(t)}(\bR^{(t)} - \bR_0^{(t)}) 
+ (\bR^{(t)} - \bR_0^{(t)})\bR_0^{(t)}
\}\bigg\|_2\\
&\leq \frac{1}{m}\sum_{t = 1}^m\left\|
\bR^{(t)} - \bR_0^{(t)}\right\|_2^2 + \frac{2}{m}\sum_{t = 1}^m\left\|\bR_0^{(t)}\right\|_2\left\|\bR^{(t)} - \bR_0^{(t)}\right\|_2\\
&\leq \frac{1}{m}\sum_{t = 1}^m\left\|
\bR^{(t)} - \bR_0^{(t)}\right\|_2^2 + 2\bigg(\frac{1}{m}\sum_{t = 1}^m\left\|\bR_0^{(t)}\right\|_2^2\bigg)^{1/2}\bigg(\frac{1}{m}\sum_{t = 1}^m\left\|\bR^{(t)} - \bR_0^{(t)}\right\|_2^2\bigg)^{1/2}\\
& = O(\eps_n).
\end{align*}
This implies that the eigenvalues of $(1/m)\sum_t(\bR^{(t)})^2$ are bounded away from $0$ and $+\infty$ uniformly over $\bphi\in\calN(\sqrt{m}n\eps_n, M)$ by Weyl's inequality and Assumption \ref{assumption:true_parameter}. Also, by Assumption \ref{assumption:edge_distribution}, there exists some constant $C_M, c_M > 0$ depending on $M$, such that for all $i\in[n]$, 
\begin{align*}
&c_M\begin{bmatrix}
  \eye_m & -\frac{1}{\sqrt{m}n}\sum_{t = 1}^m\be_t\bx_i\transpose\bR^{(t)}\\
  -\frac{1}{\sqrt{m}n}\sum_{t = 1}^m\bR^{(t)}\bx_i\be_t\transpose & \frac{1}{m}\sum_{t = 1}^m\bR^{(t)}(\eye_d - \bx_i\bx_i\transpose/n)\bR^{(t)}
\end{bmatrix}\preceq
-\bDelta_{\bz_i}^\sharp\\
&\quad\preceq C_M\begin{bmatrix}
  \eye_m & -\frac{1}{\sqrt{m}n}\sum_{t = 1}^m\be_t\bx_i\transpose\bR^{(t)}\\
  -\frac{1}{\sqrt{m}n}\sum_{t = 1}^m\bR^{(t)}\bx_i\be_t\transpose & \frac{1}{m}\sum_{t = 1}^m\bR^{(t)}(\eye_d - \bx_i\bx_i\transpose/n)\bR^{(t)}
\end{bmatrix}.
\end{align*}
Similarly,
\begin{align*}
\frac{c_M}{2}\bD_d\transpose\bD_d - O(n^{-1}) \preceq \frac{-1}{2n^2}\sum_{i\neq j}\ell''(\theta_{ij}^{(t)}; A_{ij}^{(t)})\bD_d\transpose(\bx_j\bx_j\transpose\otimes\bx_i\bx_i\transpose)\bD_d
\preceq \frac{C_M}{2}\bD_d\transpose\bD_d 
\end{align*}
This also implies that the eigenvalues of $(1/m)\sum_{t = 1}^m(\bR^{(t)})^2 - 1/(mn)\sum_{t = 1}^m\bR^{(t)}\bx_i\bx_i\transpose\bR^{(t)}$ are bounded away from $0$ and $+\infty$ uniformly over $\bphi\in\calN(\sqrt{m}n\eps_n, M)$. Denote $\bH(\tau) = \bPi\bDelta_{\bpsi}^\sharp(\bpsi(\bphi_0 + \tau(\breve{\bphi} - \bphi_0)))\bPi\transpose$. Hence, 
\begin{align*}
&\sup_{\bphi\in\calN(\sqrt{m}n\eps_n, M)}\|\bDelta_{\bz_i}^\sharp(\bpsi(\bphi))\|_2 = O(1),\; 
\sup_{\bphi\in\calN(\sqrt{m}n\eps_n, M)}\|\{\bDelta_{\bz_i}^\sharp(\bpsi(\bphi))\}^{-1}\|_2 = O(1),\\
&\sup_{\bphi\in\calN(\sqrt{m}n\eps_n, M)}\|\bDelta_{\bz_i}^\sharp(\bpsi(\bphi))\|_\infty = O(\sqrt{m}),\; 
\sup_{\bphi\in\calN(\sqrt{m}n\eps_n, M)}\|\{\bDelta_{\bz_i}^\sharp(\bpsi(\bphi))\}^{-1}\|_\infty = O(\sqrt{m}),\\
&\sup_{\bphi\in\calN(\sqrt{m}n\eps_n, M)}\|\bGamma_{\bR\bR}^\sharp(\bpsi(\bphi))\|_2 = O(1),\; 
\sup_{\bphi\in\calN(\sqrt{m}n\eps_n, M)}\|\{\bGamma_{\bR\bR}^\sharp(\bpsi(\bphi))\}^{-1}\|_2 = O(1),\\
&\sup_{\bphi\in\calN(\sqrt{m}n\eps_n, M)}\|\bGamma_{\bR\bR}^\sharp(\bpsi(\bphi))\|_\infty = O(1),\; 
\sup_{\bphi\in\calN(\sqrt{m}n\eps_n, M)}\|\{\bGamma_{\bR\bR}^\sharp(\bpsi(\bphi))\}^{-1}\|_\infty = O(1),\\
&\sup_{\tau\in[0, 1]}\|\bH(\tau)\|_2\vee\|\bH(\tau)^{-1}\|_2 = O(1),\;
\sup_{\tau\in[0, 1]}\|\bH(\tau)\|_\infty\vee\|\bH(\tau)^{-1}\|_\infty = O(\sqrt{m}).
\end{align*}
Denote by $\bD = \mathrm{diag}(\eye_{d_\balpha + d_\br}, \sqrt{m}\eye_{d_\bX})$ and $\widetilde{\bD} = \mathrm{diag}(\eye_{d_\balpha}/\sqrt{n}, \eye_{d_\br}/n, \eye_{d_\bX}/\sqrt{n})$. 
Since $\bDelta_\bpsi^\sharp
= \mathrm{diag}(\bDelta_\bZ^\sharp, \bGamma_{\bR\bR}^\sharp)$ and
$\bJ_n(\tau\bB) = \mathrm{diag}\{\eye_{d_\balpha + d_\br},
\bJ_\bX(\tau\bB)\}$, the $(\balpha, \br)$ and $(\br, \bX)$ blocks of
$\bJ_{0n}\{(\scrI\bDelta_\bphi^\sharp)(\breve{\bphi})\}^{-1}
\bJ_{0n}\transpose$ vanish, and $\widetilde{\bD}$ is a scalar multiple
of the identity on the $(\balpha, \bX)$ coordinates and on the $\br$
coordinates. Hence, $\widetilde{\bD}$ commutes with this matrix, and
since $\widetilde{\bLambda}_\sharp = \bD^{-1}\widetilde{\bD}$, we have 
\begin{align*}
\widetilde{\bLambda}_\sharp\bJ_{0n}\{(\scrI \bDelta_\bphi^\sharp)(\breve{\bphi})\}^{-1}\bJ_{0n}\transpose& = \bD^{-1}\bJ_{0n}\{(\scrI \bJ_n\transpose\bPi\bDelta_\bpsi^\sharp(\bpsi(\cdot))\bPi\transpose\bJ_n)(\breve{\bphi})\}^{-1}\bJ_{0n}\transpose\widetilde{\bD}\\
& = \bD^{-1}\bJ_{0n}\{(\scrI \bDelta_\bphi^\sharp)(\breve{\bphi})\}^{-1}\bJ_{0n}\transpose\widetilde{\bD}.
\end{align*}
Furthermore, write
\begin{align*}
-\int_{[0, 1]}\bDelta_{\bz_i}^\sharp(\bpsi(\bphi_0 + \tau(\breve{\bphi} - \bphi_0)))
\mathrm{d}\tau = \begin{bmatrix}
  \bar{\bDelta}_{\balpha_i\balpha_i} & \bar{\bDelta}_{\balpha_i\bx_i}\\
  \bar{\bDelta}_{\balpha_i\bx_i}\transpose & \bar{\bDelta}_{\bx_i\bx_i}
\end{bmatrix}
\end{align*}
in blocks of sizes $m$ and $d$. By \eqref{eqn:Deltazi_sharp}, $\bar{\bDelta}_{\balpha_i\balpha_i}$ is diagonal with diagonal elements in $[\kappa_1(n - 1)/n, \kappa_2]$, the elements of
$\bar{\bDelta}_{\balpha_i\bx_i}$ are $O(m^{-1/2})$, and the Schur complement
$[\bar{\bDelta}_{\bz_i}^\sharp]_{\bx_i\bx_i}^{-1} = \bar{\bDelta}_{\bx_i\bx_i} - \bar{\bDelta}_{\balpha_i\bx_i}\transpose\bar{\bDelta}_{\balpha_i\balpha_i}^{-1}\bar{\bDelta}_{\balpha_i\bx_i}$
satisfies 
\begin{align*}
  \lambda_{\min}([\bar{\bDelta}_{\bz_i}^\sharp]_{\bx_i\bx_i}^{-1}) = \lambda_{\min}(\bar{\bDelta}_{\bx_i\bx_i} - \bar{\bDelta}_{\balpha_i\bx_i}\transpose\bar{\bDelta}_{\balpha_i\balpha_i}^{-1}\bar{\bDelta}_{\balpha_i\bx_i})\geq
\lambda_{\min}\bigg(-\int_{[0, 1]}\bDelta_{\bz_i}^\sharp(\bpsi(\bphi_0 + \tau(\breve{\bphi} - \bphi_0)))\mathrm{d}\tau\bigg) \geq c 
\end{align*}
for some constant $c > 0$. Then, the block inverse formula
\begin{align*}
\begin{bmatrix}
\bar{\bDelta}_{\balpha_i\balpha_i} & \bar{\bDelta}_{\balpha_i\bx_i}\\
\bar{\bDelta}_{\bx_i\balpha_i} & \bar{\bDelta}_{\bx_i\bx_i}
\end{bmatrix}
^{-1} = \begin{bmatrix}
\bar{\bDelta}_{\balpha_i\balpha_i}^{-1} + \bar{\bDelta}_{\balpha_i\balpha_i}^{-1}\bar{\bDelta}_{\balpha_i\bx_i}[\bar{\bDelta}_{\bz_i}^\sharp]_{\bx_i\bx_i}^{-1}\bar{\bDelta}_{\balpha_i\bx_i}\transpose\bar{\bDelta}_{\balpha_i\balpha_i}^{-1} &
-\bar{\bDelta}_{\balpha_i\balpha_i}^{-1}\bar{\bDelta}_{\balpha_i\bx_i}[\bar{\bDelta}_{\bz_i}^\sharp]_{\bx_i\bx_i}^{-1}\\
 -[\bar{\bDelta}_{\bz_i}^\sharp]_{\bx_i\bx_i}^{-1}\bar{\bDelta}_{\balpha_i\bx_i}\transpose\bar{\bDelta}_{\balpha_i\balpha_i}^{-1}&
[\bar{\bDelta}_{\bz_i}^\sharp]_{\bx_i\bx_i}^{-1}
\end{bmatrix}
\end{align*}
 shows that
 \begin{align}
\label{eqn:Deltazi_sharp_infinity_norm_I}
\max_{i\in[n],t\in[m]}\bigg\|\be_t(m + d)\transpose\bigg\{\int_{[0, 1]}\bDelta_{\bz_i}^\sharp(\bpsi(\bphi_0 + \tau(\breve{\bphi} - \bphi_0)))\mathrm{d}\tau\bigg\}^{-1}\bigg\|_\infty &= O(1),\\
\label{eqn:Deltazi_sharp_infinity_norm_II}
\max_{i\in[n],k\in[d]}\bigg\|\be_{m + k}(m + d)\transpose\bigg\{\int_{[0, 1]}\bDelta_{\bz_i}^\sharp(\bpsi(\bphi_0 + \tau(\breve{\bphi} - \bphi_0)))\mathrm{d}\tau\bigg\}^{-1}\bigg\|_\infty &= O(\sqrt{m}),\\
\bigg\|\bigg\{\int_{[0, 1]}\bGamma_{\bR\bR}^\sharp(\bpsi(\bphi_0 + \tau(\breve{\bphi} - \bphi_0)))\mathrm{d}\tau\bigg\}^{-1} \bigg\|_\infty &= O(1).
\nonumber
\end{align}
By Lemma \ref{lemma:integrated_jacobian_perturbation} applied to $-\bH(\tau)$ because $\|\bB\|_2 = \Optilde(\eps_n)$ for all $\bB\in\calN(\sqrt{m}n\eps_n, M)$ and 
\[
\sup_{\tau}\|\bar{\bH}^{-1}\|_\infty^4\|\bH(\tau)\|_\infty^3\eps_n\to 0\quad\text{as }n\to\infty,
\] 
$\bH(\cdot)$ is positive definite, and the block diagonality of $\bD^{-1}$, we immediately obtain
\begin{align*}
\|\bD^{-1}\bJ_{0n}\{(\scrI \bDelta_\bphi^\sharp)(\breve{\bphi})\}^{-1}\bJ_{0n}\transpose\|_\infty = \Optilde(m^{3/2}),\quad
\|\widetilde{\bLambda}_\sharp\bJ_{0n}\{(\scrI \bDelta_\bphi^\sharp)(\breve{\bphi})\}^{-1}\bJ_{0n}\transpose\|_\infty = \Optilde\bigg(\frac{m^{3/2}}{\sqrt{n}}\bigg).
\end{align*}
Recall in the proof of Lemma \ref{lemma:integrated_jacobian_perturbation}, we have, for any invertible matrices $\bM,\bM^*$ of the same dimension, $\bM^{-1} = (\bM^{*})^{-1} + (\bM^{*})^{-1}(\bM^* - \bM)\bM^{-1}$. 
Then, with
\[
\bM = \bLambda_\sharp\bigg(\scrI\frac{\partial^2\calL}{\partial\bphi\partial\bphi\transpose}\bigg)(\breve{\bphi})\bLambda_\sharp,\;
\bM^* = (\scrI \bDelta_\bphi^\sharp)(\breve{\bphi}),
\]
and by Lemma \ref{lemma:off_diagonal_term_perturbation}, Lemma \ref{lemma:score_term_perturbation}, and Proposition \ref{prop:local_hessian_bound}, there exists a random matrix $\bXi(\bphi)\in\mathbb{R}^{d_\bpsi\times d_\bpsi}$ satisfying $\sup_{\bphi\in\calN(\sqrt{m}n\eps_n, M)}\|\widetilde{\bD}\bXi(\bphi)\widetilde{\bLambda}_\sharp^{-1}\|_\infty = \Optilde(\sqrt{m}\eps_n\log n)$, such that
\begin{align*}
&\bigg\|\widetilde{\bLambda}_\sharp\bJ_{0n}\bigg\{\bLambda_\sharp\bigg(\scrI\frac{\partial^2\calL}{\partial\bphi\partial\bphi\transpose}\bigg)(\breve{\bphi})\bLambda_\sharp\bigg\}^{-1}
\bJ_{0n}\transpose\bigg\|_\infty\\
&\quad \leq    
\left\|\widetilde{\bLambda}_\sharp\bJ_{0n}\{(\scrI \bDelta_\bphi^\sharp)(\breve{\bphi})\}^{-1}
\bJ_{0n}\transpose\right\|_\infty\\
&\qquad + \sqrt{d_\bpsi}\left\|\bD^{-1} \bJ_{0n}\{(\scrI \bDelta_\bphi^\sharp)(\breve{\bphi})\}^{-1}
\bJ_{0n}\transpose\right\|_\infty \left\|\widetilde{\bD}\left(\scrI \bJ_{0n}\bJ_n\transpose\bPi(\bGamma_{\bpsi\bpsi}^\sharp(\bpsi(\cdot)) - \bDelta_{\bpsi}^\sharp(\bpsi(\cdot)))\bPi\transpose\bJ_n\bJ_{0n}\transpose\right)(\breve{\bphi})\right\|_{2\to\infty}\\
&\qquad\times  \bigg\|\bJ_{0n}\bigg\{\bLambda_\sharp\bigg(\scrI\frac{\partial^2\calL}{\partial\bphi\partial\bphi\transpose}\bigg)(\breve{\bphi})\bLambda_\sharp\bigg\}^{-1}
\bJ_{0n}\transpose\bigg\|_2\\
&\qquad + \frac{\sqrt{d_\bpsi}}{2}\left\|\bD^{-1}\bJ_{0n}\{(\scrI \bDelta_\bphi^\sharp)(\breve{\bphi})\}^{-1}
\bJ_{0n}\transpose\right\|_\infty\\
&\qquad\times 
\bigg\|\widetilde{\bD}\scrI\bigg( 
\bJ_{0n}\bLambda_\sharp\sum_{t = 1}^m\sum_{i\neq j}\ell'(\theta_{ij}^{(t)}(\cdot); A_{ij}^{(t)})\frac{\partial^2\theta_{ij}^{(t)}}{\partial\bphi\partial\bphi\transpose}\bLambda_\sharp\bJ_{0n}\transpose - \bXi
\bigg)(\breve{\bphi})\bigg\|_{2\to\infty}
\\
&\qquad\times \bigg\|\bJ_{0n}\bigg\{\bLambda_\sharp\bigg(\scrI\frac{\partial^2\calL}{\partial\bphi\partial\bphi\transpose}\bigg)(\breve{\bphi})\bLambda_\sharp\bigg\}^{-1}
\bJ_{0n}\transpose\bigg\|_2\\
&\qquad + \frac{1}{2}\left\|\bD^{-1}\bJ_{0n}\{(\scrI \bDelta_\bphi^\sharp)(\breve{\bphi})\}^{-1}
\bJ_{0n}\transpose\right\|_\infty
\left\|\widetilde{\bD}(\scrI \bXi)(\breve{\bphi})\widetilde{\bLambda}_\sharp^{-1}\right\|_\infty
\bigg\|\widetilde{\bLambda}_\sharp\bJ_{0n}\bigg\{\bLambda_\sharp\bigg(\scrI\frac{\partial^2\calL}{\partial\bphi\partial\bphi\transpose}\bigg)(\breve{\bphi})\bLambda_\sharp\bigg\}^{-1}
\bJ_{0n}\transpose\bigg\|_\infty\\
&\quad = \Optilde\bigg(\frac{m^{3/2}\sqrt{m \vee \log n}}{\sqrt{n}}\bigg) + \Optilde(m^{2}\eps_n\log n)\bigg\|\widetilde{\bLambda}_\sharp\bJ_{0n}\bigg\{\bLambda_\sharp\bigg(\scrI\frac{\partial^2\calL}{\partial\bphi\partial\bphi\transpose}\bigg)(\breve{\bphi})\bLambda_\sharp\bigg\}^{-1}
\bJ_{0n}\transpose\bigg\|_\infty.
\end{align*}
Since $m^{2}\eps_n\log n  \to 0$ as $n\to\infty$, it follows that 
\[
\bigg\|\widetilde{\bLambda}_\sharp\bJ_{0n}\bigg\{\bLambda_\sharp\bigg(\scrI\frac{\partial^2\calL}{\partial\bphi\partial\bphi\transpose}\bigg)(\breve{\bphi})\bLambda_\sharp\bigg\}^{-1}
\bJ_{0n}\transpose\bigg\|_\infty = \Optilde\bigg(\frac{m^{3/2}\sqrt{m \vee \log n}}{\sqrt{n}}\bigg).
\]
By Section \ref{sub:proof_sketch_of_uniform_norm_consistency}, this further implies that
\begin{align}
\label{eqn:uniform_error_bound_breve_phi}
\|\bJ_0(\breve{\bphi} - \bphi_0)\|_\infty = \Optilde\bigg(\frac{m^{2}\sqrt{\log n} + m^{3/2}\log n}{\sqrt{n}}\bigg). 
\end{align}
Now denote $\breve{\balpha}^{(t)}$, $\vech(\breve{\bR}^{(t)})$, and $\vect(\breve{\bB})$ the corresponding components of $\breve{\bphi}$. We immediately obtain
\begin{align}
\label{eqn:uniform_error_bound_breve_psi}
\|\breve{\bX} - \bX_0\|_{\max} + 
\max_{t\in[m]}(\|\breve{\balpha}^{(t)} - \balpha_0^{(t)}\|_\infty + \|\breve{\bR}^{(t)} - \bR^{(t)}_0\|_{\max}) = \Optilde\bigg(\frac{m^{2}\sqrt{\log n} + m^{3/2}\log n}{\sqrt{n}}\bigg),
\end{align}
where $\breve{\bX} = \bX(\breve{\bB})$. 

\noindent $\blacksquare$ \textbf{Stationarity of $\breve{\bphi}$.} We next show that $\breve{\bphi}$ is a stationary point of $\calL$. By \eqref{eqn:uniform_error_bound_breve_psi}, 
\[
\|\bX_0\transpose(\breve{\bX} - \bX_0)\|_2/n \leq \sqrt{d}\|\breve{\bX} - \bX_0\|_{\max} = \Optilde\bigg(\frac{m^{2}\sqrt{\log n} + m^{3/2}\log n}{\sqrt{n}}\bigg),
\]
implying that
\begin{align*}
\|(\bX_0\transpose\breve{\bX})^{-1}\|_2
& = \frac{1}{n}\bigg\|\sum_{k = 0}^\infty\bigg\{-\frac{\bX_0\transpose(\breve{\bX} - \bX_0)}{n}\bigg\}^k\bigg\|_2 = \frac{1}{n} + \Optilde\bigg(\frac{m^{2}\sqrt{\log n} + m^{3/2}\log n}{n^{3/2}}\bigg),\\
\|\breve{\bB}\|_2
&\leq \|\bX_{0\perp}\|_2\sqrt{nd}\|\breve{\bX} - \bX_0\|_{\max}\|(\bX_0\transpose\breve{\bX})^{-1}\|_2 = \Optilde\bigg(\frac{m^{2}\sqrt{\log n} + m^{3/2}\log n}{\sqrt{n}}\bigg).
\end{align*}
Together with the uniform error bound \eqref{eqn:uniform_error_bound_breve_psi}, we obtain 
\begin{align*}
  \|\breve{\bphi} - \bphi_0\|_\sharp
  & = O(m^{1/2}n)\bigg(\max_{t\in[m]}\|\breve{\balpha}^{(t)} - \balpha_0^{(t)}\|_\infty + \max_{t\in[m]}\|\breve\bR^{(t)} - \bR_0^{(t)}\|_{\max} + \|\breve{\bB}\|_{\mathrm{F}}\bigg)\\
  & = \Optilde\left(m^{5/2}\sqrt{n\log n} + m^2\sqrt{n}\log n\right),
\end{align*}
so that $\|\breve{\bphi} - \bphi_0\|_\sharp < \sqrt{m}n\eps_n$ w.h.p.. Also, \eqref{eqn:uniform_error_bound_breve_phi} automatically implies 
\[
\|\bX_{0\perp}\breve{\bB}\|_{2\to\infty} < M/2,\quad 
\max_{t\in[m]}\|\breve{\balpha}^{(t)}\|_\infty < M,\quad 
\max_{t\in[m]}\|\vech(\breve{\bR}^{(t)})\|_2 < M,
\]
so that $\breve{\bphi}$ is inside the interior of $\calN(\sqrt{m}n\eps_n, M)$ w.h.p. We now argue that $\breve{\bphi}$ is a stationary point of $\calL$ w.h.p. by contradiction and assume otherwise. Indeed, by Proposition \ref{prop:local_hessian_bound} and inverse function theorem, $\partial\calL/\partial\bphi$ is a diffeomorphism between an open neighborhood $\calU\subset\calN(\sqrt{m}n\eps_n, M)$ of $\breve{\bphi}$ and an open neighborhood $\calV$ of $(\partial\calL/\partial\bphi)(\breve{\bphi})$ w.h.p., so that every $\bv\in\calV$ can be written as $(\partial\calL/\partial\bphi)(\bphi)$ for some $\bphi\in\calU$. Since $(\partial\calL/\partial\bphi)(\breve{\bphi}) \neq \zero$, we can pick a sufficiently small $\tau > 0$ such that $(1 - \tau)(\partial\calL/\partial\bphi)(\breve{\bphi})\in\calV$, and there exists some $\bphi'\in\calU$ such that $(1 - \tau)(\partial\calL/\partial\bphi)(\breve{\bphi}) = (\partial\calL/\partial\bphi)(\bphi')$, and hence, 
\[
\bigg\|\bD_0\bJ_0\frac{\partial\calL}{\partial\bphi}(\bphi')\bigg\|_\infty = (1 - \tau)\bigg\|\bD_0\bJ_0\frac{\partial\calL}{\partial\bphi}(\breve{\bphi})\bigg\|_\infty < \bigg\|\bD_0\bJ_0\frac{\partial\calL}{\partial\bphi}(\breve{\bphi})\bigg\|_\infty = \min_{\bphi\in\calN(\sqrt{m}n\eps_n, M)}\bigg\|\bD_0\bJ_0\frac{\partial\calL}{\partial\bphi}({\bphi})\bigg\|_\infty,
\]
where $\bD_0 = \mathrm{diag}(n^{-1/2}\eye_{d_\balpha}, n^{-1}\eye_{d_\br}, m^{-1/2}n^{-3/2}\eye_{d_\bX})$ is the weight matrix in the definition of $\breve{\bphi}$ in Section \ref{sub:proof_sketch_of_uniform_norm_consistency}. This contradicts with the minimization property of $\breve{\bphi}$ since $\bphi'\in\calU\subset\calN(\sqrt{m}n\eps_n, M)$. Hence $\breve{\bphi}$ is a stationary point of $\calL$ w.h.p.

\noindent $\blacksquare$ \textbf{$\widehat{\bphi} = \breve{\bphi}$ w.h.p.}
We first argue that
$\bpsi(\widehat{\bphi})$ is a maximizer of $\calL_\bpsi$ over $\Psi$, and that $\bphi(\widehat{\bpsi})$ is a maximizer of $\calL(\bphi)$ over $\bphi\in\{\bphi:\bpsi(\bphi)\in\Psi\}$. Indeed,
$\bpsi(\bphi(\widehat{\bpsi}))$ is obtained from $\widehat{\bpsi}$ by the rotation
$\bX\mapsto\bX\bW_{\widehat{\bX}}$, $\bR^{(t)}\mapsto
\bW_{\widehat{\bX}}\transpose\bR^{(t)}\bW_{\widehat{\bX}}$ for all $t\in[m]$, which leaves $\Psi$
and $\calL_\bpsi$ invariant, so that $\bpsi(\bphi(\widehat{\bpsi}))\in\Psi$, $\bphi(\widehat{\bpsi})\in\{\bphi:\bpsi(\bphi) \in \Psi\}$, and
\[
\calL_\bpsi(\widehat{\bpsi})\geq \calL_\bpsi(\bpsi(\widehat{\bphi})) = \calL(\widehat{\bphi})
\geq \calL(\bphi(\widehat{\bpsi})) = \calL_\bpsi(\widehat{\bpsi}).
\]
Now write
$\widehat{\balpha}^{(t)}$, $\widehat{\bR}^{(t)}$, $\widehat{\bB}$ for the
components of $\widehat{\bphi}$. Since $\bpsi(\widehat{\bphi})$ maximizes $\calL_\bpsi$ over $\Psi$ and $\bX_0\transpose\bX(\widehat{\bB})
= n(\eye_d + \widehat{\bB}\transpose\widehat{\bB})^{-1/2}$
is symmetric positive definite, so that
$\bW_{\bX(\widehat{\bB})} = \eye_d$, by Proposition
\ref{prop:consistency_native_norm} applied to $\bpsi(\widehat{\bphi})$, we obtain
\[
\|\widehat{\bphi} - \bphi_0\|_\sharp = \bigg(n\sum_{t = 1}^m\|\widehat{\balpha}^{(t)} - \balpha_0^{(t)}\|_2^2 + n^2\sum_{t = 1}^m\|\widehat{\bR}^{(t)} - \bR_0^{(t)}\|_{\mathrm{F}}^2 + mn^2\|\widehat{\bB}\|_{\mathrm{F}}^2\bigg)^{1/2} = \Optilde(\sqrt{mn}),
\]
and
\begin{align*}
\|\bX_{0\perp}\widehat{\bB}\|_{2\to\infty}
&\leq \|(\bX_0 + \bX_{0\perp}\widehat{\bB})(\eye_d + \widehat{\bB}\transpose\widehat{\bB})^{-1/2}(\eye_d + \widehat{\bB}\transpose\widehat{\bB})^{1/2}\|_{2\to\infty} + \|\bX_0\|_{2\to\infty}\\
&\leq \|\bX(\widehat{\bB})\|_{2\to\infty}\|(\eye_d + \widehat{\bB}\transpose\widehat{\bB})^{1/2}\|_2 + C_1\leq \frac{M}{4}\{1 + \Optilde(n^{-1})\} + C_1 < \frac{M}{2},
\end{align*}
so that $\widehat{\bphi}\in\calN(\sqrt{m}n\eps_n, M)$ w.h.p. because $M > 4C_1$. Since $\calL$ is strongly concave over $\calN(\sqrt{m}n\eps_n, M)$ w.h.p. by Proposition \ref{prop:local_hessian_bound}, $\breve{\bphi}$ is the unique maximizer of
$\calL$ over $\calN(\sqrt{m}n\eps_n, M)$, then immediately we have
$\calL(\widehat{\bphi})\leq\calL(\breve{\bphi})$. Moreover, $\bpsi(\breve{\bphi})\in\Psi$
w.h.p., because by \eqref{eqn:uniform_error_bound_breve_psi},
\begin{align*}
&\|\bX(\breve{\bB})\|_{2\to\infty}
\leq \|\bX_0\|_{2\to\infty} + \sqrt{d}\|\breve{\bX} - \bX_0\|_{\max}
\leq C_1 + o(1) < M/4,\\
&\max_{t\in[m]}\|\breve{\balpha}^{(t)}\|_\infty\leq C_1 + o(1) < M,\;
\max_{t\in[m]}\| \breve{\bR}^{(t)} \|_{\mathrm{F}}
\leq \max_{t\in[m]}\| \bR_0^{(t)} \|_{\mathrm{F}} + o(1) < M,
\end{align*}
so that $\breve{\bphi}\in\{\bphi:\bpsi(\bphi)\in\Psi\}$ and $\calL(\breve{\bphi})\leq\calL(\widehat{\bphi})$.
Hence $\widehat{\bphi} = \breve{\bphi}$ w.h.p. by the uniqueness of the maximizer of $\calL(\bphi)$ over $\calN(\sqrt{m}n\eps_n, M)$. Therefore, $\widehat{\bphi}$ is also unique and coincides with
$\breve{\bphi}$ w.h.p. Also, $\calL(\widehat{\bphi}) = \calL(\bphi(\widehat{\bpsi}))$ and the uniqueness of $\widehat{\bphi}$ w.h.p. implies that
$\widehat{\bphi} = \bphi(\widehat{\bpsi})$ w.h.p. Note that we have used the condition
$\max_t\|\bR_0^{(t)}\|_{\mathrm{F}}\leq\sqrt{d}\max_t\max_k|\lambda_k(\bR_0^{(t)})|
\leq\sqrt{d}C_2 < M$. This proves the stationarity and uniqueness of
$\widehat{\bphi}$ w.h.p. The proof is therefore completed.
\end{proof}

\section{Proofs of Theorems \ref{thm:two_to_infinity_norm_bound_MLE} and \ref{thm:asymptotic_normality_MLE}}
\label{sec:asymptotic_normality_of_mle}

\subsection{Proof architecture}
\label{sub:proof_architecture_of_the_asymptotic_normality}

We first outline the proof architecture of the asymptotic normality results of the MLE defined in Section \ref{sec:maximum_likelihood_and_one_step_estimator} of the main text. By the uniqueness (Proposition \ref{prop:stationarity_uniqueness_uniform_consistency_MLE}), it is sufficient to consider the MLE $\widehat{\bphi} = \argmax_{\bphi:\bpsi(\bphi)\in\Psi}\calL(\bphi)$ under the compact parameterization and take the orthogonal alignment matrix $\bW_{\widehat{\bX}} = \eye_d$. By the stationarity of $\widehat{\bphi}$ from Proposition \ref{prop:stationarity_uniqueness_uniform_consistency_MLE} and Taylor's theorem, 
\begin{align}
\label{eqn:Taylor_expansion}
\bJ_0(\widehat{\bphi} - \bphi_0)& = \bJ_0\bG(\bphi_0)^{-1}\frac{\partial\calL}{\partial\bphi}(\bphi_0) + \bgamma,\\
\label{eqn:Taylor_expansion_remainder}
\bgamma& = -\bJ_0\bigg\{\bigg(\scrI\frac{\partial^2\calL}{\partial\bphi\partial\bphi\transpose}\bigg)(\widehat{\bphi})\bigg\}^{-1}\bigg\{\bigg(\scrI\frac{\partial^2\calL}{\partial\bphi\partial\bphi\transpose} + \bG(\bphi_0)\bigg)(\widehat{\bphi})\bigg\}\bG(\bphi_0)^{-1}\frac{\partial\calL}{\partial\bphi}(\bphi_0).
\end{align}
Recall that $\bz_i = [\balpha_i\transpose,\bx_i\transpose]\transpose$ is the parameter associated with vertex $i$ with $\balpha_i = [\alpha_i^{(1)},\ldots,\alpha_i^{(m)}]\transpose$. Denote
\begin{align*}
\scrP_{\bz_i}
& = \begin{bmatrix}
  \eye_m\otimes\be_i\transpose & \mathbf{O}_{m\times d_\br} & \mathbf{O}_{m\times d_\bX} \\
  \mathbf{O}_{d\times d_\balpha} & \mathbf{O}_{d\times d_\br}& \eye_d\otimes\be_i\transpose
\end{bmatrix},\;
\scrP_{\bX} = \begin{bmatrix}
\mathbf{O}_{d_\bX\times(d_\balpha + d_\br)} & \eye_{d_\bX}
\end{bmatrix},\\
\scrP_{\bZ}
& = \begin{bmatrix}
  \eye_{d_\balpha} & \mathbf{O}_{d_\balpha \times d_\br} & \mathbf{O}_{d_\balpha \times d_\bX} \\
  \mathbf{O}_{d_\bX\times d_\balpha} & \mathbf{O}_{d_\bX\times d_\br}& \eye_{d_\bX}
\end{bmatrix},\; 
\scrP_{\br} = \begin{bmatrix}
  \mathbf{O}_{d_\br\times d_\balpha} & \eye_{d_\br} & \mathbf{O}_{d_\br \times d_\bX}
\end{bmatrix}. 
\end{align*}
Denote $\bG_0^\sharp = \bLambda_\sharp\bG(\bphi_0)\bLambda_\sharp$ and
\[
\bar{\mathscr{H}} = \bigg(\scrI\bLambda_\sharp\frac{\partial^2\calL}{\partial\bphi\partial\bphi\transpose}\bLambda_\sharp\bigg)(\widehat{\bphi}).
\]
 Let $\bgamma_{\bz_i} = \scrP_{\bz_i}\bgamma$ and $\bgamma_{\bF} = \vect(\bF\bX_0)\transpose\scrP_\bX\bgamma$. It follows immediately that
\begin{align*}
\bgamma & = -(\widetilde{\bLambda}_\sharp\bJ_{0n}\bar{\scrH}^{-1}\bJ_{0n}\transpose) \{\bJ_{0n}(\bar{\scrH} + \bG_0^\sharp) \bJ_{0n}\transpose\} \{\bJ_{0n}(\bG_0^\sharp)^{-1}\bJ_{0n}\transpose\}\bigg\{\bJ_{0n}\bLambda_\sharp\frac{\partial\calL}{\partial\bphi}(\bphi_0)\bigg\},\\
\bgamma_{\bF}
& = -\frac{1}{\sqrt{mn}}\{\vect(\bF\bX_0)\transpose\mathscr{P}_{\bX}\bJ_{0n}\bar{\scrH}^{-1}\bJ_{0n}\transpose\} \{\bJ_{0n}(\bar{\scrH} + \bG_0^\sharp) \bJ_{0n}\transpose\} \{\bJ_{0n}(\bG_0^\sharp)^{-1}\bJ_{0n}\transpose\}\bigg\{\bJ_{0n}\bLambda_\sharp\frac{\partial\calL}{\partial\bphi}(\bphi_0)\bigg\},\\
\bgamma_{\bz_i}
& = -\mathrm{diag}\bigg(\frac{\eye_{m}}{\sqrt{n}}, \frac{\eye_{d}}{\sqrt{mn}}\bigg)
\{\mathscr{P}_{\bz_i}\bJ_{0n}\bar{\scrH}^{-1}\bJ_{0n}\transpose\} \{\bJ_{0n}(\bar{\scrH} + \bG_0^\sharp) \bJ_{0n}\transpose\} \{\bJ_{0n}(\bG_0^\sharp)^{-1}\bJ_{0n}\transpose\}\bigg\{\bJ_{0n}\bLambda_\sharp\frac{\partial\calL}{\partial\bphi}(\bphi_0)\bigg\}. 
\end{align*}
Below, we will sketch the argument for controlling the remainders above. 
\begin{itemize}
  \item \textbf{Step 1. Bounding $\left\|\mathscr{P}_{\bZ}\bJ_{0n}\bar{\mathscr{H}}^{-1}\bJ_{0n}\transpose \mathrm{diag}(\eye_{d_\balpha}, \sqrt{n}\eye_{d_\br}, \eye_{d_\bX})\right\|_\infty$. } This is effectively done by bounding $\|\mathscr{P}_{\bZ}\bJ_{0n}\bar{\mathscr{H}}^{-1}\bJ_{0n}\transpose \scrP_\bZ\transpose\|_\infty$, $\|\mathscr{P}_{\br}\bJ_{0n}\bar{\mathscr{H}}^{-1}\bJ_{0n}\transpose \scrP_\br\transpose\|_\infty$ separately and applying the block matrix inversion formula. 

  \item \textbf{Step 2. Remainder decomposition. }
  We decompose the remainder $\bgamma_\bZ$ into 
  \begin{equation}
  \label{eqn:remainder_decomposition}
  \begin{aligned}
  \bgamma_\bZ&:= \scrP_\bZ\bgamma 
   = \bgamma_{\bZ}^{(1)} + \bgamma_{\bZ}^{(2)} + \bgamma_{\bZ}^{(3)},\\
  \bgamma_{\bZ}^{(1)}
  & = -\mathrm{diag}\bigg(\frac{\eye_{d_\balpha}}{\sqrt{n}}, \frac{\eye_{d_\bX}}{\sqrt{mn}}\bigg)\{\mathscr{P}_{\bZ}\bJ_{0n}\bar{\mathscr{H}}^{-1}\bJ_{0n}\transpose\mathrm{diag}(\eye_{d_\balpha},\sqrt{n}\eye_{d_\br},\eye_{d_\bX})\}\mathrm{diag}\bigg(\eye_{d_\balpha}, \frac{\eye_{d_\br}}{\sqrt{n}}, \eye_{d_\bX}\bigg)\\
  &\quad\times\left[
  \bJ_{0n}\bLambda_\sharp\int_0^1\{\bGamma(\bphi_\tau) - \bGamma(\bphi_0)\}\mathrm{d}\tau\bLambda_\sharp\bJ_{0n}\transpose + \bJ_{0n}\int_0^1(\mathscr{S}_\tau - \mathscr{S}_0)\mathrm{d}\tau \bJ_{0n}\transpose
  \right]\\
  &\quad\times \{\bJ_{0n}(\bG_0^\sharp)^{-1}\bJ_{0n}\transpose\}\bigg\{\bJ_{0n}\bLambda_\sharp\frac{\partial\calL}{\partial\bphi}(\bphi_0)\bigg\},\\
  \bgamma_{\bZ}^{(2)}
  & = -\mathrm{diag}\bigg(\frac{\eye_{d_\balpha}}{\sqrt{n}}, \frac{\eye_{d_\bX}}{\sqrt{mn}}\bigg)\{\mathscr{P}_{\bZ}\bJ_{0n}\bar{\mathscr{H}}^{-1}\bJ_{0n}\transpose\mathrm{diag}(\eye_{d_\balpha},\sqrt{n}\eye_{d_\br},\eye_{d_\bX})\}\\
  &\quad\times\mathrm{diag}\bigg(\eye_{d_\balpha}, \frac{\eye_{d_\br}}{\sqrt{n}}, \eye_{d_\bX}\bigg)(\bJ_{0n}\mathscr{S}_0\bJ_{0n}\transpose)\{\bJ_{0n}(\bG_0^\sharp)^{-1}\bJ_{0n}\transpose\}\bigg\{\bJ_{0n}\bLambda_\sharp\frac{\partial\calL}{\partial\bphi}(\bphi_0)\bigg\},\\
  \bgamma_{\bZ}^{(3)}
  & = -\mathrm{diag}\bigg(\frac{\eye_{d_\balpha}}{\sqrt{n}}, \frac{\eye_{d_\bX}}{\sqrt{mn}}\bigg)\{\mathscr{P}_{\bZ}\bJ_{0n}\bar{\mathscr{H}}^{-1}\bJ_{0n}\transpose\mathrm{diag}(\eye_{d_\balpha},\sqrt{n}\eye_{d_\br},\eye_{d_\bX})\}\\
  &\quad\times \mathrm{diag}\bigg(\eye_{d_\balpha}, \frac{\eye_{d_\br}}{\sqrt{n}}, \eye_{d_\bX}\bigg)\bJ_{0n}\bLambda_\sharp\{\bGamma(\bphi_0) + \bG(\bphi_0)\}\bLambda_\sharp\bJ_{0n}\transpose \\
  &\quad\times\{\bJ_{0n}(\bG_0^\sharp)^{-1}\bJ_{0n}\transpose\}\bigg\{\bJ_{0n}\bLambda_\sharp\frac{\partial\calL}{\partial\bphi}(\bphi_0)\bigg\},
  \end{aligned}
  \end{equation}
  where we denote $\theta_{ij}^{(t)}(\tau) = \theta_{ij}^{(t)}(\bphi_\tau)$ (recall that $\theta_{ij}^{(t)}(\bphi)$ is a function of $\bphi$), $\bphi_\tau = \bphi_0 + \tau(\widehat{\bphi} - \bphi_0)$, the score term
  \begin{align*}
  {\mathscr{S}}_\tau = \frac{1}{2}\bLambda_\sharp \sum_{t = 1}^m\sum_{i\neq j}\ell'(\theta_{ij}^{(t)}(\tau); A_{ij}^{(t)})\frac{\partial^2\theta_{ij}^{(t)}}{\partial\bphi\partial\bphi\transpose}(\bphi_\tau)\bLambda_\sharp, 
  \end{align*}
  and $\bGamma$ is defined in \eqref{eqn:Gamma_matrix}. These remainder terms can be sharply bounded by jointly applying Lemmas \ref{lemma:local_lipschitz_sample_Fisher_information_infinity_norm}--\ref{lemma:inverse_Fisher_information_matrix_infty_norm_bound}. 
\end{itemize}

\subsection{Lipschitz properties of the sample Hessian under the infinity norm}
\label{sub:lipschitz_properties_of_the_sample_hessian_under_the_infinity_norm}

This subsection establishes the Lipschitz properties of the sample Hessian under the rescaled matrix infinity norm, which are used in the proof of the asymptotic normality of the MLE. 

\begin{lemma}\label{lemma:local_lipschitz_sample_Fisher_information_infinity_norm}
Suppose Assumptions \ref{assumption:edge_distribution}--\ref{assumption:true_parameter} hold and $m^{12}(\log n)^4/n\to 0$ as $n\to\infty$. Then, we have
\begin{align*}
  \sup_{\tau\in[0, 1]}\left\|\mathrm{diag}\bigg(\eye_{d_\balpha}, \frac{1}{\sqrt{n}}\eye_{d_\br}, \eye_{d_\bX}\bigg)\bJ_{0n}\bLambda_\sharp\{\bGamma({\bphi}_\tau) - \bGamma(\bphi_0)\}\bLambda_\sharp\bJ_{0n}\transpose\right\|_\infty = \Optilde\bigg(\frac{m^{5/2}\sqrt{\log n} + m^2\log n}{\sqrt{n}}\bigg),
\end{align*}
where we denote $\bphi_\tau = \bphi_0 + \tau(\widehat{\bphi} - \bphi_0)$, $\widehat{\bphi}$ is defined in Proposition \ref{prop:stationarity_uniqueness_uniform_consistency_MLE}, $\bJ_{0n}$ is defined in \eqref{eqn:J_n(B)}, and $\bGamma(\bphi)$ is defined in \eqref{eqn:Gamma_matrix} with $\bphi_0$ replaced by $\bphi$. 
\end{lemma}

\begin{proof}[\bf Proof]
Let $\delta_n = (m^{2}\sqrt{\log n} + m^{3/2}\log n)/\sqrt{n}$ 
and $\theta_{ij}^{(t)}(\tau) = \theta_{ij}^{(t)}(\bphi_0 + \tau(\widehat{\bphi} - \bphi_0))$ (recall that $\theta_{ij}^{(t)}$ is implicitly a function of $\bphi$). 
By Proposition \ref{prop:stationarity_uniqueness_uniform_consistency_MLE}, it is clear that 
\[
\sup_\tau\max_{t\in[m],i\neq j}|{\theta}_{ij}^{(t)}(\tau) - \theta_{0ij}^{(t)}| = \Optilde(\delta_n). 
\]
By Proposition \ref{prop:Cayley_transform_jacobian}, 
\begin{align*}
\frac{\partial\vect(\bX)}{\partial\vect(\bB)\transpose}
& = (\eye_d + \bB\transpose\bB)^{-1/2}\otimes\bX_{0\perp} - \{\eye_d\otimes (\bX_0 + \bX_{0\perp}\bB)\}\{(\eye_d + \bB\transpose\bB)^{-1/2}\otimes (\eye_d + \bB\transpose\bB)^{-1/2}\}\\
&\quad\times \{\eye_d\otimes (\eye_d + \bB\transpose\bB)^{1/2} + (\eye_d + \bB\transpose\bB)^{1/2}\otimes\eye_d\}^{-1}(\eye_{d^2} + \bK_{dd})(\eye_d\otimes\bB\transpose).
\end{align*}
Since $\|\tau\widehat{\bB}\|_2 = \Optilde(n^{-1/2})$, $\|(\eye_d + \tau^2\widehat{\bB}\transpose\widehat{\bB})^{-1/2} - \eye_d\|_2 = \Optilde(n^{-1})$, and $\|\bX_{0\perp}\tau\widehat{\bB}\|_{2\to\infty} = \Optilde(\delta_n)$ uniformly over $\tau\in[0, 1]$ by Propositions \ref{prop:consistency_native_norm} and \ref{prop:stationarity_uniqueness_uniform_consistency_MLE}, it follows that
\begin{align*}
\sup_\tau\bigg\|\bigg(\eye_d\otimes \frac{1}{\sqrt{n}}\bX_{0\perp}\bigg)\frac{1}{\sqrt{n}}\bigg\{\frac{\partial\vect(\bX)}{\partial\vect(\bB)\transpose}(\tau\widehat{\bB}) - \eye_d\otimes \bX_{0\perp}\bigg\}\transpose\bigg\|_\infty
& = \Optilde(\delta_n),\\
\sup_\tau\bigg\|\frac{1}{\sqrt{n}}\bigg\{\frac{\partial\vect(\bX)}{\partial\vect(\bB)\transpose}(\tau\widehat{\bB}) - \eye_d\otimes \bX_{0\perp}\bigg\}\bigg(\eye_d\otimes \frac{1}{\sqrt{n}}\bX_{0\perp}\transpose\bigg)\bigg\|_\infty
& = \Optilde(\delta_n)
\end{align*}
by a direct computation. Indeed, writing
$\bN_\tau = (\eye_d + \tau^2\widehat{\bB}\transpose\widehat{\bB})^{-1/2}$ and
$\bS_\tau = \bN_\tau^{-2}$, the display above gives
\begin{align*}
&\bigg(\eye_d\otimes \frac{1}{\sqrt{n}}\bX_{0\perp}\bigg)\frac{1}{\sqrt{n}}
\bigg\{\frac{\partial\vect(\bX)}{\partial\vect(\bB)\transpose}
(\tau\widehat{\bB})
- \eye_d\otimes \bX_{0\perp}\bigg\}\transpose\\
&\quad = (\bN_\tau - \eye_d)\otimes\frac{\bX_{0\perp}\bX_{0\perp}\transpose}{n}
- \frac{1}{n}(\eye_d\otimes\bX_{0\perp}\tau\widehat{\bB})(\eye_{d^2} + \bK_{dd})
(\bS_\tau^{1/2}\otimes\eye_d + \eye_d\otimes\bS_\tau^{1/2})^{-1}
\{\bN_\tau\otimes\bX(\tau\widehat{\bB})\transpose\},
\end{align*}
so that its matrix $\|\cdot\|_\infty$-norm is bounded by
\begin{align*}
&\sup_{\tau\in[0, 1]}\bigg\|\bigg(\eye_d\otimes \frac{1}{\sqrt{n}}\bX_{0\perp}\bigg)\frac{1}{\sqrt{n}}
\bigg\{\frac{\partial\vect(\bX)}{\partial\vect(\bB)\transpose}
(\tau\widehat{\bB})
- \eye_d\otimes \bX_{0\perp}\bigg\}\transpose\bigg\|_\infty\\
&\quad\leq \|\bN_\tau - \eye_d\|_2\bigg\|\frac{1}{n}\bX_{0\perp}\bX_{0\perp}\transpose\bigg\|_\infty
+ d\sqrt{d}\|\bX_{0\perp}\tau\widehat{\bB}\|_{2\to\infty}
\|\bX(\tau\widehat{\bB})\|_{2\to\infty} = \Optilde(\delta_n),
\end{align*} 
where we use the fact that $\|\bX_{0\perp}\bX_{0\perp}\transpose/n\|_\infty
\leq 2 + \|\bX_0\bX_0\transpose/n\|_\infty = O(1)$. The second bound follows by a similar reasoning. It is also clear that $\sup_\tau\|\bJ_n(\tau\widehat{\bB})\bJ_{0n}\transpose\|_\infty = \Optilde(1)$, where $\bJ_n$ and $\bJ_{0n}$ are defined in \eqref{eqn:J_n(B)}. Recall $\bpsi$ is the ambient parameter. 
Then we obtain
\begin{align*}
  &\bJ_{0n}\bLambda_\sharp\{\bGamma({\bphi}_\tau) - \bGamma(\bphi_0)\}\bLambda_\sharp\bJ_{0n}\transpose\\
  &\quad = \bJ_{0n}\bJ_n(\tau\widehat{\bB})\transpose\bigg[\frac{1}{2}\sum_{t = 1}^m\sum_{i\neq j}\{\ell''({\theta}_{ij}^{(t)}(\tau); A_{ij}^{(t)}) - \ell''(\theta_{0ij}^{(t)}; A_{ij}^{(t)})\}\widetilde{\bLambda}_\sharp\frac{\partial\theta_{ij}^{(t)}}{\partial\bpsi}({\bphi}_\tau)\frac{\partial\theta_{ij}^{(t)}}{\partial\bpsi\transpose}({\bphi}_\tau)\widetilde{\bLambda}_\sharp\bigg]\bJ_n(\tau\widehat{\bB})\bJ_{0n}\transpose\\
  &\qquad + \bJ_{0n}\bJ_n(\tau\widehat{\bB})\transpose\bigg[\frac{1}{2}\sum_{t = 1}^m\sum_{i\neq j}\ell''(\theta_{0ij}^{(t)}; A_{ij}^{(t)})\widetilde{\bLambda}_\sharp\bigg\{\frac{\partial\theta_{ij}^{(t)}}{\partial\bpsi}({\bphi}_\tau)\frac{\partial\theta_{ij}^{(t)}}{\partial\bpsi\transpose}({\bphi}_\tau)
  -
  \frac{\partial\theta_{ij}^{(t)}}{\partial\bpsi}(\bphi_0)\frac{\partial\theta_{ij}^{(t)}}{\partial\bpsi\transpose}(\bphi_0)
  \bigg\}\widetilde{\bLambda}_\sharp\bigg]\bJ_n(\tau\widehat{\bB})\bJ_{0n}\transpose\\
  &\qquad + \bJ_{0n}\{\bJ_n(\tau\widehat{\bB}) - \bJ_{0n}\}\transpose\bigg[\frac{1}{2}\sum_{t = 1}^m\sum_{i\neq j}\ell''(\theta_{0ij}^{(t)}; A_{ij}^{(t)})\widetilde{\bLambda}_\sharp\bigg\{
  \frac{\partial\theta_{ij}^{(t)}}{\partial\bpsi}(\bphi_0)\frac{\partial\theta_{ij}^{(t)}}{\partial\bpsi\transpose}(\bphi_0)
  \bigg\}\widetilde{\bLambda}_\sharp\bigg]\bJ_n(\tau\widehat{\bB})\bJ_{0n}\transpose\\
  &\qquad + \bJ_{0n}\bJ_{0n}\transpose\bigg[\frac{1}{2}\sum_{t = 1}^m\sum_{i\neq j}\ell''(\theta_{0ij}^{(t)}; A_{ij}^{(t)})\widetilde{\bLambda}_\sharp\bigg\{
  \frac{\partial\theta_{ij}^{(t)}}{\partial\bpsi}(\bphi_0)\frac{\partial\theta_{ij}^{(t)}}{\partial\bpsi\transpose}(\bphi_0)
  \bigg\}\widetilde{\bLambda}_\sharp\bigg]\{\bJ_n(\tau\widehat{\bB}) - \bJ_{0n}\}\bJ_{0n}\transpose.
\end{align*}
Since 
\begin{align*}
\sup_{\tau\in[0, 1]}\left\{\|\bJ_{0n}\bJ_n(\tau\widehat{\bB})\transpose\|_\infty + \|\bJ_n(\tau\widehat{\bB})\bJ_{0n}\transpose\|_\infty\right\}
& = \Optilde(1),\\
\sup_{\tau\in[0, 1]}\left\{\|\bJ_{0n}\{\bJ_n(\tau\widehat{\bB}) - \bJ_{0n}\}\transpose\|_\infty + \|\{\bJ_n(\tau\widehat{\bB}) - \bJ_{0n}\}\bJ_{0n}\transpose\|_\infty\right\}
& = \Optilde(\delta_n)
\end{align*}
by Lemma \ref{lemma:jacobian_perturbation} and the two displays above
and matrix-$\|\cdot\|_\infty$ is invariant with regard to permutation of rows and/or columns, it is sufficient to consider
\begin{align*}
&\sup_{\tau\in[0, 1]} \left\|\sum_{t = 1}^m\sum_{i\neq j}\{\ell''({\theta}_{ij}^{(t)}(\tau); A_{ij}^{(t)}) - \ell''(\theta_{0ij}^{(t)}; A_{ij}^{(t)})\}
  \begin{bmatrix}
    \frac{\be_t\otimes(\be_i + \be_j)}{\sqrt{n}}\\
    \frac{\be_i\otimes{\bR}^{(t)}_\tau{\bx}_{\tau j} + \be_j\otimes{\bR}^{(t)}_{\tau}{\bx}_{\tau i}}{\sqrt{mn}}\\
    \frac{\be_t\otimes\bD_d\transpose({\bx}_{\tau j}\otimes{\bx}_{\tau i})}{n^{3/2}}
  \end{bmatrix}
  \begin{bmatrix}
    \frac{\be_t\otimes(\be_i + \be_j)}{\sqrt{n}}\\
    \frac{\be_i\otimes{\bR}^{(t)}_\tau{\bx}_{\tau j} + \be_j\otimes{\bR}^{(t)}_{\tau}{\bx}_{\tau i}}{\sqrt{mn}}\\
    \frac{\be_t\otimes\bD_d\transpose({\bx}_{\tau j}\otimes{\bx}_{\tau i})}{n}
  \end{bmatrix}\transpose
  \right.\\
&\quad + \left.
\sum_{t = 1}^m\sum_{i\neq j} \ell''(\theta_{0ij}^{(t)}; A_{ij}^{(t)})\left\{
  \begin{bmatrix}
    \frac{\be_t\otimes(\be_i + \be_j)}{\sqrt{n}}\\
    \frac{\be_i\otimes{\bR}^{(t)}_\tau{\bx}_{\tau j} + \be_j\otimes{\bR}^{(t)}_{\tau}{\bx}_{\tau i}}{\sqrt{mn}}\\
    \frac{\be_t\otimes\bD_d\transpose({\bx}_{\tau j}\otimes{\bx}_{\tau i})}{n^{3/2}}
  \end{bmatrix}
  \begin{bmatrix}
    \frac{\be_t\otimes(\be_i + \be_j)}{\sqrt{n}}\\
    \frac{\be_i\otimes{\bR}^{(t)}_\tau{\bx}_{\tau j} + \be_j\otimes{\bR}^{(t)}_{\tau}{\bx}_{\tau i}}{\sqrt{mn}}\\
    \frac{\be_t\otimes\bD_d\transpose({\bx}_{\tau j}\otimes{\bx}_{\tau i})}{n}
  \end{bmatrix}\transpose
  \right.
\right.\\
&\qquad - \left.\left.
\begin{bmatrix}
    \frac{\be_t\otimes(\be_i + \be_j)}{\sqrt{n}}\\
    \frac{(\be_i\otimes{\bR}_0^{(t)}{\bx}_{0j} + \be_j\otimes{\bR}^{(t)}_0{\bx}_{0i})}{\sqrt{mn}}\\
    \frac{\be_t\otimes\bD_d\transpose({\bx}_{0j}\otimes{\bx}_{0i})}{n^{3/2}}
  \end{bmatrix}
  \begin{bmatrix}
    \frac{\be_t\otimes(\be_i + \be_j)}{\sqrt{n}}\\
    \frac{(\be_i\otimes{\bR}_0^{(t)}{\bx}_{0j} + \be_j\otimes{\bR}^{(t)}_0{\bx}_{0i})}{\sqrt{mn}}\\
    \frac{\be_t\otimes\bD_d\transpose({\bx}_{0j}\otimes{\bx}_{0i})}{n}
  \end{bmatrix}\transpose
\right\}\right\|_\infty
\end{align*}
and
\begin{align*}
  \left\|\sum_{t = 1}^m\sum_{i\neq j}  \ell''(\theta_{0ij}^{(t)}; A_{ij}^{(t)})
  \begin{bmatrix}
    \frac{\be_t\otimes(\be_i + \be_j)}{\sqrt{n}}\\
    \frac{(\be_i\otimes\bR_0^{(t)}\bx_{0j} + \be_j\otimes\bR_0^{(t)}\bx_{0i})}{\sqrt{mn}}\\
    \frac{\be_t\otimes\bD_d\transpose(\bx_{0j}\otimes\bx_{0i})}{n^{3/2}}
  \end{bmatrix}
  \begin{bmatrix}
    \frac{\be_t\otimes(\be_i + \be_j)}{\sqrt{n}}\\
    \frac{(\be_i\otimes\bR_0^{(t)}\bx_{0j} + \be_j\otimes\bR_0^{(t)}\bx_{0i})}{\sqrt{mn}}\\
    \frac{\be_t\otimes\bD_d\transpose(\bx_{0j}\otimes\bx_{0i})}{n}
  \end{bmatrix}\transpose
  \right\|_\infty\Optilde(\delta_n),
\end{align*}
where $\bR^{(t)}_{\tau}:= \bR_0^{(t)} + \tau(\widehat{\bR}^{(t)} - \bR_0^{(t)})$ and $\bx_{\tau i} := \bX(\tau\widehat{\bB})\transpose\be_i$. 
Each of the two matrices in the above two terms has a $3\times 3$ block structure, and each one of these blocks can be unified as one of the following $9$ forms:
\begin{align*}
  &\frac{2}{n}\sum_{t = 1}^m\be_t\be_t\transpose\otimes\bigg\{\sum_{i = 1}^n\be_i\be_i\transpose\bigg(\sum_{j\in[n]\backslash\{i\}}\beta_{1ij}^{(t)}\bigg) + \sum_{i\neq j}\beta_{2ij}^{(t)}\be_i\be_j\transpose\bigg\},\\
  &\frac{2}{mn}\sum_{i = 1}^n\be_i\be_i\transpose\otimes\sum_{t = 1}^m\sum_{j\in[n]\backslash\{i\}}\bF_{1ij}^{(t)} + \frac{2}{mn}\sum_{i\neq j}\be_i\be_j\transpose\otimes\sum_{t = 1}^m\bF_{2ij}^{(t)},\quad
  \frac{1}{n^{5/2}}\sum_{t = 1}^m\be_t\be_t\transpose\otimes \sum_{i\neq j}\bF_{3ij}^{(t)},\\
  &\frac{2}{\sqrt{m}n}\sum_{t = 1}^m\be_t\otimes\sum_{i\neq j}(\be_i + \be_j)\be_i\transpose\otimes\bh_{1ij}^{(t)\mathrm{T}},
  \quad\frac{2}{\sqrt{m}n}\sum_{t = 1}^m\be_t\transpose\otimes\sum_{i\neq j}\be_i(\be_i + \be_j)\transpose\otimes\bh_{1ij}^{(t)},\\
  &\frac{1}{\sqrt{m}n^2}\sum_{t = 1}^m\sum_{i = 1}^n\be_t\be_i\transpose\otimes\sum_{j\in[n]\backslash\{i\}}\bF_{4ij}^{(t)},\quad 
  \frac{1}{\sqrt{m}n^{3/2}}\sum_{i = 1}^n\sum_{t = 1}^m\be_i\be_t\transpose\otimes\sum_{j\in[n]\backslash\{i\}}\bF_{4ij}^{(t)\mathrm{T}},\\
  &\frac{1}{n^{3/2}}\sum_{t = 1}^m\be_t\be_t\transpose\otimes \sum_{i\neq j}(\be_i + \be_j)\bh_{2ij}^{(t)\mathrm{T}},\quad 
  \frac{1}{n^2}\sum_{t = 1}^m\be_t\be_t\transpose \otimes \sum_{i\neq j}\bh_{2ij}^{(t)}(\be_i\transpose + \be_j\transpose),
\end{align*}
where $\beta_{1ij}^{(t)},\beta_{2ij}^{(t)},\bF_{1ij}^{(t)},\bF_{2ij}^{(t)},\bF_{3ij}^{(t)},\bF_{4ij}^{(t)},\bh_{1ij}^{(t)},\bh_{2ij}^{(t)}$ are placeholder scalars or matrices ($\tau$-dependent) whose dimensions are $O(1)$ and their spectral norms are $\Optilde(\delta_n)$ uniformly over $t\in[m],i,j\in[n]$, $\tau\in[0, 1]$ by their uniformly Lipschitz continuity and Proposition \ref{prop:stationarity_uniqueness_uniform_consistency_MLE} for the first matrix and by Proposition \ref{prop:stationarity_uniqueness_uniform_consistency_MLE} and Assumptions \ref{assumption:edge_distribution}--\ref{assumption:true_parameter} for the second matrix. The proof is completed by a direct calculation based on the above observation, where the $(\bX, \balpha)$-block (the fifth form above) contributes an extra $O(\sqrt{m})$ factor. 
\end{proof}

\begin{lemma}\label{lemma:local_lipschitz_hessian_score_term_infinity_norm}
Suppose Assumptions \ref{assumption:edge_distribution}--\ref{assumption:true_parameter} hold and $m^{12}(\log n)^4/n\to 0$ as $n\to\infty$. Then, 
we have
\begin{align*}
  &\sup_{\tau\in[0, 1]}\bigg\|\mathrm{diag}\bigg(\eye_{d_\balpha}, \frac{1}{\sqrt{n}}\eye_{d_\br}, \eye_{d_\bX}\bigg)\\
  &\quad\times\bJ_{0n}\bLambda_\sharp\sum_{t = 1}^m\sum_{i\neq j}\bigg\{\ell'({\theta}_{ij}^{(t)}(\tau); A_{ij}^{(t)})\frac{\partial^2\theta_{ij}^{(t)}}{\partial\bphi\partial\bphi\transpose}({\bphi}_\tau) - \ell'(\theta_{0ij}^{(t)}; A_{ij}^{(t)})\frac{\partial^2\theta_{ij}^{(t)}}{\partial\bphi\partial\bphi\transpose}(\bphi_0) \bigg\}\bLambda_\sharp\bJ_{0n}\transpose\bigg\|_\infty
   = \Optilde(\delta_n),
\end{align*}
where $\bJ_{0n}$ is defined in \eqref{eqn:J_n(B)}, $\delta_n = (m^2\sqrt{\log n} + m^{3/2}\log n)/\sqrt{n}$, $\bphi_\tau = \bphi_0 + \tau(\widehat{\bphi} - \bphi_0)$, and ${\theta}_{ij}^{(t)}(\tau) := \theta_{ij}^{(t)}(\bphi_\tau)$. 
\end{lemma}

\begin{proof}[\bf Proof]
We follow the notations in the proof of Lemma \ref{lemma:local_lipschitz_sample_Fisher_information_infinity_norm}. 
Since $\{|\ell'(\theta_{0ij}^{(t)}; A_{ij}^{(t)})|:j\in[n]\backslash\{i\}\}$
are independent sub-exponential random variables with
$\expect_0|\ell'(\theta_{0ij}^{(t)}; A_{ij}^{(t)})|\leq\sigma$ for each fixed
$t\in[m]$, $i\in[n]$ by Assumption \ref{assumption:true_parameter}, Lemma
\ref{lemma:matrix_Bernstein} and Lemma \ref{lemma:uniform_Optilde} imply that
\[
\max_{t\in[m],i\in[n]}\sum_{j\in[n]\backslash\{i\}}
|\ell'(\theta_{0ij}^{(t)}; A_{ij}^{(t)})| = \Optilde(n),
\]
and hence, by the Lipschitz continuity of $\ell'$ and Proposition
\ref{prop:stationarity_uniqueness_uniform_consistency_MLE},
\begin{align*}
\sup_{\tau\in[0, 1]}\max_{t\in[m],i\in[n]}\sum_{j\in[n]\backslash\{i\}}
|\ell'(\theta_{ij}^{(t)}(\tau); A_{ij}^{(t)})| = \Optilde(n).  
\end{align*}
These
two facts are used below whenever a sum weighted by
$\ell'(\theta_{0ij}^{(t)}; A_{ij}^{(t)})$ or
$\ell'(\theta_{ij}^{(t)}(\tau); A_{ij}^{(t)})$
is bounded.
By the calculations in Section \ref{sub:gradient_hessian}, 
\begin{align*}
&\mathrm{diag}\bigg(\eye_{d_\balpha}, \frac{1}{\sqrt{n}}\eye_{d_\br}, \eye_{d_\bX}\bigg)\bJ_{0n}\bLambda_\sharp\sum_{t = 1}^m\sum_{i\neq j}\bigg\{\ell'({\theta}_{ij}^{(t)}(\tau); A_{ij}^{(t)})\frac{\partial^2\theta_{ij}^{(t)}}{\partial\bphi\partial\bphi\transpose}({\bphi}_\tau)
-
\ell'(\theta_{0ij}^{(t)}; A_{ij}^{(t)})\frac{\partial^2\theta_{ij}^{(t)}}{\partial\bphi\partial\bphi\transpose}(\bphi_0)
\bigg\}\bLambda_\sharp\bJ_{0n}\transpose\\
&\quad = \mathrm{diag}\bigg(\eye_{d_\balpha}, \frac{1}{\sqrt{n}}\eye_{d_\br}, \eye_{d_\bX}\bigg)\bJ_{0n}\bJ_n(\tau\widehat{\bB})\transpose\sum_{t = 1}^m\sum_{i\neq j}
\widetilde{\bLambda}_\sharp
\Bigg\{
\ell'({\theta}_{ij}^{(t)}(\tau); A_{ij}^{(t)})\\
&\qquad\times \begin{bmatrix}
\mathbf{O} & \mathbf{O} & \mathbf{O}\\
\mathbf{O} & \mathbf{O} & \be_t\otimes\bD_d\transpose\{\eye_d\otimes({\bx}_{\tau i}\be_j\transpose + {\bx}_{\tau j}\be_i\transpose)\}\\
\mathbf{O} & \be_t\transpose\otimes\{\eye_d\otimes(\be_j{\bx}_{\tau i}\transpose + \be_i{\bx}_{\tau j}\transpose)\}\bD_d& \bR^{(t)}_{\tau}\otimes(\be_i\be_j\transpose + \be_j\be_i\transpose)
\end{bmatrix}\\
&\qquad - \ell'(\theta_{0ij}^{(t)}; A_{ij}^{(t)})\\
&\qquad\times \begin{bmatrix}
\mathbf{O} & \mathbf{O} & \mathbf{O}\\
\mathbf{O} & \mathbf{O} & \be_t\otimes\bD_d\transpose\{\eye_d\otimes({\bx}_{0i}\be_j\transpose + {\bx}_{0j}\be_i\transpose)\}\\
\mathbf{O} & \be_t\transpose\otimes\{\eye_d\otimes(\be_j{\bx}_{0i}\transpose + \be_i{\bx}_{0j}\transpose)\}\bD_d& \bR_0^{(t)}\otimes(\be_i\be_j\transpose + \be_j\be_i\transpose)
\end{bmatrix}\Bigg\}\\
&\qquad\times \widetilde{\bLambda}_\sharp\bJ_n(\tau\widehat{\bB})\bJ_{0n}\transpose\\
&\qquad + \mathrm{diag}\bigg(\eye_{d_\balpha}, \frac{1}{\sqrt{n}}\eye_{d_\br}, \eye_{d_\bX}\bigg)\bJ_{0n}\{\bJ_n(\tau\widehat{\bB}) - \bJ_{0n}\}\transpose\sum_{t = 1}^m\sum_{i\neq j}
\widetilde{\bLambda}_\sharp
\ell'(\theta_{0ij}^{(t)}; A_{ij}^{(t)})\\
&\qquad\times \begin{bmatrix}
\mathbf{O} & \mathbf{O} & \mathbf{O}\\
\mathbf{O} & \mathbf{O} & \be_t\otimes\bD_d\transpose\{\eye_d\otimes({\bx}_{0i}\be_j\transpose + {\bx}_{0j}\be_i\transpose)\}\\
\mathbf{O} & \be_t\transpose\otimes\{\eye_d\otimes(\be_j{\bx}_{0i}\transpose + \be_i{\bx}_{0j}\transpose)\}\bD_d& \bR_0^{(t)}\otimes(\be_i\be_j\transpose + \be_j\be_i\transpose) 
\end{bmatrix}\\
&\qquad\times \widetilde{\bLambda}_\sharp \bJ_n(\tau\widehat{\bB})\bJ_{0n}\transpose\\
&\qquad + \mathrm{diag}\bigg(\eye_{d_\balpha}, \frac{1}{\sqrt{n}}\eye_{d_\br}, \eye_{d_\bX}\bigg)\bJ_{0n} \bJ_{0n}\transpose\sum_{t = 1}^m\sum_{i\neq j}
\widetilde{\bLambda}_\sharp
\ell'(\theta_{0ij}^{(t)}; A_{ij}^{(t)})\\
&\qquad\times \begin{bmatrix}
\mathbf{O} & \mathbf{O} & \mathbf{O}\\
\mathbf{O} & \mathbf{O} & \be_t\otimes\bD_d\transpose\{\eye_d\otimes({\bx}_{0i}\be_j\transpose + {\bx}_{0j}\be_i\transpose)\}\\
\mathbf{O} & \be_t\transpose\otimes\{\eye_d\otimes(\be_j{\bx}_{0i}\transpose + \be_i{\bx}_{0j}\transpose)\}\bD_d& \bR_0^{(t)}\otimes(\be_i\be_j\transpose + \be_j\be_i\transpose)
\end{bmatrix}\\
&\qquad\times \widetilde{\bLambda}_\sharp\{\bJ_n(\tau\widehat{\bB}) - \bJ_{0n}\}\bJ_{0n}\transpose\\
&\qquad + \frac{2}{mn^2}\bJ_{0n}\sum_{t = 1}^m\sum_{i\neq j}\ell'({\theta}_{ij}^{(t)}(\tau); A_{ij}^{(t)})
\begin{bmatrix}
\mathbf{O} & \mathbf{O} & \mathbf{O}\\
\mathbf{O} & \mathbf{O} & \mathbf{O}\\
\mathbf{O} & \mathbf{O} & \sum_{k = 1}^d\frac{\partial^2x_{ik}(\tau\widehat{\bB})}{\partial\vect(\bB)\partial\vect(\bB)\transpose}\be_k\transpose\bR^{(t)}_{\tau}\bx_{\tau j}
\end{bmatrix}
\bJ_{0n}\transpose\\
&\qquad - \frac{2}{mn^2}\bJ_{0n}\sum_{t = 1}^m\sum_{i\neq j}\ell'(\theta_{0ij}^{(t)}; A_{ij}^{(t)})
\begin{bmatrix}
\mathbf{O} & \mathbf{O} & \mathbf{O}\\
\mathbf{O} & \mathbf{O} & \mathbf{O}\\
\mathbf{O} & \mathbf{O} & \sum_{k = 1}^d\frac{\partial^2x_{ik}(\mathbf{O})}{\partial\vect(\bB)\partial\vect(\bB)\transpose}\be_k\transpose\bR_0^{(t)}\bx_{0j}
\end{bmatrix}
\bJ_{0n}\transpose.
\end{align*}
The first term can be written as the sum of two terms: the first term is a sum over $t\in[m]$ and $i,j\in[n],i\neq j$, where each summand is $\{\ell'(\theta_{ij}^{(t)}(\tau); A_{ij}^{(t)}) - \ell'(\theta_{0ij}^{(t)}; A_{ij}^{(t)})\}$ (which is uniformly $\Optilde(\delta_n)$ over $t,i,j$ and $\tau$) times 
\[
\begin{bmatrix}
\mathbf{O} & \mathbf{O} & \mathbf{O}\\
\mathbf{O} & \mathbf{O} & \frac{\be_t\otimes\bD_d\transpose\{\eye_d\otimes({\bx}_{\tau i}\be_j\transpose + {\bx}_{\tau j}\be_i\transpose)\}}{\sqrt{m}n^2}\\
\mathbf{O} & \frac{\be_t\transpose\otimes\{\eye_d\otimes(\be_j{\bx}_{\tau i}\transpose + \be_i{\bx}_{\tau j}\transpose)\}\bD_d}{\sqrt{m}n^{3/2}}& \frac{\bR^{(t)}_{\tau}\otimes(\be_i\be_j\transpose + \be_j\be_i\transpose)}{mn}
\end{bmatrix},
\]
a matrix whose nonzero elements are uniformly $O(1)$ over $t,i,j$ and $\tau$; the second term is a sum over $t\in[m],i,j\in[n],i\neq j$ with the summand being $\ell'(\theta_{0ij}^{(t)}; A_{ij}^{(t)})$ times
\begin{align*}
&\begin{bmatrix}
\mathbf{O} & \mathbf{O} & \mathbf{O}\\
\mathbf{O} & \mathbf{O} & \frac{\be_t\otimes\bD_d\transpose\{\eye_d\otimes({\bx}_{\tau i}\be_j\transpose + {\bx}_{\tau j}\be_i\transpose)\}}{\sqrt{m}n^2}\\
\mathbf{O} & \frac{\be_t\transpose\otimes\{\eye_d\otimes(\be_j{\bx}_{\tau i}\transpose + \be_i{\bx}_{\tau j}\transpose)\}\bD_d}{\sqrt{m}n^{3/2}}& \frac{\bR^{(t)}_{\tau}\otimes(\be_i\be_j\transpose + \be_j\be_i\transpose)}{mn}
\end{bmatrix}  \\
&\quad - \begin{bmatrix}
\mathbf{O} & \mathbf{O} & \mathbf{O}\\
\mathbf{O} & \mathbf{O} & \frac{\be_t\otimes\bD_d\transpose\{\eye_d\otimes({\bx}_{0i}\be_j\transpose + {\bx}_{0j}\be_i\transpose)\}}{\sqrt{m}n^2}\\
\mathbf{O} & \frac{\be_t\transpose\otimes\{\eye_d\otimes(\be_j{\bx}_{0i}\transpose + \be_i{\bx}_{0j}\transpose)\}\bD_d}{\sqrt{m}n^{3/2}}& \frac{\bR_0^{(t)}\otimes(\be_i\be_j\transpose + \be_j\be_i\transpose) }{mn}
\end{bmatrix}
\end{align*}
Then, the matrix infinity norm of the first term can be shown to be $\Optilde(\delta_n)$ by a standard rowwise $\ell_\infty$-norm analysis. 
The matrix infinity norms of the second and the third terms can be directly bounded by 
\[
\sup_{\tau\in[0, 1]}\|\bJ_{0n}\{\bJ_n(\tau\widehat{\bB}) - \bJ_{0n}\}\transpose\|_\infty = \Optilde(\delta_n),\quad 
\sup_{\tau\in[0, 1]}\|\{\bJ_n(\tau\widehat{\bB}) - \bJ_{0n}\}\bJ_{0n}\transpose\|_\infty
= O(\|\widehat\bB\|_2) = \Optilde(n^{-1/2})
\]
times 
\begin{align*}
\left\|\sum_{t = 1}^m\sum_{i\neq j}
\ell'(\theta_{0ij}^{(t)}; A_{ij}^{(t)})\begin{bmatrix}
\mathbf{O} & \mathbf{O} & \mathbf{O}\\
\mathbf{O} & \mathbf{O} & \frac{\be_t\otimes\bD_d\transpose\{\eye_d\otimes({\bx}_{0i}\be_j\transpose + {\bx}_{0j}\be_i\transpose)\}}{\sqrt{m}n^2}\\
\mathbf{O} & \frac{\be_t\transpose\otimes\{\eye_d\otimes(\be_j{\bx}_{0i}\transpose + \be_i{\bx}_{0j}\transpose)\}\bD_d}{\sqrt{m}n^{3/2}}& \frac{\bR_0^{(t)}\otimes(\be_i\be_j\transpose + \be_j\be_i\transpose)}{mn}
\end{bmatrix}\right\|_\infty,
\end{align*}
the latter of which can be shown as $\Optilde(1)$ using $\max_{t\in[m],i\in[n]}\sum_{j\in[n]\backslash\{i\}}|\ell'(\theta_{0ij}^{(t)}; A_{ij}^{(t)})| = \Optilde(n)$. It is now sufficient to focus on the last term. Equivalently, it is sufficient to show that
\begin{align*}
&\sup_\tau\bigg\|\frac{1}{mn^2}\sum_{t = 1}^m\sum_{i\neq j}\bigg(\eye_d\otimes\frac{\bX_{0\perp}}{\sqrt{n}}\bigg)\\
&\quad\times\sum_{k = 1}^d\bigg\{\ell'(\theta_{ij}^{(t)}(\tau); A_{ij}^{(t)})\frac{\partial^2x_{ik}(\tau\widehat{\bB})}{\partial\vect(\bB)\partial\vect(\bB)\transpose}\be_k\transpose\bR^{(t)}_{\tau}\bx_{\tau j} - \ell'(\theta_{0ij}^{(t)};A_{ij}^{(t)})\frac{\partial^2x_{ik}(\mathbf{O})}{\partial\vect(\bB)\partial\vect(\bB)\transpose}\be_k\transpose\bR_0^{(t)}\bx_{0j}\bigg\}
\\
&\quad\times
\bigg(\eye_d\otimes\frac{\bX_{0\perp}\transpose}{\sqrt{n}}\bigg)\bigg\|_\infty = \Optilde(\delta_n). 
\end{align*}
Recall that the inverse Lyapunov operator $\mathscr{L}^{-1}$ is linear and satisfies $\mathscr{L}^{-1}(\bM) = \int_0^\infty e^{-u\bS_\tau^{1/2}}\bM e^{-u\bS_\tau^{1/2}}\mathrm{d}u$ for any symmetric $\bM\in\mathbb{R}^{d\times d}$ by Lemma \ref{lemma:inverse_Lyapunov_operator_integral}, where we denote ${\bS}_\tau = \eye_d + \tau^2\widehat{\bB}\transpose\widehat{\bB}$. Alternatively, one also has $\vect(\mathscr{L}^{-1}(\bM)) = (\bS^{1/2}_\tau\otimes \eye_d + \eye_d \otimes {\bS}_\tau^{1/2})^{-1}\vect(\bM)$. For notational simplicity, we hide the dependency of $\bR^{(t)}_{\tau}$ and $\bx_{\tau i}$ on $\tau$. Therefore, by the calculations in Section \ref{sub:gradient_hessian}, we have
\begin{align*}
&(\eye_d\otimes \bX_{0\perp})\sum_{k = 1}^d\frac{\partial^2x_{ik}(\tau\widehat{\bB})}{\partial \vect(\bB)\partial \vect(\bB)\transpose}\be_k\transpose{\bR}^{(t)}{\bx}_j(\eye_d\otimes \bX_{0\perp}\transpose)\\
&\quad = -(\eye_d\otimes \bX_{0\perp}\bX_{0\perp}\transpose\be_i)({\bx}_j\transpose{\bR}^{(t)}{\bS}_\tau^{-1/2}\otimes {\bS}_\tau^{-1/2})({\bS}_\tau^{1/2}\otimes\eye_d + \eye_d\otimes {\bS}_\tau^{1/2})^{-1}
(\eye_{d^2} + \bK_{dd})(\eye_d\otimes \tau\widehat{\bB}\transpose\bX_{0\perp}\transpose)\\
&\qquad - (\eye_d\otimes\tau\bX_{0\perp}\widehat\bB)(\eye_{d^2} + \bK_{dd})({\bS}_\tau^{1/2}\otimes\eye_d + \eye_d\otimes {\bS}_\tau^{1/2})^{-1}
({\bS}_\tau^{-1/2}{\bR}^{(t)}{\bx}_j\otimes {\bS}_\tau^{-1/2})(\eye_d\otimes \be_i\transpose\bX_{0\perp}\bX_{0\perp}\transpose)\\
&\qquad + (\eye_d\otimes\tau\bX_{0\perp}\widehat\bB)(\eye_{d^2} + \bK_{dd})({\bS}_\tau^{1/2}\otimes\eye_d + \eye_d\otimes {\bS}_\tau^{1/2})^{-1}
({\bx}_i{\bx}_j\transpose{\bR}^{(t)}{\bS}_\tau^{-1/2}\otimes {\bS}_\tau^{-1/2})\\
&\qquad\times ({\bS}_\tau^{1/2}\otimes\eye_d + \eye_d\otimes {\bS}_\tau^{1/2})^{-1}(\eye_{d^2} + \bK_{dd})(\eye_d\otimes \tau\widehat\bB\transpose\bX_{0\perp}\transpose)\\
&\qquad + (\eye_d\otimes\tau\bX_{0\perp}\widehat\bB)(\eye_{d^2} + \bK_{dd})({\bS}_\tau^{1/2}\otimes\eye_d + \eye_d\otimes {\bS}_\tau^{1/2})^{-1}({\bS}_\tau^{-1/2}{\bR}^{(t)}{\bx}_j{\bx}_i\transpose\otimes {\bS}_\tau^{-1/2})\\
&\qquad\times ({\bS}_\tau^{1/2}\otimes\eye_d + \eye_d\otimes{\bS}_\tau^{1/2})^{-1}(\eye_{d^2} + \bK_{dd})(\eye_d\otimes \tau\widehat\bB\transpose\bX_{0\perp}\transpose) \\
&\qquad - \int_0^\infty \bigg\{e^{-u{\bS}_\tau^{1/2}}({\bx}_i{\bx}_j\transpose{\bR}^{(t)}{\bS}_\tau^{-1/2} + {\bS}_\tau^{-1/2}{\bR}^{(t)}{\bx}_j{\bx}_i\transpose)e^{-u{\bS}_\tau^{1/2}}\otimes \bX_{0\perp}\bX_{0\perp}\transpose\bigg\}\mathrm{d}u \\
&\qquad + (\eye_d \otimes \tau\bX_{0\perp}\widehat\bB)(\eye_{d^2} + \bK_{dd})({\bS}_\tau^{1/2}\otimes \eye_d + \eye_d \otimes {\bS}_\tau^{1/2})^{-1}\\
&\qquad\times 
\int_0^\infty \bigg\{e^{-u{\bS}_\tau^{1/2}}({\bx}_i{\bx}_j\transpose{\bR}^{(t)}{\bS}_\tau^{-1/2} + {\bS}_\tau^{-1/2}{\bR}^{(t)}{\bx}_j{\bx}_i\transpose) e^{-u{\bS}_\tau^{1/2}}\otimes \eye_d\bigg\}\mathrm{d}u\\
&\qquad\times ({\bS}_\tau^{1/2}\otimes \eye_d + \eye_d \otimes {\bS}_\tau^{1/2})^{-1}
(\eye_{d^2} + \bK_{dd})(\eye_d\otimes \tau\widehat\bB\transpose\bX_{0\perp}\transpose).
\end{align*}
On the right-hand side above, only the fifth term is the dominating term that is close to the target and the remaining five terms are negligible. 
Since Proposition \ref{prop:stationarity_uniqueness_uniform_consistency_MLE} and the Lipschitz continuity of $\ell'(\cdot; A_{ij}^{(t)})$ imply that $\|\bX_{0\perp}\widehat{\bB}\|_{\max} = \Optilde(\delta_n)$ and 
\begin{align*}
\sup_{\tau\in[0, 1]}\max_{t\in[m],i,j\in[n],i\neq j}(\|{\bx}_{\tau i} - \bx_{0i}\|_2 + \|{\bR}^{(t)}_{\tau}{\bx}_{\tau j} - \bR_0^{(t)}\bx_{0j}\|_2 + |\ell'(\theta_{ij}^{(t)}(\tau); A_{ij}^{(t)}) - \ell'(\theta_{0ij}^{(t)}; A_{ij}^{(t)})|) = \Optilde(\delta_n),  
\end{align*}
it follows that
\begin{align*}
&\sup_\tau\bigg\|\frac{1}{mn^2}\sum_{t = 1}^m\sum_{i\neq j}\bigg(\eye_d\otimes\frac{\bX_{0\perp}}{\sqrt{n}}\bigg)\\
&\quad\times\sum_{k = 1}^d\bigg\{\ell'(\theta_{ij}^{(t)}(\tau); A_{ij}^{(t)})\frac{\partial^2x_{ik}(\tau\widehat{\bB})}{\partial\vect(\bB)\partial\vect(\bB)\transpose}\be_k\transpose\bR^{(t)}_{\tau}\bx_{\tau j} - \ell'(\theta_{0ij}^{(t)};A_{ij}^{(t)})\frac{\partial^2x_{ik}(\mathbf{O})}{\partial\vect(\bB)\partial\vect(\bB)\transpose}\be_k\transpose\bR_0^{(t)}\bx_{0j}\bigg\}
\\
&\quad\times
\bigg(\eye_d\otimes\frac{\bX_{0\perp}\transpose}{\sqrt{n}}\bigg)\bigg\|_\infty\\
&\quad\leq \sup_\tau\bigg\{\frac{C}{n}\bigg\|\eye_d\otimes \frac{\bX_{0\perp}\bX_{0\perp}\transpose}{n}\bigg\|_\infty \|\eye_d\otimes\tau(\bX_{0\perp}\widehat{\bB})\transpose\|_\infty + C\|\eye_d\otimes\tau\bX_{0\perp}\widehat{\bB}\|_\infty \bigg\|\eye_d\otimes \frac{\bX_{0\perp}\bX_{0\perp}\transpose}{n}\bigg\|_\infty\bigg\}\\
&\qquad  + \sup_\tau\bigg\{\frac{C}{n}\|\eye_d\otimes\tau(\bX_{0\perp}\widehat{\bB})\|_\infty\|\eye_d\otimes(\tau\bX_{0\perp}\widehat{\bB})\transpose\|_\infty\bigg\}\\
&\qquad + \frac{C}{n}\|\bX_{0\perp}\bX_{0\perp}\transpose\|_\infty \\
&\qquad \times \sup_\tau\frac{1}{mn^2}\sum_{t = 1}^m\sum_{i\neq j}\int_0^\infty\bigg\|e^{-u{\bS}_\tau^{1/2}}({\bx}_{\tau i}{\bx}_{\tau j}\transpose{\bR}^{(t)}_\tau{\bS}_\tau^{-1/2} + {\bS}^{-1/2}_\tau{\bR}^{(t)}_\tau{\bx}_{\tau j}{\bx}_{\tau i}\transpose)e^{-u{\bS}_\tau^{1/2}}\ell'(\theta_{ij}^{(t)}(\tau); A_{ij}^{(t)})\\
&\qquad - e^{-u\eye_d}(\bx_{0i}\bx_{0j}\transpose\bR_0^{(t)} + \bR_0^{(t)}\bx_{0j}\bx_{0i}\transpose)e^{-u\eye_d}\ell'(\theta_{0ij}^{(t)}; A_{ij}^{(t)})\bigg\|\mathrm{d}u  = \Optilde(\delta_n).
\end{align*}
The proof is thus completed. 
\end{proof}

\subsection{Additional remainder analysis}
\label{sub:additional_remainder_analysis}

This subsection provides additional analysis that completes the proof of the asymptotic normality of the MLE. 

\begin{lemma}
\label{lemma:fisher_information_at_truth_concentration_infintiy_norm}
Suppose Assumptions \ref{assumption:edge_distribution}--\ref{assumption:true_parameter} hold and $m = O(n^2)$. Denote $\bG_0^\sharp = \bLambda_\sharp\bG(\bphi_0)\bLambda_\sharp$. Then, 
\begin{align*}
&\left\|\mathrm{diag}\bigg(\eye_{d_\balpha}, \frac{1}{\sqrt{n}}\eye_{d_\br}, \eye_{d_\bX}\bigg)\bJ_{0n}\bLambda_\sharp\{\bGamma(\bphi_0) + \bG(\bphi_0)\}\bLambda_\sharp\bJ_{0n}\transpose\{\bJ_{0n}(\bG_0^\sharp)^{-1}\bJ_{0n}\transpose\}\bigg\{\bJ_{0n}\bLambda_\sharp\frac{\partial\calL}{\partial\bphi}(\bphi_0)\bigg\}
\right\|_\infty\\
&\quad = \Optilde\bigg(\frac{\sqrt{m}\log n}{\sqrt{n}}\bigg)
\end{align*}
\end{lemma}

\begin{proof}[\bf Proof]
For convenience, we denote $\ell''_{0tij} = \ell''(\theta_{0ij}^{(t)}; A_{ij}^{(t)})$ and 
\[
\bu_{tij} = \begin{bmatrix}
  \frac{1}{\sqrt{n}}\be_t\otimes (\be_i + \be_j)\\
  \frac{1}{n}\be_t\otimes\bD_d\transpose(\bx_{0j}\otimes\bx_{0i})\\ 
  \frac{1}{\sqrt{mn}}(\bR_0^{(t)}\bx_{0j}\otimes\be_i + \bR_0^{(t)}\bx_{0i}\otimes \be_j)
\end{bmatrix}.
\]
This allows us to write 
\[
\bJ_{0n}\bLambda_\sharp\frac{\partial}{\partial\bphi}\ell(\theta_{0ij}^{(t)}; A_{ij}^{(t)}) = \bJ_{0n}\bJ_{0n}\transpose\ell'(\theta_{0ij}^{(t)}; A_{ij}^{(t)})\bu_{tij}.
\] 
By Assumption \ref{assumption:edge_distribution} and
Assumption \ref{assumption:true_parameter},
$\bG(\bphi_0) = \expect_0\{\nabla\calL(\bphi_0)\nabla\calL(\bphi_0)\transpose\}
= -\expect_0\nabla^2\calL(\bphi_0) = -\expect_0\bGamma(\bphi_0)$, so that
\[
\bJ_{0n}\bLambda_\sharp\{\bGamma(\bphi_0) + \bG(\bphi_0)\}\bLambda_\sharp
\bJ_{0n}\transpose
= \bJ_{0n}\bJ_{0n}\transpose\bigg\{\frac{1}{2}\sum_{t = 1}^m\sum_{i\neq j}
(\ell''_{0tij} - \expect_0\ell''_{0tij})\bu_{tij}\bu_{tij}\transpose\bigg\}
\bJ_{0n}\bJ_{0n}\transpose.
\]
Also, by Proposition \ref{prop:expected_hessian_bound}, we have
\begin{align*}
\max_{t\in[m],i\in[n]}\sum_{j\in[n]\backslash\{i\}}\|\bu_{tij}\|_2^2 = O(1),\quad \bigg\|\sum_{s,a,b}\bu_{sab}\bu_{sab}\transpose\bigg\|_2 = O(1),\quad 
\max_{i\in[n]}\sum_{t = 1}^m\sum_{j\in[n]\backslash\{i\}}\|\bu_{tij}\|_2^2 = O(m).
\end{align*}
It is sufficient to show that
\begin{align*}
&\max_{t\in[m],i\in[n]}\left|
  \sum_{j\in[n]\backslash\{i\}}\frac{(\ell''_{0tij} - \expect_0\ell''_{0tij})}{\sqrt{n}}\bu_{tij}\transpose \bJ_{0n}(\bG_0^\sharp)^{-1}\bJ_{0n}\transpose \bJ_{0n}\bLambda_\sharp\frac{\partial\calL}{\partial\bphi}(\bphi_0)
\right| = \Optilde\bigg(\frac{ \log n }{\sqrt{n}}\bigg),\\
&\max_{k\in[d],i\in[n]}\left|
  \sum_{t = 1}^m\sum_{j\in[n]\backslash\{i\}}\frac{(\ell''_{0tij} - \expect_0\ell''_{0tij})}{\sqrt{mn}}\be_k\transpose\bR_0^{(t)}\bx_{0j}\bu_{tij}\transpose \bJ_{0n}(\bG_0^\sharp)^{-1}\bJ_{0n}\transpose \bJ_{0n}\bLambda_\sharp\frac{\partial\calL}{\partial\bphi}(\bphi_0)
\right| = \Optilde\bigg(\frac{\sqrt{m}\log n}{\sqrt{n}}\bigg),\\
&\max_{t\in[m],k\in[d(d + 1)/2]}\left|
  \sum_{i\neq j}\frac{(\ell''_{0tij} - \expect_0\ell''_{0tij})}{n^{3/2}}\be_k\transpose\bD_d\transpose(\bx_{0j}\otimes\bx_{0i})\bu_{tij}\transpose \bJ_{0n}(\bG_0^\sharp)^{-1}\bJ_{0n}\transpose \bJ_{0n}\bLambda_\sharp\frac{\partial\calL}{\partial\bphi}(\bphi_0)
\right| = \Optilde\bigg(\frac{\sqrt{m}\log n}{\sqrt{n}}\bigg).
\end{align*} 
By matrix Hoeffding's inequality and Lemma \ref{lemma:uniform_Optilde}, we have
\begin{align*}
  \max_{s,t\in[m],i\in[n],a,b\in[n]}\bigg|\sum_{j\in[n]\backslash\{i\}}\frac{(\ell_{0tij}'' - \expect_0\ell_{0tij}'')}{\sqrt{n}}\bu_{tij}\transpose\bJ_{0n}(\bG_0^\sharp)^{-1}\bJ_{0n}\transpose\bu_{sab}\bigg|& = \Optilde\bigg(\frac{\sqrt{\log n}}{n}\bigg)
\end{align*}
and
\begin{align*}
&  \max_{t\in[m],i\in[n]}\bigg\{\sum_{s\in[m],a,b\in[n],a\neq b}\bigg(\bu_{sab}\transpose\bJ_{0n}(\bG_0^\sharp)^{-1}\bJ_{0n}\transpose\sum_{j\in[n]\backslash\{i\}}\frac{(\ell''_{0tij} - \expect_0\ell''_{0tij})}{\sqrt{n}}\bu_{tij}\bigg)^2\bigg\}^{1/2}\\
&\quad\leq \max_{t\in[m],i\in[n]}\bigg\{\bigg\|\sum_{s\in[m],a,b\in[n],a\neq b}\bu_{sab}\bu_{sab}\transpose\bigg\|_2\|\bJ_{0n}(\bG_0^\sharp)^{-1}\bJ_{0n}\transpose\|_2^2\bigg\|\sum_{j\in[n]\backslash\{i\}}\frac{(\ell''_{0tij} - \expect_0\ell''_{0tij})}{\sqrt{n}}\bu_{tij}\bigg\|_2^2\bigg\}^{1/2} \\ 
&\quad = \Optilde\bigg(\sqrt{\frac{\log n}{n}}\bigg).
\end{align*}
For the first term, we apply a leave-one-out trick \cite{Abbe_Fan_Wang_Zhong_2020_AoS_Entrywise}, Bernstein's inequality, Hoeffding's inequality, a conditioning argument via Lemma \ref{lemma:conditioning_argument}, and a union bound to obtain
\begin{align*}
&\max_{t\in[m],i\in[n]}\left|
  \sum_{j\in[n]\backslash\{i\}}\frac{(\ell''_{0tij} - \expect_0\ell''_{0tij})}{\sqrt{n}}\bu_{tij}\transpose \bJ_{0n}(\bG_0^\sharp)^{-1}\bJ_{0n}\transpose \bJ_{0n}\bLambda_\sharp\frac{\partial\calL}{\partial\bphi}(\bphi_0)
\right|\\
&\quad\leq \max_{t,i}
\left|
  \sum_{j\in[n]\backslash\{i\}}
  \frac{(\ell''_{0tij} - \expect_0\ell''_{0tij})}{\sqrt{n}}\bu_{tij}\transpose \bJ_{0n}(\bG_0^\sharp)^{-1}\bJ_{0n}\transpose \sum_{l\in[n]\backslash\{i\}}\ell'(\theta_{0il}^{(t)}; A_{il}^{(t)})\bu_{til}
\right|\\
&\qquad + \max_{t,i}
\left|
  \sum_{j\in[n]\backslash\{i\}}
  \frac{(\ell''_{0tij} - \expect_0\ell''_{0tij})}{\sqrt{n}}
  \bu_{tij}\transpose \bJ_{0n}(\bG_0^\sharp)^{-1}\bJ_{0n}\transpose\right.\\
&\qquad\quad\times \left. \sum_{s\in[m],a,b\in[n],a\neq b}\ell'(\theta_{0ab}^{(s)}; A_{ab}^{(s)})\bu_{sab}\{1 - \mathbbm{1}(s = t,a = i \text{ or } b = i)\}
\right|\\
&\quad \leq \Optilde\bigg\{\sqrt{\frac{\log n}{n}}\max_{t,i}\bigg(\sum_{j\in[n]\backslash\{i\}}\|\bu_{tij}\|_2^2\bigg)^{1/2}\bigg\}\times \|\bJ_{0n}(\bG_0^\sharp)^{-1}\bJ_{0n}\transpose\|_2\\
&\qquad\times \Optilde\bigg\{
\max_{t,i,l}\|\bu_{til}\|_2\log n + \max_{t,i}\bigg(\sum_{l\in[n]\backslash\{i\}}\|\bu_{til}\|_2^2\bigg)^{1/2}\sqrt{\log n}
\bigg\}+
\Optilde\bigg\{\frac{(\log n)^{3/2}}{n} + \frac{\log n}{\sqrt{n}}\bigg\}\\
&\quad  = \Optilde\bigg(\frac{\log n}{\sqrt{n}}\bigg).
\end{align*}
Similarly, 
\begin{align*}
& \max_{k\in[d],i\in[n],s\in[m],a,b\in[n]}\bigg|
 \sum_{t = 1}^m\sum_{j\in[n]\backslash\{i\}}\frac{(\ell''_{0tij} - \expect_0\ell''_{0tij})}{\sqrt{mn}}\be_k\transpose\bR_0^{(t)}\bx_{0j}\bu_{tij}\transpose \bJ_{0n}(\bG_0^\sharp)^{-1}\bJ_{0n}\transpose\bu_{sab}\bigg|\\
&\quad = \Optilde\bigg(\frac{\sqrt{\log n}}{n}\bigg),\\
& \max_{i\in[n],k\in[d]}\bigg\{\sum_{s\in[m],a,b\in[n]}\bigg(
\sum_{t = 1}^m\sum_{j\in[n]\backslash\{i\}}\frac{(\ell''_{0tij} - \expect_0\ell''_{0tij})}{\sqrt{mn}}\be_k\transpose\bR_0^{(t)}\bx_{0j}\bu_{tij}\transpose \bJ_{0n}(\bG_0^\sharp)^{-1}\bJ_{0n}\transpose\bu_{sab}
\bigg)^2\bigg\}^{1/2}\\
&\quad = \Optilde\bigg(\sqrt{\frac{\log n}{n}}\bigg),\\
&\max_{t,s\in[m],k\in[d(d + 1)/2],a,b\in[n]}\bigg|\sum_{i\neq j}\frac{(\ell_{0tij}'' - \expect_0\ell''_{0tij})}{n^{3/2}}\be_k\transpose\bD_d\transpose(\bx_{0j}\otimes\bx_{0i})\bu_{tij}\transpose\bJ_{0n}(\bG_0^\sharp)^{-1}\bJ_{0n}\transpose\bu_{sab}\bigg|\\
&\quad = \Optilde\bigg(\frac{\sqrt{\log n}}{n^{3/2}}\bigg),\\
&\max_{t\in[m],k\in[d(d + 1)/2]}\bigg\{\sum_{s\in[m],a,b\in[n]}\bigg(\sum_{i\neq j}\frac{(\ell''_{0tij} - \expect_0\ell''_{0tij})}{n^{3/2}}\be_k\transpose\bD_d\transpose(\bx_{0j}\otimes\bx_{0i})\bu_{tij}\transpose\bJ_{0n}(\bG_0^\sharp)^{-1}\bJ_{0n}\transpose\bu_{sab}\bigg)^2\bigg\}^{1/2}\\
&\quad = \Optilde\bigg(\frac{\sqrt{\log n}}{n}\bigg).
\end{align*}
Following the same reasoning, we can work with the second term
\begin{align*}
&\max_{k\in[d],i\in[n]}\left|
  \sum_{t = 1}^m\sum_{j\in[n]\backslash\{i\}}\frac{(\ell''_{0tij} - \expect_0\ell''_{0tij})}{\sqrt{mn}}\be_k\transpose\bR_0^{(t)}\bx_{0j}\bu_{tij}\transpose \bJ_{0n}(\bG_0^\sharp)^{-1}\bJ_{0n}\transpose \bJ_{0n}\bLambda_\sharp\frac{\partial\calL}{\partial\bphi}(\bphi_0)
\right|\\
&\quad\leq \max_{k,i}\
\bigg\|
  \sum_{t = 1}^m\sum_{j\in[n]\backslash\{i\}}\frac{(\ell''_{0tij} - \expect_0\ell''_{0tij})}{\sqrt{mn}}\be_k\transpose\bR_0^{(t)}\bx_{0j}\bu_{tij}\transpose\bigg\|_2\|\bJ_{0n}(\bG_0^\sharp)^{-1}\bJ_{0n}\transpose\|_2\bigg\| \sum_{s = 1}^m\sum_{l\in[n]\backslash\{i\}}\ell'(\theta_{0il}^{(s)}; A_{il}^{(s)})\bu_{sil}\bigg\|_2\\
&\qquad + \max_{k,i}\left|
  \sum_{t = 1}^m\sum_{j\in[n]\backslash\{i\}}
  \frac{(\ell''_{0tij} - \expect_0\ell''_{0tij})}{\sqrt{mn}}
  \be_k\transpose\bR_0^{(t)}\bx_{0j}\bu_{tij}\transpose \bJ_{0n}(\bG_0^\sharp)^{-1}\bJ_{0n}\transpose\right.\\
  &\qquad\times \left.\sum_{s = 1}^m\sum_{a \neq b}\ell'(\theta_{0ab}^{(s)}; A_{ab}^{(s)})\bu_{sab}\{1 - \mathbbm{1}(a = i\text{ or }b = i)\}
\right|\\
&\quad\leq \Optilde\bigg(\sqrt{\frac{\log n}{n}} \times \sqrt{m\log n}\bigg) + \Optilde\bigg\{\frac{(\log n)^{3/2}}{n} + \frac{\log n}{\sqrt{n}}
 \bigg\} = \Optilde\bigg(\frac{\sqrt{m}\log n}{\sqrt{n}}\bigg)
\end{align*}
and
\begin{align*}
  &\max_{t\in[m],k\in[d(d + 1)/2]}\left|
  \sum_{i\neq j}\frac{(\ell''_{0tij} - \expect_0\ell''_{0tij})}{n^{3/2}}\be_k\transpose\bD_d\transpose(\bx_{0j}\otimes\bx_{0i})\bu_{tij}\transpose \bJ_{0n}(\bG_0^\sharp)^{-1}\bJ_{0n}\transpose \bJ_{0n}\bLambda_\sharp\frac{\partial\calL}{\partial\bphi}(\bphi_0)
\right|\\
&\quad\leq 
\max_{t,k}\bigg\|
\sum_{i\neq j}\frac{(\ell''_{0tij} - \expect_0\ell''_{0tij})}{n^{3/2}}\be_k\transpose\bD_d\transpose(\bx_{0j}\otimes\bx_{0i})\bu_{tij}\transpose\bigg\|_2 \|\bJ_{0n}(\bG_0^\sharp)^{-1}\bJ_{0n}\transpose\|_2 
\bigg\|
  \sum_{i\neq j}\ell'(\theta_{0ij}^{(t)}; A_{ij}^{(t)})\bu_{tij}
\bigg\|_2\\
&\qquad + 
\max_{t,k}\left|
\sum_{i\neq j}\frac{(\ell''_{0tij} - \expect_0\ell''_{0tij})}{n^{3/2}}\be_k\transpose\bD_d\transpose(\bx_{0j}\otimes\bx_{0i})\bu_{tij}\transpose \bJ_{0n}(\bG_0^\sharp)^{-1}\bJ_{0n}\transpose 
\sum_{s\in[m]\backslash\{t\}}  \sum_{a\neq b}\ell'(\theta_{0ab}^{(s)}; A_{ab}^{(s)})\bu_{sab}
\right|\\
&\quad\leq \Optilde\bigg(\frac{n\log n}{n^{3/2}}\bigg) + \Optilde\bigg\{
\frac{(\log n)^{3/2}}{n^{3/2}} + \frac{\log n}{n}\bigg\} = \Optilde\bigg(\frac{\log n}{\sqrt{n}}\bigg).
\end{align*}
The proof is completed by combining the above three results. 
\end{proof}

\begin{lemma}
\label{lemma:score_term_hessian_at_truth_concentration_infintiy_norm}
Suppose Assumptions \ref{assumption:edge_distribution}--\ref{assumption:true_parameter} hold and $m = O(n^2)$. Denote $\bG_0^\sharp = \bLambda_\sharp\bG(\bphi_0)\bLambda_\sharp$. Then, 
\begin{align*}
&\left\|\mathrm{diag}\bigg(\eye_{d_\balpha}, \frac{\eye_{d_\br}}{\sqrt{n}}, \eye_{d_\bX}\bigg)\bigg\{\bJ_{0n}
\bLambda_\sharp\sum_{t = 1}^m\sum_{i\neq j}\ell'(\theta_{0ij}^{(t)}; A_{ij}^{(t)})\frac{\partial^2\theta_{ij}^{(t)}}{\partial\bphi\partial\bphi\transpose}(\bphi_0)\bLambda_\sharp
\bJ_{0n}\transpose\bigg\}\{\bJ_{0n}(\bG_0^\sharp)^{-1}\bJ_{0n}\transpose\}\bigg\{\bJ_{0n}\bLambda_\sharp\frac{\partial\calL}{\partial\bphi}(\bphi_0)\bigg\}
\right\|_\infty
\\
&\quad = \Optilde\bigg(\frac{ \log n}{\sqrt{n}}\bigg).
\end{align*}
\end{lemma}

\begin{proof}[\bf Proof]
The proof is similar to that of Lemma \ref{lemma:fisher_information_at_truth_concentration_infintiy_norm}.  For convenience, we denote $\ell_{0tij}' = \ell'(\theta_{0ij}^{(t)}; A_{ij}^{(t)})$,
\[
\bu_{tij} = \begin{bmatrix}
  \frac{1}{\sqrt{n}}\be_t\otimes (\be_i + \be_j)\\
  \frac{1}{n}\be_t\otimes\bD_d\transpose(\bx_{0j}\otimes\bx_{0i})\\ 
  \frac{1}{\sqrt{mn}}(\bR_0^{(t)}\bx_{0j}\otimes\be_i + \bR_0^{(t)}\bx_{0i}\otimes \be_j)
\end{bmatrix},
\]
and
\begin{align*}
\bv_{k,tj}\transpose &= \begin{bmatrix}
  \zero_{d_\balpha}\transpose  & 
  \frac{1}{\sqrt{m}n^{3/2}}\be_t\transpose\otimes(\be_k\transpose\otimes\bx_{0j}\transpose)(\eye_{d^2} + \bK_{dd})\bD_d
  & 
  \frac{2}{mn}\be_k\transpose\bR_0^{(t)}\otimes\be_j\transpose
\end{bmatrix},\\
\bw_{k,ij}\transpose &= \begin{bmatrix}
  \zero_{d_\balpha}\transpose  & 
  \zero_{d_\br}\transpose & 
  \frac{1}{\sqrt{m}n^2}\be_k\transpose\bD_d\transpose(\eye_{d^2} + \bK_{dd})(\eye_d\otimes\bx_{0i}\be_j\transpose)
\end{bmatrix},\\
\by_{kl,tij}\transpose& = \begin{bmatrix}
  \zero_{d_\balpha}\transpose & 
  \zero_{d_\br}\transpose & 
  -\frac{2}{mn^2}\int_0^\infty e^{-\tau\eye_d}\be_k\transpose(\bx_{0i}\bx_{0j}\transpose\bR_0^{(t)} + \bR_0^{(t)}\bx_{0j}\bx_{0i}\transpose)e^{-\tau\eye_d}\mathrm{d}\tau \otimes \frac{1}{n}\be_l\transpose\bX_{0\perp}\bX_{0\perp}\transpose 
\end{bmatrix}.
\end{align*}
It is sufficient to show that
\begin{align*}
&\max_{k\in[d],i\in[n]}\bigg|
\sum_{t = 1}^m\sum_{j\in[n]\backslash\{i\}}\ell'_{0tij}\bv_{k,tj}\transpose \{\bJ_{0n}(\bG_0^\sharp)^{-1}\bJ_{0n}\transpose\} \bigg\{\bJ_{0n}\bLambda_\sharp\frac{\partial\calL}{\partial\bphi}(\bphi_0)\bigg\}
\bigg| = \Optilde\bigg(\frac{\log n}{\sqrt{n}}\bigg),\\
&\max_{k\in[d(d + 1)/2],t\in[m]}\bigg|
\sum_{i\neq j}\ell'_{0tij}\bw_{k,ij}\transpose \{\bJ_{0n}(\bG_0^\sharp)^{-1}\bJ_{0n}\transpose\}  \bigg\{\bJ_{0n}\bLambda_\sharp\frac{\partial\calL}{\partial\bphi}(\bphi_0)\bigg\}
\bigg| = \Optilde\bigg(\frac{\log n}{\sqrt{mn}}\bigg),\\
&\max_{k\in[d],l\in[n]}\bigg|
\sum_{t = 1}^m\sum_{i\neq j}\ell'_{0tij}\by_{kl,tij}\transpose \{\bJ_{0n}(\bG_0^\sharp)^{-1}\bJ_{0n}\transpose\}  \bigg\{\bJ_{0n}\bLambda_\sharp\frac{\partial\calL}{\partial\bphi}(\bphi_0)\bigg\}
\bigg| = \Optilde\bigg(\frac{\log n}{\sqrt{n}}\bigg).
\end{align*}
By Bernstein's inequality and union bounds, 
\begin{align*}
\max_{k\in[d],s\in[m],i,a,b\in[n]}\bigg|
\sum_{t = 1}^m\sum_{j\in[n]\backslash\{i\}}\ell_{0tij}'\bv_{k,tj}\transpose\bJ_{0n}(\bG_0^\sharp)^{-1}\bJ_{0n}\transpose\bu_{sab}
\bigg|& = \Optilde\bigg(\frac{\sqrt{\log n}}{\sqrt{m}n}\bigg),\\
\max_{k\in[d],i\in[n]}\bigg\{
\sum_{s\in[m],a,b\in[n]}\bigg(\sum_{t = 1}^m\sum_{j\in[n]\backslash\{i\}}\ell_{0tij}'\bv_{k,tj}\transpose\bJ_{0n}(\bG_0^\sharp)^{-1}\bJ_{0n}\transpose\bu_{sab}\bigg)^2
\bigg\}^{1/2}& = \Optilde\bigg(\sqrt{\frac{\log n}{mn}}\bigg),\\  
\max_{k\in[d(d + 1)/2],t,s\in[m],a,b\in[n]}\bigg|\sum_{i\neq j}\ell'_{0tij}\bw_{k,ij}\transpose  \bJ_{0n}(\bG_0^\sharp)^{-1}\bJ_{0n}\transpose \bu_{sab}\bigg|& = \Optilde\bigg(\frac{\sqrt{\log n}}{\sqrt{m}n^{3/2}}\bigg),\\
\max_{k\in[d(d + 1)/2],t\in[m]}\bigg\{\sum_{s\in[m],a,b\in[n]}\bigg(\sum_{i\neq j}\ell'_{0tij}\bw_{k,ij}\transpose\bJ_{0n}(\bG_0^\sharp)^{-1}\bJ_{0n}\transpose\bu_{sab}\bigg)^2\bigg\}^{1/2}
& = \Optilde\bigg(\frac{\sqrt{\log n}}{\sqrt{m}n}\bigg).
\end{align*}
For the first two target bounds, we proceed by following the same reasoning as in the proof of Lemma \ref{lemma:fisher_information_at_truth_concentration_infintiy_norm}:
\begin{align*}
&\max_{k\in[d],i\in[n]}\bigg|
\sum_{t = 1}^m\sum_{j\in[n]\backslash\{i\}}\ell'_{0tij}\bv_{k,tj}\transpose \{\bJ_{0n}(\bG_0^\sharp)^{-1}\bJ_{0n}\transpose\} \bigg\{\bJ_{0n}\bLambda_\sharp\frac{\partial\calL}{\partial\bphi}(\bphi_0)\bigg\}
\bigg|\\
&\quad\leq \max_{k,i}\bigg\|
\sum_{t = 1}^m\sum_{j\in[n]\backslash\{i\}}\ell'_{0tij}\bv_{k,tj}\transpose\bigg\|_2 \|\bJ_{0n}(\bG_0^\sharp)^{-1}\bJ_{0n}\transpose\|_2 \bigg\|\sum_{t = 1}^m\sum_{j\in[n]\backslash\{i\}}\ell'_{0tij}\bu_{tij}\bigg\|_2\\
&\qquad + \max_{k,i}\bigg|
\sum_{t = 1}^m\sum_{j\in[n]\backslash\{i\}}\ell'_{0tij}\bv_{k,tj}\transpose \{\bJ_{0n}(\bG_0^\sharp)^{-1}\bJ_{0n}\transpose\} \bigg[
\sum_{s = 1}^m\sum_{a\neq b}\ell'_{0sab}\bu_{sab}\{1 - \mathbbm{1}(a = i\text{ or }b = i)\}
\bigg]
\bigg|\\
&\quad\leq \Optilde\bigg(\frac{\log n}{\sqrt{n}}\bigg) + \Optilde\bigg\{
\frac{(\log n)^{3/2}}{\sqrt{m}n} + \frac{\log n}{\sqrt{mn}}\bigg\} = \Optilde\bigg(\frac{\log n}{\sqrt{n}}\bigg).
\end{align*}
Similarly, by a direct calculation, we can obtain
\begin{align*}
&\max_{k\in[d(d + 1)/2],t\in[m]}\bigg|
\sum_{i\neq j}\ell'_{0tij}\bw_{k,ij}\transpose \{\bJ_{0n}(\bG_0^\sharp)^{-1}\bJ_{0n}\transpose\} \bigg\{\bJ_{0n}\bLambda_\sharp\frac{\partial\calL}{\partial\bphi}(\bphi_0)\bigg\}
\bigg|\\
&\quad\leq \max_{k,t}\bigg\| 
\sum_{i\neq j}\ell'_{0tij}\bw_{k,ij}\transpose \bigg\|_2\|\bJ_{0n}(\bG_0^\sharp)^{-1}\bJ_{0n}\transpose\|_2\bigg\|\sum_{i\neq j}\ell'_{0tij}\bu_{tij}\bigg\|_2
\\
&\qquad + \max_{k,t}\bigg|
\sum_{i\neq j}\ell'_{0tij}\bw_{k,ij}\transpose \{\bJ_{0n}(\bG_0^\sharp)^{-1}\bJ_{0n}\transpose\}
\bigg(\sum_{s\in[m]\backslash\{t\}}\sum_{a\neq b}\ell_{0sab}'\bu_{sab}\bigg)
\bigg|\\
&\quad = \Optilde\bigg(\frac{\log n}{\sqrt{mn}}\bigg) + \Optilde\bigg\{
\frac{(\log n)^{3/2}}{\sqrt{m}n^{3/2}} + \frac{\log n}{\sqrt{m}n}\bigg\}
= \Optilde\bigg(\frac{\log n}{\sqrt{mn}}\bigg). 
\end{align*}
For the last term, by matrix Bernstein's inequality (Lemma \ref{lemma:matrix_Bernstein} and Theorem 6.2 in \cite{tropp2012user}), we have
\begin{align*}
&\max_{k\in[d],l\in[n]}\bigg|
\sum_{t = 1}^m\sum_{i\neq j}\ell'_{0tij}\by_{kl,tij}\transpose \{\bJ_{0n}(\bG_0^\sharp)^{-1}\bJ_{0n}\transpose\} \bigg\{\bJ_{0n}\bLambda_\sharp\frac{\partial\calL}{\partial\bphi}(\bphi_0)\bigg\}
\bigg|\\
&\quad\lesssim \frac{1}{mn^2}\bigg\|\sum_{t,i,j}\ell_{0tij}'
\int_0^\infty e^{-\tau\eye_d}(\bx_{0i}\bx_{0j}\transpose\bR_0^{(t)} + \bR_0^{(t)}\bx_{0j}\bx_{0i}\transpose)e^{-\tau\eye_d}
\bigg\|_2 \|\bJ_{0n}(\bG_0^\sharp)^{-1}\bJ_{0n}\transpose\|_2 \bigg\|\sum_{t,i,j}\ell'_{0tij}\bu_{tij}\bigg\|_2\\
&\quad = \Optilde\bigg(\frac{\log n}{\sqrt{n}}\bigg).
\end{align*}
The proof is thus completed. 
\end{proof}

\subsection{Infinity norm of the rescaled inverse Fisher information matrix}
\label{sub:infinity_norm_of_the_rescaled_inverse_fisher_information_matrix}

This subsection establishes a fundamental result that sharply bounds the infinity norm of $\bJ_{0n}(\bG_0^\sharp)^{-1}\bJ_{0n}\transpose$ block-wise. It also paves the way for showing that the asymptotic distribution of the MLE does not depend on the choice of the reparameterization (specifically, the generalized Cayley transform \eqref{eqn:Cayley_transform}). Denote
\begin{align*}
\bG_{\bZ\bZ}^\sharp(\bpsi)
& = -\frac{1}{n}\sum_{t = 1}^m\sum_{i\neq j} \expect_\bpsi\ell''(\theta_{ij}^{(t)}; A_{ij}^{(t)})\be_i\be_i\transpose\otimes 
\begin{bmatrix}
\be_t(m)\be_t(m)\transpose & \frac{1}{\sqrt{m}}\be_t(m)\bx_{j}\transpose\bR^{(t)}\\ 
\frac{1}{\sqrt{m}}\bR^{(t)}\bx_{j}\be_t(m)\transpose & \frac{1}{m}\bR^{(t)}\bx_{j}\bx_{j}\transpose\bR^{(t)}
\end{bmatrix}\\
&\quad -
\frac{1}{n}\sum_{t = 1}^m\sum_{i\neq j} \expect_\bpsi\ell''(\theta_{ij}^{(t)}; A_{ij}^{(t)})\be_i\be_j\transpose\otimes 
\begin{bmatrix}
\be_t(m)\be_t(m)\transpose & \frac{1}{\sqrt{m}}\be_t(m)\bx_{i}\transpose\bR^{(t)}\\ 
\frac{1}{\sqrt{m}}\bR^{(t)}\bx_{j}\be_t(m)\transpose & \frac{1}{m}\bR^{(t)}\bx_{j}\bx_{i}\transpose\bR^{(t)}
\end{bmatrix}
,\\
\bG_{\bZ\bR}^\sharp(\bpsi)
& = -\frac{1}{2n^{3/2}}\sum_{t = 1}^m\sum_{i\neq j} \expect_\bpsi\ell''(\theta_{ij}^{(t)}; A_{ij}^{(t)})\left\{\be_i\otimes\begin{bmatrix}
\be_t(m) \\ 
\frac{\bR^{(t)}\bx_{j}}{\sqrt{m}}
\end{bmatrix}
+
\be_j\otimes\begin{bmatrix}
\be_t(m) \\ 
\frac{\bR^{(t)}\bx_{i}}{\sqrt{m}}
\end{bmatrix}
\right\}\{\be_t(m)\transpose\otimes(\bx_{j}\otimes\bx_{i})\transpose\bD_d\},\\
\bG_{\bR\bR}^\sharp(\bpsi)
& = -\sum_{t = 1}^m
\be_t(m)\be_t(m)\transpose\otimes\frac{1}{2n^2}\sum_{i\neq j}\expect_\bpsi\ell''(\theta_{ij}^{(t)}; A_{ij}^{(t)})\bD_d\transpose(\bx_{j}\bx_{j}\transpose\otimes \bx_{i}\bx_{i}\transpose)\bD_d,\\
\bG_{\bpsi\bpsi}^\sharp(\bpsi)
& = \begin{bmatrix}
\bG_{\bZ\bZ}^\sharp(\bpsi) & \bG_{\bZ\bR}^\sharp(\bpsi)\\
\bG_{\bZ\bR}^\sharp(\bpsi)\transpose & \bG_{\bR\bR}^\sharp(\bpsi)
\end{bmatrix},\;
\scrP_{\balpha\bB} = \begin{bmatrix}
  \eye_{d_\balpha} & \mathbf{O}_{d_\balpha\times d_\br} & \mathbf{O}_{d_\balpha\times d_\bB}\\
  \mathbf{O}_{d_\bB\times d_\balpha} & \mathbf{O}_{d_\bB\times d_\br} & \eye_{d_\bB}
\end{bmatrix},\\
\widetilde{\scrP}_\br& = \begin{bmatrix}
\mathbf{O}_{d_\br\times d_\balpha} & \eye_{d_\br} & \mathbf{O}_{d_\br\times d_\bB}
\end{bmatrix}.
\end{align*}
Also denote $\bSigma_\bpsi^\sharp = -\expect_0\bDelta_\bpsi^\sharp(\bpsi_0)
= \mathrm{diag}\{\bSigma_\bZ^\sharp, \bG_{\bR\bR}^\sharp(\bpsi_0)\}$, where
$\bSigma_\bZ^\sharp = -\expect_0\bDelta_\bZ^\sharp(\bpsi_0)
= \sum_{i = 1}^n\be_i\be_i\transpose\otimes\bSigma_{\bz_i}^\sharp$ with
$\bSigma_{\bz_i}^\sharp = -\expect_0\bDelta_{\bz_i}^\sharp(\bpsi_0)$ positive
definite by Assumption \ref{assumption:true_parameter}.

\begin{lemma}
\label{lemma:inverse_Fisher_information_matrix_infty_norm_bound}
Suppose Assumptions \ref{assumption:edge_distribution}--\ref{assumption:true_parameter} hold. Denote by $\bJ_\bZ = \mathrm{diag}(\eye_{d_\balpha}, \eye_d\otimes\bX_{0\perp}/\sqrt{n})$. Then,
\begin{align*}
&\max_{s\in[d_\balpha]}\|\be_s\transpose\bJ_\bZ\{\bJ_\bZ\transpose\scrP_\bZ\bPi\bG_{\bpsi\bpsi}^\sharp(\bpsi_0)\bPi\transpose\scrP_\bZ\transpose\bJ_\bZ\}^{-1}\bJ_\bZ\transpose\|_\infty = O(m),\\
&\max_{s\in[d_\bX]}\|\be_{d_\balpha + s}\transpose\bJ_\bZ\{\bJ_\bZ\transpose\scrP_\bZ\bPi\bG_{\bpsi\bpsi}^\sharp(\bpsi_0)\bPi\transpose\scrP_\bZ\transpose\bJ_\bZ\}^{-1}\bJ_\bZ\transpose\|_\infty = O(m^{3/2}),\; \|\widetilde{\scrP}_\br(\bG_0^\sharp)^{-1}\widetilde{\scrP}_\br\transpose\|_2 = O(1),\\
&\max_{s\in[d_\balpha]}\|\be_s\transpose\bJ_\bZ\scrP_{\balpha\bB}(\bG_0^\sharp)^{-1}\scrP_{\balpha\bB}\transpose\bJ_\bZ\transpose\|_\infty = O(m),\\
&\max_{s\in[d_\bX]}\|\be_{d_\balpha + s}\transpose\bJ_\bZ\scrP_{\balpha\bB}(\bG_0^\sharp)^{-1}\scrP_{\balpha\bB}\transpose\bJ_\bZ\transpose\|_\infty = O(m^{3/2}),\\
& 
\max_{s\in[d_\balpha]}\|\be_s\transpose\bJ_\bZ\scrP_{\balpha\bB}(\bG_0^\sharp)^{-1}\widetilde{\scrP}_\br\transpose\|_\infty = O(m),\;
\max_{s\in[d_\bX]}\|\be_{d_\balpha + s}\transpose\bJ_\bZ\scrP_{\balpha\bB}(\bG_0^\sharp)^{-1}\widetilde{\scrP}_{\br}\transpose\|_\infty = O(m^{3/2}).
\end{align*}
\end{lemma}

\begin{proof}[\bf Proof]
Observe that
\[
\bG_{\bZ\bZ}^\sharp(\bpsi_0) - \bSigma_{\bZ}^\sharp(\bpsi_0)
 = 
 -\frac{1}{n}\sum_{t = 1}^m\sum_{i\neq j} \expect_0\ell''(\theta_{0ij}^{(t)}; A_{ij}^{(t)})\be_i\be_j\transpose\otimes 
\begin{bmatrix}
\be_t(m)\be_t(m)\transpose & \frac{1}{\sqrt{m}}\be_t(m)\bx_{0i}\transpose\bR_0^{(t)}\\ 
\frac{1}{\sqrt{m}}\bR_0^{(t)}\bx_{0j}\be_t(m)\transpose & \frac{1}{m}\bR_0^{(t)}\bx_{0j}\bx_{0i}\transpose\bR_0^{(t)}
\end{bmatrix},
\]
then, clearly, $\|\bG_{\bZ\bZ}^\sharp(\bpsi_0) - \bSigma_{\bZ}^\sharp\|_{2\to\infty} = O(n^{-1/2})$ (also see the proof of Lemma \ref{lemma:off_diagonal_term_perturbation}). Furthermore, since $\bX_0\transpose\one_n = \zero_d$ and $\bX_0\transpose\bX_0 = n\eye_d$, then by Assumption \ref{assumption:true_parameter}, we further have
\begin{align*}
\bSigma_{\bz_i}^\sharp(\bpsi_0) &\succeq C\begin{bmatrix}
  (1 - \frac{1}{n})\eye_m & -\frac{1}{n\sqrt{m}}\sum_{t = 1}^m\be_t(m)\bx_{0i}\transpose\bR_0^{(t)} \\
  -\frac{1}{n\sqrt{m}}\sum_{t = 1}^m\bR_0^{(t)}\bx_{0i}\be_t(m)\transpose & \frac{1}{m}\sum_{t = 1}^m\bR_0^{(t)}(\eye_d - \frac{1}{n}\bx_{0i}\bx_{0i}\transpose)\bR_0^{(t)} 
\end{bmatrix}\\
&\succeq \frac{C}{2}\mathrm{diag}\bigg\{\eye_{m}, \frac{1}{m}\sum_{t = 1}^m(\bR_0^{(t)})^2\bigg\}\quad\text{for all }i\in[n],
\end{align*} 
so that $\|(\bSigma_{\bZ}^\sharp)^{-1}\|_2 = O(1)$ and $\|\bSigma_{\bZ}^\sharp\|_2 = O(1)$. Also, we have $\|(\bSigma_{\bZ}^\sharp)^{-1}\|_\infty = O(\sqrt{m})$ from the proof of Proposition \ref{prop:stationarity_uniqueness_uniform_consistency_MLE}, or more specifically, \eqref{eqn:Deltazi_sharp_infinity_norm_I} and \eqref{eqn:Deltazi_sharp_infinity_norm_II}. In addition, 
\begin{align*}
\max_{t\in[m]}\|\be_t(m + d)\transpose(\bSigma_{\bz_i}^\sharp)^{-1}\|_\infty& = O(1),\quad
\max_{k\in[d]}\|\be_{m + k}(m + d)\transpose(\bSigma_{\bz_i}^\sharp)^{-1}\|_\infty = O(\sqrt{m}).
\end{align*}
Observe that $\bJ_\bZ\transpose\scrP_\bZ = \scrP_{\balpha\bB}\bJ_{0n}\transpose$. 
Therefore, by a direct computation, for any index $s\in[d_\balpha + d_\bX]$, 
\begin{align*}
&\left\|\be_s\transpose\bJ_\bZ\{\bJ_\bZ\transpose\scrP_\bZ\bPi\bG_{\bpsi\bpsi}^\sharp(\bpsi_0)\bPi\transpose\scrP_\bZ\transpose\bJ_\bZ\}^{-1}\bJ_\bZ\transpose\right\|_\infty\\
&\quad\leq \left\|\be_s\transpose\bJ_\bZ(\bJ_\bZ\transpose\scrP_\bZ\bPi\bSigma_{\bpsi}^\sharp \bPi\transpose\scrP_\bZ\transpose\bJ_\bZ)^{-1}\bJ_\bZ\transpose\right\|_\infty\\
&\qquad + 
\left\|\be_s\transpose\bJ_\bZ\left[\{\bJ_\bZ\transpose\scrP_\bZ\bPi\bG_{\bpsi\bpsi}^\sharp(\bpsi_0)\bPi\transpose\scrP_\bZ\transpose\bJ_\bZ\}^{-1} - (\bJ_\bZ\transpose\scrP_\bZ\bPi\bSigma_{\bpsi}^\sharp\bPi\transpose\scrP_\bZ\transpose\bJ_\bZ)^{-1}\right]\bJ_\bZ\transpose\right\|_\infty\\
&\quad = \left\|\be_s\transpose\bJ_\bZ(\bJ_\bZ\transpose\scrP_\bZ\bPi\bSigma_{\bpsi}^\sharp \bPi\transpose\scrP_\bZ\transpose\bJ_\bZ)^{-1}\bJ_\bZ\transpose\right\|_\infty\\
&\qquad + 
\left\|\be_s\transpose\bJ_\bZ(\bJ_\bZ\transpose\scrP_\bZ\bPi\bSigma_{\bpsi}^\sharp \bPi\transpose\scrP_\bZ\transpose\bJ_\bZ)^{-1}\bJ_\bZ\transpose\right\|_\infty\\
&\qquad\times O(\sqrt{mn})\left\|\scrP_\bZ
\bPi\{\bSigma_{\bpsi}^\sharp  - \bG_{\bpsi\bpsi}^\sharp(\bpsi_0)\}\bPi\transpose\scrP_\bZ\transpose\right\|_{2\to\infty}\left\|\bJ_\bZ\{\bJ_\bZ\transpose\scrP_\bZ\bPi\bG_{\bpsi\bpsi}^\sharp(\bpsi_0)\bPi\transpose\scrP_\bZ\transpose\bJ_\bZ\}^{-1}\bJ_\bZ\transpose\right\|_2\\
&\quad = \left\|\be_s\transpose\bJ_\bZ(\bJ_\bZ\transpose\scrP_\bZ\bPi\bSigma_{\bpsi}^\sharp \bPi\transpose\scrP_\bZ\transpose\bJ_\bZ)^{-1}\bJ_\bZ\transpose\right\|_\infty\\ 
&\qquad\times \left\{1 + O(\sqrt{mn})\left\|\bSigma_{\bZ}^\sharp - \bG_{\bZ\bZ}^\sharp(\bpsi_0)\right\|_{2\to\infty}\left\|\{\scrP_{\balpha\bB}\bJ_{0n}\transpose\bPi\bG_{\bpsi\bpsi}^\sharp(\bpsi_0)\bPi\transpose\bJ_{0n}\scrP_{\balpha\bB}\transpose\}^{-1}\right\|_2\right\}\\
&\quad \leq \left\|\be_s\transpose\bJ_\bZ(\bJ_\bZ\transpose\scrP_\bZ\bPi\bSigma_{\bpsi}^\sharp \bPi\transpose\scrP_\bZ\transpose\bJ_\bZ)^{-1}\bJ_\bZ\transpose\right\|_\infty\\ 
&\qquad\times \left\{1 + O(\sqrt{mn})\left\|\bSigma_{\bZ}^\sharp  - \bG_{\bZ\bZ}^\sharp(\bpsi_0)\right\|_{2\to\infty}\|(\bG_0^\sharp)^{-1}\|_2\right\}\\
&\quad = O(\sqrt{m})\left\|\be_s\transpose\bJ_\bZ(\bJ_\bZ\transpose\scrP_\bZ\bPi\bSigma_{\bpsi}^\sharp \bPi\transpose\scrP_\bZ\transpose\bJ_\bZ)^{-1}\bJ_\bZ\transpose\right\|_\infty.
\end{align*}
Next, we show that $\|\be_s\transpose\bJ_\bZ(\bJ_\bZ\transpose\scrP_\bZ\bPi\bSigma_{\bpsi}^\sharp \bPi\transpose\scrP_\bZ\transpose\bJ_\bZ)^{-1}\bJ_\bZ\transpose\|_\infty$ can be sharply bounded as needed. To this end, denote by $\bV_\bZ = [\mathbf{O}_{d(d + 1)\times d_\balpha}, \eye_d\otimes [\one_n,\bX_0]\transpose/\sqrt{n}]\transpose$, which has orthonormal columns and is orthogonal to $\bJ_\bZ$. Observe that the projection matrices onto the column spaces of $(\scrP_\bZ\bPi\bSigma_{\bpsi}^\sharp \bPi\transpose\scrP_\bZ\transpose)^{1/2}\bJ_\bZ$ and $(\scrP_\bZ\bPi\bSigma_{\bpsi}^\sharp \bPi\transpose\scrP_\bZ\transpose)^{-1/2}\bV_\bZ$ are
\begin{align*}
    \textsf{Proj}_1
    &:=( \scrP_\bZ\bPi\bSigma_{\bpsi}^\sharp \bPi\transpose\scrP_\bZ\transpose )^{1/2}\bJ_\bZ( \scrP_\bZ\bPi\bSigma_{\bpsi}^\sharp \bPi\transpose\scrP_\bZ\transpose )^{-1}\bJ_\bZ\transpose
    ( \scrP_\bZ\bPi\bSigma_{\bpsi}^\sharp \bPi\transpose\scrP_\bZ\transpose )^{1/2},\\
    \textsf{Proj}_2
    &:=(\scrP_\bZ\bPi\bSigma_{\bpsi}^\sharp \bPi\transpose\scrP_\bZ\transpose)^{-1/2}\bV_\bZ\{\bV_\bZ\transpose(\scrP_\bZ\bPi\bSigma_{\bpsi}^\sharp \bPi\transpose\scrP_\bZ\transpose)^{-1}\bV_\bZ\}^{-1}\bV_\bZ\transpose
    (\scrP_\bZ\bPi\bSigma_{\bpsi}^\sharp \bPi\transpose\scrP_\bZ\transpose)^{-1/2},
\end{align*}
and the direct sum of these two column spaces is the entire $\mathbb{R}^{d_\balpha + d_\bX}$. Hence, we obtain $\textsf{Proj}_1 = \eye_{d_\balpha + d_\bX} - \textsf{Proj}_2$, implying that
\begin{align*}
  \bJ_\bZ(\scrP_\bZ\bPi\bSigma_{\bpsi}^\sharp \bPi\transpose\scrP_\bZ\transpose)^{-1}\bJ_\bZ\transpose
  & = (\scrP_\bZ\bPi\bSigma_{\bpsi}^\sharp\bPi\transpose\scrP_\bZ\transpose)^{-1} - (\scrP_\bZ\bPi\bSigma_{\bpsi}^\sharp\bPi\transpose\scrP_\bZ\transpose)^{-1}\\
  &\quad\times 
  \bV_\bZ\{\bV_\bZ\transpose(\scrP_\bZ\bPi\bSigma_{\bpsi}^\sharp\bPi\transpose\scrP_\bZ\transpose)^{-1}\bV_\bZ\}^{-1}\bV_\bZ\transpose
  (\scrP_\bZ\bPi\bSigma_{\bpsi}^\sharp\bPi\transpose\scrP_\bZ\transpose)^{-1},
\end{align*}
and hence, for any $s\in[d_\balpha + d_\bX]$,
\begin{align*}
  &\|\be_s\transpose\bJ_\bZ(\bJ_\bZ\transpose\scrP_\bZ\bPi\bSigma_\bpsi^\sharp\bPi\transpose\scrP_\bZ\transpose\bJ_\bZ)^{-1}\bJ_\bZ\transpose\|_\infty\\
  &\quad\leq  \|\be_s\transpose( \scrP_\bZ\bPi\bSigma_\bpsi^\sharp\bPi\transpose\scrP_\bZ\transpose )^{-1}\|_\infty\\
  &\qquad + \|\be_s\transpose( \scrP_\bZ\bPi\bSigma_\bpsi^\sharp\bPi\transpose\scrP_\bZ\transpose )^{-1}\|_\infty\|\bV_\bZ\|_{\infty}
  \|\{\bV_\bZ\transpose( \scrP_\bZ\bPi\bSigma_\bpsi^\sharp\bPi\transpose\scrP_\bZ\transpose )^{-1}\bV_\bZ\}^{-1}\|_\infty\\
  &\qquad\times 
  \|\bV_\bZ\transpose\|_\infty\|( \scrP_\bZ\bPi\bSigma_\bpsi^\sharp\bPi\transpose\scrP_\bZ\transpose )^{-1}\|_\infty\\
  &\quad
  \leq \|\be_s\transpose( \scrP_\bZ\bPi\bSigma_\bpsi^\sharp\bPi\transpose\scrP_\bZ\transpose )^{-1}\|_\infty\\
  &\qquad\times \{1 + \|\bV_\bZ\|_{\infty}
  d\|\{\bV_\bZ\transpose( \scrP_\bZ\bPi\bSigma_\bpsi^\sharp\bPi\transpose\scrP_\bZ\transpose )^{-1}\bV_\bZ\}^{-1}\|_2\|\bV_\bZ\transpose\|_\infty
  \|( \scrP_\bZ\bPi\bSigma_\bpsi^\sharp\bPi\transpose\scrP_\bZ\transpose )^{-1}\|_\infty\}\\
  &\quad\leq O(\sqrt{m})\|\be_s\transpose( \scrP_\bZ\bPi\bSigma_\bpsi^\sharp\bPi\transpose\scrP_\bZ\transpose )^{-1}\|_\infty.
\end{align*}
Hence, by calculating the inverse of each diagonal block of $\bSigma_{\bZ}^\sharp$ using the block matrix inversion formula (the same reasoning as \eqref{eqn:Deltazi_sharp_infinity_norm_I} and \eqref{eqn:Deltazi_sharp_infinity_norm_II}), we obtain
\begin{align*}
\max_{s\in[d_\balpha]}\|\be_s\transpose\bJ_\bZ(\bJ_\bZ\transpose\scrP_\bZ\bPi\bSigma_{\bpsi}^\sharp\bPi\transpose\scrP_\bZ\transpose\bJ_\bZ)^{-1}\bJ_\bZ\transpose\|_\infty& = O(\sqrt{m}),\\
\max_{s\in[d_\bX]}\|\be_{d_\balpha + s}\transpose\bJ_\bZ(\bJ_\bZ\transpose\scrP_\bZ\bPi\bSigma_{\bpsi}^\sharp \bPi\transpose\scrP_\bZ\transpose\bJ_\bZ)^{-1}\bJ_\bZ\transpose\|_\infty& = O(m),
\end{align*}
and
\begin{align*}
\max_{s\in[d_\balpha]}\|\be_s\transpose\bJ_\bZ\{\bJ_\bZ\transpose\scrP_\bZ\bPi\bG_{\bpsi\bpsi}^\sharp(\bpsi_0)\bPi\transpose\scrP_\bZ\transpose\bJ_\bZ\}^{-1}\bJ_\bZ\transpose\|_\infty& = O(m),\\
\max_{s\in[d_\bX]}\|\be_{d_\balpha + s}\transpose\bJ_\bZ\{\bJ_\bZ\transpose\scrP_\bZ\bPi\bG_{\bpsi\bpsi}^\sharp(\bpsi_0)\bPi\transpose\scrP_\bZ\transpose\bJ_\bZ\}^{-1}\bJ_\bZ\transpose\|_\infty& = O(m^{3/2}).
\end{align*}
This completes the first two infinity norm bounds. For the third bound on spectral norm, note that the spectral norm of a submatrix is no larger than the original matrix, so that
$\|\widetilde{\scrP}_\br(\bG_0^\sharp)^{-1}\widetilde{\scrP}_\br\transpose\|_2
\leq \|(\bG_0^\sharp)^{-1}\|_2 = O(1)$. For the remaining bounds, we denote $\bL = \bJ_{0n}\{\bJ_{0n}\transpose\bPi
\bSigma_\bpsi^\sharp\bPi\transpose\bJ_{0n}\}^{-1}\bJ_{0n}\transpose$. By
\eqref{eqn:permutation_matrix_psi}, $\bPi$ transports the coordinates of
$\bz_1,\ldots,\bz_n$ to the corresponding ones in $(\balpha^{(1)},\ldots,\balpha^{(m)}, \vect(\bX))$ and the
coordinates of $\vech(\bR^{(t)})$ to themselves, so that
$\scrP_\br\bPi\bSigma_\bpsi^\sharp\bPi\transpose\scrP_\bZ\transpose
= \mathbf{O}$. Note that $\scrP_\bZ\bJ_{0n} = \bJ_\bZ\scrP_{\balpha\bB}$
and $\bJ_{0n}\widetilde{\scrP}_\br\transpose = \scrP_\br\transpose$, we then obtain the following direct-sum decomposition of $\bL$:
\[
\bL = \scrP_\bZ\transpose\bJ_\bZ\{\bJ_\bZ\transpose\scrP_\bZ\bPi
\bSigma_\bpsi^\sharp\bPi\transpose\scrP_\bZ\transpose\bJ_\bZ\}^{-1}
\bJ_\bZ\transpose\scrP_\bZ
+ \scrP_\br\transpose\bG_{\bR\bR}^\sharp(\bpsi_0)^{-1}\scrP_\br.
\]
In particular, $\scrP_\bZ\bL\scrP_\br\transpose = \mathbf{O}$. Also, a simple algebra shows that
\[
\bJ_{0n}(\bG_0^\sharp)^{-1}\bJ_{0n}\transpose
= \bL + \bL\bPi\{\bSigma_\bpsi^\sharp - \bG_{\bpsi\bpsi}^\sharp(\bpsi_0)\}
\bPi\transpose\bJ_{0n}(\bG_0^\sharp)^{-1}\bJ_{0n}\transpose.
\]
For each $i\in[d_\balpha + d_\bX]$, the set of indices $\calJ_i$ of the nonzero elements of the row vector $\be_i\transpose\scrP_\bZ\bL$ corresponds to the coordinates of $(\balpha^{(1)},\ldots,\balpha^{(m)}, \vect(\bX))$, and for each $j\in\calJ_i$, the row vector $\be_j\transpose\bPi\{\bSigma_\bpsi^\sharp - \bG_{\bpsi\bpsi}^\sharp(\bpsi_0)\}
\bPi\transpose$ has $\ell_2$-norm $O(n^{-1/2})$. Indeed, it is upper bounded by $\|\bSigma_\bZ^\sharp - \bG_{\bZ\bZ}^\sharp(\bpsi_0)\|_{2\to\infty} + \|\bG_{\bZ\bR}^\sharp(\bpsi_0)\|_{2\to\infty} = O(n^{-1/2})$ by the beginning of this proof and the proof of Lemma \ref{lemma:off_diagonal_term_perturbation}. 
Hence, for any $i\in [d_\balpha + d_\bX]$, 
\[
\|\be_i\transpose\scrP_\bZ\bJ_{0n}(\bG_0^\sharp)^{-1}\bJ_{0n}\transpose
\|_\infty \leq \|\be_i\transpose\scrP_\bZ\bL\|_\infty\{1
+ \sqrt{d_\bpsi}\,O(n^{-1/2})\|(\bG_0^\sharp)^{-1}\|_2\}
= O(\sqrt m)\|\be_i\transpose\scrP_\bZ\bL\|_\infty,
\]
which is $O(m)$ for $i\in[d_\balpha]$ and $O(m^{3/2})$ for
$i\in\{d_\balpha + 1,\ldots,d_\balpha + d_\bX\}$. These rows include the coordinates corresponding to the $\br$ columns, so the bounds on $\bJ_\bZ\scrP_{\balpha\bB}(\bG_0^\sharp)^{-1}
\widetilde{\scrP}_\br\transpose$ follow as well.
\end{proof}

\subsection{Completing the proof of Theorem \ref{thm:asymptotic_normality_MLE}}
\label{sub:Proof of the asymptotic normality}

\begin{proof}[\bf Proof of Theorems \ref{thm:two_to_infinity_norm_bound_MLE} and \ref{thm:asymptotic_normality_MLE}]
The Frobenius norm error bound of $\widehat{\bX}$ in Theorem \ref{thm:two_to_infinity_norm_bound_MLE} follows from Proposition \ref{prop:consistency_native_norm}: since $\widehat{\bX}\bW_{\widehat{\bX}} - \bX_0\in\mathbb{R}^{n\times d}$ has rank at most $d$, $\|\widehat{\bX}\bW_{\widehat{\bX}} - \bX_0\|_{\mathrm{F}}\leq\sqrt{d}\|\widehat{\bX}\bW_{\widehat{\bX}} - \bX_0\|_2 = \Optilde(1)$. Below, we focus on proving Theorem \ref{thm:asymptotic_normality_MLE}, and the proof itself contains the two-to-infinity-norm error bound of $\widehat{\bX}$ in Theorem \ref{thm:two_to_infinity_norm_bound_MLE} as a special case. 

\noindent$\blacksquare$ \textbf{Step 1. Bounding $\left\|\mathscr{P}_{\bZ}\bJ_{0n}\bar{\mathscr{H}}^{-1}\bJ_{0n}\transpose 
\mathrm{diag}(\eye_{d_\balpha}, \sqrt{n}\eye_{d_\br}, \eye_{d_\bX})
\right\|_\infty$. } 

\noindent \underline{We first rigorously establish an error bound for $\|\mathscr{P}_{\bZ}\bJ_{0n}\bar{\mathscr{H}}^{-1}\bJ_{0n}\transpose \scrP_\bZ\transpose\|_\infty$.}
Following the same analysis in the proof of Proposition \ref{prop:stationarity_uniqueness_uniform_consistency_MLE} and applying Lemma \ref{lemma:off_diagonal_term_perturbation} and Lemma \ref{lemma:score_term_perturbation} with $a_n = c_n = 1, b_n=0, \eps_n = K_c n^{-1/2}$ for some constant $K_c > 0$ depending on $c$, we have
\begin{align*}
&\|\scrP_\bZ\bJ_{0n}\bar{\scrH}^{-1}
\bJ_{0n}\transpose\scrP_\bZ\transpose\|_\infty\\
&\quad \leq    
\left\|\scrP_{\bZ}\bJ_{0n}\{(\scrI \bDelta_\bphi^\sharp)(\widehat{\bphi})\}^{-1}
\bJ_{0n}\transpose\scrP_{\bZ}\transpose\right\|_\infty\\
&\qquad + \sqrt{d_\bpsi}\left\|\scrP_{\bZ}\bJ_{0n}\{(\scrI \bDelta_\bphi^\sharp)(\widehat{\bphi})\}^{-1}
\bJ_{0n}\transpose\right\|_\infty\\
&\qquad\times \left\|
\mathrm{diag}(\eye_{d_\balpha},\mathbf{O}_{ d_\br \times d_\br }, \eye_{d_\bX})
\left(\scrI \bJ_{0n}\bJ_n\transpose\bPi(\bGamma_{\bpsi\bpsi}^\sharp(\bpsi(\cdot)) - \bDelta_{\bpsi}^\sharp(\bpsi(\cdot)))\bPi\transpose\bJ_n\bJ_{0n}\transpose\right)(\widehat{\bphi})\right\|_{2\to\infty}  \|\bJ_{0n}\bar{\scrH}^{-1}
\bJ_{0n}\transpose\|_2\\
&\qquad + \frac{\sqrt{d_\bpsi}}{2}\left\|\scrP_{\bZ}\bJ_{0n}\{(\scrI \bDelta_\bphi^\sharp)(\widehat{\bphi})\}^{-1}
\bJ_{0n}\transpose\right\|_\infty\\
&\qquad\times 
\bigg\|\mathrm{diag}(\eye_{d_\balpha},\mathbf{O}_{ d_\br \times d_\br }, \eye_{d_\bX})\scrI\bigg( 
\bJ_{0n}\bLambda_\sharp\sum_{t = 1}^m\sum_{i\neq j}\ell'(\theta_{ij}^{(t)}(\cdot); A_{ij}^{(t)})\frac{\partial^2\theta_{ij}^{(t)}}{\partial\bphi\partial\bphi\transpose}\bLambda_\sharp\bJ_{0n}\transpose - \bXi
\bigg)(\widehat{\bphi})\bigg\|_{2\to\infty}
\|\bJ_{0n}\bar{\scrH}^{-1}\bJ_{0n}\transpose\|_2\\
&\qquad + \frac{1}{2}\left\|\scrP_\bZ\bJ_{0n}\{(\scrI \bDelta_\bphi^\sharp)(\widehat{\bphi})\}^{-1}
\bJ_{0n}\transpose\right\|_\infty
\left\|\mathrm{diag}(\eye_{d_\balpha},\mathbf{O}_{d_\br\times d_\br},\eye_{d_\bX})(\scrI \bXi)(\widehat{\bphi})\mathrm{diag}(\eye_{d_\balpha},\mathbf{O}_{d_\br\times d_\br},\eye_{d_\bX})\right\|_\infty\\
&\qquad\times 
\|\scrP_\bZ\bJ_{0n}\bar{\scrH}^{-1}
\bJ_{0n}\transpose\scrP_\bZ\transpose\|_\infty\\
&\quad = \Optilde(m^{5/2}\vee m^2\sqrt{\log n}) + \Optilde(m^{2}n^{-1/2}\log n)\|\scrP_\bZ\bJ_{0n}\bar{\scrH}^{-1}
\bJ_{0n}\transpose\scrP_\bZ\transpose\|_\infty,
\end{align*}
so that
\begin{align*}
\|\scrP_\bZ\bJ_{0n}\bar{\scrH}^{-1}
\bJ_{0n}\transpose\scrP_\bZ\transpose\|_\infty = \Optilde(m^{5/2}\vee m^2\sqrt{\log n}),  
\end{align*}
where we have used the fact that $\|\scrP_\bZ\bJ_{0n}\{(\scrI\bDelta_\bphi^\sharp)(\widehat{\bphi})\}^{-1}
\bJ_{0n}\transpose\scrP_\bZ\transpose\|_\infty = \Optilde(m^2)$ by Lemma \ref{lemma:integrated_jacobian_perturbation} and Proposition \ref{prop:local_hessian_bound}, since $\|\widehat{\bB}\|\sup_\tau\|\bar{\bH}^{-1}\|_\infty^4\|\bH(\tau)\|_\infty^3 = \Optilde(m^{7/2}n^{-1/2})$. 

\noindent
\underline{We next rigorously establish an error bound on $\|\mathscr{P}_{\br}\bJ_{0n}\bar{\mathscr{H}}^{-1}\bJ_{0n}\transpose \scrP_\br\transpose\|_\infty$.}
Proposition \ref{prop:local_hessian_bound} also implies that 
\begin{align*}
  \|\scrP_\br\bJ_{0n}\bar{\scrH}^{-1}\bJ_{0n}\transpose\scrP_\br\transpose\|_\infty\leq \sqrt{d_\br}\|\scrP_\br\bJ_{0n}\bar{\scrH}^{-1}\bJ_{0n}\transpose\scrP_\br\transpose\|_2 = \Optilde(\sqrt{m}).
\end{align*}
Also, the proof of Lemma \ref{lemma:off_diagonal_term_perturbation} shows that $\sup_\bpsi\|\bGamma_{\bZ\bR}^\sharp(\bpsi)\|_{2\to\infty} = \Optilde(n^{-1/2})$, and Lemma \ref{lemma:score_term_perturbation} with $a_n = c_n = 1,b_n = 0,\eps_n = O(n^{-1/2})$ shows that
\begin{align*}
&\sup_{\bphi\in\calN(\sqrt{m}n\eps_n, M)}\bigg\|\scrP_{\bZ}\bigg\{\bJ_{0n}\bLambda_\sharp\sum_{t = 1}^m\sum_{i\neq j}\ell'({\theta}_{ij}^{(t)}; A_{ij}^{(t)})\frac{\partial^2\theta_{ij}^{(t)}}{\partial\bphi\partial\bphi\transpose}\bLambda_\sharp\bJ_{0n}\transpose\bigg\}\scrP_{\br}\transpose\bigg\|_{\infty}\\
&\quad\leq \sup_{\bphi\in\calN(\sqrt{m}n\eps_n, M)}\sqrt{d_\br}\bigg\|\scrP_\bZ\bigg\{\bJ_{0n}\bLambda_\sharp\sum_{t = 1}^m\sum_{i\neq j}\ell'({\theta}_{ij}^{(t)}; A_{ij}^{(t)})\frac{\partial^2\theta_{ij}^{(t)}}{\partial\bphi\partial\bphi\transpose}\bLambda_\sharp\bJ_{0n}\transpose\bigg\}\scrP_\br\transpose\bigg\|_{2\to\infty} = \Optilde\bigg(\frac{\sqrt{m}\log n}{\sqrt{n}}\bigg).
\end{align*}
These two results imply that $\|\scrP_\bZ\bJ_{0n}\bar{\scrH}\bJ_{0n}\transpose\scrP_{\br}\transpose\|_\infty = \Optilde(\log n\sqrt{m/n})$. Also, denote 
\begin{align*}
\bar{\scrH}_{11} = \scrP_{\balpha\bB}\bar{\scrH}\scrP_{\balpha\bB}\transpose,\quad 
\bJ_{\bZ} = \mathrm{diag}\bigg(\eye_{d_\balpha}, \frac{1}{\sqrt{n}}\eye_d\otimes \bX_{0\perp}\bigg).
\end{align*}
Since $\bJ_\bZ\scrP_{\balpha\bB} = \scrP_\bZ\bJ_{0n}$ and $\scrP_{\bZ} = \scrP_{\bZ}\mathrm{diag}(\eye_{d_\balpha}, \mathbf{O}_{d_\br\times d_\br}, \eye_{d_\bX})$,
the same reasoning also shows that
\begin{align*}
&\|\bJ_{\bZ}(\scrP_{\balpha\bB}\bar{\scrH}\scrP_{\balpha\bB}\transpose)^{-1}
\bJ_{\bZ}\transpose\|_\infty\\
&\quad \leq    
\left\|\bJ_{\bZ}\{\scrP_{\balpha\bB}(\scrI \bDelta_\bphi^\sharp)(\widehat{\bphi})\scrP_{\balpha\bB}\transpose\}^{-1}
\bJ_{\bZ}\transpose\right\|_\infty\\
&\qquad + \sqrt{d_\bpsi}\left\|\bJ_{\bZ}\{\scrP_{\balpha\bB}(\scrI \bDelta_\bphi^\sharp)(\widehat{\bphi})\scrP_{\balpha\bB}\transpose\}^{-1}
\bJ_{\bZ}\transpose\right\|_\infty\\
&\qquad\times \left\|\scrP_{\bZ}
\mathrm{diag}(\eye_{d_\balpha},\mathbf{O}_{ d_\br \times d_\br }, \eye_{d_\bX})
\scrI\left(\bJ_{0n}\bJ_n\transpose\bPi(\bGamma_{\bpsi\bpsi}^\sharp(\bpsi(\cdot)) - \bDelta_{\bpsi}^\sharp(\bpsi(\cdot)))\bPi\transpose\bJ_n\bJ_{0n}\transpose\right)(\widehat{\bphi})\scrP_\bZ\transpose\right\|_{2\to\infty}\\
&\qquad\times  \|\bJ_{\bZ}(\scrP_{\balpha\bB}\bar{\scrH}\scrP_{\balpha\bB}\transpose)^{-1}
\bJ_{\bZ}\transpose\|_2\\
&\qquad + \frac{\sqrt{d_\bpsi}}{2}\left\|\bJ_{\bZ}\{\scrP_{\balpha\bB}(\scrI \bDelta_\bphi^\sharp)(\widehat{\bphi})\scrP_{\balpha\bB}\transpose\}^{-1}
\bJ_{\bZ}\transpose\right\|_\infty\\
&\qquad\times 
\bigg\|
\scrP_\bZ
\mathrm{diag}(\eye_{d_\balpha},\mathbf{O}_{ d_\br \times d_\br }, \eye_{d_\bX})\scrI\bigg( 
\bJ_{0n}\bLambda_\sharp\sum_{t = 1}^m\sum_{i\neq j}\ell'(\theta_{ij}^{(t)}(\cdot); A_{ij}^{(t)})\frac{\partial^2\theta_{ij}^{(t)}}{\partial\bphi\partial\bphi\transpose}\bLambda_\sharp\bJ_{0n}\transpose - \bXi
\bigg)(\widehat{\bphi})\scrP_\bZ\transpose\bigg\|_{2\to\infty}\\
&\qquad\times
\|\bJ_{\bZ}(\scrP_{\balpha\bB}\bar{\scrH}\scrP_{\balpha\bB}\transpose)^{-1}\bJ_{\bZ}\transpose\|_2\\
&\qquad + \frac{1}{2}\left\|\bJ_{\bZ}\{\scrP_{\balpha\bB}(\scrI \bDelta_\bphi^\sharp)(\widehat{\bphi})\scrP_{\balpha\bB}\transpose\}^{-1}
\bJ_{\bZ}\transpose\right\|_\infty
\left\|\mathrm{diag}(\eye_{d_\balpha},\mathbf{O}_{d_\br\times d_\br},\eye_{d_\bX})(\scrI \bXi)(\widehat{\bphi})\mathrm{diag}(\eye_{d_\balpha},\mathbf{O}_{d_\br\times d_\br},\eye_{d_\bX})\right\|_\infty\\
&\qquad\times 
\| \bJ_{\bZ}(\scrP_{\balpha\bB}\bar{\scrH}\scrP_{\balpha\bB}\transpose)^{-1}
\bJ_{\bZ}\transpose\|_\infty\\
&\quad = \Optilde(m^{5/2}\vee m^2\sqrt{\log n}) + \Optilde(m^{2}n^{-1/2}\log n)\|\bJ_{\bZ}(\scrP_{\balpha\bB}\bar{\scrH}\scrP_{\balpha\bB}\transpose)^{-1}
\bJ_{\bZ}\transpose\|_\infty,
\end{align*}
so that
\begin{align*}
\|\bJ_{\bZ}(\scrP_{\balpha\bB}\bar{\scrH}\scrP_{\balpha\bB}\transpose)^{-1}
\bJ_{\bZ}\transpose\|_\infty = \Optilde(m^{5/2}\vee m^2\sqrt{\log n}).
\end{align*}
\underline{We then apply the block matrix inverse formula to bound $\|\scrP_\bZ\bJ_{0n}\bar{\scrH}^{-1}\bJ_{0n}\transpose\mathrm{diag}(\eye_{d_\balpha}, \sqrt{n}\eye_{d_\br}, \eye_{d_\bX})\|_\infty$. }
By the block matrix inverse formula, we further obtain
\begin{align*}
\|\scrP_\bZ\bJ_{0n}\bar{\scrH}^{-1}\bJ_{0n}\transpose\scrP_\br\transpose\|_\infty 
&\leq \|\bJ_{\bZ}\bar{\scrH}_{11}^{-1}\bJ_\bZ\transpose\|_\infty \|\scrP_\bZ\bJ_{0n}\bar{\scrH}\bJ_{0n}\transpose\scrP_\br\transpose\|_\infty\|\scrP_\br\bJ_{0n}\bar{\scrH}^{-1}\bJ_{0n}\transpose\scrP_\br\transpose\|_\infty\\
& = \Optilde\bigg\{\frac{m^{7/2}\log n + m^{3}(\log n)^{3/2}}{\sqrt{n}}\bigg\}.
\end{align*}
This enables us to write
\begin{align*}
\left\|\mathscr{P}_{\bZ}\bJ_{0n}\bar{\mathscr{H}}^{-1}\bJ_{0n}\transpose 
\mathrm{diag}(\eye_{d_\balpha}, \sqrt{n}\eye_{d_\br}, \eye_{d_\bX})
\right\|_\infty
&\leq \left\|\mathscr{P}_{\bZ}\bJ_{0n}\bar{\mathscr{H}}^{-1}\bJ_{0n}\transpose\scrP_\bZ\transpose\right\|_\infty + 
\sqrt{n}\left\| \scrP_\bZ\bJ_{0n}\bar{\scrH}^{-1}\bJ_{0n}\transpose\scrP_\br\transpose
\right\|_\infty\\
& = \Optilde\{m^{7/2}\log n + m^{3}(\log n)^{3/2}\}.
\end{align*}

\noindent
$\blacksquare$ \textbf{Step 2. Bounding $\|\mathrm{diag}(\sqrt{n}\eye_{d_\balpha}, \sqrt{mn}\eye_{d_\bX})\scrP_\bZ\bgamma\|_\infty$ and $\max_{t\in[m]}\|\widehat{\bR}^{(t)} - \bR_0^{(t)}\|_{\mathrm{F}}$. }

\noindent Recall the remainder decomposition \eqref{eqn:remainder_decomposition}. Consider
\[
  \bigg\|\bJ_{0n}(\bG_0^\sharp)^{-1}\bJ_{0n}\transpose\bJ_{0n}\bLambda_\sharp
\frac{\partial\calL}{\partial\bphi}(\bphi_0)\bigg\|_\infty.
\]
Since $\bJ_{0n}\transpose\bJ_{0n}\bJ_{0n}\transpose = \bJ_{0n}\transpose$, the 
$i$th entry of the vector inside $\|\cdot\|_\infty$ is 
\[
\frac{1}{2}\sum_{t = 1}^m\sum_{a\neq b}\ell'(\theta_{0ab}^{(t)};
A_{ab}^{(t)})\be_i\transpose\bJ_{0n}(\bG_0^\sharp)^{-1}\bJ_{0n}\transpose
\bu_{tab},
\]
so Bernstein's inequality and a union bound over $i\in[d_\bpsi]$ yield  
\begin{align}
\label{eqn:J0n_s_hat_bound}
\bigg\|\bJ_{0n}(\bG_0^\sharp)^{-1}\bJ_{0n}\transpose\bJ_{0n}\bLambda_\sharp
\frac{\partial\calL}{\partial\bphi}(\bphi_0)\bigg\|_\infty = \Optilde(\sqrt{\log n}).
\end{align}
By Lemma \ref{lemma:local_lipschitz_sample_Fisher_information_infinity_norm}, Lemma \ref{lemma:local_lipschitz_hessian_score_term_infinity_norm}, Lemma \ref{lemma:inverse_Fisher_information_matrix_infty_norm_bound}, and Lemma \ref{lemma:score_uniform_convergence}, we have
\begin{align*}
\left\|\mathrm{diag}(\sqrt{n}\eye_{d_\balpha}, \sqrt{mn}\eye_{d_\bX})\bgamma_\bZ^{(1)}\right\|_\infty
&\leq \|\mathscr{P}_{\bZ}\bJ_{0n}\bar{\mathscr{H}}^{-1}\bJ_{0n}\transpose\mathrm{diag}(\eye_{d_\balpha},\sqrt{n}\eye_{d_\br},\eye_{d_\bX})\|_\infty
\Optilde\bigg(\frac{m^{5/2}\sqrt{\log n} + m^2\log n}{\sqrt{n}}\bigg)\\
&\quad\times \bigg\|\bJ_{0n}(\bG_0^\sharp)^{-1}\bJ_{0n}\transpose\bJ_{0n}\bLambda_\sharp
\frac{\partial\calL}{\partial\bphi}(\bphi_0)\bigg\|_\infty\\
&\leq \Optilde\bigg\{\frac{m^{6}(\log n)^2 + m^{11/2}(\log n)^{5/2} + m^{5}(\log n)^3}{\sqrt{n}}\bigg\}. 
\end{align*}
By Lemma \ref{lemma:score_term_hessian_at_truth_concentration_infintiy_norm}, we have
\begin{align*}
\left\|\mathrm{diag}(\sqrt{n}\eye_{d_\balpha}, \sqrt{mn}\eye_{d_\bX})\bgamma_\bZ^{(2)}\right\|_\infty = \Optilde\bigg\{\frac{m^{7/2}(\log n)^2 + m^3(\log n)^{5/2}}{\sqrt{n}}\bigg\}.
\end{align*}
By Lemma \ref{lemma:fisher_information_at_truth_concentration_infintiy_norm}, we have
\begin{align*}
\left\|\mathrm{diag}(\sqrt{n}\eye_{d_\balpha}, \sqrt{mn}\eye_{d_\bX})\bgamma_\bZ^{(3)}\right\|_\infty = \Optilde\bigg\{\frac{m^4(\log n)^2 + m^{7/2}(\log n)^{5/2}}{\sqrt{n}}\bigg\}.
\end{align*}
We conclude that the remainder $\bgamma_\bZ$ satisfies 
\[
\|\mathrm{diag}(\sqrt{n}\eye_{d_\balpha}, \sqrt{mn}\eye_{d_\bX})\bgamma_\bZ\|_\infty = \Optilde\bigg\{\frac{m^{6}(\log n)^2 + m^{11/2}(\log n)^{5/2} + m^{5}(\log n)^3}{\sqrt{n}}\bigg\}. 
\]
To further bound $\max_t\|\widehat{\bR}^{(t)} - \bR_0^{(t)}\|_{\mathrm{F}}$, by \eqref{eqn:Taylor_expansion_remainder}--\eqref{eqn:remainder_decomposition}, 
\[
n(\widehat\br - \br_0) = \scrP_\br\bJ_{0n}(\bG_0^\sharp)^{-1}\bJ_{0n}\transpose\bJ_{0n}\bLambda_\sharp\frac{\partial\calL}{\partial\bphi}(\bphi_0) + n\scrP_\br\bgamma,
\] 
where we recall the notation $\widehat{\br} = [\vech(\widehat{\bR}^{(1)})\transpose,\ldots,\vech(\widehat{\bR}^{(m)})\transpose]\transpose$, $\br_0 = [\vech(\bR_0^{(1)})\transpose,\ldots,\vech(\bR_0^{(m)})\transpose]\transpose$, and the $\ell_\infty$-norm of the first term is
$\Optilde(\sqrt{\log n})$ by the Bernstein's inequality and a union bound. A simple algebra shows that \eqref{eqn:remainder_decomposition} holds with $\scrP_\bZ$ replaced by $\scrP_\br$ and the first factor $\mathrm{diag}\{n^{-1/2}\eye_{d_\balpha}, (mn)^{-1/2}\eye_{d_\bX}\}$ on the right-hand sides of $\bgamma_{\bZ}^{(1)},\bgamma_\bZ^{(2)},\bgamma_\bZ^{(3)}$ replaced by $n^{-1}$, thereby implying that
\begin{align*}
\|n\scrP_\br\bgamma\|_\infty&\leq\|\scrP_\br\bJ_{0n}\bar\scrH^{-1}\bJ_{0n}\transpose\mathrm{diag}(\eye_{d_\balpha}, \sqrt n\eye_{d_\br}, \eye_{d_\bX})\|_\infty\\
&\quad\times\Optilde\bigg\{\frac{(m^{5/2}\sqrt{\log n} + m^2\log n)\sqrt{\log n} + \sqrt m\log n + \log n}{\sqrt n}\bigg\},
\end{align*}
where the second factor is obtained by Lemma \ref{lemma:local_lipschitz_sample_Fisher_information_infinity_norm}--\ref{lemma:score_term_hessian_at_truth_concentration_infintiy_norm} since it does not depend on the projection. Since
$\|\scrP_\br\bJ_{0n}\bar\scrH^{-1}\bJ_{0n}\transpose\scrP_\bZ\transpose\|_\infty\leq\sqrt{d_\balpha + d_\bX}\|\bJ_{0n}\bar\scrH^{-1}\bJ_{0n}\transpose\|_2 = O(\sqrt{mn})$ and $\sqrt n\|\scrP_\br\bJ_{0n}\bar\scrH^{-1}\bJ_{0n}\transpose\scrP_\br\transpose\|_\infty = O(\sqrt{mn})$ by Proposition
\ref{prop:local_hessian_bound} and the equivalence between $\|\cdot\|_2$ and $\|\cdot\|_\infty$ for a $d_\br\times d_\br$ matrix (for the second bound), we obtain 
\begin{align}
\label{eqn:Rhat_bound}
\max_{t\in[m]}\|\widehat\bR^{(t)} - \bR_0^{(t)}\|_{\mathrm F} = \Optilde\bigg\{\frac{(m^3\log n + m^{5/2}(\log n)^{3/2})}{n}\bigg\}.  
\end{align}

\noindent
$\blacksquare$
\textbf{Step 3. Completing the proof.} 
Without loss of generality, we may assume that $\bX_0\transpose\widehat{\bX}$ is symmetric positive definite, so that $\bW_{\widehat{\bX}} = \eye_d$. Let $\widehat{\bx}_i$ denotes the $i$th row of $\widehat{\bX}$. 
We first note that for $\bX(\widehat{\bB})$, we have
\begin{align*}
\bX(\widehat{\bB}) - \bX_0 & = \bX_0\{(\eye_d + \widehat{\bB}\transpose\widehat{\bB})^{-1/2} - \eye_d\} + \bX_{0\perp}\widehat{\bB} + \bX_{0\perp}\widehat{\bB}\{(\eye_d + \widehat{\bB}\transpose\widehat{\bB})^{-1/2} - \eye_d\}.
\end{align*}
With $\|\widehat{\bB}\|_2 = \Optilde(n^{-1/2})$ from Proposition \ref{prop:consistency_native_norm} and $\|\bX_{0\perp}\widehat{\bB}\|_{2\to\infty}\leq M/2$, we have 
\begin{align*}
\|\bX(\widehat{\bB}) - \bX_0 - \bX_{0\perp}\widehat{\bB}\|_{\mathrm{F}} = \Optilde(n^{-1/2}),\quad 
\widehat{\bx}_i - \bx_{0i} = \widehat{\bB}\transpose\bX_{0\perp}\transpose\be_i + \Optilde(n^{-1}). 
\end{align*}
Denote $\bg_{ni} = [\sqrt{n}\bg_{\balpha_i}\transpose, \sqrt{mn}\bg_{\bx_i}\transpose]\transpose$. Then, our target functional in item (i) can be written as
\begin{align*}
\sum_{i\in\calI}\bg_{ni}\transpose(\widehat{\bz}_i - \bz_{0i})
& = 
\sum_{i\in\calI}\left\{\sqrt{n}\bg_{\balpha_i}\transpose(\widehat{\balpha}_i - \balpha_{0i}) + 
\sqrt{mn}\bg_{\bx_i}\transpose(\widehat{\bx}_i - \bx_{0i})\right\}  \\
& = \sum_{i\in\calI}\left\{\sqrt{n}\bg_{\balpha_i}\transpose(\widehat{\balpha}_i - \balpha_{0i}) + 
\sqrt{mn}(\bg_{\bx_i}\transpose\otimes\be_i\transpose\bX_{0\perp})\vect(\widehat{\bB})\right\} + \Optilde(\sqrt{m/n})\\
& = \sum_{i\in\calI}\left\{\bg_i\transpose\mathrm{diag}(\sqrt{n}\eye_m, \sqrt{mn}\eye_d)\scrP_{\bz_i}\bJ_0(\widehat{\bphi} - \bphi_0)\right\} + \Optilde(\sqrt{m/n})\\
& = \sum_{i\in\calI}\left\{\bg_i\transpose\scrP_{\bz_i}\bJ_{0n}(\bG_0^\sharp)^{-1}\bLambda_\sharp\frac{\partial\calL}{\partial\bphi}(\bphi_0)\right\} + \Optilde(\|\bg\transpose\|_\infty\vartheta_n)\\
& = \sum_{i\in\calI}\sum_{t = 1}^m\sum_{a < b}\ell'(\theta_{0ab}^{(t)}; A_{ab}^{(t)})\bg_i\transpose\scrP_{\bz_i}\bJ_{0n}(\bG_0^\sharp)^{-1}\bJ_{0n}\transpose\bu_{tab} + \Optilde(\|\bg\transpose\|_\infty\vartheta_n),
\end{align*}
where we denote 
\[
\vartheta_n = \frac{m^{6}(\log n)^2 + m^{11/2}(\log n)^{5/2} + m^{5}(\log n)^3}{\sqrt{n}},\quad
\bu_{tab} = \begin{bmatrix}
\frac{1}{\sqrt{n}}\be_t\otimes(\be_a + \be_b)\\
\frac{1}{n}\be_t\otimes\bD_d\transpose(\bx_{0b}\otimes\bx_{0a})\\
\frac{1}{\sqrt{mn}}(\bR_0^{(t)}\bx_{0b}\otimes\be_a + \bR_0^{(t)}\bx_{0a}\otimes \be_b)
\end{bmatrix}.
\]
For the functional of interest in item (ii), we 
denote by $\widehat{\bN} = (\eye_d + \widehat{\bB}\transpose\widehat{\bB})^{-1/2}$ and write $\bF = \sum_{i,j\in[n]}F_{ij}\be_i\be_j\transpose$, where $F_{ij}$ denotes the $(i, j)$th element of $\bF$. 
Under the conditions that $\|\bF\bX_0\|_{\mathrm{F}} = 1$, $\|\bX_0\transpose\bF\bX_0\|_2 = O(1)$, and $\|\vect(\bF)\|_1 = O(1)$, by von Neumann's trace inequality, Cauchy-Schwarz inequality and the proof of Proposition \ref{prop:stationarity_uniqueness_uniform_consistency_MLE}, we have
\begin{align*}
|\mathrm{tr}(\widehat{\bB}\transpose\bX_{0\perp}\transpose\bF\bX_{0\perp}\widehat{\bB})|
&\leq  \sum_{i,j\in[n]}|F_{ij}|\|\bX_{0\perp}\widehat{\bB}\|_{2\to\infty}^2\leq  \|\bJ_0(\widehat{\bphi} - \bphi_0)\|_\infty^2\|\vect(\bF)\|_1 = \Optilde(\delta_n^2),\\
\|\widehat{\bN} - \eye_d\|_2&\leq \|\widehat{\bB}\|_2^2 = \Optilde(n^{-1}),
\end{align*}
where $\delta_n = (m^2\sqrt{\log n} + m^{3/2}\log n)/\sqrt{n}$, and 
\begin{align*}
&\left\|(\bX_0 + \bX_{0\perp}\widehat{\bB})\transpose\bF(\bX_0 + \bX_{0\perp}\widehat{\bB})
\right\|_2\\
&\quad \leq \|\bX_0\transpose\bF\bX_0\|_2 + 2\|\bX_0\transpose\bF\bX_{0\perp}\widehat{\bB}\|_2  + \|\widehat{\bB}\transpose\bX_{0\perp}\transpose\bF\bX_{0\perp}\widehat{\bB}\|_2\\
&\quad \leq O(1) + 2 \|\bX_0\transpose\bF\|_2\|\bX_{0\perp}\|_2\|\widehat{\bB}\|_2 + d^2\|\widehat{\bB}\transpose\bX_{0\perp}\transpose\bF\bX_{0\perp}\widehat{\bB}\|_{\max}\\
&\quad \leq O(1) + 2\Optilde(n^{-1/2})\|\bX_0\transpose\bF\|_2\|\bX_{0\perp}\|_2 + d\|\bX_{0\perp}\widehat{\bB}\|_{2\to\infty}^2\|\vect(\bF)\|_1  = \Optilde(1),
\end{align*}
which implies that
\begin{align*}
\mathrm{tr}(\widehat{\bX}\transpose\bF\widehat{\bX})
& = \mathrm{tr}(\bX_0\transpose\bF\bX_0) + 2\mathrm{tr}(\bX_0\transpose\bF\bX_{0\perp}\widehat{\bB}) + \mathrm{tr}(\widehat{\bB}\transpose\bX_{0\perp}\transpose\bF\bX_{0\perp}\widehat{\bB})\\
&\quad + 2\mathrm{tr}\left\{(\widehat{\bN} - \eye_d)(\bX_0 + \bX_{0\perp}\widehat{\bB})\transpose\bF(\bX_0 + \bX_{0\perp}\widehat{\bB})
\right\}\\
&\quad  + \mathrm{tr}\left\{(\widehat{\bN} - \eye_d)(\bX_0 + \bX_{0\perp}\widehat{\bB})\transpose\bF(\bX_0 + \bX_{0\perp}\widehat{\bB})(\widehat{\bN} - \eye_d)
\right\}\\
& = \mathrm{tr}(\bX_0\transpose\bF\bX_0) + 2\mathrm{tr}(\bX_0\transpose\bF\bX_{0\perp}\widehat{\bB}) + \Optilde(\delta_n^2).
\end{align*}
Also, it is clear that $\|\vect(\bF\bX_0)\|_1\leq \sum_{k = 1}^d\|\sum_{i,j\in[n]}F_{ij}\be_i\be_j\transpose\bX_0\be_k(d)\|_1\leq d \sum_{i,j\in[n]}|F_{ij}|\|\bX_0\|_{\max} = O(1)$.
It follows that
\begin{align*}
\langle \bF, \widehat{\bX}\widehat{\bX}\transpose - \bX_0\bX_0\transpose\rangle_{\mathrm{F}}
& = 2\mathrm{tr}(\bX_0\transpose\bF\bX_{0\perp}\widehat{\bB}) + \Optilde(\delta_n^2) = 2\vect(\bF\bX_0)\transpose(\eye_d\otimes\bX_{0\perp})\vect(\widehat{\bB}) + \Optilde(\delta_n^2)\\
& = 2\vect(\bF\bX_0)\transpose \scrP_\bX\bJ_0(\widehat{\bphi} - \bphi_0) + \Optilde(\delta_n^2)\\
& = 2\vect(\bF\bX_0)\transpose \scrP_\bX\bJ_0\bG(\bphi_0)^{-1}\frac{\partial\calL}{\partial\bphi}(\bphi_0) + \Optilde\bigg(\frac{ \vartheta_n}{\sqrt{mn}}\bigg)\\
& = \frac{1}{\sqrt{mn}}2\vect(\bF\bX_0)\transpose\scrP_\bX\bJ_{0n}(\bG_0^\sharp)^{-1}\bLambda_\sharp\frac{\partial\calL}{\partial\bphi}(\bphi_0) + \Optilde\bigg(\frac{ \vartheta_n}{\sqrt{mn}}\bigg)\\
& = \frac{1}{\sqrt{mn}}\sum_{t = 1}^m\sum_{i < j}2\ell'(\theta_{0ij}^{(t)}; A_{ij}^{(t)})\vect(\bF\bX_0)\transpose\scrP_\bX\bJ_{0n}(\bG_0^\sharp)^{-1}\bJ_{0n}\transpose\bu_{tij}
 + \Optilde\bigg(\frac{ \vartheta_n}{\sqrt{mn}}\bigg).
\end{align*}
Below, we focus on establishing two arguments: the leading terms of the two functionals of interest satisfy the Lindeberg condition for the Lindeberg-Feller central limit theorem, and their asymptotic variances are independent of the choice of our reparameterization based on the generalized Cayley transform \eqref{eqn:Cayley_transform}. 

\noindent\underline{Lindeberg condition.} 
A direct calculation shows that
\begin{align*}
\var\bigg\{\sum_{i\in\calI}\bg_i\transpose\scrP_{\bz_i}\bJ_{0n}(\bG_0^\sharp)^{-1}\bLambda_\sharp\frac{\partial\calL}{\partial\bphi}(\bphi_0)\bigg\}
& = \bigg(\sum_{i\in\calI}\bg_i\transpose\scrP_{\bz_i}\bigg)\bJ_{0n}(\bG_0^\sharp)^{-1}\bJ_{0n}\transpose\bigg(\sum_{i\in\calI}\scrP_{\bz_i}\transpose\bg_i\bigg),\\
\var\bigg\{
2\vect(\bF\bX_0)\transpose\scrP_\bX\bJ_{0n}(\bG_0^\sharp)^{-1}\bLambda_\sharp\frac{\partial\calL}{\partial\bphi}(\bphi_0)
\bigg\}
& = 4\vect(\bF\bX_0)\transpose\scrP_\bX\bJ_{0n}(\bG_0^\sharp)^{-1}\bJ_{0n}\transpose\scrP_\bX\transpose\vect(\bF\bX_0)
\end{align*}
are bounded away from $0$ and $+\infty$. 
It is sufficient to consider the sums of the absolute third moments of the summands converge to zero as $n\to\infty$. By Proposition \ref{prop:hessian_bound} and its proof, $\|\bJ_{0n}\transpose\sum_{t,a,b}\bu_{tab}\bu_{tab}\transpose\bJ_{0n}\|_2 = O(1)$. Then we have
\begin{align*}
&\sum_{t = 1}^m\sum_{a < b}\expect\left|\sum_{i\in\calI}\ell'(\theta_{0ab}^{(t)}; A_{ab}^{(t)})\bg_i\transpose\scrP_{\bz_i}\bJ_{0n}(\bG_0^\sharp)^{-1}\bJ_{0n}\transpose\bu_{tab}\right|^3\\
&\quad\lesssim \bigg\|\bigg(\sum_{i\in\calI}\bg_i\transpose\scrP_{\bz_i}\bigg)\bigg\|_2^3\|\bJ_{0n}(\bG_0^\sharp)^{-1}\|_2^3\bigg(\max_{t\in[m],a,b\in[n]}\|\bu_{tab}\|_2\bigg)\bigg\|\bJ_{0n}\transpose\sum_{t = 1}^m\sum_{a \neq b}\bu_{tab}\bu_{tab}\transpose\bJ_{0n}\bigg\|_2 = O(n^{-1/2}),\\
&\sum_{t = 1}^m\sum_{i < j}\expect\left|\ell'(\theta_{0ij}^{(t)}; A_{ij}^{(t)})\vect(\bF\bX_0)\transpose\scrP_{\bX}\bJ_{0n}(\bG_0^\sharp)^{-1}\bJ_{0n}\transpose\bu_{tij}\right|^3\\
&\quad\lesssim\|\vect(\bF\bX_0)\transpose\|_2^3\|\scrP_{\bX}\bJ_{0n}(\bG_0^\sharp)^{-1}\|_2^3\bigg(\max_{t\in[m],i,j\in[n]}\|\bu_{tij}\|_2\bigg)\bigg\|\bJ_{0n}\transpose\sum_{t = 1}^m\sum_{i \neq j}\bu_{tij}\bu_{tij}\transpose\bJ_{0n}\bigg\|_2 = O(n^{-1/2}),
\end{align*}
so that the Lindeberg condition holds. Then, by Lindeberg-Feller central limit theorem (in the form of Theorem 7.1.2 in \cite{Chung_2001_ProbabilityTheory}) and Slutsky's theorem, 
\begin{align*}
\frac{\sum_{i\in\calI}\bg_{ni}\transpose(\widehat{\bz}_i - \bz_{0i})}{\sqrt{(\sum_{i\in\calI}\bg_i\transpose\scrP_{\bz_i})\bJ_{0n}(\bG_0^\sharp)^{-1}\bJ_{0n}\transpose(\sum_{i\in\calI}\scrP_{\bz_i}\transpose\bg_i)}}
&\overset{\calL}{\to}\mathrm{N}(0, 1)\quad\text{as }n\to\infty,\\
\frac{\sqrt{mn}\langle\bF, \widehat{\bX}\widehat{\bX}\transpose - \bX_0\bX_0\transpose\rangle}{\sqrt{4\vect(\bF\bX_0)\transpose\scrP_\bX\bJ_{0n}(\bG_0^\sharp)^{-1}\bJ_{0n}\transpose\scrP_\bX\transpose\vect(\bF\bX_0)}}
&\overset{\calL}{\to}\mathrm{N}(0, 1)\quad\text{as }n\to\infty.
\end{align*}
\underline{Asymptotic variances are independent of the reparameterization.} We now argue that these asymptotic variances can be approximated without introducing the generalized Cayley transform \eqref{eqn:Cayley_transform}. Recall that $\bG_0^\sharp = \bJ_{0n}\transpose\bPi\bG_{\bpsi\bpsi}^\sharp(\bpsi_0)\bPi\transpose\bJ_{0n}$ and $\bG_{\bpsi\bpsi}^\sharp(\bpsi_0)$ is independent of the reparameterization. Recall the definition of $\bDelta_{\bpsi}^\sharp(\bpsi_0)$ in \eqref{eqn:Deltapsi_sharp}, which does not depend on the reparameterization either. A simple algebra shows that
\begin{align*}
&\bJ_{0n}(\bG_0^\sharp)^{-1}\bJ_{0n}\transpose + \bJ_{0n}\{\bJ_{0n}\transpose\bPi\expect_0\bDelta_\bpsi^\sharp(\bpsi_0)\bPi\transpose\bJ_{0n}\}^{-1}\bJ_{0n}\transpose\\
&\quad = \bJ_{0n}(\bG_0^\sharp)^{-1}\bJ_{0n}\transpose\bPi\left\{\bG_{\bpsi\bpsi}^\sharp(\bpsi_0) + \expect_0\bDelta_\bpsi^\sharp(\bpsi_0)\right\}\bPi\transpose\bJ_{0n}\{\bJ_{0n}\transpose\bPi\expect_0\bDelta_\bpsi^\sharp(\bpsi_0)\bPi\transpose\bJ_{0n}\}^{-1}\bJ_{0n}\transpose
\end{align*}
By definition of $\bGamma_{\bpsi\bpsi}^\sharp$ and $\bDelta_\bpsi^\sharp$, 
\begin{align*}
\|\bG_{\bpsi\bpsi}^\sharp(\bpsi_0) + \expect_0\bDelta_\bpsi^\sharp(\bpsi_0)\|_{\max}
&\leq \max_{i,j\in[n]}\left\|\frac{1}{n}\sum_{t = 1}^m\expect_0\ell''(\theta_{0ij}^{(t)}; A_{ij}^{(t)})\begin{bmatrix}
\be_t\be_t\transpose & \frac{\be_t\bx_{0i}\transpose\bR_0^{(t)}}{\sqrt{m}} \\
\frac{\bR_0^{(t)}\bx_{0i}\be_t\transpose}{\sqrt{m}} & \frac{\bR_0^{(t)}\bx_{0j}\bx_{0i}\transpose\bR_0^{(t)}}{m}
\end{bmatrix}\right\|_{\max}\\
&\quad + \max_{t\in[m],i\in[n]}\left\|
\sum_{j\in[n]\backslash\{i\}}\frac{\expect_0\ell''(\theta_{0ij}^{(t)}; A_{ij}^{(t)})}{2n^{3/2}}\begin{bmatrix}
\be_t\\ \frac{\bR_0^{(t)}\bx_{0j}}{\sqrt{m}}
\end{bmatrix}\|\bx_{0j}\|_2\|\bx_{0i}\|_2\|\bD_d\|_2
\right\|_{\max}\\
& = O(n^{-1/2}).
\end{align*}
Then, by Lemma \ref{lemma:inverse_Fisher_information_matrix_infty_norm_bound} and its proof, we further obtain
\begin{align*}
&\max_{a,b\in[d_\balpha + d_\bX]}|\be_a\transpose[\bJ_{0n}(\bG_0^\sharp)^{-1}\bJ_{0n}\transpose + \bJ_{0n}\{\bJ_{0n}\transpose\bPi\expect_0\bDelta_\bpsi^\sharp(\bpsi_0)\bPi\transpose\bJ_{0n}\}^{-1}\bJ_{0n}\transpose]\be_b|\\
&\quad \leq \max_{a\in[d_\balpha + d_\bX]}\|\be_a\transpose\bJ_{0n}(\bG_0^\sharp)^{-1}\bJ_{0n}\transpose\|_\infty\|\bG_{\bpsi\bpsi}^\sharp(\bpsi_0) + \expect_0\bDelta_\bpsi^\sharp(\bpsi_0)\|_{\max}\|\bJ_{0n}\{\bJ_{0n}\transpose\bPi\expect_0\bDelta_\bpsi^\sharp(\bpsi_0)\bPi\transpose\bJ_{0n}\}^{-1}\bJ_{0n}\transpose\|_\infty\\
&\quad = O\bigg(\frac{m^{5/2}}{\sqrt{n}}\bigg).
\end{align*}
Furthermore, $\mathrm{Span}\{(\bPi\expect_0\bDelta_\bpsi^\sharp(\bpsi_0)\bPi\transpose)^{1/2}\bJ_{0n}\}^\perp = \mathrm{Span}\{(\bPi\expect_0\bDelta_\bpsi^\sharp(\bpsi_0)\bPi\transpose)^{-1/2}\bV_\bB\}$ where $\bV_\bB$ has orthonormal columns spanning the orthogonal complement of $\mathrm{Span}(\bJ_{0n})$. This implies that
\begin{align*}
  &\{\bPi\expect_0\bDelta_\bpsi^\sharp(\bpsi_0)\bPi\transpose\}^{1/2}\bJ_{0n}\{\bJ_{0n}\transpose\bPi\expect_0\bDelta_\bpsi^\sharp(\bpsi_0)\bPi\transpose\bJ_{0n}\}^{-1}\bJ_{0n}\transpose\{\bPi\expect_0\bDelta_\bpsi^\sharp(\bpsi_0)\bPi\transpose\}^{1/2}\\
  &\quad + \{\bPi\expect_0\bDelta_\bpsi^\sharp(\bpsi_0)\bPi\transpose\}^{-1/2}\bV_\bB[\bV_\bB\transpose\{\bPi\expect_0\bDelta_\bpsi^\sharp(\bpsi_0)\bPi\transpose\}^{-1}\bV_\bB]^{-1}\bV_\bB\transpose\{\bPi\expect_0\bDelta_\bpsi^\sharp(\bpsi_0)\bPi\transpose\}^{-1/2} = \eye_{d_\bpsi},
\end{align*}
so that
\begin{align*}
  &\|\bJ_{0n}\{\bJ_{0n}\transpose\bPi\expect_0\bDelta_\bpsi^\sharp(\bpsi_0)\bPi\transpose\bJ_{0n}\}^{-1}\bJ_{0n}\transpose - \{\bPi\expect_0\bDelta_\bpsi^\sharp(\bpsi_0)\bPi\transpose\}^{-1}\|_{\max}\\
  &\quad \leq \|\{\bPi\expect_0\bDelta_\bpsi^\sharp(\bpsi_0)\bPi\transpose\}^{-1}\|_\infty^2\|\bV_\bB\|_{2\to\infty}^2\|[\bV_\bB\transpose\{\bPi\expect_0\bDelta_\bpsi^\sharp(\bpsi_0)\bPi\transpose\}^{-1}\bV_\bB]^{-1}\|_2\\
  &\quad \leq \|\{\bPi\expect_0\bDelta_\bpsi^\sharp(\bpsi_0)\bPi\transpose\}^{-1}\|_\infty^2\|\bV_\bB\|_{2\to\infty}^2\|\bPi\expect_0\bDelta_\bpsi^\sharp(\bpsi_0)\bPi\transpose\|_2 = O\bigg(\frac{m}{n}\bigg)
\end{align*}
by the proof of Lemma \ref{lemma:inverse_Fisher_information_matrix_infty_norm_bound}. 
This concludes that $\|\bJ_{0n}(\bG_0^\sharp)^{-1}\bJ_{0n}\transpose + \{\bPi\expect_0\bDelta_\bpsi^\sharp(\bpsi_0)\bPi\transpose\}^{-1}\|_{\max} = O(m^{5/2}/\sqrt{n})$. 
Since 
\[
\bigg(\sum_{i\in\calI}\bg_i\transpose \scrP_{\bz_i}\bigg)\left\{\bPi\expect_0\bDelta_\bpsi^\sharp(\bpsi_0)\bPi\transpose\right\}^{-1}\bigg(\sum_{i\in\calI}\scrP_{\bz_i}\transpose\bg_i\bigg) = \sum_{i\in\calI}\bg_i\transpose\{\expect_0\bDelta_{\bz_i}^\sharp(\bpsi_0)\}^{-1}\bg_i = -\sum_{i\in\calI}\bg_i\transpose(\bSigma_{\bz_i}^\sharp)^{-1}\bg_i, 
\]
it follows that
\begin{align*}
\left|\bigg(\sum_{i\in\calI}\bg_i\transpose\scrP_{\bz_i}\bigg)\bJ_{0n}(\bG_0^\sharp)^{-1}\bJ_{0n}\transpose\bigg(\sum_{i\in\calI}\scrP_{\bz_i}\transpose\bg_i\bigg) - \sum_{i\in\calI}\bg_i\transpose(\bSigma_{\bz_i}^\sharp)^{-1}\bg_i\right| = O\bigg(\frac{\|\bg\transpose\|_\infty m^{5/2}}{\sqrt{n}}\bigg).
\end{align*}
This implies that for $\calI = \{i\}$,
\begin{equation}
\label{eqn:stochastic_expansion_zihat}
\begin{aligned}
\bg_{ni}\transpose(\widehat{\bz}_i - \bz_{0i})
& = \bg_i\transpose\scrP_{\bz_i}\bJ_{0n}(\bG_0^\sharp)^{-1}\bJ_{0n}\transpose\bJ_{0n}\bLambda_\sharp\frac{\partial\calL}{\partial\bphi}(\bphi_0) + \Optilde( \vartheta_n)\\
& = \bg_i\transpose\scrP_{\bz_i}\left\{\bJ_{0n}(\bG_0^\sharp)^{-1}\bJ_{0n}\transpose + \bPi(\expect_0\bDelta_\bpsi^\sharp(\bpsi_0))^{-1}\bPi\transpose\right\}\bJ_{0n}\bLambda_\sharp\frac{\partial\calL}{\partial\bphi}(\bphi_0)\\
&\quad - \bg_i\transpose\{\expect_0\bDelta_{\bz_i}^\sharp(\bpsi_0)\}^{-1}\scrP_{\bz_i}\bJ_{0n}\bLambda_\sharp\frac{\partial\calL}{\partial\bphi}(\bphi_0) + \Optilde( \vartheta_n)\\
& = -\sum_{t = 1}^m\sum_{a < b}\ell'(\theta_{0ab}^{(t)}; A_{ab}^{(t)})\bg_i\transpose\{\expect_0\bDelta_{\bz_i}^\sharp(\bpsi_0)\}^{-1}\\
&\quad\times \begin{bmatrix}
\frac{1}{\sqrt{n}}\be_t(\be_i\transpose\be_a + \be_i\transpose\be_b) \\
\frac{1}{\sqrt{mn}}\{\bR_0^{(t)}\bx_{0b}(\be_i\transpose\be_a - 1/n - \bx_{0a}\transpose\bx_{0i}/n) + \bR_0^{(t)}\bx_{0a}(\be_i\transpose\be_b - 1/n - \bx_{0i}\transpose\bx_{0b}/n)\}
\end{bmatrix}\\
&\quad  + \Optilde( \vartheta_n)\\
& = \sum_{t = 1}^m\sum_{j\in[n]\backslash\{i\}}\ell'(\theta_{0ij}^{(t)}; A_{ij}^{(t)})\bg_i\transpose(\bSigma_{\bz_i}^\sharp)^{-1}\begin{bmatrix}
\frac{1}{\sqrt{n}}\be_t \\
\frac{1}{\sqrt{mn}}\bR_0^{(t)}\bx_{0j}
\end{bmatrix} + \Optilde(\vartheta_n). 
\end{aligned}
\end{equation}
Indeed, the first term on the right-hand side of the second equality in \eqref{eqn:stochastic_expansion_zihat} is 
\[
\sum_{t = 1}^m\sum_{a < b}\ell'(\theta_{0ab}^{(t)}; A_{ab}^{(t)})c_{ti,ab}
\] 
with
$c_{ti,ab} = \bg_i\transpose\scrP_{\bz_i}\{\bJ_{0n}(\bG_0^\sharp)^{-1}\bJ_{0n}\transpose + \bPi(\expect_0\bDelta_\bpsi^\sharp(\bpsi_0))^{-1}\bPi\transpose\}\bJ_{0n}\bJ_{0n}\transpose\bu_{tab}$, so that
\[
\sum_{t = 1}^m\sum_{a < b}\ell'(\theta_{0ab}^{(t)}; A_{ab}^{(t)})c_{ti,ab}
=
  \Optilde\bigg\{\log n\max_{t\in[m], i,a,b\in[n]}|c_{ti,ab}| + \sqrt{\log n}\max_{i\in[n]}\bigg(\sum_{t\in[m], a\neq b}c_{ti,ab}^2\bigg)^{1/2}\bigg\}
\]
by Bernstein's inequality, and
\[
\max_{i\in[n]}\sum_{t\in[m], a\neq b}c_{ti,ab}^2\leq\bigg\|\sum_{t\in[m], a\neq b}\bu_{tab}\bu_{tab}\transpose\bigg\|_2\max_{i\in[n]}\left\|\bg_i\transpose\scrP_{\bz_i}\{\bJ_{0n}(\bG_0^\sharp)^{-1}\bJ_{0n}\transpose - (\bPi\bSigma_\bpsi^\sharp\bPi\transpose)^{-1}\}\right\|_2^2.
\]
Write
\[
\bJ_{0n}(\bG_0^\sharp)^{-1}\bJ_{0n}\transpose - (\bPi\bSigma_\bpsi^\sharp\bPi\transpose)^{-1} = \{\bJ_{0n}(\bG_0^\sharp)^{-1}\bJ_{0n}\transpose - \bL\} + \{\bL - (\bPi\bSigma_\bpsi^\sharp\bPi\transpose)^{-1}\}
\]
with $\bL := \bJ_{0n}\{\bJ_{0n}\transpose\bPi\bSigma_{\bpsi}^\sharp\bPi\transpose\bJ_{0n}\}^{-1}\bJ_{0n}\transpose$.
By the identity 
\[
\bJ_{0n}(\bG_0^\sharp)^{-1}\bJ_{0n}\transpose = \bL + \bL\bPi\{\bSigma_\bpsi^\sharp - \bG_{\bpsi\bpsi}^\sharp(\bpsi_0)\}\bPi\transpose\bJ_{0n}(\bG_0^\sharp)^{-1}\bJ_{0n}\transpose,
\]
for any $a\in[d_\balpha + d_\bX]$, the set of indices $\calJ_a$ of the nonzero elements of the row vector $\be_a\transpose\scrP_\bZ\bL$ corresponds to the coordinates of $(\balpha^{(1)},\ldots,\balpha^{(m)}, \vect(\bX))$, and
\[
\|\be_a\transpose\{\bJ_{0n}(\bG_0^\sharp)^{-1}\bJ_{0n}\transpose - \bL\}\|_2 \leq 
\|\be_a\transpose\scrP_\bZ\bL\|_\infty\max_{j\in\calJ_a}\|\be_j\transpose\bPi\{\bSigma_\bpsi^\sharp - \bG_{\bpsi\bpsi}^\sharp(\bpsi_0)\}\bPi\transpose\|_2\|(\bG_0^\sharp)^{-1}\|_2 = O(m/\sqrt n).
\]
By the fact that the direct sum of $\mathrm{Span}\{(\bPi\bSigma_\bpsi^\sharp\bPi\transpose)^{-1/2}\bV_\bB\}$ and $\mathrm{Span}\{(\bPi\bSigma_\bpsi^\sharp\bPi\transpose)^{1/2}\bJ_{0n}\}$ is the entire $\mathbb{R}^{d_\bpsi}$, we obtain
\[
\bL - (\bPi\bSigma_\bpsi^\sharp\bPi\transpose)^{-1} = -(\bPi\bSigma_\bpsi^\sharp\bPi\transpose)^{-1}\bV_\bB\{\bV_\bB\transpose(\bPi\bSigma_\bpsi^\sharp\bPi\transpose)^{-1}\bV_\bB\}^{-1}\bV_\bB\transpose(\bPi\bSigma_\bpsi^\sharp\bPi\transpose)^{-1},
\]
and for any $a\in[d_\balpha + d_\bX]$,
\[
\|\be_a\transpose\{\bL - (\bPi\bSigma_\bpsi^\sharp\bPi\transpose)^{-1}\}\|_2\leq 
\|(\bSigma_{\bz_i}^\sharp)^{-1}\|_\infty\|\bV_\bB\|_{2\to\infty}\|\bSigma_\bpsi^\sharp\|_2\|(\bSigma_\bpsi^\sharp)^{-1}\|_2 = O(\sqrt{m/n}). 
\]
Hence $\sum_{t, a\neq b}c_{tab}^2 = O(m^2/n)$,
$\max|c_{tab}| = O(m/n)$, and the first term on the right-hand side of the second equality of \eqref{eqn:stochastic_expansion_zihat} is $\Optilde(\sqrt{\log n/n})$. Similarly, in the third equality of \eqref{eqn:stochastic_expansion_zihat}, the coefficients of $\ell'(\theta_{0ab}^{(t)}; A_{ab}^{(t)})$ contributed by the terms $-1/n - \bx_{0a}\transpose\bx_{0i}/n$ and $-1/n - \bx_{0i}\transpose\bx_{0b}/n$ are $O(m^{-1/2}n^{-3/2})$ uniformly in $(t, a, b)$, so their total contribution is also $\Optilde(\sqrt{\log n/n})$ by Bernstein's inequality, which gives the last equality of \eqref{eqn:stochastic_expansion_zihat}. Theorem \ref{thm:two_to_infinity_norm_bound_MLE} is then proved by applying Bernstein's inequality to the above linearization with $\calI = \{i\}$, $\bg_{\balpha_i} = \zero_m$, and $\bg_{\bx_i} = \be_k(d)$, together with a union bound. 
For $\scrP_\bX\{\bPi\expect_0\bDelta_\bpsi^\sharp(\bpsi_0)\bPi\transpose\}^{-1}\scrP_\bX\transpose$, note that
\begin{align*}
&\left|\vect(\bF\bX_0)\transpose\scrP_{\bX}\bJ_{0n}(\bG_0^\sharp)^{-1}\bJ_{0n}\transpose\scrP_{\bX}\transpose\vect(\bF\bX_0) + \vect(\bF\bX_0)\transpose\scrP_\bX\{\bPi\expect_0\bDelta_{\bpsi}^\sharp(\bpsi_0)\bPi\transpose\}^{-1}\scrP_\bX\transpose\vect(\bF\bX_0)\right|\\
&\quad = O\bigg(\frac{m^{5/2}}{\sqrt{n}}\bigg)
\end{align*}
and
\begin{align*}
&-\vect(\bF\bX_0)\transpose\scrP_\bX\{\bPi\expect_0\bDelta_\bpsi^\sharp(\bpsi_0)\bPi\transpose\}^{-1}\scrP_\bX\transpose\vect(\bF\bX_0)\\
&\quad = -\sum_{i = 1}^n\vect(\bF\bX_0)\transpose\Bigg\{
 \begin{bmatrix}
\mathbf{O}_{d\times m} & \eye_d
\end{bmatrix}\{\expect_0\bDelta_{\bz_i}^\sharp(\bpsi_0)\}^{-1}\begin{bmatrix}
  \mathbf{O}_{m\times d} \\ \eye_d
\end{bmatrix}\otimes\be_i\be_i\transpose
\Bigg\}\vect(\bF\bX_0)\\
&\quad = \sum_{i = 1}^n\begin{bmatrix}
\zero_m\transpose & \be_i\transpose\bF\bX_0
\end{bmatrix}(\bSigma_{\bz_i}^\sharp)^{-1}\begin{bmatrix}
  \zero_m \\ \bX_0\transpose\bF\be_i
\end{bmatrix} = \sum_{i = 1}^n\be_i\transpose\bF\bX_0\bSigma_{\bx_i}^\sharp\bX_0\transpose\bF\be_i.
\end{align*}
The proof is therefore completed by Slutsky's theorem. 
\end{proof}

\section{Proof of Theorem \ref{thm:asymptotic_normality_R}}
\label{sec:proof_of_theorem_asymptotic_normality_R}

\subsection{Proof architecture}
\label{sub:proof_architecture_asymptotic_normality_R}

By the uniqueness in the sense of Proposition \ref{prop:stationarity_uniqueness_uniform_consistency_MLE} and the convention in Subsection \ref{sub:proof_architecture_of_the_asymptotic_normality}, we may assume without loss of generality that $\bW_{\widehat{\bX}} = \eye_d$. 
The asymptotic normality of $\widehat{\bR}^{(t)}$ and the stochastic expansion lies in the following third-order Taylor expansion of the log-likelihood:
\begin{align}
\label{eqn:third_order_taylor_expansion_loglikelihood}
\zero_{d_\bphi}& = \nabla\calL(\bphi_0) + \nabla^2\calL(\bphi_0)(\widehat{\bphi} - \bphi_0) + \frac{1}{2}\nabla^3\calL(\bphi_0)[\widehat{\bphi} - \bphi_0](\widehat{\bphi} - \bphi_0) + \bgamma_\calL.
\end{align}
Denote
\begin{align}
\label{eqn:s_hat}
\widehat{\bs}:=(\bG_0^\sharp)^{-1}\bLambda_\sharp\nabla\calL(\bphi_0)
\end{align}
so that by \eqref{eqn:Taylor_expansion}, we know that 
\[
\bgamma_\bphi:=\widehat{\bphi} - \bphi_0 - \bLambda_\sharp\widehat{\bs}
\]
satisfies $\bJ_0\bgamma_\bphi = \bgamma$, where $\bgamma$ is the remainder in \eqref{eqn:Taylor_expansion}. 
Let $\bH^\sharp := \bLambda_\sharp\nabla^2\calL(\bphi_0)\bLambda_\sharp$ and, for $t\in[m]$,
\[
  \widetilde{\scrP}_{\br^{(t)}} = 
\begin{bmatrix}
\mathbf{O}_{d_r\times d_\balpha} & \be_t(m)\transpose\otimes\eye_{d_r} & \mathbf{O}_{d_r\times d_\bB}  
\end{bmatrix},
\]
so that $\widetilde{\scrP}_{\br^{(t)}}\bv$ extracts the coordinates of $\bv\in\mathbb{R}^{d_\bphi}$ corresponding to $\vech(\bR^{(t)})$. For $\bu,\bv\in\mathbb{R}^{d_\bphi}$, we write $\nabla^3\calL(\bphi_0)[\bu, \bv] := \nabla^3\calL(\bphi_0)[\bu]\bv$, where $\nabla^3\calL(\bphi_0)[\bu] = (\mathrm{d}/\mathrm{d}\tau)\nabla^2\calL(\bphi_0 + \tau\bu)|_{\tau = 0}$ denotes the directional derivative of $\nabla^2\calL(\bphi_0)$ along the direction of $\bu$. Denote
\begin{align}
\label{eqn:R_bias_Q1}
\bQ_1 &:= (\bG_0^\sharp)^{-1}(\bH^\sharp + \bG_0^\sharp)\widehat{\bs},\\
\label{eqn:R_bias_Q2}
\bQ_2 &:= \frac{1}{2}(\bG_0^\sharp)^{-1}\bLambda_\sharp\nabla^3\calL(\bphi_0)[\bLambda_\sharp\widehat{\bs}, \bLambda_\sharp\widehat{\bs}],
\end{align}
Immediately,
$\nabla^3\calL(\bphi_0)[\bu, \bv] = (\partial^2/\partial \tau_1\partial \tau_2)\nabla\calL(\bphi_0 + \tau_1\bu + \tau_2\bv)|_{\tau_1 = \tau_2 = 0}$ is symmetric in $(\bu, \bv)$ since $\calL$ is three times continuously differentiable. Then, a direct calculation from \eqref{eqn:third_order_taylor_expansion_loglikelihood} shows that
\begin{equation}
\label{eqn:third_order_expansion_phi}
\begin{aligned}
\bLambda_\sharp^{-1}(\widehat{\bphi} - \bphi_0)
& = \widehat{\bs} + \bQ_1 + \bQ_2 + (\bG_0^\sharp)^{-1}(\bH^\sharp + \bG_0^\sharp)\bLambda_\sharp^{-1}\bgamma_\bphi \\
&\quad + (\bG_0^\sharp)^{-1}\bigg\{\bLambda_\sharp\nabla^3\calL(\bphi_0)[\bLambda_\sharp\widehat{\bs}, \bgamma_\bphi] + \frac{1}{2}\bLambda_\sharp\nabla^3\calL(\bphi_0)[\bgamma_\bphi, \bgamma_\bphi]\bigg\} + (\bG_0^\sharp)^{-1}\bLambda_\sharp\bgamma_\calL.
\end{aligned}
\end{equation}
Applying $\widetilde{\scrP}_{\br^{(t)}}$ to both sides of \eqref{eqn:third_order_expansion_phi} immediately implies
\begin{equation}
\label{eqn:third_order_expansion_vech_Rt}
\begin{aligned}
&n\{\vech(\widehat{\bR}^{(t)}) - \vech(\bR_0^{(t)})\} - \widetilde{\scrP}_{\br^{(t)}}\widehat{\bs} - \expect_0\widetilde{\scrP}_{\br^{(t)}}(\bQ_1 + \bQ_2)\\
&\quad = \widetilde{\scrP}_{\br^{(t)}}(\bQ_1 - \expect_0\bQ_1) + \widetilde{\scrP}_{\br^{(t)}}(\bQ_2 - \expect_0\bQ_2) + \widetilde{\scrP}_{\br^{(t)}}(\bG_0^\sharp)^{-1}(\bH^\sharp + \bG_0^\sharp)\bLambda_\sharp^{-1}\bgamma_\bphi \\
&\qquad + \widetilde{\scrP}_{\br^{(t)}}(\bG_0^\sharp)^{-1}\bigg\{\bLambda_\sharp\nabla^3\calL(\bphi_0)[\bLambda_\sharp\widehat{\bs}, \bgamma_\bphi] + \frac{1}{2}\bLambda_\sharp\nabla^3\calL(\bphi_0)[\bgamma_\bphi, \bgamma_\bphi]\bigg\} + \widetilde{\scrP}_{\br^{(t)}}(\bG_0^\sharp)^{-1}\bLambda_\sharp\bgamma_\calL.
\end{aligned}
\end{equation}
The fluctuation of $\widetilde{\scrP}_{\br^{(t)}}\bQ_1$ and $\widetilde{\scrP}_{\br^{(t)}}\bQ_2$ are handled by Subsections \ref{sub:controlling_the_bias_remainder_for_Rt_hat_Q1} and \ref{sub:controlling_the_bias_remainder_for_Rt_hat_Q2}, respectively. The fourth term on the right-hand side is controlled by Subsection \ref{sub:controlling_the_third_order_derivatives_of_the_log_likelihood_function}, and the last term is controlled by Subsection \ref{sub:third_order_taylor_expansion_of_the_log_likelihood}. Finally, the third term and the complete proof of Theorem \ref{thm:asymptotic_normality_R} are collected in Subsection \ref{sub:completing_the_proof_of_asymptotic_normality_Rt}. 

\subsection{Third-order Taylor expansion of the log-likelihood}
\label{sub:third_order_taylor_expansion_of_the_log_likelihood}

This subsection elaborates on a third-order Taylor expansion analysis of the log-likelihood function $\calL(\bphi)$, or equivalently, the second-order Taylor expansion analysis of the score function $\partial\calL/\partial\bphi$. 
Let $\widehat{\bphi}$ be the MLE over the compact parameterization defined in Proposition \ref{prop:stationarity_uniqueness_uniform_consistency_MLE}. 

\begin{lemma}\label{lemma:third_order_Taylor_expansion_loglikelihood}
Suppose Assumptions \ref{assumption:edge_distribution}--\ref{assumption:third_order_lipschitz_ell} hold and $m^{12}(\log n)^4/n\to 0$ as $n\to\infty$. Then, the remainder $\bgamma_\calL$ in \eqref{eqn:third_order_taylor_expansion_loglikelihood} satisfies $\|\bLambda_\sharp\bgamma_\calL\|_2 = \Optilde\{\sqrt{m}(\log n)^{5/2}/\sqrt{n}\}$. 
\end{lemma}

\begin{proof}[\bf Proof]
For notational simplicity, in this proof, we denote $\calE=\{(t,i,j)\in[m]\times[n]\times[n]:i\neq j\}$. Also, for any $\tau\in[0, 1]$ and any $e := (t,i,j)\in\calE$, denote
\begin{equation}
\begin{aligned}
&\bphi_\tau = \bphi_0 + \tau(\widehat{\bphi} - \bphi_0),\; 
\bpsi_\tau = \bpsi(\bphi_\tau),\; 
\dot{\bpsi} := \frac{\mathrm{d}}{\mathrm{d}\tau}\bpsi_\tau = \frac{\partial\bpsi}{\partial\bphi\transpose}(\bphi_\tau)(\widehat{\bphi} - \bphi_0) = \bJ(\tau\widehat{\bB})(\widehat{\bphi} - \bphi_0),\\
&\ddot{\bpsi} := \frac{\mathrm{d}^2}{\mathrm{d}\tau^2}\bpsi_\tau = \frac{\mathrm{d}}{\mathrm{d}\tau}\frac{\partial\bpsi}{\partial\bphi\transpose}(\bphi_\tau)(\widehat{\bphi} - \bphi_0) = \dot{\bJ}(\tau\widehat{\bB})[\widehat{\bphi} - \bphi_0](\widehat{\bphi} - \bphi_0),\\
&\dddot{\bpsi} := \frac{\mathrm{d}^3}{\mathrm{d}\tau^3}\bpsi_\tau = \frac{\mathrm{d}}{\mathrm{d}\tau}\dot{\bJ}(\tau\widehat{\bB})[\widehat{\bphi} - \bphi_0](\widehat{\bphi} - \bphi_0) = \ddot{\bJ}(\tau\widehat{\bB})[\widehat{\bphi} - \bphi_0, \widehat{\bphi} - \bphi_0](\widehat{\bphi} - \bphi_0),\\
&\theta_{e} :=\theta_{ij}^{(t)}(\bpsi_\tau),\;
\nabla_\bpsi\theta_e := \frac{\partial\theta_{ij}^{(t)}}{\partial\bpsi}(\bpsi_\tau),\;
\nabla^2_\bpsi\theta_e := \frac{\partial^2\theta_{ij}^{(t)}}{\partial\bpsi\partial\bpsi}(\bpsi_\tau),\;
\nabla^3_\bpsi\theta_e[\dot{\bpsi}] = \frac{\mathrm{d}}{\mathrm{d}\tau}\nabla^2_\bpsi\theta_e,\\
&\nabla^3\calL(\bphi_0)[\widehat{\bphi} - \bphi_0] = \frac{\mathrm{d}}{\mathrm{d}\tau}\nabla^2\calL(\bphi_0 + \tau(\widehat{\bphi} - \bphi_0))\mathrel{\bigg|_{\tau = 0}},\;\ell(\theta_e):=\ell(\theta_{ij}^{(t)}(\bpsi_\tau); A_{ij}^{(t)}),
\end{aligned}
\end{equation}
where we inherit the dot-notations from \eqref{eqn:dot_notations} and view $\theta_{ij}^{(t)}$ as a function of the ambient parameter $\bpsi$. 
By Taylor's theorem with an integral remainder and the Leibniz formula for higher-order derivatives, 
\begin{align*}
\bgamma_\calL& = \frac{1}{2}\int_0^1(1 - \tau)^2\frac{\mathrm{d}^3}{\mathrm{d}\tau^3}\nabla\calL(\bphi_\tau)\mathrm{d}\tau = \frac{1}{4}\int_0^1(1 - \tau)^2\sum_{e\in\calE}\frac{\mathrm{d}^3}{\mathrm{d}\tau^3}\left\{\ell'(\theta_e)\bJ(\tau\widehat{\bB})\transpose\nabla_\bpsi\theta_e\right\}\mathrm{d}\tau\\
& = \frac{1}{4}\int_0^1(1 - \tau)^2\sum_{e\in\calE}\bJ(\tau\widehat{\bB})\transpose\frac{\mathrm{d}^3}{\mathrm{d}\tau^3}\left\{\ell'(\theta_e)\nabla_\bpsi\theta_e\right\}\mathrm{d}\tau + \frac{3}{4}\int_0^1(1 - \tau)^2\sum_{e\in\calE}\dot{\bJ}(\tau\widehat{\bB})[\widehat{\bB}]\transpose\frac{\mathrm{d}^2}{\mathrm{d}\tau^2}\left\{\ell'(\theta_e)\nabla_\bpsi\theta_e\right\}\mathrm{d}\tau\\
&\quad + \frac{3}{4}\int_0^1(1 - \tau)^2\sum_{e\in\calE}\ddot{\bJ}(\tau\widehat{\bB})[\widehat{\bB}, \widehat{\bB}]\transpose
\left\{\ell''(\theta_e)\nabla_\bpsi\theta_e\nabla_\bpsi\theta_e\transpose\dot{\bpsi} + \ell'(\theta_e)\nabla^2_\bpsi\theta_e\dot{\bpsi}\right\}\mathrm{d}\tau\\
&\quad + \frac{1}{4}\int_0^1(1 - \tau)^2\sum_{e\in\calE}\dddot{\bJ}(\tau\widehat{\bB})[\widehat{\bB}, \widehat{\bB}, \widehat{\bB}]\transpose\left\{\ell'(\theta_e)\nabla_\bpsi\theta_e\right\}\mathrm{d}\tau.
\end{align*}

\medskip\noindent \textbf{$\blacksquare$ Step 1. Higher-order derivative bounds of $\bpsi$ and $\nabla_\bpsi\theta_e$.}
Without loss of generality, we may assume $\bW_{\widehat{\bX}} = \eye_d$. 
By \eqref{eqn:stochastic_expansion_zihat}, Bernstein's inequality, \eqref{eqn:Rhat_bound}, Lemma \ref{lemma:Lipschitz_Cayley_transform}, and Propositions \ref{prop:consistency_native_norm}--\ref{prop:stationarity_uniqueness_uniform_consistency_MLE}, we know that 
\begin{align*}
&\sup_{\tau\in[0, 1]}\|\dot{\bpsi}\|_\infty \leq \max\bigg(\max_{i\in[n]}\|\widehat{\balpha}_i - \balpha_{0i}\|_\infty,\max_{i\in[n]}\|\be_i\transpose\dot{\bX}(\tau\widehat{\bB})[\widehat{\bB}]\|_2, \max_{t\in[m]}\|\widehat{\bR}^{(t)} - \bR_0^{(t)}\|_{\mathrm{F}}\bigg) = \Optilde\bigg(\sqrt{\frac{\log n}{n}}\bigg),\\
&\sup_{\tau\in[0, 1]}\|\ddot{\bpsi}\|_\infty \leq \max_{i\in[n]}\|\be_i\transpose\ddot{\bX}(\tau\widehat{\bB})[\widehat{\bB}, \widehat{\bB}]\|_2 = \Optilde(n^{-1}),\\
&\sup_{\tau\in[0, 1]}\|\dddot{\bpsi}\|_\infty \leq \max_{i\in[n]}\|\be_i\transpose\dddot{\bX}(\tau\widehat{\bB})[\widehat{\bB}, \widehat{\bB}, \widehat{\bB}]\|_2 = \Optilde\bigg(\frac{\sqrt{\log n}}{\sqrt{m}n^{3/2}}\bigg),\\
&\sup_{\tau\in[0, 1]}\|\bJ(\tau\widehat{\bB})\|_2 = O(1),\; 
\sup_{\tau\in[0, 1]}\|\dot{\bJ}(\tau\widehat{\bB})[\widehat{\bB}]\|_2 = O(\sqrt{n}\|\widehat{\bB}\|_2) = \Optilde(1),\\
&\sup_{\tau\in[0, 1]}\|\ddot{\bJ}(\tau\widehat{\bB})[\widehat{\bB}, \widehat{\bB}]\|_2 = O(\sqrt{n}\|\widehat{\bB}\|_2^2) = \Optilde(n^{-1/2}),\;
\sup_{\tau\in[0, 1]}\|\dddot{\bJ}(\tau\widehat{\bB})[\widehat{\bB}, \widehat{\bB}, \widehat{\bB}]\|_2 = O(\sqrt{n}\|\widehat{\bB}\|_2^3) = \Optilde(n^{-1}).
\end{align*}
Now, for any $e = (t,i,j)\in\calE$, let $\calS_1(e)\subset[d_\bpsi]$ be the index set of the coordinates of $\bpsi$ on which $\theta_{ij}^{(t)}$ depends, namely, those of $\alpha_i^{(t)},\alpha_j^{(t)},\vech(\bR^{(t)}),\bx_i,\bx_j$. Then $|\calS_1(e)| = 2 + d(d + 1)/2 + 2d = O(1)$, and index set of the nonzero coordinates of $\nabla_\bpsi\theta_e$, $\nabla_\bpsi^2\theta_e\delta\bpsi$, and $\nabla_\bpsi^3\theta_e[\delta\bpsi_1]\delta\bpsi_2$ are subsets of $\calS_1(e)$ for all $\delta\bpsi,\delta\bpsi_1,\delta\bpsi_2\in\mathbb{R}^{d_\bpsi}$. Furthermore, we have the following inequalities:
\begin{align*}
\sup_{\tau\in[0, 1]}|\dot{\bpsi}\transpose\nabla_\bpsi\theta_e|
&\leq \sup_{\tau\in[0, 1]}\|\nabla_\bpsi\theta_e\|_1\|\dot{\bpsi}\|_\infty = O\bigg(\sup_{\tau\in[0, 1]}\|\dot{\bpsi}\|_\infty\bigg) = \Optilde\bigg(\sqrt{\frac{\log n}{n}}\bigg),\\
\sup_{\tau\in[0, 1]}|\ddot{\bpsi}\transpose\nabla_\bpsi\theta_e|
&\leq \sup_{\tau\in[0, 1]}\|\nabla_\bpsi\theta_e\|_1\|\ddot{\bpsi}\|_\infty = O\bigg(\sup_{\tau\in[0, 1]}\|\ddot{\bpsi}\|_\infty\bigg) = \Optilde(n^{-1}),\\
\sup_{\tau\in[0, 1]}|\dot{\bpsi}\transpose\nabla_\bpsi^2\theta_e\dot{\bpsi}|
&\leq \sup_{\tau\in[0, 1]}\|\vect(\nabla^2_\bpsi\theta_e)\|_1\|\dot{\bpsi}\|_\infty^2 = O\bigg(\sup_{\tau\in[0, 1]}\|\dot{\bpsi}\|_\infty^2\bigg) = \Optilde\bigg(\frac{\log n}{n}\bigg),\\
\sup_{\tau\in[0, 1]}|\dddot{\bpsi}\transpose\nabla_\bpsi\theta_e|
&\leq \sup_{\tau\in[0, 1]}\|\nabla_\bpsi\theta_e\|_1\|\dddot{\bpsi}\|_\infty = O\bigg(\sup_{\tau\in[0, 1]}\|\dddot{\bpsi}\|_\infty\bigg) = \Optilde\bigg(\frac{\sqrt{\log n}}{\sqrt{m}n^{3/2}}\bigg),\\
\sup_{\tau\in[0, 1]}|\ddot{\bpsi}\transpose\nabla_\bpsi^2\theta_e\dot{\bpsi}|
&\leq \sup_{\tau\in[0, 1]}\|\vect(\nabla_\bpsi^2\theta_e)\|_1\|\dot{\bpsi}\|_\infty\|\ddot{\bpsi}\|_\infty = O\bigg(\sup_{\tau\in[0, 1]}\|\dot{\bpsi}\|_\infty\|\ddot{\bpsi}\|_\infty\bigg) = \Optilde\bigg(\frac{\sqrt{\log n}}{n^{3/2}}\bigg),\\
\sup_{\tau\in[0, 1]}|\dot{\bpsi}\transpose\nabla^3_\bpsi\theta_e[\dot{\bpsi}]\dot{\bpsi}|
& = O\bigg(\sup_{\tau\in[0, 1]}\|\dot{\bpsi}\|_\infty^3\bigg) = \Optilde\bigg\{\frac{(\log n)^{3/2}}{n^{3/2}}\bigg\},\\
\sup_{\tau\in[0, 1]}\|\nabla_\bpsi^2\theta_e\dot{\bpsi}\|_\infty 
&\leq \sup_{\tau\in[0, 1]}\|\nabla_\bpsi^2\theta_e\|_\infty\|\dot{\bpsi}\|_\infty = O\bigg(\sup_{\tau\in[0, 1]}\|\dot{\bpsi}\|_\infty\bigg) = \Optilde\bigg(\sqrt{\frac{\log n}{n}}\bigg),\\
\sup_{\tau\in[0, 1]}\|\nabla_\bpsi^2\theta_e\ddot{\bpsi}\|_\infty 
&\leq \sup_{\tau\in[0, 1]}\|\nabla_\bpsi^2\theta_e\|_\infty\|\ddot{\bpsi}\|_\infty = O\bigg(\sup_{\tau\in[0, 1]}\|\ddot{\bpsi}\|_\infty\bigg) = \Optilde(n^{-1}),\\
\sup_{\tau\in[0, 1]}\|\nabla_\bpsi^3\theta_e[\dot{\bpsi}]\dot{\bpsi}\|_\infty 
&\leq \sup_{\tau\in[0, 1]}\|\nabla_\bpsi^3\theta_e[\dot{\bpsi}]\|_\infty\|\dot{\bpsi}\|_\infty = O\bigg(\sup_{\tau\in[0, 1]}\|\dot{\bpsi}\|_\infty^2\bigg) = \Optilde\bigg(\frac{\log n}{n}\bigg),\\
\sup_{\tau\in[0, 1]}\|\nabla_\bpsi^3\theta_e[\ddot{\bpsi}]\dot{\bpsi}\|_\infty 
&\leq \sup_{\tau\in[0, 1]}\|\nabla_\bpsi^3\theta_e[\ddot{\bpsi}]\|_\infty\|\dot{\bpsi}\|_\infty = O\bigg(\sup_{\tau\in[0, 1]}\|\dot{\bpsi}\|_\infty\|\ddot{\bpsi}\|_\infty\bigg) = \Optilde\bigg(\frac{\sqrt{\log n}}{n^{3/2}}\bigg),\\
\sup_{\tau\in[0, 1]}\|\nabla_\bpsi^3\theta_e[\dot{\bpsi}]\ddot{\bpsi}\|_\infty 
&\leq \sup_{\tau\in[0, 1]}\|\nabla_\bpsi^3\theta_e[\dot{\bpsi}]\|_\infty\|\ddot{\bpsi}\|_\infty = O\bigg(\sup_{\tau\in[0, 1]}\|\dot{\bpsi}\|_\infty\|\ddot{\bpsi}\|_\infty\bigg) = \Optilde\bigg(\frac{\sqrt{\log n}}{n^{3/2}}\bigg),\\
\sup_{\tau\in[0, 1]}\|\nabla^2_\bpsi\theta_e \dddot{\bpsi}\|_\infty
&\leq \sup_{\tau\in[0, 1]}\|\nabla^2_\bpsi\theta_e\|_\infty \|\dddot{\bpsi}\|_\infty = O\bigg(\sup_{\tau\in[0, 1]}\|\dddot{\bpsi}\|_\infty\bigg) = \Optilde\bigg(\frac{\sqrt{\log n}}{\sqrt{m}n^{3/2}}\bigg).
\end{align*}

\medskip\noindent\textbf{$\blacksquare$ Step 2. Generic sparse formula for the higher-order derivatives.}
Since $\theta_{ij}^{(t)}$ is a cubic polynomial in $\bpsi$, $\nabla^3_\bpsi\theta_e$ does not depend on
$\tau$ and $\nabla^4_\bpsi\theta_e$ vanishes (as a zero tensor), then, with
$\dot\theta_e = \dot{\bpsi}\transpose\nabla_\bpsi\theta_e$,
$\ddot\theta_e = \dot{\bpsi}\transpose\nabla^2_\bpsi\theta_e\dot{\bpsi}
+ \ddot{\bpsi}\transpose\nabla_\bpsi\theta_e$, and
$\dddot\theta_e = \dot{\bpsi}\transpose\nabla^3_\bpsi\theta_e[\dot{\bpsi}]\dot{\bpsi}
+ 3\ddot{\bpsi}\transpose\nabla^2_\bpsi\theta_e\dot{\bpsi}
+ \dddot{\bpsi}\transpose\nabla_\bpsi\theta_e$,
\begin{align*}
\frac{\mathrm{d}}{\mathrm{d}\tau}\{\ell'(\theta_e)\nabla_\bpsi\theta_e\}
& = \ell''(\theta_e)\dot\theta_e\nabla_\bpsi\theta_e + \ell'(\theta_e)\nabla^2_\bpsi\theta_e\dot{\bpsi},\\
\frac{\mathrm{d}^2}{\mathrm{d}\tau^2}\{\ell'(\theta_e)\nabla_\bpsi\theta_e\}
& = \{\ell'''(\theta_e)\dot\theta_e^2 + \ell''(\theta_e)\ddot\theta_e\}\nabla_\bpsi\theta_e
  + 2\ell''(\theta_e)\dot\theta_e\nabla^2_\bpsi\theta_e\dot{\bpsi}\\
&\quad + \ell'(\theta_e)\{\nabla^3_\bpsi\theta_e[\dot{\bpsi}]\dot{\bpsi}
  + \nabla^2_\bpsi\theta_e\ddot{\bpsi}\},\\
\frac{\mathrm{d}^3}{\mathrm{d}\tau^3}\{\ell'(\theta_e)\nabla_\bpsi\theta_e\}
& = \{\ell''''(\theta_e)\dot\theta_e^3 + 3\ell'''(\theta_e)\dot\theta_e\ddot\theta_e
  + \ell''(\theta_e)\dddot\theta_e\}\nabla_\bpsi\theta_e\\
&\quad + 3\{\ell'''(\theta_e)\dot\theta_e^2
  + \ell''(\theta_e)\ddot\theta_e\}\nabla^2_\bpsi\theta_e\dot{\bpsi}\\
&\quad + 3\ell''(\theta_e)\dot\theta_e\{\nabla^3_\bpsi\theta_e[\dot{\bpsi}]\dot{\bpsi}
  + \nabla^2_\bpsi\theta_e\ddot{\bpsi}\}
  + \ell'(\theta_e)\{3\nabla^3_\bpsi\theta_e[\dot{\bpsi}]\ddot{\bpsi}
  + \nabla^2_\bpsi\theta_e\dddot{\bpsi}\}.
\end{align*}
By Proposition \ref{prop:stationarity_uniqueness_uniform_consistency_MLE} and the definition of $\Psi$, $|\theta_e|\leq b$ almost surely,
so that $|\ell''(\theta_e)|, |\ell'''(\theta_e)|, |\ell''''(\theta_e)|$ are uniformly bounded by
Assumptions \ref{assumption:edge_distribution} and \ref{assumption:third_order_lipschitz_ell}, and the nonzero elements of
$\nabla_\bpsi\theta_e$, $\nabla^2_\bpsi\theta_e$, $\nabla^3_\bpsi\theta_e$ are $O(1)$ uniformly over $\tau\in[0, 1]$ and $e\in\calE$. Together with the
inequalities above, this implies that there exists an absolute constant $K\in\mathbb{N}_+$, scalars $(\alpha^{(1)}_{e,k}, \alpha^{(2)}_{e,k}, \alpha^{(3)}_{e,k}:e\in\calE,k\in[K])$, $(\delta^{(1)}_{e,k}, \delta^{(2)}_{e,k}, \delta^{(3)}_{e,k}:e\in\calE,k\in[K])$, and vectors $(\bbeta_{e,k}^{(1)}, \bbeta_{e,k}^{(2)}, \bbeta_{e,k}^{(3)}:e\in\calE,k\in[K])\subset\mathbb{R}^{d_\bpsi}$, such that 
\begin{align*}
&\sup_\tau\max_{e,k}|\alpha_{e,k}^{(1)}| = \Optilde\bigg\{\frac{(\log n)^{1/2}}{n^{1/2}}\bigg\},\;
\sup_\tau\max_{e,k}|\alpha_{e,k}^{(2)}| = \Optilde\bigg(\frac{\log n}{n}\bigg),\;
\sup_\tau\max_{e,k}|\alpha_{e,k}^{(3)}| = \Optilde\bigg\{\frac{(\log n)^{3/2}}{n^{3/2}}\bigg\},\\
&\sup_\tau\max_{e,k}|\delta_{e,k}^{(1)}| = \Optilde\bigg\{\frac{(\log n)^{1/2}}{n^{1/2}}\bigg\},\;
\sup_\tau\max_{e,k}|\delta_{e,k}^{(2)}| = \Optilde\bigg(\frac{\log n}{n}\bigg),\;
\sup_\tau\max_{e,k}|\delta_{e,k}^{(3)}| = \Optilde\bigg\{\frac{(\log n)^{3/2}}{n^{3/2}}\bigg\},\\
&\sup_\tau\max_{e,k}\|\bbeta_{e,k}^{(1)}\|_\infty\leq 1,\;
\sup_\tau\max_{e,k}\|\bbeta_{e,k}^{(2)}\|_\infty \leq 1,\; 
\sup_\tau\max_{e,k}\|\bbeta_{e,k}^{(3)}\|_\infty\leq 1,\\
&\{i\in[d_\bpsi]:\be_i\transpose\bbeta_{e,k}^{(1)}\neq 0\},\{i\in[d_\bpsi]:\be_i\transpose\bbeta_{e,k}^{(2)}\neq 0\},\{i\in[d_\bpsi]:\be_i\transpose\bbeta_{e,k}^{(3)}\neq 0\}\subset\calS_1(e),
\end{align*}
and for any $r\in\{1,2,3\}$,
\begin{align*}
\sum_{e\in\calE}\frac{\mathrm{d}^r}{\mathrm{d}\tau^r}\left\{
\ell'(\theta_e)\nabla_\bpsi\theta_e
\right\} = \sum_{k = 1}^K\sum_{e\in\calE}\left\{\alpha_{e,k}^{(r)} + \delta_{e,k}^{(r)}\ell'(\theta_e)\right\}\bbeta_{e,k}^{(r)}.
\end{align*}

\medskip\noindent\textbf{$\blacksquare$ Step 3. Completing the proof.}
Note that for any $l\in[d_\bpsi]$, 
\begin{align*}
\sum_e\mathbbm{1}\{l\in\calS_1(e)\} = \begin{cases}
O(n),&\text{ if }l\in[d_\balpha],\\
O(n^2),&\text{ if }l\in\{d_\balpha + 1,\ldots,d_\balpha + d_\br\},\\
O(mn),&\text{ if }l\in\{d_\balpha + d_\br + 1,\ldots,d_\bpsi\}.
\end{cases}  
\end{align*}
Then, for any unit vector $\bw = [w_1,\ldots,w_{d_\bpsi}]\transpose\in\mathbb{R}^{d_\bpsi}$ and any $r\in\{0,1,2,3\}$ (denoting $\bbeta_{e,k}^{(0)}:=\nabla_\bpsi\theta_e$, whose entries are $O(1)$), with $(\widetilde{\bLambda}_\sharp)_{ll}$ denoting the $l$th diagonal element of $\widetilde{\bLambda}_\sharp$, we obtain by Cauchy--Schwarz inequality
\begin{align*}
\sum_{k = 1}^K\sum_{e\in\calE}(\bw\transpose\widetilde{\bLambda}_\sharp\bbeta_{e,k}^{(r)})^2 &\lesssim \max_{e\in\calE}|\calS_1(e)|\sum_{k = 1}^K\sum_{e\in\calE}\sum_{l\in\calS_1(e)}(\widetilde{\bLambda}_\sharp)_{ll}^2w_l^2\\
& = \max_{e\in\calE}|\calS_1(e)|\sum_{k = 1}^K\sum_{l = 1}^{d_\bpsi}w_l^2(\widetilde{\bLambda}_\sharp)_{ll}^2\sum_{e\in\calE}\mathbbm{1}\{l\in\calS_1(e)\} = O(1).
\end{align*}
In particular, this implies that
\begin{align}
\label{eqn:frame_bound}
\bigg\|\sum_{e\in\calE}\bLambda_\sharp\frac{\partial\theta_e}{\partial\bphi}\frac{\partial\theta_e}{\partial\bphi\transpose}\bLambda_\sharp\bigg\|_2
\leq
\bigg\|\sum_{e\in\calE}\widetilde{\bLambda}_\sharp\nabla_\bpsi\theta_e\nabla_\bpsi\theta_e\transpose\widetilde{\bLambda}_\sharp\bigg\|_2 = O(1).
\end{align}
By the Lipschitz continuity of $\ell'$, the fact that $\ell'(\theta_{0ij}^{(t)}; A_{ij}^{(t)})$'s are uniformly $\Optilde(\log n)$, and Lemma \ref{lemma:uniform_Optilde}, we know that
\begin{align*}
\sup_{\tau\in[0, 1]}\max_{e\in\calE}|\ell'(\theta_e)|
&\leq \sup_{\tau\in[0, 1]}\max_{t,i,j}|\ell'(\theta_{ij}^{(t)}(\bpsi_\tau); A_{ij}^{(t)}) - \ell'(\theta_{0ij}^{(t)}; A_{ij}^{(t)})| + \max_{t,i,j}|\ell'(\theta_{0ij}^{(t)}; A_{ij}^{(t)})|\\
&\lesssim \sup_{\tau\in[0, 1]}\max_{t,i,j}|\theta_{ij}^{(t)}(\bpsi_\tau) - \theta_{0ij}^{(t)}| + \Optilde(\log n) = \Optilde(\log n).
\end{align*}
Therefore, by Cauchy-Schwarz inequality,
\begin{align*}
\sup_{\tau\in[0, 1]}\bigg\|\widetilde{\bLambda}_\sharp\sum_{e\in\calE}\ell'(\theta_e)\nabla_\bpsi\theta_e\bigg\|_2
& =
\sup_{\tau\in[0, 1]}\sup_{\bw\in\mathbb{R}^{d_\bpsi}:\|\bw\|_2 = 1}\bigg|\bw\transpose\widetilde{\bLambda}_\sharp\sum_{e\in\calE}\ell'(\theta_e)\nabla_\bpsi\theta_e\bigg|\\
&\leq \bigg[\sup_{\tau\in[0, 1]}\sum_{e\in\calE}\{\ell'(\theta_e)\}^2\bigg]^{1/2}
\bigg\{\sum_{e\in\calE}(\bw\transpose\widetilde{\bLambda}_\sharp \nabla_\bpsi\theta_e)^2\bigg\}^{1/2} = \Optilde(\sqrt{m}n\log n),
\end{align*}
which immediately implies that the fourth term in the decomposition of $\bgamma_\calL$ satisfies
\begin{align*}
&\bigg\|\bLambda_\sharp\frac{1}{4}\int_0^1(1 - \tau)^2\sum_{e\in\calE}\dddot{\bJ}(\tau\widehat{\bB})[\widehat{\bB}, \widehat{\bB}, \widehat{\bB}]\transpose\left\{\ell'(\theta_e)\nabla_\bpsi\theta_e\right\}\mathrm{d}\tau\bigg\|_2\\
&\quad\leq \frac{1}{4\sqrt{n}}\sup_{\tau\in[0, 1]}\|\dddot{\bJ}(\tau\widehat{\bB})[\widehat{\bB}, \widehat{\bB}, \widehat{\bB}]\|_2\sup_{\tau\in[0, 1]}\bigg\|\widetilde{\bLambda}_\sharp\sum_{e\in\calE}\ell'(\theta_e)\nabla_\bpsi\theta_e\bigg\|_2 = \Optilde\bigg(\frac{\sqrt{m}\log n}{\sqrt{n}}\bigg).
\end{align*}
For the first term in the decomposition of $\bgamma_\calL$, by Cauchy-Schwarz inequality again,
\begin{align*}
&\bigg\|\bLambda_\sharp \int_0^1(1 - \tau)^2\sum_{e\in\calE}\bJ(\tau\widehat{\bB})\transpose\frac{\mathrm{d}^3}{\mathrm{d}\tau^3}\left\{\ell'(\theta_e)\nabla_\bpsi\theta_e\right\}\mathrm{d}\tau \bigg\|_2\\
&\quad\leq \sup_{\tau\in[0, 1]}\|\bJ_n(\tau\widehat{\bB})\|_2\bigg\|\widetilde{\bLambda}_\sharp\sum_{e\in\calE}\frac{\mathrm{d}^3}{\mathrm{d}\tau^3}\left\{\ell'(\theta_e)\nabla_\bpsi\theta_e\right\}\bigg\|_2 \leq \Optilde(1)\sup_{\tau\in[0, 1]}\bigg\|\widetilde{\bLambda}_\sharp\sum_{k = 1}^K\sum_{e\in\calE}\left\{\alpha_{e,k}^{(3)} + \delta_{e,k}^{(3)}\ell'(\theta_e)\right\}\bbeta_{e,k}^{(3)}\bigg\|_2\\
&\quad = \Optilde(1)\sup_{\tau\in[0, 1]}\sup_{\bw:\|\bw\|_2 = 1}\bigg|\sum_{k = 1}^K\sum_{e\in\calE}\left\{\alpha_{e,k}^{(3)} + \delta_{e,k}^{(3)}\ell'(\theta_e)\right\}\bw\transpose\widetilde{\bLambda}_\sharp\bbeta_{e,k}^{(3)}\bigg|\\
&\quad\leq \Optilde(1)\sup_{\tau\in[0, 1]}\bigg[\sum_{k = 1}^K\sum_{e\in\calE}\left\{\alpha_{e,k}^{(3)} + \delta_{e,k}^{(3)}\ell'(\theta_e)\right\}^2\bigg]^{1/2}\sup_{\bw:\|\bw\|_2 = 1}\bigg\{\sum_{k = 1}^K\sum_{e\in\calE}(\bw\transpose\widetilde{\bLambda}_\sharp\bbeta_{e,k}^{(3)})^2\bigg\}^{1/2}\\
&\quad = \Optilde\bigg\{\frac{\sqrt{m}(\log n)^{5/2}}{\sqrt{n}}\bigg\}.
\end{align*}
The analysis of the second and third terms in the decomposition of $\bgamma_\calL$ can be done by the same reasoning, and the proof is thus completed.
\end{proof}

\subsection{Controlling the third-order derivatives of the log-likelihood function}
\label{sub:controlling_the_third_order_derivatives_of_the_log_likelihood_function}

Throughout this subsection, 
\[
\vartheta_n = \frac{m^{6}(\log n)^2 + m^{11/2}(\log n)^{5/2} + m^{5}(\log n)^3}{\sqrt{n}}  
\] 
denotes the remainder rate of Section \ref{sub:Proof of the asymptotic normality}.

\begin{lemma}\label{lemma:third_order_cross_terms}
Suppose Assumptions \ref{assumption:edge_distribution}--\ref{assumption:true_parameter} hold. Then:
\begin{enumerate}[(i)]
\item There exists a constant $C > 0$ depending only on $d, C_1, C_2$ and $\kappa_2(b_0), \kappa_3(b_0)$ with $b_0 = 2C_1 + C_1^2C_2$, such that for all $\bu, \bv\in\mathbb{R}^{d_\bphi}$,
\begin{align*}
\|\bLambda_\sharp\nabla^3\calL(\bphi_0)[\bu, \bv]\|_2\leq C\sqrt{m}n\bigg\{1 + \max_{t\in[m],i\neq j}|\ell'(\theta_{0ij}^{(t)}; A_{ij}^{(t)})|\bigg\}\|\bJ_0\bu\|_\infty\|\bJ_0\bv\|_\infty.
\end{align*}
\item If, in addition, $m^{12}(\log n)^4/n\to 0$ as $n\to\infty$, then
\begin{align*}
&\|\bLambda_\sharp\nabla^3\calL(\bphi_0)[\bLambda_\sharp\widehat{\bs}, \bgamma_\bphi]\|_2 = \Optilde\{\sqrt{m}(\log n)^{3/2}\vartheta_n\},\quad
\|\bLambda_\sharp\nabla^3\calL(\bphi_0)[\bgamma_\bphi, \bgamma_\bphi]\|_2 = \Optilde\{\sqrt{m}(\log n)\vartheta_n^2\},\\
&\left\|2\bLambda_\sharp\nabla^3\calL(\bphi_0)[\bLambda_\sharp\widehat{\bs}, \bgamma_\bphi]  + \bLambda_\sharp\nabla^3\calL(\bphi_0)[\bgamma_\bphi, \bgamma_\bphi]\right\|_2 = \Optilde\{\sqrt{m}(\log n)^{3/2}\vartheta_n\}.
\end{align*}
\end{enumerate}
\end{lemma}

\begin{proof}[\bf Proof]
We adopt the notations in the proof of Lemma \ref{lemma:third_order_Taylor_expansion_loglikelihood} but specialize to the case when $\tau = 0$. Let $\calE = \{(t,i,j)\in[m]\times[n]\times[n]:i\neq j\}$, $e = (t,i,j)$, $\calS_1(e)\subset[d_\bpsi]$ the indices of the coordinates of $\bpsi$ on which $\theta_{ij}^{(t)}$ depends, and we view $\theta_e = \theta_{ij}^{(t)}$ as a function of $\bpsi$. All quantities below are evaluated at the truth: $\theta_{0e} = \theta_{0ij}^{(t)}$, $\ell_e':= \ell'(\theta_{0ij}^{(t)}; A_{ij}^{(t)})$, $\ell_e'':= \ell''(\theta_{0ij}^{(t)}; A_{ij}^{(t)})$, $\ell_e''':= \ell'''(\theta_{0ij}^{(t)}; A_{ij}^{(t)})$, $\nabla_\bpsi\theta_e = \nabla_\bpsi\theta_e(\bpsi_0)$, $\nabla^2_\bpsi\theta_e := \nabla^2_\bpsi\theta_e(\bpsi_0)$, and note that $\nabla^3_\bpsi\theta_e$ does not depend on $\bpsi$. Since $|\theta_{0e}|\leq 2C_1 + C_1^2C_2 = b_0$ by Assumption \ref{assumption:true_parameter}, we have $|\ell''_e|\leq\kappa_2(b_0)$ and $|\ell'''_e|\leq\kappa_3(b_0)$ by Assumption \ref{assumption:edge_distribution}. Also, the absolute values of the elements of $\nabla_\bpsi\theta_e$, $\nabla^2_\bpsi\theta_e$, and $\nabla^3_\bpsi\theta_e$ (viewed as a tensor) are bounded by a constant $C_0$ depending only on $d, C_1, C_2$, and $\nabla_\bpsi\theta_e, \nabla_\bpsi^2\theta_e\delta\bpsi_1$, $\nabla^3_\bpsi\theta_e[\delta\bpsi_1, \delta\bpsi_2]$only have nonzero elements over $\calS_1(e)$. 

\medskip\noindent $\blacksquare$ \textbf{Step 1. Lipschitz property of $\bX(\cdot)$ at $\bB = \mathbf{O}$.} For $\bU, \bV\in\mathbb{R}^{(n - 1 - d)\times d}$, Lemma \ref{lemma:integral_formula_matrix_squarerootinverse} at $\bB = \mathbf{O}$ gives $\bR_\tau = (\tau^2 + 1)^{-1}\eye_d$, $\dot{\bS}(\mathbf{O})[\cdot] = \mathbf{O}$, and hence $\dot{\bN}(\mathbf{O})[\bU] = \mathbf{O}$, $\ddot{\bN}(\mathbf{O})[\bU, \bV] = -(\bU\transpose\bV + \bV\transpose\bU)/2$, and $\dddot{\bN}(\mathbf{O})[\cdot,\cdot,\cdot] = \mathbf{O}$, where $\bS(\bB) = \eye_d + \bB\transpose\bB$ and $\bN(\bB) = \bS(\bB)^{-1/2}$. By the formulas for $\dot{\bX}, \ddot{\bX}, \dddot{\bX}$ in the proof of Lemma \ref{lemma:Lipschitz_Cayley_transform}, for any $\bU, \bV, \bU_1, \bU_2, \bU_3\in\mathbb{R}^{(n - 1 - d)\times d}$, 
\begin{align*}
\dot{\bX}(\mathbf{O})[\bU] & = \bX_{0\perp}\bU,\quad
\ddot{\bX}(\mathbf{O})[\bU, \bV] = -\frac{1}{2}\bX_0(\bU\transpose\bV + \bV\transpose\bU),\\
\dddot{\bX}(\mathbf{O})[\bU_1, \bU_2, \bU_3] & = -\frac{1}{2}\bX_{0\perp}\bU_1(\bU_2\transpose\bU_3 + \bU_3\transpose\bU_2)
-\frac{1}{2}\bX_{0\perp}\bU_2(\bU_1\transpose\bU_3 + \bU_3\transpose\bU_1)\\
&\quad
-\frac{1}{2}\bX_{0\perp}\bU_3(\bU_1\transpose\bU_2 + \bU_2\transpose\bU_1).
\end{align*}
Since $\bJ(\bB) = \mathrm{diag}\{\eye_{d_\balpha + d_\br}, \bJ_\bX(\bB)\}$ and $\bJ_\bX(\bB)\vect(\delta\bB) = \vect\{\dot{\bX}(\bB)[\delta\bB]\}$, the directional derivatives $\dot{\bJ}(\mathbf{O})[\bU]$ and $\ddot{\bJ}(\mathbf{O})[\bU, \bV]$ are only nonzero on the third diagonal block. Since $\|\bX_0\|_2 = \|\bX_{0\perp}\|_2 = \sqrt{n}$, we obtain
\begin{align*}
\|\dot{\bJ}(\mathbf{O})[\bU]\|_2\leq\sqrt{n}\|\bU\|_2,\quad
\|\ddot{\bJ}(\mathbf{O})[\bU, \bV]\|_2\leq 3\sqrt{n}\|\bU\|_{\mathrm{F}}\|\bV\|_{\mathrm{F}}.
\end{align*}
Furthermore, for $\bu\in\mathbb{R}^{d_\bphi}$ with $[\mathbf{O},\eye_{d_\bB}]\bu = \vect(\bB_\bu)$, we have $[\mathbf{O}, \eye_{d_\bX}]\bJ_0\bu = \vect(\bX_{0\perp}\bB_\bu)$ and 
\begin{align*}
\|\bB_\bu\|_{\mathrm{F}} = n^{-1/2}\|\bX_{0\perp}\bB_\bu\|_{\mathrm{F}}\leq n^{-1/2}\sqrt{nd}\,\|\bX_{0\perp}\bB_\bu\|_{\max}\leq\sqrt{d}\|\bJ_0\bu\|_\infty.
\end{align*}

\medskip\noindent$\blacksquare$ \textbf{Step 2. Generic sparse formula for the mixed derivative.} Let $\bB_\bu, \bB_\bv\in\mathbb{R}^{(n - 1 - d)\times d}$ satisfy $\vect(\bB_\bu) = [\mathbf{O}, \eye_{d_\bB}]\bu$ and $\vect(\bB_\bv) = [\mathbf{O}, \eye_{d_\bB}]\bv$, 
and $\bff_e(\bpsi) = \ell'(\theta_e(\bpsi); A_{ij}^{(t)})\nabla_\bpsi\theta_e(\bpsi)$, so that 
\[
  \nabla\calL(\bphi ) = \frac{1}{2}\sum_{e\in\calE}\bJ(\bB)\transpose\bff_e(\bpsi(\bphi)),
\]
where we recall that $\bphi\mapsto \bpsi(\bphi)$ is the natural embedding of $\bphi$ into the ambient parameter space and $\theta_e(\cdot)$ is viewed as a function of $\bpsi$. By the Leibniz formula,
\begin{align*}
\nabla^3\calL(\bphi_0)[\bu, \bv]
& = \frac{1}{2}\sum_{e\in\calE}\bJ_0\transpose\ddot{\bff}_e(\bphi_0)[\bu, \bv]
 + \frac{1}{2}\sum_{e\in\calE}\dot{\bJ}(\mathbf{O})[\bB_\bu]\transpose\dot{\bff}_e(\bphi_0)[\bv]
 + \frac{1}{2}\sum_{e\in\calE}\dot{\bJ}(\mathbf{O})[\bB_\bv]\transpose\dot{\bff}_e(\bphi_0)[\bu]\\
&\quad + \frac{1}{2}\sum_{e\in\calE}\ddot{\bJ}(\mathbf{O})[\bB_\bu, \bB_\bv]\transpose\bff_e(\bphi_0).
\end{align*}
Following the directional derivatives notations in \eqref{eqn:dot_notations}, we have
\begin{align*}
&\dot{\bpsi}(\bphi_0)[\bu]  = \bJ_0\bu,\;  
\dot{\bpsi}(\bphi_0)[\bv] = \bJ_0\bv,\;
\ddot{\bpsi}(\bphi_0)[\bu, \bv] = [\zero_{d_\balpha + d_\br}\transpose, \vect\{\ddot{\bX}(\mathbf{O})[\bB_\bu, \bB_\bv]\}\transpose]\transpose,\\
&\dot{\theta}_e(\bpsi(\bphi_0))[\bu] = \dot{\bpsi}(\bphi_0)[\bu]\transpose\nabla_\bpsi\theta_e(\bpsi(\bphi_0)),\\
&\ddot{\theta}_e(\bpsi(\bphi_0))[\bu, \bv] = \dot{\bpsi}(\bphi_0)[\bu]\transpose\nabla^2_\bpsi\theta_e(\bpsi(\bphi_0))\dot{\bpsi}(\bphi_0)[\bv] + \ddot{\bpsi}(\bphi_0)[\bu, \bv]\transpose\nabla_\bpsi\theta_e(\bpsi(\bphi_0)).
\end{align*}
Since $\theta_e$ is a cubic polynomial in $\bpsi$, the chain rule gives (omitting the dependence on $\bphi_0$)
\begin{align*}
\dot{\bff}_e[\bu] & = \ell''_e\dot{\theta}_e[\bu]\nabla_\bpsi\theta_e + \ell'_e\nabla^2_\bpsi\theta_e\dot{\bpsi}[\bu],\\
\dot{\bff}_e[\bv] & = \ell''_e\dot{\theta}_e[\bv]\nabla_\bpsi\theta_e + \ell'_e\nabla^2_\bpsi\theta_e\dot{\bpsi}[\bv],\\
\ddot{\bff}_e[\bu, \bv]
& = (\ell'''_e\dot{\theta}_e[\bu]\dot{\theta}_e[\bv] + \ell''_e\ddot{\theta}_e[\bu, \bv])\nabla_\bpsi \theta_e
 + \ell''_e\dot{\theta}_e[\bu]\nabla^2_\bpsi\theta_e\dot{\bpsi}[\bv] + \ell''_e\dot{\theta}_e[\bv]\nabla^2_\bpsi\theta_e\dot{\bpsi}[\bu]\\
&\quad + \ell'_e\{\nabla^3_\bpsi\theta_e[\dot{\bpsi}[\bv]]\dot{\bpsi}[\bu] + \nabla^2_\bpsi\theta_e\ddot{\bpsi}[\bu, \bv]\}.
\end{align*}
By Step 1 and $\|\bX_0\|_{2\to\infty}\leq C_1$,
\[
\|\ddot{\bpsi}(\bphi_0)[\bu, \bv]\|_\infty\leq\|\ddot{\bX}(\mathbf{O})[\bB_\bu, \bB_\bv]\|_{2\to\infty}\leq C_1\|\bB_\bu\|_2\|\bB_\bv\|_2\leq C_1d\|\bJ_0\bu\|_\infty\|\bJ_0\bv\|_\infty.  
\]
Because the absolute values of the nonzero elements of $\nabla_\bpsi\theta_e, \nabla^2_\bpsi\theta_e, \nabla^3_\bpsi\theta_e$ (viewed as a tensor) are bounded by a constant and there are at most $|\calS_1(e)|$ many of them, we then have, at $\bphi = \bphi_0$,
\begin{align*}
&|\dot{\theta}_e[\bu]|\leq C\|\bJ_0\bu\|_\infty,\quad
\|\nabla^2_\bpsi\theta_e\dot{\bpsi}[\bu]\|_\infty\leq C\|\bJ_0\bu\|_\infty,\\
&|\ddot{\theta}_e[\bu, \bv]| + \|\nabla^3_\bpsi\theta_e[\dot{\bpsi}[\bv]]\dot{\bpsi}[\bu]\|_\infty + \|\nabla^2_\bpsi\theta_e\ddot{\bpsi}[\bu, \bv]\|_\infty\leq C\|\bJ_0\bu\|_\infty\|\bJ_0\bv\|_\infty.
\end{align*}
Consequently, there exist an absolute constant $K\in\mathbb{N}_+$, scalars $(\alpha_{e,k}, \alpha'_{e,k}, \alpha''_{e,k}:e\in\calE,k\in[K])$, and vectors $(\bbeta_{e,k},\bbeta'_{e,k},\bbeta''_{e,k}:e\in\calE,k\in[K])\subset\mathbb{R}^{d_\bpsi}$, where the index set of the nonzero elements of $\bbeta_{e,k},\bbeta_{e,k}',\bbeta_{e,k}''$ are contained in $\calS_1(e)$ with $\|\bbeta_{e,k}\|_\infty,\|\bbeta'_{e,k}\|_\infty,\|\bbeta''_{e,k}\|_\infty\leq 1$, such that
\begin{align*}
&\ddot{\bff}_e[\bu, \bv] = \sum_{k = 1}^K \alpha_{e,k}\bbeta_{e,k},\quad
\dot{\bff}_e[\bu] = \sum_{k = 1}^K \alpha'_{e,k}\bbeta'_{e,k},\quad
\dot{\bff}_e[\bv] = \sum_{k = 1}^K \alpha''_{e,k}\bbeta''_{e,k},
\\
&|\alpha_{e,k}|\leq C(1 + |\ell'_e|)\|\bJ_0\bu\|_\infty\|\bJ_0\bv\|_\infty,\quad
|\alpha'_{e,k}|\leq C(1 + |\ell'_e|)\|\bJ_0\bu\|_\infty,\quad
|\alpha''_{e,k}|\leq C(1 + |\ell'_e|)\|\bJ_0\bv\|_\infty.
\end{align*}

\medskip\noindent\textbf{$\blacksquare$ Step 3. Proof of item (i).} As shown in the proof of Lemma \ref{lemma:third_order_Taylor_expansion_loglikelihood}, for any scalars $(c_e)_{e\in\calE}$ and vectors $(\bbeta_e)_{e\in\calE}$ with the index of nonzero elements of $\bbeta_e$ contained in $\calS_1(e)$ and $\|\bbeta_e\|_\infty\leq 1$, Cauchy--Schwarz inequality yields
\begin{align*}
\bigg\|\widetilde{\bLambda}_\sharp\sum_{e\in\calE}c_e\bbeta_e\bigg\|_2
 = \sup_{\|\bw\|_2 = 1}\bigg|\sum_{e\in\calE}c_e\bw\transpose\widetilde{\bLambda}_\sharp\bbeta_e\bigg|
\leq\bigg(\sum_{e\in\calE}c_e^2\bigg)^{1/2}\sup_{\|\bw\|_2 = 1}\bigg\{\sum_{e\in\calE}(\bw\transpose\widetilde{\bLambda}_\sharp\bbeta_e)^2\bigg\}^{1/2}
\lesssim \bigg(\sum_{e\in\calE}c_e^2\bigg)^{1/2}.
\end{align*}
In particular, by Step 2 and $|\calE|\leq mn^2$,
\begin{align*}
&\bigg\|\widetilde{\bLambda}_\sharp\sum_{e\in\calE}\ddot{\bff}_e[\bu, \bv]\bigg\|_2\lesssim \sqrt{m}n\bigg(1 + \max_{e\in\calE}|\ell'_e|\bigg)\|\bJ_0\bu\|_\infty\|\bJ_0\bv\|_\infty,\;
\bigg\|\widetilde{\bLambda}_\sharp\sum_{e\in\calE}\dot{\bff}_e[\bv]\bigg\|_2\lesssim \sqrt{m}n\bigg(1 + \max_{e\in\calE}|\ell'_e|\bigg)\|\bJ_0\bv\|_\infty,\\
&\bigg\|\widetilde{\bLambda}_\sharp\sum_{e\in\calE}\dot{\bff}_e[\bu]\bigg\|_2\lesssim \sqrt{m}n\bigg(1 + \max_{e\in\calE}|\ell'_e|\bigg)\|\bJ_0\bu\|_\infty,\;
\bigg\|\widetilde{\bLambda}_\sharp\sum_{e\in\calE}\bff_e\bigg\|_2\lesssim \sqrt{m}n\max_{e\in\calE}|\ell'_e|.
\end{align*}
Since $\bLambda_\sharp\bJ_0\transpose = \bJ_{0n}\transpose\widetilde{\bLambda}_\sharp$ by \eqref{eqn:J_n(B)} and $\|\bJ_{0n}\|_2 = 1$, the first bound above gives 
\begin{align*}
\bigg\|\bLambda_\sharp\sum_{e\in\calE}\bJ_0\transpose\ddot{\bff}(\bphi_0)[\bu, \bv]\bigg\|_2\leq\bigg\|\widetilde{\bLambda}_\sharp\sum_{e\in\calE}\ddot{\bff}_e(\bphi_0)[\bu, \bv]\bigg\|_2\lesssim \sqrt{m}n\bigg(1 + \max_{e}|\ell'_e|\bigg)\|\bJ_0\bu\|_\infty\|\bJ_0\bv\|_\infty  
\end{align*}
Similarly, $\bLambda_\sharp\dot{\bJ}[\bB_\bu]\transpose = n^{-1/2}\dot{\bJ}[\bB_\bu]\transpose\widetilde{\bLambda}_\sharp$ and $\bLambda_\sharp\dot{\bJ}[\bB_\bv]\transpose = n^{-1/2}\dot{\bJ}[\bB_\bv]\transpose\widetilde{\bLambda}_\sharp$, so that by Step 1, we have
\begin{align*}
\bigg\|\bLambda_\sharp\sum_{e\in\calE}\dot{\bJ}(\mathbf{O})[\bB_\bv]\transpose\dot{\bff}_e(\bphi_0)[\bu]\bigg\|_2
&\leq \frac{1}{\sqrt{n}}\|\dot{\bJ}(\mathbf{O})[\bB_\bv]\|_2\bigg\|\sum_{e\in\calE}\widetilde{\bLambda}_\sharp\dot{\bff}_e(\bphi_0)[\bu]\bigg\|_2
\leq\|\bB_\bv\|_2\bigg\|\sum_{e\in\calE}\widetilde{\bLambda}_\sharp\dot{\bff}_e(\bphi_0)[\bu]\bigg\|_2\\
&\lesssim \sqrt{m}n\bigg(1 + \max_{e\in\calE}|\ell'_e|\bigg)\|\bJ_0\bu\|_\infty\|\bJ_0\bv\|_\infty,
\end{align*}
and the same upper bound for the remaining two terms appearing in the decomposition of $\nabla^3\calL(\bphi_0)[\bu, \bv]$. This proves item (i).

\medskip\noindent\textbf{$\blacksquare$ Step 4. Proof of item (ii).} By Lemma \ref{lemma:uniform_Optilde} and Assumption \ref{assumption:true_parameter}, 
\begin{align}
\label{eqn:max_ellprime_bound}
\max_{e\in\calE}|\ell'_e| = \Optilde(\log n).
\end{align}
By \eqref{eqn:J_n(B)}, $\bJ_0\bLambda_\sharp\widehat{\bs} = \widetilde{\bLambda}_\sharp\bJ_{0n}\widehat{\bs}$, and $\bJ_{0n}\widehat{\bs} = \bJ_{0n}(\bG_0^\sharp)^{-1}\bJ_{0n}\transpose\bJ_{0n}\bLambda_\sharp\nabla\calL(\bphi_0)$, so \eqref{eqn:J0n_s_hat_bound} implies $\|\bJ_{0n}\widehat{\bs}\|_\infty = \Optilde(\sqrt{\log n})$, and hence
\[
\|\bJ_0\bLambda_\sharp\widehat{\bs}\|_\infty\leq n^{-1/2}\|\bJ_{0n}\widehat{\bs}\|_\infty = \Optilde\bigg(\sqrt{\frac{\log n}{n}}\bigg).
\]
Next, $\bJ_0\bgamma_\bphi = \bgamma$ is the remainder in \eqref{eqn:Taylor_expansion}, and Step 2 of the proof in Section \ref{sub:Proof of the asymptotic normality} shows that
\[
\|\mathrm{diag}(\sqrt{n}\eye_{d_\balpha}, \sqrt{mn}\eye_{d_\bX})\scrP_\bZ\bgamma\|_\infty = \Optilde(\vartheta_n)  
\] and $\|n\scrP_\br\bgamma\|_\infty = \Optilde\{m^3\log n + m^{5/2}(\log n)^{3/2}\}$. Since $m^3\log n + m^{5/2}(\log n)^{3/2}\leq\sqrt{n}\vartheta_n$, it follows that
\[
\|\bJ_0\bgamma_\bphi\|_\infty = \|\bgamma\|_\infty\leq\max\bigg\{\frac{\|\mathrm{diag}(\sqrt{n}\eye_{d_\balpha}, \sqrt{mn}\eye_{d_\bX})\scrP_\bZ\bgamma\|_\infty}{\sqrt{n}}, \frac{\|n\scrP_\br\bgamma\|_\infty}{n}\bigg\} = \Optilde\bigg(\frac{\vartheta_n}{\sqrt{n}}\bigg).
\]
Invoking these three bounds into (i) with $(\bu, \bv) = (\bLambda_\sharp\widehat{\bs}, \bgamma_\bphi)$ and $(\bu, \bv) = (\bgamma_\bphi, \bgamma_\bphi)$ yields
\begin{align*}
\|\bLambda_\sharp\nabla^3\calL(\bphi_0)[\bLambda_\sharp\widehat{\bs}, \bgamma_\bphi]\|_2
& = \Optilde(\sqrt{m}n\log n)\Optilde\bigg(\sqrt{\frac{\log n}{n}}\bigg)\Optilde\bigg(\frac{\vartheta_n}{\sqrt{n}}\bigg) = \Optilde\{\sqrt{m}(\log n)^{3/2}\vartheta_n\},\\
\|\bLambda_\sharp\nabla^3\calL(\bphi_0)[\bgamma_\bphi, \bgamma_\bphi]\|_2
& = \Optilde(\sqrt{m}n\log n)\Optilde\bigg(\frac{\vartheta_n^2}{n}\bigg) = \Optilde\{\sqrt{m}(\log n)\vartheta_n^2\}.
\end{align*}
Finally, $\vartheta_n\to 0$ under $m^{12}(\log n)^4/n\to 0$, so that $\sqrt{m}(\log n)\vartheta_n^2\leq\sqrt{m}(\log n)^{3/2}\vartheta_n$ for all large $n$, and the last assertion follows from the triangle inequality, thereby completing the proof.
\end{proof}

\subsection{Controlling the fluctuation of $\bQ_1$}
\label{sub:controlling_the_bias_remainder_for_Rt_hat_Q1}

This subsection will establish the $\widetilde{\scrP}_{\br^{(t)}}\bQ_1$ centers around its expected value. 

\begin{lemma}\label{lemma:Q1_fluctuation}
Suppose Assumptions \ref{assumption:edge_distribution}--\ref{assumption:true_parameter} hold and $m\leq n^{\delta_0}$ for some constant $\delta_0 > 0$. Then, 
\begin{align*}
\max_{t\in[m]}\left\|\widetilde{\scrP}_{\br^{(t)}}\bQ_1 - \expect_0\widetilde{\scrP}_{\br^{(t)}}\bQ_1\right\|_2 = \Optilde\bigg(\frac{\log n}{\sqrt{n}}\bigg).
\end{align*}
\end{lemma}

\begin{proof}[\bf Proof]
We adopt the notations in the proof of Lemma \ref{lemma:third_order_cross_terms} and denote $\calE^\circ := \{e = (t, a, b): t\in[m], 1\leq a < b\leq n\}$ for the set of independent edges.
Specifically, recall that $\nabla_\bpsi\theta_e := \nabla_\bpsi\theta_e(\bpsi_0)$ and $\nabla^2_\bpsi\theta_e := \nabla^2_\bpsi\theta_e(\bpsi_0)$ have entries bounded by a constant and the index set of the nonzero elements are contained in $\calS_1(e)$, $|\calS_1(e)| = O(1)$, and $(\widetilde{\bLambda}_\sharp)_{ll}^2\sum_{e\in\calE^\circ}\mathbbm{1}\{l\in\calS_1(e)\}\leq 1$ for every $l\in[d_\bpsi]$, where $(\widetilde{\bLambda}_\sharp)_{ll}$ denotes the $l$th diagonal element of $\widetilde{\bLambda}_\sharp$.

\medskip\noindent\textbf{$\blacksquare$ Step 1. Decomposition.} 
Recall that
\begin{align}
\label{eqn:Q1_building_blocks}
\bLambda_\sharp\nabla\calL(\bphi_0) = \sum_{e\in\calE^\circ}\ell'_e\bJ_{0n}\transpose\widetilde{\bLambda}_\sharp\nabla_\bpsi\theta_e,\;
\bH^\sharp = \sum_{e\in\calE^\circ}\bigg\{\ell''_e\bJ_{0n}\transpose\widetilde{\bLambda}_\sharp\nabla_\bpsi\theta_e\nabla_\bpsi\theta_e \transpose\widetilde{\bLambda}_\sharp\bJ_{0n} + \ell'_e\bLambda_\sharp\frac{\partial^2\theta_e}{\partial\bphi\partial\bphi\transpose}(\bpsi(\bphi_0))\bLambda_\sharp\bigg\}
\end{align}
by definition, so that
\begin{align*}
\widetilde{\scrP}_{\br^{(t)}}\bQ_1
& = \widetilde{\scrP}_{\br^{(t)}}(\bG_0^\sharp)^{-1}\sum_{e\in\calE^\circ}\bigg\{(\ell''_e - \expect_0\ell''_e)\bJ_{0n}\transpose\widetilde{\bLambda}_\sharp\nabla_\bpsi\theta_e\nabla_\bpsi\theta_e \transpose\widetilde{\bLambda}_\sharp\bJ_{0n} + \ell'_e\bLambda_\sharp\frac{\partial^2\theta_e}{\partial\bphi\partial\bphi\transpose}(\bpsi(\bphi_0))\bLambda_\sharp\bigg\}\\
&\quad \times (\bG_0^\sharp)^{-1}\sum_{e'\in\calE^\circ}\ell'_{e'}\bJ_{0n}\transpose\widetilde{\bLambda}_\sharp\nabla_\bpsi\theta_{e'}\\
& = \sum_{e,e'\in\calE^\circ}(\ell''_e - \expect_0\ell''_e)\ell'_{e'}\left\{
\nabla_\bpsi\theta_e\transpose\widetilde{\bLambda}_\sharp\bJ_{0n}
(\bG_0^\sharp)^{-1}\bJ_{0n}\transpose\widetilde{\bLambda}_\sharp\nabla_\bpsi\theta_{e'}
\right\}\widetilde{\scrP}_{\br^{(t)}}(\bG_0^\sharp)^{-1}\bJ_{0n}\transpose\widetilde{\bLambda}_\sharp\nabla_\bpsi\theta_e\\
&\quad + \sum_{e,e'\in\calE^\circ}\ell'_e\ell'_{e'}\widetilde{\scrP}_{\br^{(t)}}(\bG_0^\sharp)^{-1}
\bLambda_\sharp\frac{\partial^2\theta_e}{\partial\bphi\partial\bphi\transpose}(\bpsi(\bphi_0))\bLambda_\sharp
(\bG_0^\sharp)^{-1}\bJ_{0n}\transpose\widetilde{\bLambda}_\sharp\nabla_\bpsi\theta_{e'}.
\end{align*}
This implies that
\begin{align*}
\widetilde{\scrP}_{\br^{(t)}}\bQ_1 - 
\expect_0\widetilde{\scrP}_{\br^{(t)}}\bQ_1 
& = \sum_{e \in\calE^\circ}\{(\ell''_e - \expect_0\ell''_e)\ell'_{e} - \expect_0(\ell''_e\ell'_{e})\}\left\|
(\bG_0^\sharp)^{-1/2}\bJ_{0n}\transpose\widetilde{\bLambda}_\sharp\nabla_\bpsi\theta_{e}
\right\|_2^2\widetilde{\scrP}_{\br^{(t)}}(\bG_0^\sharp)^{-1}\bJ_{0n}\transpose\widetilde{\bLambda}_\sharp\nabla_\bpsi\theta_e\\
&\quad + \sum_{e \in\calE^\circ}\left\{(\ell'_e)^2 - \expect_0(\ell'_e)^2\right\}\widetilde{\scrP}_{\br^{(t)}}(\bG_0^\sharp)^{-1}
\bLambda_\sharp\frac{\partial^2\theta_e}{\partial\bphi\partial\bphi\transpose}(\bpsi(\bphi_0))\bLambda_\sharp
(\bG_0^\sharp)^{-1}\bJ_{0n}\transpose\widetilde{\bLambda}_\sharp\nabla_\bpsi\theta_{e }\\
&\quad + \sum_{e\neq e' }(\ell''_e - \expect_0\ell''_e)\ell'_{e'}\left\{
\nabla_\bpsi\theta_e\transpose\widetilde{\bLambda}_\sharp\bJ_{0n}
(\bG_0^\sharp)^{-1}\bJ_{0n}\transpose\widetilde{\bLambda}_\sharp\nabla_\bpsi\theta_{e'}
\right\}\widetilde{\scrP}_{\br^{(t)}}(\bG_0^\sharp)^{-1}\bJ_{0n}\transpose\widetilde{\bLambda}_\sharp\nabla_\bpsi\theta_e\\
&\quad + \sum_{e\neq e'}\ell'_e\ell'_{e'}\widetilde{\scrP}_{\br^{(t)}}(\bG_0^\sharp)^{-1}
\bLambda_\sharp\frac{\partial^2\theta_e}{\partial\bphi\partial\bphi\transpose}(\bpsi(\bphi_0))\bLambda_\sharp
(\bG_0^\sharp)^{-1}\bJ_{0n}\transpose\widetilde{\bLambda}_\sharp\nabla_\bpsi\theta_{e'}.
\end{align*}
In the above decomposition, the four terms on the right-hand side correspond to the diagonal terms (the first two terms) and the off-diagonal terms (the last two terms), and they call for different bounding techniques. 

\medskip\noindent\textbf{$\blacksquare$ Step 2. Deterministic bounds.}
\begin{enumerate}[(a)]
  \item We first have 
  $\|\bJ_{0n}\transpose\widetilde{\bLambda}_\sharp\nabla_\bpsi\theta_e\|_2\leq\|\widetilde{\bLambda}_\sharp\nabla_\bpsi\theta_e\|_1\leq Cn^{-1/2}$ since there are at most $O(1)$-many nonzero elements in $\nabla_\bpsi\theta_e$ for each fixed $e\in\calE^\circ$.  
   By Proposition \ref{prop:expected_hessian_bound}, $\bG_0^\sharp$ and $(\bG_0^\sharp)^{-1}$ have bounded eigenvalues, so $\|(\bG_0^\sharp)^{-1}\|_2\leq C$, $\|(\bG_0^\sharp)^{-1/2}\bJ_{0n}\transpose\widetilde{\bLambda}_\sharp\nabla_\bpsi\theta_{e}\|_2^2 = O(n^{-1})$, and $\|\widetilde{\scrP}_{\br^{(t)}}(\bG_0^\sharp)^{-1}\bJ_{0n}\transpose\widetilde{\bLambda}_\sharp\nabla_\bpsi\theta_{e}\|_2 = O(n^{-1/2})$. Moreover, $\min_{e\in\calE^\circ}(-\ell''_e)$ is bounded away from $0$, so that by \eqref{eqn:frame_bound}, 
  \[
  \sum_{e\in\calE^\circ}\left\|\widetilde{\scrP}_{\br^{(t)}}(\bG_0^\sharp)^{-1}\bJ_{0n}\transpose\widetilde{\bLambda}_\sharp\nabla_\bpsi\theta_{e}\right\|_2^2\lesssim \mathrm{tr}\left\{\widetilde{\scrP}_{\br^{(t)}}(\bG_0^\sharp)^{-1}\widetilde{\scrP}_{\br^{(t)}}\transpose\right\} = O(1).
  \]

  \item By the calculations in Subsection \ref{sub:gradient_hessian}, we have
  \begin{align}
  \label{eqn:Ve_formula}
  \bLambda_\sharp\frac{\partial^2\theta_e}{\partial\bphi\partial\bphi\transpose}\bLambda_\sharp = \bJ_{0n}\transpose\widetilde{\bLambda}_\sharp\nabla^2_\bpsi\theta_e\widetilde{\bLambda}_\sharp\bJ_{0n} + \frac{1}{mn^2}\mathrm{diag}\Big\{\mathbf{O}_{d_\balpha + d_\br}, -\frac{1}{2}(\bM_e + \bM_e\transpose)\otimes\eye_{n - 1 - d}\Big\},
  \end{align}
  where we denote $\bM_e = \bR_0^{(t)}(\bx_{0i}\bx_{0j}\transpose + \bx_{0j}\bx_{0i}\transpose)$ for $e = (t, i, j)\in\calE^\circ$. Note that the spectral norm of the diagonal matrix in the second term above is $O(1)$. 

  \item For any $j\in[d(d + 1)/2]$, we have
  $\|\bJ_{0n}(\bG_0^\sharp)^{-1}\widetilde{\scrP}_{\br^{(t)}}\transpose\be_j\|_\infty\leq \|\bJ_{0n}(\bG_0^\sharp)^{-1}\widetilde{\scrP}_{\br^{(t)}}\transpose\be_j\|_2  = O(1)$, and the vector $\widetilde{\bLambda}_\sharp\nabla^2_\bpsi\theta_e\widetilde{\bLambda}_\sharp\bJ_{0n}(\bG_0^\sharp)^{-1}\widetilde{\scrP}_{\br^{(t)}}\transpose\be_j$ is supported on $\calS_1(e)$. Since the entries of $\nabla^2_\bpsi\theta_e$ are uniformly bounded by a constant, $|\calS_1(e)| = O(1)$, and $\sum_{l\in\calS_1(e)}(\widetilde{\bLambda}_\sharp)_{ll}^2 = O(n^{-1})$, Cauchy--Schwarz inequality implies
  \begin{align*}
  \max_{e\in\calE^\circ}\left\|\widetilde{\bLambda}_\sharp\nabla^2_\bpsi\theta_e\widetilde{\bLambda}_\sharp\bJ_{0n}(\bG_0^\sharp)^{-1}\widetilde{\scrP}_{\br^{(t)}}\transpose\be_j\right\|_2^2
  &\lesssim \max_{e\in\calE^\circ}\frac{1}{n}\sum_{l\in\calS_1(e)}(\widetilde{\bLambda}_\sharp)_{ll}^2\left\|\bJ_{0n}(\bG_0^\sharp)^{-1}\widetilde{\scrP}_{\br^{(t)}}\transpose\be_j\right\|_\infty^2 = O(n^{-2}),\\
  \sum_{e\in\calE^\circ}\left\|\widetilde{\bLambda}_\sharp\nabla^2_\bpsi\theta_e\widetilde{\bLambda}_\sharp\bJ_{0n}(\bG_0^\sharp)^{-1}\widetilde{\scrP}_{\br^{(t)}}\transpose\be_j\right\|_2^2
  &\lesssim \frac{1}{n}\sum_{l = 1}^{d_\bpsi}\left\{\be_l\transpose\bJ_{0n}(\bG_0^\sharp)^{-1}\widetilde{\scrP}_{\br^{(t)}}\transpose\be_j\right\}^2(\widetilde{\bLambda}_\sharp)_{ll}^2\sum_{e\in\calE^\circ}\mathbbm{1}\{l\in\calS_1(e)\}\\
  &\lesssim \frac{1}{n}\left\|\bJ_{0n}(\bG_0^\sharp)^{-1}\widetilde{\scrP}_{\br^{(t)}}\transpose\be_j\right\|_2^2 = O(n^{-1}),
  \end{align*}
  where we have used the fact that $\max_{l\in[d_\bpsi]}(\widetilde{\bLambda})_{ll}^2\sum_{e\in\calE}\mathbbm{1}\{l\in\calS_1(e)\} = O(1)$. 

  \item By (b), (c), and Cauchy--Schwarz inequality, we have, 
  \begin{align*}
  &\max_{e\in\calE^\circ}
  \left|\be_j\transpose\widetilde{\scrP}_{\br^{(t)}}(\bG_0^\sharp)^{-1}
    \bLambda_\sharp\frac{\partial^2\theta_e}{\partial\bphi\partial\bphi\transpose}\bLambda_\sharp
  (\bG_0^\sharp)^{-1}\bJ_{0n}\transpose\widetilde{\bLambda}_\sharp\nabla_\bpsi\theta_e
  \right|_2\\
  &\quad\leq \max_{e\in\calE^\circ}\left\|\widetilde{\bLambda}_\sharp\nabla^2_\bpsi\theta_e\widetilde{\bLambda}_\sharp\bJ_{0n}(\bG_0^\sharp)^{-1}\widetilde{\scrP}_{\br^{(t)}}\transpose\be_j\right\|_2\|(\bG_0^\sharp)^{-1}\|_2\|\bJ_{0n}\transpose\widetilde{\bLambda}_\sharp\nabla_\bpsi\theta_e\|_2\\
  &\qquad + \frac{1}{mn^2}\max_{e\in\calE^\circ}\|\bM_e\|_2\| (\bG_0^\sharp)^{-1}\|_2^2\|\bJ_{0n}\transpose\widetilde{\bLambda}_\sharp\nabla_\bpsi\theta_e\|_2 = O(n^{-3/2})
  \end{align*}
  and
  \begin{align*}
  &\sum_{e\in\calE^\circ}\left\|\widetilde{\scrP}_{\br^{(t)}}(\bG_0^\sharp)^{-1}
    \bLambda_\sharp\frac{\partial^2\theta_e}{\partial\bphi\partial\bphi\transpose}\bLambda_\sharp
  (\bG_0^\sharp)^{-1}\bJ_{0n}\transpose\widetilde{\bLambda}_\sharp\nabla_\bpsi\theta_e
  \right\|_2^2\\
  &\quad\leq 2\sum_{j = 1}^{d(d + 1)/2}\bigg\{\sum_{e\in\calE^\circ}\|(\bG_0^\sharp)^{-1}\|_2^2\left\|\widetilde{\bLambda}_\sharp\nabla^2_\bpsi\theta_e\widetilde{\bLambda}_\sharp\bJ_{0n}(\bG_0^\sharp)^{-1}\widetilde{\scrP}_{\br^{(t)}}\transpose\be_j\right\|_2^2\max_{e\in\calE^\circ}\|\bJ_{0n}\transpose\widetilde{\bLambda}_\sharp\nabla_\bpsi\theta_e\|_2^2\\
  &\qquad + \frac{|\calE^\circ|}{m^2n^4}\|(\bG_0^\sharp)^{-1}\|_2^4\max_{e\in\calE^\circ}\|\bM_e\|_2^2\|\bJ_{0n}\transpose\widetilde{\bLambda}_\sharp\nabla_\bpsi\theta_e\|_2^2\bigg\} = O(n^{-2}).
  \end{align*}
\end{enumerate}

\medskip\noindent\textbf{$\blacksquare$ Step 3. The diagonal sums (the first two terms).} The random variables $\{(\ell''_e - \expect_0\ell''_e)\ell'_e - \expect_0(\ell''_e\ell'_e):e\in\calE^\circ\}$ are independent, mean-zero sub-exponential random variables with a uniform variance proxy parameter. By Lemma \ref{lemma:matrix_Bernstein}, Step 2(a), the first diagonal sum is therefore $\Optilde(\sqrt{\log n}/n)$. For the second diagonal sum, it is sufficient to apply Theorem \ref{thm:HW} elementwise, which implies that
\begin{align*}
\bigg\|\sum_{e \in\calE^\circ}\left\{(\ell'_e)^2 - \expect_0(\ell'_e)^2\right\}\widetilde{\scrP}_{\br^{(t)}}(\bG_0^\sharp)^{-1}
\bLambda_\sharp\frac{\partial^2\theta_e}{\partial\bphi\partial\bphi\transpose}(\bpsi(\bphi_0))\bLambda_\sharp
(\bG_0^\sharp)^{-1}\bJ_{0n}\transpose\widetilde{\bLambda}_\sharp\nabla_\bpsi\theta_{e }\bigg\|_2 = \Optilde\bigg(\frac{\sqrt{\log n}}{n}\bigg)
\end{align*}
by Step 2(d). 

\medskip\noindent\textbf{$\blacksquare$ Step 4. The first off-diagonal sums (the third term).} Let $(\bar{A}_e)_{e\in\calE^\circ}$ be an independent copy of $(A_e)_{e\in\calE^\circ}$, $\bar{\ell}'_e := \ell'(\theta_{0e}; \bar{A}_e)$, and $\bar{\bs} := (\bG_0^\sharp)^{-1}\sum_{e\in\calE^\circ}\bar{\ell}'_e\bJ_{0n}\transpose\widetilde{\bLambda}_\sharp\nabla_\bpsi\theta_e$, i.e.\ $\widehat{\bs}$ computed from $\bar{A}$. By Theorem \ref{thm:decoupling}, it suffices to consider
\begin{align*}
&\sum_{e\neq e' }(\ell''_e - \expect_0\ell''_e)\bar{\ell}'_{e'}\left\{
\nabla_\bpsi\theta_e\transpose\widetilde{\bLambda}_\sharp\bJ_{0n}
(\bG_0^\sharp)^{-1}\bJ_{0n}\transpose\widetilde{\bLambda}_\sharp\nabla_\bpsi\theta_{e'}
\right\}\widetilde{\scrP}_{\br^{(t)}}(\bG_0^\sharp)^{-1}\bJ_{0n}\transpose\widetilde{\bLambda}_\sharp\nabla_\bpsi\theta_e\\
&\quad = \sum_{e\in\calE^\circ}(\ell''_e - \expect_0\ell''_e)\bigg[\sum_{e'\in\calE^\circ}\bar{\ell}'_{e'}\left\{
\nabla_\bpsi\theta_e\transpose\widetilde{\bLambda}_\sharp\bJ_{0n}
(\bG_0^\sharp)^{-1}\bJ_{0n}\transpose\widetilde{\bLambda}_\sharp\nabla_\bpsi\theta_{e'}
\right\}\bigg]\widetilde{\scrP}_{\br^{(t)}}(\bG_0^\sharp)^{-1}\bJ_{0n}\transpose\widetilde{\bLambda}_\sharp\nabla_\bpsi\theta_e\\
&\qquad - \sum_{e\in\calE^\circ}(\ell''_e - \expect_0\ell''_e)\bar{\ell}'_{e}\left\{
\nabla_\bpsi\theta_e\transpose\widetilde{\bLambda}_\sharp\bJ_{0n}
(\bG_0^\sharp)^{-1}\bJ_{0n}\transpose\widetilde{\bLambda}_\sharp\nabla_\bpsi\theta_{e}
\right\}\widetilde{\scrP}_{\br^{(t)}}(\bG_0^\sharp)^{-1}\bJ_{0n}\transpose\widetilde{\bLambda}_\sharp\nabla_\bpsi\theta_e.
\end{align*}
The argument of \eqref{eqn:J0n_s_hat_bound} applied to $\bar{A}$ yields $\|\bJ_{0n}\bar{\bs}\|_\infty = \Optilde(\sqrt{\log n})$, so by Step 2(a), we have
\[
\max_{e\in\calE^\circ}\bigg|\sum_{e'\in\calE^\circ}\bar{\ell}'_{e'}\left\{
\nabla_\bpsi\theta_e\transpose\widetilde{\bLambda}_\sharp\bJ_{0n}
(\bG_0^\sharp)^{-1}\bJ_{0n}\transpose\widetilde{\bLambda}_\sharp\nabla_\bpsi\theta_{e'}
\right\}\bigg|\leq\max_{e\in\calE^\circ}\|\widetilde{\bLambda}_\sharp\nabla_\bpsi\theta_e\|_1\|\bJ_{0n}\bar{\bs}\|_\infty = \Optilde\bigg(\sqrt{\frac{\log n}{n}}\bigg)
\]
and
\begin{align*}
&\sum_{e\in\calE^\circ}\bigg|\sum_{e'\in\calE^\circ}\bar{\ell}'_{e'}\left\{
\nabla_\bpsi\theta_e\transpose\widetilde{\bLambda}_\sharp\bJ_{0n}
(\bG_0^\sharp)^{-1}\bJ_{0n}\transpose\widetilde{\bLambda}_\sharp\nabla_\bpsi\theta_{e'}
\right\}\bigg|^2\|\widetilde{\scrP}_{\br^{(t)}}(\bG_0^\sharp)^{-1}\bJ_{0n}\transpose\widetilde{\bLambda}_\sharp\nabla_\bpsi\theta_e\|_2^2\\
&\quad\leq \max_{e\in\calE^\circ}\bigg|\sum_{e'\in\calE^\circ}\bar{\ell}'_{e'}\left\{
\nabla_\bpsi\theta_e\transpose\widetilde{\bLambda}_\sharp\bJ_{0n}
(\bG_0^\sharp)^{-1}\bJ_{0n}\transpose\widetilde{\bLambda}_\sharp\nabla_\bpsi\theta_{e'}
\right\}\bigg|^2 \sum_{e\in\calE^\circ}\|\widetilde{\scrP}_{\br^{(t)}}(\bG_0^\sharp)^{-1}\bJ_{0n}\transpose\widetilde{\bLambda}_\sharp\nabla_\bpsi\theta_e\|_2^2 = \Optilde\bigg(\frac{\log n}{n}\bigg)
\end{align*}
by Step 2(a) and Lemma \ref{lemma:uniform_Optilde}. 
Then Lemma \ref{lemma:conditioning_argument} implies that the first term is $\Optilde(\log n/\sqrt{n})$. For the second sum, 
\begin{align*}
&\max_{e\in\calE^\circ}\left\|\bar{\ell}'_{e}\left\{
\nabla_\bpsi\theta_e\transpose\widetilde{\bLambda}_\sharp\bJ_{0n}
(\bG_0^\sharp)^{-1}\bJ_{0n}\transpose\widetilde{\bLambda}_\sharp\nabla_\bpsi\theta_{e}
\right\}\widetilde{\scrP}_{\br^{(t)}}(\bG_0^\sharp)^{-1}\bJ_{0n}\transpose\widetilde{\bLambda}_\sharp\nabla_\bpsi\theta_e\right\|_2\\
&\quad\leq \max_{e\in\calE^\circ}|\bar{\ell}'_e|\left\|(\bG_0^\sharp)^{-1/2}\bJ_{0n}\transpose\widetilde{\bLambda}_\sharp\nabla_\bpsi\theta_{e}\right\|_2^2\|(\bG_0^\sharp)^{-1}\|_2 \|\bJ_{0n}\transpose\widetilde{\bLambda}_\sharp\nabla_\bpsi\theta_e\|_2 = \Optilde\bigg(\frac{\log n}{n^{3/2}}\bigg),\\
&\sum_{e\in\calE^\circ}\left\|\bar{\ell}'_{e}\left\{
\nabla_\bpsi\theta_e\transpose\widetilde{\bLambda}_\sharp\bJ_{0n}
(\bG_0^\sharp)^{-1}\bJ_{0n}\transpose\widetilde{\bLambda}_\sharp\nabla_\bpsi\theta_{e}
\right\}\widetilde{\scrP}_{\br^{(t)}}(\bG_0^\sharp)^{-1}\bJ_{0n}\transpose\widetilde{\bLambda}_\sharp\nabla_\bpsi\theta_e\right\|_2^2\\
&\quad\leq \max_{e\in\calE^\circ}|\bar{\ell}'_e|^2\left\|(\bG_0^\sharp)^{-1/2}\bJ_{0n}\transpose\widetilde{\bLambda}_\sharp\nabla_\bpsi\theta_{e}\right\|_2^4\sum_{e\in\calE^\circ}\|\widetilde{\scrP}_{\br^{(t)}}(\bG_0^\sharp)^{-1}\bJ_{0n}\transpose\widetilde{\bLambda}_\sharp\nabla_\bpsi\theta_e\|_2^2 = \Optilde\bigg\{\frac{(\log n)^2}{n^2}\bigg\}
\end{align*}
by Step 2(a) and Lemma \ref{lemma:uniform_Optilde}, so Lemma \ref{lemma:conditioning_argument} again implies that the second term is $\Optilde\{(\log n)^{3/2}/n\}$. 

\medskip\noindent\textbf{$\blacksquare$ Step 5. The second off-diagonal sum (the last term)} By Theorem \ref{thm:decoupling} again, it suffices to bound 
\begin{align*}
&\bigg\|\sum_{e\neq e'}\ell'_e\bar{\ell}'_{e'}\widetilde{\scrP}_{\br^{(t)}}(\bG_0^\sharp)^{-1}
\bLambda_\sharp\frac{\partial^2\theta_e}{\partial\bphi\partial\bphi\transpose}(\bpsi(\bphi_0))\bLambda_\sharp
(\bG_0^\sharp)^{-1}\bJ_{0n}\transpose\widetilde{\bLambda}_\sharp\nabla_\bpsi\theta_{e'}\bigg\|_2
\\
&\quad\leq 
\bigg\|\sum_{e\in\calE^\circ}\ell'_e\bigg[\sum_{e'\in\calE^\circ}\bar{\ell}'_{e'}\widetilde{\scrP}_{\br^{(t)}}(\bG_0^\sharp)^{-1}
\bLambda_\sharp\frac{\partial^2\theta_e}{\partial\bphi\partial\bphi\transpose}(\bpsi(\bphi_0))\bLambda_\sharp
(\bG_0^\sharp)^{-1}\bJ_{0n}\transpose\widetilde{\bLambda}_\sharp\nabla_\bpsi\theta_{e'}\bigg]\bigg\|_2\\
&\qquad + \bigg\|\sum_{e\in\calE^\circ}\ell'_e\bar{\ell}_e'\widetilde{\scrP}_{\br^{(t)}}(\bG_0^\sharp)^{-1}
\bLambda_\sharp\frac{\partial^2\theta_e}{\partial\bphi\partial\bphi\transpose}(\bpsi(\bphi_0))\bLambda_\sharp
(\bG_0^\sharp)^{-1}\bJ_{0n}\transpose\widetilde{\bLambda}_\sharp\nabla_\bpsi\theta_{e}\bigg\|_2
\end{align*}
For the first part, by Step 2(c) and \eqref{eqn:J0n_s_hat_bound},
\begin{align*}
&\max_{e\in\calE^\circ}\bigg\|\sum_{e'\in\calE^\circ}\bar{\ell}'_{e'}\widetilde{\scrP}_{\br^{(t)}}(\bG_0^\sharp)^{-1}
\bLambda_\sharp\frac{\partial^2\theta_e}{\partial\bphi\partial\bphi\transpose}(\bpsi(\bphi_0))\bLambda_\sharp
(\bG_0^\sharp)^{-1}\bJ_{0n}\transpose\widetilde{\bLambda}_\sharp\nabla_\bpsi\theta_{e'}\bigg\|_2\\
&\quad \lesssim \max_{e\in\calE^\circ}\bigg\|\sum_{e'\in\calE^\circ}\bar{\ell}'_{e'}\widetilde{\scrP}_{\br^{(t)}}(\bG_0^\sharp)^{-1}
\bJ_{0n}\transpose\widetilde{\bLambda}_\sharp\nabla_\bpsi^2\theta_e\widetilde{\bLambda}_\sharp\bJ_{0n}
(\bG_0^\sharp)^{-1}\bJ_{0n}\transpose\widetilde{\bLambda}_\sharp\nabla_\bpsi\theta_{e'}\bigg\|_\infty\\
&\qquad + \frac{1}{2mn^2} \max_{e\in\calE^\circ}  \bigg\|\sum_{e'\in\calE^\circ}\bar{\ell}'_{e'}\widetilde{\scrP}_{\br^{(t)}}(\bG_0^\sharp)^{-1}
\mathrm{diag}\big\{\mathbf{O}_{d_\balpha + d_\br}, (\bM_e + \bM_e\transpose)\otimes\eye_{n - 1 - d}\big\}
(\bG_0^\sharp)^{-1}\bJ_{0n}\transpose\widetilde{\bLambda}_\sharp\nabla_\bpsi\theta_{e'}\bigg\|_2\\
&\quad\leq \max_{e\in\calE^\circ,j\in[d(d + 1)/2]}
\left\|\be_j\transpose\widetilde{\scrP}_{\br^{(t)}}(\bG_0^\sharp)^{-1}
\bJ_{0n}\transpose\widetilde{\bLambda}_\sharp\nabla_\bpsi^2\theta_e\widetilde{\bLambda}_\sharp\right\|_\infty \|\bJ_{0n}\bar{\bs}\|_\infty + \frac{1}{mn^2} \max_{e\in\calE^\circ}\|\bM_e\|_2\|(\bG_0^\sharp)^{-1}\|_2 \|\bJ_{0n}\bar{\bs}\|_2\\
&\quad\lesssim \max_{e\in\calE^\circ,j\in[d(d + 1)/2]}
\left\|\be_j\transpose\widetilde{\scrP}_{\br^{(t)}}(\bG_0^\sharp)^{-1}
\bJ_{0n}\transpose\widetilde{\bLambda}_\sharp\nabla_\bpsi^2\theta_e\widetilde{\bLambda}_\sharp\right\|_2 \|\bJ_{0n}\bar{\bs}\|_\infty + \frac{\sqrt{d_\bpsi}}{mn^2} \max_{e\in\calE^\circ}\|\bM_e\|_2\|(\bG_0^\sharp)^{-1}\|_2 \|\bJ_{0n}\bar{\bs}\|_\infty\\
&\quad = \Optilde\bigg(\frac{\sqrt{\log n}}{n}\bigg) + \Optilde\bigg(\frac{\sqrt{\log n}}{\sqrt{m}n^{3/2}}\bigg) = \Optilde\bigg(\frac{\sqrt{\log n}}{n}\bigg)
\end{align*}
and
\begin{align*}
&\sum_{e\in\calE^\circ}\bigg\|\sum_{e'\in\calE^\circ}\bar{\ell}'_{e'}\widetilde{\scrP}_{\br^{(t)}}(\bG_0^\sharp)^{-1}
\bLambda_\sharp\frac{\partial^2\theta_e}{\partial\bphi\partial\bphi\transpose}(\bpsi(\bphi_0))\bLambda_\sharp
(\bG_0^\sharp)^{-1}\bJ_{0n}\transpose\widetilde{\bLambda}_\sharp\nabla_\bpsi\theta_{e'}\bigg\|_2^2\\
&\quad \lesssim \sum_{e\in\calE^\circ}\bigg\|\sum_{e'\in\calE^\circ}\bar{\ell}'_{e'}\widetilde{\scrP}_{\br^{(t)}}(\bG_0^\sharp)^{-1}
\bJ_{0n}\transpose\widetilde{\bLambda}_\sharp\nabla_\bpsi^2\theta_e\widetilde{\bLambda}_\sharp\bJ_{0n}
(\bG_0^\sharp)^{-1}\bJ_{0n}\transpose\widetilde{\bLambda}_\sharp\nabla_\bpsi\theta_{e'}\bigg\|_\infty^2\\
&\qquad + \frac{1}{m^2n^4} \sum_{e\in\calE^\circ}  \bigg\|\sum_{e'\in\calE^\circ}\bar{\ell}'_{e'}\widetilde{\scrP}_{\br^{(t)}}(\bG_0^\sharp)^{-1}
\mathrm{diag}\big\{\mathbf{O}_{d_\balpha + d_\br}, (\bM_e + \bM_e\transpose)\otimes\eye_{n - 1 - d}\big\}
(\bG_0^\sharp)^{-1}\bJ_{0n}\transpose\widetilde{\bLambda}_\sharp\nabla_\bpsi\theta_{e'}\bigg\|_2^2\\
&\quad\leq \sum_{e\in\calE^\circ,j\in[d(d + 1)/2]}
\left\|\be_j\transpose\widetilde{\scrP}_{\br^{(t)}}(\bG_0^\sharp)^{-1}
\bJ_{0n}\transpose\widetilde{\bLambda}_\sharp\nabla_\bpsi^2\theta_e\widetilde{\bLambda}_\sharp\right\|_\infty^2 \|\bJ_{0n}\bar{\bs}\|_\infty^2 + \frac{1}{m^2n^4} \sum_{e\in\calE^\circ}\|\bM_e\|_2^2\|(\bG_0^\sharp)^{-1}\|_2^2 \|\bJ_{0n}\bar{\bs}\|_2^2\\
&\quad\lesssim \sum_{e\in\calE^\circ}\max_{j\in[d(d + 1)/2]}
\left\|\be_j\transpose\widetilde{\scrP}_{\br^{(t)}}(\bG_0^\sharp)^{-1}
\bJ_{0n}\transpose\widetilde{\bLambda}_\sharp\nabla_\bpsi^2\theta_e\widetilde{\bLambda}_\sharp\right\|_2^2 \|\bJ_{0n}\bar{\bs}\|_\infty^2 + \frac{d_\bpsi}{m^2n^4} \sum_{e\in\calE^\circ}\|\bM_e\|_2^2\|(\bG_0^\sharp)^{-1}\|_2^2 \|\bJ_{0n}\bar{\bs}\|_\infty^2\\
&\quad = \Optilde\bigg(\frac{\sqrt{\log n}}{n}\bigg) + \Optilde\bigg(\frac{\log n}{n}\bigg) = \Optilde\bigg(\frac{\log n}{n}\bigg).
\end{align*}
Then, Lemma \ref{lemma:conditioning_argument} implies that the first term in the second off-diagonal sum can be bounded as
\begin{align*}
  \bigg\|\sum_{e\in\calE^\circ}\ell'_e\bigg[\sum_{e'\in\calE^\circ}\bar{\ell}'_{e'}\widetilde{\scrP}_{\br^{(t)}}(\bG_0^\sharp)^{-1}
\bLambda_\sharp\frac{\partial^2\theta_e}{\partial\bphi\partial\bphi\transpose}(\bpsi(\bphi_0))\bLambda_\sharp
(\bG_0^\sharp)^{-1}\bJ_{0n}\transpose\widetilde{\bLambda}_\sharp\nabla_\bpsi\theta_{e'}\bigg]\bigg\|_2 = \Optilde\bigg(\frac{\log n}{\sqrt{n}}\bigg).
\end{align*}
For the second term, Lemma \ref{lemma:uniform_Optilde} and Step 2(d) directly implies that
\begin{align*}
&\max_{e\in\calE^\circ}\bigg\|\bar{\ell}_e'\widetilde{\scrP}_{\br^{(t)}}(\bG_0^\sharp)^{-1}
\bLambda_\sharp\frac{\partial^2\theta_e}{\partial\bphi\partial\bphi\transpose}(\bpsi(\bphi_0))\bLambda_\sharp
(\bG_0^\sharp)^{-1}\bJ_{0n}\transpose\widetilde{\bLambda}_\sharp\nabla_\bpsi\theta_{e}\bigg\|_2 = \Optilde\bigg(\frac{\log n}{n^{3/2}}\bigg),\\
&\sum_{e\in\calE^\circ}\left\|\bar{\ell}'_e\widetilde{\scrP}_{\br^{(t)}}(\bG_0^\sharp)^{-1}
    \bLambda_\sharp\frac{\partial^2\theta_e}{\partial\bphi\partial\bphi\transpose}(\bpsi(\bphi_0))\bLambda_\sharp
  (\bG_0^\sharp)^{-1}\bJ_{0n}\transpose\widetilde{\bLambda}_\sharp\nabla_\bpsi\theta_e
  \right\|_2^2 = \Optilde\bigg\{\frac{(\log n)^2}{n^2}\bigg\}.
\end{align*}
Similarly, by Lemma \ref{lemma:conditioning_argument}, we obtain
\begin{align*}
   \bigg\|\sum_{e\in\calE^\circ}\ell'_e\bar{\ell}_e'\widetilde{\scrP}_{\br^{(t)}}(\bG_0^\sharp)^{-1}
\bLambda_\sharp\frac{\partial^2\theta_e}{\partial\bphi\partial\bphi\transpose}(\bpsi(\bphi_0))\bLambda_\sharp
(\bG_0^\sharp)^{-1}\bJ_{0n}\transpose\widetilde{\bLambda}_\sharp\nabla_\bpsi\theta_{e}\bigg\|_2 = \Optilde\bigg\{\frac{(\log n)^{3/2}}{n}\bigg\}.
\end{align*}
The proof is therefore completed by combining Steps 1--5. 
\end{proof}

\subsection{Controlling the fluctuation of $\bQ_2$}
\label{sub:controlling_the_bias_remainder_for_Rt_hat_Q2}

This subsection will next establish the $\widetilde{\scrP}_{\br^{(t)}}\bQ_2$ centers around its expected value. Compared to the fluctuation of $\widetilde{\scrP}_{\br^{(t)}}\bQ_1$, the proof below is more involved and lengthy but necessary en route to the asymptotic normality of $\widehat{\bR}^{(t)}$. 

\begin{lemma}\label{lemma:Q2_fluctuation}
Suppose Assumptions \ref{assumption:edge_distribution}--\ref{assumption:true_parameter} hold and $m\leq n$. Then
\begin{align*}
\max_{t\in[m]}\left\|\widetilde{\scrP}_{\br^{(t)}}\bQ_2 - \expect_0\widetilde{\scrP}_{\br^{(t)}}\bQ_2\right\|_2 = \Optilde\bigg(\frac{\log n}{\sqrt{n}}\bigg).
\end{align*}
\end{lemma}

\begin{proof}[\bf Proof]
We adopt the notations in the proof of Lemma \ref{lemma:Q1_fluctuation}. With a slight abuse of notation, all sums over $e, e', e''\in\calE$ in this proof are taken over the set $\calE^\circ$ of independent edges. In addition, write $\ell'''_e := \ell'''(\theta_{0e}; A_e)$ and introduce the following notations: for any $\delta\bphi_1,\delta\bphi_2,\delta\bphi_3\in\mathbb{R}^{d_\bphi}$, 
\begin{align*}
&\bV_e = \bLambda_\sharp\frac{\partial^2\theta_e}{\partial\bphi\partial\bphi\transpose}\bLambda_\sharp,\; 
\tau_e(\delta\bphi_1, \delta\bphi_2, \delta\bphi_3) = \frac{\partial^3\theta_e}{\partial\tau_1\partial\tau_2\partial\tau_3}(\bpsi(\bphi_0 + \tau_1\bLambda_\sharp\delta\bphi_1 + \tau_2\bLambda_\sharp\delta\bphi_2 + \tau_3\bLambda_\sharp\delta\bphi_3)){\mathrel{\bigg|_{\tau_1 = \tau_2 = \tau_3 = 0}}},\\
&\bw_e[\delta\bphi_1,\delta\bphi_2] = \sum_{j = 1}^{d_\bphi}\be_j \tau_e(\delta\bphi_1, \delta\bphi_2, \be_j(d_\bphi)),\;
\bT_e(\delta\bphi_3) = \sum_{j_1,j_2\in[d_\bphi]}\be_{j_1}\be_{j_2}\transpose\tau_e(\be_{j_1}, \be_{j_2}, \delta\bphi_3),\\
&\nabla_\bphi\theta_e = \frac{\partial\theta_e(\bpsi(\bphi))}{\partial\bphi}{\mathrel{\bigg|_{\bphi = \bphi_0}}} = \bJ_0\transpose\nabla_\bpsi\theta_e.
\end{align*}
In particular, $\dot{\theta}_e[\bLambda_\sharp\delta\bphi] = \delta\bphi\transpose\bLambda_\sharp\bJ_0\transpose\nabla_\bpsi\theta_e = \delta\bphi\transpose\bLambda_\sharp\nabla_\bphi\theta_e$. 
Then, following Step 2 of the proof of Lemma \ref{lemma:third_order_cross_terms}, we have
\begin{align*}
\bLambda_\sharp\nabla^3\calL(\bphi_0)[\bLambda_\sharp\delta\bphi_1, \bLambda_\sharp\delta\bphi_2]
& = \sum_{e\in\calE^\circ}\Big[\ell'''_e (\bLambda_\sharp\nabla_\bphi\theta_e)\dot{\theta}_e[\bLambda_\sharp\delta\bphi_1]\dot{\theta}_e[\bLambda_\sharp\delta\bphi_2]\\
&\qquad + \ell''_e\big\{\bLambda_\sharp\nabla_\bphi\theta_e(\delta\bphi_1\transpose\bV_e\delta\bphi_2) + \bV_e\delta\bphi_2\dot{\theta}_e[\bLambda_\sharp\delta\bphi_1] + \bV_e\delta\bphi_1\dot{\theta}_e[\bLambda_\sharp\delta\bphi_2]\big\}\\
&\qquad + \ell'_e\bw_e[\delta\bphi_1, \delta\bphi_2]\Big].
\end{align*}
In this proof, we also write $\scrP = \widetilde{\scrP}_{\br^{(t)}}$ for simplicity. 

\medskip\noindent\textbf{$\blacksquare$ Step 1. Decomposition of the target. } For $k \in \{1,2,3,4\}$, define random variable $\xi_e^{(k)}$ by
\begin{align*}
&\xi_e^{(1)} = \ell'''_e,\quad \xi_e^{(2)} = \xi_e^{(3)} = \ell''_e, \quad \xi_e^{(4)} = \ell'_e,
\end{align*}
and the following vector-valued symmetric bilinear functions $\bg_e^{(k)}(\cdot, \cdot):\mathbb{R}^{d_\bphi}\times\mathbb{R}^{d_\bphi}\to\mathbb{R}^{d(d + 1)/2}$ by
\begin{align*}
&\bg_e^{(1)}(\delta\bphi_1, \delta\bphi_2) := \scrP(\bG_0^\sharp)^{-1}\bLambda_\sharp\nabla_\bphi\theta_e\dot{\theta}_e[\bLambda_\sharp\delta\bphi_1]\dot{\theta}_e[\bLambda_\sharp\delta\bphi_2],\\ 
&\bg_e^{(2)}(\delta\bphi_1, \delta\bphi_2) := \scrP(\bG_0^\sharp)^{-1}\bLambda_\sharp\nabla_\bphi\theta_e(\delta\bphi_1\transpose\bV_e\delta\bphi_2),\\
&\bg_e^{(3)}(\delta\bphi_1, \delta\bphi_2) := \scrP(\bG_0^\sharp)^{-1}\bV_e\delta\bphi_2\dot{\theta}_e[\bLambda_\sharp\delta\bphi_1] + \scrP(\bG_0^\sharp)^{-1}\bV_e\delta\bphi_1\dot{\theta}_e[\bLambda_\sharp\delta\bphi_2],\\
&\bg_e^{(4)}(\delta\bphi_1, \delta\bphi_2) := \scrP(\bG_0^\sharp)^{-1}\bw_e[\delta\bphi_1, \delta\bphi_2]. 
\end{align*}
Similarly, define the following vector-valued function $\bg_e^{(k)}[\cdot]:\mathrm{Sym}(d_\bphi)\to\mathbb{R}^{d(d + 1)/2}$ by
\begin{align*}
&\bg_e^{(1)}[\bM] := \scrP(\bG_0^\sharp)^{-1}\bLambda_\sharp\nabla_\bphi\theta_e(\nabla_\bphi\theta_e\transpose\bLambda_\sharp\bM\bLambda_\sharp\nabla_\bphi\theta_e),\\ 
&\bg_e^{(2)}[\bM] := \scrP(\bG_0^\sharp)^{-1}\bLambda_\sharp\nabla_\bphi\theta_e \mathrm{tr}(\bV_e\bM),\\
&\bg_e^{(3)}[\bM] := 2\scrP(\bG_0^\sharp)^{-1}\bV_e\bM\bLambda_\sharp\nabla_\bphi\theta_e,\\
&\bg_e^{(4)}[\bM] := \sum_{j = 1}^{d(d + 1)/2}\be_j\mathrm{tr}\{\bT_e((\bG_0^\sharp)^{-1}\scrP\transpose\be_j)\bM\}. 
\end{align*}
Also, a direct calculation shows that if $\bM$ can be written as $\bM = \sum_s\lambda_s\bv_s\bv_s\transpose$ for scalars $(\lambda_s)_s$ and vectors $(\bv_s)_s\subset\mathbb{R}^{d_\bphi}$, then $\bg_e^{(k)}[\bM] = \sum_s\lambda_s\bg_e^{(k)}(\bv_s, \bv_s)$. This allows us to rewrite $\scrP\bQ_2$ as
\begin{align*}  
\scrP\bQ_2 
& = \frac{1}{2}\sum_{e\in\calE}\ell'''_e\scrP(\bG_0^\sharp)^{-1}\bLambda_\sharp\nabla_\bphi\theta_e(\dot{\theta}_e[\bLambda_\sharp\widehat{\bs}])^2 + \frac{1}{2}\sum_{e\in\calE}\ell''_e\scrP(\bG_0^\sharp)^{-1}\bLambda_\sharp\nabla_\bphi\theta_e(\widehat{\bs}\transpose\bV_e\widehat{\bs})\\
&\quad + \sum_{e\in\calE}\ell''_e\scrP(\bG_0^\sharp)^{-1}\bV_e\widehat{\bs}\dot{\theta}_e[\bLambda_\sharp\widehat{\bs}] + \frac{1}{2}\sum_{e\in\calE}\ell'_e\scrP(\bG_0^\sharp)^{-1}\bw_e[\widehat{\bs}, \widehat{\bs}]\\
& = \frac{1}{2}\sum_{e\in\calE}\ell'''_e\bg_e^{(1)}(\widehat{\bs}, \widehat{\bs}) + \frac{1}{2}\sum_{e\in\calE}\ell''_e\bg_e^{(2)}(\widehat{\bs}, \widehat{\bs}) + \frac{1}{2}\sum_{e\in\calE}\ell''_e\bg_e^{(3)}(\widehat{\bs}, \widehat{\bs}) + \frac{1}{2}\sum_{e\in\calE}\ell'_e\bg_e^{(4)}(\widehat{\bs}, \widehat{\bs}).
\end{align*}
Consequently, using $\widehat{\bs} = (\bG_0^\sharp)^{-1}\sum_e \ell'_e\bLambda_\sharp\nabla_\bphi\theta_e$, we obtain
\begin{align*}
2\scrP\bQ_2 - 2\expect_0\scrP\bQ_2 
& = \sum_{k = 1}^4\left\{\bq_1^{(k)} + \bq_{21}^{(k)} + \bq_{22}^{(k)} + \bq_{23}^{(k)} + \bq_{31}^{(k)} + \bq_{32}^{(k)} + \bq_{41}^{(k)} + \bq_{42}^{(k)}\right\},
\end{align*}
where
\begin{align*}
\bq_1^{(k)}
& = \sum_e \{\xi_e^{(k)}(\ell'_e)^2 - \expect_0\xi_e^{(k)}(\ell'_e)^2\}\bg_e^{(k)}((\bG_0^\sharp)^{-1}\bLambda_\sharp\nabla_\bphi\theta_e, (\bG_0^\sharp)^{-1}\bLambda_\sharp\nabla_\bphi\theta_e),\\
\bq_{21}^{(k)}
& = \sum_e (\xi_e^{(k)} - \expect_0\xi_e^{(k)})\sum_{e'\in\calE\backslash\{e\}}\expect_0(\ell'_{e'})^2\bg_e^{(k)}((\bG_0^\sharp)^{-1}\bLambda_\sharp\nabla_\bphi\theta_{e'}, (\bG_0^\sharp)^{-1}\bLambda_\sharp\nabla_\bphi\theta_{e'}),\\
\bq_{22}^{(k)}
& = \sum_{e'} \{(\ell'_{e'})^2 - \expect_0(\ell'_{e'})^2\}\sum_{e\in\calE\backslash\{e'\}}\expect_0\xi_e^{(k)}\bg_e^{(k)}((\bG_0^\sharp)^{-1}\bLambda_\sharp\nabla_\bphi\theta_{e'}, (\bG_0^\sharp)^{-1}\bLambda_\sharp\nabla_\bphi\theta_{e'}),\\
\bq_{23}^{(k)}
& = \sum_{e\neq e'}(\xi_e^{(k)} - \expect_0\xi_e^{(k)})\{(\ell'_{e'})^2 - \expect_0(\ell'_{e'})^2\}\bg_e^{(k)}((\bG_0^\sharp)^{-1}\bLambda_\sharp\nabla_\bphi\theta_{e'}, (\bG_0^\sharp)^{-1}\bLambda_\sharp\nabla_\bphi\theta_{e'}),\\
\bq_{31}^{(k)}
& = 2\sum_{e''}\ell'_{e''}\sum_{e\in\calE\backslash\{e''\}}\expect_0\xi_e^{(k)}\ell'_e\bg_e^{(k)}((\bG_0^\sharp)^{-1}\bLambda_\sharp\nabla_\bphi\theta_e, (\bG_0^\sharp)^{-1}\bLambda_\sharp\nabla_\bphi\theta_{e''})\\
\bq_{32}^{(k)}
& = 2\sum_{e\neq e''}(\xi_e^{(k)}\ell'_e - \expect_0\xi_e^{(k)}\ell'_e)\ell'_{e''}\bg_e^{(k)}((\bG_0^\sharp)^{-1}\bLambda_\sharp\nabla_\bphi\theta_e, (\bG_0^\sharp)^{-1}\bLambda_\sharp\nabla_\bphi\theta_{e''}),\\
\bq_{41}^{(k)}
& = \sum_{e'\neq e''}\ell'_{e'}\ell'_{e''}\sum_{e\in\calE\backslash\{e', e''\}}\expect_0\xi_e^{(k)}\bg_e^{(k)}((\bG_0^\sharp)^{-1}\bLambda_\sharp\nabla_\bphi\theta_{e'}, (\bG_0^\sharp)^{-1}\bLambda_\sharp\nabla_\bphi\theta_{e''}),\\
\bq_{42}^{(k)}
& = \sum_{e\neq e'\neq e''\neq e}(\xi_e^{(k)} - \expect_0\xi_e^{(k)})\ell'_{e'}\ell'_{e''}\bg_e^{(k)}((\bG_0^\sharp)^{-1}\bLambda_\sharp\nabla_\bphi\theta_{e'}, (\bG_0^\sharp)^{-1}\bLambda_\sharp\nabla_\bphi\theta_{e''})
\end{align*}

\medskip\noindent\textbf{$\blacksquare$
Step 2. Deterministic bounds. 
}We now collect several deterministic bounds that are useful for the concentration bounds of the fluctuation of $\scrP\bQ_2$. 

\begin{enumerate}[(A)]
  \item $\max_e\|(\bG_0^\sharp)^{-1}\bLambda_\sharp\nabla_\bphi\theta_e\|_2 \leq \|(\bG_0^\sharp)^{-1}\|_2\|\bJ_{0n}\|_2\|\widetilde{\bLambda}_\sharp\|_2\|\nabla_\bpsi\theta_e\|_2 = O(n^{-1/2})$; This can be seen by observing that there at most $O(1)$ nonzero elements in $\nabla_\bpsi\theta_e$ and these elements are uniformly bounded. 

  \item For any vector $\delta\bphi_1,\delta\bphi_2\in\mathbb{R}^{d_\bphi}$, 
  \begin{equation}
  \label{eqn:kernel_bound_vector}
  \begin{aligned}
  &\max_{e\in\calE}\|\bg_e^{(k)}(\delta\bphi_1, \delta\bphi_2)\|_2 \lesssim \|\bJ_{0n}\delta\bphi_1\|_\infty\|\bJ_{0n}\delta\bphi_2\|_\infty n^{-3/2},\\
  &\sum_{e\in\calE}\|\bg_e^{(k)}(\delta\bphi_1, \delta\bphi_2)\|_2^2 \lesssim\|\bJ_{0n}\delta\bphi_1\|_\infty^2\|\bJ_{0n}\delta\bphi_2\|_\infty^2 n^{-2},\\
  &\max_{e\in\calE}\|\bg_e^{(k)}((\bG_0^\sharp)^{-1}\bLambda_\sharp\nabla_\bphi\theta_e, (\bG_0^\sharp)^{-1}\bLambda_\sharp\nabla_\bphi\theta_e)\|_2
  \lesssim  n^{-5/2},\\
  &\sum_{e\in\calE}\|\bg_e^{(k)}((\bG_0^\sharp)^{-1}\bLambda_\sharp\nabla_\bphi\theta_e, (\bG_0^\sharp)^{-1}\bLambda_\sharp\nabla_\bphi\theta_e)\|_2^2 \lesssim  n^{-4},\\
  &\max_{e\in\calE}\|\bg_e^{(k)}((\bG_0^\sharp)^{-1}\bLambda_\sharp\nabla_\bphi\theta_e, \delta\bphi_2)\|_2
   \lesssim \|\bJ_{0n}\delta\bphi_2\|_\infty n^{-2},\\
  &\sum_{e\in\calE}\|\bg_e^{(k)}((\bG_0^\sharp)^{-1}\bLambda_\sharp\nabla_\bphi\theta_e, \delta\bphi_2)\|_2^2
   \lesssim \|\bJ_{0n}\delta\bphi_2\|_\infty^2 n^{-3};
  \end{aligned}
  \end{equation}
  \noindent\underline{For $k = 1$,} we use the fact that $|\calS_1(e)| = O(1)$ for all $e$ and $\|\sum_{e}\widetilde{\bLambda}_\sharp\nabla_\bpsi\theta_e\nabla_\bpsi\theta_e\transpose\widetilde{\bLambda}_\sharp\| = O(1)$ and $\sum_e\|\scrP(\bG_0^\sharp)^{-1}\widetilde{\bLambda}_\sharp\nabla_\bpsi\theta_e\|_2^2 = O(1)$ (see Step 3 of Lemma \ref{lemma:third_order_Taylor_expansion_loglikelihood}) to obtain
    \begin{align*}
    \max_{e\in\calE}\|\bg_e^{(1)}(\delta\bphi_1, \delta\bphi_2)\|_2
    & = 
    \max_{e\in\calE}\left\|\scrP(\bG_0^\sharp)^{-1}\bLambda_\sharp\nabla_\bphi\theta_e\dot{\theta}_e[\bLambda_\sharp\delta\bphi_1]\dot{\theta}_e[\bLambda_\sharp\delta\bphi_2]\right\|_2\\
    &\lesssim n^{-1/2}\max_{e\in\calE}|\delta\bphi_1\transpose\bJ_{0n}\transpose\widetilde{\bLambda}_\sharp\nabla_\bpsi\theta_e||\delta\bphi_2\transpose\bJ_0\transpose\widetilde{\bLambda}_\sharp\nabla_\bpsi\theta_e|\\
    &\lesssim n^{-3/2}\|\bJ_{0n}\delta\bphi_1\|_\infty\|\bJ_{0n}\delta\bphi_2\|_\infty,\\
    \sum_{e\in\calE}\|\bg_e^{(1)}(\delta\bphi_1, \delta\bphi_2)\|_2^2
    &\lesssim \sum_e\|\scrP(\bG_0^\sharp)^{-1}\widetilde{\bLambda}_\sharp\nabla_\bpsi\theta_e\|_2^2 \\
    &\quad\times |\delta\bphi_1\transpose\bJ_{0n}\transpose\widetilde{\bLambda}_\sharp\nabla_\bpsi\theta_e|^2|\delta\bphi_2\transpose\bJ_0\transpose\widetilde{\bLambda}_\sharp\nabla_\bpsi\theta_e|^2\\
    &\lesssim \max_{e\in\calE}|\delta\bphi_1\transpose\bJ_{0n}\transpose\widetilde{\bLambda}_\sharp\nabla_\bpsi\theta_e|^2|\delta\bphi_2\transpose\bJ_0\transpose\widetilde{\bLambda}_\sharp\nabla_\bpsi\theta_e|^2\\
    &\lesssim n^{-2}\|\bJ_{0n}\delta\bphi_1\|_\infty^2\|\bJ_{0n}\delta\bphi_2\|_\infty^2,
    \\
    \max_{e\in\calE}\|\bg_e^{(1)}((\bG_0^\sharp)^{-1}\bLambda_\sharp\nabla_\bphi\theta_e, (\bG_0^\sharp)^{-1}\bLambda_\sharp\nabla_\bphi\theta_e)\|_2
    &\lesssim n^{-3/2}\max_{e\in\calE}\|\bJ_{0n}(\bG_0^\sharp)^{-1}\bJ_{0n}\transpose\widetilde{\bLambda}_\sharp\nabla_\bpsi\theta_e\|_\infty^2\\
    &\lesssim n^{-3/2}\max_{e\in\calE}\|\bJ_{0n}(\bG_0^\sharp)^{-1}\bJ_{0n}\transpose\|_2^2\|\widetilde{\bLambda}_\sharp\|_2^2\|\nabla_\bpsi\theta_e\|_2^2\\
    &\lesssim n^{-5/2},\\
    \sum_{e\in\calE}\|\bg_e^{(1)}((\bG_0^\sharp)^{-1}\bLambda_\sharp\nabla_\bphi\theta_e, (\bG_0^\sharp)^{-1}\bLambda_\sharp\nabla_\bphi\theta_e)\|_2^2
    &\lesssim \sum_{e}\|\scrP(\bG_0^\sharp)^{-1}\widetilde{\bLambda}_\sharp\nabla_\bpsi\theta_e\|_2^2\|\bJ_{0n}\transpose\widetilde{\bLambda}_\sharp\|_2^4\|\nabla_\bpsi\theta_e\|_2^4\\
    &\quad\times \|(\bG_0^\sharp)^{-1}\bLambda_\sharp\nabla_\bphi\theta_e\|_2^4\\
    &\lesssim n^{-4}, \\
    \max_{e\in\calE}\|\bg_e^{(1)}((\bG_0^\sharp)^{-1}\bLambda_\sharp\nabla_\bphi\theta_e, \delta\bphi_2)\|_2
    &\lesssim n^{-3/2}\max_{e\in\calE}\|\bJ_{0n}(\bG_0^\sharp)^{-1}\bJ_{0n}\transpose\widetilde{\bLambda}_\sharp\nabla_\bpsi\theta_e\|_\infty\|\bJ_{0n}\delta\bphi_2\|_\infty\\
    &\lesssim n^{-3/2}\max_{e\in\calE}\|\bJ_{0n}(\bG_0^\sharp)^{-1}\bJ_{0n}\transpose\|_2 \|\widetilde{\bLambda}_\sharp\|_2 \|\nabla_\bpsi\theta_e\|_2\|\bJ_{0n}\delta\bphi_2\|_\infty \\
    &\lesssim n^{-2}\|\bJ_{0n}\delta\bphi_2\|_\infty,\\
    \sum_{e\in\calE}\|\bg_e^{(1)}((\bG_0^\sharp)^{-1}\bLambda_\sharp\nabla_\bphi\theta_e, \delta\bphi_2)\|_2^2
    &\lesssim \sum_{e\in\calE}\|\scrP(\bG_0^\sharp)^{-1}\widetilde{\bLambda}_\sharp\nabla_\bpsi\theta_e\|_2^2\|\bJ_{0n}\transpose\widetilde{\bLambda}_\sharp\|_2^2\|\nabla_\bpsi\theta_e\|_2^2\|\widetilde{\bLambda}_\sharp\|_2^4\\
    &\quad\times \|\bJ_{0n}\delta\bphi_2\|_\infty^2
    \\
    &\lesssim n^{-3}\|\bJ_{0n}\delta\bphi_2\|_\infty^2.
    \end{align*}

    \noindent\underline{For $k = 2$,} we first explore the properties of $\bV_e$. For any $\delta\bphi_1,\delta\bphi_2\in\mathbb{R}^{d_\bphi}$, we have
    \begin{align*}
    \delta\bphi_1\transpose\bV_e\delta\bphi_2
    & = (\bJ_{0n}\delta\bphi_1)\transpose\widetilde{\bLambda}_\sharp\nabla_\bpsi^2\theta_e\widetilde{\bLambda}_\sharp(\bJ_{0n}\delta\bphi_2) - \frac{1}{2mn^2}\mathrm{tr}\{\bB_{\delta\bphi_1}\transpose\bB_{\delta\bphi_2}(\bM_e + \bM_e\transpose)\},
    \end{align*}
    where $\bB_{\delta\bphi_i}$ is defined as the $(n - 1 - d)\times d$ matrix such that $\vect(\bB_{\delta\bphi_i}) = [\mathbf{O}, \eye_{d_\bB}]\delta\bphi_{i}$, $i = 1,2$, and we recall that $\bM_e = \bR_0^{(t)}(\bx_{0i}\bx_{0j}\transpose + \bx_{0j}\bx_{0i}\transpose)$ for $e = (t, i, j)$. The absolute value of the first term above is $O(\|\bJ_{0n}\delta\bphi_1\|_\infty\|\bJ_{0n}\delta\bphi_2\|_\infty/(mn))$ by the fact that $\nabla_\bpsi^2\theta_e$ has at most $O(1)$-many nonzero elements and they are uniformly bounded. Since $\|\bB_{\delta\bphi_i}\|_{\mathrm{F}} = n^{-1/2}\|\bX_{0\perp}\bB_{\delta\bphi_i}\|_{\mathrm{F}} \lesssim \|\bJ_0\delta\bphi_i\|_\infty\lesssim \sqrt{n}\|\bJ_{0n}\delta\bphi_i\|_\infty$, $i\in\{1,2\}$, the absolute value of the second term on the right-hand side of the above display is $O(\|\bJ_{0n}\delta\bphi_1\|_\infty\|\bJ_{0n}\delta\bphi_2\|_\infty/(mn))$ as well. Note that it can also be implied that
    \begin{align}
    \label{eqn:Ve_nuclear_norm}
    \max_{e\in\calE}\|\bV_e\|_* \lesssim \frac{1}{mn}
    \end{align}
    since $\nabla^2_\bpsi\theta_e$ has only nonzero blocks corresponding to $(\br, \bX)$, $(\bX, \br)$, and $(\bX, \bX)$ and has at most $O(1)$ many nonzero elements, which are uniformly bounded in absolute value.
    Therefore, 
    \begin{align*}
    \max_{e\in\calE}\|\bg_e^{(2)}(\delta\bphi_1, \delta\bphi_2)\|_2
    & \leq 
    \max_{e\in\calE}\|\scrP(\bG_0^\sharp)^{-1}\|_2\|\bJ_{0n}\transpose\|_2\|\widetilde{\bLambda}_\sharp\|_2\|\nabla_\bpsi\theta_e\|_2|\delta\bphi_1\transpose\bV_e\delta\bphi_2|\\
    & \lesssim m^{-1}n^{-3/2}\|\bJ_{0n}\delta\bphi_1\|_\infty\|\bJ_{0n}\delta\bphi_2\|_\infty,\\
    \sum_{e\in\calE}\|\bg_e^{(2)}(\delta\bphi_1, \delta\bphi_2)\|_2^2
    & \lesssim 
    \sum_{e\in\calE}\|\scrP(\bG_0^\sharp)^{-1}\widetilde{\bLambda}_\sharp \nabla_\bpsi\theta_e\|_2^2|\delta\bphi_1\transpose\bV_e\delta\bphi_2|^2\\
    & \lesssim (mn)^{-2}\|\bJ_{0n}\delta\bphi_1\|_\infty^2\|\bJ_{0n}\delta\bphi_2\|_\infty^2,\\
    \max_{e\in\calE}\|\bg_e^{(2)}((\bG_0^\sharp)^{-1}\bLambda_\sharp\nabla_\bphi\theta_e, (\bG_0^\sharp)^{-1}\bLambda_\sharp\nabla_\bphi\theta_e)\|_2
    &\lesssim m^{-1}n^{-3/2}\max_{e\in\calE}\|\bJ_{0n}(\bG_0^\sharp)^{-1}\bJ_{0n}\transpose\widetilde{\bLambda}_\sharp\nabla_\bpsi\theta_e\|_\infty^2\\
    &\lesssim m^{-1}n^{-3/2}\max_{e\in\calE}\|\bJ_{0n}(\bG_0^\sharp)^{-1}\bJ_{0n}\transpose\|_2^2\|\widetilde{\bLambda}_\sharp\|_2^2\|\nabla_\bpsi\theta_e\|_2^2\\
    &\lesssim m^{-1}n^{-5/2},\\
    \sum_{e\in\calE}\|\bg_e^{(2)}((\bG_0^\sharp)^{-1}\bLambda_\sharp\nabla_\bphi\theta_e, (\bG_0^\sharp)^{-1}\bLambda_\sharp\nabla_\bphi\theta_e)\|_2^2
    &\lesssim (mn)^{-2}\sum_{e}\|\scrP(\bG_0^\sharp)^{-1}\widetilde{\bLambda}_\sharp\nabla_\bpsi\theta_e\|_2^2 \\
    &\quad\times \|\bJ_{0n}(\bG_0^\sharp)^{-1}\bJ_{0n}\transpose\widetilde{\bLambda}_\sharp\nabla_\bpsi\theta_e\|_\infty^4\\
    &\lesssim m^{-2}n^{-4},\\
    \max_{e\in\calE}\|\bg_e^{(2)}((\bG_0^\sharp)^{-1}\bLambda_\sharp\nabla_\bphi\theta_e, \delta\bphi_2)\|_2
    &\lesssim m^{-1}n^{-3/2}\max_{e\in\calE}\|\bJ_{0n}(\bG_0^\sharp)^{-1}\bJ_{0n}\transpose\widetilde{\bLambda}_\sharp\nabla_\bpsi\theta_e\|_\infty\|\bJ_{0n}\delta\bphi_2\|_\infty\\
    &\lesssim m^{-1}n^{-2}\|\bJ_{0n}\delta\bphi_2\|_\infty,\\
    \sum_{e\in\calE}\|\bg_e^{(2)}((\bG_0^\sharp)^{-1}\bLambda_\sharp\nabla_\bphi\theta_e, \delta\bphi_2)\|_2^2
    &\lesssim (mn)^{-2}\sum_{e\in\calE}\|\scrP(\bG_0^\sharp)^{-1}\widetilde{\bLambda}_\sharp\nabla_\bpsi\theta_e\|_2^2\\
    &\quad\times \|\bJ_{0n}(\bG_0^\sharp)^{-1}\bJ_{0n}\transpose\widetilde{\bLambda}_\sharp\nabla_\bpsi\theta_e\|_2^2
    \|\bJ_{0n}\delta\bphi_2\|_\infty^2
    \\
    &\lesssim m^{-2}n^{-3}\|\bJ_{0n}\delta\bphi_2\|_\infty^2.
    \end{align*}

    \underline{For $k = 3$,} note that $\max_e\|\scrP(\bG_0^\sharp)^{-1}\bV_e\delta\bphi_i\|_2 = O(\|\bJ_{0n}\delta\bphi_i\|_\infty/(mn))$, and by Cauchy--Schwarz inequality,
    \begin{equation}
    \label{eqn:V2}
    \begin{aligned}
    \sum_{e\in\calE}(\delta\bphi_1\bV_e\delta\bphi_2)^2
    &\leq 2\sum_{e\in\calE}\left\{(\bJ_{0n}\delta\bphi_1)\transpose\widetilde{\bLambda}_\sharp\nabla_\bpsi^2\theta_e\widetilde{\bLambda}_\sharp(\bJ_{0n}\delta\bphi_2)\right\}^2 + \frac{C}{m^2n^4}\sum_{e\in\calE}\|\bB_{\delta\bphi_1}\|_{\mathrm{F}}^2\|\bB_{\delta\bphi_2}\|_{\mathrm{F}}^2\\
    &\lesssim \sum_{e\in\calE}\bigg\{\sum_{a,b\in\calS_1(e)}
    (\be_a\transpose\bJ_{0n}\delta\bphi_1)(\be_a\transpose\widetilde{\bLambda}_\sharp\nabla_\bpsi^2\theta_e\widetilde{\bLambda}_\sharp\be_b)(\be_b\transpose\bJ_{0n}\delta\bphi_2)
    \bigg\}^2\\
    &\quad + \frac{1}{mn}\|\bJ_{0n}\delta\bphi_1\|_\infty^2\|\bJ_{0n}\delta\bphi_2\|_2^2\\
    &\lesssim \sum_{a,b\in [d_\bphi]}(\be_a\transpose\bJ_{0n}\delta\bphi_1)^2(\be_b\transpose\bJ_{0n}\delta\bphi_2)^2\sum_{e:a,b\in\calS_1(e)}(\be_a\transpose\widetilde{\bLambda}_\sharp\nabla_\bpsi^2\theta_e\widetilde{\bLambda}_\sharp\be_b)^2\\
    &\quad + \frac{1}{mn}\|\bJ_{0n}\delta\bphi_1\|_\infty^2\|\bJ_{0n}\delta\bphi_2\|_2^2\\
    &\lesssim \frac{1}{mn}\|\bJ_{0n}\delta\bphi_1\|_\infty^2\|\bJ_{0n}\delta\bphi_2\|_2^2.
    \end{aligned}
    \end{equation}
    The same reasoning also implies that, if $(\delta\bphi_e)_{e\in\calE}$ is a collection of vectors indexed by $e\in\calE$, then
    \begin{align}
    \label{eqn:V2_prime}
    \sum_{e\in\calE}(\delta\bphi_e\bV_e\delta\bphi_2)^2  \lesssim \frac{1}{mn}\max_e\|\bJ_{0n}\delta\bphi_e\|_\infty^2\|\bJ_{0n}\delta\bphi_2\|_2^2.
    \end{align}
    Then, we have
    \begin{align*}
    \max_{e\in\calE}\|\bg_e^{(3)}(\delta\bphi_1, \delta\bphi_2)\|_2
    &\lesssim \frac{1}{mn}\|\bJ_{0n}\delta\bphi_1\|_\infty|\delta\bphi_2\transpose\bJ_{0n}\transpose\widetilde{\bLambda}_\sharp\nabla_\bpsi\theta_e|
    + \frac{1}{mn}\|\bJ_{0n}\delta\bphi_2\|_\infty|\delta\bphi_1\transpose\bJ_{0n}\transpose\widetilde{\bLambda}_\sharp\nabla_\bpsi\theta_e|\\
    &\lesssim \frac{1}{mn^{3/2}}\|\bJ_{0n}\delta\bphi_1\|_\infty\|\bJ_{0n}\delta\bphi_2\|_\infty,\\
    \sum_{e\in\calE}\|\bg_e^{(3)}(\delta\bphi_1, \delta\bphi_2)\|_2^2
    &\lesssim \sum_{j = 1}^{d(d + 1)/2}\sum_{e\in\calE}\left\{\be_j\transpose\scrP(\bG_0^\sharp)^{-1}\bV_e\delta\bphi_1\right\}^2\left(\delta\bphi_2\transpose\bJ_{0n}\transpose\widetilde{\bLambda}_\sharp\nabla_\bpsi\theta_e\right)^2\\
    &\quad + \sum_{j = 1}^{d(d + 1)/2}\sum_{e\in\calE}\left\{\be_j\transpose\scrP(\bG_0^\sharp)^{-1}\bV_e\delta\bphi_2\right\}^2\left(\delta\bphi_1\transpose\bJ_{0n}\transpose\widetilde{\bLambda}_\sharp\nabla_\bpsi\theta_e\right)^2\\
    &\lesssim \frac{1}{n^2}\|\bJ_{0n}\delta\bphi_1\|_\infty^2\|\bJ_{0n}\delta\bphi_2\|_\infty^2.
    \end{align*}
    Similarly,
    \begin{align*}
    \max_{e\in\calE}\|\bg_e^{(3)}((\bG_0^\sharp)^{-1}\bLambda_\sharp\nabla_\bphi\theta_e, (\bG_0^\sharp)^{-1}\bLambda_\sharp\nabla_\bphi\theta_e)\|_2
    &\lesssim n^{-5/2},\\
    \max_{e\in\calE}\|\bg_e^{(3)}((\bG_0^\sharp)^{-1}\bLambda_\sharp\nabla_\bphi\theta_e, \delta\bphi_2)\|_2
    &\lesssim n^{-2}\|\bJ_{0n}\delta\bphi_2\|_\infty. 
    \end{align*}
    For the sum of squares, by \eqref{eqn:V2_prime}, we have
    \begin{align*}
    &\sum_{e\in\calE}\|\bg_e^{(3)}((\bG_0^\sharp)^{-1}\bLambda_\sharp\nabla_\bphi\theta_e, (\bG_0^\sharp)^{-1}\bLambda_\sharp\nabla_\bphi\theta_e)\|_2^2\\
    &\quad\leq 4\sum_{e\in\calE}\|\scrP(\bG_0^\sharp)^{-1}\bV_e(\bG_0^\sharp)^{-1}\bJ_{0n}\transpose\widetilde{\bLambda}_\sharp\nabla_\bpsi\theta_e\|_2^2 
    \left\{\nabla_\bpsi\theta_e\transpose\widetilde{\bLambda}_\sharp\bJ_{0n}(\bG_0^\sharp)^{-1}\bJ_{0n}\transpose\widetilde{\bLambda}_\sharp\nabla_\bpsi\theta_e\right\}^2\\
    &\quad\lesssim \frac{1}{n}\max_{e\in\calE}\|(\bG_0^\sharp)^{-1}\bJ_{0n}\transpose\widetilde{\bLambda}_\sharp\nabla_\bpsi\theta_e\|_\infty^2 \|\widetilde{\bLambda}_\sharp\|_2^4\lesssim \frac{1}{n^4},\\
    &\sum_{e\in\calE}\|\bg_e^{(3)}((\bG_0^\sharp)^{-1}\bLambda_\sharp\nabla_\bphi\theta_e, \delta\bphi_2)\|_2^2\\
    &\quad\leq 2\sum_{e\in\calE}\|\scrP(\bG_0^\sharp)^{-1}\bV_e\delta\bphi_2\|_2^2 
    \left\{\nabla_\bpsi\theta_e\transpose\widetilde{\bLambda}_\sharp\bJ_{0n}(\bG_0^\sharp)^{-1}\bJ_{0n}\transpose\widetilde{\bLambda}_\sharp\nabla_\bpsi\theta_e\right\}^2\\
    &\qquad + 2\sum_{e\in\calE}\|\scrP(\bG_0^\sharp)^{-1}\bV_e(\bG_0^\sharp)^{-1}\bJ_{0n}\transpose\widetilde{\bLambda}_\sharp\nabla_\bpsi\theta_e\|_2^2 
    \left(\nabla_\bpsi\theta_e\transpose\widetilde{\bLambda}_\sharp\bJ_{0n}\delta\bphi_2\right)^2\\
    &\quad\lesssim 
    \frac{1}{n}\|\bJ_{0n}\delta\bphi_2\|_\infty^2\|\widetilde{\bLambda}_\sharp\|_2^4
    +
    \frac{1}{n}\max_{e\in\calE}\|(\bG_0^\sharp)^{-1}\bJ_{0n}\transpose\widetilde{\bLambda}_\sharp\nabla_\bpsi\theta_e\|_\infty^2 \|\widetilde{\bLambda}_\sharp\|_2^2\|\bJ_{0n}\delta\bphi_2\|_\infty^2\lesssim \frac{1}{n^3}\|\bJ_{0n}\delta\bphi_2\|_\infty^2.    
    \end{align*}
    
    \noindent\underline{For $k = 4$,} we need to first establish several properties related to the operator $\bT$. Recall the natural embedding of $\bphi$ into the ambient space $\bpsi(\bphi)$. For any $\delta\bphi_a\in\mathbb{R}^{d_\bphi}$, $a\in\{1,2,3\}$, denote 
    \begin{align*}
    \bpsi^{(a)} &:= \bJ_0\bLambda_\sharp\delta\bphi_a = \widetilde{\bLambda}_\sharp\bJ_{0n}\delta{\bphi}_a,\quad 
    \bpsi^{(ab)} := \frac{1}{mn^2}[\zero\transpose, \vect\{\ddot{\bX}(\mathbf{O})[\bB_{\delta\bphi_a}, \bB_{\delta\bphi_b}]\}\transpose]\transpose,\\
    \bpsi^{(123)} &:= \frac{1}{(\sqrt{m}n)^3}[\zero\transpose, \vect\{\dddot{\bX}(\mathbf{O})[\bB_{\delta\bphi_1}, \bB_{\delta\bphi_2}, \bB_{\delta\bphi_3}]\}\transpose]\transpose
    \end{align*}
    Since $\theta_e$ is a cubic polynomial in $\bpsi$, by the chain rule, we obtain
\begin{align*}
\tau_e(\delta\bphi_1, \delta\bphi_2, \delta\bphi_3) & = \nabla^3_\bpsi\theta_e[\bpsi^{(1)}, \bpsi^{(2)}, \bpsi^{(3)}] + \bpsi^{(12)\mathrm{T}}\nabla^2_\bpsi\theta_e\bpsi^{(3)} + \bpsi^{(13)\mathrm{T}}\nabla^2_\bpsi\theta_e \bpsi^{(2)} \\
&\quad + \bpsi^{(23)\mathrm{T}}\nabla^2_\bpsi\theta_e \bpsi^{(1)}  + \nabla_\bpsi\theta_e\transpose\bpsi^{(123)}
.
\end{align*}
By the calculations in Subsection \ref{sub:gradient_hessian} and Step 1 of the proof of Lemma \ref{lemma:third_order_cross_terms}, for any $\bU_1,\bU_2,\bU_3\in\mathbb{R}^{(n - 1 - d)\times d}$, we have
\begin{align*}
\ddot{\bX}(\mathbf{O})[\bU_1, \bU_2]
& = -\frac{1}{2}\bX_0(\bU_1\transpose\bU_2 + \bU_2\transpose\bU_1),\\
\dddot{\bX}(\mathbf{O})[\bU_1, \bU_2, \bU_3]
& = -\frac{1}{2}\bX_{0\perp}\bU_1(\bU_2\transpose\bU_3 + \bU_3\transpose\bU_2) -\frac{1}{2}\bX_{0\perp}\bU_2(\bU_1\transpose\bU_3 + \bU_3\transpose\bU_1)\\
&\quad -\frac{1}{2}\bX_{0\perp}\bU_3(\bU_1\transpose\bU_2 + \bU_2\transpose\bU_1).
\end{align*}
Now for any $e = (t, i, j)\in\calE$, we introduce the following edge-dependent local native norm $\langle\cdot\rangle_e$
\begin{align}
\label{eqn:local_norm}
\langle\delta\bphi\rangle_e := \frac{1}{n}\|\widetilde{\scrP}_{\br^{(t)}}\delta\bphi\|_2 + \frac{1}{\sqrt{mn}}\left(\left\|[\mathbf{O}, \eye_d\otimes\be_i(n)\transpose]\bJ_{0n}\delta\bphi\right\|_2 + \left\|[\mathbf{O}, \eye_d\otimes\be_j(n)\transpose]\bJ_{0n}\delta\bphi\right\|_2\right).
\end{align}
It is clear that $\max_e\langle\delta\bphi\rangle_e\lesssim (mn)^{-1/2}\|\bJ_{0n}\delta{\bphi}\|_\infty$. Also, it is clear that
\begin{align*}
\sum_{e\in\calE^\circ}\langle\delta\bphi\rangle_e^2\lesssim \sum_{s\in[m]}\|\widetilde{\scrP}_{\br^{(s)}}\delta\bphi\|_2^2 +  \sum_{i}\left\|[\mathbf{O}, \eye_d\otimes\be_i\transpose]\bJ_{0n}\delta{\bphi}\right\|_2^2\lesssim \min\{\|\delta\bphi\|_2^2,\ n\|\bJ_{0n}\delta{\bphi}\|_\infty^2\}.
\end{align*}
Now: 
\begin{itemize}
  \item For the first term in $\tau_e$ with $e = (t, i, j)\in\calE$, the nonzero indices of $\nabla^3_\bpsi\theta_e$ (as a three-way tensor) are only associate with the indices corresponding to $(\br^{(t)}, \bx_i, \bx_j)$ (in any order) with entries bounded by $1$ (since $\theta_e$ is a cubic polynomial of $\bpsi$), so the absolute value of the first term in $\tau_e$ is $O(\sum_{\sigma\in \calS_3}\langle\delta\bphi_{\sigma(1)}\rangle_e\langle\delta\bphi_{\sigma(2)}\rangle_e\langle\delta\bphi_{\sigma(3)}\rangle_e)$, where $\calS_3$ is the set of all permutations over $\{1,2,3\}$. 

  \item For the second, third, and fourth terms in $\tau_e$ with $e = (t, i, j)\in \calE$, $\bpsi^{(ab)}$ has only nonzero coordinates associated with $\bX$, and its block corresponding to $\bx_v$ with $v$ taking values in $\{i, j\}$ are $-(2mn)^{-1}(\bB_{\delta\bphi_a}\transpose\bB_{\delta\bphi_b} + \bB_{\delta\bphi_b}\transpose\bB_{\delta\bphi_a})\transpose\bx_{0v}$, and this block is $O(\|\bB_{\delta\bphi_a}\|_{\mathrm{F}}\|\bB_{\delta\bphi_b}\|_{\mathrm{F}}/(mn^2))$. Also, note that $\nabla^2_\bpsi\theta_e$ has nonzero elements only in indices associated with $e$, so that for $a,b,c\in\{1,2,3\}$, $a\neq b\neq c\neq a$, it is clear that $\bpsi^{(ab)\mathrm{T}}\nabla_{\bpsi}^2\theta_e\bpsi^{(c)}\lesssim (mn^2)^{-1}\|\bB_{\delta\bphi_a}\|_{\mathrm{F}}\|\bB_{\delta\bphi_b}\|_{\mathrm{F}}\langle\delta\bphi_c\rangle_e$. 

  \item For the last term, $\nabla_\bpsi\theta_e$ has nonzero elements in the coordinates associated with $\bX$ as $\bR_0^{(t)}\bx_{0b}$ and $\bR_0^{(t)}\bx_{0a}$, whose norms are uniformly bounded. Therefore, by the formula of $\bpsi^{(123)}$, it is also clear that $|\nabla_\bpsi\theta_e\transpose\bpsi^{(123)}|\lesssim (mn^2)^{-1}\sum_{\sigma\in\calS_3}\langle\delta\bphi_{\sigma(1)}\rangle_e\|\bB_{\delta\bphi_{\sigma(2)}}\|_{\mathrm{F}}\|\bB_{\delta\bphi_{\sigma(3)}}\|_{\mathrm{F}}$. 
\end{itemize}
Altogether,
\begin{align}
\label{eqn:W_local_bound}
|\tau_e(\delta\bphi_1, \delta\bphi_2, \delta\bphi_3)|\lesssim\sum_{\sigma\in S_3}\bigg\{\langle\delta\bphi_{\sigma(1)}\rangle_e\langle\delta\bphi_{\sigma(2)}\rangle_e\langle\delta\bphi_{\sigma(3)}\rangle_e + \frac{\|\bB_{\delta\bphi_{\sigma(1)}}\|_{\mathrm{F}}\|\bB_{\delta\bphi_{\sigma(2)}}\|_{\mathrm{F}}\langle\delta\bphi_{\sigma(3)}\rangle_e}{mn^2}\bigg\}.
\end{align}
In particular, this inequality implies:
\begin{itemize}
\item[(T1)] $|\tau_e(\delta\bphi_1, \delta\bphi_2, \delta\bphi_3)|\lesssim (mn)^{-3/2}\|\bJ_{0n}\delta{\bphi}_1\|_\infty\|\bJ_{0n}\delta{\bphi}_2\|_\infty\|\bJ_{0n}\delta{\bphi}_3\|_\infty$ for every $e$;
\item[(T2)] $\sum_e\tau_e(\delta\bphi_1, \delta\bphi_2, \delta\bphi_3)^2\leq C(mn)^{-2}\|\bJ_{0n}\delta{\bphi}_1\|_\infty^2\|\bJ_{0n}\delta{\bphi}_2\|_\infty^2\|\delta\bphi_3\|_2^2$. To see this, for the first summand in \eqref{eqn:W_local_bound}, $\max_e\langle\delta\bphi_1\rangle_e\langle\delta\bphi_2\rangle_e\lesssim C(mn)^{-1}\|\bJ_{0n}\delta{\bphi}_1\|_\infty\|\bJ_{0n}\delta{\bphi}_2\|_\infty$, and summing the remaining factor over $e$ results in $\sum_e\langle\delta\bphi_3\rangle_e^2\lesssim \|\delta\bphi_3\|_2^2$. For the second term in the summand of \eqref{eqn:W_local_bound}, if $\sigma(3) = 3$, then we have
\begin{align*}
\frac{1}{m^2n^4}\sum_{e\in\calE}\|\bB_{\delta\bphi_{\sigma(1)}}\|^2_{\mathrm{F}}\|\bB_{\delta\bphi_{\sigma(2)}}\|^2_{\mathrm{F}}\langle\delta\bphi_3\rangle_e^2\lesssim \frac{mn^2}{m^2n^4}\|\bJ_{0n}\delta\bphi_{\sigma(1)}\|_\infty^2\|\bJ_{0n}\delta\bphi_{\sigma(2)}\|_\infty^2\|\delta\bphi_3\|_2^2,
\end{align*}
and if $\sigma(3)\neq 3$, then for any $\sigma_1,\sigma_3\in\{1,2\},\sigma_1\neq \sigma_3$, we have
\begin{align*}
\frac{1}{m^2n^4}\sum_{e\in\calE}\|\bB_{\delta\bphi_{\sigma_1}}\|^2_{\mathrm{F}}\|\bB_{\delta\bphi_{3}}\|^2_{\mathrm{F}}\langle\delta\bphi_{\sigma_3}\rangle_e^2
&\le  \frac{mn^2}{m^2n^4} \|(\bX_{0\perp}/\sqrt{n})\bB_{\delta\bphi_{\sigma_1}}\|_{\mathrm{F}}^2\max_{e\in\calE}\langle\delta\bphi_{\sigma_3}\rangle_e^2\|\delta\bphi_3\|_2^2\\
&\lesssim \frac{1}{(mn)^2}\|\bJ_{0n}\delta\bphi_{\sigma_1}\|_\infty^2\|\bJ_{0n}\delta\bphi_{\sigma_3}\|_\infty^2\|\delta\bphi_3\|_2^2.
\end{align*}
Then, (T2) is established by combining the above cases. 

\item[(T3)] for fixed $\bkappa\in\mathbb{R}^{d_\bphi}$, the nuclear norm $\nu_e(\bkappa) := \|\bT_e(\bkappa)\|_*$ of the matrix $\bT_e(\bkappa)$ satisfies
\[
\max_{e\in\calE^\circ}\nu_e(\bkappa)\lesssim (mn)^{-3/2}\|\bkappa\|_2,\qquad
\sum_{e\in\calE^\circ}\nu_e(\bkappa)^2\lesssim (mn)^{-2}\|\bkappa\|_2^2.
\]
Consequently, $|\tau_e(\delta\bphi_1, \delta\bphi_2, \bkappa)|\leq\nu_e(\bkappa)\|\delta\bphi_1\|_2\|\delta\bphi_2\|_2$ and $|\mathrm{tr}\{\bT_e(\bkappa)\bM\}|\leq\nu_e(\bkappa)\|\bM\|_2$ for every symmetric $\bM$ by von Neumann trace inequality.

To prove (T3), fix $\delta\bphi_3 = \bkappa$ and inspect each of the five terms appearing in the formula of $\tau_e(\cdot, \cdot, \bkappa)$. For the first term, the indices of the nonzero elements of $\nabla_\bpsi^3\theta_e$ only appear in $\calS_1(e)\times\calS_1(e)\times\calS_1(e)$, so that
\begin{align*}
&\bigg\|\sum_{j_1,j_2\in[d_\bphi]}\nabla^3_\bpsi\theta_e[\widetilde{\bLambda}_\sharp\bJ_{0n}\be_{j_1}, \widetilde{\bLambda}_\sharp\bJ_{0n}\be_{j_2}, \widetilde{\bLambda}_\sharp\bJ_{0n}\bkappa]\be_{j_1}\be_{j_2}\transpose\bigg\|_*\\
&\quad = \bigg\|\sum_{j_1,j_2\in[d_\bphi]}\sum_{a,b,c\in\calS_1(e)}\nabla^3_\bpsi\theta_e[\be_a,\be_b,\be_c](\be_a\transpose\widetilde{\bLambda}_\sharp\bJ_{0n}\be_{j_1})(\be_b\transpose\widetilde{\bLambda}_\sharp\bJ_{0n}\be_{j_2})(\be_c\transpose\widetilde{\bLambda}_\sharp\bJ_{0n}\bkappa)\be_{j_1}\be_{j_2}\transpose\bigg\|_*\\
&\quad \leq \sum_{a,b,c\in\calS_1(e)}|\nabla^3_\bpsi\theta_e[\be_a,\be_b,\be_c]| \bigg\|\sum_{j_1,j_2\in[d_\bphi]}(\be_a\transpose\widetilde{\bLambda}_\sharp\bJ_{0n}\be_{j_1})(\be_b\transpose\widetilde{\bLambda}_\sharp\bJ_{0n}\be_{j_2})(\be_c\transpose\widetilde{\bLambda}_\sharp\bJ_{0n}\bkappa)\be_{j_1}\be_{j_2}\transpose\bigg\|_*\\
&\quad \lesssim \sum_{a,b,c\in\calS_1(e)}\langle\bkappa\rangle_e\bigg\|\sum_{j_1,j_2\in[d_\bphi]}\be_{j_1}\be_{j_1}\transpose\bJ_{0n}\transpose\widetilde{\bLambda}_\sharp\be_a \be_b\transpose\widetilde{\bLambda}_\sharp\bJ_{0n}\be_{j_2}\be_{j_2}\transpose\bigg\|_*\lesssim \frac{1}{n}\langle\bkappa\rangle_e.
\end{align*}
For the second term, with $e = (t, i, j)$, we have
\begin{align*}
&\bigg\|
\sum_{j_1,j_2\in[d_\bphi]} \frac{1}{mn^2}[\zero\transpose, \vect\{\ddot{\bX}(\mathbf{O})[\bB_{\be_{j_1}}, \bB_{\be_{j_2}}]\}\transpose] \nabla^2_\bpsi \theta_e\widetilde{\bLambda}_\sharp\bJ_{0n}\bkappa \be_{j_1}\be_{j_2}\transpose
\bigg\|_*\\
&\quad = \bigg\|\sum_{a,b\in\calS_1(e)}
\sum_{j_1,j_2\in[d_\bphi]} \frac{1}{mn^2}[\zero\transpose, \vect\{\ddot{\bX}(\mathbf{O})[\bB_{\be_{j_1}}, \bB_{\be_{j_2}}]\}\transpose]\be_a \be_a\transpose\nabla^2_\bpsi \theta_e\be_b\be_b\transpose\widetilde{\bLambda}_\sharp\bJ_{0n}\bkappa \be_{j_1}\be_{j_2}\transpose
\bigg\|_*\\
&\quad \leq \sum_{a,b\in\{d_{\balpha} + d_{\br} + (k - 1)n + i, d_{\balpha} + d_{\br} + (k - 1)n + j:k\in[d]\}}\\
&\qquad \bigg\|
\sum_{j_1,j_2\in[d_\bphi]} \frac{1}{mn^2}[\zero\transpose, \vect\{\ddot{\bX}(\mathbf{O})[\bB_{\be_{j_1}}, \bB_{\be_{j_2}}]\}\transpose]\be_a \be_a\transpose\nabla^2_\bpsi \theta_e\be_b\be_b\transpose\widetilde{\bLambda}_\sharp\bJ_{0n}\bkappa \be_{j_1}\be_{j_2}\transpose
\bigg\|_*\\
&\quad\lesssim \langle\bkappa\rangle_e\max_{a\in\{i, j\},k\in[d]}\bigg\|
\sum_{j_1,j_2\in[d_\bphi]} \frac{1}{mn^2}\bx_{0a}\transpose(\bB_{\be_{j_1}}\transpose\bB_{\be_{j_2}} + \bB_{\be_{j_2}}\transpose\bB_{\be_{j_1}})\be_k\be_{j_1}\be_{j_2}\transpose
\bigg\|_*\\
&\quad = \langle \bkappa\rangle_e
\max_{a\in\{i, j\}, k\in[d]}\frac{1}{mn^2}\left\|\begin{bmatrix}
  \mathbf{O} \\ \eye_{d_\bX}
\end{bmatrix}
(\bx_{0a}\be_k\transpose + \be_k\bx_{0a}\transpose)\otimes\eye_{n - 1 - d}
\begin{bmatrix}
\mathbf{O} & \eye_{d_\bX}
\end{bmatrix}
\right\|_*\lesssim \frac{\langle\bkappa\rangle_e}{mn}.
\end{align*}
For the third term, we directly write
\begin{align*}
&\bigg\|
\sum_{j_1,j_2\in[d_\bphi]} \frac{1}{mn^2}[\zero\transpose, \vect\{\ddot{\bX}(\mathbf{O})[\bB_{\be_{j_1}}, \bB_{\bkappa}]\}\transpose] \nabla^2_\bpsi \theta_e\widetilde{\bLambda}_\sharp\bJ_{0n}\be_{j_2} \be_{j_1}\be_{j_2}\transpose
\bigg\|_*\\
&\quad = \frac{1}{mn^2}\bigg\|
\sum_{j_1 \in[d_\bphi]} \be_{j_1}[\zero\transpose, \vect\{\ddot{\bX}(\mathbf{O})[\bB_{\be_{j_1}}, \bB_{\bkappa}]\}\transpose] \nabla^2_\bpsi \theta_e\widetilde{\bLambda}_\sharp\bJ_{0n} 
\bigg\|_*\\
&\quad = \frac{1}{mn^2}\left\|
\begin{bmatrix}
\mathbf{O} & \\
& -\frac{1}{2}\left\{\eye_d\otimes \bX_0\bB_{\bkappa}\transpose + (\bB_{\bkappa}\transpose\otimes\bX_0)\bK_{(n - 1 - d)d}\right\}
\end{bmatrix}\transpose\nabla_\bpsi^2\theta_e \widetilde{\bLambda}_\sharp\bJ_{0n}
\right\|_*
\lesssim \frac{\|\bkappa\|_2}{mn^{5/2}}
\end{align*}
and likewise for the fourth term. 

Finally, for the last term, it is the sum of three symmetric summands according to the formula of $\dddot{\bX}(\mathbf{O})$. The first and the second summands are symmetric with $\bU_3 = \bB_\bkappa$, so we can consider without loss of generality that
\begin{align*}
&\bigg\|
\sum_{j_1,j_2\in[d_\bphi]} \frac{1}{(\sqrt{m}n)^3}[\zero\transpose, \vect\{\bX_{0\perp}\bB_{\be_{j_1}}(\bB_{\be_{j_2}}\transpose\bB_{\bkappa} + \bB_{\bkappa}\transpose\bB_{\be_{j_2}})\}\transpose] \nabla_\bpsi \theta_e \be_{j_1}\be_{j_2}\transpose
\bigg\|_* \\
&\quad\leq 
\frac{1}{(\sqrt{m}n)^3}\left\|\eye_d\otimes\bigg(\bX_{0\perp}\transpose\frac{\partial\theta_e}{\partial\bX}\bB_\bkappa\transpose + \bB_\bkappa\frac{\partial\theta_e}{\partial\bX\transpose}\bX_{0\perp}\bigg)\right\|_*\lesssim \frac{\|\bkappa\|_2}{m^{3/2}n^{5/2}}. 
\end{align*}
The first summand with $\bU_3 = \bB_\bkappa$ is similar to the second term above, which we write
\begin{align*}
&\bigg\|
\sum_{j_1,j_2\in[d_\bphi]} \frac{1}{(\sqrt{m}n)^3}[\zero\transpose, \vect\{\bX_{0\perp}\bB_\bkappa(\bB_{\be_{j_1}}\transpose\bB_{\be_{j_2}} + \bB_{\be_{j_2}}\transpose\bB_{\be_{j_1}})\}\transpose] \nabla_\bpsi \theta_e \be_{j_1}\be_{j_2}\transpose
\bigg\|_* \\
&\quad\leq 
\frac{1}{(\sqrt{m}n)^3}\left\|\bigg(\bB_\bkappa\transpose\bX_{0\perp}\transpose\frac{\partial\theta_e}{\partial\bX} + \frac{\partial\theta_e}{\partial\bX\transpose}\bX_{0\perp}\bB_\bkappa\bigg)\right\|_*\lesssim \frac{\langle \bkappa \rangle_e}{m n }.
\end{align*}
Combining the above bounds, we see that
$\nu_e(\bkappa)\lesssim (mn)^{-1}\langle\bkappa\rangle_e + m^{-1}n^{-5/2}\|\bkappa\|_2 $, and (T3) follows from $\max_e\langle\bkappa\rangle_e\lesssim (mn)^{-1/2}\|\bkappa\|_2$, $\sum_e\langle\bkappa\rangle_e^2\leq C\|\bkappa\|_2^2$ and $|\calE^\circ|m^{-2}n^{-5}\leq m^{-1}n^{-3}\leq (mn)^{-2}$.
\end{itemize}

\underline{Now we are back with the case of $k = 4$.} By (T1), we directly have
\begin{align*}
&\max_{e\in\calE}\|\bg_e^{(4)}(\delta\bphi_1, \delta\bphi_2)\|_2\\
&\quad\lesssim\max_{e\in\calE,k}|\be_k\transpose\scrP(\bG_0^\sharp)^{-1}\bw_e[\delta\bphi_1, \delta\bphi_2]|\lesssim\max_{e,k}|\tau_e(\delta\bphi_1, \delta\bphi_2, (\bG_0^\sharp)^{-1}\scrP\transpose\be_k)|\\
&\quad\lesssim \frac{1}{(mn)^{3/2}}\|\bJ_{0n}\delta\bphi_1\|_\infty\|\bJ_{0n}\delta\bphi_2\|_\infty,\\
&\max_{e\in\calE}\|\bg_e^{(4)}((\bG_0^\sharp)^{-1}\bLambda_\sharp\bJ_0\transpose\nabla_\bpsi\theta_e, (\bG_0^\sharp)^{-1}\bLambda_\sharp\bJ_0\transpose\nabla_\bpsi\theta_e)\|_2\\
&\quad
\lesssim\max_{e,k}|\tau_e((\bG_0^\sharp)^{-1}\bLambda_\sharp\bJ_0\transpose\nabla_\bphi\theta_e, (\bG_0^\sharp)^{-1}\bLambda_\sharp\bJ_0\transpose\nabla_\bphi\theta_e, (\bG_0^\sharp)^{-1}\scrP\transpose\be_k)|\\
&\quad\lesssim \frac{1}{(mn)^{3/2}}\|\bJ_{0n}(\bG_0^\sharp)^{-1}\bLambda_\sharp\bJ_0\transpose\nabla_\bpsi\theta_e\|_\infty^2\lesssim \frac{1}{m^{3/2}n^{5/2}},\\
&\max_{e\in\calE}\|\bg_e^{(4)}((\bG_0^\sharp)^{-1}\bLambda_\sharp\bJ_0\transpose\nabla_\bpsi\theta_e, \delta\bphi_2)\|_2\\
&\quad
\lesssim\max_{e,k}|\tau_e((\bG_0^\sharp)^{-1}\bLambda_\sharp\bJ_0\transpose\nabla_\bphi\theta_e, \delta\bphi_2, (\bG_0^\sharp)^{-1}\scrP\transpose\be_k)|\\
&\quad\lesssim \frac{1}{(mn)^{3/2}}\|\bJ_{0n}(\bG_0^\sharp)^{-1}\bLambda_\sharp\bJ_0\transpose\nabla_\bpsi\theta_e\|_\infty\|\bJ_{0n}\delta\bphi_2\|_\infty\lesssim \frac{1}{m^{3/2}n^{2}}\|\bJ_{0n}\delta\bphi_2\|_\infty.
\end{align*}
For the sum of squared norms, by (T1), (T2) and (T3), we have
\begin{align*}
&\sum_{e\in\calE}\|\bg_e^{(4)}(\delta\bphi_1, \delta\bphi_2)\|_2^2\\
&\quad\lesssim\sum_{e\in\calE,k\in[d(d + 1)/2]}|\be_k\transpose\scrP(\bG_0^\sharp)^{-1}\bw_e[\delta\bphi_1, \delta\bphi_2]|^2
\lesssim\sum_{e,k}|\tau_e(\delta\bphi_1, \delta\bphi_2, (\bG_0^\sharp)^{-1}\scrP\transpose\be_k)|^2\\
&\quad\lesssim \frac{1}{(mn)^{2}}\|\bJ_{0n}\delta\bphi_1\|_\infty^2\|\bJ_{0n}\delta\bphi_2\|_\infty^2,\\
&\sum_{e\in\calE}\|\bg_e^{(4)}((\bG_0^\sharp)^{-1}\bLambda_\sharp\bJ_0\transpose\nabla_\bpsi\theta_e, (\bG_0^\sharp)^{-1}\bLambda_\sharp\bJ_0\transpose\nabla_\bpsi\theta_e)\|_2^2\\
&\quad
\lesssim\sum_{e,k}|\tau_e((\bG_0^\sharp)^{-1}\bLambda_\sharp\bJ_0\transpose\nabla_\bphi\theta_e, (\bG_0^\sharp)^{-1}\bLambda_\sharp\bJ_0\transpose\nabla_\bphi\theta_e, (\bG_0^\sharp)^{-1}\scrP\transpose\be_k)|^2\\
&\quad
\lesssim  
\max_k\sum_{e} \|\bT_e((\bG_0^\sharp)^{-1}\scrP\transpose\be_k)\|_*^2\|(\bG_0^\sharp)^{-1}\bLambda_\sharp\bJ_0\transpose\nabla_\bphi\theta_e\|_2^4\\
&\quad\lesssim \frac{1}{(mn)^2}\|\bJ_{0n}(\bG_0^\sharp)^{-1}\bLambda_\sharp\bJ_0\transpose\nabla_\bpsi\theta_e\|_2^4\lesssim \frac{1}{m^2n^{4}},\\
&\sum_{e\in\calE}\|\bg_e^{(4)}((\bG_0^\sharp)^{-1}\bLambda_\sharp\bJ_0\nabla_\bpsi\theta_e, \delta\bphi_2)\|_2^2\\
&\quad
\lesssim mn^2\max_{e,k}|\tau_e((\bG_0^\sharp)^{-1}\bLambda_\sharp\bJ_0\nabla_\bphi\theta_e, \delta\bphi_2, (\bG_0^\sharp)^{-1}\scrP\transpose\be_k)|^2\\
&\quad\lesssim \frac{mn^2}{(mn)^{3}}\|\bJ_{0n}(\bG_0^\sharp)^{-1}\bLambda_\sharp\bJ_0\nabla_\bpsi\theta_e\|_\infty^2\|\bJ_{0n}\delta\bphi_2\|_\infty^2\lesssim \frac{1}{(mn)^{2}}\|\bJ_{0n}\delta\bphi_2\|_\infty^2.
\end{align*}

  \item For any matrix $\bM\in\mathbb{R}^{d_\bphi\times d_\bphi}$, 
  \begin{equation}
  \label{eqn:kernel_bound_matrix}
  \sum_{e\in\calE}\|\bg_e^{(k)}[(\bG_0^\sharp)^{-1}]\|_2^2 \lesssim n^{-2},
  \quad 
  \max_{e\in\calE}\|\bg_e^{(k)}[\bM]\|_2 \lesssim \|\bM\|_2n^{-3/2},\quad\sum_{e\in\calE}\|\bg_e^{(k)}[\bM]\|_2^2 \lesssim \|\bM\|_2^2 n^{-1};  
  \end{equation}
  For the second bound, by \eqref{eqn:Ve_nuclear_norm} and (T3), we have
  \begin{align*}
  \max_{e\in\calE}\|\bg_e^{(1)}[\bM]\|_2
  &\lesssim \|\widetilde{\bLambda}_\sharp\|_2^3\|\bM\|_2\lesssim n^{-3/2}\|\bM\|_2,\\
  \max_{e\in\calE}\|\bg_e^{(2)}[\bM]\|_2
  &\lesssim \max_{e\in\calE}\|\widetilde{\bLambda}_\sharp\|_2\|\bV_e\|_*\|\bM\|_2\lesssim m^{-1}n^{-3/2}\|\bM\|_2,\\
  \max_{e\in\calE}\|\bg_e^{(3)}[\bM]\|_2
  &\lesssim \max_{e\in\calE}\|\widetilde{\bLambda}_\sharp\|_2\|\bM\|_2\|\bV_e\|_2\lesssim n^{-3/2}\|\bM\|_2,\\
  \max_{e\in\calE}\|\bg_e^{(4)}[\bM]\|_2
  &\lesssim \max_{e\in\calE}\bigg\{\sum_{j = 1}^{d(d + 1)/2}\|\bT_e((\bG_0^\sharp)^{-1}\scrP\transpose\be_j)\|_*^2\|\bM\|_2^2\bigg\}^{1/2}\lesssim (mn)^{-3/2}\|\bM\|_2.
  \end{align*}
  For the sum of squared norms, by \eqref{eqn:Ve_nuclear_norm}, (T3), and item (D) below (whose derivation is independent), we have
  \begin{align*}
  \sum_{e\in\calE}\|\bg_e^{(1)}[\bM]\|_2^2
  &\lesssim \sum_e\big\|\scrP(\bG_0^\sharp)^{-1}{\bLambda}_\sharp\nabla_\bphi\theta_e\big\|_2^2\|\widetilde{\bLambda}_\sharp\|_2^4\|\bM\|_2^2\lesssim n^{-2}\|\bM\|_2^2,\\
  \sum_{e\in\calE}\|\bg_e^{(2)}[\bM]\|_2^2
  &\lesssim \sum_e\big\|\scrP(\bG_0^\sharp)^{-1}{\bLambda}_\sharp\nabla_\bphi\theta_e\big\|_2^2
  \|\bV_e\|_*^2\|\bM\|_2^2\lesssim (mn)^{-2}\|\bM\|_2^2,\\
  \sum_{e\in\calE}\|\bg_e^{(3)}[\bM]\|_2^2
  &\lesssim mn^2\max_{e\in\calE}\|{\bLambda}_\sharp\nabla_\bphi\theta_e\|_2^2\|\bM\|_2^2\|\bV_e\|_2^2\lesssim (mn)^{-1}\|\bM\|_2^2,\\
  \sum_{e\in\calE}\|\bg_e^{(4)}[\bM]\|_2^2
  &\lesssim \sum_{e\in\calE} \sum_{j = 1}^{d(d + 1)/2}\|\bT_e((\bG_0^\sharp)^{-1}\scrP\transpose\be_j)\|_*^2\|\bM\|_2^2 \lesssim (mn)^{-2}\|\bM\|_2^2.
  \end{align*}
  Note that for these sum of squared norms of $\|\bg_e^{(k)}[\bM]\|_2^2$, only the case of $k = 3$ is $O((mn)^{-1}\|\bM\|_2^2)$ and the remaining bounds are $O(n^{-2}\|\bM\|_2^2)$. Therefore, when $\bM$ is specialized to $(\bG_0^\sharp)^{-1}$, which is the first bound, it is sufficient to consider $k = 3$, in which case we have
  \begin{align*}
  \sum_{e\in\calE}\|\bg_e^{(3)}[(\bG_0^\sharp)^{-1}]\|_2^2
  & = 4\sum_{e\in\calE}\left\|\scrP(\bG_0^\sharp)^{-1}\bV_e(\bG_0^\sharp)^{-1}{\bLambda}_\sharp\nabla_\bphi\theta_e\right\|_2^2\lesssim \frac{1}{n^2}
  \end{align*}
  by Step 2(d) in the proof of Lemma \ref{lemma:Q1_fluctuation}. 

  \item For any matrix $\bM\in\mathbb{R}^{d(d + 1)/2\times d_\bphi}$, 
  \[
  \max_{e\in\calE}\|\bM(\bG_0^\sharp)^{-1}\bLambda_\sharp\nabla_\bphi\theta_e\|_2\lesssim n^{-1/2}\|\bM\|_2\quad\text{and}\quad 
  \sum_{e\in\calE}\|\bM(\bG_0^\sharp)^{-1}\bLambda_\sharp\nabla_\bphi\theta_e\|_2^2\lesssim \|\bM\|_2^2.
  \]
  The first bound is immediate since $\|(\bG_0^\sharp)^{-1}\|_2 = O(1)$, $\|\widetilde{\bLambda}_\sharp\|_2 = O(n^{-1/2})$, and $\max_e\|\nabla_\bpsi\theta_e\|_2 = O(1)$. The second bound follows from von Neumann trace inequality and
  \begin{align*}
  \sum_e\|\bM(\bG_0^\sharp)^{-1}\bLambda_\sharp\nabla_\bphi\theta_e\|_2^2
  & = \mathrm{tr}\bigg\{\sum_e \bLambda_\sharp\nabla_\bphi\theta_e \nabla_\bphi\theta_e\transpose\bLambda_\sharp(\bG_0^\sharp)^{-1}\bM\transpose\bM(\bG_0^\sharp)^{-1}
  \bigg\}\\
  &\leq \bigg\|\sum_e \bLambda_\sharp\nabla_\bphi\theta_e \nabla_\bphi\theta_e\transpose\bLambda_\sharp\bigg\|_2\left\|(\bG_0^\sharp)^{-1}\bM\transpose\bM(\bG_0^\sharp)^{-1}\right\|_*\lesssim \|\bM\|_2^2.
  \end{align*}

  \item For any scalars $(\alpha_e)_{e\in\calE}$ and \underline{any $k\in\{1,2,3\}$}, the following bounds hold:
  \begin{equation}
  \label{eqn:kernel_vector_bound_scalar_multiple}
  \begin{aligned}
  \max_{e'\in\calE}\bigg\|\sum_{e\in\calE}\alpha_e\bg_e^{(k)}((\bG_0^\sharp)^{-1}\bLambda_\sharp\nabla_\bphi\theta_e, (\bG_0^\sharp)^{-1}\bLambda_\sharp\nabla_\bphi\theta_{e'})\bigg\|_2
  & \lesssim \max_e|\alpha_e|n^{-3/2},\\
  \sum_{e'\in\calE}\bigg\|\sum_{e\in\calE}\alpha_e\bg_e^{(k)}((\bG_0^\sharp)^{-1}\bLambda_\sharp\nabla_\bphi\theta_e, (\bG_0^\sharp)^{-1}\bLambda_\sharp\nabla_\bphi\theta_{e'})\bigg\|_2^2
  & \lesssim \max_e|\alpha_e|^2n^{-2},\\
  \max_{e'\in\calE}\bigg\|\sum_{e\in\calE}\alpha_e\bg_e^{(k)}((\bG_0^\sharp)^{-1}\bLambda_\sharp\nabla_\bphi\theta_{e'}, (\bG_0^\sharp)^{-1}\bLambda_\sharp\nabla_\bphi\theta_{e'})\bigg\|_2
  & \lesssim \max_e|\alpha_e|n^{-3/2},\\
  \sum_{e'\in\calE}\bigg\|\sum_{e\in\calE}\alpha_e\bg_e^{(k)}((\bG_0^\sharp)^{-1}\bLambda_\sharp\nabla_\bphi\theta_{e'}, (\bG_0^\sharp)^{-1}\bLambda_\sharp\nabla_\bphi\theta_{e'})\bigg\|_2^2
  & \lesssim \max_e|\alpha_e|^2n^{-2},
  \end{aligned}
  \end{equation}
  and for $k = 4$, 
  \begin{equation}
  \label{eqn:kernel_vector_bound_scalar_multiple_k4}
  \begin{aligned}
  \max_{e'\in\calE}\bigg\|\sum_{e\in\calE}\alpha_e\bg_e^{(4)}((\bG_0^\sharp)^{-1}\bLambda_\sharp\nabla_\bphi\theta_e, (\bG_0^\sharp)^{-1}\bLambda_\sharp\nabla_\bphi\theta_{e'})\bigg\|_2
  & \lesssim \max_e|\alpha_e|n^{-1},\\
  \sum_{e'\in\calE}\bigg\|\sum_{e\in\calE}\alpha_e\bg_e^{(4)}((\bG_0^\sharp)^{-1}\bLambda_\sharp\nabla_\bphi\theta_e, (\bG_0^\sharp)^{-1}\bLambda_\sharp\nabla_\bphi\theta_{e'})\bigg\|_2^2
  & \lesssim \max_e|\alpha_e|^2n^{-1}.
  \end{aligned}
  \end{equation}
  Furthermore, for any $\delta\bphi\in\mathbb{R}^{d_\bphi}$ and \underline{any $k\in\{1,2,3\}$},
  \begin{equation}
  \label{eqn:kernel_vector_bound_scalar_multiple_deltaphi}
  \begin{aligned}
  \max_{e'\in\calE}\bigg\|\sum_{e\in\calE}\alpha_e\bg_e^{(k)}((\bG_0^\sharp)^{-1}\bLambda_\sharp\nabla_\bphi\theta_{e'}, \delta\bphi)\bigg\|_2
  & \lesssim \max_e|\alpha_e|\|\bJ_{0n}\delta\bphi\|_\infty n^{-1},\\
  \sum_{e'\in\calE}\bigg\|\sum_{e\in\calE}\alpha_e\bg_e^{(k)}((\bG_0^\sharp)^{-1}\bLambda_\sharp\nabla_\bphi\theta_e, \delta\bphi)\bigg\|_2^2
  & \lesssim \max_e|\alpha_e|^2\|\bJ_{0n}\delta\bphi\|_2^2n^{-1}.
  \end{aligned}
  \end{equation}

  \underline{For \eqref{eqn:kernel_vector_bound_scalar_multiple} and \eqref{eqn:kernel_vector_bound_scalar_multiple_deltaphi} with $k = 1$,} write
  \begin{align*}
  &\sum_{e\in\calE}\alpha_e\bg_e^{(1)}((\bG_0^\sharp)^{-1}\bLambda_\sharp\nabla_\bphi\theta_e, (\bG_0^\sharp)^{-1}\bLambda_\sharp\nabla_\bphi\theta_{e'})\\
  &\quad = \scrP(\bG_0^\sharp)^{-1}\bigg\{\sum_{e\in\calE}\alpha_e\bLambda_\sharp\nabla_\bphi\theta_e\{(\bG_0^\sharp)^{-1}\bLambda_\sharp\nabla_\bphi\theta_e\}\transpose\bLambda_\sharp\bJ_0\transpose\nabla_\bpsi\theta_e\nabla_\bpsi\theta_e\transpose\bJ_0\bLambda_\sharp\bigg\}(\bG_0^\sharp)^{-1}\bLambda_\sharp\nabla_\bphi\theta_{e'},\\
  &\sum_{e\in\calE}\alpha_e\bg_e^{(1)}((\bG_0^\sharp)^{-1}\bLambda_\sharp\nabla_\bphi\theta_{e'}, \delta\bphi)\\
  &\quad = \scrP(\bG_0^\sharp)^{-1}\bigg\{\sum_{e\in\calE}\alpha_e\bLambda_\sharp\nabla_\bphi\theta_e
  (\delta\bphi
  \transpose\bLambda_\sharp\bJ_0\transpose\nabla_\bpsi\theta_e)
  \nabla_\bphi\theta_e\transpose\bLambda_\sharp\bigg\}(\bG_0^\sharp)^{-1}\bLambda_\sharp\nabla_\bphi\theta_{e'},\\
  &\sum_{e\in\calE}\alpha_e\bg_e^{(1)}((\bG_0^\sharp)^{-1}\bLambda_\sharp\nabla_\bphi\theta_{e'}, (\bG_0^\sharp)^{-1}\bLambda_\sharp\nabla_\bphi\theta_{e'})\\
  &\quad = \scrP(\bG_0^\sharp)^{-1} \sum_{e\in\calE}\alpha_e\bLambda_\sharp\nabla_\bphi\theta_e
  \{\nabla_\bpsi\theta_e\transpose\bJ_0\bLambda_\sharp (\bG_0^\sharp)^{-1}\bLambda_\sharp\nabla_\bphi\theta_{e'}\}^2.
  \end{align*}
  For the first and second bounds in \eqref{eqn:kernel_vector_bound_scalar_multiple}, item (D) above implies that
  \begin{align*}
  &\max_{e'\in\calE}\bigg\|\sum_{e\in\calE}\alpha_e\bg_e^{(1)}((\bG_0^\sharp)^{-1}\bLambda_\sharp\nabla_\bphi\theta_e, (\bG_0^\sharp)^{-1}\bLambda_\sharp\nabla_\bphi\theta_{e'})\bigg\|_2\\
  &\quad\lesssim n^{-1/2}\bigg\|\scrP(\bG_0^\sharp)^{-1}\bigg\{\sum_{e\in\calE}\alpha_e\bLambda_\sharp\nabla_\bphi\theta_e\{(\bG_0^\sharp)^{-1}\bLambda_\sharp\nabla_\bphi\theta_e\}\transpose\bLambda_\sharp\bJ_0\transpose\nabla_\bpsi\theta_e\nabla_\bpsi\theta_e\transpose\bJ_0\bLambda_\sharp\bigg\}\bigg\|_2\\
  &\quad\lesssim n^{-1/2}\max_e|\alpha_e|\sup_{\|\bw\|_2 = 1}
  \sum_{e\in\calE}  \nabla_\bphi\theta_e\transpose\bLambda_\sharp(\bG_0^\sharp)^{-1}\bLambda_\sharp\bJ_0\transpose\nabla_\bpsi\theta_e(\nabla_\bpsi\theta_e\transpose\bJ_0\bLambda_\sharp\bw)^2\\
  &\quad\lesssim n^{-3/2}\max_e|\alpha_e|\sup_{\|\bw\|_2 = 1}
  \sum_{e\in\calE}(\nabla_\bpsi\theta_e\transpose\bJ_0\bLambda_\sharp\bw)^2\lesssim n^{-3/2}\max_{e}|\alpha_e|,\\
  &\sum_{e'\in\calE}\bigg\|\sum_{e\in\calE}\alpha_e\bg_e^{(1)}((\bG_0^\sharp)^{-1}\bLambda_\sharp\nabla_\bphi\theta_e, (\bG_0^\sharp)^{-1}\bLambda_\sharp\nabla_\bphi\theta_{e'})\bigg\|_2^2\lesssim \max_e|\alpha_e|^2n^{-2},
  \end{align*}
  and for \eqref{eqn:kernel_vector_bound_scalar_multiple_deltaphi} with $k = 1$,
  \begin{align*}
  &\max_{e'\in\calE}\bigg\|\sum_{e\in\calE}\alpha_e\bg_e^{(1)}((\bG_0^\sharp)^{-1}\bLambda_\sharp\nabla_\bphi\theta_{e'}, \delta\bphi)\bigg\|_2\\
  &\quad\lesssim n^{-1/2}\bigg\|\scrP(\bG_0^\sharp)^{-1}\bigg\{\sum_{e\in\calE}\alpha_e\bLambda_\sharp\nabla_\bphi\theta_e
  (\delta\bphi\transpose\bLambda_\sharp\bJ_0\transpose\nabla_\bpsi\theta_e)
  \nabla_\bphi\theta_e\transpose\bLambda_\sharp\bigg\}\bigg\|_2\\
  &\quad\lesssim n^{-1/2}\max_e|\alpha_e|\|\widetilde{\bLambda}_\sharp\|_\infty\|\bJ_{0n}\delta\bphi\|_\infty
  \lesssim n^{-1}\max_e|\alpha_e|,\\
  &\sum_{e'\in\calE}\bigg\|\sum_{e\in\calE}\alpha_e\bg_e^{(1)}((\bG_0^\sharp)^{-1}\bLambda_\sharp\nabla_\bphi\theta_{e'}, \delta\bphi)\bigg\|_2^2\lesssim \max_e|\alpha_e|^2n^{-1}.
  \end{align*}
  For the third and fourth bounds in \eqref{eqn:kernel_vector_bound_scalar_multiple}, by Cauchy--Schwarz inequality and item (D), we have
  \begin{align*}
  &\max_{e'\in\calE}\bigg\|\sum_{e\in\calE}\alpha_e\bg_e^{(1)}((\bG_0^\sharp)^{-1}\bLambda_\sharp\nabla_\bphi\theta_{e'}, (\bG_0^\sharp)^{-1}\bLambda_\sharp\nabla_\bphi\theta_{e'})\bigg\|_2\\
  &\quad \leq \max_{e'\in\calE}\bigg[\sum_{e\in\calE}\alpha_e^2\left\|\scrP(\bG_0^\sharp)^{-1}\bLambda_\sharp\nabla_\bphi\theta_e\right\|_2^2
  \{\nabla_\bpsi\theta_e\transpose\bJ_0\bLambda_\sharp (\bG_0^\sharp)^{-1}\bLambda_\sharp\nabla_\bphi\theta_{e'}\}^2\bigg]^{1/2}\\
  &\qquad\times \bigg[
    \nabla_\bphi\theta_{e'}\transpose
  \bLambda_\sharp
  (\bG_0^\sharp)^{-1}
  \bigg\{
  \sum_{e\in\calE} 
  \bLambda_\sharp
  \bJ_0\transpose
  \nabla_\bpsi\theta_e
  \nabla_\bpsi\theta_e\transpose\bJ_0\bLambda_\sharp\bigg\} (\bG_0^\sharp)^{-1}\bLambda_\sharp\nabla_\bphi\theta_{e'} \bigg]^{1/2}
  \lesssim \max_e|\alpha_e|n^{-3/2},\\
  &\sum_{e'\in\calE}\bigg\|\sum_{e\in\calE}\alpha_e\bg_e^{(1)}((\bG_0^\sharp)^{-1}\bLambda_\sharp\nabla_\bphi\theta_{e'}, (\bG_0^\sharp)^{-1}\bLambda_\sharp\nabla_\bphi\theta_{e'})\bigg\|_2^2\\
  &\quad\leq \sum_{e\in\calE}\alpha_e^2\left\|\scrP(\bG_0^\sharp)^{-1}\bLambda_\sharp\nabla_\bphi\theta_e\right\|_2^2\bigg[\sum_{e'\in\calE}
  \{\nabla_\bpsi\theta_e\transpose\bJ_0\bLambda_\sharp (\bG_0^\sharp)^{-1}\bLambda_\sharp\nabla_\bphi\theta_{e'}\}^2\bigg] \\
  &\qquad\times \bigg[
    \nabla_\bphi\theta_{e'}\transpose
  \bLambda_\sharp
  (\bG_0^\sharp)^{-1}
  \bigg\{
  \sum_{e\in\calE} 
  \bLambda_\sharp
  \bJ_0\transpose
  \nabla_\bpsi\theta_e
  \nabla_\bpsi\theta_e\transpose\bJ_0\bLambda_\sharp\bigg\} (\bG_0^\sharp)^{-1}\bLambda_\sharp\nabla_\bphi\theta_{e'} \bigg]\\
  &\quad\lesssim \frac{1}{n}\max_{e\in\calE}|\alpha_e|^2\sum_{e'\in\calE}
  \{\nabla_\bpsi\theta_e\transpose\bJ_0\bLambda_\sharp (\bG_0^\sharp)^{-1}\bLambda_\sharp\nabla_\bphi\theta_{e'}\}^2\lesssim \frac{1}{n^2}\max_{e\in\calE}|\alpha_e|^2.
  \end{align*}

  \underline{For \eqref{eqn:kernel_vector_bound_scalar_multiple} and \eqref{eqn:kernel_vector_bound_scalar_multiple_deltaphi} with $k = 2$,} write
  \begin{align*}
  &\sum_{e\in\calE}\alpha_e\bg_e^{(2)}((\bG_0^\sharp)^{-1}\bLambda_\sharp\nabla_\bphi\theta_{e}, (\bG_0^\sharp)^{-1}\bLambda_\sharp\nabla_\bphi\theta_{e'})\\
  &\quad = \scrP(\bG_0^\sharp)^{-1}\bigg[\sum_{e\in\calE}\alpha_e
  \bLambda_\sharp\nabla_\bphi\theta_e\{(\bG_0^\sharp)^{-1}\bLambda_\sharp\nabla_\bphi\theta_{e}\}\transpose\bV_e\bigg](\bG_0^\sharp)^{-1}\bLambda_\sharp\nabla_\bphi\theta_{e'},\\
  &\sum_{e\in\calE}\alpha_e\bg_e^{(2)}((\bG_0^\sharp)^{-1}\bLambda_\sharp\nabla_\bphi\theta_{e'}, \delta\bphi)\\
  &\quad = \scrP(\bG_0^\sharp)^{-1}\bigg[\sum_{e\in\calE}\alpha_e
  \bLambda_\sharp\nabla_\bphi\theta_e\delta\bphi\transpose\bV_e\bigg](\bG_0^\sharp)^{-1}\bLambda_\sharp\nabla_\bphi\theta_{e'},
  \end{align*}
  so item (D) above, \eqref{eqn:Ve_nuclear_norm}, and \eqref{eqn:V2} imply that the first and the second bounds of \eqref{eqn:kernel_vector_bound_scalar_multiple} and \eqref{eqn:kernel_vector_bound_scalar_multiple_deltaphi} with $k = 2$ are
  \begin{align*}
  &\max_{e'\in\calE}\bigg\|\sum_{e\in\calE}\alpha_e\bg_e^{(2)}((\bG_0^\sharp)^{-1}\bLambda_\sharp\nabla_\bphi\theta_{e}, (\bG_0^\sharp)^{-1}\bLambda_\sharp\nabla_\bphi\theta_{e'})\bigg\|_2\\
  &\quad\lesssim n^{-1/2}\bigg\|\scrP(\bG_0^\sharp)^{-1}\bigg[\sum_{e\in\calE}\alpha_e
  \bLambda_\sharp\nabla_\bphi\theta_e\{(\bG_0^\sharp)^{-1}\bLambda_\sharp\nabla_\bphi\theta_{e}\}\transpose\bV_e\bigg]\bigg\|_2\\
  &\quad\lesssim n^{-1/2}\bigg\| \sum_{e\in\calE}\alpha_e
  \bLambda_\sharp\nabla_\bphi\theta_e
  \nabla_\bphi\theta_e\transpose\bLambda_\sharp
  (\bG_0^\sharp)^{-1} \bV_e \bigg\|_2 \lesssim n^{-1/2}\max_{e\in\calE}|\alpha_e|\|\bV_e\|_* \lesssim \max_{e\in\calE}|\alpha_e|m^{-1/2}n^{-3/2},\\
  &\sum_{e'\in\calE}\bigg\|\sum_{e\in\calE}\alpha_e\bg_e^{(2)}((\bG_0^\sharp)^{-1}\bLambda_\sharp\nabla_\bphi\theta_{e}, (\bG_0^\sharp)^{-1}\bLambda_\sharp\nabla_\bphi\theta_{e'})\bigg\|_2^2\\
  &\quad\lesssim \bigg\|\scrP(\bG_0^\sharp)^{-1}\bigg[\sum_{e\in\calE}\alpha_e
  \bLambda_\sharp\nabla_\bphi\theta_e\{(\bG_0^\sharp)^{-1}\bLambda_\sharp\nabla_\bphi\theta_{e}\}\transpose\bV_e\bigg]\bigg\|_2^2\\
  &\quad\lesssim \bigg\| \sum_{e\in\calE}\alpha_e
  \bLambda_\sharp\nabla_\bphi\theta_e
  \nabla_\bphi\theta_e\transpose\bLambda_\sharp
  (\bG_0^\sharp)^{-1} \bV_e \bigg\|_2^2 \lesssim  \max_{e\in\calE}|\alpha_e|^2\|\bV_e\|_*^2 \lesssim \max_{e\in\calE}|\alpha_e|^2m^{-1}n^{-2},\\
  &\max_{e'\in\calE}\bigg\|\sum_{e\in\calE}\alpha_e\bg_e^{(2)}( (\bG_0^\sharp)^{-1}\bLambda_\sharp\nabla_\bphi\theta_{e'}, \delta\bphi)\bigg\|_2\\
  &\quad\lesssim n^{-1/2}\bigg\|\scrP(\bG_0^\sharp)^{-1}\bigg[\sum_{e\in\calE}\alpha_e
  \bLambda_\sharp\nabla_\bphi\theta_e \delta\bphi \transpose\bV_e\bigg]\bigg\|_2
  \lesssim n^{-1/2}\bigg\| \sum_{e\in\calE}\alpha_e
  \bLambda_\sharp\nabla_\bphi\theta_e
  \delta\bphi\transpose\bV_e \bigg\|_2\\
  &
  \quad \leq n^{-1/2}\max_e|\alpha_e|\sup_{\|\bw_1\|_2 = \|\bw_2\|_2 = 1}
  \sum_e (\bw_1\transpose\bLambda_\sharp\nabla_\bphi\theta_e)(\delta\bphi\transpose\bV_e\bw_2)\\
  &\quad\leq n^{-1/2}\max_e|\alpha_e|\sup_{\bw_1,\bw_2}\bigg\{\sum_e(\bw_1\transpose\bLambda_\sharp\nabla_\bphi\theta_e)^2\bigg\}^{1/2}\bigg\{\sum_e(\delta\bphi\transpose\bV_e\bw_2)^2\bigg\}^{1/2} \\
  &\quad  \lesssim n^{-1}\max_{e\in\calE}|\alpha_e|\|\bJ_{0n}\delta\bphi\|_2,\\
  &\sum_{e'\in\calE}\bigg\|\sum_{e\in\calE}\alpha_e\bg_e^{(2)}((\bG_0^\sharp)^{-1}\bLambda_\sharp\nabla_\bphi\theta_{e'}, \delta\bphi)\bigg\|_2^2\\
  &\quad\lesssim \bigg\|\scrP(\bG_0^\sharp)^{-1}\bigg[\sum_{e\in\calE}\alpha_e
  \bLambda_\sharp\nabla_\bphi\theta_e\delta\bphi\transpose\bV_e\bigg]\bigg\|_2^2 \lesssim \max_{e\in\calE}|\alpha_e|^2\|\bJ_{0n}\delta\bphi\|_2^2n^{-1},
  \end{align*}
  For the other two bounds in \eqref{eqn:kernel_vector_bound_scalar_multiple}, for any $j\in[d(d + 1)/2]$, we have
  \begin{align*}
  &\sum_{e\in\calE}\alpha_e\be_j\transpose\bg_e^{(2)}((\bG_0^\sharp)^{-1}\bLambda_\sharp\nabla_\bphi\theta_{e'}, (\bG_0^\sharp)^{-1}\bLambda_\sharp\nabla_\bphi\theta_{e'})\\
  &\quad = \{(\bG_0^\sharp)^{-1}\bLambda_\sharp\nabla_\bphi\theta_{e'}\}\transpose\bigg\{\sum_{e\in\calE} (\alpha_e\be_j\transpose\scrP(\bG_0^\sharp)^{-1}
    \bLambda_\sharp\nabla_\bphi\theta_e\bV_e)\bigg\}(\bG_0^\sharp)^{-1}\bLambda_\sharp\nabla_\bphi\theta_{e'},
  \end{align*}
  and the spectral norm of the matrix inside the bracket can be bounded as
  \begin{align*}
  \bigg\|\sum_{e\in\calE} (\alpha_e\be_j\transpose\scrP(\bG_0^\sharp)^{-1}
    \bLambda_\sharp\nabla_\bphi\theta_e\bV_e)\bigg\|_2
  & = \sup_{\|\bw\|_2 = 1}\sum_{e\in\calE} (\alpha_e\be_j\transpose\scrP(\bG_0^\sharp)^{-1}
    \bLambda_\sharp\nabla_\bphi\theta_e)(\bw\transpose\bV_e\bw)\\
  &\leq \bigg\{\sum_{e\in\calE} (\alpha_e\be_j\transpose\scrP(\bG_0^\sharp)^{-1}
    \bLambda_\sharp\nabla_\bphi\theta_e)^2\bigg\}^{1/2}\sup_{\|\bw\|_2 = 1}\bigg\{\sum_{e\in\calE}(\bw\transpose\bV_e\bw)^2\bigg\}^{1/2}\\
  &\leq n^{-1/2}\max_e|\alpha_e|\|\bJ_{0n}\bw\|_\infty\lesssim n^{-1/2}\max_e|\alpha_e|
  \end{align*}  
  by \eqref{eqn:V2}. Then, the third bound of \eqref{eqn:kernel_vector_bound_scalar_multiple} is derived as follows:
  \begin{align*}
  &\max_{e'\in\calE,j\in[d(d + 1)/2]}\bigg|\sum_{e\in\calE}\alpha_e\be_j\transpose\bg_e^{(2)}((\bG_0^\sharp)^{-1}\bLambda_\sharp\nabla_\bphi\theta_{e'}, (\bG_0^\sharp)^{-1}\bLambda_\sharp\nabla_\bphi\theta_{e'})\bigg|
  \lesssim n^{-3/2}\max_e|\balpha_e|.
  \end{align*}
  For the fourth bound with $k = 2$, we denote $\bv_{ee'} = \{(\bG_0^\sharp)^{-1}\bLambda_\sharp\nabla_\bphi\theta_{e'}\}\transpose\bV_e\{(\bG_0^\sharp)^{-1}\bLambda_\sharp\nabla_\bphi\theta_{e'}\}$ for notational simplicity. By the definition of $\bV_e$ and the definition of the edge-dependent local native norm \eqref{eqn:local_norm}, it is clear that $|\bw\transpose\bV_e\bw|\lesssim \langle\bw\rangle_e^2 + \|\bB_\bw\|_{\mathrm{F}}^2/(mn^2)$, so that 
  \begin{align}
  \label{eqn:V3}
  \sum_e|\bw\transpose\bV_e\bw|\lesssim \|\bw\|_2^2
  \end{align}
  for any vector $\bw$. This implies that $\max_{e'}\sum_e|v_{ee'}|\lesssim n^{-1}$ and $\max_e\sum_{e'}|v_{ee'}|\lesssim \|\bV_e\|_*\lesssim (mn)^{-1}$ by \eqref{eqn:Ve_nuclear_norm}. Then, we derive by Cauchy--Schwarz inequality that
  \begin{align*}
  &\sum_{e'\in\calE}\bigg|\sum_{e\in\calE}\alpha_e\be_j\transpose\bg_e^{(2)}((\bG_0^\sharp)^{-1}\bLambda_\sharp\nabla_\bphi\theta_{e'}, (\bG_0^\sharp)^{-1}\bLambda_\sharp\nabla_\bphi\theta_{e'})\bigg|^2\\
  &\quad\leq 
  \sum_{e'\in\calE}\bigg(\sum_e|v_{ee'}|\bigg)\bigg\{\sum_e \alpha_e^2|v_{ee'}|(\be_j\transpose\scrP(\bG_0^\sharp)^{-1}\bLambda_\sharp\nabla_\bphi\theta_e)^2\bigg\}\\
  &\quad\lesssim \frac{\max_e|\alpha_e|^2}{n}\sum_{e}(\be_j\transpose\scrP(\bG_0^\sharp)^{-1}\bLambda_\sharp\nabla_\bphi\theta_e)^2\sum_{e'}|v_{ee'}|\lesssim \frac{1}{mn^2}\max_e|\alpha_e|^2.
  \end{align*}

  \underline{For \eqref{eqn:kernel_vector_bound_scalar_multiple} and \eqref{eqn:kernel_vector_bound_scalar_multiple_deltaphi} with $k = 3$,} we first write by definition that
  \begin{align*}
  &\sum_{e\in\calE}\alpha_e\bg_e^{(3)}((\bG_0^\sharp)^{-1}\bLambda_\sharp\nabla_\bphi\theta_e, (\bG_0^\sharp)^{-1}\bLambda_\sharp\nabla_\bphi\theta_{e'})\\
  &\quad = \scrP(\bG_0^\sharp)^{-1}\bigg[
  \sum_{e\in\calE}\alpha_e \{\nabla_\bphi\theta_e\transpose\bLambda_\sharp(\bG_0^\sharp)^{-1}\bLambda_\sharp\nabla_\bphi\theta_e\} \bV_e 
  + \sum_{e\in\calE}\alpha_e \{\bV_e(\bG_0^\sharp)^{-1}\bLambda_\sharp\nabla_\bphi\theta_e\}\nabla_\bphi\theta_e\transpose\bLambda_\sharp
  \bigg]\\
  &\qquad\times(\bG_0^\sharp)^{-1}\bLambda_\sharp\nabla_\bphi\theta_{e'},\\
  &\sum_{e\in\calE}\alpha_e\bg_e^{(3)}((\bG_0^\sharp)^{-1}\bLambda_\sharp\nabla_\bphi\theta_{e'}, \delta\bphi)\\
  &\quad = \scrP(\bG_0^\sharp)^{-1}\bigg[
  \sum_{e\in\calE}\alpha_e (\nabla_\bphi\theta_e\transpose\bLambda_\sharp\delta\bphi) \bV_e 
  + \sum_{e\in\calE}\alpha_e (\bV_e\delta\bphi)\nabla_\bphi\theta_e\transpose\bLambda_\sharp
  \bigg]
  (\bG_0^\sharp)^{-1}\bLambda_\sharp\nabla_\bphi\theta_{e'}.
  \end{align*}
  By item (D) above, we need to bound
  \begin{align*}
  &\bigg\|
  \sum_{e\in\calE}\alpha_e \{\nabla_\bphi\theta_e\transpose\bLambda_\sharp(\bG_0^\sharp)^{-1}\bLambda_\sharp\nabla_\bphi\theta_e\} \bV_e 
  + \sum_{e\in\calE}\alpha_e \{\bV_e(\bG_0^\sharp)^{-1}\bLambda_\sharp\nabla_\bphi\theta_e\}\nabla_\bphi\theta_e\transpose\bLambda_\sharp
  \bigg\|_2,\\
  &\bigg\|\sum_{e\in\calE}\alpha_e (\nabla_\bphi\theta_e\transpose\bLambda_\sharp\delta\bphi) \bV_e 
  + \sum_{e\in\calE}\alpha_e (\bV_e\delta\bphi)\nabla_\bphi\theta_e\transpose\bLambda_\sharp\bigg\|_2.
  \end{align*}
  By \eqref{eqn:V2}, \eqref{eqn:V2_prime}, \eqref{eqn:V3}, and Cauchy--Schwarz inequality,
  \begin{align*}
  \bigg\|
  \sum_{e\in\calE}\alpha_e \{\nabla_\bphi\theta_e\transpose\bLambda_\sharp(\bG_0^\sharp)^{-1}\bLambda_\sharp\nabla_\bphi\theta_e\} \bV_e \bigg\|_2
  & \leq \sup_{\|\bw\|_2 = 1}\sum_{e\in\calE}|\alpha_e| \{\nabla_\bphi\theta_e\transpose\bLambda_\sharp(\bG_0^\sharp)^{-1}\bLambda_\sharp\nabla_\bphi\theta_e\} |\bw\transpose\bV_e\bw|\\
  & \lesssim \frac{\max_e|\alpha_e|}{n},\\
  \bigg\|
  \sum_{e\in\calE}\alpha_e (\nabla_\bphi\theta_e\transpose\bLambda_\sharp\delta\bphi) \bV_e \bigg\|_2
  & \leq \sup_{\|\bw\|_2 = 1}\sum_{e\in\calE}|\alpha_e| (\nabla_\bphi\theta_e\transpose\bLambda_\sharp\delta\bphi) |\bw\transpose\bV_e\bw|\\
  & \lesssim \frac{\max_e|\alpha_e|}{n^{1/2}}\|\bJ_{0n}\delta\bphi\|_\infty,\\
  \bigg\|\sum_{e\in\calE}\alpha_e \{\bV_e(\bG_0^\sharp)^{-1}\bLambda_\sharp\nabla_\bphi\theta_e\}\nabla_\bphi\theta_e\transpose\bLambda_\sharp
  \bigg\|_2
  & \leq \sup_{\|\bw\|_2 = 1}\sum_{e\in\calE}|\alpha_e|\{\bw\transpose\bV_e(\bG_0^\sharp)^{-1}\bLambda_\sharp\nabla_\bphi\theta_e\}\nabla_\bphi\theta_e\transpose\bLambda_\sharp\bw\\
  & \leq \bigg[\sup_{\|\bw\|_2 = 1}\sum_e|\alpha_e|^2\{\bw\transpose\bV_e(\bG_0^\sharp)^{-1}\bLambda_\sharp\nabla_\bphi\theta_e\}^2\bigg]^{1/2}\\
  &\quad\times \bigg\{\sup_{\|\bw\|_2 = 1}\sum_e\bw\transpose\bLambda_\sharp\nabla_\bphi\theta_e\nabla_\bphi\theta_e\transpose\bLambda_\sharp\bw\bigg\}^{1/2}\\
  &\lesssim \frac{\max_e|\alpha_e|}{\sqrt{mn}}\max_e\|\bJ_{0n}(\bG_0^\sharp)^{-1}\bJ_{0n}\transpose\widetilde{\bLambda}_\sharp \nabla_\bpsi\theta_e\|_\infty\|\bJ_{0n}\bw\|_2\\
  &\lesssim \frac{\max_e|\alpha_e|}{\sqrt{m}n},\\
   \bigg\|\sum_{e\in\calE}\alpha_e (\bV_e\delta\bphi)\nabla_\bphi\theta_e\transpose\bLambda_\sharp
  \bigg\|_2
  & \leq \sup_{\|\bw\|_2 = 1}\sum_{e\in\calE}|\alpha_e| |\bw\transpose\bV_e\delta\bphi||\nabla_\bphi\theta_e\transpose\bLambda_\sharp\bw|\\
  & \leq \bigg[\sup_{\|\bw\|_2 = 1}\sum_e|\alpha_e|^2(\bw\transpose\bV_e\delta\bphi)^2\bigg]^{1/2}\\
  &\quad\times \bigg\{\sup_{\|\bw\|_2 = 1}\sum_e\bw\transpose\bLambda_\sharp\nabla_\bphi\theta_e\nabla_\bphi\theta_e\transpose\bLambda_\sharp\bw\bigg\}^{1/2}\\
  &\lesssim \frac{\max_e|\alpha_e|}{\sqrt{mn}}\|\bJ_{0n}\delta\bphi\|_\infty\|\bJ_{0n}\bw\|_2 \lesssim \frac{\max_e|\alpha_e|}{\sqrt{mn}}\|\bJ_{0n}\delta\bphi\|_\infty.
  \end{align*}
  Then, the first and the second bounds of \eqref{eqn:kernel_vector_bound_scalar_multiple} and \eqref{eqn:kernel_vector_bound_scalar_multiple_deltaphi} with $k = 3$ follow from item (D) above. For the third and fourth bounds of \eqref{eqn:kernel_vector_bound_scalar_multiple}, we have
  \begin{align*}
  &\sum_{e\in\calE}\alpha_e\bg_e^{(3)}((\bG_0^\sharp)^{-1}\bLambda_\sharp\nabla_\bphi\theta_{e'}, (\bG_0^\sharp)^{-1}\bLambda_\sharp\nabla_\bphi\theta_{e'})\\
  &\quad = 2\sum_{e\in\calE}\{\scrP(\bG_0^\sharp)^{-1}\bV_e(\bG_0^\sharp)^{-1}\bLambda_\sharp\nabla_\bphi\theta_{e'}\}\{\nabla_\bphi\theta_e\transpose\bLambda_\sharp(\bG_0^\sharp)^{-1}\bLambda_\sharp\nabla_\bphi\theta_{e'}\}.
  \end{align*}
  For any $j\in[d(d + 1)/2]$, by Cauchy--Schwarz inequality, 
  \begin{align*}
  &\bigg|\be_j\transpose\sum_{e\in\calE}\alpha_e\bg_e^{(3)}((\bG_0^\sharp)^{-1}\bLambda_\sharp\nabla_\bphi\theta_{e'}, (\bG_0^\sharp)^{-1}\bLambda_\sharp\nabla_\bphi\theta_{e'})\bigg| \\
  &\quad\leq 2\max_{e\in\calE}|\alpha_e|\bigg[\sum_{e\in\calE}\{\be_j\transpose\scrP(\bG_0^\sharp)^{-1}\bV_e(\bG_0^\sharp)^{-1}\bLambda_\sharp\nabla_\bphi\theta_{e'}\}^2\bigg]^{1/2}
  \bigg[\sum_{e\in\calE}\{\nabla_\bphi\theta_e\transpose\bLambda_\sharp(\bG_0^\sharp)^{-1}\bLambda_\sharp\nabla_\bphi\theta_{e'}\}^2\bigg]^{1/2}.
  \end{align*}
  For any vector $\delta\bphi\in\mathbb{R}^{d_\bphi}$, by the derivation of Step 2(d) in the proof of Lemma \ref{lemma:Q1_fluctuation}, $\sum_{e}\|\bV_e\delta\bphi\|_2^2 \lesssim n^{-1}\|\delta\bphi\|_2^2$,
  so that by \eqref{eqn:V2},
  \begin{align*}
  &\max_{e'\in\calE}\bigg|\be_j\transpose\sum_{e\in\calE}\alpha_e\bg_e^{(3)}((\bG_0^\sharp)^{-1}\bLambda_\sharp\nabla_\bphi\theta_{e'}, (\bG_0^\sharp)^{-1}\bLambda_\sharp\nabla_\bphi\theta_{e'})\bigg|\\
  &\quad\leq 2\max_{e\in\calE}|\alpha_e|\max_{e'\in\calE}\bigg[\sum_{e\in\calE}\{\be_j\transpose\scrP(\bG_0^\sharp)^{-1}\bV_e(\bG_0^\sharp)^{-1}\bLambda_\sharp\nabla_\bphi\theta_{e'}\}^2\bigg]^{1/2}\\
  &\qquad\times 
  \bigg[\{(\bG_0^\sharp)^{-1}\bLambda_\sharp\nabla_\bphi\theta_{e'}\}\transpose\bigg\{\sum_{e\in\calE}\bLambda_\sharp\nabla_\bphi\theta_e \nabla_\bphi\theta_e\transpose\bLambda_\sharp\bigg\}(\bG_0^\sharp)^{-1}\bLambda_\sharp\nabla_\bphi\theta_{e'} \bigg]^{1/2}\\
  &\quad\lesssim \frac{\max_{e\in\calE}|\alpha_e|}{\sqrt{m}n}\|\bJ_{0n}(\bG_0^\sharp)^{-1}\scrP\transpose\be_j\|_\infty \|\bJ_{0n}(\bG_0^\sharp)^{-1}\bJ_{0n}\transpose \widetilde{\bLambda}_\sharp\nabla_\bpsi\theta_{e'}\|_\infty\lesssim \frac{\max_e|\alpha_e|}{\sqrt{m}n^{3/2}},\\
  &\sum_{e'\in\calE}\bigg|\be_j\transpose\sum_{e\in\calE}\alpha_e\bg_e^{(3)}((\bG_0^\sharp)^{-1}\bLambda_\sharp\nabla_\bphi\theta_{e'}, (\bG_0^\sharp)^{-1}\bLambda_\sharp\nabla_\bphi\theta_{e'})\bigg|^2\\
  &\quad\leq 4\max_{e\in\calE}|\alpha_e|^2\sum_{e'\in\calE}\bigg[\sum_{e\in\calE}\{\be_j\transpose\scrP(\bG_0^\sharp)^{-1}\bV_e(\bG_0^\sharp)^{-1}\bLambda_\sharp\nabla_\bphi\theta_{e'}\}^2\bigg]\\
  &\qquad\times 
  \bigg[\{(\bG_0^\sharp)^{-1}\bLambda_\sharp\nabla_\bphi\theta_{e'}\}\transpose\bigg\{\sum_{e\in\calE}\bLambda_\sharp\nabla_\bphi\theta_e \nabla_\bphi\theta_e\transpose\bLambda_\sharp\bigg\}(\bG_0^\sharp)^{-1}\bLambda_\sharp\nabla_\bphi\theta_{e'} \bigg]\\
  &\quad\lesssim \frac{\max_{e\in\calE}|\alpha_e|^2}{n}\sum_{e\in\calE}
  \be_j\transpose\scrP(\bG_0^\sharp)^{-1}\bV_e(\bG_0^\sharp)^{-1}\bigg\{\sum_{e'\in\calE}\bLambda_\sharp\nabla_\bphi\theta_{e'}\nabla_\bphi\theta_{e'}\transpose\bLambda_\sharp\bigg\}
  (\bG_0^\sharp)^{-1}\bV_e(\bG_0^\sharp)^{-1}\scrP\transpose\be_j\\
  &\quad\lesssim \frac{\max_{e\in\calE}|\alpha_e|^2}{n}\sum_{e\in\calE}\|\bV_e(\bG_0^\sharp)^{-1}\scrP\transpose\be_j\|_2^2\lesssim \frac{\max_{e\in\calE}|\alpha_e|^2}{n^2}.
  \end{align*}
  
  \underline{For the \eqref{eqn:kernel_vector_bound_scalar_multiple_k4} with $k = 4$,} by definition, for any $j\in[d(d + 1)/2]$,
  \begin{align*}
  &\sum_{e\in\calE}\alpha_e\be_j\transpose\bg_e^{(4)}((\bG_0^\sharp)^{-1}\bLambda_\sharp\nabla_\bphi\theta_{e}, (\bG_0^\sharp)^{-1}\bLambda_\sharp\nabla_\bphi\theta_{e'})\\
  &\quad = \sum_{e\in\calE}\alpha_e \tau_e((\bG_0^\sharp)^{-1}\bLambda_\sharp\nabla_\bphi\theta_{e}, (\bG_0^\sharp)^{-1}\bLambda_\sharp\nabla_\bphi\theta_{e'}, (\bG_0^\sharp)^{-1}\scrP\transpose\be_j)\\
  &\quad = \left\{(\bG_0^\sharp)^{-1}\bLambda_\sharp\nabla_\bphi\theta_{e'}\right\}\transpose\sum_{e\in\calE}\alpha_e\bT_e((\bG_0^\sharp)^{-1}\scrP\transpose\be_j)(\bG_0^\sharp)^{-1}\bLambda_\sharp\nabla_\bphi\theta_e,
  \end{align*}
  so that by (T3) and Cauchy--Schwarz inequality, 
  \begin{align*}
  &\max_{e',j}\bigg|\sum_{e\in\calE}\alpha_e\be_j\transpose\bg_e^{(4)}((\bG_0^\sharp)^{-1}\bLambda_\sharp\nabla_\bphi\theta_{e}, (\bG_0^\sharp)^{-1}\bLambda_\sharp\nabla_\bphi\theta_{e'})\bigg|\\
  &\quad \leq \max_{e'}\|\bLambda_\sharp\nabla_\bphi\theta_{e'}\|_2\sum_{e\in\calE}|\alpha_e|\nu_e((\bG_0^\sharp)^{-1}\scrP\transpose\be_j)\|(\bG_0^\sharp)^{-1}\bLambda_\sharp\nabla_\bphi\theta_e\|_2\\
  &\quad\lesssim \frac{\max_e|\alpha_e|}{\sqrt{n}}\sqrt{m}n\bigg\{\sum_{e\in\calE}\nu_e((\bG_0^\sharp)^{-1}\scrP\transpose\be_j)^2\bigg\}^{1/2}\max_{e\in\calE}\|(\bG_0^\sharp)^{-1}\bLambda_\sharp\nabla_\bphi\theta_e\|_2\\
  &\quad\lesssim \frac{\max_e|\alpha_e|}{\sqrt{m}n}
  \end{align*}
  and by item (D) above,
  \begin{align*}
  &\sum_{e',j}\bigg|\sum_{e\in\calE}\alpha_e\be_j\transpose\bg_e^{(4)}((\bG_0^\sharp)^{-1}\bLambda_\sharp\nabla_\bphi\theta_{e}, (\bG_0^\sharp)^{-1}\bLambda_\sharp\nabla_\bphi\theta_{e'})\bigg|^2\\
  &\quad \lesssim \max_j \sum_{e'}\bigg[\bigg\{\sum_{e\in\calE}\alpha_e\bT_e((\bG_0^\sharp)^{-1}\scrP\transpose\be_j)(\bG_0^\sharp)^{-1}\bLambda_\sharp\nabla_\bphi\theta_e\bigg\}\transpose(\bG_0^\sharp)^{-1}\bLambda_\sharp\nabla_\bphi\theta_{e'}\bigg]^2\\
  &\quad \lesssim \bigg\|\sum_{e\in\calE}\alpha_e\bT_e((\bG_0^\sharp)^{-1}\scrP\transpose\be_j)(\bG_0^\sharp)^{-1}\bLambda_\sharp\nabla_\bphi\theta_e\bigg\|_2^2\\
  &\quad \lesssim \max_e|\alpha_e|^2mn^2\bigg\{\sum_{e\in\calE}\nu_e((\bG_0^\sharp)^{-1}\scrP\transpose\be_j)^2\bigg\}\max_e\|(\bG_0^\sharp)^{-1}\bLambda_\sharp\nabla_\bphi\theta_e\|_2^2\lesssim \frac{\max_e|\alpha_e|^2}{mn}.
  \end{align*}
\end{enumerate}

\medskip\noindent\textbf{$\blacksquare$
Step 3. A truncation argument.
} For any $c > 0$ and $\delta > 0$, let $K_c > 0$ be a sufficient large constant that may depend on $c$, such that $\prob(\max_e|\ell'_e|\leq K_c\log n)\geq 1 - O(n^{-2c - 6\delta - 19})$ by \eqref{eqn:max_ellprime_bound}. For any sum of the form $\sum_{e,e',e''\in\calE}(\chi_e - \expect_0\chi_e)(\chi_{e'} - \expect_0\chi_{e'})(\chi_{e''} - \expect_0\chi_{e''})\bbeta_{e,e',e''}$ with deterministic $\max_{e,e',e''}\|\bbeta_{e,e',e''}\|_2\leq n^{\delta}$ and $\chi_e\in\{\ell'''_e(\ell'_e)^2, \ell''_e(\ell'_e)^2, (\ell'_e)^3, \ell'''_e\ell'_e, \ell''_e\ell'_e, (\ell'_e)^2\}$ appears below, we replace $\chi_e$ by $\chi_e\mathbbm{1}(|\ell'_e|\leq K_c\log n)$ and likewise for $\chi_{e'}$ and $\chi_{e''}$. Over the event $\{\max_e|\ell'_e|\leq K_c\log n\}$, the random variable $\chi_e$ is not affected as $\chi_e = \chi_e\mathbbm{1}(|\ell'_e|\leq K_c\log n)$. By Cauchy--Schwarz inequality, over such event, the truncation error satisfies
\begin{align*}
&\sum_{e_1,e_2,e_3\in\calE}\bigg[\prod_{k = 1}^3|\{\chi_{e_k} - \expect_0\chi_{e_k}\mathbbm{1}(|\ell'_{e_k}| \leq K_c\log n)\}|^{1 - \delta_k} |\expect_0\{\chi_{e_k}\mathbbm{1}(|\ell'_{e_k}| > K_c\log n)\}|^{\delta_k}\bigg]\left\|\bbeta_{e_1e_2e_3}\right\|_2^3\\
&\quad\leq (mn^2)^{3}n^{3\delta}(2K_c\log n)^3\max_{e\in\calE}(\expect_0\chi_e^2)^{(\delta_1 + \delta_2 + \delta_3)/2}\prob(|\ell'_e| > K_c\log n)^{(\delta_1 + \delta_2 + \delta_3)/2} = O(n^{-c}).
\end{align*}
for $\delta_1,\delta_2,\delta_3\in\{0,1\}$ and $\delta_1 + \delta_2 + \delta_3\in\{1,2,3\}$. 
The truncated and centered variables are independent, mean zero, and bounded by $CK_c^3(\log n)^3$ for some constant $C > 0$, so they satisfy the moment condition of Lemma \ref{lemma:matrix_Bernstein} with $K = C(K_c\log n)^3$ and $v_k = C$. 

\medskip\noindent\textbf{$\blacksquare$
Step 4. Bounding $\bq_1^{(k)},\bq_{21}^{(k)},\bq_{22}^{(k)},\bq_{23}^{(k)}$. 
} For $\bq_1^{(k)}$, it can be bounded using the matrix Bernstein's inequality (Lemma \ref{lemma:matrix_Bernstein} and Theorem 6.2 in \cite{tropp2012user}), together with the truncation argument in Step 3 and \eqref{eqn:kernel_bound_matrix} in Step 2, as
\begin{align*}
\|\bq_1^{(k)}\|_2
& = \Optilde\bigg[\bigg\{\sum_e\|\bg_e^{(k)}((\bG_0^\sharp)^{-1}\bLambda_\sharp\nabla_\bphi\theta_e, (\bG_0^\sharp)^{-1}\bLambda_\sharp\nabla_\bphi\theta_e)\|_2^2\bigg\}^{1/2}\sqrt{\log n}\\
&\quad + \max_e\|\bg_e^{(k)}((\bG_0^\sharp)^{-1}\bLambda_\sharp\nabla_\bphi\theta_e, (\bG_0^\sharp)^{-1}\bLambda_\sharp\nabla_\bphi\theta_e)\|_2(\log n)^3\bigg] = \Optilde\bigg(\frac{1}{n^{3/2}}\bigg).
\end{align*}
For $\bq_{21}^{(k)}$, note that by \eqref{eqn:kernel_bound_vector} and \eqref{eqn:kernel_bound_matrix}, 
\begin{align*}
&\sum_e\bigg\|
\sum_{e'\in\calE\backslash\{e\}}\expect_0(\ell'_{e'})^2\bg_e^{(k)}((\bG_0^\sharp)^{-1}\bLambda_\sharp\nabla_\bphi\theta_{e'}, (\bG_0^\sharp)^{-1}\bLambda_\sharp\nabla_\bphi\theta_{e'})
\bigg\|_2^2\\
&\quad\lesssim \sum_{e\in\calE}\bigg\|
\bg_e^{(k)}\bigg[(\bG_0^\sharp)^{-1}\bLambda_\sharp
\sum_{e'\in\calE\backslash\{e\}}\expect_0(\ell'_{e'})^2\nabla_\bphi\theta_{e'}\nabla_\bphi\theta_{e'}\transpose \bLambda_\sharp(\bG_0^\sharp)^{-1}
\bigg]
\bigg\|_2^2\\
&\qquad + \sum_{e\in\calE}\left\|\expect_0(\ell'_{e})^2\bg_e^{(k)}((\bG_0^\sharp)^{-1}\bLambda_\sharp\nabla_\bphi\theta_{e}, (\bG_0^\sharp)^{-1}\bLambda_\sharp\nabla_\bphi\theta_{e})\right\|_2^2\\
&\quad\lesssim \sum_{e\in\calE}\left\|\bg_e^{(k)}[(\bG_0^\sharp)^{-1}]\right\|_2^2 + \sum_{e\in\calE}\left\| \bg_e^{(k)}((\bG_0^\sharp)^{-1}\bLambda_\sharp\nabla_\bphi\theta_{e}, (\bG_0^\sharp)^{-1}\bLambda_\sharp\nabla_\bphi\theta_{e})\right\|_2^2 \lesssim n^{-2},\\
&\max_e\bigg\|
\sum_{e'\in\calE\backslash\{e\}}\expect_0(\ell'_{e'})^2\bg_e^{(k)}((\bG_0^\sharp)^{-1}\bLambda_\sharp\nabla_\bphi\theta_{e'}, (\bG_0^\sharp)^{-1}\bLambda_\sharp\nabla_\bphi\theta_{e'})
\bigg\|_2 \\
&\quad\leq \max_{e\in\calE}\bigg\|
\bg_e^{(k)}\bigg[(\bG_0^\sharp)^{-1}\bLambda_\sharp
\max_{e'\in\calE\backslash\{e\}}\expect_0(\ell'_{e'})^2\nabla_\bphi\theta_{e'}\nabla_\bphi\theta_{e'}\transpose \bLambda_\sharp(\bG_0^\sharp)^{-1}
\bigg]
\bigg\|_2 \\
&\qquad + \max_{e\in\calE}\left\|\expect_0(\ell'_{e})^2\bg_e^{(k)}((\bG_0^\sharp)^{-1}\bLambda_\sharp\nabla_\bphi\theta_{e}, (\bG_0^\sharp)^{-1}\bLambda_\sharp\nabla_\bphi\theta_{e})\right\|_2 \\
&\quad\lesssim \max_{e\in\calE}\left\|\bg_e^{(k)}[(\bG_0^\sharp)^{-1}]\right\|_2 + \max_{e\in\calE}\left\| \bg_e^{(k)}((\bG_0^\sharp)^{-1}\bLambda_\sharp\nabla_\bphi\theta_{e}, (\bG_0^\sharp)^{-1}\bLambda_\sharp\nabla_\bphi\theta_{e})\right\|_2  \lesssim n^{-2}.
\end{align*} 
Then, by matrix Bernstein, we obtain
\begin{align*}
\|\bq_{21}^{(k)}\|_2
& = \Optilde\bigg[\bigg\{\sum_e\bigg\|
\sum_{e'\in\calE\backslash\{e\}}\expect_0(\ell'_{e'})^2\bg_e^{(k)}((\bG_0^\sharp)^{-1}\bLambda_\sharp\nabla_\bphi\theta_{e'}, (\bG_0^\sharp)^{-1}\bLambda_\sharp\nabla_\bphi\theta_{e'})
\bigg\|_2^2\bigg\}^{1/2}\sqrt{\log n}\\
&\quad + \max_e\bigg\|
\sum_{e'\in\calE\backslash\{e\}}\expect_0(\ell'_{e'})^2\bg_e^{(k)}((\bG_0^\sharp)^{-1}\bLambda_\sharp\nabla_\bphi\theta_{e'}, (\bG_0^\sharp)^{-1}\bLambda_\sharp\nabla_\bphi\theta_{e'})
\bigg\|_2(\log n)^4\bigg] = \Optilde\bigg(\frac{\sqrt{\log n}}{n}\bigg),
\end{align*}
and the same reasoning implies that $\|\bq_{22}^{(k)}\|_2 = \Optilde(n^{-1}\sqrt{\log n})$. For $\bq_{23}^{(k)}$, by Theorem \ref{thm:decoupling}, it is sufficient to consider
\begin{align*}
&\bigg\|\sum_{e\neq e'}(\xi_e - \expect_0\xi_e)\{(\bar{\ell}'_{e'})^2 - \expect_0(\bar{\ell}'_{e'})^2\}\bg_e^{(k)}((\bG_0^\sharp)^{-1}\bLambda_\sharp\nabla_\bphi\theta_{e'}, (\bG_0^\sharp)^{-1}\bLambda_\sharp\nabla_\bphi\theta_{e'})\bigg\|_2\\
&\quad = \bigg\|\sum_{e\in\calE}(\xi_e - \expect_0\xi_e)
\bigg\{
\bg_e^{(k)}\bigg[\sum_{e'\in\calE}\{(\bar{\ell}'_{e'})^2 - \expect_0(\bar{\ell}'_{e'})^2\}
(\bG_0^\sharp)^{-1}\bLambda_\sharp\nabla_\bphi\theta_{e'}\nabla_\bphi\theta_{e'}\transpose\bLambda_\sharp(\bG_0^\sharp)^{-1}
\bigg]\\
&\qquad - \{(\bar{\ell}'_{e})^2 - \expect_0(\bar{\ell}'_{e})^2\}\bg_e^{(k)}((\bG_0^\sharp)^{-1}\bLambda_\sharp\nabla_\bphi\theta_{e}, (\bG_0^\sharp)^{-1}\bLambda_\sharp\nabla_\bphi\theta_{e})\bigg\}
\bigg\|_2,
\end{align*}
where $(\bar{\ell}'_e:e\in\calE)$ is an independent copy of $(\ell'_e:e\in\calE)$. 
Let $\bar{\bs}$ be an independent copy of $\widehat{\bs}$ computed based on $(\bar{\ell}'_e:e\in\calE)$, namely, $\bar{\bs} = (\bG_0^\sharp)^{-1}\bLambda_\sharp\sum_e\bar{\ell}'_e\nabla_\bphi\theta_e$.
By \eqref{eqn:max_ellprime_bound}, \eqref{eqn:kernel_bound_vector}, \eqref{eqn:kernel_bound_matrix}, and matrix Bernstein, we have
\begin{align*}
&\max_{e\in\calE}\bigg\|
\bg_e^{(k)}\bigg[\sum_{e'\in\calE}\{(\bar{\ell}'_{e'})^2 - \expect_0(\bar{\ell}'_{e'})^2\}
(\bG_0^\sharp)^{-1}\bLambda_\sharp\nabla_\bphi\theta_{e'}\nabla_\bphi\theta_{e'}\transpose\bLambda_\sharp(\bG_0^\sharp)^{-1}
\bigg]\\
&\quad - \{(\bar{\ell}'_{e})^2 - \expect_0(\bar{\ell}'_{e})^2\}\bg_e^{(k)}((\bG_0^\sharp)^{-1}\bLambda_\sharp\nabla_\bphi\theta_{e}, (\bG_0^\sharp)^{-1}\bLambda_\sharp\nabla_\bphi\theta_{e})\bigg\|_2\\
&\quad\lesssim n^{-3/2}\bigg\|
\sum_{e'\in\calE}\{(\bar{\ell}'_{e'})^2 - \expect_0(\bar{\ell}'_{e'})^2\}
(\bG_0^\sharp)^{-1}\bLambda_\sharp\nabla_\bphi\theta_{e'}\nabla_\bphi\theta_{e'}\transpose\bLambda_\sharp(\bG_0^\sharp)^{-1}\bigg\|_2\\
&\qquad + \Optilde\{(\log n)^2\}
 \max_{e\in\calE}\left\|\bg_e^{(k)}((\bG_0^\sharp)^{-1}\bLambda_\sharp\nabla_\bphi\theta_{e}, (\bG_0^\sharp)^{-1}\bLambda_\sharp\nabla_\bphi\theta_{e})\right\|_2 = \Optilde\bigg(\frac{\sqrt{\log n}}{n^2}\bigg),\\
&\sum_{e\in\calE}\bigg\|
\bg_e^{(k)}\bigg[\sum_{e'\in\calE}\{(\bar{\ell}'_{e'})^2 - \expect_0(\bar{\ell}'_{e'})^2\}
(\bG_0^\sharp)^{-1}\bLambda_\sharp\nabla_\bphi\theta_{e'}\nabla_\bphi\theta_{e'}\transpose\bLambda_\sharp(\bG_0^\sharp)^{-1}
\bigg]\\
&\quad - \{(\bar{\ell}'_{e})^2 - \expect_0(\bar{\ell}'_{e})^2\}\bg_e^{(k)}((\bG_0^\sharp)^{-1}\bLambda_\sharp\nabla_\bphi\theta_{e}, (\bG_0^\sharp)^{-1}\bLambda_\sharp\nabla_\bphi\theta_{e})\bigg\|_2^2 = \Optilde\bigg(\frac{\log n}{n^2}\bigg).
\end{align*}
This further implies by Lemma \ref{lemma:conditioning_argument} that
\begin{align*}
  \bigg\|\sum_{e\neq e'}(\xi_e - \expect_0\xi_e)\{(\bar{\ell}'_{e'})^2 - \expect_0(\bar{\ell}'_{e'})^2\}\bg_e^{(k)}((\bG_0^\sharp)^{-1}\bLambda_\sharp\nabla_\bphi\theta_{e'}, (\bG_0^\sharp)^{-1}\bLambda_\sharp\nabla_\bphi\theta_{e'})\bigg\|_2 = \Optilde\bigg(\frac{\log n}{n}\bigg)
\end{align*}

\medskip\noindent\textbf{$\blacksquare$ Step 5. Bounding $\bq_{31}^{(k)}$ and $\bq_{32}^{(k)}$.}
For $\bq_{31}^{(k)}$, by matrix Bernstein, \eqref{eqn:kernel_bound_vector}, \eqref{eqn:kernel_vector_bound_scalar_multiple}, and \eqref{eqn:kernel_vector_bound_scalar_multiple_k4}, we have
\begin{align*}
\|\bq_{31}^{(k)}\|_2
& = 2\bigg\|\sum_{e''}\ell'_{e''}\sum_{e\in\calE\backslash\{e''\}}\expect_0\xi_e\ell'_e\bg_e^{(k)}((\bG_0^\sharp)^{-1}\bLambda_\sharp\nabla_\bphi\theta_e, (\bG_0^\sharp)^{-1}\bLambda_\sharp\nabla_\bphi\theta_{e''})\bigg\|_2\\
&\leq 2\bigg\|\sum_{e''}\ell'_{e''}\bigg\{\sum_{e\in\calE}\expect_0\xi_e\ell'_e\bg_e^{(k)}((\bG_0^\sharp)^{-1}\bLambda_\sharp\nabla_\bphi\theta_e, (\bG_0^\sharp)^{-1}\bLambda_\sharp\nabla_\bphi\theta_{e''})\bigg\}\bigg\|_2\\
&\quad + 2\bigg\|\sum_{e }\ell'_{e } (\expect_0\xi_e\ell'_e)\bg_e^{(k)}((\bG_0^\sharp)^{-1}\bLambda_\sharp\nabla_\bphi\theta_e, (\bG_0^\sharp)^{-1}\bLambda_\sharp\nabla_\bphi\theta_{e''})\bigg\|_2\\
& = \Optilde\bigg[
\bigg\{\sum_{e''\in\calE}\bigg\|\sum_{e\in\calE}\expect_0\xi_e\ell'_e\bg_e^{(k)}((\bG_0^\sharp)^{-1}\bLambda_\sharp\nabla_\bphi\theta_e, (\bG_0^\sharp)^{-1}\bLambda_\sharp\nabla_\bphi\theta_{e''})\bigg\|_2^2\bigg\}^{1/2}\sqrt{\log n}\\
&\quad + \max_{e''\in\calE} \bigg\|\sum_{e\in\calE}\expect_0\xi_e\ell'_e\bg_e^{(k)}((\bG_0^\sharp)^{-1}\bLambda_\sharp\nabla_\bphi\theta_e, (\bG_0^\sharp)^{-1}\bLambda_\sharp\nabla_\bphi\theta_{e''})\bigg\|_2 (\log n)^4
\bigg]\\
&\quad + \Optilde\bigg[
\bigg\{\sum_{e\in\calE}\left\|(\expect_0\xi_e\ell'_e)\bg_e^{(k)}((\bG_0^\sharp)^{-1}\bLambda_\sharp\nabla_\bphi\theta_e, (\bG_0^\sharp)^{-1}\bLambda_\sharp\nabla_\bphi\theta_{e''})\right\|_2^2\bigg\}^{1/2}\sqrt{\log n}\\
&\qquad + \max_e\left\|(\expect_0\xi_e\ell'_e)\bg_e^{(k)}((\bG_0^\sharp)^{-1}\bLambda_\sharp\nabla_\bphi\theta_e, (\bG_0^\sharp)^{-1}\bLambda_\sharp\nabla_\bphi\theta_{e''})\right\|_2(\log n)^2
\bigg] = \Optilde\bigg(\sqrt{\frac{\log n}{n}}\bigg). 
\end{align*}
For $\bq_{32}^{(k)}$, by Theorem \ref{thm:decoupling}, with $(\bar{\ell}'_e:e\in\calE)$ being an independent copy of $(\ell'_e:e\in\calE)$, it is sufficient to consider
\begin{align*}
&\bigg\|\sum_{e\neq e''}(\xi_e^{(k)}\ell'_e - \expect_0\xi_e^{(k)}\ell'_e)\bar{\ell}'_{e''}\bg_e^{(k)}((\bG_0^\sharp)^{-1}\bLambda_\sharp\nabla_\bphi\theta_e, (\bG_0^\sharp)^{-1}\bLambda_\sharp\nabla_\bphi\theta_{e''})\bigg\|_2\\
&\quad\leq 
\bigg\|\sum_{e\in\calE}(\xi_e^{(k)}\ell'_e - \expect_0\xi_e^{(k)}\ell'_e)\bigg\{\sum_{e''\in\calE}\bar{\ell}'_{e''}\bg_e^{(k)}((\bG_0^\sharp)^{-1}\bLambda_\sharp\nabla_\bphi\theta_e, (\bG_0^\sharp)^{-1}\bLambda_\sharp\nabla_\bphi\theta_{e''})\bigg\}\bigg\|_2\\
&\qquad + \bigg\|\sum_{e\in\calE}(\xi_e^{(k)}\ell'_e - \expect_0\xi_e^{(k)}\ell'_e)\bar{\ell}'_{e}\bg_e^{(k)}((\bG_0^\sharp)^{-1}\bLambda_\sharp\nabla_\bphi\theta_e, (\bG_0^\sharp)^{-1}\bLambda_\sharp\nabla_\bphi\theta_{e})\bigg\|\\
&\quad\leq 
\bigg\|\sum_{e\in\calE}(\xi_e^{(k)}\ell'_e - \expect_0\xi_e^{(k)}\ell'_e)\bg_e^{(k)}((\bG_0^\sharp)^{-1}\bLambda_\sharp\nabla_\bphi\theta_e, \bar{\bs})\bigg\|_2\\
&\qquad + \bigg\|\sum_{e\in\calE}(\xi_e^{(k)}\ell'_e - \expect_0\xi_e^{(k)}\ell'_e)\bar{\ell}'_{e}\bg_e^{(k)}((\bG_0^\sharp)^{-1}\bLambda_\sharp\nabla_\bphi\theta_e, (\bG_0^\sharp)^{-1}\bLambda_\sharp\nabla_\bphi\theta_{e})\bigg\|.
\end{align*}
Then, by matrix Bernstein, Lemma \ref{lemma:conditioning_argument},  \eqref{eqn:J0n_s_hat_bound}, and \eqref{eqn:kernel_bound_vector}, we have
\begin{align*}
\|\bq_{32}^{(k)}\|_2
& = \Optilde\bigg[\bigg\{\sum_{e\in\calE}\|\bg_e^{(k)}((\bG_0^\sharp)^{-1}\bLambda_\sharp\nabla_\bphi\theta_e, \bar{s})\|_2^2\bigg\}^{1/2}\sqrt{\log n} + \max_e\|\bg_e^{(k)}((\bG_0^\sharp)^{-1}\bLambda_\sharp\nabla_\bphi\theta_e, \bar{s})\|_2(\log n)^3\bigg]\\
&\quad + \Optilde\bigg[\bigg\{\sum_{e\in\calE}\bigg\|
\bar{\ell}'_{e}\bg_e^{(k)}((\bG_0^\sharp)^{-1}\bLambda_\sharp\nabla_\bphi\theta_e, (\bG_0^\sharp)^{-1}\bLambda_\sharp\nabla_\bphi\theta_{e})
\bigg\|_2^2\bigg\}^{1/2}\sqrt{\log n}\\
&\qquad + \max_{e\in\calE}\bigg\|\bar{\ell}'_{e}\bg_e^{(k)}((\bG_0^\sharp)^{-1}\bLambda_\sharp\nabla_\bphi\theta_e, (\bG_0^\sharp)^{-1}\bLambda_\sharp\nabla_\bphi\theta_{e})\bigg\|(\log n)^3\bigg] = \Optilde\bigg(\frac{\log n}{n^{3/2}}\bigg).
\end{align*}

\medskip\noindent\textbf{Step 6. Bounding $\bq_{41}^{(k)}$. }
Note that $\bq_{41}^{(4)}$ vanishes so it is sufficient to consider $k\in\{1,2,3\}$. By Theorem \ref{thm:decoupling}, it is sufficient to consider
\begin{align*}
&\sum_{e'\in\calE}\ell'_{e'}\bigg\{\sum_{e''\in\calE\backslash\{e'\}}\bar\ell'_{e''}\sum_{e\in\calE\backslash\{e', e''\}}\expect_0\xi_e^{(k)}\bg_e^{(k)}((\bG_0^\sharp)^{-1}\bLambda_\sharp\nabla_\bphi\theta_{e'}, (\bG_0^\sharp)^{-1}\bLambda_\sharp\nabla_\bphi\theta_{e''})  \bigg\}\\
&\quad = \sum_{e'\in\calE}\ell'_{e'}\bigg[
\sum_{e\in\calE}\expect_0\xi_e^{(k)}\bg_e^{(k)}((\bG_0^\sharp)^{-1}\bLambda_\sharp\nabla_\bphi\theta_{e'}, \bar{\bs}) - \sum_{e \in\calE}\bar{\ell}'_e\expect_0\xi_e^{(k)}\bg_e^{(k)}((\bG_0^\sharp)^{-1}\bLambda_\sharp\nabla_\bphi\theta_{e}, (\bG_0^\sharp)^{-1}\bLambda_\sharp\nabla_\bphi\theta_{e'})\\
&\qquad - \bar{\ell}'_{e'}\sum_{e\in\calE}\expect_0\xi_e^{(k)}\bg_e^{(k)}((\bG_0^\sharp)^{-1}\bLambda_\sharp\nabla_\bphi\theta_{e'}, (\bG_0^\sharp)^{-1}\bLambda_\sharp\nabla_\bphi\theta_{e'})\\
&\qquad + (\expect_0\xi_{e'}^{(k)})\{2\bar{\ell}'_{e'}\bg_{e'}^{(k)}((\bG_0^\sharp)^{-1}\bLambda_\sharp\nabla_\bphi\theta_{e'}, (\bG_0^\sharp)^{-1}\bLambda_\sharp\nabla_\bphi\theta_{e'}) - \bg_{e'}^{(k)}((\bG_0^\sharp)^{-1}\bLambda_\sharp\nabla_\bphi\theta_{e'}, \bar{\bs})\}
\bigg].
\end{align*}
By \eqref{eqn:max_ellprime_bound}, matrix Bernstein, \eqref{eqn:kernel_bound_vector}, \eqref{eqn:kernel_vector_bound_scalar_multiple}, and \eqref{eqn:kernel_vector_bound_scalar_multiple_deltaphi}, we know that the sum of the squared (Euclidean) norms of the vectors inside the square bracket above over $e'\in\calE$ is $\Optilde(n^{-1}\log n)$, and the maximum of these norms over $e'$ is $\Optilde(n^{-1}\sqrt{\log n})$ for $k \in \{1,2,3\}$. Then, by matrix Bernstein and Lemma \ref{lemma:conditioning_argument}, we have $\|\bq_{41}^{(k)}\|_2 = \Optilde(n^{-1/2}\log n)$. 

\medskip\noindent\textbf{Step 7. Bounding $\bq_{42}^{(k)}$. }
Now let $(\tilde{\ell}'_e:e\in\calE)$ be a copy of $(\ell'_e:e\in\calE)$ that is also independent of $(\ell'_e:e\in\calE)$ and $(\bar{\ell}'_e:e\in\calE)$ and take $\tilde{\bs} = (\bG_0^\sharp)^{-1}\bLambda_\sharp\sum_{e\in\calE}\tilde{\ell}'_e\nabla_\bphi\theta_e$.
By definition and Theorem \ref{thm:decoupling}, it is sufficient to consider
\begin{align*}
\sum_{e\in\calE}(\xi_e^{(k)} - \expect_0\xi_e^{(k)})\bh_e^{(k)},
\end{align*}
where
\begin{align*}
\bh_e^{(k)}& = \bg_e^{(k)}(\bar{\bs}, \tilde{\bs}) - \bg_e^{(k)}\bigg[
(\bG_0^\sharp)^{-1}
\sum_{e'\in\calE}\bar{\ell}'_{e'}\tilde{\ell}'_{e'}\bLambda_\sharp\nabla_\bphi\theta_{e'}\nabla_\bphi\theta_{e'}\transpose\bLambda_\sharp
(\bG_0^\sharp)^{-1}
\bigg] - \bar{\ell}'_{e}\bg_{e}^{(k)}((\bG_0^\sharp)^{-1}\bLambda_\sharp\nabla_\bphi\theta_e, \tilde{\bs})\\
&\quad - \tilde{\ell}'_{e}\bg_{e}^{(k)}((\bG_0^\sharp)^{-1}\bLambda_\sharp\nabla_\bphi\theta_e, \bar{\bs}) + 2\bar{\ell}'_e\tilde{\ell}'_e\bg_{e}^{(k)}((\bG_0^\sharp)^{-1}\bLambda_\sharp\nabla_\bphi\theta_e,
(\bG_0^\sharp)^{-1}\bLambda_\sharp\nabla_\bphi\theta_e). 
\end{align*}
By \eqref{eqn:kernel_bound_vector} and \eqref{eqn:J0n_s_hat_bound}, we have
\begin{align*}
\max_{e\in\calE}\|\bg_e^{(k)}(\bar{\bs}, \widetilde{\bs})\|_2 = \Optilde(n^{-3/2}\log n),\quad
\sum_{e\in\calE}\|\bg_e^{(k)}(\bar{\bs}, \widetilde{\bs})\|_2^2 = \Optilde\{n^{-2}(\log n)^2\};
\end{align*}
By \eqref{eqn:kernel_bound_matrix} and matrix Bernstein, 
\begin{align*}
&\bigg\|(\bG_0^\sharp)^{-1}
\sum_{e'\in\calE}\bar{\ell}'_{e'}\tilde{\ell}'_{e'}\bLambda_\sharp\nabla_\bphi\theta_{e'}\nabla_\bphi\theta_{e'}\transpose\bLambda_\sharp
(\bG_0^\sharp)^{-1}\bigg\|_2 = \Optilde\bigg(\sqrt{\frac{\log n}{n}}\bigg),\\
&\max_{e\in\calE}\bigg\|\bg_e^{(k)}\bigg[(\bG_0^\sharp)^{-1}
\sum_{e'\in\calE}\bar{\ell}'_{e'}\tilde{\ell}'_{e'}\bLambda_\sharp\nabla_\bphi\theta_{e'}\nabla_\bphi\theta_{e'}\transpose\bLambda_\sharp
(\bG_0^\sharp)^{-1}\bigg]\bigg\|_2 = \Optilde\bigg(\frac{\sqrt{\log n}}{n^2}\bigg),\\
&\sum_{e\in\calE}\bigg\|\bg_e^{(k)}\bigg[(\bG_0^\sharp)^{-1}
\sum_{e'\in\calE}\bar{\ell}'_{e'}\tilde{\ell}'_{e'}\bLambda_\sharp\nabla_\bphi\theta_{e'}\nabla_\bphi\theta_{e'}\transpose\bLambda_\sharp
(\bG_0^\sharp)^{-1}\bigg]\bigg\|_2^2 = \Optilde\bigg(\frac{\log n}{n^2}\bigg).
\end{align*}
By \eqref{eqn:J0n_s_hat_bound}, \eqref{eqn:max_ellprime_bound}, and \eqref{eqn:kernel_bound_vector},
\begin{align*}
&\max_{e\in\calE}\left\|\bar{\ell}'_{e}\bg_{e}^{(k)}((\bG_0^\sharp)^{-1}\bLambda_\sharp\nabla_\bphi\theta_e, \tilde{\bs})\right\|_2 = \Optilde\bigg\{\frac{(\log n)^{3/2}}{n^2}\bigg\},\;
\sum_{e\in\calE}\left\|\bar{\ell}'_{e}\bg_{e}^{(k)}((\bG_0^\sharp)^{-1}\bLambda_\sharp\nabla_\bphi\theta_e, \tilde{\bs})\right\|_2^2 = \Optilde\bigg\{\frac{(\log n)^{3}}{n^3}\bigg\},\\
&\max_{e\in\calE}\left\|\tilde{\ell}'_{e}\bg_{e}^{(k)}((\bG_0^\sharp)^{-1}\bLambda_\sharp\nabla_\bphi\theta_e, \bar{\bs})\right\|_2 = \Optilde\bigg\{\frac{(\log n)^{3/2}}{n^2}\bigg\},\;
\sum_{e\in\calE}\left\|\tilde{\ell}'_{e}\bg_{e}^{(k)}((\bG_0^\sharp)^{-1}\bLambda_\sharp\nabla_\bphi\theta_e, \bar{\bs})\right\|_2^2 = \Optilde\bigg\{\frac{(\log n)^{3}}{n^3}\bigg\}.
\end{align*}
By \eqref{eqn:max_ellprime_bound} and \eqref{eqn:kernel_bound_vector}, 
\begin{align*}
&\max_{e\in\calE}\left\|\bar{\ell}'_e\tilde{\ell}'_e\bg_{e}^{(k)}((\bG_0^\sharp)^{-1}\bLambda_\sharp\nabla_\bphi\theta_e,
(\bG_0^\sharp)^{-1}\bLambda_\sharp\nabla_\bphi\theta_e)\right\|_2 = \Optilde\bigg\{\frac{(\log n)^2}{n^{5/2}}\bigg\},\\
&\sum_{e\in\calE}\left\|\bar{\ell}'_e\tilde{\ell}'_e\bg_{e}^{(k)}((\bG_0^\sharp)^{-1}\bLambda_\sharp\nabla_\bphi\theta_e,
(\bG_0^\sharp)^{-1}\bLambda_\sharp\nabla_\bphi\theta_e)\right\|_2^2
 = \Optilde\bigg\{\frac{(\log n)^4}{n^4}\bigg\}.
\end{align*}
Hence, we obtain
\begin{align*}
\max_{e\in\calE}\|\bh_e^{(k)}\|_2 = \Optilde\bigg(\frac{\log n}{n^{3/2}}\bigg),\quad
\sum_{e\in\calE}\|\bh_e^{(k)}\|_2^2 = \Optilde\bigg\{\frac{(\log n)^2}{n^{2}}\bigg\},
\end{align*}
so that Lemma \ref{lemma:conditioning_argument} implies that $\|\bq_{42}^{(k)}\|_2 = \Optilde\{(\log n)^{3/2}/n\}$. The proof is completed by combining the error bounds for $\|\bq_{1}^{(k)}\|_2$ through $\|\bq_{42}^{(k)}\|_2$. 
\end{proof}

\subsection{Completing the proof of Theorem \ref{thm:asymptotic_normality_R}}
\label{sub:completing_the_proof_of_asymptotic_normality_Rt}

\begin{proof}[\bf Proof of Theorem \ref{thm:asymptotic_normality_R}]
By Subsection \ref{sub:proof_architecture_asymptotic_normality_R}, it is sufficient to control $\|(\bG_0^\sharp)^{-1}(\bH^\sharp + \bG_0^\sharp)\bLambda_\sharp^{-1}\bgamma_\bphi\|_2$, show that $\expect\widetilde{\scrP}_{\br^{(t)}}(\bQ_1 + \bQ_2) = \mathbf{b}_n^{(t)}(\bpsi_0) + o(1)$, and verify that the leading term satisfies the Lindeberg condition. We also record where the growth condition of Theorem \ref{thm:asymptotic_normality_R} enters: by Lemma \ref{lemma:third_order_cross_terms} (ii), the fourth term on the right-hand side of \eqref{eqn:third_order_expansion_vech_Rt} is $\Optilde\{\sqrt{m}(\log n)^{3/2}\vartheta_n\}$, and
\[
\sqrt{m}(\log n)^{3/2}\vartheta_n = \bigg\{\frac{m^{13}(\log n)^7}{n}\bigg\}^{1/2} + \bigg\{\frac{m^{12}(\log n)^8}{n}\bigg\}^{1/2} + \bigg\{\frac{m^{11}(\log n)^9}{n}\bigg\}^{1/2}\leq 3\bigg\{\frac{m^{13}(\log n)^9}{n}\bigg\}^{1/2}\to 0.
\]

\medskip\noindent\textbf{Step 1. Bounding $\|(\bG_0^\sharp)^{-1}(\bH^\sharp + \bG_0^\sharp)\bLambda_\sharp^{-1}\bgamma_\bphi\|_2$. }
Multiplying $(\bG_0^\sharp)^{-1}(\bH^\sharp + \bG_0^\sharp)$ to both sides of \eqref{eqn:third_order_expansion_phi} with $\widehat{\bs}$ moved to the left-hand side leads to
\begin{align*}
\big\|(\bG_0^\sharp)^{-1}(\bH^\sharp + \bG_0^\sharp)\bLambda_\sharp^{-1}\bgamma_\bphi\big\|_2
& \leq\big\|(\bG_0^\sharp)^{-1}(\bH^\sharp + \bG_0^\sharp)\big\|_2\bigg\{\|\bQ_1\|_2 + \|\bQ_2\|_2 + \|(\bG_0^\sharp)^{-1}\|_2\|\bLambda_\sharp\bgamma_\calL\|_2\\
&\quad + \|(\bG_0^\sharp)^{-1}\|_2\bigg\|\bLambda_\sharp\nabla^3\calL(\bphi_0)[\bLambda_\sharp\widehat{\bs}, \bgamma_\bphi] + \frac{1}{2}\bLambda_\sharp\nabla^3\calL(\bphi_0)[\bgamma_\bphi, \bgamma_\bphi]\bigg\|_2\bigg\}\\
&\quad + \big\|(\bG_0^\sharp)^{-1}(\bH^\sharp + \bG_0^\sharp)\big\|_2^2\|\bLambda_\sharp^{-1}\bgamma_\bphi\|_2.
\end{align*}

\begin{itemize}
  \item \underline{We show that $\|(\bG_0^\sharp)^{-1}(\bH^\sharp + \bG_0^\sharp)\|_2 = \Optilde\{\sqrt{(\log n)/n}\}$.} We inherit the notations in the proofs of Lemma \ref{lemma:Q1_fluctuation} and Lemma \ref{lemma:Q2_fluctuation} and write by \eqref{eqn:Q1_building_blocks} that 
  \[
  \bH^\sharp + \bG_0^\sharp = \sum_{e\in\calE^\circ}(\ell''_e - \expect_0\ell_e'')\bJ_{0n}\transpose\widetilde{\bLambda}_\sharp\nabla_\bpsi\theta_e\nabla_\bpsi\theta_e\transpose\widetilde{\bLambda}_\sharp\bJ_{0n} + \sum_{e\in\calE^\circ}\ell'_e\bV_e  
  \]
  The second sum is $\bLambda_\sharp\bDelta\bLambda_\sharp$ with $\bDelta$ defined in \eqref{eqn:Gamma_matrix}, and $\|\bLambda_\sharp\bDelta\bLambda_\sharp\|_2 = \Optilde(\sqrt{\log n/n})$ by the proof of Proposition \ref{prop:hessian_bound}. The first sum is a sum of independent, mean-zero, symmetric matrices, and this sum has spectral norm being $\Optilde\{\sqrt{(\log n)/n}\}$ by matrix Bernstein. 

  \item \underline{We bound the terms inside the bracket.} By \eqref{eqn:J0n_s_hat_bound}, $\|\bJ_{0n}\widehat{\bs}\|_\infty = \Optilde(\sqrt{\log n})$, so that 
  \[
  \|\widehat{\bs}\|_2 = \|\bJ_{0n}\widehat{\bs}\|_2\leq\sqrt{d_\bpsi}\|\bJ_{0n}\widehat{\bs}\|_\infty = \Optilde(\sqrt{mn\log n}),
  \]
  and hence, $\|\bQ_1\|_2 = \Optilde(\sqrt{m}\log n)$. By Lemma \ref{lemma:third_order_cross_terms} (i) with $\bu = \bv = \bLambda_\sharp\widehat{\bs}$, 
  \[
  \|\bJ_0\bLambda_\sharp\widehat{\bs}\|_\infty = \|\widetilde{\bLambda}_\sharp\bJ_{0n}\widehat{\bs}\|_\infty\leq n^{-1/2}\|\bJ_{0n}\widehat{\bs}\|_\infty = \Optilde\bigg(\sqrt{\frac{\log n}{n}}\bigg),
  \]
  and $\max_{t,i\neq j}|\ell'(\theta_{0ij}^{(t)}; A_{ij}^{(t)})| = \Optilde(\log n)$ by \eqref{eqn:max_ellprime_bound}, we have
  \[
  \|\bQ_2\|_2\lesssim \|\bLambda_\sharp\nabla^3\calL(\bphi_0)[\bLambda_\sharp\widehat{\bs}, \bLambda_\sharp\widehat{\bs}]\|_2 = \sqrt{m}n\,\Optilde(\log n)\,\Optilde\bigg(\frac{\log n}{n}\bigg) = \Optilde\{\sqrt{m}(\log n)^2\}.
  \]
  By Lemma \ref{lemma:third_order_cross_terms} (ii), 
  \[
  \bigg\|\bLambda_\sharp\nabla^3\calL(\bphi_0)[\bLambda_\sharp\widehat{\bs}, \bgamma_\bphi] + \frac{1}{2}\bLambda_\sharp\nabla^3\calL(\bphi_0)[\bgamma_\bphi, \bgamma_\bphi]\bigg\|_2 = \Optilde\{\sqrt{m}(\log n)^{3/2}\vartheta_n\}.
  \]
  By Lemma \ref{lemma:third_order_Taylor_expansion_loglikelihood}, $\|\bLambda_\sharp\bgamma_\calL\|_2 = \Optilde\{\sqrt{m}(\log n)^{5/2}/\sqrt{n}\}$. Since $\bJ_0\bgamma_\bphi = \bgamma$ is the remainder of \eqref{eqn:Taylor_expansion} and $\bJ_{0n}\bLambda_\sharp^{-1} = \widetilde{\bLambda}_\sharp^{-1}\bJ_0$, we have $\|\bLambda_\sharp^{-1}\bgamma_\bphi\|_2 = \|\bJ_{0n}\bLambda_\sharp^{-1}\bgamma_\bphi\|_2 = \|\widetilde{\bLambda}_\sharp^{-1}\bgamma\|_2$. Step 2 of the proof in Section \ref{sub:Proof of the asymptotic normality} shows that $\|\mathrm{diag}(\sqrt{n}\eye_{d_\balpha}, \sqrt{mn}\eye_{d_\bX})\scrP_\bZ\bgamma\|_\infty = \Optilde(\vartheta_n)$ and $\|n\scrP_\br\bgamma\|_\infty = \Optilde\{m^3\log n + m^{5/2}(\log n)^{3/2}\}$. Then, by the fact that $d_\balpha + d_\bX = O(mn)$ and $d_\br = md(d + 1)/2$, we obtain
  \begin{align*}
  \|\bLambda_\sharp^{-1}\bgamma_\bphi\|_2&\lesssim \sqrt{mn}\|\mathrm{diag}(\sqrt{n}\eye_{d_\balpha}, \sqrt{mn}\eye_{d_\bX})\scrP_\bZ\bgamma\|_\infty + \sqrt{m}\|n\scrP_\br\bgamma\|_\infty = \Optilde(\sqrt{mn}).
  \end{align*}
\end{itemize}
Combining the above mini-bounds, we obtain
\begin{align*}
&\big\|(\bG_0^\sharp)^{-1}(\bH^\sharp + \bG_0^\sharp)\bLambda_\sharp^{-1}\bgamma_\bphi\big\|_2\\
&\quad = \Optilde\bigg(\sqrt{\frac{\log n}{n}}\bigg)\Optilde\Big\{\sqrt{m}\log n + \sqrt{m}(\log n)^2 + \sqrt{m}(\log n)^{3/2}\vartheta_n + \sqrt{m/n}(\log n)^{5/2}\Big\}\\
&\qquad + \Optilde\bigg(\frac{\log n}{n}\bigg)\Optilde(\sqrt{mn})
 = \Optilde\bigg\{\frac{\sqrt{m}(\log n)^{5/2}}{\sqrt{n}}\bigg\}.
\end{align*}

\medskip\noindent\textbf{Step 2. We show that $\expect_0\widetilde{\scrP}_{\br^{(t)}}(\bQ_1 + \bQ_2) = \mathbf{b}_n^{(t)}(\bpsi_0) + o(1)$. }By definition and the facts that $\bOmega(\bpsi_0) = \bJ_{0n}(\bG_0^\sharp)^{-1}\bJ_{0n}\transpose$ (since $\widetilde{\bLambda}_\sharp \scrJ(\bpsi_0) = \widetilde{\bLambda}_\sharp\bJ_{0n} = \bJ_0\bLambda_\sharp$) and $\widetilde{\scrP}_{\br^{(t)}} = \scrP_{\br^{(t)}}\bJ_{0n}$, we have
\begin{align*}
\expect_0\widetilde{\scrP}_{\br^{(t)}}\bQ_1
& = \sum_{e\in\calE^\circ}
\expect_0\ell''_e\ell'_e
\widetilde{\scrP}_{\br^{(t)}}(\bG_0^\sharp)^{-1}
\bJ_{0n}\transpose\widetilde{\bLambda}_\sharp\nabla_\bpsi\theta_e\nabla_\bpsi\theta_e\transpose\widetilde{\bLambda}_\sharp\bJ_{0n}(\bG_0^\sharp)^{-1}\bJ_{0n}\transpose\widetilde{\bLambda}_\sharp\nabla_\bpsi\theta_e\\
&\quad + \sum_{e\in\calE^\circ}
\expect_0 (\ell'_e)^2
\widetilde{\scrP}_{\br^{(t)}}(\bG_0^\sharp)^{-1}\bigg[\bLambda_\sharp\frac{\partial^2\theta_e}{\partial\bphi\partial\bphi\transpose}\bLambda_\sharp\bigg]
(\bG_0^\sharp)^{-1}\bJ_{0n}\transpose\widetilde{\bLambda}_\sharp\nabla_\bpsi\theta_e\\
& = \sum_{e\in\calE^\circ}
\expect_0\ell''_e\ell'_e
\bigg\{\frac{\partial\theta_e}{\partial\bpsi\transpose}\widetilde{\bLambda}_\sharp\bOmega(\bpsi_0)\widetilde{\bLambda}_\sharp\frac{\partial\theta_e}{\partial\bpsi}\bigg\}{\scrP}_{\br^{(t)}}\bOmega(\bpsi_0)
\widetilde{\bLambda}_\sharp\frac{\partial\theta_e}{\partial\bpsi}\\
&\quad + \sum_{e\in\calE^\circ}
\expect_0 (\ell'_e)^2
\widetilde{\scrP}_{\br^{(t)}}(\bG_0^\sharp)^{-1}\bigg[\bLambda_\sharp\frac{\partial^2\theta_e}{\partial\bphi\partial\bphi\transpose}\bLambda_\sharp\bigg]
(\bG_0^\sharp)^{-1}\bJ_{0n}\transpose\widetilde{\bLambda}_\sharp\nabla_\bpsi\theta_e.
\end{align*}
The formula for $\expect_0\widetilde{\scrP}_{\br^{(t)}}\bQ_2$ is slightly more involved. We inherit the notations in the proof of Lemma \ref{lemma:Q2_fluctuation} and write
\begin{align*}
\expect_0\widetilde{\scrP}_{\br^{(t)}}\bQ_2
& = 
\frac{1}{2}\sum_{k = 1}^4\sum_{e\in\calE}\expect_0\xi_e^{(k)}\bg_e^{(k)}(\widehat{\bs}, \widehat{\bs})\\
& = \frac{1}{2}\sum_{k = 1}^4\sum_{e,e',e''\in\calE^\circ}\expect_0\xi_e^{(k)}\ell'_{e'}\ell'_{e''}\bg_e^{(k)}\left(
(\bG_0^\sharp)^{-1}\bJ_{0n}\transpose\widetilde{\bLambda}_\sharp\nabla_\bpsi\theta_{e'},
(\bG_0^\sharp)^{-1}\bJ_{0n}\transpose\widetilde{\bLambda}_\sharp\nabla_\bpsi\theta_{e''}
\right)\\
& = \frac{1}{2}\sum_{k = 1}^4\bigg\{\sum_{e = e' = e''} + \sum_{e' = e'' \neq e} + \sum_{e = e'\neq e''} + \sum_{e = e'' \neq e'} + \sum_{e\neq e'\neq e''\neq e}\bigg\}\expect_0\xi_e^{(k)}\ell'_{e'}\ell'_{e''}\\
&\quad\times \bg_e^{(k)}\left(
(\bG_0^\sharp)^{-1}\bJ_{0n}\transpose\widetilde{\bLambda}_\sharp\nabla_\bpsi\theta_{e'},
(\bG_0^\sharp)^{-1}\bJ_{0n}\transpose\widetilde{\bLambda}_\sharp\nabla_\bpsi\theta_{e''}
\right)\\
& = \frac{1}{2}\sum_{k = 1}^4\sum_{e\in\calE^\circ}\expect_0\xi_e^{(k)}(\ell'_e)^2\bg_e^{(k)}\left(
(\bG_0^\sharp)^{-1}\bJ_{0n}\transpose\widetilde{\bLambda}_\sharp\nabla_\bpsi\theta_{e},
(\bG_0^\sharp)^{-1}\bJ_{0n}\transpose\widetilde{\bLambda}_\sharp\nabla_\bpsi\theta_{e}
\right)\\
&\quad + \frac{1}{2}\sum_{k = 1}^4\sum_{e,e'\in\calE^\circ,e\neq e'}\expect_0\xi_e^{(k)}(\ell'_{e'})^2\bg_e^{(k)}\left(
(\bG_0^\sharp)^{-1}\bJ_{0n}\transpose\widetilde{\bLambda}_\sharp\nabla_\bpsi\theta_{e'},
(\bG_0^\sharp)^{-1}\bJ_{0n}\transpose\widetilde{\bLambda}_\sharp\nabla_\bpsi\theta_{e'}
\right)\\
& = \frac{1}{2}\sum_{k = 1}^4\sum_{e\in\calE^\circ}\left\{\expect_0\xi_e^{(k)}(\ell'_e)^2 - (\expect_0\xi_e^{(k)})\expect_0(\ell'_e)^2\right\}\bg_e^{(k)}\left(
(\bG_0^\sharp)^{-1}\bJ_{0n}\transpose\widetilde{\bLambda}_\sharp\nabla_\bpsi\theta_{e},
(\bG_0^\sharp)^{-1}\bJ_{0n}\transpose\widetilde{\bLambda}_\sharp\nabla_\bpsi\theta_{e}
\right)\\
&\quad + \frac{1}{2}\sum_{k = 1}^4\sum_{e\in\calE^\circ}\expect_0\xi_e^{(k)}\bg_e^{(k)}[(\bG_0^\sharp)^{-1}].
\end{align*}
By the formulas of $\xi_e^{(k)}$ and $\bg_e^{(k)}$, we obtain
\begin{align*}
\expect_0\widetilde{\scrP}_{\br^{(t)}}\bQ_2 
& = \frac{1}{2}\sum_{e\in\calE^\circ}\bigg[
\expect_0\ell'''_e {\scrP}_{\br^{(t)}}\bOmega(\bpsi_0)\widetilde{\bLambda}_\sharp\frac{\partial\theta_e}{\partial\bpsi}\bigg\{
\frac{\partial\theta_e}{\partial\bpsi\transpose}\widetilde{\bLambda}_\sharp\bOmega(\bpsi_0)\widetilde{\bLambda}_\sharp\frac{\partial\theta_e}{\partial\bpsi}
\bigg\}\\
&\quad
+ \expect_0\ell''_e{\scrP}_{\br^{(t)}}\bOmega(\bpsi_0)\widetilde{\bLambda}_\sharp\frac{\partial\theta_e}{\partial\bpsi}\mathrm{tr}\{\bV_e(\bG_0^\sharp)^{-1}\}
\bigg]\\
&\quad + \sum_{e\in\calE^\circ} \expect_0\ell''_e \widetilde{\scrP}_{\br^{(t)}}(\bG_0^\sharp)^{-1}\bigg[\bLambda_\sharp\frac{\partial^2\theta_e}{\partial\bphi\partial\bphi\transpose}\bLambda_\sharp\bigg](\bG_0^\sharp)^{-1}\bJ_{0n}\transpose\widetilde{\bLambda}_\sharp\nabla_\bpsi\theta_e\\
&\quad + \frac{1}{2}\sum_{e\in\calE^\circ}\left\{\expect_0\ell'''_e(\ell'_e)^2 - (\expect_0\ell'''_e)\expect_0(\ell'_e)^2\right\}
\left\{\nabla_\bpsi\theta_e\transpose\widetilde{\bLambda}_\sharp\bJ_{0n}(\bG_0^\sharp)^{-1}\bJ_{0n}\transpose\widetilde{\bLambda}_\sharp\nabla_\bpsi\theta_e\right\}^2\\
&\quad\times \widetilde{\scrP}_{\br^{(t)}}(\bG_0^\sharp)^{-1}\bJ_{0n}\transpose\widetilde{\bLambda}_\sharp\nabla_\bpsi\theta_e\\
&\quad + \frac{1}{2}\sum_{e\in\calE^\circ}\left\{\expect_0\ell''_e(\ell'_e)^2 - (\expect_0\ell''_e)\expect_0(\ell'_e)^2\right\}\widetilde{\scrP}_{\br^{(t)}}(\bG_0^\sharp)^{-1}\bJ_{0n}\transpose\widetilde{\bLambda}_\sharp\nabla_\bpsi\theta_e\\
&\quad\times\left\{
\nabla_\bpsi\theta_e\transpose\widetilde{\bLambda}_\sharp \bJ_{0n}(\bG_0^\sharp)^{-1}\bigg[\bLambda_\sharp\frac{\partial^2\theta_e}{\partial\bphi\partial\bphi\transpose}\bLambda_\sharp\bigg](\bG_0^\sharp)^{-1}\bJ_{0n}\transpose\widetilde{\bLambda}_\sharp\nabla_\bpsi\theta_e
\right\}\\
&\quad + \sum_{e\in\calE^\circ}\expect_0(\ell''_e - \expect_0\ell''_e)(\ell'_e)^2\widetilde{\scrP}_{\br^{(t)}}(\bG_0^\sharp)^{-1}\bigg[\bLambda_\sharp\frac{\partial^2\theta_e}{\partial\bphi\partial\bphi\transpose}\bLambda_\sharp\bigg](\bG_0^\sharp)^{-1}\bJ_{0n}\transpose\widetilde{\bLambda}_\sharp\nabla_\bpsi\theta_e\\
&\quad\times \left\{\nabla_\bpsi\theta_e\transpose\widetilde{\bLambda}_\sharp\bJ_{0n}(\bG_0^\sharp)^{-1}\bJ_{0n}\transpose \widetilde{\bLambda}_\sharp \nabla_\bpsi\theta_e\right\}\\
&\quad + \frac{1}{2}\sum_{e\in\calE^\circ}\sum_{j = 1}^{d(d + 1)/2} \expect_0(\ell'_e)^3 \be_j\tau_e\left(
(\bG_0^\sharp)^{-1}\bJ_{0n}\transpose\widetilde{\bLambda}_\sharp\nabla_\bpsi\theta_e,
(\bG_0^\sharp)^{-1}\bJ_{0n}\transpose\widetilde{\bLambda}_\sharp\nabla_\bpsi\theta_e,
(\bG_0^\sharp)^{-1}\widetilde{\scrP}_{\br^{(t)}}\transpose\be_j
\right).
\end{align*}
Note that in the above formulas, the second term in $\expect_0\widetilde{\scrP}_{\br^{(t)}}\bQ_2$ cancels with the second term in $\expect_0\widetilde{\scrP}_{\br^{(t)}}\bQ_1$ since $\expect_0\ell''_e = - \expect_0(\ell'_e)^2$. For the term inside the bracket of the first term in $\expect_0\widetilde{\scrP}_{\br^{(t)}}\bQ_2$ involving $\bV_e$, we recall in the proof of Lemma \ref{lemma:Q1_fluctuation} that
\begin{align*}
\bV_e = \bJ_{0n}\transpose\widetilde{\bLambda}_\sharp\nabla^2_\bpsi\theta_e\widetilde{\bLambda}_\sharp\bJ_{0n} + \frac{1}{mn^2}\mathrm{diag}\Big\{\mathbf{O}_{d_\balpha + d_\br}, -\frac{1}{2}(\bM_e + \bM_e\transpose)\otimes\eye_{n - 1 - d}\Big\},
\end{align*}
where $\bM_e = \bR_0^{(s)}(\bx_{0a}\bx_{0b}\transpose + \bx_{0b}\bx_{0a}\transpose)$ for $e = (s, a, b)\in\calE^\circ$. Then, by a direct calculation,
\begin{align*}
\mathrm{tr}\left\{\bJ_{0n}\transpose\widetilde{\bLambda}_\sharp\nabla^2_\bpsi\theta_e\widetilde{\bLambda}_\sharp\bJ_{0n}\right\} = \mathrm{tr}\bigg\{\widetilde{\bLambda}_\sharp\frac{\partial^2\theta_e}{\partial\bpsi\partial\bpsi\transpose}\widetilde{\bLambda}_\sharp\bOmega(\bpsi_0)
\bigg\}
\end{align*}
and
\begin{align*}
&\mathrm{tr}\bigg[
\mathrm{diag}\Big\{\mathbf{O}_{d_\balpha + d_\br}, -\frac{1}{2}(\bM_e + \bM_e\transpose)\otimes\eye_{n - 1 - d}\Big\}(\bG_0^\sharp)^{-1}
\bigg]\\
&\quad = \mathrm{tr}\bigg[
\mathrm{diag}\Big\{\mathbf{O}_{d_\balpha + d_\br}, -\frac{1}{2}(\bM_e + \bM_e\transpose)\otimes\eye_n\Big\}\bJ_{0n}(\bG_0^\sharp)^{-1}\bJ_{0n}\transpose
\bigg]\\
&\quad = \mathrm{tr}\bigg[\sum_{c = 1}^n
\mathrm{diag}\Big\{\mathbf{O}_{d_\balpha + d_\br}, -\frac{1}{2}(\bM_e + \bM_e\transpose)\otimes\eye_n\Big\}\scrP_{\bx_c}\transpose\scrP_{\bx_c}\bOmega(\bpsi_0)
\bigg]\\
&\quad = \mathrm{tr}\bigg[\sum_{c = 1}^n\scrP_{\bx_c}\bOmega(\bpsi_0)
\mathrm{diag}\Big\{\mathbf{O}_{d_\balpha + d_\br}, -\frac{1}{2}(\bM_e + \bM_e\transpose)\otimes\eye_n\Big\}\scrP_{\bx_c}\transpose
\bigg]\\
&\quad = -\frac{1}{2}\mathrm{tr}\bigg[\sum_{c = 1}^n\scrP_{\bx_c}\bOmega(\bpsi_0)\scrP_{\bx_c}\transpose\bR_0^{(s)}(\bx_{0a}\bx_{0b}\transpose + \bx_{0b}\bx_{0a}\transpose)\Big\}
\bigg] = -\bx_{0b}\transpose\{\bT(\bpsi_0)\bR_0^{(s)} + \bR_0^{(s)}\bT(\bpsi_0)\}\bx_{0a}\\
&\quad = - \vech\left\{\bR_0^{(s)}\bT(\bpsi_0) + \bT(\bpsi_0)\bR_0^{(s)}\right\}\transpose\frac{\partial\theta_e}{\partial\vech(\bR^{(s)})}\\
&\quad = - \sum_{s' = 1}^m\vech\left\{\bR_0^{(s')}\bT(\bpsi_0) + \bT(\bpsi_0)\bR_0^{(s')}\right\}\transpose\frac{\partial\theta_e}{\partial\vech(\bR^{(s')})}\\
&\quad = - \sum_{s' = 1}^mn\begin{bmatrix}
\zero_{d_\balpha}\transpose &  \be_{s'}(m)\transpose\otimes\vech\left\{\bR_0^{(s')}\bT(\bpsi_0) + \bT(\bpsi_0)\bR_0^{(s')}\right\}\transpose & \zero_{d_\bB}\transpose
\end{bmatrix}
\bJ_{0n}\transpose \widetilde{\bLambda}_\sharp \frac{\partial\theta_e}{\partial\bpsi},
\end{align*}
so that
\begin{align*}
&\frac{1}{2mn^2}\sum_{e\in\calE^\circ}\expect_0\ell''_e\scrP_{\br^{(t)}}\bOmega(\bpsi_0)\widetilde{\bLambda}_\sharp\frac{\partial\theta_e}{\partial\bpsi}\mathrm{tr}\bigg[
\mathrm{diag}\Big\{\mathbf{O}_{d_\balpha + d_\br}, -\frac{1}{2}(\bM_e + \bM_e\transpose)\otimes\eye_{n - 1 - d}\Big\}(\bG_0^\sharp)^{-1}
\bigg]\\
&\quad = \frac{\widetilde{\scrP}_{\br^{(t)}}(\bG_0^\sharp)^{-1}}{2mn}
\bigg(
\sum_{e\in\calE^\circ}\expect_0\ell''_e\bJ_{0n}\transpose\widetilde{\bLambda}_\sharp\frac{\partial\theta_e}{\partial\bpsi}\frac{\partial\theta_e}{\partial\bpsi\transpose}\widetilde{\bLambda}_\sharp\bJ_{0n}
\bigg)
\sum_{s' = 1}^m\begin{bmatrix}
\zero_{d_\balpha}\\ 
\be_{s'}\otimes\vech\left\{\bR_0^{(s')}\bT(\bpsi_0) + \bT(\bpsi_0)\bR_0^{(s')}\right\} \\ 
\zero_{d_\bB}
\end{bmatrix}\\
&\quad = \frac{1}{2mn}\widetilde{\scrP}_{\br^{(t)}}\sum_{s' = 1}^m\begin{bmatrix}
\zero_{d_\balpha}\\ 
\be_{s'}\otimes\vech\left\{\bR_0^{(s')}\bT(\bpsi_0) + \bT(\bpsi_0)\bR_0^{(s')}\right\} \\ 
\zero_{d_\bB}
\end{bmatrix} = \frac{\vech\{\bR_0^{(t)}\bT(\bpsi_0) + \bT(\bpsi_0)\bR_0^{(t)}\}}{2mn}.
\end{align*}
Therefore, it is sufficient to show that the third term through the sixth term on the right-hand side of the formula of $\expect_0\widetilde{\scrP}_{\br^{(t)}}\bQ_2$ are $o(1)$. For the third term, by Cauchy--Schwarz inequality and Step 2 item (D) in the proof of Lemma \ref{lemma:Q2_fluctuation}, we have 
\begin{align*}
&\bigg\|\sum_{e\in\calE^\circ}\left\{\expect_0\ell'''_e(\ell'_e)^2 - (\expect_0\ell'''_e)\expect_0(\ell'_e)^2\right\}
\left\{\nabla_\bpsi\theta_e\transpose\widetilde{\bLambda}_\sharp\bJ_{0n}(\bG_0^\sharp)^{-1}\bJ_{0n}\transpose\widetilde{\bLambda}_\sharp\nabla_\bpsi\theta_e\right\}^2\widetilde{\scrP}_{\br^{(t)}}(\bG_0^\sharp)^{-1}\bJ_{0n}\transpose\widetilde{\bLambda}_\sharp\nabla_\bpsi\theta_e\bigg\|_2\\
&\quad\lesssim \sqrt{m}n\max_{e\in\calE}\left\{\nabla_\bpsi\theta_e\transpose\widetilde{\bLambda}_\sharp\bJ_{0n}(\bG_0^\sharp)^{-1}\bJ_{0n}\transpose\widetilde{\bLambda}_\sharp\nabla_\bpsi\theta_e\right\}^2\bigg\{\sum_{e\in\calE^\circ}\|\widetilde{\scrP}_{\br^{(t)}}(\bG_0^\sharp)^{-1}\bJ_{0n}\transpose\widetilde{\bLambda}_\sharp\nabla_\bpsi\theta_e\|_2^2\bigg\}^{1/2} = O\bigg(\frac{\sqrt{m}}{n}\bigg).
\end{align*}
For the fourth term, by Cauchy--Schwarz inequality, Step 2 item (D), and \eqref{eqn:Ve_nuclear_norm}, we have 
\begin{align*}
&\bigg\|\frac{1}{2}\sum_{e\in\calE^\circ}\left\{\expect_0\ell''_e(\ell'_e)^2 - (\expect_0\ell''_e)\expect_0(\ell'_e)^2\right\}\widetilde{\scrP}_{\br^{(t)}}(\bG_0^\sharp)^{-1}\bJ_{0n}\transpose\widetilde{\bLambda}_\sharp\nabla_\bpsi\theta_e\\
&\quad\times\left\{
\nabla_\bpsi\theta_e\transpose\widetilde{\bLambda}_\sharp \bJ_{0n}(\bG_0^\sharp)^{-1}\bigg[\bLambda_\sharp\frac{\partial^2\theta_e}{\partial\bphi\partial\bphi\transpose}\bLambda_\sharp\bigg](\bG_0^\sharp)^{-1}\bJ_{0n}\transpose\widetilde{\bLambda}_\sharp\nabla_\bpsi\theta_e
\right\}\bigg\|_2\\
&\quad\lesssim \max_{e\in\calE}\|\bV_e\|_2\|(\bG_0^\sharp)^{-1}\bJ_{0n}\transpose\widetilde{\bLambda}_\sharp\nabla_\bpsi\theta_e\|_2^2 \sqrt{m}n \bigg\{\sum_{e\in\calE^\circ}\|\widetilde{\scrP}_{\br^{(t)}}(\bG_0^\sharp)^{-1}\bJ_{0n}\transpose\widetilde{\bLambda}_\sharp\nabla_\bpsi\theta_e\|_2^2\bigg\}^{1/2}\lesssim \frac{1}{\sqrt{m}n}. 
\end{align*}
For the fifth term, by Cauchy--Schwarz inequality and \eqref{eqn:V2_prime},
\begin{align*}
&\bigg\|\sum_{e\in\calE^\circ}\expect_0(\ell''_e - \expect_0\ell''_e)(\ell'_e)^2\widetilde{\scrP}_{\br^{(t)}}(\bG_0^\sharp)^{-1}\bigg[\bLambda_\sharp\frac{\partial^2\theta_e}{\partial\bphi\partial\bphi\transpose}\bLambda_\sharp\bigg](\bG_0^\sharp)^{-1}\bJ_{0n}\transpose\widetilde{\bLambda}_\sharp\nabla_\bpsi\theta_e\\
&\quad\times \left\{\nabla_\bpsi\theta_e\transpose\widetilde{\bLambda}_\sharp\bJ_{0n}(\bG_0^\sharp)^{-1}\bJ_{0n}\transpose \widetilde{\bLambda}_\sharp \nabla_\bpsi\theta_e\right\}\bigg\|_2\\
&\quad\lesssim \frac{1}{n}\sqrt{m}n\bigg\{\sum_{e\in\calE^\circ}\|\widetilde{\scrP}_{\br^{(t)}}(\bG_0^\sharp)^{-1}\bV_e(\bG_0^\sharp)^{-1}\bJ_{0n}\transpose\widetilde{\bLambda}_\sharp\nabla_\bpsi\theta_e\|_2^2\bigg\}^{1/2}
\lesssim \frac{\sqrt{m}}{\sqrt{mn}}\max_{e\in\calE}\|\bJ_{0n}\transpose\widetilde{\bLambda}_\sharp\nabla_\bpsi\theta_e\|_2\lesssim \frac{1}{n}.
\end{align*}
For the last term, by Cauchy--Schwarz inequality and Step 2 item (B) property (T2), 
\begin{align*}
&\max_{j\in[d(d + 1)/2]}\bigg|\sum_{e\in\calE^\circ}\expect_0(\ell'_e)^3 \tau_e\left(
(\bG_0^\sharp)^{-1}\bJ_{0n}\transpose\widetilde{\bLambda}_\sharp\nabla_\bpsi\theta_e,
(\bG_0^\sharp)^{-1}\bJ_{0n}\transpose\widetilde{\bLambda}_\sharp\nabla_\bpsi\theta_e,
(\bG_0^\sharp)^{-1}\widetilde{\scrP}_{\br^{(t)}}\transpose\be_j
\right)\bigg|\\
&\quad\lesssim \sqrt{m}n\bigg\{\sum_{e\in\calE^\circ}\tau_e\left(
(\bG_0^\sharp)^{-1}\bJ_{0n}\transpose\widetilde{\bLambda}_\sharp\nabla_\bpsi\theta_e,
(\bG_0^\sharp)^{-1}\bJ_{0n}\transpose\widetilde{\bLambda}_\sharp\nabla_\bpsi\theta_e,
(\bG_0^\sharp)^{-1}\widetilde{\scrP}_{\br^{(t)}}\transpose\be_j
\right)^2\bigg\}^{1/2}\lesssim \frac{1}{\sqrt{m}n}.
\end{align*}

\medskip\noindent\textbf{Step 3. Lindeberg condition of the leading term.} Since $\calT$ and $\bC$ are fixed and the bounds in Steps 1--2 and in Subsections \ref{sub:third_order_taylor_expansion_of_the_log_likelihood}--\ref{sub:controlling_the_bias_remainder_for_Rt_hat_Q2} hold uniformly in $t\in[m]$, multiplying \eqref{eqn:third_order_expansion_vech_Rt} by $\bC_t$ and summing over $t\in\calT$ shows that $\sum_{t\in\calT}\bC_t[n\{\vech(\widehat{\bR}^{(t)}) - \vech(\bR_0^{(t)})\} - \bb_n^{(t)}(\bpsi_0)]$ equals its leading term up to $o_p(1)$. The leading term has the form
\begin{align*}
\{\scrP_\calT(\bC)\bOmega(\bpsi_0)\scrP_\calT(\bC)\transpose\}^{-1/2}\sum_{t\in\calT}\sum_{e\in\calE^\circ}\bC_t\widetilde{\scrP}_{\br^{(t)}}(\bG_0^\sharp)^{-1}\bJ_{0n}\transpose\widetilde{\bLambda}_\sharp\ell'_e\frac{\partial\theta_e}{\partial\bpsi},
\end{align*}
where $\scrP_\calT(\bC)\bOmega(\bpsi_0)\scrP_\calT(\bC)\transpose = \{\sum_{t\in\calT}\bC_t\widetilde{\scrP}_{\br^{(t)}}\}(\bG_0^\sharp)^{-1}\{\sum_{t\in\calT}\bC_t\widetilde{\scrP}_{\br^{(t)}}\}\transpose$ has eigenvalues bounded away from $0$ and $\infty$ since $(\sum_{t}\bC_t\widetilde{\scrP}_{\br^{(t)}})(\sum_{t}\bC_t\widetilde{\scrP}_{\br^{(t)}})\transpose = \sum_{t\in\calT}\bC_t\bC_t\transpose$ is nondegenerate and the eigenvalues of $\bG_0^\sharp$ are bounded away from $0$ and $\infty$ by Proposition \ref{prop:expected_hessian_bound}, 
and
\begin{align*}
\sum_{e\in\calE^\circ}\expect_0\left\|\sum_{t\in\calT}\bC_t\widetilde{\scrP}_{\br^{(t)}}(\bG_0^\sharp)^{-1}\bJ_{0n}\transpose\widetilde{\bLambda}_\sharp\ell'_e\frac{\partial\theta_e}{\partial\bpsi}\right\|_2^3
&\lesssim \frac{1}{\sqrt{n}}\sum_{e\in\calE^\circ}\left\|\sum_{t\in\calT}\bC_t\widetilde{\scrP}_{\br^{(t)}}(\bG_0^\sharp)^{-1}\bJ_{0n}\transpose\widetilde{\bLambda}_\sharp\frac{\partial\theta_e}{\partial\bpsi}\right\|_2^2\\
& = \frac{1}{\sqrt{n}}\sum_{e\in\calE^\circ}\sum_j\bigg\{\be_j\transpose\scrP_{\calT}(\bC)(\bG_0^\sharp)^{-1}\bJ_{0n}\transpose\widetilde{\bLambda}_\sharp\frac{\partial\theta_e}{\partial\bpsi}\bigg\}^2\\
&\lesssim \frac{1}{\sqrt{n}}\sum_j\be_j\transpose\scrP_{\calT}(\bC)(\bG_0^\sharp)^{-1}\scrP_{\calT}(\bC)\transpose\be_j \lesssim \frac{1}{\sqrt{n}}\to 0
\end{align*}
as $n\to\infty$ by Step 2 item (D) in the proof of Lemma \ref{lemma:Q2_fluctuation}. The proof is thus completed. 
\end{proof}

\section{Plug-in inference}
\label{sec:plug_in_inference}

\subsection{Consistency of the plug-in estimators}
\label{sub:consistency_of_the_plug_in_estimators}

This subsection establishes that the plug-in estimators defined in Subsection \ref{sub:plugin_inference} and the plug-in estimated bias $\mathbf{b}_n^{(t)}(\widehat{\bpsi})$ are consistent for their respective estimands. We first prove a Lipschitz continuity property of the moments regarding the derivatives of $\ell(\theta; x)$. 

\begin{lemma}
\label{lemma:edge_moment_lipschitz}
Suppose Assumptions \ref{assumption:edge_distribution}--\ref{assumption:true_parameter} hold and $\max_{t\in[m],i,j\in[n]}|\theta_{0ij}^{(t)}|\leq b$. Then, there exist constants $c_0\in(0, 1]$ and $C_0 > 0$ depending only on $\kappa_2(b + 1)$, $\kappa_3(b + 1)$, and $\sigma$, such that for every $e = (t, i, j)\in\calE^\circ$ and every $\theta$ with $|\theta - \theta_{0ij}^{(t)}|\leq c_0$, with $A\sim p_A(\cdot; \theta)$,
\begin{align*}
\big|\expect_\theta\ell''(\theta; A) - \expect_0\ell''(\theta_{0ij}^{(t)}; A_{ij}^{(t)})\big| + \big|\expect_\theta\{\ell'(\theta; A)\ell''(\theta; A)\} - \expect_0\{\ell'(\theta_{0ij}^{(t)}; A_{ij}^{(t)})\ell''(\theta_{0ij}^{(t)}; A_{ij}^{(t)})\}\big|\leq C_0|\theta - \theta_{0ij}^{(t)}|.
\end{align*}
If in addition Assumption \ref{assumption:third_order_lipschitz_ell} holds, then 
\[ 
|\expect_\theta\ell'''(\theta; A) - \expect_0\ell'''(\theta_{0ij}^{(t)}; A_{ij}^{(t)})|\leq C_0|\theta - \theta_{0ij}^{(t)}|,
\]
with $C_0$ being a constant depending on $\kappa_2(b + 1)$, $\kappa_3(b + 1)$, and $\kappa_4(b + 1)$.
\end{lemma}

\begin{proof}[\bf Proof]
For a given and fixed $e = (t, i, j)\in\calE^\circ$, we write $\theta_0 = \theta_{0ij}^{(t)}$ for simplicity
 $\delta = |\theta - \theta_0|\leq c_0\leq 1$, $\kappa_k = \kappa_k(b + 1)$, and let $h \in \{\ell'', \ell'\ell'', \ell'''\}$. Denote $\Delta(x) := \ell(\theta; x) - \ell(\theta_0; x)$ the log-likelihood ratio so that $p_A(x; \theta) = p_A(x; \theta_0)\exp\{\Delta(x)\}$ for all $x\in\calX$. By Assumption \ref{assumption:edge_distribution}, we then have
\begin{align*}
\expect_\theta h(\theta; A) - \expect_0h(\theta_0; A_{ij}^{(t)}) = \expect_0\{h(\theta; A_{ij}^{(t)}) - h(\theta_0; A_{ij}^{(t)})\} + \expect_0 h(\theta; A_{ij}^{(t)})\{e^{\Delta(A_e)} - 1\}.
\end{align*}
Let $L := |\ell'(\theta_0; A_{ij}^{(t)})| + \kappa_2(b + 1)$. By the mean value theorem, Assumption \ref{assumption:edge_distribution}, the facts that $\max(|\theta|, |\theta_0|)\leq b + 1$ and $|\ell'(\bar\theta; A_{ij}^{(t)})|\leq L$ for every $\bar\theta$ between $\theta_0$ and $\theta$, we then have $|\Delta(A_{ij}^{(t)})|\leq\delta L$ and 
$|e^{\Delta(A_{ij}^{(t)})} - 1|\leq\delta Le^{\delta L}$. Also, 
\[
|h(\theta; A_{ij}^{(t)})|\leq\max\{\kappa_2(b + 1), \kappa_2(b + 1)L, \kappa_3(b + 1)\}
\] 
and 
\[
|h(\theta; A_{ij}^{(t)}) - h(\theta_0; A_{ij}^{(t)})|\leq\delta\max\{\kappa_3(b + 1), \kappa_2(b + 1)^2 + \kappa_3(b + 1)L, \kappa_4(b + 1)\}
\] 
for $h \in \{\ell'', \ell'\ell'', \ell'''\}$, respectively, using $(\ell'\ell'')' = (\ell'')^2 + \ell'\ell'''$ and Assumption \ref{assumption:third_order_lipschitz_ell} for the last case. 
Therefore, 
\[
|\expect_\theta h(\theta; A) - \expect_0h(\theta_0; A_{ij}^{(t)})|\leq C\delta\,\expect_0\{(1 + L)^2e^{\delta L}\}
\] 
for some constant $C$ that only depend on $\kappa_2(b + 1)$, $\kappa_3(b + 1)$, $\kappa_4(b + 1)$. 
By Assumption \ref{assumption:true_parameter}, $\expect_0e^{|\ell'(\theta_{0}; A_{ij}^{(t)})|/\sigma}\leq 2$ and $\expect_0(1 + L)^4\leq C$, where we can take $C$ to be larger than $\expect_0(1 + L)^4$. Now take $c_0 := \min\{1, 1/(2\sigma)\}$, so that we have $e^{\delta L}\leq e^{\kappa_2(b + 1)/(2\sigma)}e^{|\ell'(\theta_0; A_{ij}^{(t)})|/(2\sigma)}$. Then, by Cauchy--Schwarz inequality, we have 
\[
\expect_0\{(1 + L)^2e^{\delta L}\}\leq e^{\kappa_2(b + 1)/(2\sigma)}\{\expect_0(1 + L)^4\}^{1/2}\{\expect_0e^{|\ell'(\theta_{0}; A_{ij}^{(t)})|/\sigma}\}^{1/2}\leq C
\]
by picking $C$ large enough, and the proof is done.
\end{proof}

Next, we argue that the plug-in estimators are consistent. 
\begin{lemma}
\label{lemma:plug_in_consistency}
Suppose Assumptions \ref{assumption:edge_distribution}--\ref{assumption:true_parameter} hold, $m^{12}(\log n)^4/n\to 0$, and let $\delta_n := (m^2\sqrt{\log n} + m^{3/2}\log n)/\sqrt n$. Then, with $M$ as in Theorem \ref{thm:asymptotic_normality_MLE},
\begin{enumerate}[(i)]
\item $\max_{e\in\calE^\circ}|\theta_e(\widehat\bpsi) - \theta_{0e}| = \Optilde(\delta_n)$, where $\theta_e(\widehat{\bpsi}):=\widehat{\theta}_{ij}^{(t)}$ for $e = (t, i, j)\in\calE^\circ$ and $\theta_{0e} := \theta_{0ij}^{(t)}$;
\item $\max_{i\in[n]}\{\|\widehat\bSigma_{\bz_i}^\sharp - \bSigma_{\bz_i}^\sharp \|_2 + \|(\widehat\bSigma_{\bz_i}^\sharp)^{-1} - (\bSigma_{\bz_i}^\sharp )^{-1}\|_2 + \|\widehat\bSigma_{\bx_i}^\sharp - \bSigma_{\bx_i}^\sharp\|_2\} = \Optilde(\delta_n)$;
\item $\|\bOmega(\widehat\bpsi) - \bOmega(\bpsi_0)\|_2 = \Optilde(\delta_n)$;
\item if in addition, Assumption \ref{assumption:third_order_lipschitz_ell} holds, then $\max_{t\in[m]}\|\bb_n^{(t)}(\widehat\bpsi) - \bb_n^{(t)}(\bpsi_0)\|_2 = \Optilde(\sqrt m\,\delta_n)$.
\end{enumerate}
\end{lemma}

\begin{proof}[\bf Proof]
Throughout, $C$ denotes a constant depending only on $d, C_1, C_2, M$, $\{\kappa_k(b_0 + 1)\}_{k = 1}^4$ and $\sigma$. We also follow the convention that $\bW_{\widehat{\bX}} = \eye_d$ without loss of generality by the uniqueness in the sense of Proposition \ref{prop:stationarity_uniqueness_uniform_consistency_MLE}. 

\medskip\noindent\textbf{Step 1. Convergence rate preparations and item (i).} By Proposition \ref{prop:stationarity_uniqueness_uniform_consistency_MLE}, Theorem \ref{thm:two_to_infinity_norm_bound_MLE} and \eqref{eqn:Rhat_bound},
\begin{align}
\label{eqn:plug_in_rates}
&\max_{t\in[m]}\|\widehat\balpha^{(t)} - \balpha_0^{(t)}\|_\infty = \Optilde(\delta_n),\quad
\|\widehat\bX - \bX_0\|_{2\to\infty} = \Optilde\bigg(\sqrt{\frac{\log n}{mn}}\bigg),\nonumber\\
&\varepsilon := \max_{t\in[m]}\|\widehat\bR^{(t)} - \bR_0^{(t)}\|_{\mathrm F} + \|\widehat\bX - \bX_0\|_{2\to\infty} = \Optilde(\delta_n).
\end{align}
Since $\widehat\bpsi\in\Psi$, $\|\widehat\bX\|_{2\to\infty}\leq M/4$, $\|\widehat\bR^{(t)}\|_{\mathrm F}\leq M$ and $\|\widehat\balpha^{(t)}\|_\infty\leq M$, then, for $e = (t, i, j)$,
\begin{align*}
\max_{e\in\calE}|\theta_e(\widehat\bpsi) - \theta_{0e}|
&\leq 2\|\widehat\balpha^{(t)} - \balpha_0^{(t)}\|_\infty + \max_{i\in[n]}\|\widehat\bx_i - \bx_{0i}\|_2\|\widehat\bR^{(t)}\|_2\max_{j\in[n]}\|\widehat\bx_j\|_2\\
&\quad + \max_{i,j\in[n]}\|\bx_{0i}\|_2\|\widehat\bR^{(t)} - \bR_0^{(t)}\|_2\|\widehat\bx_j\|_2 + \max_{i\in[n]}\|\bx_{0i}\|_2\|\bR_0^{(t)}\|_2\max_{j\in[n]}\|\widehat\bx_j - \bx_{0j}\|_2 = \Optilde(\delta_n)
\end{align*}
uniformly in $e$, which is (i). In particular, $\max_e|\theta_e(\widehat\bpsi) - \theta_{0e}|\leq c_0$  w.h.p. with $c_0$ from Lemma \ref{lemma:edge_moment_lipschitz}, which implies
\begin{align}
\label{eqn:plug_in_moments}
\max_{e\in\calE^\circ}\big|\expect_{\widehat\bpsi}h(\theta_e(\widehat\bpsi); A_e) - \expect_0h(\theta_{0e}; A_e)\big| = \Optilde(\delta_n),h\in\{\ell'', \ell'\ell''\}\text{, and for }h = \ell'''\text{ under Assumption \ref{assumption:third_order_lipschitz_ell}}.
\end{align}

\medskip\noindent\textbf{Step 2. Deterministic bounds.} For $\bpsi\in\Psi$ and $\delta\bpsi\in\mathbb{R}^{d_\bpsi}$, write $\delta\alpha_{a}^{(t)}$, $\delta\br^{(t)}$, $\delta\bx_a$ denotes the $\alpha_a^{(t)}$, $\br^{(t)}$, $\bx_a$ blocks of $\delta\bpsi$, respectively.  
For $e = (t, a, b)$ and $\delta\bR^{(t)}\in\mathrm{Sym}(d)$ such that $\vech(\delta\bR^{(t)}) = \delta\br^{(t)}$, we have
\[
\frac{\partial\theta_e}{\partial\bpsi\transpose}(\bpsi)\widetilde{\bLambda}_\sharp\delta\bpsi = \frac{\delta \alpha_a^{(t)}}{\sqrt{n}} + \frac{\delta \alpha_b^{(t)}}{\sqrt{n}} + \frac{\bx_a\transpose\delta\bR^{(t)}\bx_b}{n} + \frac{(\bR^{(t)}\bx_b)\transpose\delta\bx_a}{\sqrt{mn}} + \frac{(\bR^{(t)}\bx_a)\transpose\delta\bx_b}{\sqrt{mn}}. 
\]
By Cauchy--Schwarz inequality, uniformly over $\bpsi\in\Psi$, we have
\begin{align}
\label{eqn:frame_bound_psi}
\sum_{e\in\calE^\circ}\bigg\{\frac{\partial\theta_e}{\partial\bpsi\transpose}(\bpsi)\widetilde{\bLambda}_\sharp\delta\bpsi\bigg\}^2
&\lesssim \bigg\{\sum_{t, a}(\delta \alpha_a^{(t)})^2 + \sum_t\|\delta\bR^{(t)}\|_2^2 + \frac{1}{mn}\sum_{t, a}\delta\bx_a\transpose(\bR^{(t)})^2\delta\bx_a\bigg\}\lesssim \|\delta\bpsi\|_2^2.
\end{align}
Then, by the norm duality and Cauchy--Schwarz inequality, $\|\sum_{e\in\calE^\circ}\alpha_e\widetilde{\bLambda}_\sharp(\partial\theta_e/\partial\bpsi)(\bpsi)\|_2\lesssim (\sum_{e\in\calE^\circ}\alpha_e^2)^{1/2}$ for any scalars $(\alpha_e)_{e\in\calE^\circ}$. The same computation applies to $(\partial\theta_e/\partial\bpsi)(\widehat{\bpsi}) - (\partial\theta_e/\partial\bpsi)(\bpsi_0)$, whose $\balpha$-block vanishes, whose $\vech(\bR^{(t)})$-block has Euclidean norm bounded by $O(\|\widehat\bX - \bX_0\|_{2\to\infty})$, and whose $\bx_a$- and $\bx_b$-blocks are $\widehat\bR^{(t)}\widehat\bx_b - \bR_0^{(t)}\bx_{0b}$ and $\widehat\bR^{(t)}\widehat\bx_a - \bR_0^{(t)}\bx_{0a}$, and the Euclidean norms of the $\bx_a$- and $\bx_b$-blocks are bounded by $O(\varepsilon)$. Therefore, the same reasoning leads to
\begin{equation}
\label{eqn:frame_bound_difference}
\begin{aligned}
\sum_{e\in\calE^\circ}\bigg[\bigg\{\frac{\partial\theta_e}{\partial\bpsi}(\widehat{\bpsi}) - \frac{\partial\theta_e}{\partial\bpsi}(\bpsi_0)\bigg\}\transpose\widetilde{\bLambda}_\sharp\delta\bpsi\bigg]^2&\lesssim \varepsilon^2\|\delta\bpsi\|_2^2,\\
\bigg\|\sum_{e\in\calE^\circ}\beta_e\widetilde{\bLambda}_\sharp\bigg\{\frac{\partial\theta_e}{\partial\bpsi}(\widehat{\bpsi}) - \frac{\partial\theta_e}{\partial\bpsi}(\bpsi_0)\bigg\}\bigg\|_2&\lesssim \varepsilon\bigg(\sum_{e\in\calE^\circ}\beta_e^2\bigg)^{1/2},
\end{aligned}
\end{equation}
and also $\sup_{\bpsi\in\Psi}\|\widetilde{\bLambda}_\sharp(\partial\theta_e/\partial\bpsi)(\bpsi)\|_2\lesssim n^{-1/2}$ and $\|\widetilde{\bLambda}_\sharp\{(\partial\theta_e/\partial\bpsi)(\widehat\bpsi) - (\partial\theta_e/\partial\bpsi)(\bpsi_0)\}\|_2\lesssim \varepsilon n^{-1/2}$ for all $e$.

\medskip\noindent\textbf{Step 3. Proof of item (ii).} For any $\bpsi\in\Psi$ and scalars $(\alpha_{tj}:t\in[m],j\in[n])$, it is clear that 
\begin{align*}
\sup_{\bpsi\in\Psi}\Bigg\|\frac{1}{n}\sum_{t = 1}^m\sum_{j\in[n]\backslash\{i\}}\alpha_{tj} \begin{bmatrix}
\be_t \be_t\transpose & \frac{1}{\sqrt{m}}\be_t\bx_j\transpose\bR^{(t)}\\
\frac{1}{\sqrt{m}}\bR^{(t)}\bx_j\be_t\transpose & \frac{1}{m}\bR^{(t)}\bx_j\bx_j\transpose\bR^{(t)}
\end{bmatrix}\Bigg\|_2\lesssim \max_{t\in[m],j\in[n]}|\alpha_{tj}|.
\end{align*}
Then, by Cauchy--Schwarz inequality, $\bM = \sup_{\|\bw\|_2 = 1}\bw\transpose\bM$ (Rayleigh--Ritz theorem), \eqref{eqn:plug_in_moments}, \eqref{eqn:frame_bound}, and \eqref{eqn:frame_bound_difference}
\begin{align*}
\|\widehat{\bSigma}_{\bz_i}^\sharp - \bSigma_{\bz_i}^\sharp \|_2
& \leq \Bigg\|\frac{1}{n}\sum_{t = 1}^m\sum_{j\in[n]\backslash\{i\}}
\left\{
\expect_{\widehat{\bpsi}}\ell''(\widehat{\theta}_{ij}^{(t)}; A_{ij}^{(t)}) - \expect_0\ell''(\theta_{0ij}^{(t)}; A_{ij}^{(t)})\right\}
\begin{bmatrix}
\be_t \be_t\transpose & \frac{1}{\sqrt{m}}\be_t\widehat\bx_j\transpose\widehat\bR^{(t)}\\
\frac{1}{\sqrt{m}}\widehat\bR^{(t)}\widehat\bx_j\be_t\transpose & \frac{1}{m}\widehat\bR^{(t)}\widehat\bx_j\widehat\bx_j\transpose\widehat\bR^{(t)}
\end{bmatrix}
\Bigg\|_2\\
&\quad + \kappa_2(b)\Bigg\|\frac{1}{n}\sum_{t = 1}^m\sum_{j\in[n]\backslash\{i\}}
\begin{bmatrix}
\be_t \be_t\transpose & \frac{1}{\sqrt{m}}\be_t(\widehat\bx_j\transpose\widehat\bR^{(t)} - \bx_{0j}\transpose\bR_0^{(t)})\\
\frac{1}{\sqrt{m}}(\widehat\bR^{(t)}\widehat\bx_j - \bR_0^{(t)}\bx_{0j})\be_t\transpose & \frac{1}{m}(\widehat\bR^{(t)}\widehat\bx_j\widehat\bx_j\transpose\widehat\bR^{(t)} - \bR_0^{(t)}\bx_{0j}\bx_{0j}\transpose\bR_0^{(t)})
\end{bmatrix}
\Bigg\|_2\\
&\lesssim \max_{t,j}\left|\expect_{\widehat{\bpsi}}\ell''(\widehat{\theta}_{ij}^{(t)}; A_{ij}^{(t)}) - \expect_0\ell''(\theta_{0ij}^{(t)}; A_{ij}^{(t)})\right|\\
&\quad + \Bigg\|\frac{1}{n}\sum_{t = 1}^m\sum_{j\in[n]\backslash\{i\}}
\begin{bmatrix}
\zero_m \\ \frac{1}{\sqrt{m}}(\widehat{\bR}^{(t)}\widehat{\bx}_j - \bR_0^{(t)}\bx_{0j})
\end{bmatrix}\begin{bmatrix}
\be_t\\ \frac{1}{\sqrt{m}}\widehat{\bR}^{(t)}\widehat{\bx}_{j}
\end{bmatrix}\transpose\Bigg\|_2\\
&\quad + \Bigg\|\frac{1}{n}\sum_{t = 1}^m\sum_{j\in[n]\backslash\{i\}}
\begin{bmatrix}
\be_t\\ \frac{1}{\sqrt{m}}\bR_0^{(t)}\bx_{0j}
\end{bmatrix}
\begin{bmatrix}
\zero_m \\ \frac{1}{\sqrt{m}}(\widehat{\bR}^{(t)}\widehat{\bx}_j - \bR_0^{(t)}\bx_{0j})
\end{bmatrix}\transpose\Bigg\|_2\\
&\lesssim \Optilde(\delta_n) + \Bigg\|\frac{1}{n}\sum_{t,j}
\begin{bmatrix}
\be_t\\ \frac{1}{\sqrt{m}}(\widehat{\bR}^{(t)}\widehat{\bx}_j - \bR_0^{(t)}\bx_{0j})
\end{bmatrix}
\begin{bmatrix}
\be_t\\ \frac{1}{\sqrt{m}}(\widehat{\bR}^{(t)}\widehat{\bx}_j - \bR_0^{(t)}\bx_{0j})
\end{bmatrix}\transpose\Bigg\|_2^{1/2}\\
&\lesssim \Optilde(\delta_n) + \eps = \Optilde(\delta_n). 
\end{align*}
Since the eigenvalues of $\bSigma_{\bz_i}^\sharp$ are bounded away from $0$ and $\infty$ uniformly in $i$, then so are the eigenvalues of $\widehat\bSigma_{\bz_i}^\sharp$ for all $i$ w.h.p., and $\|(\widehat\bSigma_{\bz_i}^\sharp)^{-1} - (\bSigma_{\bz_i}^\sharp)^{-1}\|_2\leq\|(\widehat\bSigma_{\bz_i}^\sharp)^{-1}\|_2\| \widehat\bSigma_{\bz_i}^\sharp - \bSigma_{\bz_i}^\sharp \|_2\|(\widehat\bSigma_{\bz_i}^\sharp)^{-1}\|_2 = \Optilde(\delta_n)$. We also note that $\bSigma_{\bx_i}^\sharp$ and $\widehat{\bSigma}_{\bx_i}^\sharp$ are submatrices of the two matrix inverses so that their eigenvalues are bounded accordingly.

\medskip\noindent\textbf{Step 4. Proof of (iii).} Recall the permutation matrix $\bPi$ defined in \eqref{eqn:permutation_matrix_psi} and the definition of $\bG_{\bpsi\bpsi}^\sharp(\bpsi)$ in Subsection \ref{sub:infinity_norm_of_the_rescaled_inverse_fisher_information_matrix}. 
Let $A_e = A_{ij}^{(t)}$ for $e = (t, i, j)\in\calE^\circ$ and
\[
\scrA(\bpsi) := \bPi\bG_{\bpsi\bpsi}^\sharp(\bpsi)\bPi\transpose = \sum_{e\in\calE^\circ}\{-\expect_\bpsi\ell''(\theta_e(\bpsi); A_e)\}\widetilde{\bLambda}_\sharp\frac{\partial\theta_e}{\partial\bpsi}(\bpsi)\frac{\partial\theta_e}{\partial\bpsi\transpose}(\bpsi)\widetilde{\bLambda}_\sharp  
\]
Recall from Subsection \ref{sub:R_inference} of the main text that $\bOmega(\bpsi) = \scrJ(\bpsi)\{\scrJ(\bpsi)\transpose\scrA(\bpsi)\scrJ(\bpsi)\}^{-1}\scrJ(\bpsi)\transpose$, where $\scrJ(\bpsi) = \mathrm{diag}(\eye_{d_\balpha + d_\br}, \eye_d\otimes\bU_\perp)$ and $\bU_\perp$ has orthonormal columns spanning $\mathrm{Span}(\one_n, \bX)^\perp$, and that $\scrJ(\bpsi_0)\transpose\scrA(\bpsi_0)\scrJ(\bpsi_0) = \bG_0^\sharp$ with $\scrJ(\bpsi_0) = \bJ_{0n}$. We write $\widehat\bU_\perp$ for the matrix $\bU_\perp$ associated with $\widehat\bpsi$.
By \eqref{eqn:frame_bound_psi}, 
\[
\|\scrA(\bpsi)\|_2\leq\kappa_2(b_0 + 1)\bigg\|\sum_e\widetilde{\bLambda}_\sharp\frac{\partial\theta_e}{\partial\bpsi}(\bpsi)\frac{\partial\theta_e}{\partial\bpsi\transpose}(\bpsi)\widetilde{\bLambda}_\sharp\bigg\|_2 = O(1)
\] 
for $\bpsi\in\{\bpsi_0, \widehat\bpsi\}$. Then, a simple algebra shows that
\begin{align*}
&\scrA(\widehat\bpsi) - \scrA(\bpsi_0)\\
&\quad = \sum_{e\in\calE^\circ}\{\expect_0\ell''(\theta_{0e}; A_e) - \expect_{\widehat{\bpsi}}\ell''(\theta_e(\widehat\bpsi); A_e)\}\widetilde{\bLambda}_\sharp\frac{\partial\theta_e}{\partial\bpsi}(\widehat\bpsi)\frac{\partial\theta_e}{\partial\bpsi\transpose}(\widehat\bpsi)\widetilde{\bLambda}_\sharp\\
&\qquad - \sum_{e\in\calE^\circ}\expect_0\ell''(\theta_{0e}; A_e)\widetilde{\bLambda}_\sharp\bigg[\bigg\{
\frac{\partial\theta_e}{\partial\bpsi}(\widehat\bpsi) - \frac{\partial\theta_e}{\partial\bpsi}(\bpsi_0)\bigg\}\frac{\partial\theta_e}{\partial\bpsi\transpose}(\widehat\bpsi) + \frac{\partial\theta_e}{\partial\bpsi}(\bpsi_0)\bigg\{
\frac{\partial\theta_e}{\partial\bpsi}(\widehat\bpsi) - \frac{\partial\theta_e}{\partial\bpsi}(\bpsi_0)\bigg\}\transpose\bigg]\widetilde{\bLambda}_\sharp.
\end{align*}
The first term has spectral norm $\Optilde(\delta_n)$ by Step 3 above, \eqref{eqn:plug_in_moments} and \eqref{eqn:frame_bound_psi}, and the second term has spectral norm $\Optilde(\delta_n)$ by the Cauchy--Schwarz inequality of Step 3, \eqref{eqn:frame_bound_psi} and \eqref{eqn:frame_bound_difference}. Hence $\|\scrA(\widehat\bpsi) - \scrA(\bpsi_0)\|_2 = \Optilde(\delta_n)$. We next work on $\scrJ(\widehat{\bpsi})$ and $\scrJ(\bpsi_0)$. Without loss of generality, we assume that $\widehat{\bU}_\perp\transpose\bU_{0\perp}$ is symmetric, where $\bU_{0\perp}:=\bX_{0\perp}/\sqrt{n}$. 
Denote by $\widehat\bP := n^{-1}\one_n\one_n\transpose + n^{-1}\widehat\bX\widehat\bX\transpose$ and $\bP_0 := n^{-1}\one_n\one_n\transpose + n^{-1}\bX_0\bX_0\transpose$ the orthogonal projections onto $\mathrm{Span}([\one_n, \widehat\bX])$ and $\mathrm{Span}([\one_n, \bX_0])$, respectively. By Theorem \ref{thm:two_to_infinity_norm_bound_MLE}, $\|\widehat\bP - \bP_0\|_2 = \Optilde(n^{-1/2})$, so that,
\begin{align*}
\|\scrJ(\widehat{\bpsi}) - \scrJ(\bpsi_0)\|_2
& = \|\widehat\bU_\perp - \bU_{0\perp}\|_2
\lesssim \|\sin\Theta(\widehat{\bU}_\perp, \bU_{0\perp})\|_2
\lesssim \|(\eye_n - \widehat{\bU}_\perp\widehat{\bU}_\perp\transpose) - (\eye_n - \bU_{0\perp}\bU_{0\perp}\transpose)\|_2\\
&\leq C\|\widehat\bP - \bP_0\|_2 = \Optilde(n^{-1/2}).
\end{align*}
Since $\|\scrJ(\widehat{\bpsi})\|_2 = \|\scrJ(\bpsi_0)\|_2 = 1$, then
\begin{align*}
\|\scrJ(\widehat{\bpsi})\transpose\scrA(\widehat{\bpsi})\scrJ(\widehat{\bpsi}) - \bG_0^\sharp\|_2& = \|\scrJ(\widehat{\bpsi})\transpose\scrA(\widehat{\bpsi})\scrJ(\widehat{\bpsi}) - \scrJ(\bpsi_0)\transpose\scrA(\bpsi_0)\scrJ(\bpsi_0)\|_2\\
&\leq 2\|\scrJ(\widehat{\bpsi}) - \scrJ(\bpsi_0)\|_2\|\scrA(\widehat\bpsi)\|_2 + \|\scrA(\widehat\bpsi) - \scrA(\bpsi_0)\|_2 = \Optilde(\delta_n). 
\end{align*}
Since $ \bG_0^\sharp$ has eigenvalues bounded away from $0$ by Proposition \ref{prop:expected_hessian_bound}, then the eigenvalues of $\scrJ(\widehat{\bpsi})\transpose\scrA(\widehat{\bpsi})\scrJ(\widehat{\bpsi})$ are bounded away from $0$ w.h.p., and 
\begin{align*}
&\left\|\left\{\scrJ(\widehat{\bpsi})\transpose\scrA(\widehat{\bpsi})\scrJ(\widehat{\bpsi})\right\}^{-1} - (\bG_0^\sharp)^{-1}\right\|_2\\
&\quad \leq\left\|\left\{\scrJ(\widehat{\bpsi})\transpose\scrA(\widehat{\bpsi})\scrJ(\widehat{\bpsi})\right\}^{-1}\right\|_2\left\|\scrJ(\widehat{\bpsi})\transpose\scrA(\widehat{\bpsi})\scrJ(\widehat{\bpsi}) - \bG_0^\sharp\right\|_2\|(\bG_0^\sharp)^{-1}\|_2 = \Optilde(\delta_n).  
\end{align*}
Therefore, we obtain
\begin{align*}
\|\bOmega(\widehat{\bpsi}) - \bOmega(\bpsi_0)\|_2& \leq \left\|\scrJ(\widehat{\bpsi}) - \scrJ(\bpsi_0)\right\|_2\left\|\left\{\scrJ(\widehat{\bpsi})\transpose\scrA(\widehat{\bpsi})\scrJ(\widehat{\bpsi})\right\}^{-1}\right\|_2 \left\|\scrJ(\widehat\bpsi)\right\|_2\\
&\quad + \left\|\left\{\scrJ(\widehat{\bpsi})\transpose\scrA(\widehat{\bpsi})\scrJ(\widehat{\bpsi})\right\}^{-1} - (\bG_0^\sharp)^{-1}\right\|_2 + \|(\bG_0^\sharp)^{-1}\|_2\|\scrJ(\widehat{\bpsi}) - \scrJ(\bpsi_0)\|_2 = \Optilde(\delta_n).
\end{align*}

\medskip\noindent\textbf{Step 5. Proof of (iv).} 
Denote by 
\begin{align*}
\bb_{n1}^{(t)}(\bpsi)
& = \sum_{s = 1}^m\sum_{1\leq i < j\leq n}\bigg\{
\expect_\bpsi \ell''(\theta_{ij}^{(s)}; A_{ij}^{(s)})\ell'(\theta_{ij}^{(s)}; A_{ij}^{(s)}) + \frac{1}{2}\expect_\bpsi \ell'''(\theta_{ij}^{(s)}; A_{ij}^{(s)})
\bigg\}\\
&\quad\times \bigg\{\frac{\partial\theta_{ij}^{(s)}}{\partial\bpsi\transpose}(\bpsi)\widetilde{\bLambda}_\sharp\bOmega(\bpsi)\widetilde{\bLambda}_\sharp\frac{\partial\theta_{ij}^{(s)}}{\partial\bpsi}(\bpsi)\bigg\}\scrP_{\br^{(t)}}\bOmega(\bpsi)\widetilde{\bLambda}_\sharp\frac{\partial\theta_{ij}^{(s)}}{\partial\bpsi}(\bpsi),\\
\bb_{n2}^{(t)}(\bpsi)
& = \frac{1}{2}\sum_{s = 1}^m\sum_{1\leq i < j\leq n}\expect_\bpsi\ell''(\theta_{ij}^{(s)}; A_{ij}^{(s)})\mathrm{tr}\bigg\{\widetilde{\bLambda}_\sharp\frac{\partial^2\theta_{ij}^{(s)}}{\partial\bpsi\partial\bpsi\transpose}(\bpsi)\widetilde{\bLambda}_\sharp\bOmega(\bpsi)\bigg\}
\scrP_{\br^{(t)}}\bOmega(\bpsi)\widetilde{\bLambda}_\sharp\frac{\partial\theta_{ij}^{(s)}}{\partial\bpsi}(\bpsi),\\
\bb_{n3}^{(t)}(\bpsi)
& = \frac{1}{2mn}\vech\left\{\bT(\bpsi)\bR^{(t)} + \bR^{(t)}\bT(\bpsi)\right\}.
\end{align*}
By Step 2 and (iii),
\begin{align*}
&\max_{t\in[m],i,j\in[n]}\bigg|
\frac{\partial\theta_{ij}^{(t)}}{\partial\bpsi\transpose}(\widehat\bpsi)\widetilde{\bLambda}_\sharp\bOmega(\widehat\bpsi)\widetilde{\bLambda}_\sharp\frac{\partial\theta_{ij}^{(t)}}{\partial\bpsi}(\widehat\bpsi) - 
\frac{\partial\theta_{ij}^{(t)}}{\partial\bpsi\transpose}(\bpsi_0)\widetilde{\bLambda}_\sharp\bOmega(\bpsi_0)\widetilde{\bLambda}_\sharp\frac{\partial\theta_{ij}^{(t)}}{\partial\bpsi}(\bpsi_0)
\bigg|\\
&\quad\leq\|\bOmega(\widehat{\bpsi}) - \bOmega(\bpsi_0)\|_2
\bigg\|\widetilde{\bLambda}_\sharp\frac{\partial\theta_{ij}^{(t)}}{\partial\bpsi}(\widehat\bpsi)\bigg\|_2^2\\
&\qquad + \|\bOmega(\bpsi_0)\|_2\bigg\{\bigg\|\widetilde{\bLambda}_\sharp\frac{\partial\theta_{ij}^{(t)}}{\partial\bpsi\transpose}(\widehat\bpsi)\bigg\|_2 + \bigg\|\widetilde{\bLambda}_\sharp\frac{\partial\theta_{ij}^{(t)}}{\partial\bpsi\transpose}(\bpsi_0)\bigg\|_2\bigg\}\bigg\|\widetilde{\bLambda}_\sharp\bigg\{\frac{\partial\theta_{ij}^{(t)}}{\partial\bpsi\transpose}(\widehat\bpsi) - \frac{\partial\theta_{ij}^{(t)}}{\partial\bpsi\transpose}(\bpsi_0)\bigg\}\bigg\|_2 = \Optilde\bigg(\frac{\delta_n}{n}\bigg).
\end{align*}
Next, for $e = (t, a, b)\in\calE^{\circ}$, $\nabla^2_\bpsi\theta_e(\bpsi)$ has nonzero entries only in the $(\bx_a, \bx_b)$-, $(\bx_b, \bx_a)$-blocks equaling to $\bR^{(t)}$ and in the $(\bx_a, \vech(\bR^{(t)}))$-, $(\vech(\bR^{(t)}), \bx_a)$-, $(\bx_b, \vech(\bR^{(t)}))$-, $(\vech(\bR^{(t)}), \bx_b)$-blocks, so $\widetilde{\bLambda}_\sharp\nabla^2_\bpsi\theta_e(\bpsi)\widetilde{\bLambda}_\sharp$ has rank at most $2d + 2d_r$ (with $d_r = d(d + 1)/2$) and entries bounded by $O(1/(mn))$ (the $\bx\bx$-blocks) and $O(1/(n\sqrt{mn}))$ (the $\bx\br$-blocks). Therefore, $\|\widetilde{\bLambda}_\sharp\nabla^2_\bpsi\theta_e(\bpsi)\widetilde{\bLambda}_\sharp\|_* = O(1/(mn))$, and the same holds for $\|\widetilde{\bLambda}_\sharp\{\nabla^2_\bpsi\theta_e(\widehat\bpsi) - \nabla^2_\bpsi\theta_e(\bpsi_0)\}\widetilde{\bLambda}_\sharp \|_* = \Optilde(\varepsilon/(mn))$ by the same reasoning.

For simplicity, for $e = (t, i, j)\in\calE^\circ$, denote 
\begin{align*}
&\expect_{\widehat{\bpsi}}\widehat{\ell}'_e = \expect_{\widehat{\bpsi}}\ell'(\theta_e(\hat{\bpsi}); A_{ij}^{(t)}),\; 
\expect_0{\ell}'_e = \expect_0\ell'(\theta_{0ij}^{(t)}; A_{ij}^{(t)}),\;
\expect_{\widehat{\bpsi}}\widehat{\ell}''_e = \expect_{\widehat{\bpsi}}\ell''(\theta_e(\hat{\bpsi}); A_{ij}^{(t)}),\;
\expect_0{\ell}''_e = \expect_0\ell''(\theta_{0ij}^{(t)}; A_{ij}^{(t)}),\\
&\expect_{\widehat{\bpsi}}\widehat{\ell}''_e\widehat{\ell}'_e = \expect_{\widehat{\bpsi}}
\ell''(\theta_e(\hat{\bpsi}); A_{ij}^{(t)})
\ell'(\theta_e(\hat{\bpsi}); A_{ij}^{(t)}),\;
\expect_0{\ell}''_e{\ell}'_e = \expect_0
\ell''(\theta_{0ij}^{(t)}; A_{ij}^{(t)})
\ell'(\theta_{0ij}^{(t)}; A_{ij}^{(t)}),\\
& \expect_{\widehat{\bpsi}}\widehat{\ell}'''_e = \expect_{\widehat{\bpsi}}\ell'''(\theta_e(\hat{\bpsi}); A_{ij}^{(t)}),\; 
\expect_0{\ell}'_e = \expect_0\ell'(\theta_{0ij}^{(t)}; A_{ij}^{(t)}),\\
&\widehat{\alpha}_e = \expect_{\widehat{\bpsi}}\widehat{\ell}''_e\widehat{\ell}'_e + \frac{1}{2}\expect_{\widehat{\bpsi}}\widehat\ell'''_e,\;
\alpha_e = \expect_0\ell''_e\ell'_e + \frac{1}{2}\expect_0\ell'''_e,\;
\widehat\delta_e = \frac{\partial\theta_{ij}^{(t)}}{\partial\bpsi\transpose}(\widehat{\bpsi})\widetilde{\bLambda}_\sharp\bOmega(\widehat{\bpsi})\widetilde{\bLambda}_\sharp\frac{\partial\theta_{ij}^{(t)}}{\partial\bpsi}(\widehat{\bpsi}),\\
&\delta_e = \frac{\partial\theta_{ij}^{(t)}}{\partial\bpsi\transpose}(\bpsi_0)\widetilde{\bLambda}_\sharp\bOmega(\bpsi_0)\widetilde{\bLambda}_\sharp\frac{\partial\theta_{ij}^{(t)}}{\partial\bpsi}(\bpsi_0).
\end{align*}

\medskip\noindent
\underline{Consistency of $\bb_{n1}^{(t)}(\widehat{\bpsi})$:}
Now write
\begin{align*}
\bb_{n1}^{(t)}(\widehat\bpsi) - \bb_{n1}^{(t)}(\bpsi_0)
& = \scrP_{\br^{(t)}}\{\bOmega(\widehat{\bpsi}) - \bOmega(\bpsi_0)\}\sum_e\widehat \alpha_e\widehat \delta_e\widetilde{\bLambda}_\sharp
\frac{\partial\theta_e}{\partial\bpsi}(\widehat{\bpsi})
 + \scrP_{\br^{(t)}}\bOmega(\bpsi_0)\sum_e(\widehat \alpha_e\widehat \delta_e - \alpha_e\delta_e)\widetilde{\bLambda}_\sharp\frac{\partial\theta_e}{\partial\bpsi}(\widehat{\bpsi})\\
&\quad + \scrP_{\br^{(t)}}\bOmega(\bpsi_0)\sum_e\alpha_e\delta_e\widetilde{\bLambda}_\sharp\bigg\{\frac{\partial\theta_e}{\partial\bpsi}(\widehat{\bpsi}) - \frac{\partial\theta_e}{\partial\bpsi}(\bpsi_0)\bigg\}.
\end{align*}
By \eqref{eqn:frame_bound_psi} at $\bpsi = \widehat\bpsi$ and Cauchy--Schwarz inequality, 
\[
\bigg\|\sum_e\widehat \alpha_e\widehat \delta_e\widetilde{\bLambda}_\sharp\frac{\partial\theta_e}{\partial\bpsi}(\widehat{\bpsi})\bigg\|_2\lesssim \bigg(\sum_e\widehat \alpha_e^2\widehat \delta_e^2\bigg)^{1/2}\lesssim (mn^2\cdot n^{-2})^{1/2} = O(\sqrt{m})
\]
so the first term is $\Optilde(\sqrt m\delta_n)$. Since $|\widehat \alpha_e\widehat \delta_e - \alpha_e \delta_e|\leq|\widehat \alpha_e - \alpha_e|\widehat \delta_e + |\alpha_e||\widehat \delta_e - \delta_e| = \Optilde(\delta_n/n)$, by the same reasoning, the second term is $\Optilde\{(mn^2\cdot\delta_n^2/n^2)^{1/2}\} = \Optilde(\sqrt m\delta_n)$, and the third term is $\Optilde\{\varepsilon(mn^2\cdot n^{-2})^{1/2}\} = \Optilde(\sqrt m\delta_n)$ by \eqref{eqn:frame_bound_difference}. 

\medskip\noindent\underline{Consistency of $\bb_{n2}^{(t)}(\widehat{\bpsi})$.}
Write
\begin{align*}
\bb_{n2}^{(t)}(\widehat{\bpsi}) - \bb_{n2}^{(t)}(\bpsi_0)
& = \frac{1}{2}\scrP_{\br^{(t)}}\{\bOmega(\widehat\bpsi) - \bOmega(\bpsi_0)\}\\
&\quad\times\sum_e\expect_{\widehat{\bpsi}}\widehat{\ell}''_e \mathrm{tr} \bigg\{\widetilde{\bLambda}_\sharp\frac{\partial^2\theta_e}{\partial\bpsi\partial\bpsi\transpose}(\widehat{\bpsi})\widetilde{\bLambda}_\sharp\bOmega(\widehat{\bpsi})\bigg\} \widetilde{\bLambda}_\sharp\frac{\partial\theta_e}{\partial\bpsi}(\widehat\bpsi)\\
&\quad + \frac{1}{2}\scrP_{\br^{(t)}}\bOmega(\bpsi_0)\\
&\quad\times\sum_e\bigg[\expect_{\widehat{\bpsi}}\widehat{\ell}''_e\mathrm{tr} \bigg\{\widetilde{\bLambda}_\sharp\frac{\partial^2\theta_e}{\partial\bpsi\partial\bpsi\transpose}(\widehat{\bpsi})\widetilde{\bLambda}_\sharp\bOmega(\widehat{\bpsi})\bigg\}\\
&\qquad - \expect_0 {\ell}''_e\mathrm{tr} \bigg\{\widetilde{\bLambda}_\sharp\frac{\partial^2\theta_e}{\partial\bpsi\partial\bpsi\transpose}(\bpsi_0)\widetilde{\bLambda}_\sharp\bOmega(\bpsi_0)\bigg\}\bigg] \widetilde{\bLambda}_\sharp\frac{\partial\theta_e}{\partial\bpsi}(\widehat\bpsi)\\
&\quad + \frac{1}{2}\scrP_{\br^{(t)}}\bOmega(\bpsi_0)\sum_e\expect_0 {\ell}''_e\mathrm{tr} \bigg\{\widetilde{\bLambda}_\sharp\frac{\partial^2\theta_e}{\partial\bpsi\partial\bpsi\transpose}(\bpsi_0)\widetilde{\bLambda}_\sharp\bOmega(\bpsi_0)\bigg\}\widetilde{\bLambda}_\sharp\bigg\{\frac{\partial\theta_e}{\partial\bpsi}(\widehat\bpsi) - \frac{\partial\theta_e}{\partial\bpsi}(\bpsi_0)\bigg\}.
\end{align*}
By von Neumann's trace inequality, 
\begin{align*}
&\sup_{\bpsi\in\Psi}\bigg|\mathrm{tr}\bigg\{\widetilde{\bLambda}_\sharp\frac{\partial^2\theta_e}{\partial\bpsi\partial\bpsi\transpose}(\widehat{\bpsi})\widetilde{\bLambda}_\sharp\bOmega(\widehat{\bpsi})\bigg\}\bigg|\leq \sup_{\bpsi\in\Psi}\bigg\|\widetilde{\bLambda}_\sharp\frac{\partial^2\theta_e}{\partial\bpsi\partial\bpsi\transpose}(\widehat{\bpsi})\widetilde{\bLambda}_\sharp\bigg\|_2\|\bOmega(\widehat{\bpsi})\|_2 \lesssim \frac{1}{mn},\\
&\bigg|\mathrm{tr}\bigg\{\widetilde{\bLambda}_\sharp\frac{\partial^2\theta_e}{\partial\bpsi\partial\bpsi\transpose}(\widehat{\bpsi})\widetilde{\bLambda}_\sharp\bOmega(\widehat{\bpsi}) - \widetilde{\bLambda}_\sharp\frac{\partial^2\theta_e}{\partial\bpsi\partial\bpsi\transpose}({\bpsi}_0)\widetilde{\bLambda}_\sharp\bOmega({\bpsi}_0)\bigg\}\bigg|\\
&\quad \leq \bigg|\mathrm{tr}\bigg[\bigg\{\widetilde{\bLambda}_\sharp\frac{\partial^2\theta_e}{\partial\bpsi\partial\bpsi\transpose}(\widehat{\bpsi})\widetilde{\bLambda}_\sharp - \widetilde{\bLambda}_\sharp\frac{\partial^2\theta_e}{\partial\bpsi\partial\bpsi\transpose}({\bpsi}_0)\widetilde{\bLambda}_\sharp\bigg\}\bOmega(\widehat{\bpsi}) \bigg]\bigg| + \bigg|\mathrm{tr}\bigg[ \widetilde{\bLambda}_\sharp\frac{\partial^2\theta_e}{\partial\bpsi\partial\bpsi\transpose}({\bpsi}_0)\widetilde{\bLambda}_\sharp\{\bOmega(\widehat{\bpsi}) - \bOmega({\bpsi}_0)\}\bigg]\bigg|\\
&\quad = \Optilde\bigg(\frac{\eps + \delta_n}{mn}\bigg).
\end{align*}
Then, by the same reasoning as in the consistency of $\bb_{n1}^{(t)}(\widehat{\bpsi})$, the three terms in the decomposition of $\bb_{n2}^{(t)}(\widehat{\bpsi}) - \bb_{n2}^{(t)}(\bpsi_0)$ are $\Optilde(\delta_n/\sqrt{m})$, $\Optilde(\delta_n/\sqrt{m})$, and $\Optilde(\eps/\sqrt{m})$, so that $\|\bb_{n2}^{(t)}(\widehat\bpsi) - \bb_{n2}^{(t)}(\bpsi_0)\|_2 = \Optilde(m^{-1/2}\delta_n)$. 

\medskip\noindent\underline{Consistency of $\bb_{n3}^{(t)}(\widehat{\bpsi})$}
For $\bb_{n3}^{(t)}(\cdot)$, $\|\bT(\bpsi)\|_2\leq n\|\bOmega(\bpsi)\|_2 = O(n)$ and $\|\bT(\widehat\bpsi) - \bT(\bpsi_0)\|_2= \Optilde(n\delta_n)$, so that
\begin{align*}
\|\bb_{n3}^{(t)}(\widehat\bpsi) - \bb_{n3}^{(t)}(\bpsi_0)\|_2\lesssim \frac{1}{mn}\Big\{ \|\bT(\widehat\bpsi) - \bT(\bpsi_0)\|_2\|\widehat\bR^{(t)}\|_2 +  \|\bT(\bpsi_0)\|_2\|\widehat\bR^{(t)} - \bR_0^{(t)}\|_2\Big\} = \Optilde\bigg(\frac{\delta_n + \varepsilon}{m}\bigg),
\end{align*}
and the proof is done. 
\end{proof}

\subsection{Proofs of Corollaries \ref{corr:plug_in_vertex} and \ref{corr:plug_in_R}}
\label{sub:proofs_of_corollaries_plug_in}

Let $\bQ\in\mathbb{O}(d)$ and let $\bpsi\mapsto\bpsi_\bQ := (\balpha, \bQ\transpose\bR^{(t)}\bQ, \bX\bQ)$ denote the simultaneous rotation that leaves the likelihood invariant. Since $\theta_e(\bpsi_\bQ) = \theta_e(\bpsi)$, we have
\begin{align*}
\frac{\partial\theta_e}{\partial\bpsi}(\bpsi_\bQ) &= 
\mathrm{diag}(\eye_{d_\balpha}, \eye_m\otimes\bM_\bQ, \bQ\transpose\otimes\eye_n)^{-\mathrm T}\frac{\partial\theta_e}{\partial\bpsi}(\bpsi),\\
\frac{\partial^2\theta_e}{\partial\bpsi\partial\bpsi\transpose}(\bpsi_\bQ) &
= \mathrm{diag}(\eye_{d_\balpha}, \eye_m\otimes\bM_\bQ, \bQ\transpose\otimes\eye_n)^{-\mathrm T}\frac{\partial^2\theta_e}{\partial\bpsi\partial\bpsi\transpose}(\bpsi)\mathrm{diag}(\eye_{d_\balpha}, \eye_m\otimes\bM_\bQ, \bQ\transpose\otimes\eye_n)^{-1},
\end{align*}
where $\bM_\bQ := \bD_d^\dagger(\bQ\transpose\otimes\bQ\transpose)\bD_d$ is the matrix with $\vech(\bQ\transpose\bR\bQ) = \bM_\bQ\vech(\bR)$ and $\bD_d^\dagger$ denotes the Moore--Penrose inverse of $\bD_d$. Since $\mathrm{Span}([\one_n, \bX\bQ]) = \mathrm{Span}([\one_n, \bX])$ and
\begin{align*}
\mathrm{diag}(\eye_{d_\balpha}, \eye_m\otimes\bM_\bQ, \bQ\transpose\otimes\eye_n)\scrJ(\bpsi) = \scrJ(\bpsi) \mathrm{diag}(\eye_{d_\balpha}, \eye_m\otimes\bM_\bQ, \bQ\transpose\otimes\eye_{n - 1 - d}),
\end{align*}
then, a direct computation shows that
\begin{align*}
\bDelta_{\bz_i}^\sharp(\bpsi_\bQ)& = \mathrm{diag}(\eye_m, \bQ)\transpose\bDelta_{\bz_i}^\sharp(\bpsi)\mathrm{diag}(\eye_m, \bQ),\\
\bOmega(\bpsi_\bQ)& = \mathrm{diag}(\eye_{d_\balpha}, \eye_m\otimes\bM_\bQ, \bQ\transpose\otimes\eye_n)\bOmega(\bpsi)\mathrm{diag}(\eye_{d_\balpha}, \eye_m\otimes\bM_\bQ, \bQ\transpose\otimes\eye_n)\transpose,\\
\bT(\bpsi_\bQ)& = \bQ\transpose\bT(\bpsi)\bQ,\qquad
\bb_n^{(t)}(\bpsi_\bQ) = \bM_\bQ\bb_n^{(t)}(\bpsi).
\end{align*}
Consequently, the two corollaries can be recovered by observing that we can denote $\widehat\bpsi^{\,\circ}$ as the maximum likelihood estimator as computed so that $\widehat\bpsi = \widehat\bpsi^{\,\circ}_{\bW_{\widehat\bX}} = \widehat{\bpsi}^a$.

\begin{proof}[\bf Proof of Corollary \ref{corr:plug_in_vertex}]
For item (i), by Theorem \ref{thm:asymptotic_normality_MLE}(i), 
\begin{align*}
\frac{\sum_{i\in\calI}\{\sqrt{n}\bg_{\balpha_i}\transpose(\widehat{\balpha}_i - \balpha_{0i})
+ \sqrt{mn}\bg_{\bx_i}\transpose(\bW_{\widehat{\bX}}\transpose\widehat{\bx}_i - \bx_{0i})\} }{\sqrt{\sum_{i\in\calI}\bg_i\transpose (\bSigma_{\bz_i}^\sharp)^{-1}\bg_i}}\overset{\calL}{\to}\mathrm{N}(0, 1)\quad\text{as }n\to\infty.
\end{align*}
Since the eigenvalues of $ \bSigma_{\bz_i}^\sharp $ are bounded away from $0$ and $\infty$ uniformly over $i\in[n]$ and $\|\bg_i\|_2 = 1$, then
\begin{align*}
\bigg|\sum_{i\in\calI}\bg_i\transpose \{(\widehat\bSigma_{\bz_i}^\sharp)^{-1} - (\bSigma_{\bz_i}^\sharp)^{-1}\}\bg_i\bigg|\leq\sum_{i\in\calI}\|\bSigma_{\bz_i}^\sharp(\widehat\bpsi)^{-1} - \bSigma_{\bz_i}^\sharp(\bpsi_0)^{-1}\|_2 = \Optilde(\delta_n)
\end{align*}
by Lemma \ref{lemma:plug_in_consistency} (ii), so that item (i) holds by Slutsky's theorem. For item (ii), we know that
\[
4\sum_{i = 1}^n\be_i\transpose\bF\bX_0\bSigma_{\bx_i}^\sharp \bX_0\transpose\bF\be_i\geq 4\min_{i\in[n]}\lambda_{\min}(\bSigma_{\bx_i}^\sharp)\|\bF\bX_0\|_{\mathrm{F}}^2 = 4\min_{i\in[n]}\lambda_{\min}(\bSigma_{\bx_i}^\sharp)
\]
and it is bounded away from $0$. 
Writing 
\begin{align*}
\widehat\bX\widehat{\bSigma}_{\bx_i}^\sharp \widehat\bX\transpose - \bX_0\bSigma_{\bx_i}^\sharp \bX_0\transpose = (\widehat\bX - \bX_0)\widehat\bSigma_{\bx_i}^\sharp \widehat\bX\transpose + \bX_0(\widehat\bSigma_{\bx_i}^\sharp - \bSigma_{\bx_i}^\sharp)\widehat\bX\transpose + \bX_0\bSigma_{\bx_i}^\sharp(\widehat\bX - \bX_0)\transpose  
\end{align*}
and 
\begin{align*}
\|\bF(\widehat\bX - \bX_0)\|_{\mathrm F} = \bigg\|\sum_{a = 1}^n\sum_{b = 1}^nF_{ab}\be_a(\widehat\bx_b - \bx_{0b})\transpose\bigg\|_{\mathrm F}\leq\|\vect(\bF)\|_1\|\widehat\bX - \bX_0\|_{2\to\infty} = \Optilde\bigg(\sqrt{\frac{\log n}{mn}}\bigg),  
\end{align*}
then, by $\|\bF\widehat\bX\|_{\mathrm F}\leq 1 + \|\bF(\widehat\bX - \bX_0)\|_{\mathrm F}$, $\max_i\|\widehat\bSigma_{\bx_i}^\sharp \|_2 = \Optilde(1)$, Cauchy--Schwarz inequality, and Lemma \ref{lemma:plug_in_consistency} (ii), we obtain
\begin{align*}
&\bigg|4\sum_{i = 1}^n\be_i\transpose\bF(\widehat{\bX}\widehat{\bSigma}^\sharp_{\bx_i}\widehat{\bX}\transpose - \bX_0\bSigma_{\bx_i}^\sharp \bX_0\transpose)\bF\be_i\bigg|\\
&\quad\leq 4\sum_{i = 1}^n \left\{\|\be_i\transpose\bF(\widehat{\bX} - \bX_0)\|_2 (\|\widehat{\bSigma}_{\bx_i}^\sharp\|_2 \|\widehat{\bX}\transpose\bF\be_i\|_2 + \|\bSigma_{\bx_i}^\sharp\|_2\|\bX_0\transpose\bF\be_i\|_2) + \|\be_i\transpose\bF\bX_0\|_2\|\widehat{\bSigma}_{\bx_i}^\sharp  - \bSigma_{\bx_i}^\sharp\|_2 \|\widehat{\bX}\transpose\bF\be_i\|_2\right\} \\
&\quad\leq 4\Big\{\Optilde(1)\|\bF(\widehat\bX - \bX_0)\|_{\mathrm F}\big(\|\bF\widehat\bX\|_{\mathrm F} + \|\bF\bX_0\|_{\mathrm F}\big) + \max_i\|\widehat\bSigma_{\bx_i}^\sharp  - \bSigma_{\bx_i}^\sharp \|_2\|\bF\bX_0\|_{\mathrm F}\|\bF\widehat\bX\|_{\mathrm F}\Big\} = \Optilde(\delta_n).
\end{align*}
The conclusion follows from Theorem \ref{thm:asymptotic_normality_MLE} (ii) and Slutsky's theorem as in (i).
\end{proof}

\begin{proof}[\bf Proof of Corollary \ref{corr:plug_in_R}]
By the rotation equivariance above we may take $\bW_{\widehat\bX} = \eye_d$, so that $\widehat\bpsi^{\rm a} = \widehat\bpsi$. Let $\bSigma_\calT := \scrP_\calT(\bC)\bOmega(\bpsi_0)\scrP_\calT(\bC)\transpose$ and $\widehat{\bSigma}_\calT := \scrP_\calT(\bC)\bOmega(\widehat\bpsi)\scrP_\calT(\bC)\transpose$. Since $\|\scrP_\calT(\bC)\|_2 = \|\sum_{t\in\calT}\bC_t\bC_t\transpose\|_2^{1/2}$, Lemma \ref{lemma:plug_in_consistency} (iii)--(iv) implies
\[
\|\widehat{\bSigma}_\calT - \bSigma_\calT\|_2\leq\bigg\|\sum_{t\in\calT}\bC_t\bC_t\transpose\bigg\|_2\|\bOmega(\widehat\bpsi) - \bOmega(\bpsi_0)\|_2 = \Optilde(\delta_n),\quad
\bigg\|\sum_{t\in\calT}\bC_t\{\bb_n^{(t)}(\widehat\bpsi) - \bb_n^{(t)}(\bpsi_0)\}\bigg\|_2 = \Optilde(\sqrt m\,\delta_n),
\]
and $\sqrt m\,\delta_n\to 0$ under $m^{13}(\log n)^9/n\to 0$. Since the eigenvalues of $\bSigma_\calT$ are bounded away from $0$ and $\infty$ by Theorem \ref{thm:asymptotic_normality_R}, $\widehat{\bSigma}_\calT^{-1/2}\bSigma_\calT^{1/2} = \eye_{d'} + o_p(1)$. Writing $\bv_n := \sum_{t\in\calT}\bC_t[n\{\vech(\widehat\bR^{(t)}) - \vech(\bR_0^{(t)})\} - \bb_n^{(t)}(\bpsi_0)]$, we have
\begin{align*}
&\widehat{\bSigma}_\calT^{-1/2}\sum_{t\in\calT}\bC_t\big[n\{\vech(\widehat\bR^{(t)}) - \vech(\bR_0^{(t)})\} - \bb_n^{(t)}(\widehat\bpsi)\big]\\
&\quad = (\widehat{\bSigma}_\calT^{-1/2}\bSigma_\calT^{1/2})\bSigma_\calT^{-1/2}\bv_n
 + \widehat{\bSigma}_\calT^{-1/2}\sum_{t\in\calT}\bC_t\{\bb_n^{(t)}(\bpsi_0) - \bb_n^{(t)}(\widehat\bpsi)\}\overset{\calL}{\to}\mathrm{N}(\zero_{d'}, \eye_{d'})
\end{align*}
by Theorem \ref{thm:asymptotic_normality_R} and Slutsky's theorem, and the proof is done.
\end{proof}

\section{Asymptotic normality of the one-step estimator}
\label{sec:asymptotic_normality_of_the_one_step_estimator}

\subsection{Stochastic expansion of the preliminary estimators}
\label{sub:asymptotic_normality_of_the_preliminary_estimators}

This subsection first establishes the first-order stochastic expansion of the preliminary estimators generated by aggregating the layer-wise maximum likelihood estimators. Specifically, denote $\calL^{(t)}$ the log-likelihood of the $t$th layer defined by
\[
\calL^{(t)}(\balpha^{(t)},\bX,\bR^{(t)}) = \sum_{i < j}\ell(\alpha_i^{(t)} + \alpha_j^{(t)} + \bx_i\transpose\bR^{(t)}\bx_j)
\]
with the corresponding parameter space defined by
\begin{align*}
\Psi^{(t)} = \left\{(\balpha^{(t)},\bR^{(t)},\bX):\|\balpha^{(t)}\|_\infty \leq M, \|\bR^{(t)}\|_{\mathrm{F}}\leq M, \|\bX\|_{2\to\infty}\leq M/4, \bX\transpose\bX = n\eye_d, \bX\transpose\one_n = \zero_d\right\},
\end{align*}
and define
\begin{align}
\label{eqn:layer_wise_MLE}
(\widetilde{\balpha}^{(t)}, \widetilde{\bX}^{(t)}, \overline{\bR}^{(t)}) = \argmax_{(\balpha^{(t)},\bX,\bR^{(t)})\in\Psi^{(t)}}\calL^{(t)}(\balpha^{(t)},\bX,\bR^{(t)}).
\end{align}
Let $\widetilde{\bX}/\sqrt{n}$ be the eigenvector matrix with orthonormal columns corresponding to the $d$-largest eigenvalues of $(1/m)\sum_{t = 1}^m\widetilde{\bX}^{(t)}\widetilde{\bX}^{(t)\mathrm{T}}$ and $\widetilde{\bR}^{(t)} = (1/n^2)\widetilde{\bX}\transpose\widetilde{\bX}^{(t)}\overline{\bR}^{(t)}\widetilde{\bX}^{(t)\mathrm{T}}\widetilde{\bX}$. Let $\widetilde{\alpha}_i^{(t)}$ denote the $i$th element of $\widetilde{\balpha}^{(t)}$ and let $\widetilde{\balpha}_i = [\widetilde{\alpha}_i^{(1)},\ldots,\widetilde{\alpha}_i^{(m)}]\transpose$. 

\begin{proposition}\label{prop:stochastic_expansion_aggregated_estimators}
Suppose Assumptions \ref{assumption:edge_distribution}--\ref{assumption:true_parameter} hold and $m(\log n)^6/n = O(1)$. 
Let $\widetilde{\bX}$ be the aggregated estimator defined as above. Denote by 
\[
\bOmega_i^{(t)} = -\frac{1}{n}\sum_{j\in[n]\backslash\{i\}}\expect_0\ell''(\theta_{0ij}^{(t)}; A_{ij}^{(t)})\begin{bmatrix}
1 & \bx_{0j}\transpose\bR_0^{(t)}\\
\bR_0^{(t)}\bx_{0j} & \bR_0^{(t)}\bx_{0j}\bx_{0j}\transpose\bR_0^{(t)}
\end{bmatrix}.
\]
Then, there exists an orthogonal matrix $\bW\in\mathbb{O}(d)$, such that
\begin{align*}
\begin{bmatrix}
\widetilde{\alpha}_i^{(1)}- \alpha_{0i}^{(1)}\\
\vdots\\
\widetilde{\alpha}_i^{(m)} - \alpha_{0i}^{(m)}\\
\bW\transpose\widetilde{\bx}_i - \bx_{0i}
\end{bmatrix}
& = \frac{1}{n}\sum_{t = 1}^m\sum_{j\in[n]\backslash\{i\}}\ell'(\theta_{0ij}^{(t)}; A_{ij}^{(t)})
\begin{bmatrix}
  \be_t(m)\be_1\transpose(\bOmega_i^{(t)})^{-1}\\
  (1/m)\begin{bmatrix}
  \zero_d & \eye_d
  \end{bmatrix}(\bOmega_i^{(t)})^{-1}
\end{bmatrix}\begin{bmatrix}
1 \\ \bR_0^{(t)}\bx_{0j}
\end{bmatrix} + \bgamma_{\widetilde{\bz}_i},
\end{align*}
where $\max_{i\in[n]}\|\bgamma_{\widetilde{\bz}_i}\|_\infty = \Optilde\{(\log n)^3/n\}$. In particular, 
\begin{align*}
\|\widetilde{\bX}\bW - \bX_0\|_{2\to\infty} = \Optilde\bigg(\sqrt{\frac{\log n}{mn}}\bigg),\;
\|\widetilde{\bX}\bW - \bX_0\|_{\mathrm{F}} = \Optilde(1),\; 
\max_{t\in[m]}\|\bW\transpose\widetilde{\bR}^{(t)}\bW - \bR_0^{(t)}\|_{\mathrm{F}} = \Optilde\bigg\{\frac{(\log n)^{3/2}}{n}\bigg\}.
\end{align*} 
\end{proposition}

\begin{proof}[\bf Proof]
By Theorem \ref{thm:asymptotic_normality_MLE} and its proof in Section \ref{sub:Proof of the asymptotic normality} applied to each layer $t\in[m]$ separately and treating each layer as a single-layer network, there exists a random orthogonal matrix 
\[
\bW_t = (\bX_0\transpose\widetilde{\bX}^{(t)})^{-1}\{(\widetilde{\bX}^{(t)\mathrm{T}}\bX_0)^{-1}(\bX_0\transpose\widetilde{\bX}^{(t)})^{-1}\}^{-1/2}\in\mathbb{O}(d),
\] 
such that $\max_{t\in[m]}\|\widetilde{\bX}^{(t)}\bW_t - \bX_0\|_{\mathrm{F}} = \Optilde(1)$, $\max_{t\in[m]}\|\bW_t\transpose\overline{\bR}^{(t)}\bW_t - \bR_0^{(t)}\|_{\mathrm{F}} = \Optilde(n^{-1/2})$ (by Proposition \ref{prop:consistency_native_norm}), and 
\begin{align*}
\begin{bmatrix}
\widetilde{\alpha}_i^{(t)} - \alpha_{0i}^{(t)}\\
\bW_t\transpose\widetilde{\bx}_i^{(t)} - \bx_{0i}
\end{bmatrix} = \frac{1}{n}\sum_{j\in[n]\backslash\{i\}}\ell'(\theta_{0ij}^{(t)}; A_{ij}^{(t)})(\bOmega_i^{(t)})^{-1}\begin{bmatrix}
1\\ \bR_0^{(t)}\bx_{0j}
\end{bmatrix} + \bgamma_{i}^{(t)},\quad t\in[m],i\in[n],
\end{align*}
where $(\bgamma_{i}^{(t)}:t\in[m],i\in[n])$ are uniformly $\Optilde\{(\log n)^3/n\}$. Clearly, this implies that
\begin{align*}
\widetilde{\alpha}_i^{(t)} - \alpha_{0i}^{(t)}
& = \frac{1}{n}\sum_{j\in[n]\backslash\{i\}}\ell'(\theta_{0ij}^{(t)}; A_{ij}^{(t)})\be_1(d + 1)\transpose(\bOmega_i^{(t)})^{-1}\begin{bmatrix}
1\\ \bR_0^{(t)}\bx_{0j}
\end{bmatrix} + \be_1(d + 1)\transpose\bgamma_i^{(t)},\\
\bW_t\transpose\widetilde{\bx}_i^{(t)} - \bx_{0i}
& = \frac{1}{n}\sum_{j\in[n]\backslash\{i\}}\ell'(\theta_{0ij}^{(t)}; A_{ij}^{(t)})\begin{bmatrix}
0 & \eye_d
\end{bmatrix}(\bOmega_i^{(t)})^{-1}\begin{bmatrix}
1\\ \bR_0^{(t)}\bx_{0j}
\end{bmatrix} + \begin{bmatrix}
0 & \eye_d
\end{bmatrix}\bgamma_{i}^{(t)}.
\end{align*}
Applying Lemma \ref{lemma:zt-common} (Theorem 1 in \cite{Zheng_Tang_2026_COSIELimit}) with 
\begin{align*}
&\bT_0^{(t)\mathrm{T}}\be_i = \frac{1}{n^{3/2}}\sum_{j\in[n]\backslash\{i\}}\ell'(\theta_{0ij}^{(t)}; A_{ij}^{(t)})\begin{bmatrix}
0 & \eye_d
\end{bmatrix}(\bOmega_i^{(t)})^{-1}\begin{bmatrix}
1\\ \bR_0^{(t)}\bx_{0j}
\end{bmatrix},\quad \bT^{(t)\mathrm{T}}\be_i = \frac{1}{\sqrt{n}}\begin{bmatrix}
0 & \eye_d
\end{bmatrix}\bgamma_{i}^{(t)},\\
&\|\bT^{(t)}\|_{2\to\infty} = \Optilde\bigg\{\frac{(\log n)^3}{n^{3/2}}\bigg\},\quad 
\|\bT_0^{(t)}\|_{2\to\infty} = \Optilde\bigg(\frac{\sqrt{\log n}}{n}\bigg),\\
&\zeta_\bU = \frac{\|\bX_0\|_{2\to\infty}}{\sqrt{n}} = O(n^{-1/2}),\quad 
\epsilon_\star = \max_{t\in[m]}\frac{1}{\sqrt{n}}\|\bX_0\transpose\bT_0^{(t)}\|_2 = \Optilde\bigg(\frac{\sqrt{\log n}}{n}\bigg),\\
&\epsilon_{\bT_0} = \max_{t\in[m]}\|\bT_0^{(t)}\|_2\leq \max_{t\in[m]}\sqrt{n}\|\bT_0^{(t)}\|_{2\to\infty}  = \sqrt{n}\zeta_{\bT_0} = \Optilde\bigg(\sqrt{\frac{\log n}{n}}\bigg),\\
&\epsilon_{\bT} = \max_{t\in[m]}\|\bT^{(t)}\|_2\leq \max_{t\in[m]}\sqrt{n}\|\bT^{(t)}\|_{2\to\infty}  = \sqrt{n}\zeta_{\bT} = \Optilde\bigg\{\frac{(\log n)^3}{n}\bigg\},
\end{align*}
we further obtain
\begin{align*}
\bW\transpose\widetilde{\bx}_i - \bx_{0i}
& = \frac{1}{mn}\sum_{t = 1}^m\sum_{j\in[n]\backslash\{i\}}\ell'(\theta_{0ij}^{(t)}; A_{ij}^{(t)})\begin{bmatrix}
0 & \eye_d
\end{bmatrix}(\bOmega_i^{(t)})^{-1}\begin{bmatrix}
1\\ \bR_0^{(t)}\bx_{0j}
\end{bmatrix} + \bgamma_{\widetilde{\bx}_i}
\end{align*}
for some (random) orthogonal matrix $\bW\in\mathbb{O}(d)$,
and $\bgamma_{\widetilde{\bx}_i}$ satisfies $\max_{i\in[n]}\|\bgamma_{\widetilde{\bx}_i}\|_\infty = \Optilde\{{(\log n)^3}/{n}\}$. Here, $\eps_\star$ is obtained by applying a Bernstein's inequality to a sum of $O(mn^2)$ independent mean-zero $d\times d$ matrices associated with $\bX_0\transpose\bT_0^{(t)}$.
Note that Bernstein's inequality and a union bound over $i\in[n]$ imply that $\|\widetilde{\bX}\bW - \bX_0\|_{2\to\infty} = \Optilde\{\sqrt{(\log n)/(mn)}\}$. 
Rewriting $\widetilde{\balpha}_i - \balpha_{0i} = \sum_{t = 1}^m\be_t(\widetilde{\alpha}_i^{(t)} - \alpha_{0i}^{(t)})$ gives
\begin{align*}
\begin{bmatrix}
\widetilde{\balpha}_i - \balpha_{0i}\\
\bW\transpose\widetilde{\bx}_i - \bx_{0i}
\end{bmatrix}
& = \frac{1}{n}\sum_{t = 1}^m\sum_{j\in[n]\backslash\{i\}}\ell'(\theta_{0ij}^{(t)}; A_{ij}^{(t)})
\begin{bmatrix}
  \be_t(m)\be_1\transpose(\bOmega_i^{(t)})^{-1}\\
  (1/m)\begin{bmatrix}
  \zero_d & \eye_d
  \end{bmatrix}(\bOmega_i^{(t)})^{-1}
\end{bmatrix}\begin{bmatrix}
1 \\ \bR_0^{(t)}\bx_{0j}
\end{bmatrix} + \bgamma_{\widetilde{\bz}_i}.
\end{align*}

Since $\max_{t\in[m]}\|\widetilde\bX^{(t)}\bW_t - \bX_0\|_{\mathrm F} = \Optilde(1)$, the Davis--Kahan theorem gives 
\begin{align*}
\|\widetilde{\bX}\bW - \bX_0\|_{\mathrm{F}}
&\leq \sqrt{2dn}\bigg\|\sin\Theta\bigg(\frac{\widetilde{\bX}}{\sqrt{n}}, \frac{\bX_0}{\sqrt{n}}\bigg)\bigg\|_2\leq 2\sqrt{2dn}\bigg\|\frac{1}{mn}\sum_{t = 1}^m\widetilde{\bX}^{(t)}\widetilde{\bX}^{(t)\mathrm{T}} - \frac{1}{n}\bX_0\bX_0\transpose\bigg\|_2\\
&\leq 2\sqrt{2dn}\max_{t\in[m]}\bigg\|\sin\Theta\bigg(\frac{\widetilde{\bX}^{(t)}}{\sqrt{n}}, \frac{\bX_0}{\sqrt{n}}\bigg)\bigg\|_2\leq 2\sqrt{2d}\max_{t\in[m]}\|\widetilde{\bX}^{(t)}\bW_t - \bX_0\|_2
 = \Optilde(1).
\end{align*}
Note that $\bW$ is constructed as $\bW_1\bW_2\transpose$, where $\bW_1\bS\bW_2\transpose$ is the singular value decomposition (SVD) of $\widetilde{\bX}\transpose\bX_0$ with $\bW_1,\bW_2\in\mathbb{O}(d)$, and $\|n^{-1}\bS - \eye_d\|_2\leq \|\sin\Theta(n^{-1/2}\bX_0, n^{-1/2}\widetilde{\bX})\|_2^2 = \Optilde(n^{-1})$. This implies that 
$\|(\widetilde\bX\bW - \bX_0)\transpose\bX_0\|_2 = \Optilde(1)$, and likewise
$\max_{t\in[m]}\|\bX_0\transpose(\widetilde\bX^{(t)}\bW_t - \bX_0)\|_2 = \Optilde(1)$ since $\bW_t$ can be equivalently constructed, so that 
\begin{align*}
\max_{t\in[m]}\left\|n^{-1}(\widetilde{\bX}\bW)\transpose\widetilde{\bX}^{(t)}\bW_t - \eye_d\right\|_2  
&\leq \max_{t\in[m]}\left\|n^{-1}(\widetilde{\bX}\bW - \bX_0)\transpose(\widetilde{\bX}^{(t)}\bW_t - \bX_0)\right\|_2\\
&\quad +  \max_{t\in[m]}\left\|n^{-1}(\widetilde{\bX}\bW - \bX_0)\transpose\bX_0\right\|_2 +  \max_{t\in[m]}\left\|n^{-1}\bX_0\transpose(\widetilde{\bX}^{(t)}\bW_t - \bX_0)\right\|_2\\
& = \Optilde(n^{-1}).
\end{align*}
Therefore,
\begin{align*}
\bW\transpose\widetilde\bR^{(t)}\bW - \bR_0^{(t)}
& = \bigg\{\eye_d + \frac{1}{n}(\widetilde{\bX}\bW)\transpose\widetilde{\bX}^{(t)}\bW_t - \eye_d\bigg\}\{\bR_0^{(t)} + (\bW_t\transpose\overline\bR^{(t)}\bW_t - \bR_0^{(t)})\}\\
&\quad\times \bigg\{\eye_d + \frac{1}{n}(\widetilde{\bX}\bW)\transpose\widetilde{\bX}^{(t)}\bW_t - \eye_d\bigg\}\transpose - \bR_0^{(t)}\\
& = (\bW_t\transpose\overline\bR^{(t)}\bW_t - \bR_0^{(t)}) + \bigg\{\frac{1}{n}(\widetilde{\bX}\bW)\transpose\widetilde{\bX}^{(t)}\bW_t - \eye_d\bigg\}\bR_0^{(t)} + \bR_0^{(t)}\bigg\{\frac{1}{n}(\widetilde{\bX}\bW)\transpose\widetilde{\bX}^{(t)}\bW_t - \eye_d\bigg\}\transpose\\
&\quad + 
\bigg\{\frac{1}{n}(\widetilde{\bX}\bW)\transpose\widetilde{\bX}^{(t)}\bW_t - \eye_d\bigg\}(\bW_t\transpose\overline\bR^{(t)}\bW_t - \bR_0^{(t)})\\
&\quad + (\bW_t\transpose\overline\bR^{(t)}\bW_t - \bR_0^{(t)})\bigg\{\frac{1}{n}(\widetilde{\bX}\bW)\transpose\widetilde{\bX}^{(t)}\bW_t - \eye_d\bigg\}\transpose\\
&\quad
+ \bigg\{\frac{1}{n}(\widetilde{\bX}\bW)\transpose\widetilde{\bX}^{(t)}\bW_t - \eye_d\bigg\}\bR_0^{(t)}\bigg\{\frac{1}{n}(\widetilde{\bX}\bW)\transpose\widetilde{\bX}^{(t)}\bW_t - \eye_d\bigg\}\transpose\\
&\quad + \bigg\{\frac{1}{n}(\widetilde{\bX}\bW)\transpose\widetilde{\bX}^{(t)}\bW_t - \eye_d\bigg\}(\bW_t\transpose\overline\bR^{(t)}\bW_t - \bR_0^{(t)})\bigg\{\frac{1}{n}(\widetilde{\bX}\bW)\transpose\widetilde{\bX}^{(t)}\bW_t - \eye_d\bigg\}\transpose\\
& = (\bW_t\transpose\overline\bR^{(t)}\bW_t - \bR_0^{(t)}) + \Optilde(n^{-1}),
\end{align*}
and $\max_{t\in[m]}\|\bW\transpose\widetilde\bR^{(t)}\bW - \bR_0^{(t)}\|_{\mathrm F} = \Optilde\{(\log n)^{3/2}/n\}$ by \eqref{eqn:Rhat_bound} applied to each layer. The proof is therefore completed. 
\end{proof}

\subsection{Proof of the vertex-wise asymptotic normality}
\label{sub:proof_of_the_vertex_wise_asymptotic_normality}

\begin{proof}[\bf Proof of Theorem \ref{thm:asymptotic_normality_OSE}]
Denote by
\begin{align}
\label{eqn:Sigmazi}
\bSigma_{\bz_i}(\bpsi) & = -\frac{1}{n}\sum_{t = 1}^m\sum_{j\in[n]\backslash\{i\}}\ell''(\theta_{ij}^{(t)}; A_{ij}^{(t)})\begin{bmatrix}
\be_t(m)\be_t(m)\transpose & \be_t(m)\bx_j\transpose\bR^{(t)}\\
\bR^{(t)}\bx_j\be_t(m)\transpose & \bR^{(t)}\bx_j\bx_j\transpose\bR^{(t)}
\end{bmatrix},\quad \overline{\bSigma}_{\bz_i} = \expect_0\bSigma_{\bz_i}(\bpsi_0).
\end{align}
In the analysis below in this section, we omit the superscript $(\textsf{OS})$ for notation conciseness and denote $\widehat{\balpha}_i = [(\widehat{\alpha}_i^{(1)})^{\textsf{(OS)}},\ldots, (\widehat{\alpha}_i^{(m)})^{\textsf{(OS)}}]\transpose$. Denote $\widehat{\bz}_i = [\widehat{\balpha}_i\transpose, \widetilde{\bx}_i^{(\textsf{OS})\mathrm{T}}]\transpose$. Also, denote $\widetilde{\balpha}_i = [\widetilde{\alpha}_i^{(1)},\ldots,\widetilde{\alpha}_i^{(m)}]\transpose$ and $\widetilde{\bz}_i = [\widetilde{\balpha}_i\transpose, \widetilde{\bx}_i\transpose]\transpose$, and 
\[
\widetilde{\bpsi} = [\widetilde{\balpha}^{(1)\mathrm{T}},\ldots,\widetilde{\balpha}^{(m)\mathrm{T}},\vech(\widetilde{\bR}^{(1)})\transpose,\ldots,\vech(\widetilde{\bR}^{(m)})\transpose,\vect(\widetilde{\bX})]\transpose. 
\]
Write $\widetilde{\bX} = [\widetilde{\bx}_1,\ldots,\widetilde{\bx}_n]\transpose$. 
Without loss of generality, we can pick a rotation of the rows of $\widetilde{\bX}$ such that $\bX_0\transpose\widetilde{\bX}$ is symmetric positive definite, so that $\bW = \argmin_{\mathbf{V}\in\mathbb{O}(d)}\|\widetilde{\bX}\mathbf{V} - \bX_0\|_{\mathrm{F}} = \eye_d$. 
By construction and a direct calculation, we have
\begin{equation}
\label{eqn:OSE_decomposition}
\begin{aligned}
&\mathrm{diag}(\sqrt{n}\eye_m, \sqrt{mn}\eye_d)(\widehat{\bz}_i - \bz_{0i})\\
&\quad = \sqrt{n}(\bSigma_{\bz_i}^\sharp)^{-1}\mathrm{diag}\bigg(\eye_m, \frac{\eye_d}{\sqrt{m }}\bigg)\frac{1}{n}\frac{\partial\calL}{\partial\bz_i}(\bpsi_0) + \mathrm{diag}(\sqrt{n}\eye_m, \sqrt{mn}\eye_d)(\widetilde{\bz}_i - \bz_{0i})\\
&\qquad + 
\sqrt{n}(\bSigma_{\bz_i}^\sharp)^{-1}\mathrm{diag}\bigg(\eye_m, \frac{\eye_d}{\sqrt{m }}\bigg)\frac{1}{n}\bigg\{\frac{\partial\calL}{\partial\bz_i}(\widetilde{\bpsi}) - \frac{\partial\calL}{\partial\bz_i}(\bpsi_0)\bigg\}\\
&\qquad + \sqrt{n}\mathrm{diag}( \eye_m, \sqrt{m }\eye_d)\left\{\bSigma_{\bz_i}(\widetilde{\bpsi})^{-1} - \overline{\bSigma}_{\bz_i}^{-1}\right\}\mathrm{diag}( \eye_m, \sqrt{m }\eye_d)\\
&\qquad\times \mathrm{diag}\bigg(\eye_m, \frac{\eye_d}{\sqrt{m }}\bigg)\frac{1}{n}\bigg\{\frac{\partial\calL}{\partial\bz_i}(\widetilde{\bpsi}) - \frac{\partial\calL}{\partial\bz_i}(\bpsi_0)\bigg\}\\
&\qquad + \sqrt{n}\mathrm{diag}( \eye_m, \sqrt{m }\eye_d)\left\{\bSigma_{\bz_i}(\widetilde{\bpsi})^{-1} - \overline{\bSigma}_{\bz_i}^{-1}\right\}\mathrm{diag}( \eye_m, \sqrt{m }\eye_d)\\
&\qquad\times \mathrm{diag}\bigg(\eye_m, \frac{\eye_d}{\sqrt{m }}\bigg)\frac{1}{n}\frac{\partial\calL}{\partial\bz_i}(\bpsi_0).
\end{aligned}
\end{equation}

\noindent$\blacksquare$ \textbf{Bounding $\bSigma_{\bz_i}(\widetilde{\bpsi})^{-1} - \overline{\bSigma}_{\bz_i}^{-1}$. }Following the same reasoning as in the beginning of the proof of Proposition \ref{prop:stationarity_uniqueness_uniform_consistency_MLE} (right after \eqref{eqn:integrated_hessian_inverse_infinity_norm_bound}), by Proposition \ref{prop:stochastic_expansion_aggregated_estimators}, we have
\begin{align*}
\bigg\|\frac{1}{m}\sum_{t = 1}^m\{(\widetilde\bR^{(t)})^2 - (\bR_0^{(t)})^2\}\bigg\|_2
& = \bigg\|\frac{1}{m}\sum_{t = 1}^m\{
(\widetilde\bR^{(t)} - \bR_0^{(t)})^2 + 
\bR_0^{(t)}(\widetilde\bR^{(t)} - \bR_0^{(t)}) 
+ (\widetilde\bR^{(t)} - \bR_0^{(t)})\bR_0^{(t)}
\}\bigg\|_2\\
&\leq \frac{1}{m}\sum_{t = 1}^m\left\|
\widetilde\bR^{(t)} - \bR_0^{(t)}\right\|_2^2 + \frac{2}{m}\sum_{t = 1}^m\left\|\bR_0^{(t)}\right\|_2\left\|\widetilde\bR^{(t)} - \bR_0^{(t)}\right\|_2\\
&\leq \frac{1}{m}\sum_{t = 1}^m\left\|
\widetilde\bR^{(t)} - \bR_0^{(t)}\right\|_2^2 + 2\bigg(\frac{1}{m}\sum_{t = 1}^m\left\|\bR_0^{(t)}\right\|_2^2\bigg)^{1/2}\bigg(\frac{1}{m}\sum_{t = 1}^m\left\|\widetilde\bR^{(t)} - \bR_0^{(t)}\right\|_2^2\bigg)^{1/2}\\
& = \Optilde(n^{-1/2}).
\end{align*}
This implies that the eigenvalues of $(1/m)\sum_t(\widetilde\bR^{(t)})^2$ are bounded away from $0$ and $+\infty$ by Weyl's inequality and Assumption \ref{assumption:true_parameter}. Also, since $\widetilde{\bX}\transpose\one_n = \zero_d$ by its construction, then, by Assumption \ref{assumption:edge_distribution}, there exists some constant $C_M, c_M > 0$ depending on $M$, such that for all $i\in[n]$, 
\begin{align*}
&c_M\begin{bmatrix}
  \eye_m & -\frac{1}{\sqrt{m}n}\sum_{t = 1}^m\be_t\widetilde\bx_i\transpose\widetilde\bR^{(t)}\\
  -\frac{1}{\sqrt{m}n}\sum_{t = 1}^m\widetilde\bR^{(t)}\widetilde\bx_i\be_t\transpose & \frac{1}{m}\sum_{t = 1}^m\widetilde\bR^{(t)}(\eye_d - \widetilde\bx_i\widetilde\bx_i\transpose/n)\widetilde\bR^{(t)}
\end{bmatrix}\preceq
-\bDelta_{\bz_i}^\sharp(\widetilde{\bpsi})\\
&\quad\preceq C_M\begin{bmatrix}
  \eye_m & -\frac{1}{\sqrt{m}n}\sum_{t = 1}^m\be_t\widetilde\bx_i\transpose\widetilde\bR^{(t)}\\
  -\frac{1}{\sqrt{m}n}\sum_{t = 1}^m\widetilde\bR^{(t)}\widetilde\bx_i\be_t\transpose & \frac{1}{m}\sum_{t = 1}^m\widetilde\bR^{(t)}(\eye_d - \widetilde\bx_i\widetilde\bx_i\transpose/n)\widetilde\bR^{(t)}
\end{bmatrix},
\end{align*}
where $\bDelta_{\bz_i}^\sharp(\bpsi)$ is defined in \eqref{eqn:Deltazi_sharp}. Note that $\mathrm{diag}(\eye_m, m^{-1/2}\eye_d)\bSigma_{\bz_i}(\bpsi)\mathrm{diag}(\eye_m, m^{-1/2}\eye_d) = -\bDelta_{\bz_i}^\sharp(\bpsi)$ and $\mathrm{diag}(\eye_m, m^{-1/2}\eye_d)\overline{\bSigma}_{\bz_i}\mathrm{diag}(\eye_m, m^{-1/2}\eye_d) = -\expect_0\bDelta_{\bz_i}^\sharp(\bpsi_0) = \bSigma_{\bz_i}^\sharp$. Since 
\[
\max_{t\in[m],j\in[n]}\|\widetilde{\bR}^{(t)}\widetilde{\bx}_j - \bR_0^{(t)}\bx_{0j}\|_2 = \Optilde\bigg(\frac{1}{\sqrt{n}} + \sqrt{\frac{\log n}{mn}}\bigg),\quad \max_{t\in[m],i\in[n]} |\widetilde{\alpha}_i^{(t)} - \alpha_{0i}^{(t)}| = \Optilde\bigg(\sqrt{\frac{\log n}{n}}\bigg),
\]
it follows from Proposition \ref{prop:stochastic_expansion_aggregated_estimators} that
\begin{align*}
\max_{t\in[m],i,j\in[n]}|\widetilde{\theta}_{ij}^{(t)} - \theta_{0ij}^{(t)}|
&:= \max_{t\in[m],i,j\in[n]}|(\widetilde{\alpha}_i^{(t)} + \widetilde{\alpha}_j^{(t)} + \widetilde{\bx}_i\transpose\widetilde{\bR}^{(t)}\widetilde{\bx}_j) - \theta_{0ij}^{(t)}|\\
&\leq 2\max_{t\in[m],i\in[n]}|\widetilde{\alpha}_i^{(t)} - \alpha_{0i}^{(t)}| + 2\max_{t\in[m],i,j\in[n]}\|\widetilde{\bx}_i - \bx_{0i}\|_2\|\widetilde{\bR}^{(t)}\widetilde{\bx}_j\|_2\\
&\quad + \max_{t\in[m]}\|\widetilde{\bX}\|_{2\to\infty}^2\|\widetilde{\bR}^{(t)} - \bR_0^{(t)}\|_2 = \Optilde\bigg(\sqrt{\frac{\log n}{n}}\bigg)
\end{align*}
and the Lipschitz continuity of $\ell''$ that $\|\bDelta_{\bz_i}^\sharp(\widetilde{\bpsi}) - \bDelta_{\bz_i}^\sharp(\bpsi_0)\|_2 = \Optilde\{\sqrt{(\log n)/n}\}$. 
Also, by Bernstein's inequality, we have
$\|\bDelta_{\bz_i}^\sharp(\bpsi_0) - \expect_0\bDelta_{\bz_i}^\sharp(\bpsi_0)\|_2 = \Optilde\{\sqrt{(\log n)/n}\}$. 
Namely,
\begin{align*}
&\left\|\mathrm{diag}(\eye_m, \sqrt{m}\eye_d)\left\{
\bSigma_{\bz_i}(\widetilde{\bpsi})^{-1} - \overline{\bSigma}_{\bz_i}^{-1}
\right\}\mathrm{diag}(\eye_m, \sqrt{m}\eye_d)\right\|_2\\
&\quad = \left\|\bDelta^\sharp_{\bz_i}(\widetilde{\bpsi})^{-1}\left\{\expect_0\bDelta_{\bz_i}^\sharp(\bpsi_0)
 - \bDelta_{\bz_i}^\sharp(\widetilde{\bpsi})
\right\}(\bSigma^\sharp_{\bz_i})^{-1}\right\|_2 = \Optilde\bigg(\sqrt{\frac{\log n}{n}}\bigg).
\end{align*}
This immediately implies that the fifth term in the decomposition \eqref{eqn:OSE_decomposition} has the following error bound:
\begin{align*}
&\sqrt{n}\bigg\|\mathrm{diag}(\eye_m, \sqrt{m}\eye_d)\left\{\bSigma_{\bz_i}(\widetilde{\bpsi})^{-1} - \overline{\bSigma}_{\bz_i}^{-1}\right\}\frac{1}{n}\frac{\partial\calL}{\partial\bz_i}(\bpsi_0)\bigg\|_2\\
&\quad\leq \sqrt{n}\left\|\mathrm{diag}(\eye_m, \sqrt{m}\eye_d)\left\{\bSigma_{\bz_i}(\widetilde{\bpsi})^{-1} - \overline{\bSigma}_{\bz_i}^{-1}\right\}\mathrm{diag}(\eye_m, \sqrt{m}\eye_d)\right\|_2\\
&\qquad\times \Bigg\|\frac{1}{n}\sum_{t = 1}^m\sum_{j\in[n]\backslash\{i\}}\ell'(\theta_{0ij}^{(t)}; A_{ij}^{(t)})\begin{bmatrix}
\be_t\\ m^{-1/2}\bR_0^{(t)}\bx_{0j}
\end{bmatrix}\Bigg\|_2 = \Optilde\bigg(\frac{\sqrt{m}\log n}{\sqrt{n}}\bigg)
\end{align*}
by matrix Bernstein's inequality (Lemma \ref{lemma:matrix_Bernstein} and Theorem 6.2 in \cite{tropp2012user}). 

\noindent $\blacksquare$ \textbf{Bounding $(\partial\calL/\partial\bz_i)(\widetilde{\bpsi}) - (\partial\calL/\partial\bz_i)(\bpsi_0)$. }By Proposition \ref{prop:stochastic_expansion_aggregated_estimators}, there exist deterministic coefficients $(\omega_{ij}^{(t)}, \bzeta_{ij}^{(t)}:i,j\in[n],t\in[m])$, such that
\begin{align*}
\widetilde{\alpha}_i^{(t)} - \alpha_{0i}^{(t)}& = \frac{1}{n}\sum_{j\in[n]\backslash\{i\}}\ell'(\theta_{0ij}^{(t)}; A_{ij}^{(t)})\omega_{ij}^{(t)} + \gamma_{\alpha_i^{(t)}},\\
\widetilde{\bx}_j  - \bx_{0j}& = \frac{1}{mn}\sum_{s = 1}^m\sum_{a \in [n]\backslash\{j\}}\ell'(\theta_{0ja}^{(s)}; A_{ja}^{(s)})\bzeta_{ja}^{(s)} + \bgamma_{\widetilde{\bx}_j},\\
\widetilde{\bR}^{(t)}\widetilde{\bx}_j  - \bR_0^{(t)}\bx_{0j}& = \frac{1}{mn}\sum_{s = 1}^m\sum_{a \in [n]\backslash\{j\}}\ell'(\theta_{0ja}^{(s)}; A_{ja}^{(s)})\bR_0^{(t)}\bzeta_{ja}^{(s)} + \bgamma_{\by_j^{(t)}},
\end{align*}
where $\max_{t,i,j}(\omega_{ij}^{(t)},\|\bzeta_{ij}^{(t)}\|_2) = O(1)$ and $\max_{t,i,j}(|\gamma_{\alpha_i^{(t)}}| + \|\bgamma_{\widetilde{\bx}_j}\|_2 + \|\bgamma_{\by_j^{(t)}}\|_2) = \Optilde\{(\log n)^3/n\}$. This implies that
\begin{align*}
\widetilde{\theta}_{ij}^{(t)} - \theta_{0ij}^{(t)}
& = (\widetilde{\alpha}_i^{(t)} - \alpha_{0i}^{(t)}) + (\widetilde{\bx}_i - \bx_{0i})\transpose\bR_0^{(t)}\bx_{0j} + (\widetilde{\alpha}_j^{(t)} - \alpha_{0j}^{(t)}) + \bx_{0i}\transpose(\widetilde{\bR}^{(t)}\widetilde{\bx}_j - \bR_0^{(t)}\bx_{0j})\\
&\quad + (\widetilde{\bx}_i - \bx_{0i})\transpose(\widetilde{\bR}^{(t)}\widetilde{\bx}_j - \bR_0^{(t)}\bx_{0j})\\
& = \begin{bmatrix}
\be_t(m)\transpose & \bx_{0j}\transpose\bR_0^{(t)}
\end{bmatrix}(\widetilde{\bz}_i - \bz_{0i})\\
&\quad + \frac{1}{n}\sum_{b\in[n]\backslash\{j\}}\ell'(\theta_{0jb}^{(t)}; A_{jb}^{(t)})\omega_{jb}^{(t)} + \frac{1}{mn}\sum_{s = 1}^m\sum_{b\in[n]\backslash\{j\}}\ell'(\theta_{0jb}^{(s)}; A_{jb}^{(s)})\bzeta_{jb}^{(s)\mathrm{T}}\bR_0^{(t)}\bx_{0i} + \gamma_{\widetilde{\theta}_{ij}^{(t)}},
\end{align*}
where $\max_{t,i,j}|\gamma_{\widetilde{\theta}_{ij}^{(t)}}| = \Optilde\{(\log n)^3/n\}$. By Assumption \ref{assumption:edge_distribution} and Taylor's theorem, this further implies that there exists some $\bar{\theta}_{ij}^{(t)}$ between $\theta_{0ij}^{(t)}$ and $\widetilde{\theta}_{ij}^{(t)}$, such that
\begin{align*}
&\mathrm{diag}\bigg(\eye_m, \frac{\eye_d}{\sqrt{m }}\bigg)\Bigg\{\ell'(\widetilde{\theta}_{ij}^{(t)}; A_{ij}^{(t)})\begin{bmatrix}
\be_t\\  \widetilde{\bR}^{(t)}\widetilde{\bx}_j
\end{bmatrix} - \ell'(\theta_{0ij}^{(t)}; A_{ij}^{(t)})\begin{bmatrix}
\be_t\\  \bR_0^{(t)}\bx_{0j}
\end{bmatrix}\Bigg\}\\
&\quad = \mathrm{diag}\bigg(\eye_m, \frac{\eye_d}{\sqrt{m }}\bigg)\left\{\ell'(\widetilde{\theta}_{ij}^{(t)}; A_{ij}^{(t)}) - \ell'(\theta_{0ij}^{(t)}; A_{ij}^{(t)})\right\}
\left\{
\begin{bmatrix}
\be_t\\  \widetilde{\bR}^{(t)}\widetilde{\bx}_j
\end{bmatrix} - 
\begin{bmatrix}
\be_t\\  \bR_0^{(t)}\bx_{0j}
\end{bmatrix}
\right\}\\
&\qquad + \mathrm{diag}\bigg(\eye_m, \frac{\eye_d}{\sqrt{m }}\bigg)\left\{\ell'(\widetilde{\theta}_{ij}^{(t)}; A_{ij}^{(t)}) - \ell'(\theta_{0ij}^{(t)}; A_{ij}^{(t)})\right\}\begin{bmatrix}
\be_t\\  \bR_0^{(t)}\bx_{0j}
\end{bmatrix}\\
&\qquad + \mathrm{diag}\bigg(\eye_m, \frac{\eye_d}{\sqrt{m }}\bigg)\ell'(\theta_{0ij}^{(t)}; A_{ij}^{(t)})\left\{
\begin{bmatrix}
\be_t\\  \widetilde{\bR}^{(t)}\widetilde{\bx}_j
\end{bmatrix} - 
\begin{bmatrix}
\be_t\\  \bR_0^{(t)}\bx_{0j}
\end{bmatrix}
\right\}\\
&\quad = \left\{\ell''(\theta_{0ij}^{(t)}; A_{ij}^{(t)})(\widetilde{\theta}_{ij}^{(t)} - \theta_{0ij}^{(t)}) + \frac{1}{2}\ell'''(\bar{\theta}_{ij}^{(t)}; A_{ij}^{(t)})(\widetilde{\theta}_{ij}^{(t)} - \theta_{0ij}^{(t)})^2\right\} 
\begin{bmatrix}
\be_t\\  m^{-1/2}\bR_0^{(t)}\bx_{0j}
\end{bmatrix}\\
&\qquad + \frac{1}{m n}\sum_{s = 1}^m\sum_{b\in[n]\backslash\{j\}}\ell'(\theta_{0ij}^{(t)}; A_{ij}^{(t)})\ell'(\theta_{0jb}^{(s)}; A_{jb}^{(s)})
\begin{bmatrix}
\zero_m\\ m^{-1/2}\bR_0^{(t)}\bzeta_{jb}^{(s)}
\end{bmatrix}
 + \bar{\bgamma}_{ij}^{(t)}\\
&\quad = \ell''(\theta_{0ij}^{(t)}; A_{ij}^{(t)})\begin{bmatrix}
\be_t(m)\be_t(m)\transpose & m^{-1/2}\be_t(m)\bx_{0j}\transpose\bR_0^{(t)}\\
 m^{-1/2}\bR_0^{(t)}\bx_{0j}\be_t(m)\transpose &  m^{-1}\bR_0^{(t)}\bx_{0j}\bx_{0j}\transpose\bR_0^{(t)}
\end{bmatrix}\mathrm{diag}(\eye_m, \sqrt{m}\eye_d)(\widetilde{\bz}_i - \bz_{0i})\\
&\qquad + \frac{1}{n}\sum_{s = 1}^m\sum_{b\in[n]\backslash\{j\}}\ell'(\theta_{0jb}^{(s)}; A_{jb}^{(s)})\expect_0\ell''(\theta_{0ij}^{(t)}; A_{ij}^{(t)}) \bigg(\be_t\transpose\be_s\omega_{jb}^{(s)} + \frac{1}{m}\bzeta_{jb}^{(s)\mathrm{T}}\bR_0^{(t)}\bx_{0i}\bigg)\begin{bmatrix}
\be_t\\  m^{-1/2}\bR_0^{(t)}\bx_{0j}
\end{bmatrix}
\\
&\qquad + \frac{1}{n}\sum_{s = 1}^m\sum_{b\in[n]\backslash\{j\}}\ell'(\theta_{0jb}^{(s)}; A_{jb}^{(s)})\left\{\ell''(\theta_{0ij}^{(t)}; A_{ij}^{(t)}) - \expect_0\ell''(\theta_{0ij}^{(t)}; A_{ij}^{(t)})\right\}\\
&\qquad\times \bigg(\be_t\transpose\be_s\omega_{jb}^{(s)} + \frac{1}{m}\bzeta_{jb}^{(s)\mathrm{T}}\bR_0^{(t)}\bx_{0i}\bigg)\begin{bmatrix}
\be_t\\  m^{-1/2}\bR_0^{(t)}\bx_{0j}
\end{bmatrix}\\
&\qquad + \frac{1}{m^{3/2} n}\sum_{s = 1}^m\sum_{b\in[n]\backslash\{j\}}\ell'(\theta_{0ij}^{(t)}; A_{ij}^{(t)})\ell'(\theta_{0jb}^{(s)}; A_{jb}^{(s)})
\begin{bmatrix}
\zero_m\\ \bR_0^{(t)}\bzeta_{jb}^{(s)}
\end{bmatrix} + \bgamma_{ij}^{(t)},
\end{align*}
where $\max_{t,i,j}(\|\bar{\bgamma}_{ij}^{(t)}\|_2 + \|\bgamma_{ij}^{(t)}\|_2) = \Optilde\{(\log n)^3/n\}$. Hence, we further obtain
\begin{align*}
& \mathrm{diag}\bigg(\eye_m, \frac{\eye_d}{\sqrt{m}}\bigg)\frac{1}{n}\bigg\{\frac{\partial\calL}{\partial\bz_i}(\widetilde{\bpsi}) - \frac{\partial\calL}{\partial\bz_i}( \bpsi_0 )\bigg\}\\
&\quad = \mathrm{diag}\bigg(\eye_m, \frac{\eye_d}{\sqrt{m}}\bigg)\Bigg\{\frac{1}{n}\sum_{t = 1}^m\sum_{j\in[n]\backslash\{i\}}\ell'(\widetilde{\theta}_{ij}^{(t)}; A_{ij}^{(t)})\begin{bmatrix}
\be_t\\ \widetilde{\bR}^{(t)}\widetilde{\bx}_{j}
\end{bmatrix}  - \frac{1}{n}\sum_{t = 1}^m\sum_{j\in[n]\backslash\{i\}}\ell'(\theta_{0ij}^{(t)}; A_{ij}^{(t)})\begin{bmatrix}
\be_t\\ \bR_0^{(t)}\bx_{0j}
\end{bmatrix}\Bigg\} \\
&\quad = -\bSigma_{\bz_i}^\sharp \mathrm{diag}(\eye_m, \sqrt{m}\eye_d)(\widetilde{\bz}_i - \bz_{0i}) + \{\bDelta_{\bz_i}^\sharp(\bpsi_0) - \expect_0\bDelta_{\bz_i}^\sharp(\bpsi_0)\}\mathrm{diag}(\eye_m, \sqrt m\eye_d)(\widetilde\bz_i - \bz_{0i})\\
&\qquad + \frac{1}{n^2}\sum_{t,s\in[m]}\sum_{j\in[n]\backslash\{i\}}\sum_{b\in[n]\backslash\{j\}}\ell'(\theta_{0jb}^{(s)}; A_{jb}^{(s)})\expect_0\ell''(\theta_{0ij}^{(t)}; A_{ij}^{(t)}) \bigg(\be_t\transpose\be_s\omega_{jb}^{(s)} + \frac{1}{m}\bzeta_{jb}^{(s)\mathrm{T}}\bR_0^{(t)}\bx_{0i}\bigg)\begin{bmatrix}
\be_t\\ m^{-1/2}\bR_0^{(t)}\bx_{0j}
\end{bmatrix}\\
&\qquad + \frac{1}{n^2}\sum_{s,t\in[m]}\sum_{j\in[n]\backslash\{i\}}\sum_{b\in[n]\backslash\{j\}}\ell'(\theta_{0jb}^{(s)}; A_{jb}^{(s)})\left\{\ell''(\theta_{0ij}^{(t)}; A_{ij}^{(t)}) - \expect_0\ell''(\theta_{0ij}^{(t)}; A_{ij}^{(t)})\right\}\\
&\qquad\times \bigg(\be_t\transpose\be_s\omega_{jb}^{(s)} + \frac{1}{m}\bzeta_{jb}^{(s)\mathrm{T}}\bR_0^{(t)}\bx_{0i}\bigg)\begin{bmatrix}
\be_t\\ m^{-1/2}\bR_0^{(t)}\bx_{0j}
\end{bmatrix}\\
&\qquad + \frac{1}{m^{3/2} n^2}\sum_{t,s\in[m]}\sum_{j\in[n]\backslash\{i\}}\sum_{b\in[n]\backslash\{j\}}\ell'(\theta_{0ij}^{(t)}; A_{ij}^{(t)})\ell'(\theta_{0jb}^{(s)}; A_{jb}^{(s)})
\begin{bmatrix}
\zero_m\\ \bR_0^{(t)}\bzeta_{jb}^{(s)}
\end{bmatrix} + \frac{1}{n}\sum_{t = 1}^m\sum_{j\in[n]\backslash\{i\}}\bgamma_{ij}^{(t)}.
\end{align*}
The second term is $\Optilde\{\sqrt{(\log n)/n}\}\Optilde\{\sqrt{(m\log n)/n}\} = \Optilde\{\sqrt m(\log n)/n\}$. 
By matrix Bernstein's inequality (Lemma \ref{lemma:matrix_Bernstein} and Theorem 6.2 in \cite{tropp2012user}), the third term is $\Optilde(\sqrt{m\log n}/n)$. For the fourth and fifth terms, 
we apply the decoupling inequality (Theorem \ref{thm:decoupling} and Theorem 1 in \cite{delaPena_MontgomerySmith_1995_AoP_DecouplingInequalities}) and let $(\bar{A}_{ij}^{(t)}:t\in[m],1\leq i\leq j\leq n)$ be an independent copy of $({A}_{ij}^{(t)}:t\in[m],1\leq i\leq j\leq n)$. 
In these two terms, the summands with
$s = t, b = i$ involve a single edge $(i, j)$ are
\begin{align*}
&\left\|\frac{1}{n^2}\sum_{t\in[m]}\sum_{j\in[n]\backslash\{i\}}\ell'(\theta_{0ij}^{(t)}; A_{ij}^{(t)})\left\{\ell''(\theta_{0ij}^{(t)}; A_{ij}^{(t)}) - \expect_0\ell''(\theta_{0ij}^{(t)}; A_{ij}^{(t)})\right\}
\bigg(\omega_{ji}^{(t)} + \frac{1}{m}\bzeta_{ji}^{(t)\mathrm{T}}\bR_0^{(t)}\bx_{0i}\bigg)\begin{bmatrix}
\be_t\\\frac{\bR_0^{(t)}\bx_{0j}}{\sqrt{m}}
\end{bmatrix}\right\|_2\\
&\quad\lesssim  \frac{1}{n^2}\sum_{t\in[m]}\sum_{j\in[n]\backslash\{i\}}\left\{|\ell'(\theta_{0ij}^{(t)}; A_{ij}^{(t)})| - \expect_0 |\ell'(\theta_{0ij}^{(t)}; A_{ij}^{(t)})|\right\} + O\bigg(\frac{m}{n}\bigg) = \Optilde\bigg(\frac{\sqrt{m\log n}}{n^{3/2}}\bigg) + O\bigg(\frac{m}{n}\bigg) = \Optilde\bigg(\frac{m}{n}\bigg),\\
&\left\|\frac{1}{m^{3/2} n^2}\sum_{t\in[m]}\sum_{j\in[n]\backslash\{i\}}\{\ell'(\theta_{0ij}^{(t)}; A_{ij}^{(t)})\}^2
\begin{bmatrix}
\zero_m\\ \bR_0^{(t)}\bzeta_{ij}^{(t)}
\end{bmatrix}\right\|_2\lesssim \frac{1}{m^{3/2}n^2}\sum_{t\in[m]}\sum_{j\in[n]\backslash\{i\}}\{\ell'(\theta_{0ij}^{(t)}; A_{ij}^{(t)})\}^2 = \Optilde\bigg(\frac{\log n}{m^{1/2}n}\bigg)
\end{align*}
by Bernstein's inequality and the fact that $\max_{t,i,j}|\ell'(\theta_{0ij}^{(t)}; A_{ij}^{(t)})| = \Optilde(\log n)$ by Lemma \ref{lemma:uniform_Optilde} (note that $\{\ell'(\theta_{0ij}^{(t)}; A_{ij}^{(t)}):1\leq i < j\leq n,t\in[m]\}$ are uniformly sub-exponential random variables). 
Therefore, the remaining summands involve two distinct edges. Consider the following two functions:
\begin{align*}
f_{(t,k,j),(s,a,b)}^{(1)}(A_{kj}^{(t)}, A_{ab}^{(s)})
& = \frac{1}{n^2}\mathbbm{1}(s\neq t\text{ or }b\neq i)\mathbbm{1}(k = i, a = j)\ell'(\theta_{0jb}^{(s)}; A_{jb}^{(s)})\left\{\ell''(\theta_{0kj}^{(t)}; A_{kj}^{(t)}) - \expect_0\ell''(\theta_{0kj}^{(t)}; A_{kj}^{(t)})\right\}\\
&\quad\times \bigg(\be_t\transpose\be_s\omega_{jb}^{(s)} + \frac{1}{m}\bzeta_{jb}^{(s)\mathrm{T}}\bR_0^{(t)}\bx_{0k}\bigg)\begin{bmatrix}
\be_t\\m^{-1/2} \bR_0^{(t)}\bx_{0j}
\end{bmatrix},\\
f_{(t,k,j),(s,a,b)}^{(2)}(A_{kj}^{(t)}, A_{ab}^{(s)})
& = \frac{1}{m^{3/2} n^2}\mathbbm{1}(s\neq t\text{ or }b\neq i)\mathbbm{1}(k = i, a = j)\ell'(\theta_{0kj}^{(t)}; A_{kj}^{(t)})\ell'(\theta_{0jb}^{(s)}; A_{jb}^{(s)})
\begin{bmatrix}
\zero_m\\ \bR_0^{(t)}\bzeta_{jb}^{(s)}
\end{bmatrix}.
\end{align*}
For the fourth and fifth terms, it is sufficient to bound
\begin{align*}
&\frac{1}{n^2}\sum_{s,t\in[m]}\sum_{j\in[n]\backslash\{i\}}\sum_{b\in[n]\backslash\{j\}}\mathbbm{1}(s\neq t\text{ or }b\neq i)\ell'(\theta_{0jb}^{(s)}; \bar{A}_{jb}^{(s)})\left\{\ell''(\theta_{0ij}^{(t)}; A_{ij}^{(t)}) - \expect_0\ell''(\theta_{0ij}^{(t)}; A_{ij}^{(t)})\right\}\\
&\quad\times \bigg(\be_t\transpose\be_s\omega_{jb}^{(s)} + \frac{1}{m}\bzeta_{jb}^{(s)\mathrm{T}}\bR_0^{(t)}\bx_{0i}\bigg)\begin{bmatrix}
\be_t\\ m^{-1/2}\bR_0^{(t)}\bx_{0j}
\end{bmatrix},\\
&\frac{1}{m^{3/2} n^2}\sum_{s,t\in[m]}\sum_{j\in[n]\backslash\{i\}}\sum_{b\in[n]\backslash\{j\}}\mathbbm{1}(s\neq t\text{ or }b\neq i)\ell'(\theta_{0ij}^{(t)}; A_{ij}^{(t)})\ell'(\theta_{0jb}^{(s)}; \bar{A}_{jb}^{(s)})
\begin{bmatrix}
\zero_m\\ \bR_0^{(t)}\bzeta_{jb}^{(s)}
\end{bmatrix}.
\end{align*}
Since matrix Bernstein's inequality and matrix Hoeffding's inequality imply that
\begin{align*}
&\max_{t,i,j}\Bigg\|\frac{1}{n^2}\sum_{s = 1}^m\sum_{b\in[n]\backslash\{j\}}\mathbbm{1}(s\neq t\text{ or }b\neq i)\ell'(\theta_{0jb}^{(s)}; \bar{A}_{jb}^{(s)})\bigg(\be_t\transpose\be_s\omega_{jb}^{(s)} + \frac{1}{m}\bzeta_{jb}^{(s)\mathrm{T}}\bR_0^{(t)}\bx_{0i}\bigg)\begin{bmatrix}
\be_t\\ \frac{\bR_0^{(t)}\bx_{0j}}{\sqrt{m}}
\end{bmatrix}\Bigg\|_2\\
&\quad = \Optilde\bigg(\frac{\sqrt{m\log n}}{n^{3/2}}\bigg),\\
&\max_{t,i,j}\Bigg\|\frac{1}{m^{3/2}n^2}\sum_{s\in[m]}\sum_{b\in[n]\backslash\{j\}}\mathbbm{1}(s\neq t\text{ or }b\neq i)\ell'(\theta_{0jb}^{(s)}; \bar{A}_{jb}^{(s)})
\begin{bmatrix}
\zero_m\\ \bR_0^{(t)}\bzeta_{jb}^{(s)}
\end{bmatrix}\Bigg\|_2 = \Optilde\bigg(\frac{\sqrt{\log n}}{m n^{3/2}}\bigg). 
\end{align*}
Then, we apply Lemma \ref{lemma:conditioning_argument} to obtain
\begin{align*}
&\Bigg\|\frac{1}{n^2}\sum_{s,t\in[m]}\sum_{j\in[n]\backslash\{i\}}\sum_{b\in[n]\backslash\{j\}}\mathbbm{1}(s\neq t\text{ or }b\neq i)\ell'(\theta_{0jb}^{(s)}; \bar{A}_{jb}^{(s)})\left\{\ell''(\theta_{0ij}^{(t)}; A_{ij}^{(t)}) - \expect_0\ell''(\theta_{0ij}^{(t)}; A_{ij}^{(t)})\right\}\\
&\quad\times \bigg(\be_t\transpose\be_s\omega_{jb}^{(s)} + \frac{1}{m}\bzeta_{jb}^{(s)\mathrm{T}}\bR_0^{(t)}\bx_{0i}\bigg)\begin{bmatrix}
\be_t\\ m^{-1/2}\bR_0^{(t)}\bx_{0j}
\end{bmatrix}\Bigg\| = \Optilde\bigg(\frac{m\log n}{n}\bigg),\\
&\Bigg\|\frac{1}{m^{3/2} n^2}\sum_{s,t\in[m]}\sum_{j\in[n]\backslash\{i\}}\sum_{b\in[n]\backslash\{j\}}\mathbbm{1}(s\neq t\text{ or }b\neq i)\ell'(\theta_{0ij}^{(t)}; A_{ij}^{(t)})\ell'(\theta_{0jb}^{(s)}; \bar{A}_{jb}^{(s)})
\begin{bmatrix}
\zero_m\\ \bR_0^{(t)}\bzeta_{jb}^{(s)}
\end{bmatrix}\Bigg\|_2 = \Optilde\bigg(\frac{\log n}{\sqrt{m}n}\bigg). 
\end{align*}
Then, the decoupling inequality directly implies that the fourth term and fifth terms are $\Optilde\{(m\log n)/n\}$. Hence, we obtain
\[
\bigg\|\mathrm{diag}(\eye_m, \sqrt{m} \eye_d)(\widetilde{\bz}_i - \bz_{0i}) + (\bSigma_{\bz_i}^\sharp)^{-1}\mathrm{diag}\bigg(\eye_m, \frac{\eye_d}{\sqrt{m}}\bigg)\frac{1}{n}\bigg\{\frac{\partial\calL}{\partial\bz_i}(\widetilde{\bpsi}) - \frac{\partial\calL}{\partial\bz_i}(\bpsi_0)\bigg\}\bigg\|_2 = \Optilde\bigg\{\frac{m\log n + \sqrt{m}(\log n)^3}{n}\bigg\}. 
\]
This further implies that 
\[
\Bigg\|(\bSigma_{\bz_i}^\sharp)^{-1}\mathrm{diag}\bigg(\eye_m, \frac{\eye_d}{\sqrt{m}}\bigg)\frac{1}{n}\bigg\{\frac{\partial\calL}{\partial\bz_i}(\widetilde{\bpsi}) - \frac{\partial\calL}{\partial\bz_i}(\bpsi_0)\bigg\}\Bigg\|_2 = \Optilde\bigg(\sqrt{\frac{m\log n}{n}}\bigg)
\] 
under the condition that $m^2\log n = O(n)$ by Proposition \ref{prop:stochastic_expansion_aggregated_estimators}. Consequently, the sum of the second and third terms in \eqref{eqn:OSE_decomposition} is $\Optilde[\{m\log n + \sqrt{m}(\log n)^3\}/\sqrt{n}]$ and the fourth term there satisfies
\begin{align*}
&\Bigg\|\sqrt{n}\mathrm{diag}( \eye_m, \sqrt{m }\eye_d)\left\{\bSigma_{\bz_i}(\widetilde{\bpsi})^{-1} - \overline{\bSigma}_{\bz_i}^{-1}\right\}\mathrm{diag}( \eye_m, \sqrt{m }\eye_d)\mathrm{diag}\bigg(\eye_m, \frac{\eye_d}{\sqrt{m }}\bigg)\frac{1}{n}\bigg\{\frac{\partial\calL}{\partial\bz_i}(\widetilde{\bpsi}) - \frac{\partial\calL}{\partial\bz_i}(\bpsi_0)\bigg\}\Bigg\|_2\\
&\quad\leq \Optilde\bigg(\frac{\sqrt{m}\log n}{\sqrt{n}}\bigg).
\end{align*}

\noindent$\blacksquare$ \textbf{Completing the proof.} By combining the error bounds for the second terms through the fifth terms in \eqref{eqn:OSE_decomposition}, we conclude that
\begin{align*}
\mathrm{diag}(\sqrt{n}\eye_m, \sqrt{mn}\eye_d)(\widehat{\bz}_i - \bz_{0i})
& = (\bSigma_{\bz_i}^\sharp)^{-1}\frac{1}{\sqrt{n}}\sum_{t = 1}^m\sum_{j\in[n]\backslash\{i\}}\ell'(\theta_{0ij}^{(t)}; A_{ij}^{(t)})\begin{bmatrix}
\be_t(m)\\
m^{-1/2}\bR_0^{(t)}\bx_{0j}
\end{bmatrix}  + \bgamma_i^{(\textsf{OS})},
\end{align*}
where 
\[
\max_{i\in[n]}\|\bgamma_i^{(\textsf{OS})}\|_2 = \Optilde\bigg\{\frac{m(\log n) + \sqrt{m}(\log n)^3}{\sqrt{n}}\bigg\}.
\]
We now argue that the projection step does not change the stochastic expansion of $\widehat{\bX}$. First, note that
$\widehat\bX^{(\mathrm{OS})} = \widetilde\bX^{(\mathrm{OS})} - \one_n\bar\bx\transpose$ with $\bar\bx = (1/n)\sum_{j = 1}^n(\widetilde\bx_j^{(\mathrm{OS})} - \bx_{0j})$ because $\bX_0\transpose\one_n = \zero$. By the above stochastic expansion,
$\sqrt{mn}\,\bar\bx$ is a sum of $O(mn^2)$ independent mean-zero terms with coefficients $O(n^{-3/2})$ plus $(1/n)\sum_j[\mathbf{O},\eye_d]\bgamma_j^{(\mathrm{OS})}$, so that
$\sqrt{mn}\|\bar\bx\|_2 = \Optilde\{\sqrt{m(\log n)/n} + (m\log n + \sqrt m(\log n)^3)/\sqrt n\}$, and the projection does not change the expansion of $\sqrt{mn}(\widehat\bx_i - \bx_{0i})$.
Then, item (i) follows from the Lindeberg--Feller argument in
Section \ref{sub:Proof of the asymptotic normality} with $\vartheta_n$ replaced by $\{m\log n + \sqrt m(\log n)^3\}/\sqrt n$. Next, for item (ii), 
\[
\langle\bF, \widehat\bX\widehat\bX\transpose - \bX_0\bX_0\transpose\rangle_{\mathrm F} = 2\sum_{i = 1}^n\be_i\transpose\bF\bX_0(\widehat\bx_i - \bx_{0i}) + \langle\bF, (\widehat\bX - \bX_0)(\widehat\bX - \bX_0)\transpose\rangle_{\mathrm F},
\] 
where the first term is obtained by inserting $\eye_n = \sum_{i = 1}^n\be_i\be_i\transpose$ between $\bX_0\transpose\bF$ and $\widehat{\bX} - \bX_0$ and the cyclic property of the trace operation, and the second term is bounded by
$\|\vect(\bF)\|_1\|\widehat\bX - \bX_0\|_{2\to\infty}^2 = \Optilde\{(\log n)/(mn)\}$ and, by the expansion above and $\|\vect(\bF\bX_0)\|_1 = O(1)$,
\begin{align*}
\sqrt{mn}\sum_{i = 1}^n\be_i\transpose\bF\bX_0(\widehat\bx_i - \bx_{0i})
& = \sum_{i = 1}^n\be_i\transpose\bF\bX_0\big[\zero_{d\times m}\;\eye_d\big](\bSigma_{\bz_i}^\sharp)^{-1}\frac{1}{\sqrt n}\sum_{t = 1}^m\sum_{j\in[n]\backslash\{i\}}\ell'(\theta_{0ij}^{(t)}; A_{ij}^{(t)})\begin{bmatrix}\be_t(m)\\ m^{-1/2}\bR_0^{(t)}\bx_{0j}\end{bmatrix}\\
&\quad + \Optilde\bigg\{\frac{m\log n + \sqrt m(\log n)^3}{\sqrt n}\bigg\}.
\end{align*}
Checking Lindeberg condition is similar to those of Section \ref{sub:Proof of the asymptotic normality} and is routine, and the proof is completed by Slutsky's theorem.
\end{proof}

\section{Technical lemmas}
\label{sec:technical_lemmas}

\paragraph*{Matrix Bernstein's inequality for sub-exponential random matrices}

\begin{lemma}[Theorem 6.2 in \cite{tropp2012user}]
\label{lemma:matrix_Bernstein}
Let $(\bX_k)_k$ be independent symmetric $n\times n$ sub-exponential random matrices satisfying 
\[
\expect\bX_k = \mathbf{O}\quad\text{and}\quad \expect\bX_k^p\preceq \frac{p!}{2} R^{p - 2}\bA_k^2\text{ for }p = 2, 3, 4, \ldots
\]
for deterministic matrices $(\bA_k)_k$. 
Let $\sigma^2:=\|\sum_k\bA_k^2\|_2$ denote the variance parameter. Then, the following inequality holds for all $t \geq 0$:
\begin{align*}
\prob\bigg\{\lambda_{\max}\bigg(\sum_k\bX_k\bigg) \geq t\bigg\}\leq n\exp\bigg(\frac{-t^2/2}{\sigma^2 + Rt}\bigg). 
\end{align*}
Furthermore, if $(\bX_k)_k$ are $p_1\times p_2$ rectangular random matrices such that $\bX_k = \xi_k\bY_k$ for a scalar random variable $\xi_k$ satisfying
\[
\expect\xi_k = 0,\quad \expect|\xi_k|^p\leq \frac{p!}{2}K^{p - 2}v_k\text{ for }p = 2, 3, 4, \ldots
\]
for deterministic $K, v_k > 0$ and a deterministic matrix $\bY_k$ (and $(\xi_k)_k$ are independent), then $R$, $\sigma^2$, and $n$ above should be replaced by
\[
R = K\max_k\|\bY_k\|_2,\quad \sigma^2 = \max\bigg\{
\bigg\|\sum_kv_k\bY_k\bY_k\transpose\bigg\|_2,
\bigg\|\sum_kv_k\bY_k\transpose\bY_k\bigg\|_2\bigg\},\quad n = p_1 + p_2
\]
\end{lemma}

\begin{proof}[\bf Proof]
The symmetric case is Theorem 6.2 in \cite{tropp2012user}, so 
we will prove the rectangular case with a scalar random multiplier using the symmetric case. Denote the symmetric dilation of $\bY_k$ by
\[
  \bS_k = \begin{bmatrix}
   & \bY_k\\ \bY_k\transpose & 
  \end{bmatrix}.
\]
Suppose $\bY_k = \sum_i\sigma_i\bu_i\bv_i\transpose$ is the singular value decomposition of $\bY_k$, where $(\bu_i)_i$'s are orthonormal vectors, and so $(\bv_i)_i$. Then, the spectral decomposition of $\bS_k$ is given by
\[
\bS_k = \sum_i\sigma_i\frac{1}{\sqrt{2}}\begin{bmatrix}
  \bu_i\\
  \bv_i
\end{bmatrix}
\frac{1}{\sqrt{2}}
\begin{bmatrix}
  \bu_i\transpose & \bv_i\transpose
\end{bmatrix}
+ 
\sum_i(-\sigma_i)\frac{1}{\sqrt{2}}\begin{bmatrix}
  \bu_i\\
  -\bv_i
\end{bmatrix}
\frac{1}{\sqrt{2}}
\begin{bmatrix}
  \bu_i\transpose & -\bv_i\transpose
\end{bmatrix}.
\]
Then, with $|\bS_k|$ defined by applying absolute value to the eigenvalues, we obtain
\[
|\bS_k| = \sum_i\sigma_i\begin{bmatrix}
\bu_i\bu_i\transpose & \\ & \bv_i\bv_i\transpose
\end{bmatrix} = \begin{bmatrix}
  (\bY_k\bY_k\transpose)^{1/2} & \\ & (\bY_k\transpose\bY_k)^{1/2}
\end{bmatrix},
\]
where the matrix square root is applied to the eigenvalues. Since $\bS_k^p \preceq \bS_k|\bS_k|^{p - 2}\bS_k\preceq \|\bY_k\|_2^{p - 2}\bS_k^2$ for any $p\geq 2$, we obtain $\expect (\xi_k\bS_k)^p\preceq \expect |\xi_k|^p \|\bY_k\|_2^{p - 2}\bS_k^2\preceq (p!/2)(K\|\bY_k\|_2)^{p - 2}v_k\bS_k^2$, so that
\[
\bigg\|\sum_kv_k\bS_k^2\bigg\|_2 = \left\|
  \begin{bmatrix}
  \sum_k v_k\bY_k\bY_k\transpose & \\ 
  & \sum_k v_k\bY_k\transpose\bY_k
  \end{bmatrix}
\right\|_2\leq \max\bigg\{\bigg\|\sum_kv_k\bY_k\bY_k\transpose\bigg\|_2,
\bigg\|\sum_kv_k\bY_k\transpose\bY_k\bigg\|_2\bigg\},
\]
which completes the proof.
\end{proof}

\paragraph*{Properties of the Lyapunov operator}
\begin{lemma}[Theorem VII.2.3 in \cite{bhatia2013matrix}]
\label{lemma:inverse_Lyapunov_operator_integral}
Let $\bM_1$ and $\bM_2$ be operators whose spectra are contained in the open right half-plane and the open left half-plane, respectively. Then the solution of the equation $\bM_1\bX - \bX\bM_2 = \bY$ can be expressed as
\begin{align*}
	\bX = \int_0^\infty e^{-\tau\bM_1}\bY e^{\tau\bM_2}\mathrm{d}\tau. 
\end{align*}
\end{lemma}

\paragraph*{Properties related to $\Optilde(\cdot)$ notation}

\begin{lemma}\label{lemma:uniform_Optilde}
If $N$ sequences of random matrices $\{(\bX_{\alpha n})_{n = 1}^\infty: \alpha\in[N]\}$ are uniformly $\Optilde(\eps_n)$ for a deterministic nonnegative sequence $(\eps_n)_{n = 1}^\infty$ and $N\leq n^{\delta}$ for some constant $\delta > 0$, then 
\begin{align*}
\max_{\alpha\in[N]}\|\bX_{\alpha n}\|_2 = \Optilde(\eps_n). 
\end{align*}
\end{lemma}

\begin{proof}[\bf Proof]
Since $\{(\bX_{\alpha n})_{n = 1}^\infty: \alpha\in[N]\}$ are uniformly $\Optilde(\eps_n)$, then for any $c > 0$, there exist constants $N_c,K_c,M_c > 0$ that only depend on $c$ but do not depend on $\alpha\in[N]$, such that 
\[
\prob(\|\bX_{\alpha n}\|_2 > M_c\eps_n)\leq K_c n^{-(c + \delta)}
\] for any $n\geq N_c$ and any $\alpha\in[N]$. Therefore, by a union bound over $\alpha\in[N]$, 
\begin{align*}
\prob\bigg(\max_{\alpha\in[N]}\|\bX_{\alpha n}\|_2 > M_c\eps_n\bigg)  
&\leq \sum_{\alpha = 1}^N\prob(\|\bX_{\alpha n}\|_2 > M_c\eps_n)\leq K_c Nn^{-(c + \delta)} \leq K_cn^{-c}. 
\end{align*}
This proves that $\max_{\alpha\in[N]}\|\bX_{\alpha n}\|_2 = \Optilde(\eps_n)$. 
\end{proof}

\begin{lemma}\label{lemma:conditioning_argument}
Let $(\xi_{\alpha n})_{\alpha, n = 1}^\infty$ be an array of (infinitely many) random variables, and suppose $(N_n)_{n = 1}^\infty$ is a diverging sequence in $\mathbb{N}_+$, such that 
$\sum_{\alpha = 1}^{N_n}\xi_{\alpha n}^2 = \Optilde(\sigma_n^2)$.
For each $n\in\mathbb{N}_+$, let $(\bX_{\alpha n})_{\alpha = 1}^{N_n}$ be $N_n$ independent mean-zero random matrices of dimension $p_n\times q_n$ that are also independent of $(\xi_{\alpha n})_{\alpha, n = 1}^\infty$, where $p_n,q_n\leq n^{\delta}$ for some constant $\delta > 0$. 
\begin{itemize}
  \item[(i)] If for each $n\in\mathbb{N}_+$, there exist $N_n$ deterministic matrices $(\bY_{\alpha n})_{\alpha = 1}^{N_n}$, such that $\bX_{\alpha n}\bX_{\alpha n}\transpose\preceq \bY_{\alpha n}\bY_{\alpha n}\transpose$ and $\bX_{\alpha n}\transpose\bX_{\alpha n}\preceq \bY_{\alpha n}\transpose\bY_{\alpha n}$ almost surely (a.s.), then
  \begin{align*}
  \sum_{\alpha = 1}^{N_n}\xi_{\alpha n}\bX_{\alpha n} = \Optilde\bigg(\sigma_n \max_{\alpha\in[N_n]}\|\bY_{\alpha n}\|_2 \sqrt{\log n}\bigg). 
  \end{align*}

  \item[(ii)] If $\max_{\alpha\in[N_n]}|\xi_{\alpha n}| = \Optilde(r_n)
$ and for each $n\in\mathbb{N}_+$, there exist $N_n$ deterministic matrices $(\bY_{\alpha n})_{\alpha = 1}^{N_n}$ and random variables $(\eta_{\alpha n})_{\alpha = 1}^{N_n}$, such that $\bX_{\alpha n} = \eta_{\alpha n}\bY_{\alpha n}$ and $\expect \eta_{\alpha n} = 0$, $\expect|\eta_{\alpha n}|^p\leq (p!/2)K^{p - 2}v_{\alpha n}$ for $p = 2,3,4,\ldots$, then 
  \begin{align*}
    \sum_{\alpha = 1}^{N_n}\xi_{\alpha n}\bX_{\alpha n} = \Optilde\bigg(
    Kr_n\max_{\alpha\in[N_n]}\|\bY_{\alpha n}\|_2\log n + \max_{\alpha\in[N_n]}|v_{\alpha n}|^{1/2}\|\bY_{\alpha n}\|_2\sigma_n\sqrt{\log n}
    \bigg).
  \end{align*}
\end{itemize}
\end{lemma}

\begin{proof}[\bf Proof]
For item (i), for any $c > 0$, there exist constants $K_c,M_c,N_c > 0$, such that the event
\[
\calE_n := \bigg\{\sum_{\alpha = 1}^{N_n}\xi_{\alpha n}^2\leq M_c \sigma^2_n \bigg\}
\]
occurs with probability at least $1 - K_cn^{-c}$ for all $n\geq N_c$. Conditioning on this event, by matrix Hoeffding's inequality (see Theorem 1.3 in \cite{tropp2012user}), we have, for any $t > 0$,
\begin{align*}
&\prob\bigg\{
\bigg\|\sum_{\alpha = 1}^{N_n}\xi_{\alpha n}\bX_{\alpha n}\bigg\|_2 > t \bigg| (\xi_{\alpha n})_{\alpha = 1}^{N_n}
\bigg\}\mathbbm{1}(\calE_n)\\
&\quad\leq 2(p_n + q_n)\exp\bigg\{-\frac{t^2}{8\max(\|\sum_\alpha\xi_{\alpha n}^2\bY_{\alpha n}\bY_{\alpha n}\transpose\|_2, \|\sum_\alpha\xi_{\alpha n}^2\bY_{\alpha n}\transpose\bY_{\alpha n}\|_2)}\bigg\}\mathbbm{1}(\calE_n)\\
&\quad\leq 2(p_n + q_n)\exp\bigg(-\frac{t^2}{8\max_{\alpha}\|\bY_{\alpha n}\|_2^2\sum_\alpha\xi_{\alpha n}^2}\bigg)\leq 4n^\delta\exp\bigg(-\frac{t^2}{8\max_{\alpha}\|\bY_{\alpha n}\|_2^2M_c\sigma^2_n}\bigg) 
\end{align*}
almost surely with respect to the distribution of $(\xi_{\alpha n})_{\alpha = 1}^{N_n}$. Then, 
\begin{align*}
  \prob\bigg(\bigg\|\sum_{\alpha = 1}^{N_n}\xi_{\alpha n}\bX_{\alpha n}\bigg\|_2 > t\bigg)
  & = \expect_{(\xi_{\alpha n})_{\alpha = 1}^{N_n}}\bigg[\prob\bigg\{
\bigg\|\sum_{\alpha = 1}^{N_n}\xi_{\alpha n}\bX_{\alpha n}\bigg\|_2 > t \bigg| (\xi_{\alpha n})_{\alpha = 1}^{N_n}
\bigg\}\bigg]\\
&\leq \expect_{(\xi_{\alpha n})_{\alpha = 1}^{N_n}}\bigg[\prob\bigg\{
\bigg\|\sum_{\alpha = 1}^{N_n}\xi_{\alpha n}\bX_{\alpha n}\bigg\|_2 > t \bigg| (\xi_{\alpha n})_{\alpha = 1}^{N_n}
\bigg\}\mathbbm{1}(\calE_n)\bigg] + \prob(\calE_n^c)\\
&\leq 4n^\delta\exp\bigg(-\frac{t^2}{8\max_{\alpha}\|\bY_{\alpha n}\|_2^2M_c\sigma^2_n}\bigg) + K_cn^{-c}
\end{align*}
for all $n\geq N_c$. 
Now take $t = \{8\max_\alpha\|\bY_{\alpha n}\|_2^2M_c\sigma^2_n(c + \delta)\log n\}^{1/2}$. Then, 
\begin{align*}
  \prob\bigg\{\bigg\|\sum_{\alpha = 1}^{N_n}\xi_{\alpha n}\bX_{\alpha n}\bigg\|_2 > \{8\max_\alpha\|\bY_{\alpha n}\|_2^2M_c\sigma^2_n(c + \delta)\log n\}^{1/2}\bigg\}\leq (4 + K_c)n^{-c}
\end{align*}
for all $n\geq N_c$. This completes the proof of item (i). For item (ii), we modify the event $\calE_n$ to include the additional constraint:
\[
\calF_n := \bigg\{ \max_{\alpha\in[N_n]}|\xi_{\alpha n}| \leq M_c r_n,\sum_{\alpha = 1}^{N_n}\xi_{\alpha n}^2\leq M_c \sigma^2_n \bigg\},
\]
which occurs with probability at least $1 - 2K_cn^{-c}$ for all $n\geq N_c$ without loss of generality. Then, following the same reasoning, by matrix Bernstein's inequality (Lemma \ref{lemma:matrix_Bernstein}), for any $t > 0$,
\begin{align*}
&\prob\bigg\{
\bigg\|\sum_{\alpha = 1}^{N_n}\xi_{\alpha n}\bX_{\alpha n}\bigg\|_2 > t \bigg| (\xi_{\alpha n})_{\alpha = 1}^{N_n}
\bigg\}\mathbbm{1}(\calF_n)\\
&\quad\leq 4n^\delta\exp\bigg\{-\frac{t^2/2}{\max(\|\sum_\alpha v_{\alpha n}\xi_{\alpha n}^2\bY_{\alpha n}\bY_{\alpha n}\transpose\|_2, \|\sum_\alpha v_{\alpha n}\xi_{\alpha n}^2\bY_{\alpha n}\transpose\bY_{\alpha n}\|_2) + K\max_{\alpha} |\xi_{\alpha n}|\|\bY_{\alpha n}\|_2t }\bigg\}\mathbbm{1}(\calF_n)\\
&\quad\leq 4n^\delta\exp\bigg(-\frac{t^2/2}{\max_{\alpha}|v_{\alpha n}|\|\bY_{\alpha n}\|_2^2M_c\sigma_n^2 + K\max_{\alpha}\|\bY_{\alpha n}\|_2t M_cr_n}\bigg)
\end{align*}
almost surely with respect to the distribution of $(\xi_{\alpha n})_{\alpha = 1}^{N_n}$, so that
\begin{align*}
  \prob\bigg(\bigg\|\sum_{\alpha = 1}^{N_n}\xi_{\alpha n}\bX_{\alpha n}\bigg\|_2 > t\bigg)
  &\leq 4n^\delta\exp\bigg(-\frac{t^2/2}{\max_{\alpha}|v_{\alpha n}|\|\bY_{\alpha n}\|_2^2M_c\sigma_n^2 + K\max_{\alpha}\|\bY_{\alpha n}\|_2t M_cr_n}\bigg) + K_cn^{-c}
\end{align*}
for all $n\geq N_c$. Now taking 
\[
t = \max_\alpha|v_{\alpha n}|^{1/2}\|\bY_{\alpha n}\|_2M_c^{1/2}\sigma_n\sqrt{2(c + \delta)\log n} + K\max_\alpha\|\bY_{\alpha n}\|_2M_cr_n2(c + \delta)\log n
\]
completes the proof.  
\end{proof}

\paragraph*{Decoupling inequality for multivariate $U$-statistic}
\begin{theorem}[Theorem 1 of \cite{delaPena_MontgomerySmith_1995_AoP_DecouplingInequalities}]
\label{thm:decoupling}
Let $\{X_i\}$ be a sequence of independent random variables on a measurable
space $(\mathscr{S}, S)$ and let $\{X_i^{(j)}\}$, $j = 1, \ldots, k$, be $k$
independent copies of $\{X_i\}$. Let $f_{i_1 i_2 \cdots i_k}$ be families of
functions of $k$ variables taking $(S \times \cdots \times S)$ into a Banach
space $(B, \|\cdot\|)$. Then, for all $n \ge k \ge 2$, $t > 0$, there exists a numerical constant $C_k$ depending on $k$ only so that
\begin{align*}
  &P\Biggl( \Biggl\| \sum_{1 \le i_1 \ne i_2 \ne \cdots \ne i_k \le n}
      f_{i_1 \cdots i_k}\bigl(X_{i_1}^{(1)}, X_{i_2}^{(1)}, \ldots, X_{i_k}^{(1)}\bigr)
    \Biggr\| \ge t \Biggr) \\
  &\qquad \le C_k \, P\Biggl( C_k \Biggl\| \sum_{1 \le i_1 \ne i_2 \ne \cdots \ne i_k \le n}
      f_{i_1 \cdots i_k}\bigl(X_{i_1}^{(1)}, X_{i_2}^{(2)}, \ldots, X_{i_k}^{(k)}\bigr)
    \Biggr\| \ge t \Biggr).
\end{align*}
The notation $\{i_1 \ne i_2 \ne \cdots \ne i_k\}$
denotes that all of $i_1, \ldots, i_k$ are different.
\end{theorem}

\paragraph*{Properties of the aggregated subspace}
\begin{lemma}[Theorem 1 of \cite{Zheng_Tang_2026_COSIELimit}, v6]\label{lemma:zt-common}
Let $\mathbf{U}$ be an $n \times d $ matrix with orthonormal columns. For each $t \in [m]$, suppose that we have an estimate $\widehat{\mathbf{U}}^{(t)} \in \mathbb{R}^{n \times d }$ of $\mathbf{U} $ such that
\[
  \widehat{\mathbf{U}}^{(t)}\mathbf{W}_{\mathbf{U}}^{(t)} - \mathbf{U} 
  = \mathbf{T}_0^{(t)} + \mathbf{T}^{(t)}
\]
for some orthogonal matrix $\mathbf{W}_{\mathbf{U}}^{(t)}$, where $\mathbf{T}_0^{(t)}$ and $\mathbf{T}^{(t)}$ satisfy
\[
  \max_{t \in [m]} \Bigl( 2\|\mathbf{T}_0^{(t)}\|_2 + 2\|\mathbf{T}^{(t)}\|_2 + \|\mathbf{T}_0^{(t)} + \mathbf{T}^{(t)}\|^2_2 \Bigr) \leq \delta
\]
for some constant $\delta < \frac{1}{2}$. Define
\[
\begin{aligned}
  \zeta_{\mathbf{U}} &= \|\mathbf{U} \|_{2\to\infty}, &
  \epsilon_{\star} &= \max_{t \in [m]} \|\mathbf{U}\transpose\mathbf{T}_0^{(t)}\|_2, &
  \epsilon_{\mathbf{T}_0} &= \max_{t \in [m]} \|\mathbf{T}_0^{(t)}\|_2, &
  \zeta_{\mathbf{T}_0} &= \max_{t \in [m]} \|\mathbf{T}_0^{(t)}\|_{2\to\infty}, \\
  \epsilon_{\mathbf{T}} &= \max_{t \in [m]} \|\mathbf{T}^{(t)}\|_2, &
  \zeta_{\mathbf{T}} &= \max_{t \in [m]} \|\mathbf{T}^{(t)}\|_{2\to\infty}.
\end{aligned}
\]
Now let $\widehat{\mathbf{U}}$ denote the eigenvector matrix with orthonormal columns corresponding to the $d$ largest eigenvalues of $(1/m)\sum_{t=1}^{m} \widehat{\mathbf{U}}^{(t)}\widehat{\mathbf{U}}^{(t)\mathrm{T}}$. Let $\mathbf{W}_{\mathbf{U}}$ be the minimizer of $\|\widehat{\mathbf{U}} \mathbf{W} - \mathbf{U} \|_F$ over $\mathbf{W}\in\mathbb{O}(d)$. We then have
\[
  \widehat{\mathbf{U}} \mathbf{W}_{\mathbf{U} } - \mathbf{U} 
  = \frac{1}{m}\sum_{t=1}^{m} \mathbf{T}_0^{(t)} + \mathbf{Q}_{\mathbf{U} },
\]
where $\mathbf{Q}_{\mathbf{U} }$ is a matrix satisfying
\[
\begin{aligned}
  \|\mathbf{Q}_{\mathbf{U} }\|_2 &\lesssim \epsilon_{\star} + \epsilon_{\mathbf{T}_0}^2 + \epsilon_{\mathbf{T}}, \quad
  \|\mathbf{Q}_{\mathbf{U} }\|_{2\to\infty} \lesssim
  \zeta_{\mathbf{U}}\bigl(\epsilon_{\star} + \epsilon_{\mathbf{T}_0}^2 + \epsilon_{\mathbf{T}}\bigr)
  + \zeta_{\mathbf{T}_0}\bigl(\epsilon_{\star} + \epsilon_{\mathbf{T}_0} + \epsilon_{\mathbf{T}}\bigr)
  + \zeta_{\mathbf{T}}.
\end{aligned}
\]
\end{lemma}

\paragraph*{Concentration inequality for quadratic form of sub-exponential random vectors}

\begin{theorem}[Proposition 1.1 in \cite{Gotze_Sambale_Sinulis_2021_EJP_HWSubexp}]
\label{thm:HW}
Let $\alpha\in(0,1]$ and $K>0$. Let $X_1,\dots,X_n $ be independent random variables with $\mathbb{E} X_i=0$, $\expect X_i^2 = \sigma_i^2$, and $\|X_i\|_{\psi_\alpha}:=\inf\{\tau > 0:\expect\exp(|X_i|^\alpha/\tau^\alpha)\leq 2\}\le K$ for all $i$, and
let $\bM = [M_{ij}]\in\mathbb{R}^{n\times n}$ be symmetric. Then, there exist constants
$c_\alpha,C_\alpha\in(0,\infty)$ depending only on $\alpha$, such that,
for every $u\ge 0$, with probability at least $1-2e^{-u}$,
\[
\bigg|\sum_{i,j\in[n]}X_iX_jM_{ij} - \sum_{i = 1}^n\sigma_i^2M_{ii}\bigg|
\le C_\alpha K^2\bigl(\sqrt{u}\,\|\bM\|_{\mathrm{F}}+u^{2/\alpha}\|\bM\|_2\bigr).
\]
\end{theorem}

\section{Likelihood derivatives and computation}
\label{sec:implementation_audit}

\subsection{Logit and probit forms of the bias for binary edges}
For the logit link function, $\ell(\theta; A) = A \theta - \log(1 + e^\theta)$, $\ell'(\theta; A) = A - 1/(1 + e^{-\theta})$, $\ell''(\theta; A) = -e^{-\theta}/(1 + e^{-\theta})^2$, and $\ell'''(\theta; A) = e^{-\theta}(1 - e^{-\theta})/(1 + e^{-\theta})^3$, so $\mathbb E\ell''\ell'=0$ since $\ell''(\theta; A)$ does not depend on $A$ and $\mathbb E\ell' = 0$, and the two edge coefficients in the bias formula defined in Subsection \ref{sub:R_inference} of the main text are computed explicitly. For the probit model with $p=\Phi(\theta)$, $q=1-p$, $h=\phi(\theta)/p$, and $g=\phi(\theta)/q$, the scores at $A=1,0$ are $h$ and $-g$, the second derivatives are $-h(\theta+h)$ and $g(\theta-g)$, and the third derivatives are $h\{(\theta+h)(\theta+2h)-1\}$ and $g(1-\theta^2+3\theta g-2g^2)$. The variance of the score is $ph^2+qg^2$, and the first coefficient $\mathbb E(\ell''\ell')+\mathbb E(\ell''')/2$ is obtained by averaging over $A$. $\expect\ell''\ell'$ is generally nonzero. 

\subsection{Schur-complement algorithm}
\label{sec:schur}
Order the compact parameter as $\bphi = [\balpha^{(1)\mathrm{T}},\ldots,\balpha^{(m)\mathrm{T}},\bphi_D\transpose]\transpose$, where $\bphi_D$ collects the $q_D=md(d+1)/2+d(n-1-d)$ interaction and subspace coordinates. Then the Fisher information matrix has the form
\[
-\expect_\bphi\frac{\partial^2\calL}{\partial\bphi\partial\bphi\transpose}=\begin{bmatrix}
\mathrm{diag}(\bH_{\balpha^{(1)}},\ldots,\bH_{\balpha^{(m)}})&\bH_{\balpha \bphi_D}\\
\bH_{\balpha\bphi_D}\transpose & \bH_{\bphi_D\bphi_D}
\end{bmatrix},\qquad \bH_{\balpha^{(t)}}=-\expect_\bphi\ell''(\bTheta^{(t)})-\mathrm{diag}\{\expect_\bphi\ell''(\bTheta^{(t)})\one_n\},
\]
where $\ell''(\bTheta^{(t)})$ is defined after \eqref{eqn:ell_prime_Theta_t} and 
$\bH_{\balpha\bphi_D}=[\bH_{\balpha^{(1)}\bphi_D}\transpose,\ldots,\bH_{\balpha^{(m)}\bphi_D}\transpose]\transpose$. With $\bG_t = \bH_{\balpha^{(t)}}^{-1}\bH_{\balpha^{(t)}\bphi_D}$ and $\bS = \bH_{\bphi_D\bphi_D} - \sum_{t = 1}^m\bH_{\balpha^{(t)}\bphi_D}\transpose \bG_t$, we obtain
\begin{align*}
\bigg\{-\expect_\bphi\frac{\partial^2\calL}{\partial\bphi\partial\bphi\transpose}\bigg\}^{-1}
& = \begin{bmatrix} 
\bH_{\balpha^{(1)}}^{-1} + \bG_1\bS^{-1}\bG_1\transpose & \bG_1\bS^{-1}\bG_2\transpose & \ldots & \bG_1\bS^{-1}\bG_m\transpose & -\bG_1\bS^{-1}\\
\bG_2\bS^{-1}\bG_1\transpose & \bH_{\balpha^{(2)}}^{-1} + \bG_2\bS^{-1}\bG_2\transpose & \ldots & \bG_2\bS^{-1}\bG_m\transpose & - \bG_2\bS^{-1}\\
\vdots & \vdots & \ddots & \vdots & \vdots \\
\bG_m\bS^{-1}\bG_1\transpose & \bG_m\bS^{-1}\bG_2\transpose & \ldots & \bH_{\balpha^{(m)}}^{-1} + \bG_m\bS^{-1}\bG_m\transpose & -\bG_m\bS^{-1}\\
-\bS^{-1}\bG_1\transpose & -\bS^{-1}\bG_2\transpose & \ldots & -\bS^{-1}\bG_m\transpose & \bS^{-1}
\end{bmatrix}
\end{align*}
Each edge gradient $\partial\theta_{ij}^{(t)}/\partial\bpsi$ with respect to the ambient parameter $\bpsi$ has $2+d(d+1)/2+2d$ nonzero coordinates, so the terms $(\partial\theta_{ij}^{(t)}/\partial\bpsi)\transpose\widetilde{\bLambda}_\sharp \bOmega \widetilde{\bLambda}_\sharp (\partial\theta_{ij}^{(t)}/\partial\bpsi)$ and $\mathrm{tr}(\widetilde{\bLambda}_\sharp\partial^2\theta_{ij}^{(t)}/\partial\bpsi\partial\bpsi\transpose\widetilde{\bLambda}_\sharp\bOmega)$ need only the corresponding entries of these blocks, which are formed one layer at a time by the above formula. The first two terms of the bias $\bb_n^{(t)}(\bpsi)$ require computing the quantity of the form
\begin{align*}
\sum_{s = 1}^m\sum_{i < j}\bigg[c_{ij}^{(s)}\bigg\{\frac{\partial\theta_{ij}^{(s)}}{\partial\bpsi\transpose}(\bpsi)\widetilde{\bLambda}_\sharp \bOmega(\bpsi) \widetilde{\bLambda}_\sharp \frac{\partial\theta_{ij}^{(s)}}{\partial\bpsi}(\bpsi)\bigg\} + \frac{1}{2}w_{ij}^{(s)}\mathrm{tr}\bigg\{\widetilde{\bLambda}_\sharp\frac{\partial^2\theta_{ij}^{(s)}}{\partial\bpsi\partial\bpsi\transpose}(\bpsi)\widetilde{\bLambda}_\sharp\bOmega(\bpsi)\bigg\}\bigg]\widetilde{\bLambda}_\sharp\frac{\partial\theta_{ij}^{(s)}}{\partial\bpsi},
\end{align*} 
where $(c_{ij}^{(s)}, w_{ij}^{(s)}:i,j\in[n],s\in[m])$ are scalars depending on $\bpsi$. The computation of this quantity
is accumulated in the same pass as $(s,i,j)$ ranges over $\calE^\circ = [m]\times\{(i, j)\in[n]\times[n]:i < j\}$ and multiplied by $\bOmega(\bpsi)$ through the same block formulas. The sandwich covariance takes the form $\bOmega(\bpsi)\widetilde{\bLambda}_\sharp\bar{\bS}(\bpsi)\widetilde{\bLambda}_\sharp\bOmega(\bpsi)$, where $\bar{\bS}(\bpsi) = \sum_{(s, i, j)\in\calE^\circ}\{\ell'(\theta_{ij}^{(s)}; A_{ij}^{(s)})\}^2(\partial\theta_{ij}^{(s)}/\partial\bpsi)(\partial\theta_{ij}^{(s)}/\partial\bpsi\transpose)$, and is computed by building $\bar{\bS}$ with the same block routine and applying the block formulas on both sides. Table~\ref{tab:schur_check} compares the implementation with the dense reference implementation.

\begin{table}[htbp]\centering\small
\caption{Maximum relative difference between the Schur-complement implementation (\texttt{code/fisher\_block.py}) and default dense implementations, at randomly generated parameters. Columns: bias of all $\vech(R^{(t)})$, their covariance, the full sandwich covariance matrix of the interaction and subspace coordinates, and its interaction block.}\label{tab:schur_check}
\begin{tabular}{lccccc}
\toprule
Family & $(n,m,d)$ & Bias & Covariance & Sandwich (all) & Sandwich ($\bR$-component) \\
\midrule
Bernoulli & (30,2,3) & 5.5e-16 & 4.6e-16 & 1.7e-15 & 3.7e-15 \\
Bernoulli & (40,4,2) & 6.4e-16 & 4.1e-16 & 1.4e-15 & 1.2e-15 \\
Probit & (30,2,3) & 9.2e-16 & 8.7e-16 & 1.3e-15 & 3.1e-15 \\
Probit & (40,4,2) & 6.9e-16 & 2.9e-16 & 2.1e-15 & 1.1e-15 \\
Poisson & (30,2,3) & 3.5e-16 & 6.0e-16 & 1.7e-15 & 3.9e-15 \\
Poisson & (40,4,2) & 1.5e-15 & 3.8e-16 & 2.3e-15 & 1.3e-15 \\
Gaussian & (30,2,3) & 3.0e-16 & 5.3e-16 & 1.6e-15 & 2.1e-15 \\
Gaussian & (40,4,2) & 2.3e-16 & 4.9e-16 & 2.0e-15 & 7.0e-16 \\
\bottomrule
\end{tabular}

\end{table}

\subsection{Further computation details}
\label{sub:computation}

\paragraph*{Maximum likelihood.}
We compute the MLE by block Fisher scoring that exploits the structure of Proposition~\ref{prop:expected_hessian_bound}. Each iteration (1) updates every $\bR^{(t)}$ given the other parameters, a concave $d(d+1)/2$-dimensional problem; (2) updates all vertex blocks $\bz_i=(\alpha_i^{(1)},\ldots,\alpha_i^{(m)},\bx_i)$ simultaneously, which requires only the $(m+d)\times(m+d)$ matrices $\bG_i(\bpsi)$ in Section \ref{sec:maximum_likelihood_and_one_step_estimator} of the main text and the vertex scores; and (3) applies the exact centering and whitening transformation of Section~\ref{sec:multilayer_latent_space_network_with_an_invariant_subspace}, which restores the normalization without changing any $\theta_{ij}^{(t)}$. Step halving makes the log-likelihood increase monotonically, and each iteration costs $O(mn^2d^2)$ operations. Because the likelihood is not concave, the Fisher scoring algorithm can be sensitive to initializations.
We therefore run the algorithm from several starting points and retain the stationary point with the largest log-likelihood. The first starting point is the link-transformed spectral estimate defined next; the others are perturbations of the best solution found so far.

\emph{Link-transformed spectral estimate.} For each layer $t$, universal singular value thresholding \citep{chatterjee2015matrix} keeps the eigenvalues of $\bA^{(t)}$ whose modulus is at least $2.02(n\widehat\sigma_t^2)^{1/2}$, where $\widehat\sigma_t^2$ is $\bar p_t(1-\bar p_t)$ for Bernoulli edges, $\bar p_t$ for Poisson edges, $\sigma^2$ for Gaussian edges, and $\bar p_t$ is the mean off-diagonal entry of $\bA^{(t)}$. The resulting estimate of $\expect\bA^{(t)}$ is truncated entrywise to $[1/(2n),1-1/(2n)]$ (Bernoulli) or $[1/(2n),\infty)$ (Poisson), transformed entrywise by the canonical link (the identity for Gaussian edges), and double centered, which gives $\widetilde\bTheta_{\rm c}^{(t)} = \bJ_n\widetilde\bTheta^{(t)}\bJ_n$ with $\bJ_n = \eye_n - \one_n\one_n\transpose/n$. Under model~\eqref{eqn:model}, $\bJ_n\bTheta^{(t)}\bJ_n = \bX\bR^{(t)}\bX\transpose$ for every $t$, so the initial $\bX$ is $\sqrt n$ times the matrix of the $d$ leading orthonormal eigenvectors of $\sum_{t=1}^m(\widetilde\bTheta_{\rm c}^{(t)})^2$. The initial $(\balpha^{(t)},\bR^{(t)})$ are then obtained from the same identity and refined by eight Fisher scoring iterations with $\bX$ held fixed. This is the initialization referred to as ``spectral'' in Sections~\ref{sub:sim_mouse} and~\ref{sec:additional_simulations}.

\emph{Perturbations of the best fit.} After the algorithm has converged from the spectral estimate, it is restarted from perturbed copies of the current best solution $(\widehat\balpha,\widehat\bR,\widehat\bX)$, and Steps~2--5 below are run to convergence from each of them. If a restart reaches a larger log-likelihood, it becomes the new best solution. Agreement of the restarts with the best fit, as in Section~\ref{sec:mouse} of the main text, indicates that the reported fit is not an artifact of the starting point.
In our numerical examples, we enforce the boundary constraints $\max|\alpha_i^{(t)}|\leq15$, $\max_t\|\bR^{(t)}\|_{\mathrm F}\leq10$, and $\max_i\|\bx_i\|_2\leq6$ that resemble the boundary conditions in $\Psi$ in \eqref{eqn:psi_parameter_space}. A fit is declared successful if no bound is active and the largest tangent score, divided by the number of edges, is below a tolerance \texttt{tol} (which is taken as $10^{-6}$ in our numerical examples). Every reported numerical example records this diagnostic.
In summary, the algorithm proceeds as follows.
\begin{enumerate}[1.]
\item Initialize $\bpsi$ at the link-transformed spectral estimate.
\item For each $t\in[m]$, update $\bR^{(t)}$ by Fisher scoring with the remaining parameters fixed.
\item Update all vertex blocks $\bz_i$, $i\in[n]$, simultaneously by one Fisher scoring step, each of which solves an $(m+d)\times(m+d)$ linear system with matrix $\bG_i(\bpsi)$ and the vertex score as the right-hand side; halve the step until the log-likelihood increases.
\item Center and whiten $\bX$ and transform $(\balpha^{(t)},\bR^{(t)})_{t = 1}^m$ accordingly, so that $\bX\transpose\one_n = \zero_d$, $\bX\transpose\bX = n\eye_d$, and every $\theta_{ij}^{(t)}$ is unchanged.
\item Repeat Steps 2--4 until the largest tangent score, divided by the number of edges, is below \texttt{tol}.
\item Restart Steps 2--5 from perturbations of the best solution found, as described above, and retain the fit with the largest log-likelihood.
\end{enumerate}

\paragraph*{Debiasing and covariance.}
The bias $\bb_n^{(t)}$ and covariance in Corollary~\ref{corr:plug_in_R} require the inverse of the tangent Fisher information, of dimension $mn+q_D$ with $q_D=md(d+1)/2+d(n-1-d)$, together with 
$(\partial\theta_{ij}^{(t)}/\partial\bpsi)\transpose\widetilde{\bLambda}_\sharp \bOmega \widetilde{\bLambda}_\sharp (\partial\theta_{ij}^{(t)}/\partial\bpsi)$ and $\mathrm{tr}(\widetilde{\bLambda}_\sharp\partial^2\theta_{ij}^{(t)}/\partial\bpsi\partial\bpsi\transpose\widetilde{\bLambda}_\sharp\bOmega)$.
Two features of the model make their computation efficient. First, $\partial\theta_{ij}^{(t)}/\partial\bpsi$ has only $2+d(d+1)/2+2d$ nonzero coordinates, so each term touches a small submatrix of $\bOmega$. Second, the $\balpha$-block of the Fisher information matrix is block diagonal across layers, with one $n\times n$ block per layer. We therefore eliminate the degree effects block layer by layer through a Schur complement onto the $q_D$ interaction and subspace coordinates, form $\bOmega$ blockwise from the $m$ layer inverses and the $q_D\times q_D$ Schur inverse, and never store the full matrix.
 See Subsection \ref{sec:schur} above for the detailed implementation.
Storage complexity is $O(mn^2+mnq_D+q_D^2)$ and time complexity is $O(mn^3+mn^2q_D+q_D^3)$, against $O\{(mn+q_D)^2\}$ and $O\{(mn+q_D)^3\}$, respectively, for a dense inverse. For the mouse connectome data ($n=332$, $m=32$, $d=3$, 11{,}800 parameters), the computation, including the sandwich covariance matrix, takes about 
$115$
seconds on one processor core. The sandwich covariance matrix $\bOmega(\widehat{\bpsi})\widetilde{\bLambda}_\sharp\bar{\bS}(\widehat{\bpsi})\widetilde{\bLambda}_\sharp\bOmega(\widehat{\bpsi})$ of Subsection~\ref{sec:schur}, which we use as a check on the working variance in Section~\ref{sec:mouse}, is assembled from the same blocks with weights equal to the squared edge scores. 

\section{Simulation studies}
\label{sec:numerical_examples}

In this section, the simulation studies address three questions: whether the vertex-level confidence sets of Theorem~\ref{thm:asymptotic_normality_MLE} are calibrated, whether debiased inference based on Corollary~\ref{corr:plug_in_R} is needed and sufficient for the interaction matrices, and when the one-step estimator is a reliable substitute for the MLE. All code, seeds, parameter arrays, and replicate-level records are in the reproducibility package.

\subsection{Design}
We set $d=3$ and draw the entries of $\bX$ independently from the uniform distribution on $[-1,1]$, then center its columns via $\eye_n - \one_n\one_n\transpose/n$, and orthonormalize its columns so that $\bX\transpose\bX=n\eye_d$ and $\bX\transpose\one_n = \zero_d$. Each $\bR^{(t)}$ is generated by randomly sampling an orthogonal eigenbasis uniformly from $\mathbb{O}(d)$ and eigenvalue magnitudes uniformly from $[0.6,1.2]$, with the smallest eigenvalue negative in alternate layers. We further multiply $\bR^{(t)}$ by $1/2$ for the probit, Poisson, and Gaussian families. Degree effects take the form $\alpha_i^{(t)}=c+a_i+b_t+e_{it}$, with $a_i$, $b_t$, and $e_{it}$ independently and uniformly distributed on $[-0.5,0.5]$, $[-0.25,0.25]$, and $[-0.25,0.25]$, respectively. The intercept $c$ takes values $-1$ for logistic edges ($-1.5$ in a lower-density design, referred to below as the comparatively sparse design), $-0.5$ for probit edges, and $0$ for Poisson and unit-variance Gaussian edges, giving edge densities of about 0.18 (0.09 in the lower-density design) for logistic and 0.21 for probit edges. Parameters are fixed across replicates within a configuration, so the results describe sampling variability given the parameters. Estimates are aligned to the truth by orthogonal Procrustes rotation before errors are computed. Unsuccessful fits, as defined in Section~\ref{sub:computation}, are counted and reported.

\subsection{Vertex-level inference}
Table~\ref{tab:vertex} reports the MLE with $m=8$ logistic layers. For every replicate, we compute the 95\% confidence intervals of Section~\ref{sub:plugin_inference} (a) for each of the $mn$ degree effects and the 95\% Wald ellipsoids (b) for each of the $n$ latent positions, and record the fraction that covers the truth. The table averages these fractions over replicates. The coverage for all vertices, which is computed based on the average coverages of the confidence ellipsoids over all vertices, is between $0.942$ and $0.951$. These confidence ellipsoids under-cover slightly at $n=200$ ($0.942$, Monte Carlo standard error $0.001$) and reach $0.951$ at $n=800$. The coverage of the confidence ellipsoid for a single fixed vertex is noisier because it rests on one interval per replication: with $40$ replications at $n=800$, its Monte Carlo standard error is about $0.035$. The subspace error decreases approximately at the rate $n^{-1/2}$, consistent with Theorem~\ref{thm:two_to_infinity_norm_bound_MLE}.

\begin{table}[t]
\caption{Vertex-level inference for the MLE, logistic edges, $m=8$, $d=3$. Errors are averages over replications of the Frobenius $\sin\Theta$ distance and of the root mean squared error of the degree effects ($\alpha$ RMSE). Coverage columns are averages over replications of the fraction of the $mn$ degree-effect intervals and of the $n$ latent-position regions (nominal level 0.95) that contain the truth, and coverage for the first vertex alone. }
\label{tab:vertex}
\centering\small
\begin{tabular}{rrccccccc}
\toprule
& & \multicolumn{2}{c}{Error} & \multicolumn{2}{c}{Coverage for all vertices} & \multicolumn{2}{c}{Coverage for vertex 1} & Median \\
\cmidrule(lr){3-4}\cmidrule(lr){5-6}\cmidrule(lr){7-8}
$n$ & Fits & $\sin\Theta$ & $\alpha$ RMSE & $\alpha_i^{(t)}$ CI & $\bx_i$ region & $\alpha_1^{(1)}$ CI & $\bx_1$ region & seconds \\
\midrule
200 & 200/200 & 0.148 & 0.224 & 0.947 & 0.942 & 0.96 & 0.94 & 4 \\
400 & 200/200 & 0.115 & 0.159 & 0.948 & 0.946 & 0.94 & 0.94 & 15 \\
800 & 40/40 & 0.081 & 0.112 & 0.949 & 0.951 & 1.00 & 0.88 & 55 \\
\bottomrule
\end{tabular}

\end{table}

\subsection{Interaction matrices}
\label{sub:sim_R}
For each family, we use $m=2$ and 100 replications ($40$ at $n=800$), and compute the plug-in estimated bias and covariance of Corollary~\ref{corr:plug_in_R} after alignment. We compare three 95\% intervals for each of the twelve entries of $\vech(\bR^{(1)})$ and $\vech(\bR^{(2)})$: the undebiased interval, the debiased interval, and an oracle interval that uses $\bb_n^{(t)}(\bpsi_0)$ and $\bOmega(\bpsi_0)$ at the true parameter. 
The purpose of the oracle interval is to validate the theory of the asymptotic normality in Theorem \ref{thm:asymptotic_normality_R} and is not feasible in practice.

\begin{table}[t]
\caption{Interaction-matrix inference, $m=2$, $d=3$. ``Fits'' counts successful fits. Coverage columns average the twelve entrywise 95\% intervals over successful replications; ``Joint'' is the coverage of the six-dimensional debiased Wald region for a layer. The last two columns give the median over entries of the absolute empirical bias of $\widehat\bR$, and of the residual bias after subtracting $\bb_n(\bpsi_0)/n$, divided by the empirical standard deviation.}
\label{tab:R}
\centering\small
\begin{tabular}{lrrcccccc}
\toprule
& & & \multicolumn{3}{c}{Average coverage of 95\% intervals} & Joint & \multicolumn{2}{c}{Bias / SD} \\
\cmidrule(lr){4-6}\cmidrule(lr){8-9}
Family & $n$ & Fits & Naive & Debiased & Oracle & Debiased & Raw & after $\bb_n^{(t)}(\psi_0)$ \\
\midrule
Logit & 100 & 97/100 & 0.454 & 0.976 & 0.742 & 0.985 & 2.27 & 0.62 \\
Logit & 200 & 100/100 & 0.473 & 0.964 & 0.882 & 0.975 & 2.28 & 0.19 \\
Logit & 400 & 100/100 & 0.518 & 0.942 & 0.921 & 0.945 & 1.99 & 0.12 \\
Logit & 800 & 40/40 & 0.527 & 0.954 & 0.946 & 0.975 & 1.93 & 0.09 \\
Probit & 100 & 90/100 & 0.452 & 0.974 & 0.731 & 0.994 & 2.11 & 0.58 \\
Probit & 200 & 100/100 & 0.472 & 0.961 & 0.896 & 0.960 & 2.43 & 0.30 \\
Poisson & 100 & 100/100 & 0.900 & 0.948 & 0.939 & 0.945 & 0.42 & 0.07 \\
Poisson & 200 & 100/100 & 0.893 & 0.956 & 0.953 & 0.945 & 0.59 & 0.08 \\
Poisson & 400 & 100/100 & 0.917 & 0.950 & 0.949 & 0.940 & 0.40 & 0.07 \\
Gaussian & 100 & 100/100 & 0.912 & 0.948 & 0.945 & 0.960 & 0.50 & 0.05 \\
Gaussian & 200 & 100/100 & 0.911 & 0.948 & 0.948 & 0.955 & 0.45 & 0.11 \\
\bottomrule
\end{tabular}

\end{table}

\begin{figure}[t]
\centering\includegraphics[width=\linewidth]{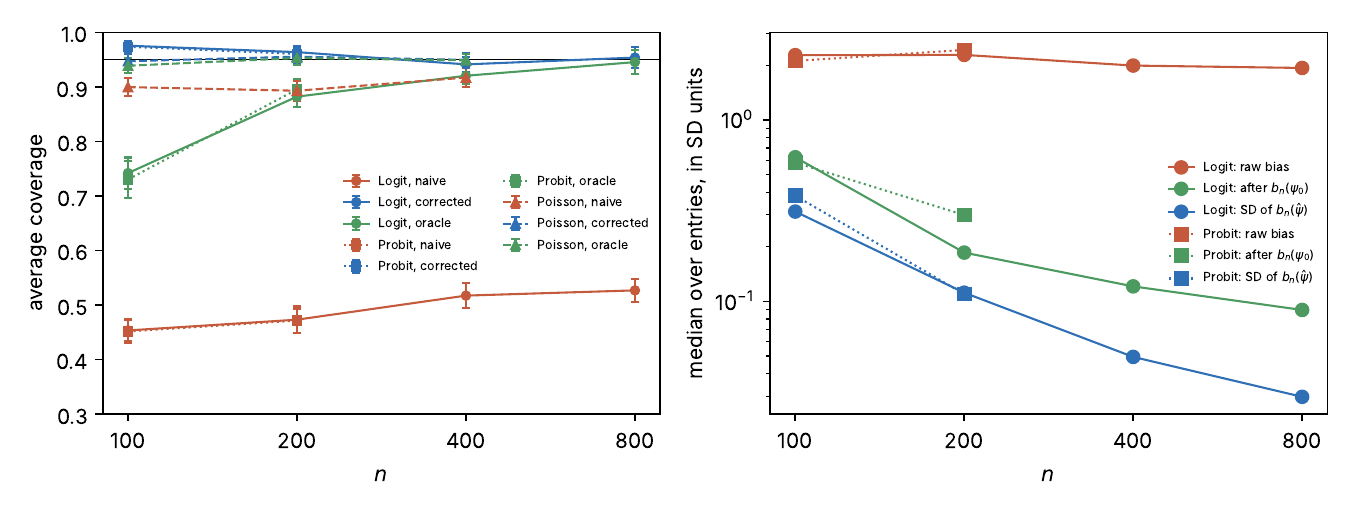}
\caption{Interaction-matrix inference across $n$ ($m=2$). Left: average coverage of undebiased (naive), debiased (corrected), and oracle 95\% intervals, with 1.96 Monte Carlo standard errors clustered by replicates. Right: median absolute empirical bias of $\widehat\bR$ (raw), median absolute residual bias after subtracting the oracle bias, and median standard deviation of the plug-in estimated bias, all in units of the empirical standard deviation of $\widehat\bR$.}
\label{fig:R_calibration}
\end{figure}

Table~\ref{tab:R} and Figure~\ref{fig:R_calibration} show three findings. First, the undebiased intervals fail for binary edges: their coverage stays near one half as $n$ grows, because the bias of $\widehat\bR^{(t)}$ for logistic edges is about 
$1.9$--$2.3$
standard deviations at every $n$, exactly as the $n^{-1}$ order of both quantities predicts. For Poisson and Gaussian edges, the bias is about half a standard deviation, and the undebiased intervals have coverage about 0.90. Second, the oracle debiasing removes the bias: the residual falls from 
$0.62$
standard deviations at $n=100$ to 
$0.12$
at $n=400$ and 
$0.09$
at $n=800$, consistent with Theorem~\ref{thm:asymptotic_normality_R}. Third, the feasible debiasing is close to nominal throughout, with coverage 
$0.942$--$0.976$
for logistic edges. At $n=100$ it over-covers while the oracle under-covers due to a finite-sample effect of plug-in estimation: the plug-in estimated bias is positively correlated with the estimation error, so subtracting it also absorbs part of the error. The standard deviation of the plug-in estimated bias relative to that of the error falls from 
$0.31$
at $n=100$ to 
$0.03$
at $n=800$, and the oracle and plug-in coverages stabilize near $95\%$. The probit family, whose bias includes the term involving $\expect(\ell''\ell')$, behaves like the logistic family, although $10$ of $100$ probit fits at $n=100$ reached a safeguard bound (operational coverage, which counts unsuccessful fits as noncoverage, 0.877; Section~\ref{sec:additional_simulations}).

\subsection{Calibration at the dimensions of the application}
\label{sub:sim_mouse}
Our growth conditions are far from satisfied with the mouse connectome dataset with $m=32$ layers of $n=332$ nodes, so we check calibration directly at these dimensions. We generated 
$60$
data sets from the Gaussian model fitted to the mouse connectomes in Section~\ref{sec:mouse}, with the fitted $(\balpha,\bR,\bX,\sigma^2)$ as truth, and repeated the complete analysis on each, including the link-transformed spectral initialization of Section~\ref{sub:computation}, the restarts, and the estimation of $\sigma^2$. Average coverage of the 192 entrywise 95\% intervals for the interaction matrices was 
$0.951$
without and 
$0.951$
with debiasing, and the six-dimensional Wald regions covered in 
$0.948$
of layers. The $95\%$ confidence interval of the hemispheric coefficient of Section~\ref{sec:mouse}, which requires the estimated subspace, has empirical coverage rate 
$0.947$
and that of the strain contrast is
$0.967$.
The $mn=10{,}624$ degree-effect intervals of Theorem~\ref{thm:asymptotic_normality_MLE} have empirical coverage rate 
$0.949$
on average. With Gaussian log counts  and $n = 332$, the bias is small, about 
$0.02$
standard errors, so the main value of the likelihood theory in Section~\ref{sec:mouse} is calibrated uncertainty for the subspace, degree, and interaction parameters jointly.
The intervals for $\rho_t$ and for the strain contrast are computed by the delta method described in Section~\ref{sub:mouse_hemi} of the main text, with variance $(\partial \rho_t/\partial\bpsi\transpose)\widetilde{\bLambda}_\sharp\bOmega(\widehat\bpsi)\widetilde{\bLambda}_\sharp(\partial \rho_t/\partial\bpsi)$ (and the corresponding gradient for the strain contrast), which uses the blocks of $\bOmega(\widehat\bpsi)$ for both the interaction and the subspace coordinates. The contribution of the subspace coordinates is not covered by Corollary~\ref{corr:plug_in_R}; Table~\ref{tab:mouse_calibration} shows that the coverage of the strain contrast is unchanged when the subspace uncertainty is ignored.

\subsection{Joint MLE and the one-step estimator}
For logistic layers, we vary $n\in\{200,400,800\}$ at $m=8$ and $m\in\{1,2,4,8,16\}$ at $n=400$, and repeat the first design with an additional setup where the intercept value results in comparatively sparse networks. Each configuration corresponds to 20 independent Monte Carlo replicates. Table~\ref{tab:accuracy_reconstruction} compares the MLE with the one-step estimate of Subsection \ref{sub:one_step_estimator}. In the dense network design case, the subspace errors of both estimates decrease with $m$ approximately at an order of $O(m^{-1/2})$, and at $n\geq400$, the two estimates are indistinguishable. At $n=200$, however, $17$ of $20$ replicates have an unsuccessful layer-wise fit even in the dense design, and the degree-effect errors of the one-step estimates are three times those of the MLE. With comparatively sparse networks, the one-step estimate inherits the instability of its single-layer fits: at $n=200$, every replicate has at least one unsuccessful layer-wise fit, and a single undamped Newton step cannot repair such a poor fit, so its degree-effect errors explode, whereas the joint MLE remains accurate. The one-step estimate is therefore useful when every layer separately supports a stable likelihood fit, but the joint MLE  is still necessary in sparse layers.

\begin{table}[t]
\caption{Accuracy of the MLE and the one-step estimate (OS), logistic edges, $d=3$, 20 replicates per row, measured in terms of the averages of the subspace estimation errors and root mean-squared errors for the degree effects ($\alpha$ RMSE). Averages include all fits. MLE failures are unsuccessful joint fits; OS failures are replicates with at least one unsuccessful layer-wise fit. The very large OS errors in the comparatively sparse scenario are undamped-update breakdowns, not outliers removed elsewhere.}
\label{tab:accuracy_reconstruction}
\centering\small
\begin{tabular}{rrlrrrrrrr}\toprule
&&&&\multicolumn{2}{c}{Subspace error}&\multicolumn{2}{c}{$\alpha$ RMSE}&\multicolumn{2}{c}{Failures}\\
\cmidrule(lr){5-6}\cmidrule(lr){7-8}\cmidrule(lr){9-10}
$n$&$m$&Density&$B$&MLE&OS&MLE&OS&MLE&OS\\\midrule
200 & 8 & Dense & 20 & 0.155 & 0.161 & 0.230 & 0.739 & 0 & 17 \\
400 & 1 & Dense & 20 & 0.389 & 0.389 & 0.199 & 0.199 & 0 & 0 \\
400 & 2 & Dense & 20 & 0.224 & 0.224 & 0.182 & 0.179 & 0 & 0 \\
400 & 4 & Dense & 20 & 0.153 & 0.153 & 0.169 & 0.168 & 0 & 0 \\
400 & 8 & Dense & 20 & 0.118 & 0.118 & 0.156 & 0.156 & 0 & 0 \\
400 & 16 & Dense & 20 & 0.079 & 0.079 & 0.156 & 0.158 & 0 & 0 \\
800 & 8 & Dense & 20 & 0.078 & 0.078 & 0.108 & 0.108 & 0 & 0 \\
200 & 8 & Sparse & 20 & 0.204 & 0.268 & 0.329 & 649.059 & 2 & 20 \\
400 & 8 & Sparse & 20 & 0.156 & 0.156 & 0.221 & 4.528 & 0 & 14 \\
800 & 8 & Sparse & 20 & 0.110 & 0.110 & 0.158 & 0.160 & 0 & 0 \\
\bottomrule\end{tabular}
\end{table}

\section{Additional simulation results}
\label{sec:additional_simulations}

\subsection{Interaction matrices}
\begin{figure}[htbp]
\centering\includegraphics[width=\linewidth]{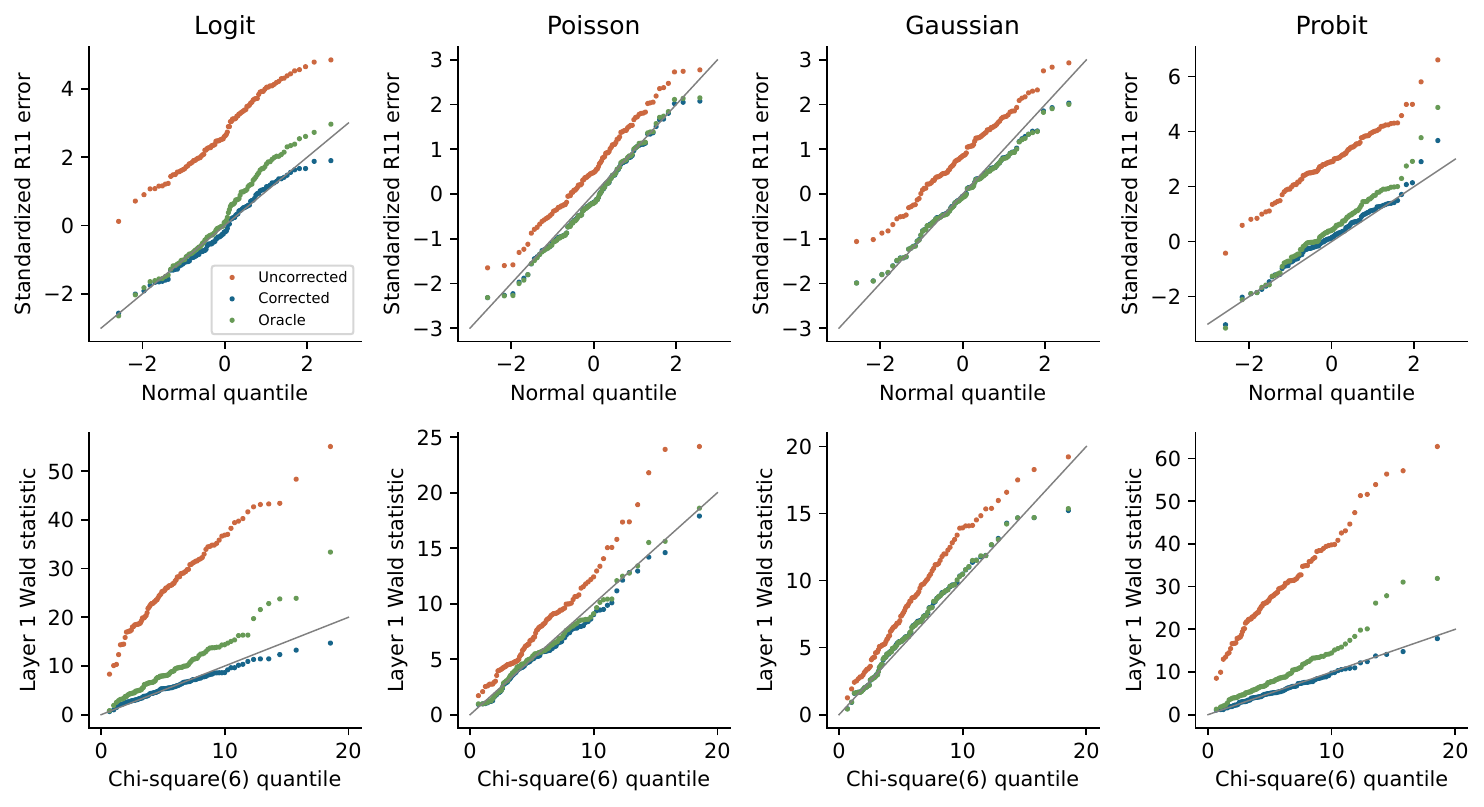}
\caption{Quantile diagnostics from 100 replications per family at $n=200$, $m=2$. Top: standardized error of $R^{(1)}_{11}$. Bottom: within-layer Wald statistics against the $\chi^2_6$ distribution. Orange: undebiased (uncorrected); blue: plug-in debiased (corrected); green: oracle.}\label{fig:R_qq}
\end{figure}

\begin{figure}[htbp]
\centering\includegraphics[width=\linewidth]{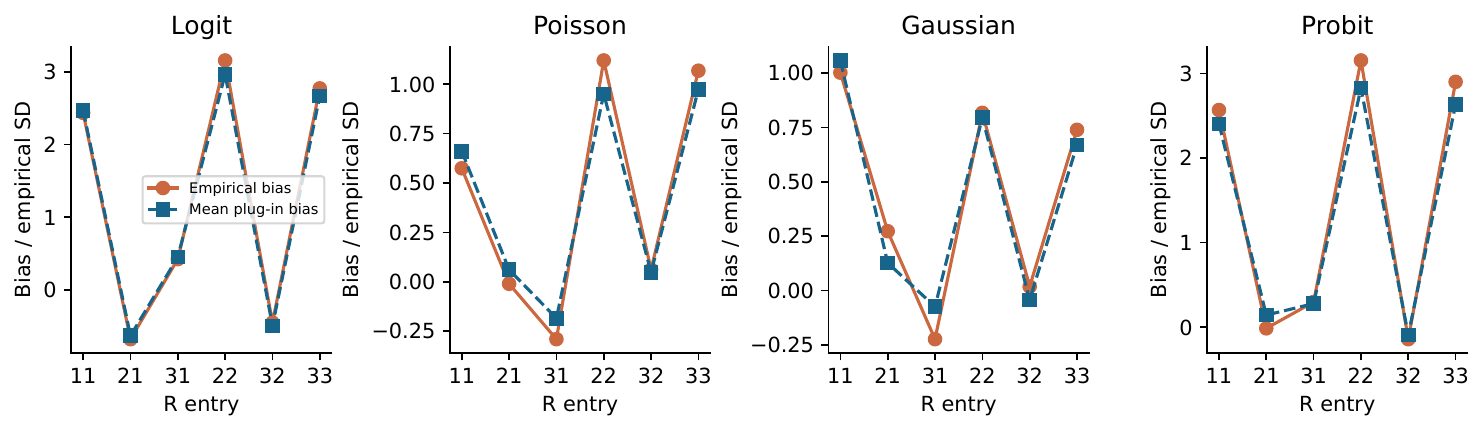}
\caption{Empirical bias and mean estimated bias, divided by the empirical standard deviation, for the six entries of $\vech(\bR^{(1)})$ at $n=200$, in column-major order $11,21,31,22,32,33$.}\label{fig:R_bias_entries}
\end{figure}

Table~\ref{tab:R_operational} counts unsuccessful fits as noncoverage. Figure~\ref{fig:R_qq} shows quantile plots of one standardized entry and of the six-dimensional Wald statistic, and Figure~\ref{fig:R_bias_entries} shows the empirical bias and mean estimated bias for every entry at $n=200$. A single-entry coverage near 0.95 from 100 replications has a Monte Carlo standard error about 0.022.

\begin{table}[htbp]
\centering\small
\caption{Empirical coverage rates of debiased confidence intervals for the twelve entries of $\vech(\bR^{(1)})$ and $\vech(\bR^{(2)})$, $m=2$, $d=3$, 100 replications. The conditional coverage rate averages over successful fits only; the operational coverage rate counts every interval from an unsuccessful fit as a miss. The Monte Carlo standard error (MCSE) of the conditional rate is clustered by replication: the twelve intervals of one replication share a data set and are dependent, so the MCSE is the standard deviation of the per-replication average coverage divided by the square root of the number of replications. The last column counts fits at which a safeguard bound of Section~\ref{sub:computation} is active.}\label{tab:R_operational}
\begin{tabular}{lccccc}\toprule
Family & $n$ & Conditional coverage rate & MCSE & Operational coverage rate & Fits at a bound\\
\midrule
Logit & 100 & 0.976 & 0.0046 & 0.947 & 3\\
Logit & 200 & 0.964 & 0.0053 & 0.964 & 0\\
Poisson & 100 & 0.948 & 0.0065 & 0.948 & 0\\
Poisson & 200 & 0.956 & 0.0065 & 0.956 & 0\\
Gaussian & 100 & 0.948 & 0.0058 & 0.948 & 0\\
Gaussian & 200 & 0.948 & 0.0062 & 0.948 & 0\\
Probit & 100 & 0.974 & 0.0052 & 0.877 & 10\\
Probit & 200 & 0.961 & 0.0056 & 0.961 & 0\\
\bottomrule
\end{tabular}
\end{table}

\subsection{Initializations and the one-step estimate}
\begin{figure}[htbp]
\centering\includegraphics[width=\linewidth]{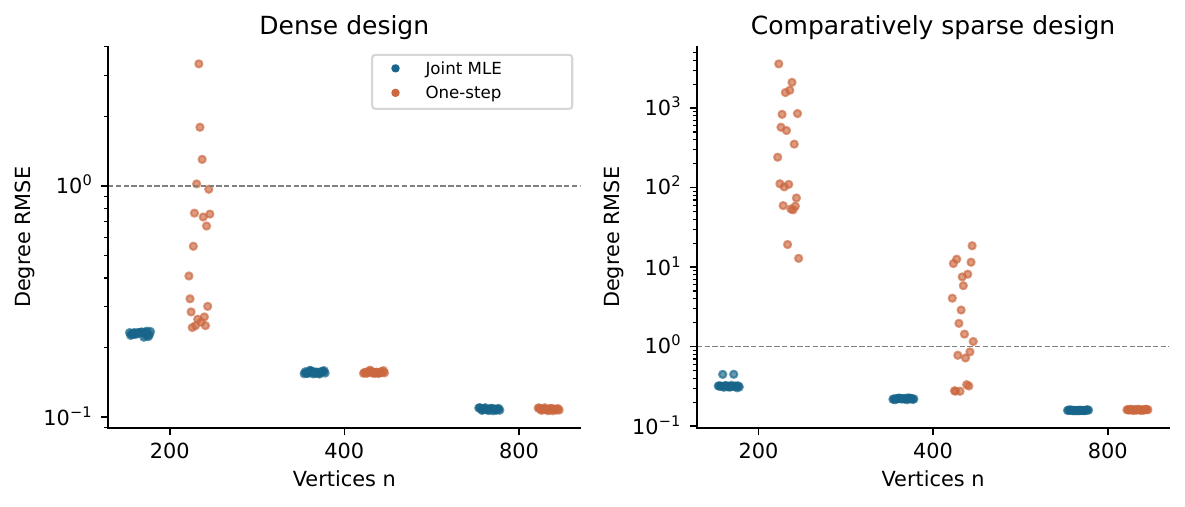}
\caption{Degree-effect RMSE of the joint MLE and the one-step estimate for all 20 replications at each $n$, $m=8$, logistic edges, in the dense design (left) and the comparatively sparse design (right). The logarithmic scale shows undamped-update breakdowns without trimming.}\label{fig:one_step_stability}
\end{figure}

The one-step estimate needs a preliminary estimate of the subspace. Table~\ref{tab:initializers} compares two candidates: the aggregate $\widetilde\bX$ of the layer-wise MLEs used in Subsection~\ref{sub:one_step_estimator} of the main text (``Layer'') and the link-transformed spectral estimate of Section~\ref{sub:computation} (``Spectral''). The table reports the average Frobenius $\sin\Theta$ distance of each to the true subspace, and the median and maximum over replications of the degree-effect RMSE of the one-step estimate. The layer-wise aggregate is the more accurate of the two in every configuration except $n=200$ in the comparatively sparse design, where every replicate has an unsuccessful layer-wise fit (Table~\ref{tab:accuracy_reconstruction}). The same unsuccessful fits produce the heavy tail of the one-step errors. Figure~\ref{fig:one_step_stability} displays every replication.

\begin{table}[htbp]
\centering\small
\caption{Preliminary subspace estimates and one-step tail errors, logistic edges, $d=3$, 20 replications per row. ``Layer'' is the aggregate of the layer-wise MLEs and ``Spectral'' the link-transformed spectral estimate (Section~\ref{sub:computation}); both columns give the average Frobenius $\sin\Theta$ distance to the true subspace. ``OS median'' and ``OS maximum'' are the median and maximum over replications of the degree-effect RMSE of the one-step estimate. All returned fits are retained.}\label{tab:initializers}
\begin{tabular}{lrrrrrr}\toprule
$n$ & $m$ & Design & Layer $\sin\Theta$ & Spectral $\sin\Theta$ & OS median & OS maximum\\
\midrule
200 & 8 & Dense & 0.190 & 0.238 & 0.478 & 3.369\\
400 & 1 & Dense & 0.389 & 0.397 & 0.201 & 0.218\\
400 & 2 & Dense & 0.235 & 0.258 & 0.178 & 0.189\\
400 & 4 & Dense & 0.167 & 0.212 & 0.167 & 0.173\\
400 & 8 & Dense & 0.132 & 0.190 & 0.156 & 0.160\\
400 & 16 & Dense & 0.090 & 0.181 & 0.158 & 0.160\\
800 & 8 & Dense & 0.084 & 0.167 & 0.108 & 0.110\\
200 & 8 & Sparse & 0.287 & 0.245 & 176.286 & 3599.444\\
400 & 8 & Sparse & 0.184 & 0.229 & 1.698 & 18.537\\
800 & 8 & Sparse & 0.126 & 0.209 & 0.160 & 0.163\\
\bottomrule
\end{tabular}
\end{table}

\subsection{Calibration at the dimensions of the mouse application}
Table~\ref{tab:mouse_calibration} gives the full results of the parametric-bootstrap study summarized in Section~\ref{sub:sim_mouse}.

\begin{table}[htbp]\centering\small
\caption{Parametric-bootstrap calibration with $n=332$, $m=32$, $d=3$, Gaussian edges, truth equal to the fit to the mouse connectomes. Coverage of nominal 95\% sets, averaged over data sets.}\label{tab:mouse_calibration}
\begin{tabular}{lcc}
\toprule
Quantity & Coverage & Monte Carlo SE \\
\midrule
Entries of $\vech(\bR^{(t)})$, undebiased (192 per data set) & 0.951 & -- \\
Entries of $\vech(\bR^{(t)})$, debiased & 0.951 & -- \\
Six-dimensional Wald regions (32 per data set) & 0.948 & -- \\
Hemispheric coefficients $\rho_t$ (32 per data set) & 0.947 & -- \\
BTBR-minus-other contrast of $\rho_t$ & 0.967 & 0.023 \\
Same, ignoring subspace uncertainty & 0.967 & 0.023 \\
Degree effects $\alpha_i^{(t)}$ (10{,}624 per data set) & 0.949 & -- \\
\midrule
Data sets & \multicolumn{2}{c}{60 (successful fits: 60)} \\
SD of standardized contrast error & \multicolumn{2}{c}{0.88} \\
\bottomrule
\end{tabular}

\end{table}

\section{Mouse connectomes: additional details}
\label{sec:mouse_details}
The data are the 32 diffusion-MRI connectomes of \citet{wang2020variability} as distributed in version 3.4.4 of \texttt{graspologic} \citep{chung2019graspy} (MIT license); file checksums are in \texttt{data/mouse/manifest.json}. Regions 1--166 are in the left hemisphere and 167--332 are their right-hemisphere counterparts. Connectivities between regions are streamline counts, and we use $\log(1 + \text{streamline counts})$ as edge weights, retaining every region and mouse. Table~\ref{tab:mouse_cv_splits} gives the held-out errors for each split. Table~\ref{tab:mouse_regions} lists the regions with the largest strain differences in degree effects.

\begin{table}[htbp]\centering\small
\caption{Held-out mean squared error for each of the three random holdouts (splits) of the region pairs (10\% of region pairs removed from all mice).}\label{tab:mouse_cv_splits}
\begin{tabular}{lccc}
\toprule
Model & Split 1 & Split 2 & Split 3 \\
\midrule
Degree only & 4.826 & 4.811 & 4.725 \\
Shared $d=2$ & 2.514 & 2.505 & 2.547 \\
Shared $d=3$ & 2.272 & 2.274 & 2.313 \\
Shared $d=4$ & 2.089 & 2.086 & 2.121 \\
Shared $d=5$ & 1.932 & 1.942 & 1.966 \\
Shared $d=6$ & 1.835 & 1.846 & 1.858 \\
Separate $d=3$ & 1.914 & 1.915 & 1.953 \\
One-step $d=3$ & 2.318 & 2.319 & 2.360 \\
\bottomrule
\end{tabular}

\end{table}

\begin{table}[htbp]\centering\small
\caption{Regions with the largest absolute values of the $t$-statistics in Welch's test of the BTBR-minus-other contrast of degree effects across mice, $d=3$. ``Contrast'' is $\bar\alpha_i^{\rm BTBR}-\bar\alpha_i^{\rm other}$; $q$ is the Benjamini--Hochberg adjusted $p$-value over 332 regions.}\label{tab:mouse_regions}
\begin{tabular}{llccc}
\toprule
Region & Atlas division & Contrast & Welch $t$-statistic & $q$ \\
\midrule
L Brachium of Superior Colliculus & white matter & 0.99 & 20.6 & 1.0e-16 \\
L Primary Visual Cortex Binocular Area & isocortex & -1.63 & -18.2 & 4.9e-14 \\
R Lateral Ventricle & white matter & -1.17 & -16.3 & 3.3e-13 \\
L Lateral Ventricle & white matter & -1.03 & -15.9 & 5.7e-09 \\
L Primary Visual Cortex Monocular Area & isocortex & -2.16 & -15.9 & 1.4e-07 \\
L Corpus Callosum & white matter & -1.12 & -14.6 & 5.4e-07 \\
L Cingulate Cortex Area 24a prime & isocortex & 1.49 & 14.5 & 2.4e-11 \\
R Corpus Callosum & white matter & -0.93 & -14.3 & 4.6e-11 \\
R Primary Visual Cortex & isocortex & -1.63 & -14.1 & 1.5e-12 \\
R Cingulate Cortex Area 29a & isocortex & 1.46 & 13.6 & 1.5e-11 \\
L Primary Somatosensory CortexForelimb Region & isocortex & -1.16 & -12.5 & 1.5e-11 \\
L Primary Somatosensory Cortex & isocortex & -1.72 & -12.2 & 1.5e-10 \\
\bottomrule
\end{tabular}

\end{table}

\section{Additional applications}
\label{sec:additional_applications}

\subsection{A neuronal connectome with sparse layers}
\label{sub:connectome}
We analyze the public three-layer \emph{Caenorhabditis elegans} connectome \citep{chen2006wiring,dedomenico2015muxviz}: 279 neurons and electrical, monadic chemical, and polyadic chemical junction layers. After removing three loop rows and merging duplicate and reverse encodings, the undirected, unweighted layers have 514, 888, and 1{,}703 edges. Ten random holdouts (splits) each reserve 20\% of neuron pairs, which are excluded from the likelihood in all three layers and used only for prediction, and we compare degree-only (model that only includes degree effects), shared (shared subspace with layer-specific degree effects and interactions, $d=1,2$), separate (all parameters are layer-specific $d=2$), one-step ($d=2$) fits, and a latent distance model in the sense of \citet{hoff2002latent}, with unrestricted node-layer degree effects. The distance model replaces the bilinear term $\bx_i\transpose\bR^{(t)}\bx_j$ of model~\eqref{eqn:model} by a term that decreases in the Euclidean distance $\|\bx_i - \bx_j\|_2$ between two-dimensional latent positions, and is fitted by maximizing the corresponding likelihood. Its degree effects are specified as in the shared $d=2$ fit, so the comparison isolates the form of the latent term.

\begin{table}[htbp]\centering\small
\caption{Observed connectome: mean validation metrics over ten dyad-blocked splits, averaging the three layers equally within each split. All candidates are numerical fits; boundary/iteration failures preclude interpreting this as a comparison of certified MLEs. For a layer with held-out pairs $\calH$ and fitted edge probabilities $\widehat p_{ij}^{(t)}$, the log loss is $-|\calH|^{-1}\sum_{(i,j)\in\calH}\{A_{ij}^{(t)}\log\widehat p_{ij}^{(t)} + (1-A_{ij}^{(t)})\log(1-\widehat p_{ij}^{(t)})\}$ and the Brier loss is the mean squared error of the probabilities, $|\calH|^{-1}\sum_{(i,j)\in\calH}(A_{ij}^{(t)}-\widehat p_{ij}^{(t)})^2$ \citep{brier1950verification}; AUC is the area under the ROC curve. Smaller log loss and Brier loss are preferable.}\label{tab:connectome}
\begin{tabular}{lrrr}\toprule
Method & AUC & Log loss & Brier loss\\\midrule
Degree only & 0.727 & 0.121 & 0.0247 \\
Shared, $d=1$ & 0.827 & 0.104 & 0.0240 \\
Shared, $d=2$ & 0.848 & 0.102 & 0.0240 \\
Layer-specific, $d=2$ & 0.815 & 0.116 & 0.0246 \\
One-step, $d=2$ & 0.676 & 4.863 & 0.2846 \\
Distance, $d=2$ & 0.860 & 0.100 & 0.0241 \\\bottomrule\end{tabular}\end{table}

\begin{figure}[htbp]\centering
\includegraphics[width=\linewidth]{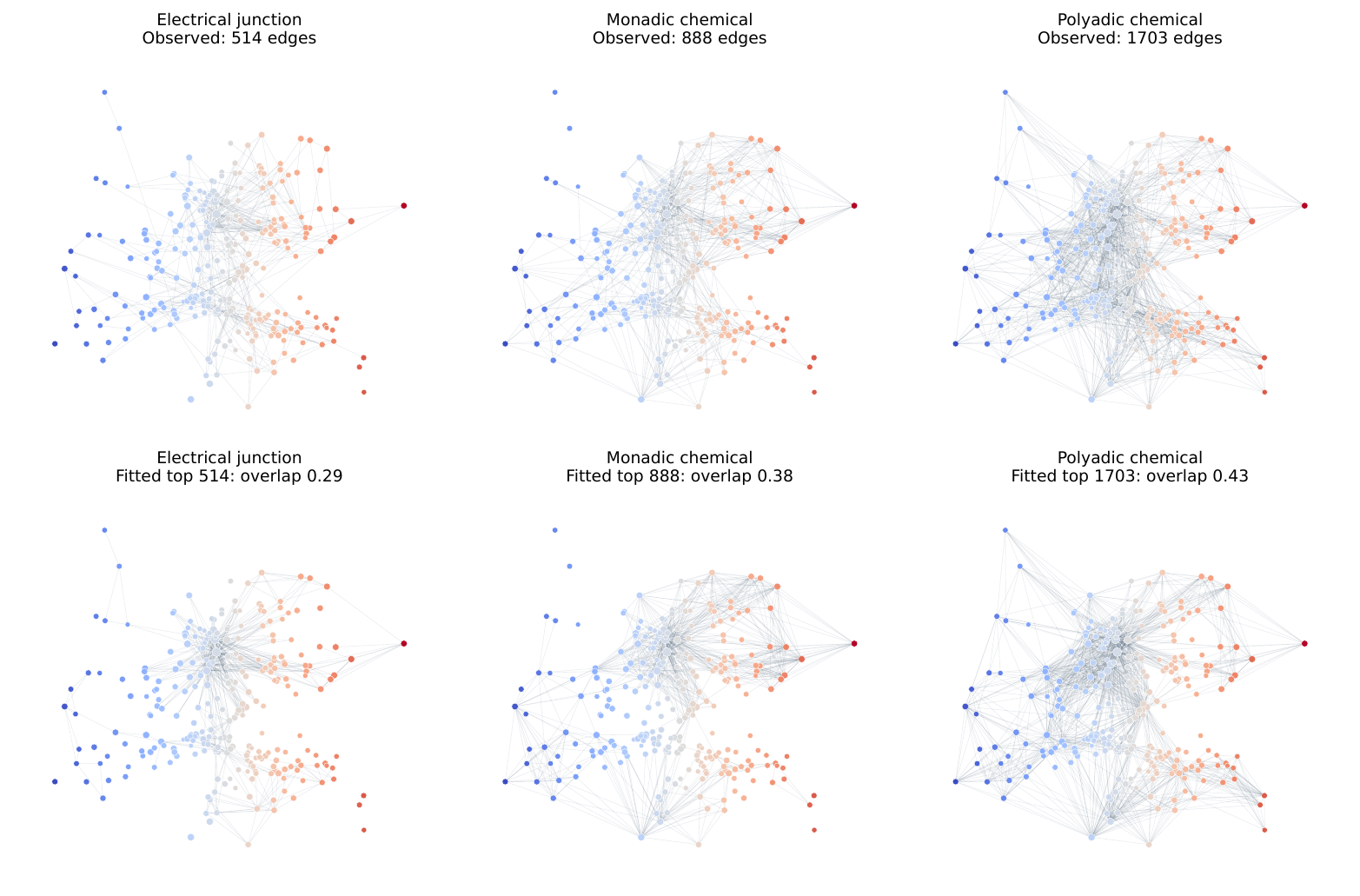}
\caption{Observed \emph{C.~elegans} connectome edges (top) and the strongest fitted probabilities (bottom), using the pooled $d=2$ layout throughout. Fitted panels show as many edges as the observed layer; the overlap is in-sample.}\label{fig:connectome_networks}
\end{figure}

The shared fits improve on degree-only prediction, and a shared subspace structure improves on layer-specific subspaces (Table~\ref{tab:connectome}). The distance model has slightly better AUC and log loss than the shared bilinear $d=2$ fit, and the one-step estimate is unreliable on these sparse layers. The layers contain 26, 19, and 1 isolated neurons. For an isolated neuron $i$ in layer $t$, $\partial L/\partial\alpha_i^{(t)}=-\sum_{j\ne i}\mathrm{logit}^{-1}(\theta_{ij}^{(t)})<0$ for every finite parameter value, so the likelihood increases without bound as $\alpha_i^{(t)}\to-\infty$ and no finite MLE exists. The bounded fits reach the safeguard. This example shows that the unrestricted-degree model needs penalization or a sparse-regime theory before it can be applied to sparse binary layers. Across the ten holdouts, the standard deviations of AUC and log loss are below 0.012 and 0.007 for every method except the one-step estimate (0.027 and 0.73). Figure~\ref{fig:connectome_networks} shows observed and fitted edges in the common layout: every panel places the neurons at their fitted positions from the pooled $d=2$ fit; the top row draws the observed edges of each layer, and the bottom row draws the same number of neuron pairs with the largest fitted probabilities.

\subsection{Surveillance profile networks}
\label{sub:cdc}
We downloaded the public mirror of CDC primary and secondary syphilis counts by jurisdiction, year, sex, and age \citep{cdcwonder2015std} (\url{https://github.com/best-dave/HEALTH_STD_DATA}, commit \texttt{39712c6d0842}). With reported cases $O_{iyk}$ and population $P_{iyk}$ in jurisdiction $i$, year $y$, and sex-by-age cell $k$ (fourteen cells), let $r_{yk}=\sum_iO_{iyk}/\sum_iP_{iyk}$, $B_{iyk}=P_{iyk}r_{yk}$, $E_{iyk}=B_{iyk}\sum_kO_{iyk}/\sum_kB_{iyk}$, and $Z_{iyk}=(O_{iyk}-E_{iyk})/(E_{iyk}+0.5)^{1/2}$. Here $r_{yk}$ is the national case rate of cell $k$ in year $y$, $B_{iyk}$ is the number of cases expected in jurisdiction $i$ at the national rates, $E_{iyk}$ rescales $B_{iyk}$ so that the expected and observed annual totals of jurisdiction $i$ agree (indirect standardization), and $Z_{iyk}$ is the resulting standardized residual, with $0.5$ added to stabilize small expected counts. The profile $(Z_{iyk})_k$ thus describes the demographic composition of the cases in jurisdiction $i$ after its overall level is removed. Within each of the periods 1996--2000, 2001--2005, 2006--2010, and 2011--2014, we concatenate the profiles over years, correlate them between the 42 retained jurisdictions, and use $\sqrt{K-3}\,\mathrm{atanh}(\rho_{ij})$ as the edge, with $K=70$ or $56$ profile entries. Edges measure similarity of demographic case profiles, not contacts between people. The resulting multilayer network has $n=42$ vertices (jurisdictions) and $m=4$ layers, one per period. Because edges sharing a jurisdiction reuse the same counts, the independent-edge theory does not apply, and the analysis is descriptive.

\begin{figure}[htbp]
\centering\includegraphics[width=\linewidth]{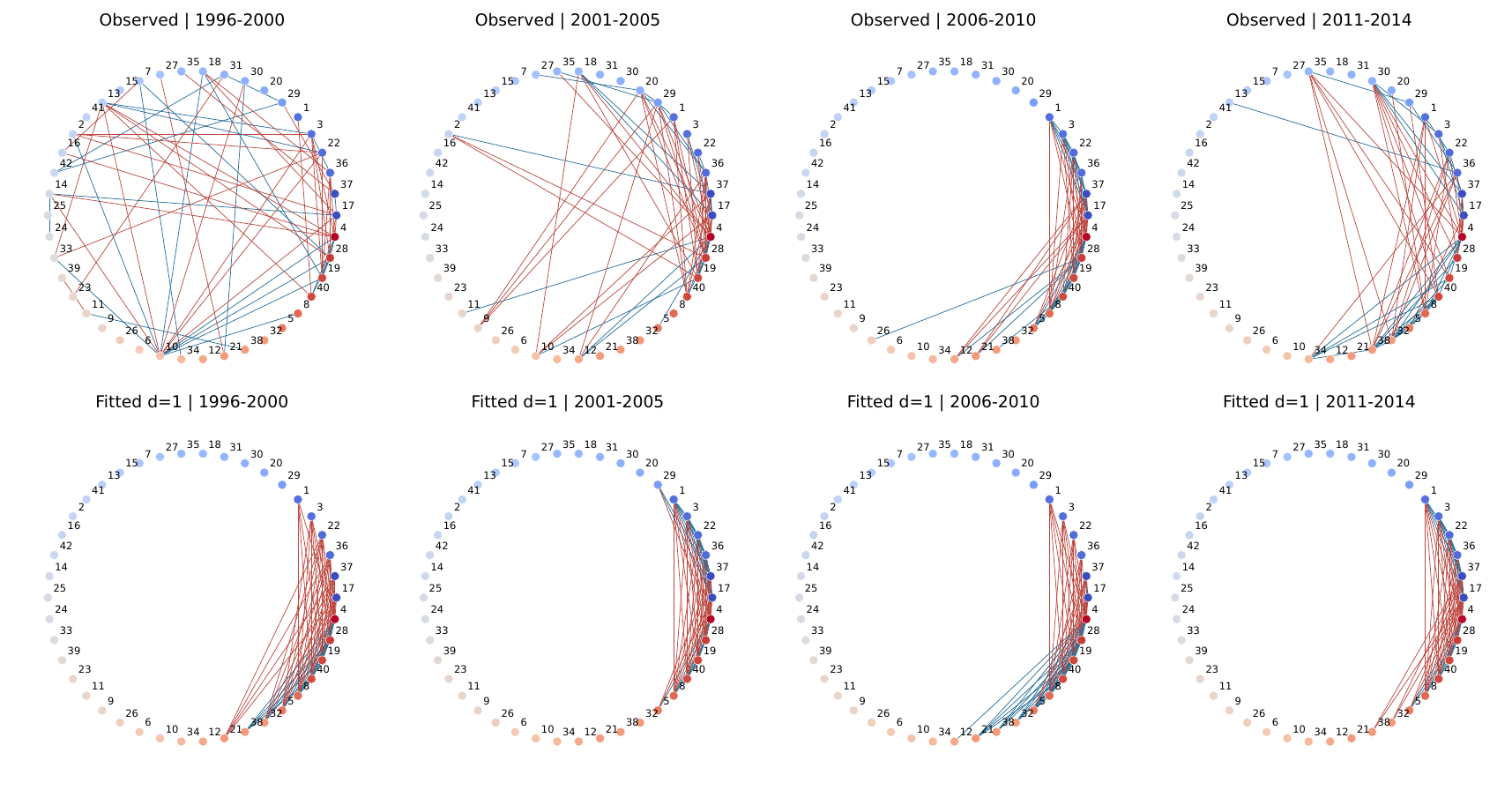}
\caption{CDC demographic-profile networks and fitted counterparts in a common layout; each panel shows the 60 largest absolute off-diagonal weights (blue positive, red negative). Numbers identify jurisdictions (Table~\ref{tab:cdc_key}); positions follow the common latent coordinate, not geography.}\label{fig:CDC_networks}
\end{figure}

\begin{figure}[htbp]
\centering\includegraphics[width=\linewidth]{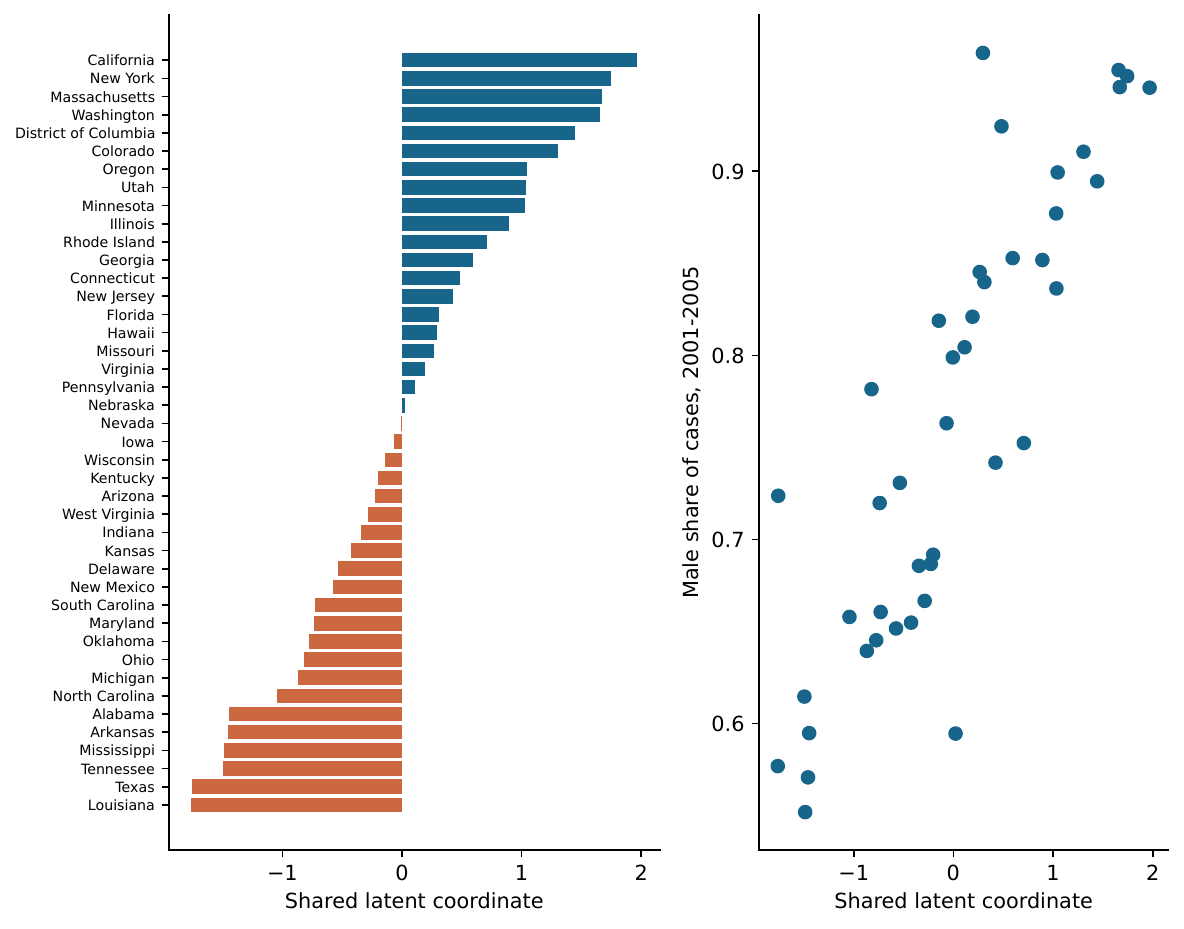}
\caption{Shared one-dimensional subspace for the 42 jurisdictions, oriented to correlate positively with the male case share in 2001--2005 (right).}\label{fig:CDC_latent}
\end{figure}

\begin{figure}[htbp]
\centering\includegraphics[width=\linewidth]{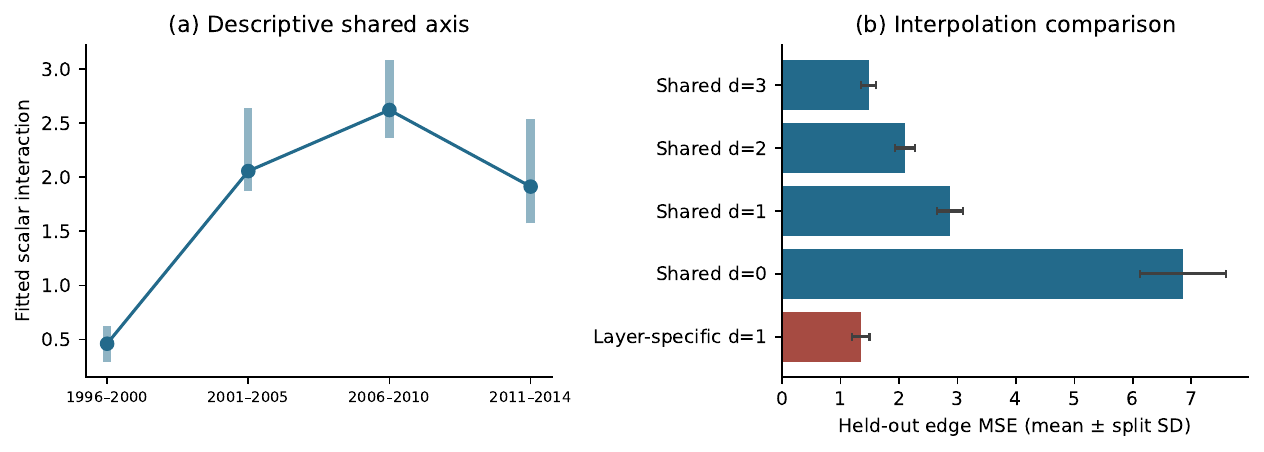}
\caption{Left: fitted scalar interactions $\bR^{(t)}$ at $d=1$ for the four periods, with 2.5--97.5\% ranges from 200 within-period year resamples. Right: mean squared prediction error on the held-out edges for each model (the panel title ``interpolation'' refers to this edge-holdout comparison).}\label{fig:CDC_descriptive}
\end{figure}

\begin{figure}[htbp]
\centering\includegraphics[width=\linewidth]{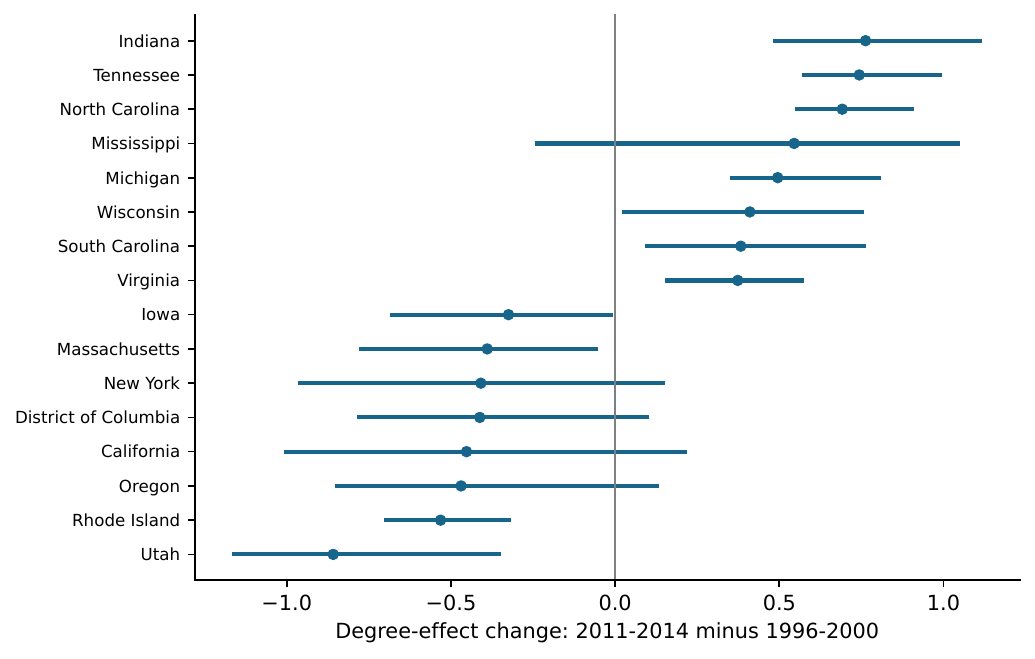}
\caption{Largest eight decreases and increases in fitted degree effects between the first and last periods, with year-resampling ranges; the selection is post hoc and descriptive.}\label{fig:CDC_contrast}
\end{figure}

At $d=1$ the fitted interaction weights are $0.461$, $2.057$, $2.622$, and $1.913$, and the shared axis correlates with the male share of cases at $0.577$, $0.862$, $0.866$, and $0.719$ (Figures~\ref{fig:CDC_networks}--\ref{fig:CDC_latent}, Table~\ref{tab:CDC_periods}). Twenty random holdouts, each removing 10\% of the edges, give mean squared errors $6.854$, $2.875$, $2.099$, and $1.484$ for shared subspace dimension ranging from zero to three, and $1.342$ for a layer-specific $d=1$ fit (where all parameters are layer-specific). Ten holdouts of two source years per period give unscaled Fisher-$z$ errors $0.0892$ and $0.0870$ for shared subspace dimension $d=1,2$ and $0.0842$ for layer-specific $d=1$ (Table~\ref{tab:CDC_prediction}). Sharing the subspace therefore gives a compact description but no predictive advantage in these data. Year-resampling ranges in Figures~\ref{fig:CDC_descriptive}--\ref{fig:CDC_contrast} are sensitivity summaries, not confidence intervals.

\begin{table}[htbp]
\centering\small
\caption{CDC descriptive summaries for the 42 included jurisdictions. Resampling SD is a within-period year sensitivity measure, not a validated standard error.}\label{tab:CDC_periods}
\begin{tabular}{lrrrr}\toprule
Period & Cases & $\widehat R$ & Axis correlation & Resampling SD\\
\midrule
1996-2000 & 39450 & 0.461 & 0.577 & 0.093\\
2001-2005 & 36639 & 2.057 & 0.862 & 0.202\\
2006-2010 & 62142 & 2.622 & 0.866 & 0.213\\
2011-2014 & 66419 & 1.913 & 0.719 & 0.235\\
\bottomrule
\end{tabular}
\end{table}

\begin{table}[htbp]
\centering\small
\caption{CDC validation summaries. Edge holdouts use scaled Fisher correlations; year holdouts use unscaled Fisher correlations. Their MSEs are not on a common scale. Split SD describes variability across overlapping splits, not independent sampling uncertainty.}\label{tab:CDC_prediction}
\begin{tabular}{llrrrrr}\toprule
Holdout & Model & $d$ & Splits & MSE & Split SD & Success\\
\midrule
Edge & Layer-specific & 1 & 20 & 1.3425 & 0.1517 & 1.00\\
Edge & Layer-specific & 2 & 20 & 26.4471 & 53.2392 & 0.80\\
Edge & Shared & 0 & 20 & 6.8540 & 0.7369 & 1.00\\
Edge & Shared & 1 & 20 & 2.8752 & 0.2261 & 1.00\\
Edge & Shared & 2 & 20 & 2.0993 & 0.1721 & 1.00\\
Edge & Shared & 3 & 20 & 1.4837 & 0.1273 & 1.00\\
Year & Layer-specific & 1 & 10 & 0.0842 & 0.0069 & 1.00\\
Year & Layer-specific & 2 & 10 & 0.0859 & 0.0059 & 1.00\\
Year & Shared & 0 & 10 & 0.1433 & 0.0063 & 1.00\\
Year & Shared & 1 & 10 & 0.0892 & 0.0070 & 1.00\\
Year & Shared & 2 & 10 & 0.0870 & 0.0065 & 1.00\\
\bottomrule
\end{tabular}
\end{table}

\begin{figure}[htbp]
\centering\includegraphics[width=\linewidth]{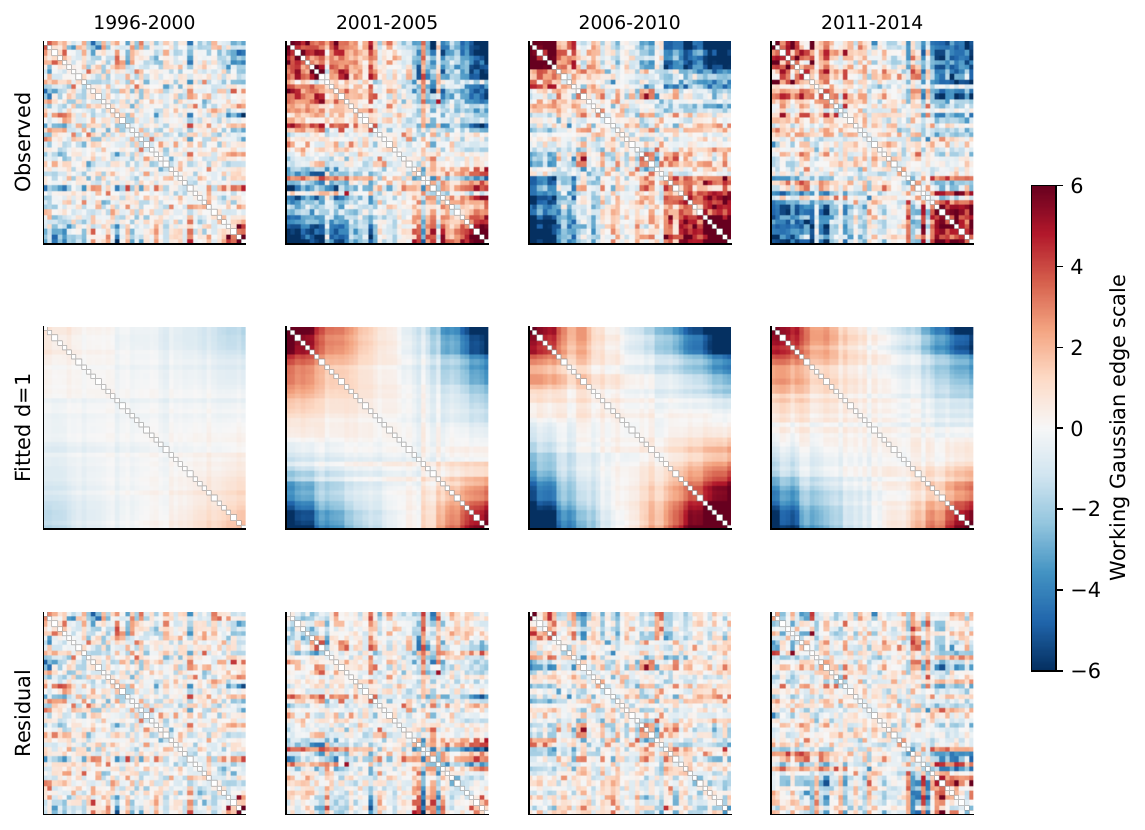}
\caption{Observed edges, fitted rank-one natural parameters, and residuals, ordered by the common coordinate, scale $[-6,6]$.}\label{fig:CDC_matrices}
\end{figure}

\begin{table}[htbp]\centering\small
\caption{Key to the jurisdiction numbers in Figure~\ref{fig:CDC_networks}.}\label{tab:cdc_key}
\begin{tabular}{rlrl}\toprule
1 & Alabama & 22 & Mississippi \\
2 & Arizona & 23 & Missouri \\
3 & Arkansas & 24 & Nebraska \\
4 & California & 25 & Nevada \\
5 & Colorado & 26 & New Jersey \\
6 & Connecticut & 27 & New Mexico \\
7 & Delaware & 28 & New York \\
8 & District of Columbia & 29 & North Carolina \\
9 & Florida & 30 & Ohio \\
10 & Georgia & 31 & Oklahoma \\
11 & Hawaii & 32 & Oregon \\
12 & Illinois & 33 & Pennsylvania \\
13 & Indiana & 34 & Rhode Island \\
14 & Iowa & 35 & South Carolina \\
15 & Kansas & 36 & Tennessee \\
16 & Kentucky & 37 & Texas \\
17 & Louisiana & 38 & Utah \\
18 & Maryland & 39 & Virginia \\
19 & Massachusetts & 40 & Washington \\
20 & Michigan & 41 & West Virginia \\
21 & Minnesota & 42 & Wisconsin \\
\bottomrule\end{tabular}
\end{table}

\subsection{A synthetic partnership network}
\label{sub:kfla}
\citet{zhao2025edge} modeled the 2019--2023 syphilis outbreak in the Kingston, Frontenac and Lennox \& Addington (KFL\&A) public health region of Ontario, Canada, by edge-based compartmental models on random graphs. Their repository releases an estimated six-month degree distribution (a truncated power law with mean 6.16) and fitted epidemic parameters, but not patient contacts. We use the released degree distribution, capped at 60, to calibrate the heterogeneity of a synthetic cohort generated from model~\eqref{eqn:model} with Bernoulli edges: $n=500$ individuals, eight half-year training layers, and a ninth half-year held out for prediction.

\emph{Degree effects.} The activity $k_i$ of individual $i$ is drawn from the released distribution and enters as $\alpha_i^{(t)} = c_t + \log k_i - \overline{\log k} + e_i^{(t)}$, with $e_i^{(t)}$ independent and uniform on $[-0.25,0.25]$. The intercepts $c_t$ are chosen so that the density of every layer, the fraction of the $n(n-1)/2$ pairs that are edges, is 0.036 on average. This cohort is much more densely connected than the KFL\&A population, so the calibration preserves the relative heterogeneity of activity, not its level.

\emph{Latent positions and interaction matrices.} The term ``three latent groups'' refers to how $\bX$ is generated; the model itself imposes no community structure. Each individual belongs to one of three groups, men who have sex with men (30\%, one fifth of whom are bisexual), heterosexual men (35\%), and women (35\%). Each $\bx_i\in\mathbb{R}^2$ is the centroid of its group, with the three centroids at the vertices of an equilateral triangle and the bisexual men placed between the first group and the women, plus independent $\mathrm{N}(\zero,0.25^2\eye_2)$ noise. The rows of $\bX$ are therefore distinct but clustered around three points. In the centroid coordinates, $\bR^{(t)}=\mathrm{diag}(r_{1t},r_{2t})$, with $r_{1t}$ decreasing from $2.0$ to $1.2$ and $r_{2t}$ from $-0.6$ to $-1.6$ over the half-years. The matrices are indefinite because heterosexual partnerships are disassortative with respect to sex.

\begin{figure}[htbp]
\centering\includegraphics[width=\linewidth]{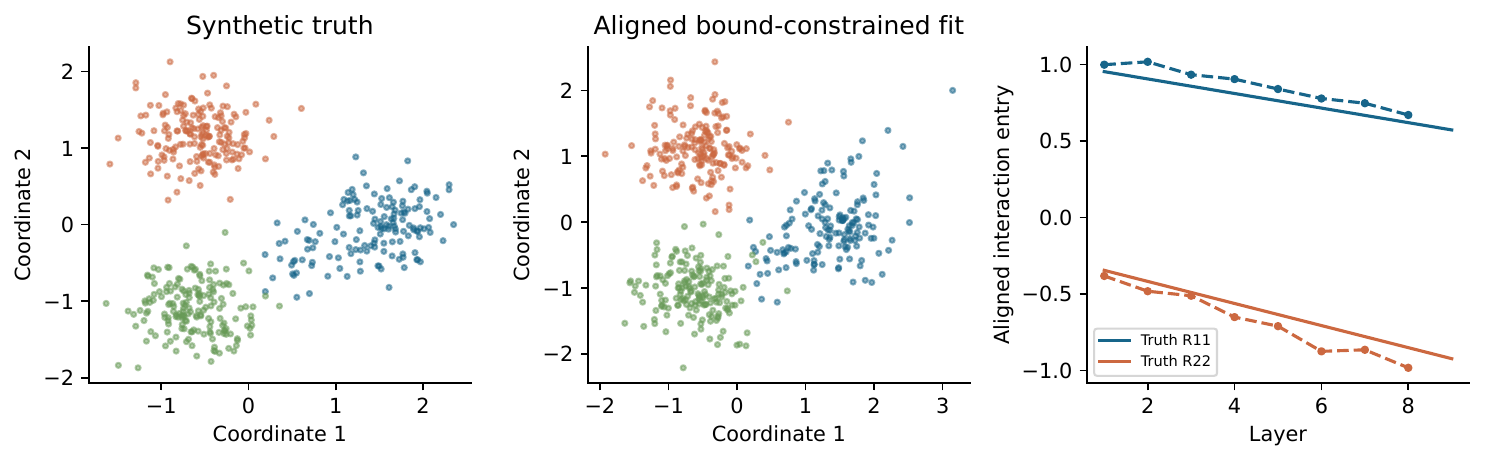}
\caption{One synthetic realization (seed 410000): true and aligned fitted positions colored by simulated group, and true and fitted interaction diagonals over the training layers.}\label{fig:KFLA_restored}
\end{figure}

Across 20 replications, the held-out AUC is 0.906 for the shared fit, 0.869 for degree effects only, and 0.915 for the true probabilities (Table~\ref{tab:KFLA_prediction}, Figure~\ref{fig:KFLA_restored}). All 20 fits reach a safeguard bound of Section~\ref{sub:computation} (here $|\alpha_i^{(t)}|\leq 8$), so this example illustrates prediction in a sparse design rather than the interior-MLE theory.

\begin{table}[htbp]
\centering\small
\caption{Synthetic partnership network calibrated to the degree distribution of the Kingston, Frontenac and Lennox \& Addington (KFL\&A) syphilis outbreak \citep{zhao2025edge} (Section~\ref{sub:kfla}). Held-out AUC over 20 replications for the shared model (``latent fit''), degree effects only, and the true edge probabilities (``oracle''), with Monte Carlo standard errors. Every latent fit reaches a safeguard bound. These results describe bounded optimization, not validated interior-MLE inference.}\label{tab:KFLA_prediction}
\begin{tabular}{lrr}\toprule
Method & AUC & MCSE\\
\midrule
Latent fit & 0.9064 & 0.0017\\
Degree only & 0.8688 & 0.0022\\
Oracle & 0.9150 & 0.0013\\
\bottomrule
\end{tabular}
\end{table}

\section{Reproducibility}
\label{sec:reproducibility}
The package contains Python code (\texttt{code/}), public data with download records and checksums (\texttt{data/}), replication-level results (\texttt{results/}), and the manuscript sources. \texttt{code/README} lists the script that produces each table, figure, and number; \texttt{code/make\_paper\_numbers.py} writes every number quoted in Section~\ref{sec:mouse} of the main text and in the numerical sections of this supplement to \texttt{manuscript/numbers.tex}. Simulations store fixed parameter arrays and seeds, so every replication can be regenerated. Unsuccessful fits are retained in all records. The model code implements the likelihoods, the block Fisher scoring algorithm, the one-step estimator, the dense reference and Schur-complement bias and covariance computations, and the vertex covariance; unit tests check derivatives, normalization invariance, vectorization conventions, rotation equivariance, and agreement of the implementations.

\bibliographystyle{plainnat}
{\sloppy\bibliography{references_theory}}

@book{higham2008functions,
  author    = {Higham, Nicholas J.},
  title     = {Functions of Matrices: Theory and Computation},
  publisher = {Society for Industrial and Applied Mathematics (SIAM)},
  address   = {Philadelphia, PA},
  year      = {2008}
}

@article{abbe2020entrywise,
  author  = {Abbe, Emmanuel and Fan, Jianqing and Wang, Kaizheng and Zhong, Yiqiao},
  title   = {Entrywise eigenvector analysis of random matrices with low expected rank},
  journal = {The Annals of Statistics},
  volume  = {48},
  number  = {3},
  pages   = {1452--1474},
  year    = {2020}
}

@article{hoff2002latent,
  author  = {Hoff, Peter D. and Raftery, Adrian E. and Handcock, Mark S.},
  title   = {Latent space approaches to social network analysis},
  journal = {Journal of the American Statistical Association},
  volume  = {97},
  number  = {460},
  pages   = {1090--1098},
  year    = {2002}
}

@article{ma2020universal,
  author  = {Ma, Zhuang and Ma, Zongming and Yuan, Hongsong},
  title   = {Universal latent space model fitting for large networks with edge covariates},
  journal = {Journal of Machine Learning Research},
  volume  = {21},
  number  = {4},
  pages   = {1--67},
  year    = {2020}
}

@article{macdonald2022latent,
  author  = {MacDonald, Peter W. and Levina, Elizaveta and Zhu, Ji},
  title   = {Latent space models for multiplex networks with shared structure},
  journal = {Biometrika},
  volume  = {109},
  number  = {3},
  pages   = {683--706},
  year    = {2022}
}

@article{arroyo2021inference,
  author  = {Arroyo, Jes{\'u}s and Athreya, Avanti and Cape, Joshua and Chen, Guodong and Priebe, Carey E. and Vogelstein, Joshua T.},
  title   = {Inference for multiple heterogeneous networks with a common invariant subspace},
  journal = {Journal of Machine Learning Research},
  volume  = {22},
  number  = {142},
  pages   = {1--49},
  year    = {2021}
}

@misc{jones2020multilayer,
  author        = {Jones, Andrew and Rubin-Delanchy, Patrick},
  title         = {The multilayer random dot product graph},
  howpublished  = {arXiv preprint arXiv:2007.10455},
  eprint        = {2007.10455},
  archivePrefix = {arXiv},
  primaryClass  = {stat.ML},
  year          = {2020}
}

@article{lei2020consistent,
  author  = {Lei, Jing and Chen, Kehui and Lynch, Brian},
  title   = {Consistent community detection in multi-layer network data},
  journal = {Biometrika},
  volume  = {107},
  number  = {1},
  pages   = {61--73},
  year    = {2020}
}

@article{lei2023bias,
  author  = {Lei, Jing and Lin, Kevin Z.},
  title   = {Bias-adjusted spectral clustering in multi-layer stochastic block models},
  journal = {Journal of the American Statistical Association},
  volume  = {118},
  number  = {544},
  pages   = {2433--2445},
  year    = {2023}
}

@article{paul2020spectral,
  author  = {Paul, Subhadeep and Chen, Yuguo},
  title   = {Spectral and matrix factorization methods for consistent community detection in multi-layer networks},
  journal = {The Annals of Statistics},
  volume  = {48},
  number  = {1},
  pages   = {230--250},
  year    = {2020}
}

@article{agterberg2025joint,
  author  = {Agterberg, Joshua and Lubberts, Zachary and Arroyo, Jes{\'u}s},
  title   = {Joint spectral clustering in multilayer degree-corrected stochastic blockmodels},
  journal = {Journal of the American Statistical Association},
  volume  = {120},
  number  = {551},
  pages   = {1607--1620},
  year    = {2025}
}

@article{xie2023efficient,
  author  = {Xie, Fangzheng and Xu, Yanxun},
  title   = {Efficient estimation for random dot product graphs via a one-step procedure},
  journal = {Journal of the American Statistical Association},
  volume  = {118},
  number  = {541},
  pages   = {651--664},
  year    = {2023}
}

@article{xie2023euclidean,
  author  = {Xie, Fangzheng},
  title   = {{E}uclidean representation of low-rank matrices and its geometric properties},
  journal = {SIAM Journal on Matrix Analysis and Applications},
  volume  = {44},
  number  = {2},
  pages   = {822--866},
  year    = {2023}
}

@article{xie2024entrywise,
  author  = {Xie, Fangzheng},
  title   = {Entrywise limit theorems for eigenvectors of signal-plus-noise matrix models with weak signals},
  journal = {Bernoulli},
  volume  = {30},
  number  = {1},
  pages   = {388--418},
  year    = {2024}
}

@article{athreya2018statistical,
  author  = {Athreya, Avanti and Fishkind, Donniell E. and Tang, Minh and Priebe, Carey E. and Park, Youngser and Vogelstein, Joshua T. and Levin, Keith and Lyzinski, Vince and Qin, Yichen and Sussman, Daniel L.},
  title   = {Statistical inference on random dot product graphs: A survey},
  journal = {Journal of Machine Learning Research},
  volume  = {18},
  number  = {226},
  pages   = {1--92},
  year    = {2018}
}

@article{cape2019two,
  author  = {Cape, Joshua and Tang, Minh and Priebe, Carey E.},
  title   = {The two-to-infinity norm and singular subspace geometry with applications to high-dimensional statistics},
  journal = {The Annals of Statistics},
  volume  = {47},
  number  = {5},
  pages   = {2405--2439},
  year    = {2019}
}

@article{rubindelanchy2022statistical,
  author  = {Rubin-Delanchy, Patrick and Cape, Joshua and Tang, Minh and Priebe, Carey E.},
  title   = {A statistical interpretation of spectral embedding: The generalised random dot product graph},
  journal = {Journal of the Royal Statistical Society: Series B (Statistical Methodology)},
  volume  = {84},
  number  = {4},
  pages   = {1446--1473},
  year    = {2022}
}

@article{chatterjee2015matrix,
  author  = {Chatterjee, Sourav},
  title   = {Matrix estimation by universal singular value thresholding},
  journal = {The Annals of Statistics},
  volume  = {43},
  number  = {1},
  pages   = {177--214},
  year    = {2015}
}

@article{bickel1975one,
  author  = {Bickel, Peter J.},
  title   = {One-step {H}uber estimates in the linear model},
  journal = {Journal of the American Statistical Association},
  volume  = {70},
  number  = {350},
  pages   = {428--434},
  year    = {1975}
}

@inproceedings{lecam1956asymptotic,
  author    = {Le Cam, Lucien},
  title     = {On the asymptotic theory of estimation and testing hypotheses},
  booktitle = {Proceedings of the Third {B}erkeley Symposium on Mathematical Statistics and Probability, Volume 1: Contributions to the Theory of Statistics},
  pages     = {129--156},
  publisher = {University of California Press},
  address   = {Berkeley, CA},
  year      = {1956}
}

@book{vandervaart1998asymptotic,
  author    = {van der Vaart, A. W.},
  title     = {Asymptotic Statistics},
  series    = {Cambridge Series in Statistical and Probabilistic Mathematics},
  volume    = {3},
  publisher = {Cambridge University Press},
  address   = {Cambridge},
  year      = {1998}
}

@article{edelman1998geometry,
  author  = {Edelman, Alan and Arias, Tom{\'a}s A. and Smith, Steven T.},
  title   = {The geometry of algorithms with orthogonality constraints},
  journal = {SIAM Journal on Matrix Analysis and Applications},
  volume  = {20},
  number  = {2},
  pages   = {303--353},
  year    = {1998}
}

@book{absil2008optimization,
  author    = {Absil, P.-A. and Mahony, R. and Sepulchre, R.},
  title     = {Optimization Algorithms on Matrix Manifolds},
  publisher = {Princeton University Press},
  address   = {Princeton, NJ},
  year      = {2008}
}

@article{wen2013feasible,
  author  = {Wen, Zaiwen and Yin, Wotao},
  title   = {A feasible method for optimization with orthogonality constraints},
  journal = {Mathematical Programming},
  volume  = {142},
  number  = {1--2},
  pages   = {397--434},
  year    = {2013}
}

@article{holland1981exponential,
  author  = {Holland, Paul W. and Leinhardt, Samuel},
  title   = {An exponential family of probability distributions for directed graphs},
  journal = {Journal of the American Statistical Association},
  volume  = {76},
  number  = {373},
  pages   = {33--50},
  year    = {1981}
}

@article{graham2017econometric,
  author  = {Graham, Bryan S.},
  title   = {An econometric model of network formation with degree heterogeneity},
  journal = {Econometrica},
  volume  = {85},
  number  = {4},
  pages   = {1033--1063},
  year    = {2017}
}

@article{yan2016asymptotics,
  author  = {Yan, Ting and Leng, Chenlei and Zhu, Ji},
  title   = {Asymptotics in directed exponential random graph models with an increasing bi-degree sequence},
  journal = {The Annals of Statistics},
  volume  = {44},
  number  = {1},
  pages   = {31--57},
  year    = {2016}
}

@article{portnoy1988asymptotic,
  author  = {Portnoy, Stephen},
  title   = {Asymptotic behavior of likelihood methods for exponential families when the number of parameters tends to infinity},
  journal = {The Annals of Statistics},
  volume  = {16},
  number  = {1},
  pages   = {356--366},
  year    = {1988}
}

@article{he2000parameters,
  author  = {He, Xuming and Shao, Qi-Man},
  title   = {On parameters of increasing dimensions},
  journal = {Journal of Multivariate Analysis},
  volume  = {73},
  number  = {1},
  pages   = {120--135},
  year    = {2000}
}

@article{xie2024biascorrected,
  author = {Xie, Fangzheng},
  title = {Bias-Corrected Joint Spectral Embedding for Multilayer Networks With Invariant Subspace},
  journal = {IEEE Transactions on Information Theory},
  year = {2024},
  volume = {70},
  number = {12},
  pages = {9036--9083}
}

@article{10.1214/11-AOP734,
  author  = {Erd{\H o}s, L{\'a}szl{\'o} and Knowles, Antti and Yau, Horng-Tzer and Yin, Jun},
  title   = {Spectral statistics of Erd{\H o}s--R{\'e}nyi graphs {I}: {L}ocal semicircle law},
  journal = {The Annals of Probability},
  volume  = {41},
  number  = {3B},
  pages   = {2279--2375},
  year    = {2013},
  doi     = {10.1214/11-AOP734}
}

@article{Bendokat_Zimmermann_Absil_2024_ACM,
  title={A {G}rassmann manifold handbook: basic geometry and computational aspects},
  author={Bendokat, Thomas and Zimmermann, Ralf and Absil, P-A},
  journal={Advances in Computational Mathematics},
  volume={50},
  number={1},
  pages={6},
  year={2024},
  publisher={Springer}
}

@article{Gotze_Sambale_Sinulis_2021_EJP_HWSubexp,
author = {Friedrich G{\"o}tze and Holger Sambale and Arthur Sinulis},
title = {Concentration inequalities for polynomials in $\alpha$-sub-exponential random variables},
volume = {26},
journal = {Electronic Journal of Probability},
number = {none},
publisher = {Institute of Mathematical Statistics and Bernoulli Society},
pages = {1--22},
year = {2021},
doi = {10.1214/21-EJP606},
URL = {https://doi.org/10.1214/21-EJP606}
}

@article{delaPena_MontgomerySmith_1995_AoP_DecouplingInequalities,
author = {Victor H. {de la Pena} and S. J. Montgomery-Smith},
title = {Decoupling Inequalities for the Tail Probabilities of Multivariate $U$-Statistics},
volume = {23},
journal = {The Annals of Probability},
number = {2},
publisher = {Institute of Mathematical Statistics},
pages = {806--816},
year = {1995},
doi = {10.1214/aop/1176988291},
URL = {https://doi.org/10.1214/aop/1176988291}
}

@article{Zheng_Tang_2026_COSIELimit,
  title={Limit results for distributed estimation of invariant subspaces in multiple networks inference and {PCA}},
  author={Zheng, Runbing and Tang, Minh},
  journal={arXiv preprint arXiv:2206.04306},
  year={2026}
}

@book{Chung_2001_ProbabilityTheory,
  author    = {Chung, Kai Lai},
  title     = {A Course in Probability Theory},
  edition   = {Third},
  publisher = {Academic Press},
  address   = {San Diego, CA},
  year      = {2001},
  pagetotal = {xvi+419},
  isbn      = {978-0-12-174151-8}
}

@article{Abbe_Fan_Wang_Zhong_2020_AoS_Entrywise,
author = {Emmanuel Abbe and Jianqing Fan and Kaizheng Wang and Yiqiao Zhong},
title = {Entrywise eigenvector analysis of random matrices with low expected rank},
volume = {48},
journal = {The Annals of Statistics},
number = {3},
publisher = {Institute of Mathematical Statistics},
pages = {1452--1474},
year = {2020},
doi = {10.1214/19-AOS1854},
URL = {https://doi.org/10.1214/19-AOS1854}
}

@article{lei_zhang_zhu_2024_AoS_computational,
  title={Computational and statistical thresholds in multi-layer stochastic block models},
  author={Lei, Jing and Zhang, Anru R and Zhu, Zihan},
  journal={The Annals of Statistics},
  volume={52},
  number={5},
  pages={2431--2455},
  year={2024},
  publisher={Institute of Mathematical Statistics}
}

@article{Tian_Sun_He_2026_Bridging,
  title={Bridging Theory and Practice: Statistical Inference for Latent Space Models of Networks},
  author={Tian, Yuang and Sun, Jiajin and He, Yinqiu},
  journal={arXiv preprint arXiv:2605.08677},
  year={2026}
}

@article{Rosenblum1956,
  author  = {Rosenblum, Marvin},
  title   = {On the operator equation {$BX-XA=Q$}},
  journal = {Duke Mathematical Journal},
  volume  = {23},
  number  = {2},
  pages   = {263--269},
  year    = {1956},
  doi     = {10.1215/S0012-7094-56-02324-9},
}

@book{Bhatia2007,
  author    = {Bhatia, Rajendra},
  title     = {Positive Definite Matrices},
  series    = {Princeton Series in Applied Mathematics},
  publisher = {Princeton University Press},
  address   = {Princeton, NJ},
  year      = {2007},
  isbn      = {978-0-691-12918-1},
  note      = {Positive linear maps $X \mapsto AX+XA$ and their integral representations},
}

@article{tropp2012user,
  title={User-friendly tail bounds for sums of random matrices},
  author={Tropp, Joel A},
  journal={Foundations of computational mathematics},
  volume={12},
  number={4},
  pages={389--434},
  year={2012},
  publisher={Springer}
}

@article{chen_chi_fan_ma_2021spectral,
  title={Spectral methods for data science: A statistical perspective},
  author={Chen, Yuxin and Chi, Yuejie and Fan, Jianqing and Ma, Cong},
  journal={Foundations and Trends in Machine Learning},
  volume={14},
  number={5},
  pages={566--806},
  year={2021},
  publisher={Emerald Publishing Limited}
}

@book{bhatia2013matrix,
  title={Matrix analysis},
  author={Bhatia, Rajendra},
  volume={169},
  year={2013},
  publisher={Springer Science \& Business Media}
}

@article{Jauch_Hoff_Dunson_2020_Bernoulli,
author = {Michael Jauch and Peter D. Hoff and David B. Dunson},
title = {Random orthogonal matrices and the {C}ayley transform},
volume = {26},
journal = {Bernoulli},
number = {2},
publisher = {Bernoulli Society for Mathematical Statistics and Probability},
pages = {1560--1586},
year = {2020},
doi = {10.3150/19-BEJ1176}
}

@article{Shepard_Brozell_Gidofalvi_2015_JPCA,
  title={The representation and parametrization of orthogonal matrices},
  author={Shepard, Ron and Brozell, Scott R and Gidofalvi, Gergely},
  journal={The Journal of Physical Chemistry A},
  volume={119},
  number={28},
  pages={7924--7939},
  year={2015},
  publisher={ACS Publications}
}

@book{Jeffreys_1961_Theory,
  author    = {Jeffreys, Harold},
  title     = {Theory of Probability},
  edition   = {Third},
  series    = {International Series of Monographs on Physics},
  publisher = {Oxford University Press},
  address   = {Oxford},
  year      = {1961},
  pagetotal = {viii+447}
}

@article{Chernozhukov_Chetverikov_Demirer_Duflo_Hansen_Newey_Robins_2018_DML_EconometricsJ,
    author = {Chernozhukov, Victor and Chetverikov, Denis and Demirer, Mert and Duflo, Esther and Hansen, Christian and Newey, Whitney and Robins, James},
    title = {Double/debiased machine learning for treatment and structural parameters},
    journal = {The Econometrics Journal},
    volume = {21},
    number = {1},
    pages = {C1--C68},
    year = {2018},
    month = {02},
    issn = {1368-4221},
    doi = {10.1111/ectj.12097},
    url = {https://doi.org/10.1111/ectj.12097},
    eprint = {https://academic.oup.com/ectj/article-pdf/21/1/C1/27684918/ectj00c1.pdf},
}

@incollection{Neyman_1959_Optimal,
  author    = {Neyman, Jerzy},
  title     = {Optimal Asymptotic Tests of Composite Statistical Hypotheses},
  booktitle = {Probability and Statistics: The {Harald Cram{\'e}r} Volume},
  editor    = {Grenander, Ulf},
  publisher = {Almqvist \& Wiksell},
  address   = {Uppsala},
  year      = {1959},
  pages     = {213--234}
}

@article{Cox_Reid_1987_Orthogonality_JRSSB,
    author = {Cox, D. R. and Reid, N.},
    title = {Parameter Orthogonality and Approximate Conditional Inference},
    journal = {Journal of the Royal Statistical Society: Series B (Methodological)},
    volume = {49},
    number = {1},
    pages = {1--18},
    year = {1987},
    month = {09},
    issn = {0035-9246},
    doi = {10.1111/j.2517-6161.1987.tb01422.x},
    url = {https://doi.org/10.1111/j.2517-6161.1987.tb01422.x},
    eprint = {https://academic.oup.com/jrsssb/article-pdf/49/1/1/49098062/jrsssb_49_1_1.pdf},
}

@article{He_Sun_Tian_Ying_Feng_2025_AoS,
author = {Yinqiu He and Jiajin Sun and Yuang Tian and Zhiliang Ying and Yang Feng},
title = {Semiparametric modeling and analysis for longitudinal network data},
volume = {53},
journal = {The Annals of Statistics},
number = {4},
publisher = {Institute of Mathematical Statistics},
pages = {1406--1430},
year = {2025},
doi = {10.1214/25-AOS2506},
URL = {https://doi.org/10.1214/25-AOS2506}
}

@article{tian2026efficient,
  title={Efficient analysis of latent spaces in heterogeneous networks},
  author={Tian, Yuang and Sun, Jiajin and He, Yinqiu},
  journal={Journal of the American Statistical Association},
  volume={\textmd{(in press)}},
  year={2026},
  publisher={Taylor \& Francis}
}

@article{zhao2025edge,
  author  = {Zhao, Sicheng and Saeed, Sahar and Carter, Megan and Stoner, Bradley and Hoover, Maggie and Guan, Hugh and Magpantay, Felicia M. G.},
  title   = {Edge-based modelling for disease transmission on random graphs: an application to mitigate a syphilis outbreak},
  journal = {Royal Society Open Science},
  year    = {2025},
  doi     = {10.1098/rsos.250032},
  note    = {Data repository: \url{https://github.com/RichardSichengZhao/Data-Repository-for-EBMforDisTransOnRandGraphs-An-Application-to-Mitigate-a-Syphilis-Outbreak}}
}

@misc{cdcwonder2015std,
  author       = {{Centers for Disease Control and Prevention}},
  title        = {Sexually Transmitted Disease Morbidity 1996--2014, by sex, age group and race/ethnicity},
  howpublished = {CDC WONDER Online Database, National Center for HIV/AIDS, Viral Hepatitis, STD, and TB Prevention, Division of STD Prevention. Reports and corrections received through June 10, 2015. \url{https://wonder.cdc.gov/std.html}},
  year         = {2015}
}

@misc{liu2026unfolding,
 author={Liu, Zhaozhe and Xu, Gongjun and Zhang, Haoran},
 title={Unfolding and Fusion: Debiased Inference for Generalized Multilayer Latent Space Models},
 year={2026}, note={arXiv:2602.19129, version 3, August 21, 2026},
 url={https://arxiv.org/abs/2602.19129v3}
}

@inproceedings{zhang2020flexible,
 author={Zhang, Xuefei and Xue, Songkai and Zhu, Ji},
 title={A Flexible Latent Space Model for Multilayer Networks},
 booktitle={Proceedings of the 37th International Conference on Machine Learning},
 volume={119}, pages={11288--11297}, year={2020}, publisher={PMLR},
 url={https://proceedings.mlr.press/v119/zhang20aa.html}
}

@article{chen2006wiring,
 author={Chen, Beth L. and Hall, David H. and Chklovskii, Dmitri B.}, title={Wiring optimization can relate neuronal structure and function}, journal={Proceedings of the National Academy of Sciences}, year={2006}, volume={103}, number={12}, pages={4723--4728}}

@article{dedomenico2015muxviz,
 author={De Domenico, Manlio and Porter, Mason A. and Arenas, Alex}, title={MuxViz: A Tool for Multilayer Analysis and Visualization of Networks}, journal={Journal of Complex Networks}, year={2015}, volume={3}, number={2}, pages={159--176}}

@article{wang2020variability,
  author  = {Wang, Nian and Anderson, Robert J. and Ashbrook, David G. and Gopalakrishnan, Vivek and Park, Youngser and Priebe, Carey E. and Qi, Yi and Laoprasert, Rick and Vogelstein, Joshua T. and Williams, Robert W. and Johnson, G. Allan},
  title   = {Variability and heritability of mouse brain structure: {M}icroscopic {MRI} atlases and connectomes for diverse strains},
  journal = {NeuroImage},
  volume  = {222},
  pages   = {117274},
  year    = {2020},
  doi     = {10.1016/j.neuroimage.2020.117274}
}

@article{wahlsten2003survey,
  author  = {Wahlsten, Douglas and Metten, Pamela and Crabbe, John C.},
  title   = {Survey of 21 inbred mouse strains in two laboratories reveals that {BTBR} {T/+} tf/tf has severely reduced hippocampal commissure and absent corpus callosum},
  journal = {Brain Research},
  volume  = {971},
  number  = {1},
  pages   = {47--54},
  year    = {2003},
  doi     = {10.1016/S0006-8993(03)02354-0}
}

@article{chung2019graspy,
  author  = {Chung, Jaewon and Pedigo, Benjamin D. and Bridgeford, Eric W. and Varjavand, Bijan K. and Helm, Hayden S. and Vogelstein, Joshua T.},
  title   = {{GraSPy}: Graph Statistics in {P}ython},
  journal = {Journal of Machine Learning Research},
  volume  = {20},
  number  = {158},
  pages   = {1--7},
  year    = {2019}
}

@inproceedings{levin2017central,
  author    = {Levin, Keith and Athreya, Avanti and Tang, Minh and Lyzinski, Vince and Priebe, Carey E.},
  title     = {A central limit theorem for an omnibus embedding of multiple random dot product graphs},
  booktitle = {2017 IEEE International Conference on Data Mining Workshops (ICDMW)},
  pages     = {964--967},
  year      = {2017}
}

@article{neyman1948consistent,
  author  = {Neyman, Jerzy and Scott, Elizabeth L.},
  title   = {Consistent estimates based on partially consistent observations},
  journal = {Econometrica},
  volume  = {16},
  number  = {1},
  pages   = {1--32},
  year    = {1948}
}

@article{hahn2004jackknife,
  author  = {Hahn, Jinyong and Newey, Whitney},
  title   = {Jackknife and analytical bias reduction for nonlinear panel models},
  journal = {Econometrica},
  volume  = {72},
  number  = {4},
  pages   = {1295--1319},
  year    = {2004}
}

@article{fernandezval2016individual,
  author  = {Fern{\'a}ndez-Val, Iv{\'a}n and Weidner, Martin},
  title   = {Individual and time effects in nonlinear panel models with large {N}, {T}},
  journal = {Journal of Econometrics},
  volume  = {192},
  number  = {1},
  pages   = {291--312},
  year    = {2016}
}

@article{brier1950verification,
  author  = {Brier, Glenn W.},
  title   = {Verification of forecasts expressed in terms of probability},
  journal = {Monthly Weather Review},
  volume  = {78},
  number  = {1},
  pages   = {1--3},
  year    = {1950}
}
\end{document}